# A wearable electrical hemodynamic imaging ring

Gia-Bao Ha[1], Lucas Takanori Sanchez Shiromizu[2], Jaehyeon Song[3], Zhuyun Xie[2], Henry Crandall[1], Dinali Assylbek[1], Alexandra Boyadzhiev[4], Huanan Zhang[4], Fernando Guevara Vasquez[5], Ramakrishna Mukkamala[6,7], Michael Widlansky[8], Shamim Nemati[9], Jesse Capecelatro[3,10], C. Alberto Figueroa[3,11], Benjamin Sanchez[2,12]

[1]Department of Electrical and Computer Engineering, University of Utah, Salt Lake City, UT, USA
[2]Department of Electrical and Computer Engineering, University of Illinois Chicago, Chicago, IL, USA
[3]Department of Mechanical Engineering, University of Michigan, Ann Arbor, MI, USA
[4]Department of Chemical Engineering, University of Utah, Salt Lake City, UT, USA
[5]Department of Mathematics, University of Utah, Salt Lake City, UT, USA
[6]Department of Bioengineering, University of Pittsburgh, PA, USA
[7]Department of Anesthesiology & Perioperative Medicine, University of Pittsburgh, PA, USA
[8]Department of Medicine, Division of Cardiovascular Medicine, Milwaukee, WI, USA
[9]Division of Biomedical Informatics, UC San Diego, San Diego, CA, USA
[10]Department of Aerospace Engineering, University of Michigan, Ann Arbor, MI, USA
[11]Department of Surgery, University of Michigan, Ann Arbor, MI, USA
[12]Richard and Loan Hill Department of Biomedical Engineering, University of Illinois Chicago, Chicago, IL, USA

Corresponding author: Benjamin Sanchez, 851 S. Morgan St., Office 1104 SEO, Chicago, IL 60607.
Email: bst@uic.edu. Phone: 312-996-5847.

Main manuscript includes:
- Number of Figures: 4
- Number of Extended Data Tables: 3
- Number of Extended Data Figures: 1
- Number of References: 38

Supplementary Information includes:
- Number of Supplementary Pages: 248
- Number of Supplementary Figures: 119
- Number of Supplementary Tables: 15
- Number of Supplementary Videos: 7
- Number of Supplementary References: 273

# Data Availability

Restrictions apply to the availability of the data that support the findings of this study, which contained information that could compromise research participant privacy and consent, and are not publicly available.

# Code Availability

Access to parts of the code may be granted upon reasonable request and subject to relevant licensing agreements.

# Acknowledgments

Benjamin Sanchez acknowledges the direct financial support for the research reported in this publication provided by NSF under Award number 2529648 and 2534572; the National Cancer Institute of the National Institutes of Health (NIH) under Award Number 1R21CA273984-01A1, 1P01CA285249-01A1, and 1R21CA289101-01A1. The content is solely the responsibility of the authors and does not necessarily represent the official views of the NIH or NSF agencies.

# Author Contributions

Conceptualization - BS Data curation - GBH, LTSS, JS, ZX, HC, DA, AB Formal Analysis - HZ, FGV, RM, SN, JC, AF, BS Funding acquisition - BS Investigation - all Methodology - HZ, FGV, JC, AF, BS Project administration - BS Resources - JC, AF, BS Software - GBH, LTSS, JS, ZX, HC, DA Supervision - JC, AF, BS Validation - all Visualization - GBH, LTSS, JS Writing – original draft - all Writing – review & editing – all

# Competing Interests

Dr. Sanchez is co-founder and holds equity in Haystack Diagnostics, Inc. He holds equity and serves as Scientific Advisor to B-Secur, Ltd, and Sobr Safe, Inc. He holds equity and serves as a Chief Scientific Officer of Hemodynamiq, Inc. He serves as Chief Scientific Officer to First Capital Ventures, LLC, and Promptus, Inc. He has equity and serves as Head of Biosensing and Product Development at NeuralPoint AI, Inc. The other authors have no conflicts of interest to declare.

# Abstract

Continuous ambulatory monitoring of peripheral vascular perfusion could enable earlier detection of vascular dysfunction in individuals with diabetes mellitus and more timely management of cardiovascular disease. Clinical imaging modalities provide high-fidelity vascular information but are impractical for ambulatory use, whereas most wearable devices are limited to single-modality sensing and do not provide imaging. Electrical bioimpedance has the potential to bridge this gap by enabling rapid spatial and temporal imaging while remaining sensitive to hemodynamic changes. Here, we introduce a wearable ring with 8 electrodes and 32-channel bioimpedance sensing for finger blood flow imaging. In 96 healthy participants measured at rest and during autonomic maneuvers, we resolve conductivity images in the digital arteries associated with pulsatile blood flow and train neural network models for continuous cuffless blood pressure waveform estimation. We demonstrate the feasibility of bioimpedance imaging in a ring form factor, supporting its potential for ambulatory cuffless hemodynamic monitoring.

# Introduction

Diabetes mellitus (DM) is recognized as one of the most powerful risk factors for coronary artery disease and other forms of cardiovascular disease in the United States,[1] affecting 34.2 million individuals with nearly 1.5 million new cases diagnosed annually.[2,3] DM morbidity is largely driven by its profound effects on the vascular system (Supplementary Discussion 1). Microvascular dysfunction in patients with DM contributes to the development of retinopathy, nephropathy, and neuropathy, while macrovascular disease increases the risk of cardiovascular morbidity and mortality.[4] Accumulating evidence indicates that vascular dysfunction in DM exhibits distinct structural and functional phenotypes affecting both large arteries and microcirculation.[5–7] Early and continuous monitoring of vascular function is therefore imperative in tracking disease progression and enabling timely therapeutic response. Existing non-invasive methods for monitoring vascular health rely on conventional imaging modalities such as ultrasound and magnetic resonance angiography, which provide detailed hemodynamic and anatomical information (Supplementary Discussion 2),[8,9] but the use of these approaches is restricted to clinical settings due to their cost and highly-specialized operator training. Wearable devices offer a pathway towards continuous and ambulatory monitoring of cuffless hemodynamic vital signs such as blood pressure (BP) and blood velocity.[10] Here, we present a wearable ring imager with 8 electrodes capable of sensing 32 independent channels of bioimpedance (BioZ) data and test the technology for monitoring cuffless BP on a healthy cohort performing dynamic BP challenges (Fig. 1a and Supplementary Discussion 3).

# Wearable ring device

We first built a wearable ring device that integrates multi-channel BioZ sensing via an 8-electrode array distributed around the ring circumference, providing 32 independent channels of data (Fig. 1b i,ii, Supplementary Discussion 4 and Supplementary Fig. 1). To ensure compatibility with established scalable manufacturing processes, we fabricated the electrodes with electroless nickel immersion gold (ENIG) finish. Because BioZ performance is sensitive to electrode–skin contact, we benchmarked ENIG electrodes against an otherwise identical ENIG electrode coated with poly(3,4-ethylenedioxythiophene) (PEDOT). We inspected the surface morphology of both ENIG and PEDOT-coated electrodes through scanning electron microscopy and found that ENIG surface had scratches due to manufacturing imperfection, while PEDOT layer contained porous clusters that increase the effective electrode surface area (Fig. 1c i,ii). We then conducted electrochemical impedance spectroscopy experiments from 10 Hz to 100 kHz and found that, at the imaging frequency of 50 kHz, the impedance magnitude and phase (mean$\pm$s.d.) for untreated ENIG electrodes were 13.33$\pm$4.31 k$\Omega$ and -13.58$\pm$9.53 degrees, respectively, an acceptable impedance range that is expected to result in negligible loading of the measurement circuitry. These results were similar to low and stable contact impedance obtained with PEDOT-coated electrodes that exhibited 8.75$\pm$1.33 k$\Omega$ and -11.57$\pm$22.17 degrees, respectively (Fig. 1d i). We further performed cyclic voltammetry (CV) and found that PEDOT-coated electrodes exhibited strong capacitive properties with capacitance 331.69$\pm$576.28 µF compared to ENIG capacitance of 5.93$\pm$3.61 µF (Fig. 1d ii). Finally, to evaluate temporal electrode durability, we performed accelerated CV for 100 cycles to simulate electrode degradation and observed that both electrodes exhibit minimal current-voltage distortion across cycles, indicating robust electrode integrity under repeated electrochemical stress and suggesting robust stability required for long term cuffless BP monitoring (Fig. 1d iii).

Fig. 1 **Wearable ring for finger vascular impedance imaging**. **a**, Underlying principles of the technology for imaging blood flow in the digital arteries with a ring; **b**, Computer-aided design and manufacturing. **i**. The ring uses eight individual printed circuit boards (PCB) as sensors, connected to current multiplexer through coaxial cables; **ii**. A prototype of the ring (size 10 North America), printed with resin stereolithography. The electrodes are finished with electroless nickel immersion gold (ENIG) during the fabrication process; **c**, Scanning electron microscopic (SEM) images: **i**. ENIG surface before poly(3,4-ethylenedioxythiophene) (PEDOT) deposition, **ii**. PEDOT layer after deposition; **d**, Characterization of eight representative electrodes before and after PEDOT deposition (mean): **i**. Electrical impedance spectroscopy, **ii**. Cyclic voltammetry curves, **iii**. Electrochemical stability test. Solid line and shaded regions, mean and standard deviation, respectively. Scale bar in **b i** inset, 5 mm; scale bars in SEM images, 5 µm.

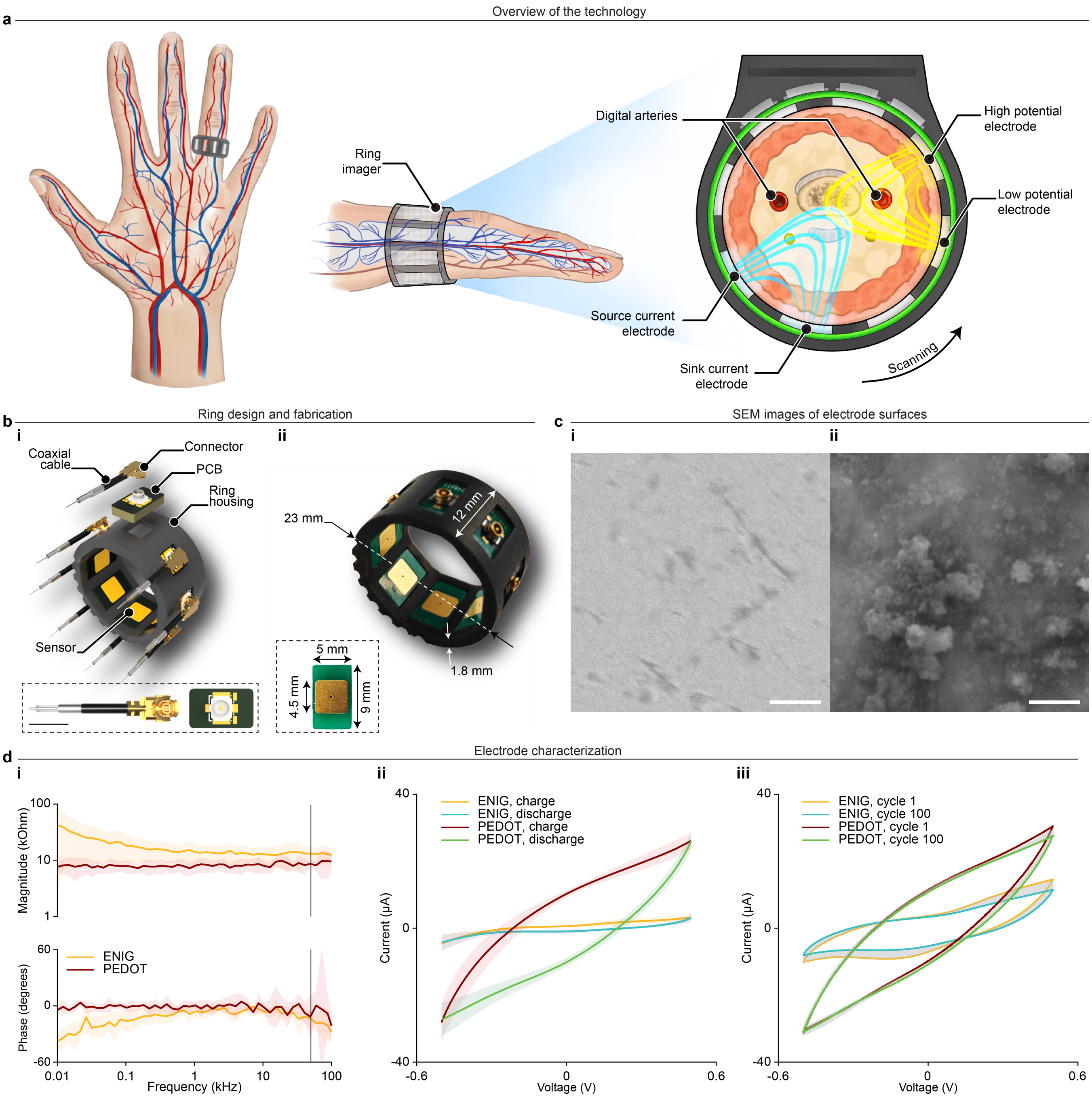

## Particle-laden blood flow simulations in the digital arteries

Chronic hyperglycemia in DM drives microvascular injury resulting in vascular leakage, luminal narrowing and reduced hand tissue perfusion.[11,12] To model how these changes alter hemodynamics in the finger, where our wearable ring images blood flow, we then generated high-fidelity simulations of the normal hemodynamics and blood cell transport from the ulnar artery to the palmar digital arteries using a coupled finite element – discrete element (FEM–DEM) framework (Fig. 2a, b, Methods, Supplementary Discussion 5, Supplementary Table 1, 2, and Supplementary Fig. 2, 3). We found that particle velocities range from near-zero during diastole to ≈80 mm/s during peak systole (Fig. 2c i, d i and Supplementary Video 1). During systole, the pulsatile inflow generated strong acceleration of the fluid phase, resulting in high particle velocities in the proximal artery (Supplementary Video 2). In contrast, during diastole, particle velocities decreased throughout the network as the pressure gradient diminished. At the bifurcation where the flow from the common palmar digital artery is divided into two proper palmar digital arteries, we observed that particle velocities decrease near the inner wall of the curvature due to viscous shear and local pressure distribution, leading to distinct transport patterns at the two downstream arteries (Fig. 2c ii,d ii). Moving along the two proper palmar arteries at the level of the ring finger, we found the particle velocity distribution exhibits a radial profile characteristic of laminar Poiseuille flow (Fig. 2c iii,d iii, Supplementary Video 3).

Given that the clinical manifestation of arterial disease in DM is often obscured by atypical symptoms and peripheral neuropathy,[13] we modeled two degrees of proximal narrowing to represent borderline and severe flow-limiting DM. Specifically, we reduced the diameter of the parent vessel proximal to the digital arterial bifurcation to generate 75% and 96% arterial area occlusion. Fig. 2e,f shows that the 75% occlusion produced only modest changes to pressure and flow rate in both arteries, whereas the 96% occlusion reduced pressure and flow rate significantly in artery 4, eliminating pulsatility. This reduction is accompanied by a compensatory increase in pulse pressure and peak flow in artery 5 (Supplementary Video 4). Together, these simulations can yield synthetic blood flow waveforms and spatially resolved conductivity distributions and serve as inputs to the electrostatic forward model for evaluating the imaging capability of the proposed ring device.

Fig. 2 **Diabetic simulations of particle-laden blood flow in the ring digital arteries**. **a,** Patient-specific palmar arterial network with ring device. **b,** Flow modeling and multiphase boundary conditions for palmar arterial simulations: **i.** Schematic diagram of boundary conditions of particle-laden flows, with arteries identified by numbers, injection site, and 0% (baseline), 75%, and 96% arterial stenosis; **ii.** Inlet waveform of fluid flow and blood cell injection. **c** and **d**, Particle distributions at representative systolic and diastolic phases, respectively: **i.** Overview of particle distributions; **ii.** Bifurcation from common palmar digital arteries to two proper palmar digital arteries (3 and 4); **iii.** Two proper palmar digital arteries (4 and 5) with the ring overlaid. **e,** Pressure waveform measured at the outlet surface of ring finger arteries. **f,** Flow waveform measured at the outlet surface of ring finger arteries. Scale bars in **a**, **b i**, **c i**, and **d i**, 10 mm; scale bars in **c ii** and **d ii**, 2 mm; scale bars in **c iii** and **d iii**, 1 mm; scale bars in **b ii**, **e**, and **f**, a quarter period.

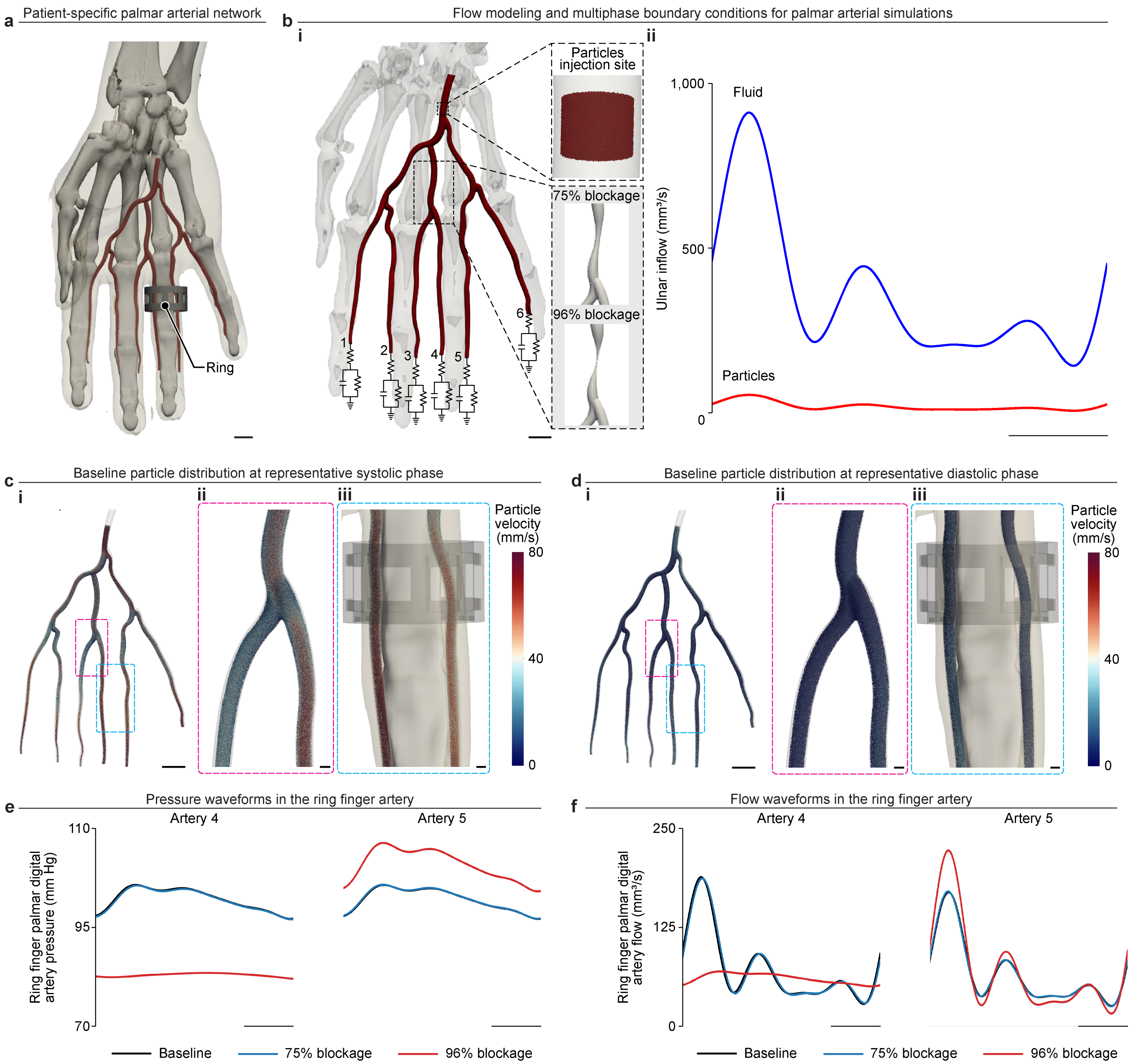
a
Patient-specific palmar arterial network
Ring
b
Flow modeling and multiphase boundary conditions for palmar arterial simulations
i
Particles injection site
75% blockage
96% blockage
1
2
3
4
5
6
ii
Ulnar inflow (mm³/s)
1,000
500
0
Fluid
Particles
c
Baseline particle distribution at representative systolic phase
i
ii
iii
Particle velocity (mm/s)
80
40
0
d
Baseline particle distribution at representative diastolic phase
i
ii
iii
Particle velocity (mm/s)
80
40
0
e
Pressure waveforms in the ring finger artery
Artery 4
Artery 5
Ring finger palmar digital artery pressure (mm Hg)
110
95
70
Baseline
75% blockage
96% blockage
f
Flow waveforms in the ring finger artery
Artery 4
Artery 5
Ring finger palmar digital artery flow (mm³/s)
250
125
0
Baseline
75% blockage
96% blockage

## Electroquasistatic simulations in the finger

The BioZ ring images finger blood flow arising from two coupled hemodynamic processes: pulsatile changes in arterial volume and changes in blood conductivity caused by red blood cell transport.[10] To determine how these sources contribute to the electrical images across the finger vasculature, we simulated ring configurations with 8 and 16 electrodes, as imaging spatial resolution increases with electrode number (Fig. 3a). We performed electroquasistatic (EQS) simulations for all bipolar current injection patterns and visualized the field distribution inside the finger through isopotential lines (Supplementary Fig. 4–13). We found strong distortion to the isopotential lines at the tissue interfaces, e.g., between skin and subcutaneous adipose tissue (SAT), a result expected due to sharp voltage gradients and electrical field discontinuity caused by high electrical conductivity contrast between adjacent tissues. Using lead-field theory,[10] we computed the sensitivity distribution and volume impedance density for selected electrode combinations (Fig. 3b i, Supplementary Fig. 14–16). We found the strongest sensitivity in the SAT and the digital arteries (Fig. 3b ii), with the region accounting for 95% of the total impedance magnitude concentrated near the electrodes while still enclosing the digital arteries (Fig. 3b iv), confirming sufficient measurement depth for ulnar and radial digital arterial detection. Quantitatively, we found that the arteries contributed 0.29% to the total absolute resistance and the total static baseline impedance magnitude, small fractions attributable to their depth and limited volume relative to surrounding tissues (Fig. 3b iii, v, Supplementary Table 3 and Supplementary Fig. 17).

Despite the small static contribution, the arteries are nonetheless expected to be detectable under time-difference imaging, as they are the only compartment exhibiting cardiosynchronous blood volume and conductivity fluctuations. To further evaluate this hypothesis, we performed simulations to reconstruct a ground-truth synthetic 100-point blood conductivity waveform prescribed at the digital arteries while keeping constant their volume (Fig. 2, Fig. 3c i and Supplementary Fig. 18).[10] At each time point, we performed simulations while sequentially switching the current-injection electrode pair around the ring, yielding a total of 800 and 1,600 simulations for the 8- and 16-electrode ring configuration studied, respectively. From the simulated multi-channel data, we applied a coupled forward–inverse reconstruction framework to recover vascular impedance images containing the spatiotemporal distribution of the internal conductivity at the finger's cross section (Fig. 3c ii, Supplementary Discussion 6, Supplementary Fig. 19–23, and Supplementary Video 5). To quantify numerical fidelity of our imaging ring pipeline, we decomposed the images and extracted the conductivity at the region of interest and compared it with the ground-truth input reference blood conductivity waveform, achieving a determination coefficient $r^2$ = 0.99 and relative error (mean$\pm$s.d.) of 0.89%$\pm$0.68% (Fig. 3c iii, iv). These results demonstrate the potential of our wearable ring to image flow driven blood conductivity changes in the digital arteries, with concurrent blood volume variations also contributing to its imaging capability.

Fig. 3 **Electroquasistatic simulations in an anatomically realistic finger model demonstrate the ability to image conductivity changes in the digital arteries**. **a**, Finite element model of the left hand, with the ring finger extracted including the digital arteries. Eight ring electrodes were modeled and placed at the medial phalanx; **b**, Impedance sensitivity and density analysis for $(\ell_1, \ell_4)$-injection pair and $(\ell_8, \ell_2)$-measurement pair: **i**. Electrode configuration, **ii**. Cross-sectional view of slices with highest real sensitivity, **iii**. Contribution of individual tissues to absolute resistance, **iv**. Impedance density of region accounting for 95% of impedance magnitude, and **v**. Contribution of individual tissues to impedance magnitude; **c**, Reconstruction of synthetic conductivity waveform: **i**. Full pipeline from data generation to image reconstruction and signal extraction, **ii**. Reconstructed waveform with representative image frames, **iii**. Correlation between reconstructed and reference conductivity, and **iv**. Histogram of absolute errors

between reconstructed and reference conductivity. $I^{(+,-)}$, current injection electrodes; $V^{(+,-)}$, voltage sensing electrodes; MRI, magnetic resonance imaging; PVI, peripheral vascular impedance; SAT, subcutaneous adipose tissue; SD, standard deviation; VID, volume impedance density; Green and blue patches in **b iv**, injection and measurement electrodes, respectively; White bars in **c ii**, region of interest; Scale bars, 5 mm.

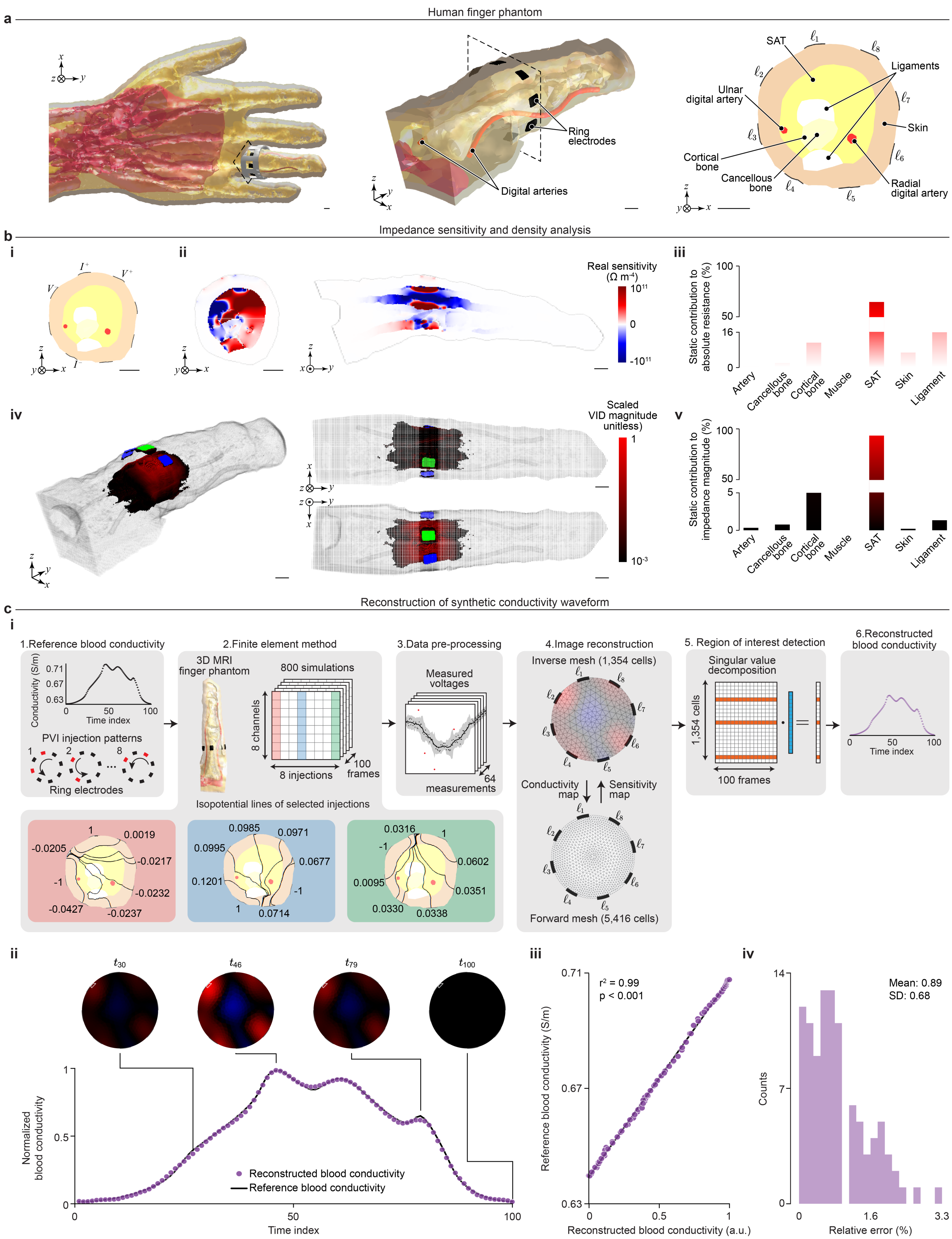

a
Human finger phantom
Ring electrodes
Digital arteries
SAT
Ligaments
Ulnar digital artery
Skin
Cortical bone
Cancellous bone
Radial digital artery
b
Impedance sensitivity and density analysis
i
ii
iii
Real sensitivity (Ω m⁻⁴)
Static contribution to absolute resistance (%)
Artery
Cancellous bone
Cortical bone
Muscle
SAT
Skin
Ligament
iv
v
Scaled VID magnitude unitless)
Static contribution to impedance magnitude (%)
c
Reconstruction of synthetic conductivity waveform
i
1.Reference blood conductivity
Conductivity (S/m)
Time index
PVI injection patterns
Ring electrodes
2.Finite element method
3D MRI finger phantom
800 simulations
8 channels
8 injections
100 frames
3.Data pre-processing
Measured voltages
64 measurements
4.Image reconstruction
Inverse mesh (1,354 cells)
Conductivity map
Sensitivity map
Forward mesh (5,416 cells)
5. Region of interest detection
Singular value decomposition
1,354 cells
100 frames
6.Reconstructed blood conductivity
Isopotential lines of selected injections
ii
Normalized blood conductivity
Reconstructed blood conductivity
Reference blood conductivity
Time index
iii
r² = 0.99
p < 0.001
Reference blood conductivity (S/m)
Reconstructed blood conductivity (a.u.)
iv
Mean: 0.89
SD: 0.68
Counts
Relative error (%)

# Hemodynamic imaging of the finger

We conducted a cross-sectional study on N = 96 healthy subjects to evaluate the feasibility of continuously imaging finger blood flow and used these data for cuffless brachial BP monitoring, followed by a pilot longitudinal study on N = 5 subjects after a gap between 3 and 12 months for recalibration evaluation (Extended Data Fig. 1a, Supplementary Discussion 7 and Supplementary Fig. 24). The cohort consisted of a male:female ratio of 40:56, age of 28.3±8.8 years, weight of 68.4±11.2 kg, height of 169.8±8.6 cm, and body mass index of 23.7±3.6 kg/m$^2$ (Extended Data Fig. 1b).

Our cross-sectional study consisted of initial Doppler flowmetry imaging (Supplementary Fig. 25 and Supplementary Video 6), followed by simultaneous recording of reference continuous BP and 32-channel BioZ data at 50 frames per second, during which participants performed Valsalva and cold pressor maneuvers to induce autonomic changes in BP. We applied an offline processing pipeline to synchronize both data streams and segment them into individual periods (Supplementary Fig. 26). In total, we collected 315,886 pairs of raw BP and BioZ periods from the cross-sectional study, of which 263,929 pairs were classified as clean (Supplementary Table 4). The dataset features diastolic BP of 82.4±11 mm Hg, systolic BP of 128.6±15.2 mm Hg, pulse pressure of 46.2±9.6 mm Hg, heart rate of 74.1±12.8 beats per minute, and peak-to-peak resistance of 6.2±4.8 mΩ (Extended Data Fig. 1c and Supplementary Fig. 27, 28).

To visualize blood flow in the digital arteries, we included the forward-inverse imaging algorithm in the data pipeline, which reconstructed sequences of images with 40×40 pixels showing the spatiotemporal fluctuations of electrical conductivity distribution inside the finger (Extended Data Fig. 1d and Supplementary Video 7). During Valsalva trials (Extended Data Fig. 1d i), the pixel-averaged conductivity signal exhibited low-frequency drift but preserved the characteristic response similar to BP: a sharp initial rise driven by positive intrathoracic pressure, followed by a sustained reduction reflecting diminished cardiac output. We applied a temporal filter to decompose the image sequences into high- and low-frequency components and found similar spatial patterns across all representative time points within each period (Extended Data Fig. 1d ii). Notably, the peak conductivity changes are consistently localized on the palmar side of the high-pass images across all periods. The low-pass images, by contrast, attribute the slow drift entirely to the image center, indicating that baseline drift is absorbed as a spatially unresolved reconstruction artifact.

# Cuffless blood pressure monitoring

To estimate reference brachial BP from the ring images, we explored five neural network architectures with increasing complexity (Methods, Supplementary Discussion 8 and Supplementary Fig. 29): linear regression (LR), multilayer perceptron (MLP), convolutional neural network (CNN), hybrid network with transformer layers (CRT),[14] and hybrid network with Samba layers (CRS).[15] As input to the models, we contrasted between raw multi-channel BioZ measured data and reconstructed images, thereby enabling the assessment of whether imaging improves cuffless brachial BP estimation accuracy. As output, we compared between estimating only the fiducial points (systolic and diastolic) against estimating the full BP waveform period; the latter presents a more challenging task. We trained and evaluated the models for N = 91 datasets under three different partitioning approaches: subject-specific (SS), in which individual models were developed and tested separately for each subject, and more challenging population-within (PW) and population-disjoint (PD), where models were trained across participants; each having a train:test ratio of 9:1 (Supplementary Fig. 30). In total, we trained 1,820 SS models, 20 PW models, and 20 PD models (Supplementary Table 5 and Supplementary Fig. 31, 32).

## Cross-sectional results

Overall, CRT and CRS models achieved optimal performance across all input and output modalities. We further found that image inputs and BioZ inputs yielded similar results. In the SS approach, we trained separate models for each individual subject's dataset and aggregated the performance metrics (Supplementary Table 6 and Supplementary Fig. 33–52). We found the CRS architecture with image input and waveform output (denoted as SS17) to be the best performer. This model achieved high regression metrics for both systolic BP (SBP) and diastolic BP (DBP), with aggregated determination coefficients ($r_\mathrm{a}^2$) of 0.81 and 0.83, respectively; and aggregated concordance correlation coefficients ($\hat{\rho}_{c,\mathrm{a}}$) of 0.9 in both cases. The model also exhibited low bias and tight agreement bounds, with mean errors and 95%-limits of agreement (ME, LOA) of 0.28, [-12.98, 12.65] mm Hg and 0.29, [-9.09, 9.23] mm Hg for SBP and DBP, respectively. We characterized the sample-wise discrepancy using absolute errors (MAE±SDAE) and found that SS17 achieved 4.73±4.46 mm Hg for SBP and 3.39±3.05 mm Hg for DBP. We also analyzed the error distribution of our models and computed the portion of estimations with absolute errors within 5, 10, and 15 mm Hg error thresholds, and found that our SS17 model achieved superior error distribution of 64%, 90%, and 97% for SBP, and 77%, 96%, and 99% for DBP, respectively. Finally, we evaluated the waveform reconstruction metrics such as average mean absolute error (AMAE) and average root mean square error (ARMSE) and found that SS17 achieved AMAE and ARMSE of 4.09 and 4.47 mm Hg, respectively.

In the PW approach, we partitioned each subject's dataset with a consistent train:test ratio before pooling the cohort-wide train and test set (Extended Data Table 1, 2; Supplementary Table 7 and Supplementary Fig. 53–72). We found the CRT architecture with image input and waveform output to yield the best estimation results (Fig. 4a). This model, denoted as PW13, achieved $r_\mathrm{a}^2$ of 0.83 and 0.82, and $\hat{\rho}_{c,\mathrm{a}}$ of 0.89 and 0.89 for SBP and DBP, respectively (Fig. 4b, c). The PW13 model also achieved ME, LOA of -1.61, [-14.38, 10.49] mm Hg for SBP and 0.23, [-9.1, 9.52] mm Hg for DBP, and MAE±SDAE of 4.7±4.39 mm Hg for SBP and 3.44±3.14 mm Hg for DBP. The cumulative AE portions of PW13 below 5, 10, and 15 mm Hg were 64%, 90%, and 97% for SBP, and 77%, 96%, and 99% for DBP, respectively. Finally, the PW13 model achieved AMAE of 3.95 mm Hg and ARMSE of 4.27 mm Hg (Fig. 4d).

After training all PW models, we evaluated their performance on holdout datasets to assess generalizability to unseen subjects. Across all architectural choices and input–output configurations, PW models exhibited performance degradation (Supplementary Table 8 and Supplementary Fig. 73–92), similarly to their PD counterparts (Supplementary Table 9 and Supplementary Fig. 93–112). Although our training dataset is the largest among academic studies, its coverage of ≈225 hours of data across 91 subjects is relatively small compared to industry led large-scale datasets comprising hundreds of thousands of individuals and millions of hours of data,[16–18] which enable generalizability of state-of-the-art models. The limited generalizability observed here suggests that the holdout dataset might lie outside the training dataset distribution, which we investigated next.

## Model robustness analysis

To characterize the distribution shift underlying the generalizability results, we defined the label gap as a dataset property independent of model choice (Methods). Applying this to the PW setting, we found that the label gap between the PW training dataset and the holdout dataset was approximately 2.6 times larger than the typical PW train-test label gap (Supplementary Table 10), confirming that holdout subjects' BP distributions fall outside the training distribution. To further characterize how model performance depends on this shift, we turned to the SS models, where both label gap and dataset quality varied across the 91 subject-specific datasets, enabling a correlation analysis (Supplementary Fig. 113). For LR and MLP models, we found strong correlation between AMAE and data quality index, with -0.51 $\leq r \leq$ -0.44; as well as between AMAE and label gap, with $0.44 \leq r \leq 0.61$. For CRS models, we found weak to moderate dependency on both dataset metrics. For CNN models, AMAE varied widely across subjects (between 4.77 and 56.8 mm Hg), with no statistically significant correlation with either dataset metric. CRT models similarly showed no statistically significant correlation to either dataset metric, but achieved lower AMAE with substantially smaller variation (between 2.46 and 16.64 mm Hg), indicating the efficacy of the CRT architecture in learning subject-specific representations. These results suggest that the limited generalizability of CRT models likely arises from insufficient data breadth, which prevents the learned representations from extending to out-of-distribution subjects.

## Pilot longitudinal recalibration study

Cuffless BP estimation models with wearable data are inherently susceptible to short-term perturbations, such as sensor position and pressure, body posture, and long-term physiological changes, including weight loss, circadian BP variation, and arterial remodeling due to aging.[19–21] However, most academic cuffless BP studies have focused on cross-sectional performance evaluation, capturing inter-subject variability only over short-term intervals.[22,23] As a result, whether pre-trained models degrade over time because of longitudinal intra-subject variability, and whether recalibration can recover performance, remains largely unexplored.

To address this question, we evaluated the pre-trained SS models on the pilot longitudinal datasets, followed by progressive fine-tuning of the models with new daily longitudinal data (Methods and Supplementary Discussion 8). For this analysis, we intentionally selected the models SS03 (LR), SS07 (MLP), SS15 (CRT), and SS17 (CRS), which achieved the highest score among their respective model class in the cross-sectional study. We evaluated the models' ARMSE while sequentially fine-tuned each model on the new daily dataset for each subject, resulting in 100 longitudinal training sessions in total (Supplementary Table 11). While all models exhibited longitudinal degradation when evaluated on unseen data, we found that feedforward models (SS03 and SS07) deteriorated far more severely than hybrid transformer models (SS15 and SS17). For example, without recalibration, the ARMSE deteriorated by 21.21 mm Hg (from 12.3 to 33.51 mm Hg) for SS03 across five longitudinal days, while the ARMSE for SS15 degraded only by 1.21 mm Hg (from 12.21 to 13.42 mm Hg) over the same time period. Furthermore, even after recalibration, feedforward models either marginally exceeded or failed to match the baseline performance of transformer-based models. For example, when evaluated on the third longitudinal day, the models SS15 and SS17 achieved ARMSE of 10.6 and 11.1 mm Hg, respectively, without fine-tuning. In contrast, SS03 and SS07, after fine-tuning with the data from the previous day, achieved ARMSE of 17.27 and 13.62 mm Hg, respectively. These findings suggest that the temporal modeling capabilities of CRS and CRT architectures may confer increased robustness to longitudinal experimental and intra-subject variability, which is not readily recovered through recalibration alone using simpler models, and may therefore support longer recalibration time intervals and improved long-term BP accuracy.

## Model ablation study

To investigate whether the our models operate near their optimal capacity, we performed an ablation study on PW15 and PW19, the best performers from the CRT and CRS classes, respectively (Methods, Extended Data Table 3, Supplementary Table 12, 13, and Supplementary Fig. 114–119). For PW15 (7.5M parameters), depth ablation (A1 and A4 variants) compressed the model to 2.12M parameters while yielding slight improvements across all metrics, indicating that the multi-stage MLP head and the additional convolutional depth might be redundant at this scale. For PW19, already one of the smallest models at 2.5M parameters, all ablations produced metric deterioration, indicating that the reference architecture might be operating near its representation limit.

Fig. 4 **Machine learning architecture for blood pressure estimation from finger electrical impedance images**. **a**, Architecture of the hybrid convolutional-recurrent-transformer (CRT) model class. **b, c**, Systolic and diastolic brachial blood pressure (BP) estimation results, respectively, from the population-within PW13 model: **i**. Correlation plot, **ii**. Bland–Altman limits of agreement, **iii**. Distribution of absolute errors (AE), **iv**. Distributions of true and estimated BP; **d**, Ensemble of BP waveforms: **i**. Estimated BP waveforms from population-within PW13 model; **ii**. True BP waveforms. $r_{\mathrm{a}}^2$, aggregated coefficient of determination; $\hat{\rho}_{c,\mathrm{a}}$, aggregated coefficient of concordance; $p$, p value; $\mathcal{P}_5$, $\mathcal{P}_{10}$, and $\mathcal{P}_{15}$, cumulative percentage of estimations with AE within 5, 10, and 15 mm Hg, respectively. $\mathcal{W}_{\mathrm{pred}}$, Wasserstein distance between true and estimated distribution. AMAE, average mean absolute error; ARMSE, average root mean square error; BiLSTM, bidirectional long short-term memory; DBP, diastolic BP; MAE and SDAE, mean and standard deviation of AE, respectively; MLP, multi-layer perceptron; SBP, systolic BP; solid and dashed lines in **d**, ensemble average and standard deviations of all periods, respectively; scale bars, a quarter period.

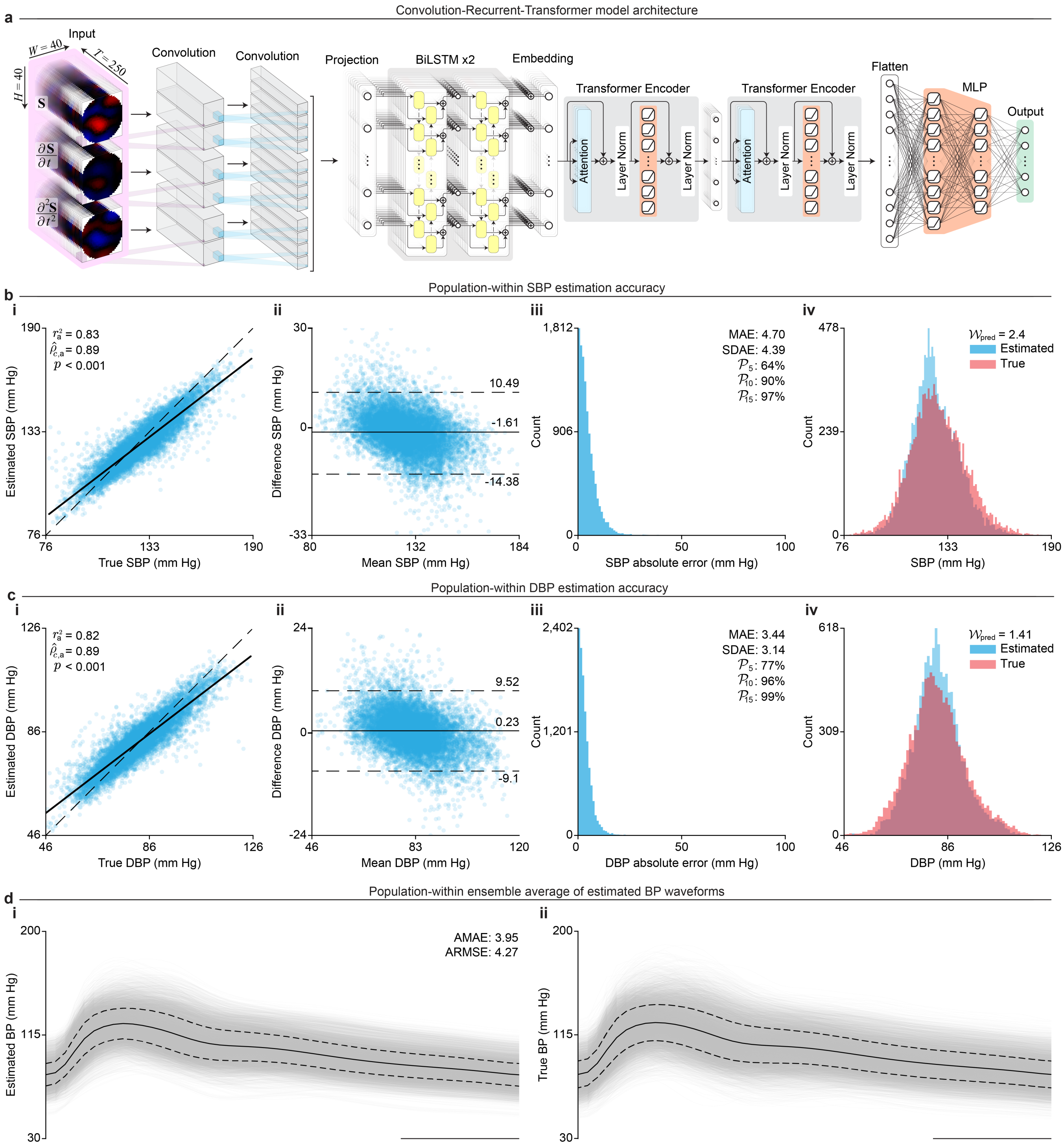

a
Convolution-Recurrent-Transformer model architecture
Input
W = 40
T = 250
H = 40
Convolution
Convolution
Projection
BiLSTM x2
Embedding
Transformer Encoder
Attention
Layer Norm
Layer Norm
Flatten
MLP
Output
b
Population-within SBP estimation accuracy
$r_a^2$ = 0.83
$\hat{\rho}_{c,a}$ = 0.89
$p$ < 0.001
Estimated SBP (mm Hg)
True SBP (mm Hg)
Difference SBP (mm Hg)
Mean SBP (mm Hg)
10.49
-1.61
-14.38
MAE: 4.70
SDAE: 4.39
$\mathcal{P}_5$: 64%
$\mathcal{P}_{10}$: 90%
$\mathcal{P}_{15}$: 97%
Count
SBP absolute error (mm Hg)
$\mathcal{W}_{pred}$ = 2.4
Estimated
True
SBP (mm Hg)
c
Population-within DBP estimation accuracy
$r_a^2$ = 0.82
$\hat{\rho}_{c,a}$ = 0.89
$p$ < 0.001
Estimated DBP (mm Hg)
True DBP (mm Hg)
Difference DBP (mm Hg)
Mean DBP (mm Hg)
9.52
0.23
-9.1
MAE: 3.44
SDAE: 3.14
$\mathcal{P}_5$: 77%
$\mathcal{P}_{10}$: 96%
$\mathcal{P}_{15}$: 99%
Count
DBP absolute error (mm Hg)
$\mathcal{W}_{pred}$ = 1.41
Estimated
True
DBP (mm Hg)
d
Population-within ensemble average of estimated BP waveforms
AMAE: 3.95
ARMSE: 4.27
Estimated BP (mm Hg)
True BP (mm Hg)

# Discussion

Continuous and ambulatory monitoring of peripheral vascular perfusion is essential for the early detection of vascular dysfunction and the timely management of DM, sepsis, and many cardiovascular diseases; however, such capabilities remain limited outside clinical settings. In this study, we developed a wearable ring with multi-channel BioZ sensing for peripheral vascular flow imaging and cuffless BP monitoring (Supplementary Discussion 9, 10). We successfully tested our approach on a healthy cohort of ninety-six subjects, representing the largest academic dataset for ring-based cuffless hemodynamic monitoring using BioZ (Supplementary Table 14). This work is also the first to acquire multi-channel BioZ data at the finger and to perform electrical imaging at this anatomical site, establishing a new paradigm for continuous and ambulatory monitoring of finger vasculature on a convenient wearable form factor. Additionally, we conducted a pilot longitudinal study tracking five subjects over multiple consecutive days to assess temporal performance stability and potential recalibration frequency requirements (Supplementary Table 11).

The selection of electrode materials critically determines the accuracy and practicality of BioZ measurements, and prior work has shown a sharp deterioration in performance when switching from wet gel electrodes to metal electrodes.[24,25] Here, we used sensing electrodes with ENIG finish and benchmarked their performance against PEDOT-coated electrodes, a popular research-grade treatment to improve contact impedance of small metal electrodes.[26] We found that ENIG exhibited impedance characteristics comparable to PEDOT at the imaging frequency of 50 kHz. Visually, we found PEDOT coatings degraded substantially due to finger insertion after minimal use, whereas ENIG electrodes remained stable throughout the entire data collection phase. Furthermore, unlike PEDOT which requires specialized post-processing techniques, ENIG finishing is integrated into standard circuit board fabrication processes. These combined advantages might position ENIG as another viable low-cost candidate, along with other metal electrodes such as silver or gold used in wearable devices.

Our coupled FEM–DEM blood flow simulation framework enables the generation of realistic blood flow to inform wearable device design. The simulations showed high-fidelity particle transport within the palmar digital arterial network under baseline condition and also at 75% and 96% area occlusion. Our results carry three implications for hemodynamic sensing at the finger. First, the propagation of pulsatile pressure and flow signatures from the upstream vasculature to the proper palmar digital arteries confirms that the finger is a physiologically viable imaging site to detect peripheral flow. Second, the near-identical pressure and flow between baseline and 75% occlusion might expose a fundamental limitation of single-modality wearables for early detection of stenosis. This finding motivates the use of imaging approaches capable of providing spatial information that single-channel modalities like photoplethysmography cannot. Third, the severe redistribution observed at 96% occlusion, with pressure collapsing at the affected artery accompanied by compensatory increase at adjacent vessels, illustrates the self-reinforcing nature of unmanaged stenosis, since the compensatory stress elevation in neighboring vessels may lead to further vascular injury, underscoring the value of continuous flow monitoring in a population at risk.

For imaging flow through the digital arteries, we developed a custom imaging algorithm to provide an adaptable path for future embedded implementation. A major challenge in BioZ imaging is the degradation of image fidelity caused by unknown boundary shapes.[27] By leveraging the rigid and predefined geometry of our ring and electrodes, we effectively eliminate artifacts due to boundary modeling error. We conducted EQS simulations to generate synthetic BioZ measurements from a normal blood conductivity waveform, and successfully demonstrated the ability to recover the radial and digital blood conductivity waveform with high morphological fidelity. Beyond validating our reconstruction algorithm, our simulation study provided a framework for choosing the optimal electrode configuration, thereby guiding the design of future BioZ-based imaging ring technology. Crucially, our study offered a detailed look at the internal potential distribution, the BioZ measurement sensitivity and volume impedance density, and the specific contribution of individual tissues with an accurate computable model, which were lacking in existing literature and necessary to determine hemodynamic and confounder physiological sources and their interaction with BioZ sensing.

Current finger- and wrist-based wearables for cuffless hemodynamic monitoring rely mainly on BP with single- or dual-channel sensing and typically estimate only SBP and DBP values (Supplementary Table 14, 15).[22,24,28–30] In instances where multi-channel BioZ data was collected, this was used to derive pulse transit time and estimate BP.[29] By contrast, as in recent ultrasound wearables,[22] we exploit 32-channel BioZ data to generate electrical images of the finger. However, ultrasound-based cuffless studies infer SBP and DBP values from arterial wall motion measurements and this approach has been shown to offer limited cross-sectional accuracy for continuous BP waveform with ARMSE of 11.65 mm Hg.[10] Here, instead, we leveraged the full electrical images to estimate continuous cuffless BP waveform and achieved better cross-sectional (ARMSE of 4.27 mm Hg for PW13 model) and longitudinal (ARMSE of 10.6 mm Hg for SS15 model without fine-tuning) performance than wall motion based ultrasound.

Specifically to ring form factor wearables, Sel et al.[30] achieved results broadly comparable to ours, with $r^2$ and ME, LOA of 0.58 and 0.11, [-10.22, 10.44] mm Hg for SBP and 0.66, and 0.11, [-7.48, 7.77] mm Hg for DBP. However, their evaluation was limited to a small, demographically narrow cohort of 10 subjects (9 male, 1 female). Furthermore, the authors trained two separate models to infer SBP and DBP independently and reduced the variability of their target SBP/DBP values through time-window averaging, thus representing a simpler learning task than ours (Supplementary Table 14). For wrist-based cuffless BP sensors, Crandall et al.[10] proposed a physics-informed neural network embedding Navier–Stokes governing equations that produces spatiotemporally BP and blood velocity field estimations from single-channel BioZ. With a network size of 1M parameters, their PW model achieved baseline results $r^2$ and MAE$\pm$SDAE of 0.77 and 7.26$\pm$8.46 mm Hg for SBP and 0.81 and 3.89$\pm$4.63 mm Hg for N = 86 patients with hypertension, cardiovascular disease, and other conditions (Supplementary Table 15). In comparison, our ablated model PW15-A4 (2.1M parameters) achieved more accurate cuffless BP estimation with $r^2$ and MAE$\pm$SDAE of 0.86 and 4.22$\pm$3.96 mm Hg for SBP and 0.86 and 3.21$\pm$2.95 mm Hg for DBP in a healthy cohort while including autonomic BP challenges (Supplementary Table 13). Our findings thus suggest that it might be possible to improve estimation performance without embedding physics-based constraints, provided that multi-channel data and more sophisticated architectures such as transformers are used. Collectively, our ML approach outperforms existing ring wearables while addressing a more challenging cuffless BP estimation task in a larger and more diverse cohort. We also conducted a small exploratory study of recalibration frequency and found that our best-performing models, particularly those incorporating transformer or Samba blocks, provide greater temporal robustness to intra-subject variability and could support longer recalibration intervals. Finally, we found that the optimal model complexity for estimating the full BP waveform lies between 2 to 3 million parameters, a relatively modest size compared to modern neural networks, while preserving estimation fidelity without overparameterization.

We identified several limitations with our study and directions for future work (Supplementary Discussion 11). First, the ring images pulsatile blood flow in the digital arteries, whereas ground-truth continuous BP waveforms were referenced at the brachial artery. This measurement-reference site mismatch is a limitation, as BP waveforms are influenced by peripheral amplification and hydrostatic effects. Although brachial BP remains the clinical standard for cardiovascular risk assessment and evaluation of distal hemodynamics with the ankle–brachial index, future work should explicitly model the complementary endpoint of local finger blood flow imaged with our ring. Second, our particle-laden blood flow simulation assumes rigid arterial wall and only one-way coupling between particles and fluid.

Future work should consider vascular deformation corresponding to pulsatile blood flow, two-way or four-way coupling to account for particle-fluid and particle-particle interactions, and use these data to model diseased flow dependency of the electrical conductivity of blood. Third, our imaging algorithm prioritizes computational efficiency over image quality through simplified regularization schemes; future work should improve reconstruction fidelity, which may yield more feature-rich inputs for downstream ML tasks. Fourth, our EQS simulations consider only changes in blood conductivity and do not account for arterial dilation driven by the pressure pulse, and thus the results may therefore be an underestimation of the real capability of the ring device to image experimental blood flow in the finger. Fifth, our experimental study was conducted in healthy participants, thus limiting generalizability of our results to broader and clinically relevant populations. Specifically, future work should evaluate real-world performance and generalizability in the target population following established protocols, for example, in patients with uncontrolled hypertension and those with periodic or orthostatic hypertension. Sixth, although we observe comparable performance between multi-channel BioZ and image-derived image inputs, with marginal improvements for image-based inputs in the CRS model; more extensive cross-validation across diverse cohorts and conditions is needed to fully assess model robustness. Finally, while our preliminary longitudinal analyses suggest potential for temporal robustness, further work is required in a larger cohort to define the limits of performance over extended monitoring periods and recalibration intervals. Despite these limitations, the convergence of vascular imaging in a ring form factor positions this new wearable technology with potential for peripheral hemodynamic evaluation in patients with DM and cardiovascular disease broadly.

# Methods

## Ring design, fabrication and characterization

We designed the individual sensing electrodes as solid printed circuit boards (PCBs) with side dimensions of 9 mm × 5 mm and a thickness of 1.2 mm (Fig. 1b, Supplementary Discussion 4). Each PCB contains a conductive region with dimension 4.5 mm × 4.5 mm on the finger-facing side to provide electrical contact, and a coaxial cable connector on the opposite side. To enclose the electrodes, we designed the ring housings with inner circumference based on the ring sizes (Supplementary Fig. 1). Each ring has eight rectangular inserts on its inner wall to contain the sensing electrodes. The rings were 3D-printed with resin stereolithography, while the electrodes were manufactured using a standard PCB fabrication process with electroless nickel immersion gold (ENIG) surface finish.

We characterized the electrodes through electrochemical impedance spectroscopy (EIS) and cyclic voltammetry (CV). We used a three-electrode system for the characterization, with silver chloride reference electrode and a platinum wire as the counter electrode. As the electrolyte, we used phosphate-buffered saline solution, which was diluted with de-ionized water to simulate the electrical conductivity of human skin (10 µS/cm). We performed EIS between 100 Hz and 100 kHz with a sinusoidal voltage of 20 mV and measured the electrodes' magnitude and phase. We then performed CV by sweeping the voltage between -0.5 and 0.5 V at a rate of 100 mV/s for five cycles. We also characterized the electrochemical stability of the sensors by performing CV for 100 cycles at a rate of 1 V/s.

As part of the characterization, we deposited poly(3,4-ethylenedioxythiophene) doped with tetrafluoroborate (PEDOT:$BF_4$) on the ENIG surface and compared them with the uncoated electrodes. For the deposition, we used the same three-electrode system in the ENIG characterization. To increase surface area and enhance adhesion, we deposited a seeding layer of gold nanoparticles (NPs) prior to PEDOT deposition.[31] We prepared the PEDOT:$BF_4$ electrolyte by dissolving tetraethylammonium tetrafluoroborate (TEA$BF_4$) in anhydrous propylene carbonate solution and adding 3,4-ethylenedioxythiophene (EDOT). We stirred the electrolyte continuously for 24 hours to thoroughly disperse the EDOT molecules in the solution and then filtered the electrolyte. Finally, we deposited PEDOT:$BF_4$ on the gold NPs seeding layer. We inspected the deposition quality with scanning electron microscopy (SEM) images acquired at 15 kV accelerating voltages. We then characterized the PEDOT:$BF_4$-coated electrodes following the same procedure in the ENIG characterization.

## Computational fluid dynamics simulations

We used a coupled finite element–discrete element method to simulate blood cell transport from the ulnar artery to the palmar digital arteries (Supplementary Discussion 5, Supplementary Fig. 2). We resolved the pulsatile blood flow within a patient-specific arterial geometry while explicitly tracking the motion of suspended blood cells. We modeled blood cells as rigid spheres with translational and rotational degrees of freedom. We included nonlinear drag, lift, added mass, and buoyancy as hydrodynamic forces acting on each particle. Particle–particle and particle–wall interactions were described using a soft-sphere contact model with elastic and dissipative components. The fluid phase, governed by the incompressible Navier-–Stokes equations, was solved in an Eulerian frame and coupled to the particulate phase through local volume fraction effects and interphase momentum exchange terms.[32] Simulations were performed within the CRIMSON framework,[33] which enabled image-based geometry reconstruction, unstructured mesh generation, and physiologically realistic boundary conditions. To account for pulsatile inflow conditions, particles were injected dynamically in time to maintain a prescribed local hematocrit near the inlet (Supplementary Fig. 3). Outlet boundary conditions were imposed using a Windkessel model, with parameters calibrated from patient-specific vessel dimensions and clinically measured ulnar artery pressure waveforms (Supplementary Table 1, 2).[10]

## Finger vascular impedance imaging

### Solver implementation and validation

We developed a modular package in MATLAB R2024a (The MathWorks Inc., Natick, MA, USA) to reconstruct finger vasculature impedance images (Supplementary Discussion 6). We used the finite element method (FEM) to formulate the complete electrode model (CEM) as a discrete forward operator, and used a one-step Gauss–Newton method to invert the linearized operator. The package has three core modules: meshing, forward solver, and inverse solver (Supplementary Fig. 19). The meshing module generates two-dimensional triangulated meshes using distmesh[34] as backend with dimensional parameters from the ring and electrodes design (Supplementary Fig. 20). The module also implements operations such as mesh refinement, discrete Laplacian, and rasterization employed in the other modules. The forward module uses first-order Lagrange polynomials as nodal bases to transform the CEM boundary value problem to its weak form and assembles the FEM stiffness matrix. The module represents injection and measurement pairs as matrices, which are combined with the stiffness matrix to form the forward operator mapping the elements' conductivity to electrode voltages for a given injection and measurement pattern. The inverse module performs difference imaging to compute the conductivity change between two time instances. The module linearizes the forward operator and uses the discrete Laplacian to formulate a regularized cost function, which is solved using Gauss–Newton method with one iteration. The inverse module applies the reciprocity theorem to compute the forward Jacobian required in the optimal solution. The refinement operation is used to compute a fine mesh used in the forward solver, while the original coarse mesh is used for the inverse solver. The reconstruction yields a vector representing the conductivity value at each element of the unstructured mesh, which is then rasterized to form a $40 \times 40$ image. The forward module was validated using a theoretical resistor model; and the inverse module was validated using an open-source dataset (Supplementary Fig. 21 and 22). We also studied the effect of regularization using a water tank experiment (Supplementary Fig. 23 and Supplementary Video 5).

### Electroquasistatic simulations in a finger phantom model

We investigated the field distribution inside the finger by analyzing the isopotential lines, the impedance sensitivity and density for different ring electrode configurations. We conducted simulations in Sim4Life v7.3 (ZurichMedTech AG, Zürich, Switzerland) using a computable finger phantom (Fig. 3a) excised from the Fats v3.1 model (male, age 37 years, height 182 cm, weight 119 kg).[35] We assigned nominal electrical parameters at 50 kHz to all tissues.[36] We projected the ring electrodes on the skin region surrounding the proximal phalanx and modeled the electrodes as perfect electrical conductors. We assigned Dirichlet boundary conditions $\pm 1$ V at the injection pair and discretized the models using a rectilinear grid with 5.2M voxels. As convergence criteria we set the relative tolerance of $10^{-12}$ with a limit of 100K iterations. We extracted the electrode voltages and computed the corresponding isopotential lines (Supplementary Fig. 4–13). For the sensitivity and impedance density analysis, we selected three electrode configurations representative of an image with 8 electrodes (Fig. 3b). We computed the 3D sensitivity distribution and impedance density based on the inner product between the lead and reciprocal electrical fields. Finally, we calculated the 3D region inside the finger that accounts for 95% of the measured impedance magnitude at the surface (Supplementary Fig. 14–16), and also quantified the contribution of each individual tissue to the absolute resistance, absolute reactance, and total impedance magnitude (Supplementary Fig. 17 and Supplementary Table 3).

To compare reconstruction quality between 8 and 16 electrodes, we also performed BioZ simulations in Sim4Life using a reference blood conductivity waveform which was generated from a multiphysics model based on real BP data (Supplementary Discussion 6).[10] We resampled the conductivity waveform to 100 points and used them to simulate conductivity variations at the arteries, while assigning nominal values at 50 kHz for the electrical properties of all other tissues. For each point in the conductivity waveform, we generated a BioZ measurement frame by cycling the boundary conditions across all feasible injection pairs (Fig. 3c). We used our imaging algorithm to reconstruct finger impedance images from the simulated electrode voltages, and detected the region of interest (ROI) with singular value decomposition. We then extracted the conductivity variation at ROI and compared it with the reference conductivity waveform (Supplementary Fig. 18).

## Experimental study

### Study protocol

We recruited N = 99 healthy subjects for the cross-sectional study (Extended Data Fig. 1a), of which N = 3 subjects dropped out due to inability to complete the protocol. Out of the 96 total subjects, N = 5 subjects were then recruited for a pilot longitudinal follow up study (Supplementary Fig. 24). The study protocol adheres to ethical standards for experiments on humans and was approved by the Institutional Review Board at the University of Utah (#00162369). All subjects were screened for eligibility and provided informed consent prior to data collection (Supplementary Discussion 7). The cross-sectional study included a 15-minute session of ultrasound imaging, followed by simultaneous recording of physiological signals (ECG, BP, and BioZ) in three phases: static (45 minutes), Valsalva (25 minutes), and cold pressor (25 minutes). Each phase consisted of multiple trials lasting between 3 minutes and 10 minutes, interleaved with 2 minutes of break. During the static trials, subjects were seated comfortably and breathed normally. During the dynamic trials (Valsalva and cold pressor), they also performed a 20-second maneuver to induce BP variation after a period of baseline recording. The longitudinal protocol took place between 3 months and 12 months after the subject's completion of the cross-sectional protocol and included five 40-minute sessions on five consecutive days. In each session, physiological signals were recorded in multiple trials lasting between 8 minutes and 10 minutes, interleaved with 2 minutes of break. During each trial, subjects were instructed to perform the 20-second Valsalva maneuver three times after at least 5 minutes of baseline data recording.

### Data collection

We collected reference ultrasound images of the subject's left index finger using the SuperSonic Mach 30 system (Hologic, Marlborough, MA, USA) with a LH20-6 transducer (Supplementary Fig. 25). We recorded three-lead ECG with gel electrodes and recorded BP using the Nano Core device (Finapres Medical Systems, Enschede, Netherlands), with the finger cuff wrapped around the middle phalanx of the left middle finger. The ECG cable and the Nano Core were connected to the NOVA Plus system (Finapres) for data acquisition. Simultaneously to ECG and BP, we recorded 32-channel BioZ at 50 frames per second as raw data using the Sciospec EIT32 system (Sciospec Scientific Instruments GmbH, Bennewitz, Germany). We placed the ring at the proximal phalanx, and connected the electrodes to the EIT32 system through coaxial cables. We used the skip-2 injection at 50 kHz, and skip-1 measurement pattern. Prior to data collection, we sanitized the subject's left index finger with 70% isopropyl alcohol and then applied 0.9% saline solution to moisturize the skin.

### Data processing

To prepare the datasets for machine learning, we passed raw experimental data through a processing pipeline for filtering, synchronization, image reconstruction, temporal segmentation and resampling, and quality assessment (Supplementary Fig. 26). Our pipeline exported 121 datasets, with 96 from the cross-sectional study and 25 from the longitudinal study. Each dataset contains tensors for BP periods ($\mathbf{P}$), high-passed BioZ periods ($\mathbf{Z}^{(\mathrm{HP})}$), low-passed BioZ periods ($\mathbf{Z}^{(\mathrm{LP})}$), high-passed image periods ($\mathbf{S}^{(\mathrm{HP})}$), and low-passed image periods ($\mathbf{S}^{(\mathrm{LP})}$). All periods were temporally aligned and resampled to 50 points. In addition, each dataset includes an auxiliary feature tensor containing the period duration and peak time of each BioZ period; and a masking vector indicating the clean periods suitable for model training.

## Machine learning

### Model configurations

We considered five model classes: linear regression (LR), multilayer perceptron (MLP), convolutional neural network (CNN), hybrid convolutional–recurrent-transformer architecture (CRT),[14] and hybrid convolutional-recurrent-samba architecture (CRS).[15] All models were implemented and trained in Python 3.11. We built the LR models to estimate BP directly from the flattened input, establishing performance baseline. We then considered the MLP class with five fully-connected (FC) hidden layers, each having 100 neurons with ReLU activations. The CNN class includes a multi-stage convolutional encoder to leverage the spatial and temporal structure of the data. The encoder output is then flattened and fed through two FC layers for BP estimation. Our last two classes (CRT and CRS) complement the convolutional layers with sequence-modeling architecture to capture long-range temporal dependency in the data. Both classes include a bidirectional long short-term memory (BiLSTM) layer after the convolutional layers to embed positional information. In the CRT class, the embedded sequential data is passed through two transformer layers with standard attention mechanism and finally mapped to the output through two FC layers. In the CRS class, the transformer and FC blocks are replaced with two Samba layers. We further configured each model to accept either BioZ data ($\mathbf{Z}$) or images ($\mathbf{S}$) as input, and to reconstruct either full BP period or only the fiducial points (SBP and DBP). Of note, our models did not incorporate any subject demographic data as input data. We formed the input samples by stacking five consecutive periods to capture long-range temporal context, with the last period associated with the BP output. We treated the high-pass and low-pass data as separate input channels; with the low-pass channels consisting of the first and second temporal derivatives, i.e. $[\mathbf{Z}^{(\mathrm{HP})}, \dot{\mathbf{Z}}^{(\mathrm{LP})}, \ddot{\mathbf{Z}}^{(\mathrm{LP})}]$ and $[\mathbf{S}^{(\mathrm{HP})}, \dot{\mathbf{S}}^{(\mathrm{LP})}, \ddot{\mathbf{S}}^{(\mathrm{LP})}]$, effectively tripling the input features. We avoided averaging the outputs across periods to preserve the full range of BP variability. Across all input-output combinations, we obtained 20 unique configurations in total, with model sizes between 96K and 154M parameters (Supplementary Table 5).

### Data partition strategy

From the N = 96 cross-sectional subject datasets, we randomly selected 5 as test-exclusive subject datasets for validation, leaving the remaining 91 datasets for model development. We used three different approaches to partition the datasets, and trained our models with all three approaches (Supplementary Discussion 8). In the subject-specific (SS) approach, we trained a separate model for each of the 91 datasets and for each of the 20 configurations. We split each dataset into train set (90%) and test set (10%). Since our samples include multiple consecutive periods, we implemented a graph-based optimization algorithm to eliminate overlapping samples between train and test sets, effectively avoiding temporal leakage (Supplementary Fig. 30). In the population-within (PW) approach, we aggregated the individual train and test sets across all 91 subjects, preserving the 9:1 split, with the test set containing unseen windows from seen subjects. In the population-disjoint (PD) approach, we assigned whole subjects exclusively to either train or test, ensuring that the test set contains only unseen subjects.

### Training cross-sectional models

As loss function we combined mean square error (MSE) with cosine distance (CD), with tunable coefficients (Supplementary Discussion 8). The MSE penalizes point-wise discrepancy while the CD penalizes morphological dissimilarity. For LR models, we used a pure MSE loss to reflect conventional linear regression, while for all other models we used 0.2×MSE + 0.8×CD. We trained the models using the AdamW optimizer,[37] with an initial learning rate of 0.0005. We multiplied the learning rate by 0.8 when the model's accuracy reached a plateau. We defined the accuracy as the average correlation between estimated and true fiducial values. We also implemented a mechanism to stop training when the accuracy stops improving or when the maximum number of epochs $\tau_{\text{max}}$ was reached. We set $\tau_{\text{max}} = 5,000$ when training SS models, and $\tau_{\text{max}} = 500$ for PW and PD models. We used minibatch size of 32 when training the models. For SS datasets, we randomly shuffled all samples before minibatch formation to avoid learning bias due to sample order. However, for PW and PD datasets, this global randomization approach incurred significant data transfer overhead. We thus implemented a stratified randomization approach with a tunable cache capacity to meet memory constraint while ensuring minibatch diversity (Supplementary Discussion 8 and Supplementary Fig. 30).

### Fine-tuning the models on longitudinal data

We selected the four best-performing SS model configurations from the cross-sectional study and extracted the five models corresponding to the longitudinal subjects. For each subject, we first evaluated how well the cross-sectional model held up over the five follow-up days without any retraining, establishing a stability baseline. We then recalibrated each model sequentially: for each follow-up day, we fine-tuned the previous day's checkpoint, and evaluated the updated model on all subsequent days to assess the benefit of progressive daily recalibration. We constructed the fine-tuning dataset for each new day by accumulating that day's data split evenly between train and test, alongside with all previously available data. We used the AdamW optimizer with 0.0001, and set the epoch limit to $\tau_{\text{max}} = 1,000$, with all other hyperparameters unchanged from the cross-sectional training. In total, we trained 100 longitudinal models.

### Ablation study

We selected two best-performing PW model configurations from the cross-sectional study, representative of the CRT and CRS class. We distinguished between three ablation variants: depth reduction, positional encoder replacement, and training strategy adjustment. In the first ablation variant, we removed layers from the convolutional encoder block and the MLP block, leaving the sequential blocks (BiLSTM + Transformer/Samba) unchanged. In the second variant, we replaced the BiLSTM with the sinusoidal encoder. The third variant departs from structural ablation and instead targets the training strategy, applied exclusively to CRT models. Here, we retrained the reference architecture using the cosine annealing learning rate scheduler instead of the default plateau scheduler.

### Evaluation metrics

We employed a variety of metrics to capture the models' goodness of fit from multiple complementary facets (Supplementary Discussion 9). First, to evaluate how well the estimated values align with the true values, we used regression metrics such as the determination coefficient ($r^2$) and concordance coefficient ($\hat{\rho}_c$). We further distinguished between inter-subject coefficients ($r^2_{\mathrm{a}}$ and $\hat{\rho}_{c,\mathrm{a}}$) and intra-subject coefficients ($r^2_{\mathrm{w}}$ and $\hat{\rho}_{c,\mathrm{w}}$), with the former computed on the aggregated test set and the latter weighted from the individual SS test sets. Second, we quantified sample-wise discrepancy through mean errors (ME) and the 95% limits of agreement (LOA). While ME is helpful in portraying the model's bias and precision in the context of LOA analysis, it is susceptible to mutual cancellation between positive and negative errors. As such, we also computed the mean and standard deviation of absolute error (MAE±SDAE). We further evaluated the cumulative percentages $\mathcal{P}_5$, $\mathcal{P}_{10}$, and $\mathcal{P}_{15}$ of estimations with absolute errors (AE) below or equal to 5, 10, and 15 mm Hg, respectively. To complement the sample-wise metrics, we evaluated the discrepancy $\mathcal{W}_{\mathrm{pred}}$ between the estimated and true BP distributions, which assesses the model's ability to capture the underlying statistical properties. Furthermore, for models estimating full BP waveform, we also reported the average mean absolute error (AMAE) and average root-mean-square error (ARMSE). Finally, to analyze how the models' performance depended on dataset properties, we analyzed the correlation between the models' AMAE and dataset quality index (DQI) and train-test label gap ($\mathcal{W}_{\mathrm{label}}$). We defined the DQI as the Wilson's lower confidence bound of the ratio between the number of samples classified as clean and the total number of samples in a dataset; and defined the label gap $\mathcal{W}_{\mathrm{label}}$ as the Wasserstein-1 distance between the train BP distribution and the test BP distribution.[38]

## Statistical analyses

Data were analyzed in MATLAB R2024a (Supplementary Discussion 12).

# Extended Data Figures

Extended Data Fig. 1 **Healthy participant study overview and synchronized reference blood pressure–conductivity tracings**. **a**, Cross-sectional study design. **b**, Population demographics. **c**, Cross-sectional population dataset features after data processing. **d**, Representative signal and images during Valsalva maneuver: **i**. Brachial blood pressure and pixel-averaged reconstructed finger conductivity change, **ii**. Representative conductivity periods before, during, and after Valsalva maneuver. The periods were interpolated to 50 points each, with representative vascular impedance images highlighted. SD, standard deviation. PVI, peripheral vascular impedance. Scale bars in **d i**, 1 s; scale bars in **d ii**, a quarter period.

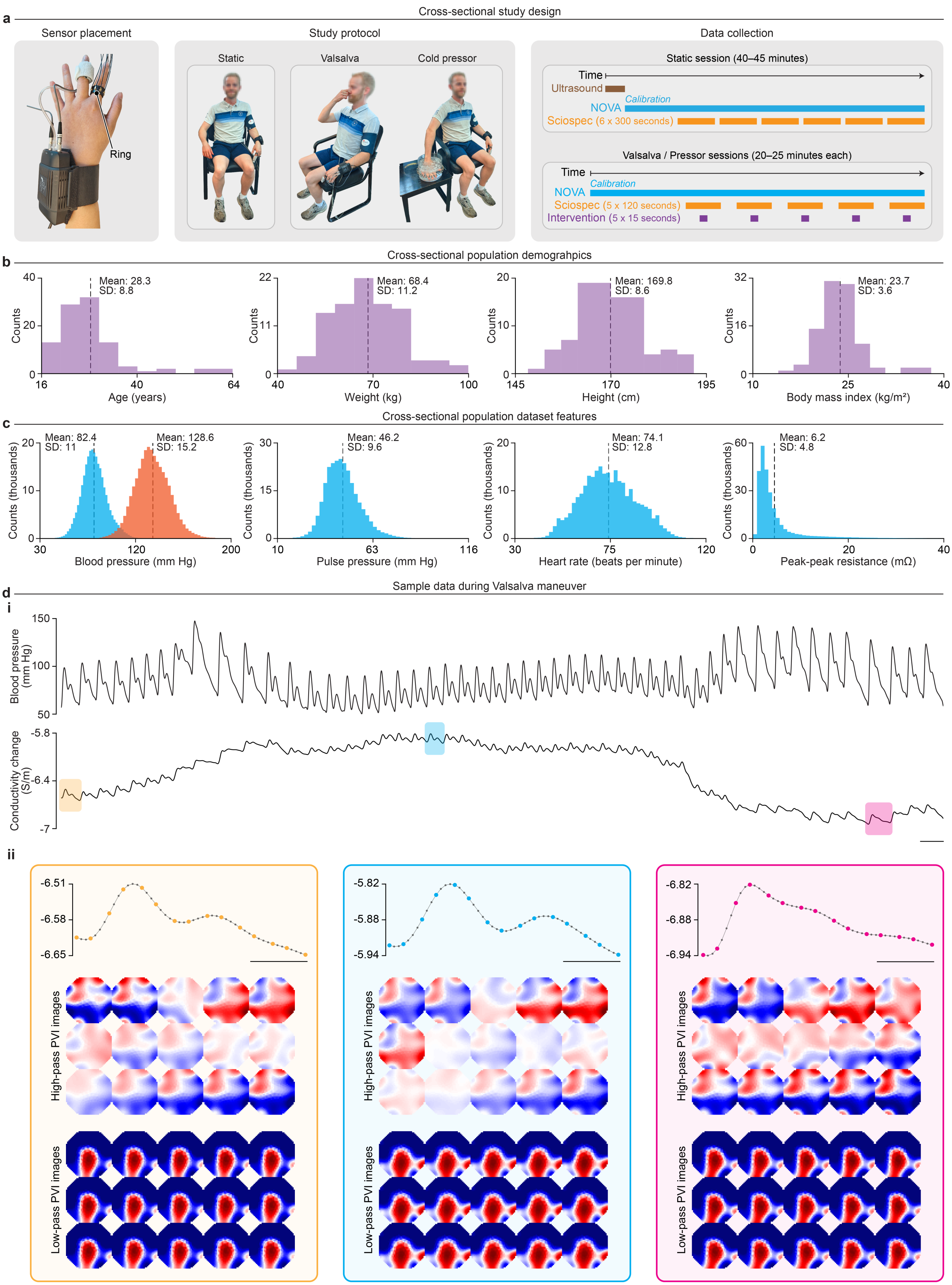

a
Cross-sectional study design
Sensor placement
Ring
Study protocol
Static
Valsalva
Cold pressor
Data collection
Static session (40–45 minutes)
Time
Ultrasound
NOVA
Calibration
Sciospec (6 x 300 seconds)
Valsalva / Pressor sessions (20–25 minutes each)
Time
NOVA
Calibration
Sciospec (5 x 120 seconds)
Intervention (5 x 15 seconds)
b
Cross-sectional population demograhpics
Mean: 28.3
SD: 8.8
Counts
Age (years)
Mean: 68.4
SD: 11.2
Weight (kg)
Mean: 169.8
SD: 8.6
Height (cm)
Mean: 23.7
SD: 3.6
Body mass index (kg/m²)
c
Cross-sectional population dataset features
Mean: 82.4
SD: 11
Mean: 128.6
SD: 15.2
Counts (thousands)
Blood pressure (mm Hg)
Mean: 46.2
SD: 9.6
Pulse pressure (mm Hg)
Mean: 74.1
SD: 12.8
Heart rate (beats per minute)
Mean: 6.2
SD: 4.8
Peak-peak resistance (mΩ)
d
Sample data during Valsalva maneuver
i
Blood pressure (mm Hg)
Conductivity change (S/m)
ii
High-pass PVI images
Low-pass PVI images

# Extended Data Tables

Extended Data Table 1. **Summary of population–within fiducial models.** Model id., Model identifier; LR, linear regression class; MLP, multilayer perceptron class; CNN, convolutional neural network class; CRT, convolutional recurrent transformer class; CRS, convolutional recurrent samba class; BioZ, multi-channel bioimpedance; BP, brachial blood pressure; AMAE, average mean absolute error; ARMSE, average root mean square error; $r_{\mathrm{a}}^2$, aggregated determination coefficient; $\hat{\rho}_{c,\mathrm{a}}$, aggregated concordance correlation coefficient; ME and LOA, mean and limits of agreement of errors, respectively; MAE and SDAE, mean and standard deviation of absolute errors (AE), respectively; $\mathcal{P}_5$, $\mathcal{P}_{10}$, and $\mathcal{P}_{15}$, cumulative percentage of estimations with AE within 5, 10, and 15 mm Hg, respectively; †, waveform metrics not applicable to fiducial models estimating SBP and DBP only.

| Model class | Input type | Model id. | BP waveform | | Systolic BP | | | | | | | | Diastolic BP | | | | | | | |
|---|---|---|---|---|---|---|---|---|---|---|---|---|---|---|---|---|---|---|---|---|
| | | | AMAE (mm Hg) | ARMSE (mm Hg) | $r_{\mathrm{a}}^2$ | $\hat{\rho}_{c,\mathrm{a}}$ | ME (mm Hg) | LOA (mm Hg) | MAE±SDAE (mm Hg) | $\mathcal{P}_5$ | $\mathcal{P}_{10}$ | $\mathcal{P}_{15}$ | $r_{\mathrm{a}}^2$ | $\hat{\rho}_{c,\mathrm{a}}$ | ME (mm Hg) | LOA (mm Hg) | MAE±SDAE (mm Hg) | $\mathcal{P}_5$ | $\mathcal{P}_{10}$ | $\mathcal{P}_{15}$ |
| LR | Image | PW02 | † | † | 0.05 | 0.20 | -3.99 | [-41.10, 27.07] | 13.39±11.47 | 27% | 48% | 64% | 0.02 | 0.14 | -2.40 | [-30.23, 21.11] | 9.89±8.58 | 34% | 61% | 78% |
| | BioZ | PW04 | † | † | 0.01 | 0.10 | -3.84 | [-43.66, 30.37] | 15.00±11.81 | 22% | 42% | 59% | 0.02 | -0.11 | -2.04 | [-31.54, 22.81] | 10.68±8.49 | 30% | 56% | 75% |
| MLP | Image | PW06 | † | † | 0.72 | 0.83 | -0.73 | [-17.05, 15.05] | 5.87±5.43 | 55% | 83% | 94% | 0.71 | 0.83 | -0.43 | [-12.41, 11.15] | 4.42±3.93 | 66% | 92% | 98% |
| | BioZ | PW08 | † | † | 0.70 | 0.82 | -1.02 | [-17.26, 15.31] | 6.31±5.37 | 49% | 80% | 93% | 0.64 | 0.73 | 2.93 | [-9.82, 16.38] | 5.61±4.60 | 55% | 84% | 96% |
| CNN | Image | PW10 | † | † | 0.28 | 0.41 | -5.89 | [-31.79, 18.58] | 10.92±8.95 | 32% | 56% | 72% | 0.31 | 0.47 | -2.04 | [-21.28, 15.58] | 7.27±5.90 | 45% | 73% | 88% |
| | BioZ | PW12 | † | † | 0.67 | 0.77 | -2.81 | [-20.69, 14.71] | 6.96±5.99 | 46% | 76% | 90% | 0.67 | 0.78 | -1.01 | [-14.24, 11.45] | 4.83±4.16 | 62% | 89% | 97% |
| CRT | Image | PW14 | † | † | 0.36 | 0.52 | -3.81 | [-27.74, 20.08] | 9.83±7.91 | 33% | 59% | 78% | 0.38 | 0.59 | -2.07 | [-19.50, 15.29] | 6.94±5.62 | 46% | 75% | 90% |
| | BioZ | PW16 | † | † | 0.80 | 0.88 | -1.09 | [-14.62, 12.10] | 4.79±4.75 | 65% | 89% | 96% | 0.80 | 0.89 | -0.04 | [-10.13, 9.76] | 3.53±3.40 | 77% | 95% | 99% |
| CRS | Image | PW18 | † | † | 0.82 | 0.90 | -0.51 | [-13.10, 11.91] | 4.63±4.47 | 66% | 90% | 97% | 0.81 | 0.90 | 0.29 | [-9.12, 9.64] | 3.48±3.14 | 77% | 96% | 99% |
| | BioZ | PW20 | † | † | 0.83 | 0.89 | -1.20 | [-14.46, 10.99] | 4.65±4.43 | 65% | 90% | 97% | 0.81 | 0.89 | -0.75 | [-10.12, 9.19] | 3.57±3.24 | 75% | 96% | 99% |

Extended Data Table 2. **Summary of population-within waveform models.** Model id., Model identifier; LR, linear regression class; MLP, multilayer perceptron class; CNN, convolutional neural network class; CRT, convolutional recurrent transformer class; CRS, convolutional recurrent samba class; BioZ, multi-channel bioimpedance; BP, brachial blood pressure; AMAE, average mean absolute error; ARMSE, average root mean square error; $r_{\mathrm{a}}^2$, aggregated determination coefficient; $\hat{\rho}_{c,\mathrm{a}}$, aggregated concordance correlation coefficient; ME and LOA, mean and limits of agreement of errors, respectively; MAE and SDAE, mean and standard deviation of absolute errors (AE), respectively; $\mathcal{P}_5$, $\mathcal{P}_{10}$, and $\mathcal{P}_{15}$, cumulative percentage of estimations with AE within 5, 10, and 15 mm Hg, respectively.

| Model class | Input type | Model id. | **BP waveform** | | **Systolic BP** | | | | | | | | **Diastolic BP** | | | | | | | |
|---|---|---|---|---|---|---|---|---|---|---|---|---|---|---|---|---|---|---|---|---|
| | | | AMAE (mm Hg) | ARMSE (mm Hg) | $r_{\mathrm{a}}^2$ | $\hat{\rho}_{c,\mathrm{a}}$ | ME (mm Hg) | LOA (mm Hg) | MAE±SDAE (mm Hg) | $\mathcal{P}_5$ | $\mathcal{P}_{10}$ | $\mathcal{P}_{15}$ | $r_{\mathrm{a}}^2$ | $\hat{\rho}_{c,\mathrm{a}}$ | ME (mm Hg) | LOA (mm Hg) | MAE±SDAE (mm Hg) | $\mathcal{P}_5$ | $\mathcal{P}_{10}$ | $\mathcal{P}_{15}$ |
| LR | Image | PW01 | 11.23 | 11.73 | 0.11 | 0.26 | -9.11 | [-40.49, 19.31] | 13.92±11.28 | 24% | 46% | 62% | 0.13 | 0.31 | -4.49 | [-26.46, 15.74] | 9.15±7.48 | 35% | 63% | 81% |
| | BioZ | PW03 | 12.75 | 13.34 | 0.01 | 0.08 | -6.36 | [-46.70, 28.28] | 15.48±12.37 | 22% | 42% | 58% | 0.02 | -0.11 | -2.56 | [-32.27, 22.56] | 10.64±8.59 | 31% | 57% | 75% |
| MLP | Image | PW05 | 4.62 | 5.08 | 0.75 | 0.85 | -2.08 | [-17.94, 12.49] | 5.62±5.39 | 57% | 85% | 94% | 0.77 | 0.87 | 0.17 | [-10.21, 11.28] | 3.87±3.64 | 73% | 94% | 98% |
| | BioZ | PW07 | 5.07 | 5.57 | 0.71 | 0.83 | 0.01 | [-16.42, 15.47] | 6.01±5.29 | 52% | 83% | 94% | 0.72 | 0.84 | 0.14 | [-11.10, 11.51] | 4.28±3.83 | 68% | 93% | 98% |
| CNN | Image | PW09 | 6.45 | 7.13 | 0.68 | 0.59 | -8.37 | [-27.83, 8.44] | 9.84±7.51 | 31% | 58% | 78% | 0.71 | 0.76 | -2.16 | [-15.13, 9.68] | 4.96±4.25 | 61% | 88% | 97% |
| | BioZ | PW11 | 6.58 | 7.16 | 0.69 | 0.59 | -5.96 | [-27.70, 10.67] | 8.71±7.50 | 38% | 67% | 83% | 0.71 | 0.72 | -1.35 | [-16.33, 10.64] | 5.13±4.36 | 59% | 88% | 96% |
| CRT | Image | PW13 | 3.95 | 4.27 | 0.83 | 0.89 | -1.61 | [-14.38, 10.49] | 4.70±4.39 | 64% | 90% | 97% | 0.82 | 0.89 | 0.23 | [-9.10, 9.52] | 3.44±3.14 | 77% | 96% | 99% |
| | BioZ | PW15 | 3.69 | 4.00 | 0.85 | 0.91 | -1.79 | [-14.06, 9.50] | 4.44±4.27 | 67% | 91% | 97% | 0.85 | 0.91 | -0.21 | [-8.55, 8.45] | 3.14±2.92 | 81% | 97% | 99% |
| CRS | Image | PW17 | 4.17 | 4.56 | 0.81 | 0.89 | -0.15 | [-13.29, 12.48] | 4.82±4.47 | 63% | 89% | 97% | 0.82 | 0.90 | 0.24 | [-8.86, 9.98] | 3.52±3.11 | 76% | 96% | 99% |
| | BioZ | PW19 | 3.80 | 4.11 | 0.84 | 0.90 | -0.47 | [-13.28, 10.96] | 4.44±4.30 | 67% | 92% | 97% | 0.83 | 0.90 | 0.40 | [-8.54, 9.22] | 3.28±3.04 | 79% | 97% | 99% |

Extended Data Table 3. **Summary of ablation results considering top performance population-within (PW) waveform models PW15 and PW19.** The models belong to the class convolutional recurrent transformer (CRT) and convolutional recurrent samba (CRS), respectively, and were trained to estimate full blood pressure (BP) waveform using 32-channel bioimpedance as input. N.A.$^{\dagger}$, not applicable (original model); A1, depth ablation; A2, positional encoder replacement; A4, depth ablation trained with cosine annealing scheduler; AMAE, average mean absolute error; ARMSE, average root mean square error; $r_{\mathrm{a}}^2$, aggregated determination coefficient; $\hat{\rho}_{c,\mathrm{a}}$, aggregated concordance correlation coefficient; ME and LOA, mean and limits of agreement of errors, respectively; MAE and SDAE, mean and standard deviation of absolute errors (AE), respectively; $\mathcal{P}_5$, $\mathcal{P}_{10}$, and $\mathcal{P}_{15}$, cumulative percentage of estimations with AE within 5, 10, and 15 mm Hg, respectively.

| Reference model | Ablation type | Model size | **BP waveform** | | **Systolic BP** | | | | | | | | **Diastolic BP** | | | | | | | |
|---|---|---|---|---|---|---|---|---|---|---|---|---|---|---|---|---|---|---|---|---|
| | | | AMAE (mm Hg) | ARMSE (mm Hg) | $r_{\mathrm{a}}^2$ | $\hat{\rho}_{c,\mathrm{a}}$ | ME (mm Hg) | LOA (mm Hg) | MAE±SDAE (mm Hg) | $\mathcal{P}_5$ | $\mathcal{P}_{10}$ | $\mathcal{P}_{15}$ | $r_{\mathrm{a}}^2$ | $\hat{\rho}_{c,\mathrm{a}}$ | ME (mm Hg) | LOA (mm Hg) | MAE±SDAE (mm Hg) | $\mathcal{P}_5$ | $\mathcal{P}_{10}$ | $\mathcal{P}_{15}$ |
| PW15 | N.A.$^{\dagger}$ | 7,545,030 | 3.69 | 4.00 | 0.85 | 0.91 | -1.79 | [-14.06, 9.50] | 4.44±4.27 | 67% | 91% | 97% | 0.85 | 0.91 | -0.21 | [-8.55, 8.45] | 3.14±2.92 | 81% | 97% | 99% |
| | A1 | 2,129,226 | 3.75 | 4.06 | 0.85 | 0.92 | 0.26 | [-11.64, 11.89] | 4.26±4.09 | 69% | 92% | 98% | 0.85 | 0.91 | 1.03 | [-7.62, 9.81] | 3.27±2.97 | 80% | 96% | 99% |
| | A2 | 7,301,986 | 4.19 | 4.50 | 0.85 | 0.88 | -3.61 | [-16.29, 7.53] | 5.28±4.59 | 58% | 87% | 96% | 0.84 | 0.90 | -1.58 | [-10.63, 7.06] | 3.52±3.08 | 76% | 96% | 99% |
| | A4 | 2,129,226 | 3.69 | 3.97 | 0.86 | 0.92 | 1.21 | [-10.34, 12.33] | 4.22±3.96 | 69% | 92% | 98% | 0.86 | 0.92 | 1.19 | [-7.02, 9.81] | 3.21±2.95 | 80% | 97% | 99% |
| PW19 | N.A.$^{\dagger}$ | 2,513,286 | 3.80 | 4.11 | 0.84 | 0.90 | -0.47 | [-13.28, 10.96] | 4.44±4.30 | 67% | 92% | 97% | 0.83 | 0.90 | 0.40 | [-8.54, 9.22] | 3.28±3.04 | 79% | 97% | 99% |
| | A1 | 2,328,390 | 4.16 | 4.52 | 0.82 | 0.90 | -0.04 | [-13.86, 12.32] | 4.69±4.48 | 65% | 90% | 97% | 0.81 | 0.89 | 0.93 | [-8.44, 10.70] | 3.63±3.28 | 75% | 95% | 99% |
| | A2 | 2,270,242 | 3.81 | 4.13 | 0.84 | 0.91 | 0.15 | [-12.65, 11.76] | 4.38±4.28 | 68% | 91% | 97% | 0.83 | 0.91 | 0.56 | [-8.40, 9.76] | 3.31±3.10 | 79% | 96% | 99% |

Supplementary Information for

# A wearable electrical hemodynamic imaging ring

Gia-Bao Ha[1], Lucas Takanori Sanchez Shiromizu[2], Jaehyeon Song[3], Zhuyun Xie[2], Henry Crandall[1], Dinali Assylbek[1], Alexandra Boyadzhiev[4], Huanan Zhang[4], Fernando Guevara Vasquez[5], Ramakrishna Mukkamala[6,7], Michael Widlansky[8], Shamim Nemati[9], Jesse Capecelatro[3,10], C. Alberto Figueroa[3,11], Benjamin Sanchez[2,12]

[1]Department of Electrical and Computer Engineering, University of Utah, Salt Lake City, UT, USA
[2]Department of Electrical and Computer Engineering, University of Illinois Chicago, Chicago, IL, USA
[3]Department of Mechanical Engineering, University of Michigan, Ann Arbor, MI, USA
[4]Department of Chemical Engineering, University of Utah, Salt Lake City, UT, USA
[5]Department of Mathematics, University of Utah, Salt Lake City, UT, USA
[6]Department of Bioengineering, University of Pittsburgh, PA, USA
[7]Department of Anesthesiology & Perioperative Medicine, University of Pittsburgh, PA, USA
[8]Department of Medicine, Division of Cardiovascular Medicine, Milwaukee, WI, USA
[9]Division of Biomedical Informatics, UC San Diego, San Diego, CA, USA
[10]Department of Aerospace Engineering, University of Michigan, Ann Arbor, MI, USA
[11]Department of Surgery, University of Michigan, Ann Arbor, MI, USA
[12]Richard and Loan Hill Department of Biomedical Engineering, University of Illinois Chicago, Chicago, IL, USA

Corresponding author: Benjamin Sanchez, 851 S. Morgan St., Office 1104 SEO, Chicago, IL 60607. Email: bst@uic.edu. Phone: 312-996-5847.

**Note:** This document contains embedded videos. Using Adobe Reader is required.

# Contents

# Supplementary discussions

## 1 Supplementary Discussion 1. Clinical significance of monitoring peripheral vasculature in diabetes mellitus

Diabetes mellitus (DM) is a leading cause of morbidity and mortality worldwide, driven largely by its profound effects on both the microvasculature and macrovasculature. In the United States alone, approximately 34.2 million individuals are affected by DM, with nearly 1.5 million new cases diagnosed annually.[1,2] Vascular complications represent the dominant clinical burden of the disease. Microvascular dysfunction contributes to the development of diabetic retinopathy, nephropathy, and neuropathy, and remains the leading cause of non-genetic blindness, end-stage renal disease, and non-traumatic limb loss.[3] In parallel, macrovascular disease markedly increases the risk of cardiovascular morbidity and mortality, with diabetes recognized as one of the most powerful risk factors for coronary artery disease and other forms of cardiovascular disease in the United States.[4]

Despite advances in pharmacologic and interventional therapies, vascular disease in individuals with DM remains difficult to treat and outcomes remain inferior compared with non-diabetic populations.[5,6] Microvascular complications occur approximately 25-fold more frequently in individuals with diabetes, and rates have declined only modestly despite improvements in glycemic control and cardiovascular risk management.[5] Similarly, the incidence of death from coronary artery disease remains more than twice that observed in well-matched non-diabetic controls. Although newer therapeutic approaches—including contemporary lipid-lowering strategies, SGLT2 inhibitors, and GLP-1 receptor agonists—have improved some cardiovascular outcomes, substantial residual risk persists.[7–9]

Accumulating evidence indicates that vascular dysfunction in diabetes exhibits distinct structural and functional phenotypes affecting both large arteries and microcirculation.[10–16] These observations suggest that diabetes-associated vascular disease is driven by unique pathophysiologic mechanisms that are not fully captured by current diagnostic approaches or therapeutic paradigms. Accordingly, technologies capable of detecting early changes in vascular and microvascular function may provide critical insight into disease progression and therapeutic response. Continuous physiological monitoring approaches, including emerging wearable sensors capable of measuring bioelectrical impedance within peripheral tissues, offer a promising strategy to non-invasively track dynamic changes in tissue perfusion and microvascular blood flow.[17] Identifying and quantifying these changes may ultimately help define pathways central to the development and progression of diabetic vascular disease and enable improved risk stratification and therapeutic targeting.[18]

# 2 Supplementary Discussion 2. Diagnostic challenges of monitoring finger microvasculature

## 2.1 Conventional imaging methods

The current gold standard for monitoring vascular health is intravascular imaging methods, as they can provide great anatomical details about the vascular system with high spatial accuracy. However, these invasive methods carry inherent risks and are only reserved for critical care situations or when planning for surgery.[19,20] Non-invasive imaging modalities such as ultrasound, magnetic resonance angiography, and computed tomography (CT) angiography have demonstrated very high diagnostic accuracy, with median sensitivity between 88–95% and median specificity between 91–97%.[21,22] They are, therefore, the recommended diagnostic methods by the American Heart Association (AHA) and European Society of Cardiology (ESC).[23,24] However, these methods are inappropriate in routine clinical examinations because they require large and expensive equipment operated by trained specialists with detailed knowledge of the vascular system.[25,26] These limitations in conventional imaging approaches motivate the development of alternative and more accessible methods that can significantly improve micro- and macrovascular monitoring.

## 2.2 Blood pressure tests

Blood pressure (BP) is a reliable measure of cardiovascular health, as the pressure exerted by blood on arterial walls during each cardiac cycle can be detected at the skin surface.[27] In the presence of atherosclerotic plaques, this pressure decreases distal to the congestion site. This relationship has led to the development of several non-invasive tests for peripheral arterial disease (PAD) diagnosis based on segmental BP measurement. The AHA and ESC recommend the ankle–brachial index (ABI) method for initial PAD screening in primary care.[23,24,28] The ABI method calculates the ratio between systolic blood pressure (SBP) measured at the ankle and at the bicep while the patient is in a supine position. The reflection of blood flow is typically stronger at the ankle, so an ABI between 1.0 and 1.3 is considered normal, while ABI $\leq 0.9$ is an indicator of PAD.[29,30]

Compared to imaging, the ABI method is nonexpensive and quick while offering sufficiently high accuracy, with both sensitivity and specificity above 80%.[31–33] However, this method suffers from several limitations. Firstly, BP at the ankle arteries (posterior tibial and dorsalis pedis) is not easy to measure and thus requires Doppler ultrasound, which can lead to false readings when operated by untrained personnel.[34] Secondly, even with oscillometry devices designed and calibrated for ankle BP measurement, the ABI method is confounded by arterial stiffness unrelated to atherosclerosis, such as in elderly or diabetic patients. For these patients that are at higher risk of PAD, the ABI offers little diagnostic value, with sensitivity ranging between 53 and 79%.[33] In these situations the toe–brachial index (TBI) is preferred, as digital arteries are less susceptible to medial calcification.[29,35] At a diagnostic threshold of TBI $\leq 0.7$,[36,37] this method has demonstrated high sensitivity (91–100%) but inconsistent specificity (51–100%).[19]

## 2.3 Bioimpedance

Bioimpedance (BioZ) is a continuous sensing modality with applications in cardiovascular monitoring. BioZ sensors distribute a harmless alternating current into the body and record the voltage corresponding to the underlying physiological activities. BioZ signals are modulated by the volumetric expansion of the arteries and the electrical properties of blood at the sensing site, making them a reliable surrogate measure for BP.[38,39] For this reason, BioZ has been adopted to diagnose PAD, though in a limited research context.[40] Early investigations showed similar patterns between BioZ and mechanical plethysmography waveforms recorded at the finger at positions above and below heart level.[41,42] Later, BioZ was used to diagnose PAD in diabetic patients with calcified arteries.[43,44] Ratio indices similar to the pressure ABI were derived from the BioZ waveform and demonstrated strong correlation ($r \geq 0.8$, $p < 0.0001$).[45] Analyzing the Fourier spectrum of BioZ signals could predict arterial occlusion with sensitivity and specificity above 90%.[46] Unlike the ABI and TBI methods that require inflatable cuffs and clinical visits, BioZ offers significant advantages for continuous BP monitoring with potential for wearable applications. This is particularly valuable for early detection of PAD in the upper limbs, which currently remains underdiagnosed.[47]

# 3 Supplementary Discussion 3. Overview of electrical impedance imaging

Electrical impedance imaging (EII), also known as electrical impedance tomography (EIT), is an imaging technique based on BioZ measurements. Instead of a single tetrapolar BioZ measurement, EII uses an array of electrodes in touch with the body and records BioZ signals simultaneously from all electrodes, providing more spatial information.[48,49] The multi-channel data are then used to reconstruct images of the body's internal electrical conductivity distribution with temporal resolution, reflecting localized physiological changes.

## 3.1 Mathematical model

The goal of EII is to estimate the internal conductivity distribution of an object based on voltages measured on the surface. This relationship between internal conductivity and surface voltages can be described mathematically by the Complete Electrode Model (CEM),[50] which consists of the following governing equations:

$$\nabla \cdot (\sigma(\mathbf{r})\nabla u(\mathbf{r})) = 0 \text{ in } \Omega, \tag{1a}$$

$$\sigma(\mathbf{r})\nabla u(\mathbf{r}) \cdot \hat{\mathbf{n}} = 0 \text{ at } \partial\Omega \setminus \left(\bigcup_{l=1}^{N_L} e_l\right), \tag{1b}$$

$$\int_{e_l} \sigma(\mathbf{r})\nabla u(\mathbf{r}) \cdot \hat{\mathbf{n}} \mathrm{d}S = I_l \text{ at } e_l,\ l = 1, 2, ...N_L, \tag{1c}$$

$$\sigma(\mathbf{r})\nabla u(\mathbf{r}) \cdot \hat{\mathbf{n}} = \frac{1}{z_l}(v_l - u(\mathbf{r})) \text{ at } e_l,\ l = 1, 2, ...N_L, \tag{1d}$$

$$\sum_{l=1}^{N_L} I_l = 0 \text{ and } \sum_{l=1}^{N_L} v_l = 0; \tag{1e}$$

in which $\Omega \subset \mathbb{R}^2$ is a closed domain in real two-dimensional space; $\mathbf{r} \in \Omega$ denotes a point within the domain; $\sigma(\mathbf{r}) : \mathbb{R}^2 \to \mathbb{R}$ and $u(\mathbf{r}) : \mathbb{R}^2 \to \mathbb{R}$ are the spatial distribution of conductivity and voltage, respectively; $\partial\Omega$ is a closed curve serving as the boundary of $\Omega$; $\hat{\mathbf{n}}$ is the normal vector on $\partial\Omega$ pointing outwards; $\mathrm{d}S$ is the length element on $\partial\Omega$; $e_l \subset \partial\Omega$ are disjoint segments serving as electrodes on the boundary; $N_L$ is the number of electrodes; and $I_l$, $v_l$ and $z_l$ are respectively the electrical current, voltage and contact impedance at each electrode $e_l$. Here, equation (1a) is the homogeneous Laplace equation, (1b) is the Neumann boundary condition, which requires the current density field to only have tangential components at the boundary regions not covered by the electrodes. While these equations can be derived from Maxwell's equations under the electroquasistatic approximation, the next two equations are specific to EII and model the field's behavior at the electrodes. Equation (1c) describes the injection current at each electrode and equation (1d) accounts for the electrode impedance at the contact regions. Finally, the constraints in equation (1e) ensure the conservation of charge as well as the uniqueness of the solution.

Equations (1a)–(1e) constitute a boundary value problem (BVP) with its solution being the voltage distribution $u(\mathbf{r})$, given the knowledge of conductivity distribution $\sigma(\mathbf{r})$, the injection currents $I_l$, and the electrodes' contact impedance $z_l$. For practical applications of EII, we consider the following two problems associated with these equations. The first is to estimate the voltages $v_l$ at the electrodes, given the knowledge of $\sigma(\mathbf{r})$, $I_l$ and $z_l$. This is also known as the forward problem in EII and it satisfies the three criteria for being a well-posed problem: it has a solution (1), that is unique (2), and that varies continuously with the initial conditions (3).[51,52] The other practical problem associated with equations (1a)–(1e) is to estimate the internal conductivity distribution $\sigma(\mathbf{r})$, given knowledge of the $I_l$, $v_l$ and $z_l$. This is the inverse problem and is what produces the images in EII. In most cases, the inverse problem does not meet at least one of the three criteria above.

## 3.2 Image reconstruction

Image reconstruction in EII refers to the estimation of snapshots of the conductivity distribution; this process includes formulating (and solving) the forward and inverse problems. The forward problem involves solving the BVP for the electrode voltages given a pre-defined current injection pattern and a known (or assumed) conductivity distribution. In a broader sense, the forward problem also encompasses formulating the appropriate BVP for a particular domain with given geometry and electrode model. The inverse problem seeks the internal conductivity distribution given the knowledge of the voltage and current at the boundary. In what follows, we briefly review the available electrode models besides the CEM, as well as the methods to solve the forward and inverse problem of EII.

### 3.2.1 Electrode models

Although the CEM is the current standard EII formulation, other electrode models also exist, for which unique analytical solutions were proven to exist and partially provided.[50,53,54] These include the continuum model, the gap model, and the shunt model,[55] all of which share the Laplace equation (1a) as the principal governing equation while differing in how the fields are modeled at the domain boundary. The continuum model is the simplest formulation, prescribing a continuous distribution of current density on the entire boundary. This model is based on Calderón's foundational paper on inverse boundary problem[56] and underpins most theoretical results regarding conductivity resolution and optimal current patterns.[57–60] The continuum model ignores electrode geometry, contact impedance, and shunting effects and therefore does not match realistic measurements, with experiments showing its overestimation of the body's resistivity by 25%.[54] The gap electrode model introduces electrodes with finite size, formulates electrical currents as uniform density at the electrodes and zero at the gaps between electrodes. This model represents the measured voltage as the average potential at each electrode.[55] While more realistic than the continuum model, the gap model still underestimates the body's resistivity since it ignores the shunting effects and contact impedance of electrodes.[54] The shunt electrode model takes into account the shorting (or shunting) effect of electrodes by modeling them as perfect conductors. Essentially, this model is the theoretical limit of the CEM when the contact impedance $z_l$ approaches null.[61] The shunting boundary condition enforces a constant voltage at each electrode, leading to an overdetermined equation system. This model has been shown to also underestimate the body's resistivity compared to experiments, with accuracy degrading when more electrodes are included.[54] This degrading behavior is attributed to the lack of contact impedance between the electrode and the boundary, which separates the measured voltage from the boundary potential. When accounting for the size and shape of electrodes as well as their shunting effect and contact impedance, we obtained the CEM as in the equation system (1), which is the most accurate electrode model in EII.[55]

### 3.2.2 The forward problem

The BVP (1) defines the forward operator $F : \sigma(\mathbf{r}) \mapsto \{v_1, v_2, \ldots, v_l\}$ that maps the conductivity distribution to a set of voltages at discrete locations. Solutions to this problem can be obtained analytically or approximated through discrete numerical methods. The analytical approach seeks closed-form expressions for the solution, which exist for domains with simple geometry and conductivity distribution, such as a cylindrical domain with concentric contrast regions or a half-space domain with embedded spherical contrast regions.[62] Seagar's thesis serves as the foundational work on low-frequency boundary measurements of internal conductivity, in which the author developed analytical methods using the continuum and point electrode models.[57] The author used these methods to solve for the boundary measurements of several scenarios in 2D, thereby establishing the theoretical limits on sensitivity and resolution and deriving the optimal injection pattern. Later, Pidcock et al. used the methods developed by Seagar and provided solutions for the gap and shunt models on several 2D and 3D models such as stacked layers or concentric spheres.[63,64] Demidenko was the first to provide an explicit solution for the CEM on a 2D unit circle with homogeneous conductivity and used the solution to estimate contact impedance in EII experiments for breast cancer screening.[65]

Although the analytical methods are useful in establishing theoretical bounds for EII measurement scenarios and systems, they employ domains with simple geometry and symmetrical properties, and assume noiseless measured data. For real-world applications in which the test body has irregular geometry, or when rapid solution is required for multiple different test bodies, numerical methods are oftentimes preferred. Among many available methods, the finite element method (FEM) remains the most popular due to its seamless integration of the CEM and its adaptability to an arbitrary domain geometry.[66–75] The FEM discretizes the solution domain into small triangular (2D) or polyhedral (3D) elements, each having a uniform conductivity inside its area (or volume) and discrete voltages at its nodes. This discretization makes the FEM suitable to represent domains with heterogeneous conductivity distribution or with irregular internal boundaries.[76,77] A drawback of the FEM is its discontinuity in field components at the interface between two media of different conductivity.[78] However, this can be circumvented by using higher-order basis functions or by adaptively increasing the mesh resolution near the interfaces.

Beside FEM, the boundary-element method (BEM) has also found applications in EII because it only discretizes the domain boundary, leading to a smaller system of equation compared to the FEM.[79–83] The reason for this apparent advantage of the BEM is that it assumes a homogeneous conductivity distribution or heterogeneous distributions with fixed internal boundaries, thereby limiting its flexibility compared to the FEM.[78,81,84] Nonetheless, a hybrid approach combining both BEM and FEM formulation is attractive when the conductivity is known or assumed homogeneous for some subdomains with fixed boundaries.[85] This hybrid approach is especially powerful when the homogeneous subdomains are near the boundary, e.g. the scalp and cerebrospinal fluid layers for brain imaging[86] or the subcutaneous adipose tissue (SAT) layers in thoracic imaging,[87,88] and allows for flexible representation of the heterogeneous conductivity deeper in the domain. However, a major disadvantage of the BEM is that it yields dense and often non-symmetric system matrices, which are expensive to solve compared to the sparse, symmetric positive definite (SPD) matrices from FEM.[76,78] Therefore, care is required in hybrid BEM/FEM implementation to yield symmetric matrices which enable the use of efficient solvers. BEM formulations exist (using combinations of single-layer and double-layer potentials) that restore symmetry, though the matrices remain dense.[78]

The discretization of the forward problem yields linear equation systems solvable numerical solvers. The choice of solver depends on the matrix properties, i.e. sparse or dense, symmetric or non-symmetric, and size; as well as the available computational resource. Direct solvers based on matrix factorization techniques are suitable for small to medium systems with dense structure, while indirect solvers are preferred for large and sparse systems.[89] LU factorization is the standard direct method for BEM and hybrid BEM/FEM approaches yielding non-symmetric dense matrices, while Cholesky factorization is optimal for sparse SPD systems from FEM formulation.[68] Specifically for sparse FEM systems, the matrices are permuted prior to factorization to reduce the total number of non-zero entries, thereby reducing both memory requirements and computational cost.[90,91] However, the combined cost of permutation, factorization and back-substitution in the direct methods is $\mathcal{O}(n^3)$, with $n$ being the size of the matrix;[90] this becomes prohibitive for large-scale systems. For this reason, indirect methods based on Krylov subspace iterations are preferred as long as they converge linearly with the matrix size.[92] Another advantage of indirect forward solvers is that their convergence tolerance can be arbitrarily set to match the signal resolution of the EII hardware.[68]

### 3.2.3 The inverse problem

The inverse problem of EII is defined as formulating and computing the inverse operator $F^{-1}$ : $\{v_1, v_2, \ldots, v_l\} \mapsto \sigma(\mathbf{r})$ mapping the discrete voltages to the conductivity distribution. While the forward problem maps from infinite-dimensional conductivity space to a finite-dimensional measurement manifold on the boundary, the inverse (or imaging) problem reverses this process, attempting to recover infinite degrees of freedom from finite boundary data.[93] This dimensional mismatch contributes to the ill-posedness of the inverse problem. In general, two imaging modalities are recognized for EII: absolute imaging and time-difference imaging; each with distinct requirements and having different applications. Absolute imaging reconstructs the static conductivity distribution from a single measurement set, requires high-fidelity data acquisition and compensation for systematic error, with main application in repetitive screening for breast cancer[94–98] and prostate cancer.[99] In contrast, time-difference imaging reconstructs conductivity changes between two time points, requires efficient real-time implementations, with applications in monitoring brain perfusion[100–102] and cardiopulmonary function.[103,104]

Reconstruction methods can be categorized into iterative and direct approaches based on how they invert the forward operator. Iterative methods define an optimization problem based on the discrepancy between the measured voltages and the forward solution, and compute the forward problem repeatedly to find the optimal conductivity distribution. Iterative methods can either be nonlinear and linearized, with the former approach used for absolute imaging with large conductivity contrasts, while the latter assuming small perturbations from a known background distribution and suitable for difference imaging in real time. Both nonlinear and linear iterative approaches rely on variants of Newton's method to compute a reconstruction matrix $\mathbf{R}$ at each iteration, representing the inverse operator $F^{-1}$. Examples of iterative linearized method include the Sheffield filtered back-projection,[49,105,106] NOSER (Newton one-step error reconstruction),[107,108] and GREIT (Graz Reconstruction consensus for Electrical Impedance Tomography) algorithm,[109,110] each defining the reconstruction matrix under different assumptions and constraints. In real-time deployment for difference imaging, the reconstruction matrix of linearized methods can be computed only once at the first iteration and reused in all subsequent iterations, thereby reducing the total computational cost.[77]

The reconstruction matrix requires formulating and inverting the first derivative of the forward operator.[71,111–113] The derivative, also known as the Jacobian, has a condition number in the order of $10^6$ due to the ill-posedness of the inverse problem, thereby necessitating the use of regularization techniques to stabilize the inverse solution.[93] Regularization is implemented by including a penalty term in the cost function, a standard technique used in most reconstruction methods, except for the Sheffield algorithm which applies spatial averaging to stabilize the inversion without an explicit penalty term.[106] The NOSER algorithm uses a variant of Tikhonov regularization, with the penalty term weighting the conductivity values by their contribution to the Jacobian.[107] The GREIT algorithm samples test regions in the reconstruction domain and optimizes the entries of the reconstruction matrix to meet specific performance criteria.[109] Another popular penalty term is the total variation (TV) constraint,[114] which has different interpretations and effects depending on applications. For absolute imaging, the TV constraint prioritizes faithful reconstruction of sharp boundaries between internal organs;[74] while for difference imaging, it assumes localized conductivity change thus reduces memory consumption.[115]

A major drawback of iterative reconstruction methods is they are susceptible to local optima, which is not the case for direct inverse methods, most notable of which is the D-bar method. This method is based on special solutions to the Schrödinger equation and the use of nonlinear Fourier transform. The theory was developed by Calderón[56] and Faddeev,[116] and later formulated for EII by Sylvester and Uhlman.[117] The D-bar method was first implemented on simulated data,[118] and later found applications in thoracic imaging.[119] For circular domain, the D-bar method was demonstrated to be robust against errors in electrode positions.[120]

## 3.3 Emerging trends in electrical impedance imaging

EII was historically developed and optimized for monitoring lung ventilation, which remains its most clinically mature application.[121–123] However, due to its core advantages compared to other imaging modalities, such as high scanning and reconstruction rate, radiation free, and low-cost equipment, EII has been explored in novel applications beyond lung ventilation. Two trends are particularly relevant to the present work. First, the high temporal resolution of EII has motivated its application in monitoring cardiovascular activity.[122] Second, advances in miniaturized analog front-ends, and dry electrode material, and flexible electrode arrays, have enabled EII to be deployed on wearable devices across a variety of anatomical sites.[124] Here, we discuss recent developments in each of these directions.

### 3.3.1 Applications in hemodynamic and cardiovascular monitoring

Besides ventilation, the periodic cardiac activities also induce blood volume and conductivity changes observable in EII images.[125] Early investigations by McArdle et al. and Eyüboğlu et al. demonstrated that cardiosynchronous conductivity variations could be resolved from the respiratory signal.[126,127] Vonk Noordegraaf et al. subsequently provided the first quantitative evidence that EII can be used to estimate right atrial emptying volume and stroke volume,[128,129] thereby motivating a series of investigations using EII to track stroke volume and cardiac output.[130–137] A concurrent line of work pursued the estimation of central BP through pulse transit time (PTT) derived from EII images, with Solà et al. first demonstrating a strong correlation between EII-derived aortic PTT and invasively measured aortic BP.[138,139] This approach was subsequently extended and experimentally validated by Braun et al. and Proença et al. for aortic BP and pulmonary artery pressure estimation, respectively, across a range of controlled physiological conditions.[140–143] Beyond the thorax, Ouypornkochagorn et al. explored scalp-mounted EII for detecting cerebral perfusion changes, demonstrating its potential to monitor hemorrhagic stroke.[144–146]

### 3.3.2 Wearable implementations

The developments of custom application-specific integrated circuits (ASICs) for EII have made its integration into wearable devices increasingly feasible in recent years.[147–152] Following the established clinical application of EII, flexible printed circuit boards (PCBs) with active electrodes were developed and integrated into chest-worn belts for continuously monitoring lung ventilation,[153–157] achieving high frame rates and multi-frequency operation suitable for ambulatory use. Concurrently, custom ASICs with low-power instrument amplifiers for wearable EII systems were developed for cardiac activity monitoring[158–161]. Beyond the thoracic region, scalp-mounted EII systems with absolute 3D imaging techniques were developed for functional brain imaging[100,162] and later extended for brain perfusion monitoring.[144–146] Beside physiological monitoring, wearable EII found a prominent application in gesture recognition due to its capability of detecting conductivity changes associated with deep muscle activation not accessible to surface electromyography. Cornelius et al.[163] developed a wrist-worn device for user identification using multi-channel BioZ data. While the system did not perform image reconstruction, it serves as a precursor to the wearable EII approach. Zhang et al.[164,165] developed the first systems using reconstructed conductivity images for gesture recognition, established the use of wearable EII for human-machine interface. Later developments focused on simplifying EII hardware,[166] integrating flexible electrodes;[167] or improving classification accuracy through 3D reconstruction,[168] contact impedance modeling,[169] deep learning algorithms,[170] and automated calibration.[167]

Taken together, this body of literature demonstrates two converging trends. First, EII can access hemodynamically meaningful information with sufficient prediction accuracy with poten-tial to be clinically relevant. Second, EII systems have been migrating away from the central thoracic region and successfully integrated into other wearable form factors, made possible by the continued advances in miniaturized hardware. However, the peripheral arterial tree has not been targeted for continuous hemodynamic imaging. Therefore, a wearable EII system implemented as a ring is a novel application that can leverage the conductivity contrast of digital arteries at the finger and enable continuous monitoring of peripheral hemodynamic parameters.

# 4 Supplementary Discussion 4. Design and characterization of the ring

## 4.1 Design and manufacturing

We designed the individual sensing electrodes as solid PCBs with side dimensions of 9 mm × 5 mm and a thickness of 1.2 mm. Each PCB contains a conductive region with dimensions 4.5 mm × 4.5 mm on the finger-facing side to provide electrical contact, and UMCC connector solder pads on the opposite side (Fig. 1, Supplementary Fig. 1). To contain the electrodes, we also designed the ring housings with inner circumference based on their sizes, ranging between 6 and 15. All rings have a height of 12 mm and wall thickness of 1.2 mm for structural rigidity. Each ring has eight rectangular inserts on its inner wall to contain sensing electrodes. Each insert has dimensions of 9.1 mm × 5.1 mm and depth of 1.25 mm. Here, we included a clearance 0.05 mm on each contact side to account for manufacturing imperfections. We designed all components in Autodesk Fusion (Autodesk Inc., San Francisco, CA, USA) and fabricated them at JLCPCB (Shenzhen, Guangdong, China). The rings were 3D-printed via stereolithography method with Imagine Black resin, while the electrode PCBs were manufactured using standard FR-4 substrate and Electroless Nickel Immersion Gold (ENIG) surface finish.

## 4.2 Electrochemical deposition and characterization

### 4.2.1 Poly(3,4-ethylenedioxythiophene) deposition

We used the SP-200 potentiostat (BioLogic, Seyssinet-Pariset, France) with a three-electrode system to deposit PEDOT doped with tetrafluoroborate (PEDOT:$BF_4$) on the manufactured ENIG sensors. We used a silver chloride reference electrode (Ag/AgCl/KCl(1M)) and a platinum (Pt) wire as the counter electrode. To increase the surface area and enhance PEDOT adhesion, we deposited a seeding layer of gold nanoparticles (NPs) prior to PEDOT deposition.[171] We first submerged the sensors in an electrolyte solution containing 0.2 mM of 99.999% tetrachloroaurate hydrate ($HAuCl_4$, trace metal basis, Thermo Fisher Scientific Inc., Waltham, MA, USA) and 0.1 M of 99.5% potassium chloride (KCl, Sigma-Aldrich, Burlington, MA, USA). We then deposited gold NPs on the ENIG surface with a constant voltage of 2V for 10 minutes. Next, we prepared the PEDOT:$BF_4$ electrolyte by dissolving 0.12 mol/kg of 99% tetraethylammonium tetrafluoroborate ($TEABF_4$) in 99.7% anhydrous propylene carbonate (PC) solution and adding 0.01 mol/kg of 97% 3,4-ethylenedioxythiophene (EDOT). We stirred the electrolyte continuously for 24 hours to thoroughly disperse the EDOT molecules in the solution and then filtered the electrolyte with a 200 µm sieve. Finally, we deposited PEDOT:$BF_4$ on the gold NPs seeding layer, using the same three-electrode system with a constant voltage of 1.4 V for 5 minutes. We inspected the deposition quality using the TM3030 scanning electron microscope (SEM, Hitachi Ltd., Tokyo, Japan). We acquired SEM images at 15 kV accelerating voltages. The high contrast between PEDOT and ENIG layers enabled clear differentiation between areas of full coverage and spotty coverage.

### 4.2.2 Sensor characterization

We performed electrochemical impedance spectroscopy (EIS) and cyclic voltammetry (CV) on both the PEDOT sensors as well as the bare ENIG sensors. We used the SP-150 potentiostat (BioLogic) for the characterization, with the same three-electrode system for PEDOT deposition. As electrolyte we used the 0.01 M phosphate-buffered saline (PBS, Cytiva Life Sciences, Marlborough, MA, USA) solution, which we diluted with de-ionized (DI) water to simulate the electrical conductivity of human skin (10 µS/cm). Throughout the dilution process, we monitored the conductivity using a SevenDirect SD30 probe (Mettler-Toledo International Inc., Greifensee, Switzerland). To ensure we only characterize the square contact regions, we insulated other metal parts of the sensors with polydimethylsiloxane (PDMS). We performed EIS between 100 Hz and 100 kHz with a sinusoidal voltage of 20 mV to measure the sensors' impedance magnitude and phase. We then performed CV by sweeping the voltage between -0.5 and 0.5 V at a rate of 100 mV/s for five cycles. To characterize the electrochemical stability of the sensors, we also performed CV for 100 cycles at a rate of 1 V/s.

# 5 Supplementary Discussion 5. Particle-laden flow simulation

We simulated the red blood cell transport from the ulnar artery to the palmar digital arteries using a coupled finite element method – discrete element method (FEM–DEM) framework. We resolved pulsatile blood flow through a patient-specific arterial geometry while tracking the motion of suspended particles representing blood cells. We modeled the fluid phase using the stabilized finite element method (FEM) to solve the incompressible Navier–Stokes equations. Concurrently, we modeled the particle phase using the discrete element method (DEM) that tracks point particles through a resolved flow field. Within this framework, particle motion was driven by the local fluid velocity and simulated hydrodynamic forces, effectively capturing particle transport throughout the upper extremity arterial network from the ulnar artery to the proper digital palmar arteries running along the phalanges.

## 5.1 Patient-specific model

We obtained the 3D patient-specific geometry by reconstructing the computed tomography images at the upper extremity of the patient. The model reconstruction, consisting of finger vascular network, was carried out using the open-source cardiovascular simulation code CRIMSON by defining center lines of the arterial structures and cross sectional geometries of finite layers. The resulting model included the distal end of the ulnar artery, proximal palmar arteries, and bifurcations into palmar and proper palmar digital arteries including two at the ring finger (Fig. 2a). Furthermore, to study the effect stenosis on particle transport, we imposed 75% and 96% of area occlusion on the upstream of the bifurcation between artery 3 and 4 (Fig. 2b).

## 5.2 Spatial discretizations

The diameter of the palmar arteries $D_j$ spans between 1.36 and 3.35 mm, where $j$ is the boundary face index. The mesh consisting of ≈2.9M isotropic tetrahedral elements was generated with local refinement to capture detailed fluid flow in the region covered by the ring device (Supplementary Fig. 2). The representative element size $\delta x$ is 0.12 mm for the region of interest, 0.2 mm for other regions, ensuring 11 to 28 elements across the diameters of all branches in the system. With the particle diameter $D_\mathrm{p}$ defined as 50 µm, the particle-to-element length scale $D_\mathrm{p}/\delta x$ ranged between 0.25 and 0.42.

## 5.3 Governing equation of the fluid and particle phase

We modeled the particle phase as Lagrangian point particles with properties such as diameter, density, and velocity. For the fluid phase, we solved the incompressible Navier–Stokes equations in the Eulerian frame of reference. For an incompressible Newtonian fluid, the filtered Eulerian governing equations[172,173] are

$$\rho \left( \frac{\partial \phi}{\partial t} + \nabla \cdot (\phi \mathbf{u}) \right) = 0,$$
$$\rho\phi \left( \frac{\partial \mathbf{u}}{\partial t} + (\mathbf{u} \cdot \nabla)\mathbf{u} \right) = \phi \left( -\nabla p + \nabla \cdot \boldsymbol{\tau} + \rho \mathbf{g} \right) + \mathbf{f}_{\mathsf{p}},$$

where $\rho \in \mathbb{R}$, $\phi \in \mathbb{R}$, and $\mathbf{u} : \mathbb{R}^3 \to \mathbb{R}^3$ are the mass density, volume fraction, and filtered velocity distribution of the fluid, respectively; $p : \mathbb{R}^3 \to \mathbb{R}$ is the pressure distribution; $\mathbf{g} \in \mathbb{R}^3$ is the gravitational acceleration; and $\mathbf{f}_{\mathsf{p}} : \mathbb{R}^3 \to \mathbb{R}^3$ is the hydrodynamic interaction of the particles exerted onto the fluid grid. Here, the shear stress tensor $\boldsymbol{\tau} : \mathbb{R}^3 \to \mathbb{R}^{3\times 3}$ is defined as $\boldsymbol{\tau} := \mu \left( \nabla \mathbf{u} + (\nabla \mathbf{u})^{\mathsf{T}} \right)$, with $\mu \in \mathbb{R}$ being the dynamic viscosity of the fluid.

We modeled the particles as rigid spheres with translational and rotational degrees of freedom. For each particle p, we applied Newton's second law of motion as

$$m \frac{\mathrm{d}\mathbf{v}}{\mathrm{d}t} = \underbrace{V_{\mathsf{p}} \left( -\nabla p + \nabla \cdot \boldsymbol{\tau} \right)_{\mathbf{r}=\mathbf{r}_{\mathsf{p}}}}_{\text{resolved stress}} + \mathbf{F}^{\text{(drag)}} + \mathbf{F}^{\text{(added mass)}} + \mathbf{F}^{\text{(lift)}} + m\mathbf{g}, \tag{2a}$$

$$I_{\mathsf{p}} \frac{\mathrm{d}\boldsymbol{\omega}}{\mathrm{d}t} = \mathbf{T}_{\mathsf{d}}; \tag{2b}$$

where $m \in \mathbb{R}$, $V_{\mathsf{p}} \in \mathbb{R}$, and $\mathbf{v} : \mathbb{R}^3 \to \mathbb{R}^3$ are the particle's mass, volume, and translational velocity, respectively; $I_{\mathsf{p}} \in \mathbb{R}$ and $\boldsymbol{\omega} : \mathbb{R}^3 \to \mathbb{R}^3$ are the particle's moment of inertia the angular velocity, respectively. In equation (2a), the resolved stress term is the point-particle approximation of the surface traction, i.e.

$$\int_{\partial V_{\mathsf{p}}} \boldsymbol{\sigma} \cdot \hat{\mathbf{n}} \mathrm{d}S,$$

where $\partial V_{\mathsf{p}}$ is the particle surface; $\hat{\mathbf{n}}$ is the outward normal vector; and $\boldsymbol{\sigma} : \mathbb{R}^3 \to \mathbb{R}^{3\times 3}$ is the Cauchy stress tensor defined as $\boldsymbol{\sigma} := -p\mathbf{I} + \boldsymbol{\tau}$, with $\mathbf{I} \in \mathbb{R}^{3\times 3}$ being the identity matrix. Also in equation (2a), the term $\mathbf{F}^{\text{(drag)}}$ is the quasi-steady drag force which takes into account the particle Reynolds number and volume fraction effect;[174] $\mathbf{F}^{\text{(added mass)}}$ is the force associated with the acceleration of the surrounding fluid; and $\mathbf{F}^{\text{(lift)}}$ is the Saffman lift force arising from velocity gradients.[175] These closure forces represent subgrid scale hydrodynamic effects not captured by the resolved-stress term acting on the particle surface. In equation (2b), $\mathbf{T}_{\mathsf{d}} : \mathbb{R}^3 \to \mathbb{R}^3$ is the drag torque exerted by the surrounding fluid, given by the Stokes solution as $\mathbf{T}_{\mathsf{d}} = 6V_{\mathsf{p}}\mu \left( \boldsymbol{\omega}_{\mathsf{f}} - \boldsymbol{\omega} \right)$, with $\boldsymbol{\omega}_{\mathsf{f}}$ being the fluid's angular velocity interpolated at the center of the particle p.

The equations (2) were integrated using a second order accurate scheme, implemented within the CRIMSON framework.[176] The fluid-particle interaction was described using one-way coupling scheme, with the fluid exerting hydrodynamic forces on the particles, while the particles did not not affect the fluid. This is assumed in dilute flow regimes where the particle volume fraction is small and the particles have small impacts on the fluid flow (Supplementary Table 2).

## 5.4 Boundary conditions

We used a discrete waveform to model the inflow at the ulnar artery.[177] The original waveform included oscillatory features that may introduce nonphysical instabilities in the solutions.

To mitigate these effects, we first applied cubic spline interpolation to resample the waveform to 1,001 points per cardiac cycle, corresponding to a temporal resolution of $\delta t = 1$ ms. We then applied a Fourier-based smoothing procedure to retain the first five Fourier modes, yielding a temporally smooth inflow profile. The Fourier reconstruction effectively removed high-frequency components, thereby improving numerical stability while preserving the dominant pulsatile features of the waveform. We then prescribed the reconstructed waveform as temporally dynamic boundary condition for both fluid inflow and particle injection rate. Given the inlet diameter $D^{\text{(in)}}$ and the characteristic velocity $U^{\text{(in)}}$, the corresponding Reynolds number is estimated as $Re = \rho U^{\text{(in)}} D^{\text{(in)}}/\mu \approx 39$, indicating that the flow remains within the laminar regime throughout the cardiac cycle. For particle injection, we generated particles at random position bounded by their cylindrical subdomain to ensure no intersection occurs. We synchronized the particle injection rate $\dot{V}_{\text{p}}$ with the fluid inflow rate $Q_{\text{f}}$, ensuring a constant hematocrit Hct:$= \dot{V}_{\text{p}}/Q_{\text{f}}$ of 6.25% throughout a cardiac cycle (Supplementary Fig. 3). The particle injection site was offset by 3 mm downstream of the ulnar artery inlet (Fig. 2b i).

The arterial network consists of six outlets representing the palmar digital arteries running along the fingers. Each outlet was terminated with a three-element Windkessel model, coupled to pressure and flow rate at the surface (Fig. 2b i). To achieve clinically relevant boundary conditions that reflect downstream vasculature resistance and deformability, we calibrated the Windkessel parameters based on the characteristic dimensions and pressure waveform obtained from clinical data in the ulnar artery.[125] The calibration was achieved using 1D nonlinear hemodynamic theory to ensure physiological impedance. The total resistance $R_T$ and outlet resistance $R_k$ distribution were calibrated as

$$R_T = \frac{\bar{P}}{\bar{Q}} \quad \text{and} \quad R_k = R_T \frac{\bar{Q}}{Q_k},$$

where $\bar{P}$ is the mean pressure of the arterial system; $\bar{Q} = 358$ mm$^3$/s is the mean inflow rate; $k \in \{1, 2, \ldots, 6\}$ is the outlet index and $Q_k$ is the corresponding outflow rate. For the same outlet $k$, the resistance distribution at the proximal and distal site were calibrated as

$$R_k^{\text{(proximal)}} = \frac{\rho c_k^{\text{(dia)}}}{A_k^{\text{(dia)}}} \quad \text{and} \quad R_k^{\text{(distal)}} = R_k - R_k^{\text{(proximal)}},$$

where $A_k^{\text{(dia)}}$ is the cross-sectional area at diastole and $c_k^{\text{(dia)}}$ is the diastolic wave speed computed as $c_k^{\text{(dia)}} = 13.3 \cdot \left(D_k^{\text{(dia)}}\right)^{-0.3}$, with $D_k^{\text{(dia)}}$ being the diastolic diameter. Finally, the compliance distribution is computed as

$$C_k = \underbrace{\frac{\Delta Q \Delta t}{\Delta P}}_{C^{\text{(peripheral)}}} \cdot \left( \frac{R_k^{\text{(proximal)}} + R_k^{\text{(distal)}}}{R_k^{\text{(distal)}}} \right) \frac{Q_k}{\bar{Q}},$$

where $\Delta Q := Q_{\text{max}} - Q_{\text{min}}$ are the difference between the maximum and minimum inflow rate; $\Delta t$ is the corresponding time interval between $Q_{\text{min}}$ and $Q_{\text{max}}$; and $\Delta P := P_{\text{max}} - P_{\text{min}}$ is the difference between maximum (systolic) and minimum (diastolic) pressure of the arterial system. Here, the peripheral compliance $C^{\text{(peripheral)}}$ represents the effective deformability of downstream arterioles and capillaries. We assumed the peripheral compliance distribution to be directly proportional to flow distribution. The resulting values after calibration is listed in Supplementary Table 1.

# 6 Supplementary Discussion 6. Peripheral vascular impedance imaging algorithm

To capture conductivity images, we developed an image reconstruction package using MATLAB R2024a (The MathWorks Inc., Natick, MA, USA) consisting of three modules: a meshing module, a forward solver, and an inverse solver. The architectural design of our package is also shown in Supplementary Fig. 19. Here, we describe the implementation and validation of the package.

## 6.1 Meshing module

The meshing module utilizes distmesh as backend[178,179] to generate 2D meshes of $N_K$ disjoint triangle elements, effectively approximating the solution domain as

$$\Omega \approx \tilde{\Omega} := \bigcup_{k=1}^{N_K} \Omega_k, \tag{3a}$$

$$\sigma(\mathbf{r}) \approx \tilde{\sigma}(\mathbf{r}) := \sum_{k=1}^{N_K} \sigma_k \chi_k(\mathbf{r}), \text{ with} \tag{3b}$$

$$\chi_k(\mathbf{r}) := \begin{cases} 1 & \text{if } \mathbf{r} \in \Omega_k, \\ 0 & \text{everywhere else;} \end{cases} \tag{3c}$$

where $\Omega_k$ denotes the element $k$, and $\sigma_k$ and $\chi_k(\mathbf{r})$ are the corresponding conductivity and indicator function, respectively. Mesh generation requires a set of boundary nodes and a constraint for adjusting the mesh resolution.[179] We used the CAD parameters of the ring to define the boundary shape conformal to the ring. We then generated a fixed number of nodes and distributed them evenly on the boundary. Next, we defined as adaptation rule a quadratic function to increase the edge length for nodes closer to the center, with the base length being the node spacing on the boundary. Finally, we assigned disjoint subsets on the boundary as electrodes using electrode dimensions from our design. Overall, we generated 19 distinct meshes corresponding to our rings with North American sizes between 6 and 15 (Supplementary Fig. 20).

The meshing module also includes efficient operations to determine the association among mesh entities (nodes, edges, and triangles). We represented these associative relationships with sparse binary matrices, which are used in three mesh operations essential to our reconstruction pipeline. The first is the discrete Laplacian matrix $\mathbf{L}$, which approximates the spatial derivative operator on unstructured meshes.[180–182] For a mesh with $N_K$ triangles, $\mathbf{L}$ is a sparse symmetric $N_K \times N_K$ matrix with the diagonal filled with $3$'s, and the rows and columns containing $-1$'s at the three adjacent triangles. Secondly, we also utilized the connectivity matrices for iterative mesh refinement. We added new nodes at the midpoints of existing edges and connected them, effectively dividing each triangle into four smaller congruent ones, resulting in a new mesh with $4N_K$ triangles.[183–185] To transfer data between the two meshes, we defined a sparse matrix $\mathbf{P}$ of dimension $4N_K \times N_K$ that represents the overlapping area between the smaller triangles of the fine mesh and the large triangle of the coarse mesh.[77] In our refinement scheme, the fine triangles are fully contained in the coarse triangle, resulting in $\mathbf{P}$ being binary. Both the Laplacian $\mathbf{L}$ and the coarse-to-fine mapping $\mathbf{P}$ are required in the inverse module, with the former being used as regularization and the latter used in the dual-mesh approach (Supplementary Discussion 6.3). The final major functionality of the meshing module is rasterization, i.e. transforming unstructured meshes into $40 \times 40$-images for training machine learning (ML) models (Supplementary Discussion 8). To this end, we first defined a structured $40 \times 40$-grid enclosing the PVI mesh. We then used the bounding box method to quickly identify pairs of overlapping triangles and pixels,[186] and finally applied the Sutherland-–Hodgman clipping algorithm to compute the area of the overlapping regions.[187]

## 6.2 Forward module

We implemented the forward module based on the Complete Electrode Model (Supplementary Discussion 3). The chained partial derivatives in equation (1a) imply that $u(\mathbf{r})$ must be twice differentiable. Often, this requirement cannot be satisfied by approximated solutions such as those from finite element methods. For this reason, equations (1a)–(1e) are called the strong formulation of the BVP, and any $u(\mathbf{r})$ that satisfies them is called a strong solution. Correspondingly, the transformation of these equations to finite element problems is called the weak formulation of the BVP, and their solutions are called weak solutions. The qualifications *strong* and *weak* mean that every strong solution is also guaranteed to satisfy the weak formulation, but not vice versa.

The solution $u(\mathbf{r})$ to the BVP (1) gives the voltage at infinitely many points in the domain $\Omega$. In the finite element approximation, the voltage is sampled at a finite number of nodes and interpolated at everywhere else in the domain. Here, we used as sample nodes the vertices of the elements $\Omega_k$, and approximate $u(\mathbf{r})$ as

$$u(\mathbf{r}) \approx \tilde{u}(\mathbf{r}) := \sum_{i=1}^{N_N} u_i \ell_i(\mathbf{r}), \tag{4}$$

where $N_N$ is the total number of nodes; $u_i$ is the voltage at each node, and $\ell_i(\mathbf{r})$ is the interpolating function assigned to each node. Taken together, the vector $\mathbf{u} = [u_1, u_2, u_3, \dots]^\intercal \in \mathbb{R}^{N_N}$ is the discrete approximation of $u(\mathbf{r})$ at the given nodes, while the choice of $\ell_i(\mathbf{r})$ determines the voltage everywhere else. To ensure the existence and uniqueness of the linear combination in (4), the interpolating functions must satisfy two requirements. First, we require $\ell_i$ to be 1 on node $i$ and 0 on other nodes, i.e. $\ell_i(\mathbf{r}_j) \equiv \delta_{ij}$ with $\delta_{ij}$ being the Kronecker delta. Secondly, $\ell_i$ and $\ell_j$ must be linearly independent for any two nodes $i, j$. Because of these requirements, the interpolating functions are also called nodal basis functions. The accuracy of the approximation in equations (3) and (4), i.e. how closely $\tilde{u}(\mathbf{r})$ approaches the true solution $u(\mathbf{r})$, depends on the resolution of the finite element mesh, as well as the choice of nodal basis functions. A domain with complex geometries and sharp boundaries can be better approximated with more elements of smaller sizes. On the other hand, the smoothness of the approximation can be achieved not only with finer meshes but also by using high-order polynomials as basis functions. The trade-off of these refinements is high computational cost, as more parameters are required. A typical choice for nodal basis functions is the set of Lagrange polynomials.[188] In our implementation, we used triangle elements to discretize the domain, and for each element we used first-order Lagrange polynomials as nodal basis functions

$$\ell_\mu(\mathbf{r}) := 1 - \sum_{\nu=1}^{3} \frac{(\mathbf{r} - \mathbf{r}_\mu) \cdot \hat{\mathbf{n}}}{(\mathbf{r}_\nu - \mathbf{r}_\mu) \cdot \hat{\mathbf{n}}}, \quad \text{for } \mu = 1, \ldots, 3; \tag{5}$$

where $\mu$ and $\nu$ are the local node indices of a triangle, and $\hat{\mathbf{n}}$ is the normal vector to the side that contains both $\mathbf{r}_\mu$ and $\mathbf{r}_\nu$. This choice offers the lowest number of unknowns while also preserving the continuity of the solution.

Due to discretization, the approximation $\tilde{u}(\mathbf{r})$ does not satisfy the twice-differentiability condition imposed by the Laplace equation (1a). To circumvent this, we transform the Laplace equation to

$$\int_{\tilde{\Omega}} w(\mathbf{r}) \nabla \cdot (\sigma \nabla \tilde{u}(\mathbf{r})) \mathrm{d}V = 0 \quad \text{for all } w(\mathbf{r}) \in W \tag{6}$$

where $w(\mathbf{r})$ is a test function in the set $W$. By introducing $w(\mathbf{r})$, we make use of the fundamental lemma of the calculus of variations, where (6) is the sufficient condition for $\nabla \cdot (\sigma(\mathbf{r}) \nabla \tilde{u}(\mathbf{r})) = 0$. By applying integration by part

$$\int_{\tilde{\Omega}} w \nabla \cdot (\sigma \nabla \tilde{u}) \mathrm{d}V = \int_{\tilde{\Omega}} \nabla \cdot (w \sigma \nabla \tilde{u}) \mathrm{d}V - \int_{\tilde{\Omega}} \nabla w (\sigma \nabla \tilde{u}) \mathrm{d}V,$$

the divergence theorem

$$\int_{\tilde{\Omega}} \nabla \cdot (w \sigma \nabla \tilde{u}) \mathrm{d}V = \int_{\partial \tilde{\Omega}} w \sigma \nabla \tilde{u} \cdot \hat{\mathbf{n}} \mathrm{d}S,$$

and the Neumann boundary conditions (1b and 1c), we acquire

$$\int_{\tilde{\Omega}} \sigma \nabla \tilde{u} \cdot \nabla w \mathrm{d}V = \sum_{l=1}^{N_L} \int_{e_l} \sigma w \nabla \tilde{u} \cdot \hat{\mathbf{n}} \mathrm{d}S. \tag{7}$$

By substituting the integrand in the right-hand side of (7) by the electrode condition (1d), we get the weak form of the EII forward problem

$$\int_{\tilde{\Omega}} \sigma \nabla \tilde{u} \cdot \nabla w \mathrm{d}V = \sum_{l=1}^{N_L} \int_{e_l} \frac{1}{z_l} (v_l - \tilde{u}) w \mathrm{d}S \tag{8}$$

The solution to the weak form is defined as $\tilde{\phi} := \{\tilde{u}(\mathbf{r}), \mathbf{v}\}$ with $\mathbf{v} = [v_1, v_2, v_3, \ldots]^\intercal \in \mathbb{R}^{N_L}$ containing the electrode voltages. For $\tilde{\phi}$ to be a solution, equation (8) must hold for all $w(\mathbf{r})$ in the set $W$. The choice of test functions $w(\mathbf{r})$ decides whether a solution exists and is unique. In our implementation, we chose $w(\mathbf{r})$ from the same set as the nodal basis function, i.e. $w_j(\mathbf{r}) \equiv \ell_j(\mathbf{r})$, with $\ell_j(\mathbf{r})$ defined in (5). Equation (8) must be solved for all $w_j$. By substituting (5) into (4), we can expand the weak form as

$$\sum_{i=1}^{N_N} \left( \int_{\tilde{\Omega}} \sigma \nabla \ell_i \nabla \ell_j \mathrm{d}V \right) u_i + \sum_{i=1}^{N_N} \left( \sum_{l=1}^{N_L} \int_{e_l} \frac{1}{z_l} \ell_i \ell_j \mathrm{d}S \right) u_i - \sum_{l=1}^{N_L} \left( \int_{e_l} \frac{1}{z_l} \ell_j \mathrm{d}S \right) v_l = 0,$$

$$- \sum_{i=1}^{N_N} \left( \int_{e_l} \frac{1}{z_l} \ell_i \mathrm{d}S \right) u_i + \int_{e_l} \frac{1}{z_l} v_l \mathrm{d}S = I_l.$$

These equations can be summarized as the following $(N_N + N_L) \times (N_N + N_L)$ matrix system

$$\underbrace{\begin{bmatrix} \mathbf{A}_m + \mathbf{A}_z & \mathbf{A}_q \\ \mathbf{A}_q^\intercal & \mathbf{A}_d \end{bmatrix}}_{\mathbf{A}_\sigma} \cdot \begin{bmatrix} \mathbf{u} \\ \mathbf{v} \end{bmatrix} = \begin{bmatrix} \mathbf{0} \\ \mathbf{q} \end{bmatrix} \cdot I, \tag{9}$$

with the submatrices $\mathbf{A}_m, \mathbf{A}_z \in \mathbb{R}^{N_N \times N_N}$, $\mathbf{A}_q \in \mathbb{R}^{N_N \times N_L}$, and $\mathbf{A}_d \in \mathbb{R}^{N_L \times N_L}$ defined as

$$\mathbf{A}_m(i,j) = \sum_{k=1}^{N_K} \sigma_k \int_{\Omega_k} \nabla \ell_i \nabla \ell_j \mathrm{d}V, \tag{10a}$$

$$\mathbf{A}_z(i,j) = \sum_{l=1}^{N_L} \frac{1}{z_l} \int_{e_l} \ell_i \ell_j \mathrm{d}S, \tag{10b}$$

$$\mathbf{A}_q(i) = -\frac{1}{z_l} \int_{e_l} \ell_i \mathrm{d}S, \tag{10c}$$

$$\mathbf{A}_d(i,j) = \frac{1}{z_l} |e_l| \, \delta_{ij}, \tag{10d}$$

and the vector $\mathbf{q} := [I_1, I_2, I_3, \ldots]^\intercal \in \mathbb{R}^{N_L}$ contains the injection current at each electrodes. For bipolar injection patterns, $\mathbf{q}$ contains only $1$ and $-1$ at the injection electrodes and $0$ everywhere else. Since the injection pair is cycled through all electrodes, we can extend the vectors $\mathbf{u}$, $\mathbf{v}$, and $\mathbf{q}$ to $\mathbf{U} := \left[\mathbf{u}^{(1)}, \mathbf{u}^{(2)}, \ldots, \mathbf{u}^{(N_L)}\right] \in \mathbb{R}^{N_N \times N_L}$, $\mathbf{V} := \left[\mathbf{v}^{(1)}, \mathbf{v}^{(2)}, \ldots, \mathbf{v}^{(N_L)}\right] \in \mathbb{R}^{N_L \times N_L}$ and $\mathbf{Q} := \left[\mathbf{q}^{(1)}, \mathbf{q}^{(2)}, \ldots, \mathbf{q}^{(N_L)}\right] \in \mathbb{R}^{N_L \times N_L}$, where the superscript $(l)$ denotes the $l$-th injection pair. Equation (9) is the finite element formulation of the forward problem, with the solution being the electrode voltages contained in $\mathbf{V}$. Solving the forward problem involves inverting the block matrix $\mathbf{A}_{\boldsymbol{\sigma}}$ and extracting the electrode voltages. We can thus summarize these steps as

$$F(\boldsymbol{\sigma}) := \mathbf{M} \cdot \mathbf{A}_{\boldsymbol{\sigma}}^{-1} \cdot \mathbf{Q} \cdot I \tag{11}$$

where $\boldsymbol{\sigma} := [\sigma_1, \sigma_2, \sigma_3, \ldots]^\intercal \in \mathbb{R}^{N_K}$ is the vector representing the conductivity of all mesh triangles, and $\mathbf{M}$ represents the measurement operator that computes the voltage drop across the pairs of measurement electrodes. Here, $F(\boldsymbol{\sigma}) : \mathbb{R}^{N_K} \to \mathbb{R}^{N_L \times N_L}$ is the forward operator representing the discretization of the original BVP (1). Our implementation included operations to assemble the matrix $\mathbf{A}_{\boldsymbol{\sigma}}$ for any given mesh, and to compute $\mathbf{Q}$ and $\mathbf{M}$ for any bipolar electrode configuration. We solved equation (11) directly using the Cholesky solver in MATLAB, since $\mathbf{A}_{\boldsymbol{\sigma}}$ is positive definite.

## 6.3 Inverse module

We developed the inverse module to reconstruct PVI images that capture the change of conductivity distribution over time, also known as differential imaging.[77] This is equivalent to seeking the operator $R(\cdot) := F^{-1}(\cdot)$, such that $\Delta\boldsymbol{\sigma} = R(\Delta\mathbf{V})$, where $\Delta\mathbf{V} := \mathbf{V}_t - \mathbf{V}_0$ represent voltage changes detected at the boundary electrodes, and $\Delta\boldsymbol{\sigma} := \boldsymbol{\sigma}_t - \boldsymbol{\sigma}_0$ is the estimated conductivity change of the mesh triangles. Assuming that $\Delta\mathbf{V}$ and $\Delta\boldsymbol{\sigma}$ are small over time, we implemented a one-step Gauss–Newton algorithm to solve for $\Delta\boldsymbol{\sigma}$. We first formulated an optimization problem with Tikhonov regularization[189] and defined the cost function as

$$\mathcal{E}(\Delta\boldsymbol{\sigma}) := \|\Delta\mathbf{V} - \Delta F\|^2 + \lambda^2 \|\mathcal{R}(\Delta\boldsymbol{\sigma})\|^2 \tag{12}$$

where $\Delta F := F(\boldsymbol{\sigma}_t) - F(\boldsymbol{\sigma}_0)$ is the voltage change estimated by the forward solver, $\mathcal{R}(\cdot)$ and $\lambda$ are the regularization function and hyperparameter, respectively; and $\|\cdot\|$ denotes the Euclidean norm. Here, we used the discrete Laplacian matrix for regularization, i.e. $\mathcal{R}(\Delta\boldsymbol{\sigma}) \equiv \mathbf{L}(\Delta\boldsymbol{\sigma})$, with $\mathbf{L}$ computed by our meshing module (Supplementary Discussion 6.1). We determined the optimal solution of the least square problem (12) to be

$$\Delta\boldsymbol{\sigma} = \left(\mathbf{J}^\intercal\mathbf{J} + \lambda^2\mathbf{L}^\intercal\mathbf{L}\right)^{-1} \mathbf{J}^\intercal (\Delta\mathbf{V} - \Delta F), \tag{13}$$

where $\mathbf{J} := \partial F/\partial \boldsymbol{\sigma}$ is a matrix of dimension $L^2 \times K$ representing the first derivative (Jacobian) of the forward operator. Instead of differentiating equation (11) directly, we leveraged the reciprocity theorem to express the Jacobian matrix as[71,111,113]

$$\mathbf{J}_k^{(p,q)} := -\frac{1}{I} \int_{\Omega_k} \nabla u^{(p)} \nabla u^{(q)} \mathrm{d}V, \tag{14}$$

where $u^{(p)}$ and $u^{(q)}$ are the lead and reciprocal field, respectively.[190] These fields arise when the injection pair and measurement pair are swapped, i.e. when the current $I$ is applied to the $p$-th electrode pair while measurements are taken at the $q$-th pair, and later when the same current is applied to the $q$-th pair while measurements are taken at the $p$-th pair. For bipolar injection and measurement patterns, the reciprocal forward problem is defined as

$$F^{(r)}(\boldsymbol{\sigma}) := \mathbf{Q}^\intercal \cdot \mathbf{A}_{\boldsymbol{\sigma}}^{-1} \cdot \mathbf{M}^\intercal \cdot I,$$

where $\mathbf{Q}$ and $\mathbf{M}$ are the original injection and measurement matrices from the lead forward problem (11). Using the reciprocity method, we solved the forward problem twice and assembled the Jacobian in a similar manner as $\mathbf{A}_m$ in equation (10a).

The last step to complete (13) is to determine the initial values $\boldsymbol{\sigma}_0$ and $F(\boldsymbol{\sigma}_0)$. With little prior information, we assumed a homogeneous conductivity distribution $\boldsymbol{\sigma}_0 := \sigma_0\mathbf{1}$, where $\mathbf{1}$ is a vector of all 1's. The corresponding forward solution is thus $F(\sigma_0\mathbf{1}) = F(\mathbf{1})/\sigma_0$, where $F(\mathbf{1})$ can be computed from (11). We then determined $\sigma_0$ by minimizing $\|\mathbf{V}_0 - F(\sigma_0\mathbf{1})\|^2$, giving

$$\sigma_0 = \frac{F(\mathbf{1})^\intercal F(\mathbf{1})}{\mathbf{V}_0^\intercal F(\mathbf{1})}.$$

To avoid the inverse crime of using the same mesh for both solvers,[76,93] we employed a dual-mesh approach[185] with a fine mesh for the forward solver (computing the electrode voltages and the Jacobian); and a coarse mesh for reconstruction (updating the conductivity). The mesh refinement scheme is carried out by the meshing module (Supplementary Discussion 6.1).

## 6.4 Validation

### 6.4.1 Forward module validation with theoretical model

We validated the forward solver module in two scenarios. The first is theoretical validation using a rectangular resistor model with a closed-form solution. We used the meshing module to generate the resistor with cross-section $A = 0.1$ m² and length $L = 0.5$ m, discretized with $475$ nodes and $846$ elements (Supplementary Fig. 21a). We assigned the resistor a homogeneous conductivity $\sigma = 0.25$ S/m and included an electrode with contact impedance $z_l = 0.1$ Ω m² at both ends. We performed a forward simulation with an excitation current $I = 0.1$ A and compared the computed nodal voltages $u_i$ as well as the electrode voltages $v_l$ with the theoretical values. In the second scenario, we used a disc model with diameter $D = 20$ cm to validate our implementation of the Jacobian (Supplementary Fig. 21b). We generated the mesh with $3,034$ nodes and $5,746$ elements. We assigned the disc a homogeneous conductivity $\sigma = 1$ S/m and generated 16 electrodes with contact impedance $z_l = 0.01$ Ω m². We distributed the electrodes evenly on the disc's boundary, with each electrode occupying $1/20$ the disc's circumference. We performed forward simulations with a current $I = 500$ µA using the skip-2 pattern for injection and adjacent pattern for measurement. We then computed the derivative $\mathbf{J}$ using equation (14).

### 6.4.2 Inverse module validation with experimental data

To validate the inverse solver module, we used the ventilation EII dataset collected from a healthy horse during normal breathing (Supplementary Fig. 22).[191] The dataset includes $T = 1,784$ measurement frames recorded over 187 seconds using the skip-4 injection pattern. We performed image reconstruction on a unit circle mesh with $N_N = 1,063$ nodes and $N_K = 1,964$ elements, and used $\lambda = 0.1$ for smoothness regularization. We also investigated the effects of regularization parameters on image quality (Supplementary Fig. 23).

## 6.5 Finger isopotential lines and sensitivity analysis

To visualize field patterns within the PVI ring geometry, we conducted electroquasistatic simulations on a computable human finger model. We performed simulations in Sim4Life v7.3 (ZurichMedTech AG, Zürich, Switzerland) using the left index finger of Fats v3.1 (male, age 37 years, height 182 cm, weight 119 kg).[192,193] We modeled the digital arteries as cylindrical splines running along the phalanges, with diameters of 1.8 mm and 1.2 mm for the radial and ulnar digital arteries, respectively. Next, we modeled the sensing electrodes as rectangular patches and distributed them evenly around the finger's circumference at the proximal phalanx. We assigned nominal values at 50 kHz for the electrical parameters of all tissues,[194] while modeling the electrodes as perfect electrical conductor. We discretized the model using a structured grid with an adaptive scheme to balance simulation time and accuracy. The grid resolutions are 0.4 mm for the finger, 0.2 mm for the electrodes, and 1 mm for the region outside the finger, resulting in 5.2M total voxels. We assigned $\pm 1$ V Dirichlet boundary condition to the injection pair, and cycled the pair across all unique bipolar combinations for 8 electrodes ($32$ pairs) and 16 electrodes ($128$ pairs), performing $160$ simulations in total. As convergence criteria, we used a relative tolerance of $1 \cdot 10^{-12}$ and maximum of 100k iterations. We then extracted the voltages at each electrodes and computed the isopotential lines matching these voltages for visualization (Supplementary Fig. 4–13).

While isopotential lines reveal the fields created by different injection configurations, they do not indicate how the fields interacts with different measurement configurations. To capture this interaction, we computed the sensitivity $S^{(p,q)}$ that quantifies the spatial coupling between the injection pair $p$ and measurement pair $q$. The sensitivity $S^{(p,q)} : \mathbb{R}^3 \to \mathbb{C}$ is a scalar field defined as

$$S^{(p,q)}(\mathbf{r}) := \frac{|\gamma(\mathbf{r})|^2}{I^{(p)} I^{(q)}} \left\langle \mathbf{E}^{(p)}(\mathbf{r}), \mathbf{E}^{(q)}(\mathbf{r}) \right\rangle \tag{15}$$

where $\mathbf{r} \in \mathbb{R}^3$ is a point in three-dimensional space; $\gamma : \mathbb{R}^3 \to \mathbb{C}$ is the local admittivity; $\mathbf{E}^{(p)}, \mathbf{E}^{(q)} : \mathbb{R}^3 \to \mathbb{C}^3$ are the lead and reciprocal electric fields, respectively; $I^{(p)}, I^{(q)} \in \mathbb{C}$ are the corresponding lead and reciprocal currents, respectively;[195,196] and $\langle \cdot, \cdot \rangle$ denotes the inner product between two vector fields. Equation (15) is the complex extension to the usual definition for sensitivity[190] that are applicable only to real vector fields. However, we computed the inner product without complex conjugation to ensure symmetry between lead and reciprocal fields.[197] The volume impedance density (VID) $\psi^{(p,q)} : \mathbb{R}^3 \to \mathbb{C}$ is then defined as[198]

$$\psi^{(p,q)}(\mathbf{r}) := \frac{S^{(p,q)}(\mathbf{r})}{\gamma(\mathbf{r})}.$$

The VID transforms the injection–measurement coupling into point-wise impedance contribution expressed in units of $\Omega/\mathrm{m}^3$, which enables the computation of total impedance as the volume integral

$$Z^{(p,q)} := \int_{\Omega} \mathrm{d}Z^{(p,q)} \equiv \int_{\mathbf{r}\in\Omega} \psi^{(p,q)}(\mathbf{r})\mathrm{d}V, \tag{16}$$

where $\Omega \in \mathbb{R}^3$ is the simulation domain with $\gamma \neq 0$, and $\mathrm{d}Z^{(p,q)} := \psi^{(p,q)}(\mathbf{r})\mathrm{d}V$ is the impedance differential. This integral representation of the impedance is particularly powerful, as it can be applied to arbitrary subregions $\Omega_t \subset \Omega$, allowing us to quantify impedance contributions from individual tissues or anatomical structures.

For computational implementation, we discretized the continuous scalar fields $\gamma$, $S^{(p,q)}$, and $\psi^{(p,q)}$ on a three-dimensional grid with dimensions $M$, $N$ and $K$ along the $x$, $y$ and $z$ axes, respectively; yielding the tensor representations $\mathbf{\Gamma}$, $\mathbf{S}^{(p,q)}$, $\mathbf{\Psi}^{(p,q)} \in \mathbb{C}^{M\times N\times K}$. Similarly, the impedance differential $\mathrm{d}Z^{(p,q)} : \mathbb{R}^3 \to \mathbb{C}$ becomes the tensor $\mathbf{Z}^{(p,q)} \in \mathbb{C}^{M\times N\times K}$, containing the impedance contribution of all voxels. With this discretization, equation (16) becomes a summation over voxels:

$$Z^{(p,q)} := \sum_k \mathbf{Z}_k^{(p,q)} \equiv \sum_k \mathbf{\Psi}_k^{(p,q)}|\omega_k|,$$

where $\omega_k \subset \Omega$ is a voxel, $|\omega_k|$ denotes its volume, and $\mathbf{Z}_k^{(p,q)} := \mathbf{\Psi}_k^{(p,q)}|\omega_k|$ denotes its impedance content.

To determine tissue contribution, we first extracted the subregions $\Omega_t \subset \Omega$ associated with the tissue of interest and computed the tissue-specific impedance $Z_t^{(p,q)}$. We then computed the tissue contribution to total resistance, total reactance, and total impedance as

$$\mathrm{PR}_t^{(p,q)} := \frac{\left|R_t^{(p,q)}\right|}{\sum_t \left|R_t^{(p,q)}\right|}, \quad \mathrm{PX}_t^{(p,q)} := \frac{\left|X_t^{(p,q)}\right|}{\sum_t \left|X_t^{(p,q)}\right|}, \quad \mathrm{PZ}_t^{(p,q)} := \frac{\left|R_t^{(p,q)}\right| + \left|X_t^{(p,q)}\right|}{\sum_t \left(\left|R_t^{(p,q)}\right| + \left|X_t^{(p,q)}\right|\right)},$$

respectively; where $R_t^{(p,q)} := \mathfrak{Re}(Z_t^{(p,q)})$ and $X_t^{(p,q)} := \mathfrak{Im}(Z_t^{(p,q)})$ are real and imaginary components of $Z_t^{(p,q)}$. Here, we used the absolute value of both components to ensure that all tissue-specific contributions are nonnegative and sum to 100% (Supplementary Fig. 17).

The discretization of $\mathrm{d}Z^{(p,q)}$ into voxel-specific impedance content $\mathbf{Z}^{(p,q)}$ also allows us to find the subregion in the finger that accounts for most of the total impedance content, e.g. $95\%$, which we used to quantify the penetration depth of the $(p,q)$-configuration. To this end, we first ranked the voxels based on their frequency content. We then computed the cumulative sum of the ranked voxels until reaching the $95\%$ of the total impedance (Supplementary Fig. 14–16).

## 6.6 Finite element analysis with synthetic ground-truth pulsatile blood electrical conductivity data

We further used the finger model to investigate the effects of electrode configurations on PVI reconstruction by generating synthetic PVI measurements from a reference blood conductivity waveform corresponding to a full cardiac cycle (Supplementary Fig. 18). We resampled the conductivity waveform, which was generated from a multiphysics model using real BP data,[125] to 100 points and used them to simulate blood conductivity variation at the arteries, while assigning nominal parameters for other tissues. A full measurement frame for any specific injection pattern would require the simulation to be repeated with the injection pair shifted to the next electrodes. With $100$ temporal data points, there are $16,000$ PVI simulations in total, with $3,200$ for 8 electrodes and $12,800$ for 16 electrodes. Here we only focused on skip-2 injection pattern for 8 electrodes ($800$ simulations) and 16 electrodes ($1,600$ simulations). We performed the simulations using the same parameters as in Supplementary Discussion 6.5.

We extracted the electrode voltages with the skip-1 measurement pattern, and passed the data through our PVI solver for image reconstruction. We perform difference reconstruction on a mesh with $N_N = 654$ nodes and $N_K = 1,354$ elements. The reconstructed data is a matrix $\mathbf{C} \in \mathbb{R}^{N_K \times T}$, with each row containing the conductivity value at each element, and each column corresponding to a time instance. We then detected the region of interest (ROI) corresponding to the arteries and extracted its conductivity waveform. To detect the ROI, we used an unsupervised algorithm based on the singular value decomposition (SVD). We first applied SVD to the reconstructed conductivity matrix and find its rank-1 approximation, i.e. $\mathbf{C} \approx s_1 \mathbf{a}_1 \mathbf{b}_1^{\mathsf{T}}$, where $s_1$ is the first singular value of $\mathbf{C}$, and $\mathbf{a}_1 \in \mathbb{R}^{N_K}$, $\mathbf{b}_1 \in \mathbb{R}^{T}$ are its corresponding left and right singular vector, respectively. We then computed the product $\mathbf{k} = \mathbf{C}\mathbf{b}$, effectively projecting the conductivity matrix onto the dimension of $\mathbf{b}$. We defined the ROI as the rows in $\mathbf{C}$, i.e., the element indices, that correspond to the largest elements in $\mathbf{k}$. In other words, the ROI is the set of elements that are most aligned with the right singular vector of $\mathbf{C}$.

# 7 Supplementary Discussion 7. Experimental study

## 7.1 Standard protocol approvals, registration, and informed consent

The study protocol adheres to ethical standards for experiments on humans and was approved by the Institutional Review Board at the University of Utah (#00162369). Potential subjects learned about the study through the University of Utah online study locator as well as through advertisement flyers and word of mouth. During registration, subjects answered questions about their cardiovascular health to evaluate their participation eligibility. Eligible subjects were then invited to the lab at the University of Utah, where they provided written informed consent before any measurements.

## 7.2 Study enrollment

Our enrollment process is shown in Supplementary Fig. 24a. To participate in the study, subjects must be: (1) at least 18 years old, (2) not pregnant, (3) not having an implanted electronic cardiac device, and (4) not diagnosed with cardiac arrhythmia, hypertension, or hypotension. In total, $N = 99$ subjects provided informed consent, of which $N = 96$ completed the study. Although we initially planned only a cross-sectional study, we also called back $N = 5$ subjects for an additional pilot longitudinal study to evaluate the frequency of recalibration required by the models when doing inference and accounting for intra-subject variability over time.

## 7.3 Study protocol

### 7.3.1 Cross-sectional study

Supplementary Fig. 24b shows the overall design of our cross-sectional study, which aims to assess the feasibility of using our ring to monitor BP of healthy subjects in controlled lab settings. The study involves a single 90-minute lab visit, during which physiological data was collected in three phases: static (45 minutes), Valsalva (25 minutes), and cold pressor (25 minutes). During the static phase, subjects were seated comfortably with back and arms supported. Data was then collected in multiple short trials, lasting between 3 minutes and 10 minutes. The trials were interleaved with 2-minute breaks during which subjects could adjust their posture to get comfortable again. The dynamic phases (Valsalva and cold pressor) were included in the protocol to induce dynamic changes in BP. Each phase included between 3 and 5 trials, interleaved with 2-minute breaks. During each trial, subjects were instructed to perform a dynamic maneuver after a period of baseline data. In the Valsalva trials, subjects inhaled deeply, pinched their nose with the right hand, closed their mouth, and attempted to exhale against the closed airways for 20 seconds. Subjects then released their hand and exhaled through their nose. During the cold pressor phase, subjects submerged their right hand in a bucket of ice water for 20 seconds.

### 7.3.2 Pilot longitudinal study

The pilot longitudinal study was designed to evaluate the efficacy of our BP estimation models on a small cohort of subjects after an extended period of time (Supplementary Fig. 24b). The study took place between 3 months and 12 months after the initial session and included five 40-minute sessions on consecutive days. In each session, data was collected in multiple short trials, each lasting between 8 minutes and 10 minutes and interleaved with 2-minute breaks. During each trial, participants was asked to perform Valsalva maneuver twice after at least 5 minutes of baseline measurement. Each Valsalva maneuver lasted 20 seconds.

## 7.4 Skin preparation

Before measurements, we used 70% isopropyl alcohol to sanitize the skin region at the PVI collection site. We then applied 0.9% saline solution to moisturize the skin.

## 7.5 Data collection

The cross-sectional study included 15 minutes of ultrasound imaging, followed by simultaneous recording of physiological data. We captured ultrasound images of the subjects' left index finger in B-mode and color Doppler mode using a SuperSonic Mach 30 system (Hologic, Marlborough, MA, USA) with a LH20-6 transducer (Supplementary Fig. 25 and Supplementary Video 6). We did not conduct ultrasound imaging for the longidudinal study sessions.

We collected ECG, PPG, BP, and PVI data simultaneously. We used the Nano Core system (Finapres Medical Systems, Enschede, Netherlands) to collect BP, with the finger cuff wrapped around the middle phalanx of the left middle finger and the actuator secured on the dorsum of the left wrist. We recorded ECG using a three-lead cable system (LifeSync Corporation, Coral Springs, FL, USA) and 3M Red Dot electrodes (3M Company, Maplewood, MN, USA). For data acquisition, we connected the Nano Core and the ECG cable to a NOVA Plus system (Finapres). We recorded PVI data at the left index finger by connecting the our ring to a Sciospec EIT32 system (Sciospec Scientific Instruments GmbH, Bennewitz, Germany) through micro coaxial cables with a characteristic impedance of 50 Ω and length of 1 m. We applied an alternating current at 50 kHz, sampled at 50 frames per second (fps). We used the skip-2 injection pattern and the skip-1 measurement pattern.

## 7.6 Data preprocessing

After data collection, we extracted the ECG and brachial BP signals from the NOVA Plus, and BioZ signals from the EIT32 system. Henceforth, we use brachial BP as the estimation target, and refer to it as BP. We developed a pipeline in MATLAB R2024a to prepare the experimental data for ML. Since the data was time series collected from two different sensor systems, they underwent separate processing steps, with the only shared components being the four main stages of the pipeline: (1) filtering, (2) alignment, (3) segmentation, and (4) data quality assessment, as shown in Supplementary Fig. 26.

### 7.6.1 Filtering

We filtered all experimental data using a third order Butterworth filter, with cutoff frequencies specific to each signal group. For ECG and BP signals, we applied a low-pass filter with 10 Hz cutoff. The raw PVI signals, which included BioZ recorded from multiple channels with the Sciospec system, were filtered with a low-pass filter with 5 Hz cutoff.

After the initial filtering round, the signals underwent intermediate steps to aid in signal alignment (Supplementary Discussion 7.6.2) and quality assessment (Supplementary Discussion 7.6.4). First, the BioZ signals were band-pass filter with cutoff frequencies at 0.9 Hz and 5 Hz to remove the baseline drift. To unify the BioZ signals, we averaged them across the channels to yield a single-channel time series, denoted by the subscript avg in $\text{BioZ}_{\text{avg}}$. We then resampled all signals to their nominal sampling rates using piecewise cubic interpolation. ECG and BP signals were resampled at 300 fps, while the $\text{BioZ}_{\text{avg}}$ signal was resampled at 50 fps. This step ensures consistent timestamps across the entire signals while preserving their original morphology, which facilitate accurate peak detection.

### 7.6.2 Alignment

The alignment stage involves detecting signal peaks, computing the inter-beat intervals (IBI), and minimizing the deviation between the IBI signals. The goal is to align the systolic peaks of BP and BioZ signals. However, we found that the frequent re-calibration of BP signal presented a challenge in direct alignment of the BP and BioZ signals. Instead, we aligned the $\text{BioZ}_{\text{avg}}$ systolic peaks with the ECG R peaks and then corrected for the delay between the R peaks and the BP systolic peaks. This approach leveraged the fact that ECG and BP signals were synchronously recorded with the same system.

We identified the peaks as the zero-crossing of the first derivative from positive to negative. We included the minimum peak distance (MPD) and minimum peak prominence (MPP) as two constraints to aid in peak detection. We defined peak distance as the interval between two adjacent peaks, and peak prominence as the peak's relative height within a window of 1.5 seconds. We required MPD $=$ 0.5 seconds for all signals, MPP $=$ 0.2 mV for ECG R peaks, and MPP $=$ 10 mm Hg for BP systolic peaks. For $\text{BioZ}_{\text{avg}}$ systolic peaks, we set the MPP $=$ 2·MAD, with MAD being the median absolute deviation of the band-passed signal. From the detected peaks, we calculated the IBI as the time difference between two adjacent peaks.

We implemented a brute force algorithm to offset the $\text{BioZ}_{\text{avg}}$ IBI along the fixed ECG IBI and compute the alignment cost $\mathcal{C}(k)$ at each shift index $k \in \mathbb{Z}$. We defined the cost as the weighted sum of least square errors in the IBI domain and the peak-time domain, both having the unit of time,

$$\mathcal{C}(k) = \alpha \cdot \underbrace{\sum_{i=0}^{P} \left(\text{IBI}_i^{(\text{ECG})} - \text{IBI}_i^{(\text{BioZ}_{(k)})}\right)^2}_{\text{IBI error}} + (1-\alpha) \cdot \underbrace{\sum_{i=0}^{P} \left(t_i^{(\text{ECG})} - t_i^{(\text{BioZ}_{(k)})}\right)^2}_{\text{peak-time error}},$$

where $\alpha \in [0,1]$ is a hyper-parameter weighting the two criteria, $P$ is the number of peaks, $i$ denotes the peak index, and the superscripts denote which signal the quantities $t$ and IBI were derived from. We used $\alpha = 0.7$ for our cost function, which we found to be more robust than a cost function with only the IBI error ($\alpha = 1$). We determined the optimal shift index $\hat{k}$ that minimizes $\mathcal{C}$ and computed the corresponding time offset $\hat{t}$. In the final alignment step, we added to $\hat{t}$ an amount equal to the pulse arrival time (PAT) to shift the $\text{BioZ}_{\text{avg}}$ systolic peaks to the BP systolic peaks. We defined the PAT as the delay between the ECG R peaks and the BP systolic peaks and computed the delay values for all ECG-BP pairs. We then used Gaussian kernels to estimate the delay distribution and determined the PAT as its mode.

### 7.6.3 Image reconstruction and temporal segmentation

After alignment, we segmented the BP signal at the its diastolic peaks. We identified these segmentation points with the same peak-detection method and parameters used on the aligned BP signal. The segmented BP periods were then resampled to a uniform length of $T = 50$ using cubic interpolation.

For the BioZ data, we first performed image reconstruction on the original BioZ signals (Supplementary Discussion 7.6.1), producing a sequence of $40 \times 40$ conductivity images. We then filtered the image sequences along the time dimension using a moving-average filter with a window of 100 consecutive frames. This step decomposed the reconstructed image sequences into a high-pass and low-pass channel, which were processed separately in our ML approach (Supplementary Discussion 8). Both image channels were then segmented along the time dimension using the BP diastolic peaks as segmentation points to ensure one-to-one association between BP data points and PVI image frames. Finally, we resampled all image segments to $T = 50$ frames per sample.

To compare ML performance using different input modality, we also segmented and stored the BioZ signals. The signals were decomposed, segmented, and resampled similar to the PVI images.

### 7.6.4 Signal quality assessment

All segmented periods were passed through an unsupervised algorithm to extract clean periods suitable for training neural networks. For BP periods, we defined SBP, DBP, pulse pressure (PP), heart rate (HR), total variation (TV), and median variation (MV) as

$$\begin{aligned}
\text{SBP}_n &:= \max_T (\mathbf{P}_n), \\
\text{DBP}_n &:= \min_T (\mathbf{P}_n), \\
\text{PP}_n &:= \text{SBP}_n - \text{DBP}_n, \\
\text{HR}_n &:= \frac{60}{\mathbf{t}_n[T] - \mathbf{t}_n[1]}, \\
\text{TV}_n &:= \sum_{k=2}^{T} |\mathbf{P}_n[k] - \mathbf{P}_n[k-1]|, \\
\text{MV}_n &:= \text{median}\left(|\mathbf{P}_n[k] - \mathbf{P}_n[k-1]|\right), \text{ for } 2 \le k \le T;
\end{aligned}$$

where $\mathbf{P}_n \in \mathbb{R}^T$ and $\mathbf{t}_n \in \mathbb{R}^T$ are the interpolated BP period and the interpolated time vector, respectively, $n$ is the period index; $k \in \{1, \ldots, T\}$ is the time index; and the functions $\max_T(\cdot)$ and $\min_T(\cdot)$ extract the extrema along the time dimension. We classified the periods as "clean" if they meet all of the following criteria: 40 mm Hg $\leq$ DBP $\leq$ 110 mm Hg, 90 mm Hg $\leq$ SBP $\leq$ 190 mm Hg, 10 mm Hg $\leq$ PP $\leq$ 100 mm Hg, 30 beats per minute (bpm) $\leq$ HR $\leq$ 200 bpm, 40 mm Hg $\leq$ TV, and 0.1 mm Hg $\leq$ MV. We found the TV and MV criteria effective in detecting anomalous BP values during NOVA Plus re-calibration phases.

We used the band-pass, channel-averaged $\text{BioZ}_{\text{avg}}$ as proxy to assess the quality of conductivity and BioZ data. This approach established a uniform quality metrics for both types of data while minimizing biases introduced by the reconstruction algorithm. We decomposed, segmented and resampled the $\text{BioZ}_{\text{avg}}$ data following the procedure in Supplementary Discussion 7.6.2. We then classified the samples in two stages. First, we computed the samples' Mahalanobis distance to the ensemble mean. All samples having distance beyond 2 standard deviations from the mean distance were excluded.[199] In the second stage, we classified the remaining samples based on systolic peak location. Samples were rejected if their systolic peaks occurred beyond the $15^{\text{th}}$ index, or if the delay between their systolic peaks and corresponding BP systolic peaks was beyond 0.1 seconds. Finally, we stacked the periods into homogeneous tensors and exported for long term storage.

# 8 Supplementary Discussion 8. Machine learning

## 8.1 Overview of our machine learning approach

Our goal is to estimate brachial BP from the ring BioZ and PVI data. We considered five model classes with increasing implementation complexity (Supplementary Fig. 29): linear regression, multilayer perceptron (MLP), convolutional neural network (CNN), hybrid architecture with convolutional and transformer layers,[200] and hybrid architecture with convolutional and Samba layers.[201] The linear regression model class serves as performance baseline for other classes. Supplementary Discussion 8.4 describes the model architecture in more details.

We further configured each model class according to different input and output modalities. As input, we contrasted BioZ signals and PVI images, enabling the assessment of image reconstruction benefits for BP estimation accuracy. As output, we compared between estimating only the fiducial points (SBP and DBP) against estimating a full BP period. The latter presents a more challenging task, requiring the accurate reconstruction of the signal morphology. These input and output combinations affect the number of parameters in our first and last layer, respectively. Therefore, we treated them as separate model configurations, with $20$ distinct configurations in total ($5$ model types $\times$ $2$ input types $\times$ $2$ output types).

For each configuration, we categorized the models based on which datasets were used to train them. Subject-specific (SS) models used datasets from individual subjects, while population models were trained on an aggregated dataset from multiple subjects. For population models, we further distinguished between two variants, based on how the aggregated dataset was partitioned during training: population-within (PW) models, and population-disjoint (PD) models. For PW models, all constituent datasets were partitioned with a consistent train:test ratio, while for (PD) models, whole subjects were assigned either in the train or test set without resampling. Supplementary Discussion 8.5 describes the partition approaches are described in more details.

We implemented and trained all models in Python (version 3.12.11) coupled with the PyTorch package (version 2.8.0).[202] We also used the following packages for support functionalities: numpy (version 2.2.6),[203] scipy (version 1.16.1),[204] scikit-learn (version 1.6.1),[205] pandas (version 2.2.3),[206] and h5py (version 3.15.1).[207,208] After training, we evaluated the models based on their performance on their respective test sets.

## 8.2 Data leakage

In ML, data leakage is defined as the incorporation of information about the test set into the training process, thereby producing overly optimistic learning outcomes. In the literal sense, data leakage occurs when test samples appear in the train set which directly influences gradient descent. This interpretation serves as the current definition of data leakage in the field of cuffless BP monitoring.[209] However, it is also important to note that the determination of whether a particular data partitioning strategy constitutes leakage depends not solely on the technical implementation of the train:test split, but fundamentally on the claims made about model generalizability and the target distribution to which the results are intended to apply.[210]

For example, a partition strategy that would constitute leakage when making claims about universal, cross-subject generalization may be entirely appropriate when the research objective is comparative evaluation of modeling approaches within a defined population.

In this work, our primary comparative analyses employed the PW partition strategy, where each subject contributes data to both the train and test set with a consistent train:test ratio. Based on the current definition, our approach does not constitute data leakage given the following considerations. Our objective is to investigate whether multi-channel BioZ data and conductivity image data collected with our ring can estimate BP accurately; and how performance depends on different input and output modalities as well as model architecture. Furthermore, we do not aim to establish our models as universal BP estimation models that can generalize to unseen subjects or new data of already seen subjects. To address generalizability explicitly, we conducted three separate analyses. First, we evaluated PW-trained model weights on the holdout datasets that were not seen during training (Supplementary Discussion 9.5.10), thereby quantifying the performance of our models in out-of-distribution scenarios. Second, we trained the same model configurations from scratch using the PD partition strategy, where entire subjects are contained wholly within either the train or test set, and report these results separately (Supplementary Discussion 9.5.11). Finally, we investigated the longitudinal stability of selected SS model configurations by evaluating them on new data from the same subjects collected at future time points (Supplementary Discussion 9.5.13). Together, these evaluations provide a complete and transparent picture of the capabilities and limitations or our approach.

## 8.3 Data sources

Our processing pipeline (Supplementary Discussion 7.6) built $96$ unique datasets from the available subject pool. Each dataset contains 5 data tensors: $\mathbf{P} \in \mathbb{R}^{N_P \times T}$, $\mathbf{S}^{(\text{HP})} \in \mathbb{R}^{N_P \times H \times W \times T}$, $\mathbf{S}^{(\text{LP})} \in \mathbb{R}^{N_P \times H \times W \times T}$, $\mathbf{Z}^{(\text{HP})} \in \mathbb{R}^{N_P \times 2C \times T}$, $\mathbf{Z}^{(\text{LP})} \in \mathbb{R}^{N_P \times 2C \times T}$, which represents brachial BP, high-pass conductivity images, low-pass conductivity images, high-pass BioZ, and low-pass BioZ, respectively. Here, $N_P$ is the total number of raw periods from each subject and can vary between subjects, from as low as $810$ to as high as $6,081$; $T = 50$ is the resampled length of each period; $H = W = 40$ are the width and height of the conductivity images; and $C = 32$ is the number of independent BioZ channels, which was derived from the skip-2 injection pattern and skip-1 measurement pattern used for PVI data collection (Supplementary Discussion 7.5). The BioZ tensors contain both real and imaginary parts, which are stacked together and treated as independent data channels in our models. In addition to the data tensors, each dataset also includes an auxiliary feature tensor $\mathbf{F} \in \mathbb{R}^{N_P \times 2}$ containing the period durations and peak time of all BioZ periods; and a binary mask $\mathbf{B} \in \{0, 1\}^{N_P}$ indicating the clean periods suitable for training.

We used the masking vector $\mathbf{B}$ to identify sequences of $m \in \mathbb{N}_{\geq 1}$ consecutive clean periods. This allows our models to capture long-term trends in the input data $\mathbf{S}$ and $\mathbf{Z}$. It is worth noting that using $m > 1$ reduces the number of available samples while increasing the number of parameters at the input layer. Furthermore, sequences with $m > 1$ periods can overlap and thus requires additional steps when constructing train and test subset to avoid leakage. In this study, we used $m = 5$ in all cases.

## 8.4 Model description

All model classes share a unified interface for processing input data and generating output values, which we outline here before describing their architectural differences. As input to our models, we used either the PVI image sequences ($\mathbf{S}^{\text{(LP)}}$ and $\mathbf{S}^{\text{(HP)}}$) or the multi-channel BioZ signals ($\mathbf{Z}^{\text{(LP)}}$ and $\mathbf{Z}^{\text{(HP)}}$). To process the data, we first stacked $m$ consecutive periods along the time dimension for all samples. We then augmented the data by including its first and second temporal derivative, effectively tripling the number of features. To avoid amplifying high-frequency noise, we only used the derivatives of the low-passed components ($\mathbf{S}^{\text{(LP)}}$ and $\mathbf{Z}^{\text{(LP)}}$). We then stacked the original data and its derivatives, producing the following tensors:

$$\mathbf{X}_n^{\text{(3D)}} := \begin{bmatrix} \mathbf{S}^{\text{(HP)}}_{[(n-m+1):n]} \\ \dot{\mathbf{S}}^{\text{(LP)}}_{[(n-m+1):n]} \\ \ddot{\mathbf{S}}^{\text{(LP)}}_{[(n-m+1):n]} \end{bmatrix} \in \mathbb{R}^{H\times W\times 3\times mT}, \tag{17a}$$

$$\mathbf{X}_n^{\text{(1D)}} := \begin{bmatrix} \mathbf{Z}^{\text{(HP)}}_{[(n-m+1):n]} \\ \dot{\mathbf{Z}}^{\text{(LP)}}_{[(n-m+1):n]} \\ \ddot{\mathbf{Z}}^{\text{(LP)}}_{[(n-m+1):n]} \end{bmatrix} \in \mathbb{R}^{6C\times mT}; \tag{17b}$$

where $n \in \{1, \ldots, N_P\}$ is the period index being processed, and the subscript $[a : b]$ denotes the combined operation of slicing the tensors from period $a$ to period $b$ and stacking along the time dimension. Besides the temporal data, we also included the period duration and peak time as auxiliary features $\mathbf{f}_n := \mathbf{F}_{[(n-m+1):n]} \in \mathbb{R}^{2\times m}$, where $\mathbf{F}$ is the original feature tensor (Supplementary Discussion 8.3). To summarize, the complete forms of the input data are

$$\tilde{\mathbf{X}}_n^{\text{(3D)}} := \mathbf{X}_n^{\text{(3D)}} \oplus \mathbf{f}_n, \tag{18a}$$

$$\tilde{\mathbf{X}}_n^{\text{(1D)}} := \mathbf{X}_n^{\text{(1D)}} \oplus \mathbf{f}_n; \tag{18b}$$

with $\oplus$ denoting the concatenation of tensors with different processing paths through the intermediate layers. With $C = 32$, $H = W = 40$, $T = 50$ and $m = 5$, the total number of input features per sample is $1,200,010$ for $\tilde{\mathbf{X}}_n^{\text{(3D)}}$ and $48,010$ for $\tilde{\mathbf{X}}_n^{\text{(1D)}}$.

As output, our models estimated either the full BP waveform $\hat{\mathbf{Y}}_n^{\text{(wave)}} \in \mathbb{R}^T$, or only the fiducial points (SBP and DBP) $\hat{\mathbf{Y}}_n^{\text{(fp)}} \in \mathbb{R}^2$. The corresponding ground truth is defined as

$$\mathbf{Y}_n^{\text{(wave)}} := \mathbf{P}_n \in \mathbb{R}^T, \tag{19a}$$

$$\mathbf{Y}_n^{\text{(fp)}} := [\min(\mathbf{P}_n), \max(\mathbf{P}_n)] \in \mathbb{R}^2. \tag{19b}$$

Without loss of generality in the subsequent discussions, we use $\hat{\mathbf{Y}}_n \in \mathbb{R}^{T_{\text{out}}}$ to denote either output type, with $T_{\text{out}} = T = 50$ for waveform output and $T_{\text{out}} = 2$ for fiducial output. Similarly, we denote the ground truth of either type as $\mathbf{Y}_n \in \mathbb{R}^{T_{\text{out}}}$. For input, we use the tensor $\mathbf{X}_n \in \mathbb{R}^{H_{\text{in}}\times W_{\text{in}}\times C_{\text{in}}\times T_{\text{in}}}$ to represent both input types, with $H_{\text{in}}$, $W_{\text{in}}$, $C_{\text{in}}$, and $T_{\text{in}}$ denoting the generic dimensions for height, width, channel size, and sequence length, respectively. Both input types have the same sequence length $T_{\text{in}} = mT = 250$ and differ in their spatial and channel dimensions, with $H_{\text{in}} = H = 40$, $W_{\text{in}} = W = 40$, and $C_{\text{in}} = 3$ for image tensors, while $H_{\text{in}} = 1$, $W_{\text{in}} = 1$, and $C_{\text{in}} = 6C = 192$ for impedance tensors. Similarly, we use $\tilde{\mathbf{X}}_n$ to denote the augmented input data. In instances where $\tilde{\mathbf{X}}$, $\mathbf{X}$, $\hat{\mathbf{Y}}$ and $\mathbf{Y}$ appear without the subscript, they represent the tensors containing all samples. Finally, we define the mapping $f_{\boldsymbol{\theta}} : \tilde{\mathbf{X}} \mapsto \hat{\mathbf{Y}}$ to represent any of the following model configurations: images to waveform ($\tilde{\mathbf{X}}^{(\text{3D})} \mapsto \hat{\mathbf{Y}}^{(\text{wave})}$), images to fiducial values ($\tilde{\mathbf{X}}^{(\text{3D})} \mapsto \hat{\mathbf{Y}}^{(\text{fp})}$), BioZ to waveform ($\tilde{\mathbf{X}}^{(\text{1D})} \mapsto \hat{\mathbf{Y}}^{(\text{wave})}$), and BioZ to fiducial values ($\tilde{\mathbf{X}}^{(\text{1D})} \mapsto \hat{\mathbf{Y}}^{(\text{fp})}$); with $\boldsymbol{\theta}$ in $f_{\boldsymbol{\theta}}$ denoting the model weights. We now describe the concrete implementation of our models.

### 8.4.1 Linear regression networks

To establish a performance baseline, we first performed multivariate linear regression (LR) to estimate BP directly from multi-dimensional input data $\hat{\mathbf{X}}$. Conventional LR requires a full data ensemble to compute the optimal loss gradient in a single step, which was infeasible in for study due to dataset size and sample dimensionality. Instead, we approximated true LR solution through a class of shallow neural networks (NN). This class first flattened the spatial-temporal tensors $\mathbf{X}_n$ to a 1D vector with dimension $D_{\text{in}} = H_{\text{in}} W_{\text{in}} C_{\text{in}} T_{\text{in}}$, which was then concatenated with the auxiliary features $\mathbf{f}_n$. The augmented input was then projected directly onto the output space through a single fully-connected (FC) layer: $(D_{\text{in}} + 2m) \to T_{\text{out}}$, without nonlinear activation, as shown in Supplementary Fig. 29a.

### 8.4.2 Multilayer perceptron networks

Our next class of NN introduces nonlinearity to the baseline LR model through hidden layers with nonlinear activation units. We used a multilayer perceptron (MLP) with five FC hidden layers, each with $100$ neurons followed by rectified linear units (ReLU), as shown in Supplementary Fig. 29b. Similar to the LR model class, MLP models still operate on flattened tensors and assume independence among features.

### 8.4.3 Convolutional networks

To capture inherent spatial and temporal patterns of the data, we implemented the Convolutional Neural Network (CNN) class consisting of a convolutional signal encoder, followed by a MLP. Supplementary Fig. 29c shows an example of the CNN model class with image inputs. The encoder stacks two convolution blocks sequentially, each having the standard architecture of convolutional filter followed by minibatch normalization, and ReLU activation. The shape of the convolutional filter adapts to the shape of the input tensors. For BioZ-based models, we used one-dimensional (1D) filters with kernel size $k_T = 7$ and $k_T = 5$ in the first and second encoder block, respectively; while for image-based models, we used three-dimensional (3D) filters with kernel size ($k_H \times k_W \times k_T$) of $3 \times 3 \times 7$ and $3 \times 3 \times 5$. Another addition to the image-based models is the maxpool filter with kernel size ($k_H \times k_W$) of $2 \times 2$ for spatial compression after each encoder block.

A common design choice for all models of this class is the twofold channel expansion through each encoder. In other words, the number of independent channels is $192 \to 384 \to 768$ for BioZ-based models and $3 \to 6 \to 12$ for image-based models. Essentially, the convolutional encoder maps the input to a feature-rich latent space while compressing the spatial dimensions. We denote this mapping as $g : \mathbb{R}^{H_{\text{in}} \times W_{\text{in}} \times C_{\text{in}} \times T_{\text{in}}} \to \mathbb{R}^{H_\ell \times W_\ell \times C_\ell \times T_\ell}$, where $(H_\ell, W_\ell, C_\ell, T_\ell)$ are the dimensions of the latent space, respectively. These dimensions are $H_\ell = H_{\text{in}}/4$, $W_\ell = W_{\text{in}}/4$, and $C_\ell = C_{\text{in}} \cdot 4$ for image-based models, while $H_\ell = H_{\text{in}} = 1$, $W_\ell = W_{\text{in}} = 1$, and $C_\ell = C_{\text{in}} \cdot 4$ for impedance-based models. In both cases, we included padding along the temporal dimension to preserve the sequence length ($T_\ell = T_{\text{in}}$).

The output from the convolutional encoder was flattened and concatenated with the auxiliary features $\mathbf{f}_n$, producing a vector $\mathbf{r} \in \mathbb{R}^{(D_\ell+2m)}$, where $D_\ell = H_\ell W_\ell H_\ell T_\ell$ is the number of latent features. The augmented data was then fed through a MLP consisting of two FC layers with ReLU activation and gradual dimensional reduction $((D_\ell + 2m) \to 512 \to 256)$. Finally, the MLP output was passed through a linear layer $(256 \to T_{\text{out}})$ to produce the estimated BP vector $\hat{\mathbf{Y}}_n$.

While CNN models leverage the spatiotemporal structure of the data, their convolutional filters struggle at capturing long-range temporal dependencies. To address this limitation, our next two classes of NN complement the convolutional layers with sequence modeling layers.

### 8.4.4 Hybrid networks with transformer

We implemented a three-stage hybrid model class: Convolutional Recurrent Transformer (CRT), as shown in Supplementary Fig. 29d. The first stage employed the same convolutional encoder as in our CNN models, which represented local spatiotemporal patterns in the latent space $\mathbb{R}^{H_\ell \times W_\ell \times C_\ell \times T_\ell}$. However, unlike our CNN models which flattened all dimensions, the CRT models only flattened the channel and spatial dimensions, producing a tensor with shape $(H_\ell W_\ell C_\ell) \times T_\ell$. This preserved the temporal structure required for sequence modeling in subsequent stages.

The second stage is a residual recurrent positional encoder (RRPE) for injecting information about temporal ordering, which is necessary since transformers are permutation invariant. The RRPE begins with a FC layer projects the flattened features from dimension $D_\ell = H_\ell W_\ell C_\ell$ to a $100$-dimensional embedding subspace, producing a tensor $\mathbf{T} \in \mathbb{R}^{100 \times T_\ell}$. This projected sequence is then processed by two bi-directional long short-term memory (BiLSTM) layers, each with $64$ hidden states capturing temporal dependencies in both forward and backward directions. The BiLSTM output $\mathbf{R} \in \mathbb{R}^{128 \times T_\ell}$ was then fed through a MLP with two FC layers: the first with ReLU activation and channel expansion $(128 \to 256)$, and the second projecting back to the embedding subspace $(256 \to 100)$. Here, a residual connection was included to combine the MLP output with the original projected features $\mathbf{T}$.

The third stage consists of a standard multi-head transformer encoder that models long-range dependencies through self-attention. We used $2$ encoder layers, each containing self-attention with $4$ heads followed by an MLP $(100 \to 512 \to 100)$ with ReLU activation. The transformer output was flattened and concatenated with the auxiliary features $\mathbf{f}_n$, producing the vector $\mathbf{r} \in \mathbb{R}^{(100T_\ell+2m)}$. The augmented output was fed through a two-layer MLP with ReLU activation, compressing the dimension $((100T_\ell + 2m) \to 256 \to 128)$. Finally, the MLP output was passed through a linear layer $(128 \to T_{\text{out}})$ to produce the estimated BP vector $\hat{\mathbf{Y}}_n$.

### 8.4.5 Hybrid networks with state-space models

As an alternative to standard transformer-based network, we implemented the Convolutional Recurrent Samba (CRS) model class, which replaces the conventional transformer blocks with Samba blocks[201,211] (Supplementary Fig. 29e). This model class shares the first two stages with CRT models: a convolutional encoder followed by RRPE for positional embedding, producing a tensor with dimension $100 \times T_\ell$.

However, the third stage differs fundamentally from the CRT class by employing Samba blocks,[201] each consisting four components stacked consecutively: (1) two Mamba layers with state-space modeling,[211,212] (2) a two-stage MLP, (3) multi-head sliding window attention (SWA) for memory retrieval,[213] and (4) a second MLP with the same architecture for additional feature sharing. We used two Samba blocks stacked sequentially, with each block maintaining the embedding dimension $100$ throughout. We used SWA with $4$ heads, each with a window size of $64$ to capture local temporal patterns. Each of the interleaving MLP block contains a FC layer $(100 \to 400)$ with sigmoid linear unit (SiLU) activation,[214] followed by another FC layer $(400 \to 100)$ mapping the data back to the embedding subspace $\mathbb{R}^{100\times T_\ell}$. The Samba output was flattened and concatenated with the auxiliary features $\mathbf{f}_n$ to yield the vector $\mathbf{r} \in \mathbb{R}^{(100T_\ell+2m)}$, which was then fed through a linear layer $(100T_\ell + 2m) \to T_{\text{out}}$, producing the estimated BP vector $\hat{\mathbf{Y}}_n$. Unlike the MLP used at the end of the CRT models, here we used a single linear mapping, since the interleaved MLP layers within Samba blocks already provide sufficient feature abstraction.

## 8.5 Dataset partition

### 8.5.1 Partition for subject-specific models

From the original pool of $96$ subjects, we randomly selected $5$ subjects (004, 018, 036, 043, and 053) exclusively for generalizability validation, representing $5.21\%$ of the datasets. We refer to their datasets as holdout (or validation) datasets. For each of the remaining $91$ datasets, we implemented a graph procedure to split the samples into train and test subsets while avoiding overlaps. Formally, we represented the train:test partition as a bipartite graph $\mathcal{G}(\mathcal{U}, \mathcal{V}, \mathcal{E})$, where $\mathcal{U}$ and $\mathcal{V}$ are the sets of nodes representing the train and test samples, respectively; and $\mathcal{E}$ is the set of edges representing overlaps between train and test samples. The node degree $d : (\mathcal{U} \cup \mathcal{V}) \to \{0, 1, \ldots, 2m-2\}$, i.e. number of incident edges, corresponds to the number of samples from the other set overlapping with it, where $m$ is the number of consecutive periods.

We defined $\mathcal{C} := \{s \in (\mathcal{U} \cup \mathcal{V}) \mid d(s) \geq 1\}$ as the set of samples with overlaps. Eliminating sample overlaps is equivalent to finding a subset $\tilde{\mathcal{C}} \subseteq \mathcal{C}$ such that once $\tilde{\mathcal{C}}$ is removed, the remaining sets $\tilde{\mathcal{U}} := \mathcal{U} \setminus \tilde{\mathcal{C}}$ and $\tilde{\mathcal{V}} := \mathcal{V} \setminus \tilde{\mathcal{C}}$ are completely disconnected. There are multiple solutions, including the trivial one in which $\tilde{\mathcal{C}} \equiv \mathcal{C}$, i.e. all samples with overlaps are removed at once. However, this would significantly reduce the number of samples and, in the extreme case where $\mathcal{C} \equiv (\mathcal{U} \cup \mathcal{V})$, would leave no remaining samples. Instead, we seek the optimal solution $\hat{\mathcal{C}}$ that satisfies the following two conditions. First, to minimize data loss, $\hat{\mathcal{C}}$ must contain the least number of samples. Second, to achieve the desired train:test ratio $r$, the remaining sets must satisfy $|\hat{\mathcal{U}}| / |\hat{\mathcal{V}}| \approx r$, where $|\cdot|$ denotes the set's cardinality, i.e. the number of samples it contains. We implemented the following iterative algorithm to approximate the optimal solution.

1. At initialization, we shuffled all samples in the original dataset $\mathcal{D}$ and split them into preliminary train and test sets $\mathcal{U}$ and $\mathcal{V}$, with the desired train:test ratio of $r := |\mathcal{U}|/|\mathcal{V}| = 9$. This step ensured that the samples in both sets were randomly distributed across the entire duration of the experiment. From the initial partition, we computed the overlaps $\mathcal{E}$ and the corresponding node degrees.

2. To construct the optimal set $\hat{\mathcal{C}}$, we targeted the samples with the most overlaps and accumulated them one after another. In other words, we constructed the set $\mathcal{C}_k := \mathcal{C}_{k-1} \cup \{c_k\}$ at each iteration $k \geq 0$, where $\mathcal{C}_0 := \varnothing$, and $c_k$ was the node with the highest degree, chosen from either $\mathcal{U}$ or $\mathcal{V}$. To decide which set to choose from, we computed the train:test ratio $r_k := |\mathcal{U}_k|/|\mathcal{V}_k|$, where $\mathcal{U}_k := \mathcal{U} \setminus \mathcal{C}_k$ and $\mathcal{V}_k := \mathcal{V} \setminus \mathcal{C}_k$. We selected $c_k \in \mathcal{U}$ if $r_k \geq r$, and $c_k \in \mathcal{V}$ otherwise. Finally, We recalculated the overlaps $\mathcal{E}_k$ of the remaining sets.

3. We repeated step 2 until the remaining sets had no overlaps, or equivalently, $\mathcal{E}_k = \varnothing$. The resulting sets satisfied the two optimality conditions and were used as train and test set in our ML models. The pseudocode for our approach is given in Algorithm 1 and visualized in Supplementary Fig. 30a.

**Algorithm 1** Dataset partition without train-test overlaps

**Input:** $\mathcal{D}, r$     */* $\mathcal{D}$ is the complete dataset */*

**Output:** Optimal partition $(\hat{\mathcal{U}}, \hat{\mathcal{V}})$

$(\mathcal{U}_k, \mathcal{V}_k) \leftarrow \texttt{split}(\mathcal{D}, r)$     */* Naive partition */*

$\mathcal{E}_k \leftarrow \texttt{compute_overlaps}(\mathcal{U}_k, \mathcal{V}_k)$     */* Also compute node degrees */*

**while** $\mathcal{E}_k \neq \varnothing$ **do**

    $r_k \leftarrow |\mathcal{U}_k|/|\mathcal{V}_k|$

    **if** $r_k \geq r$ **then**

        $c_k \leftarrow \{s \mid d(s) \geq d(u)\ \forall u \in \mathcal{U}_k\}$

        $\mathcal{U}_k \leftarrow \mathcal{U}_k \setminus \{c_k\}$

    **else**

        $c_k \leftarrow \{s \mid d(s) \geq d(v)\ \forall v \in \mathcal{V}_k\}$

        $\mathcal{V}_k \leftarrow \mathcal{V}_k \setminus \{c_k\}$

    **end if**

    $\mathcal{E}_k \leftarrow \texttt{compute_overlaps}(\mathcal{U}_k, \mathcal{V}_k)$

**end while**

$\hat{\mathcal{U}} \leftarrow \mathcal{U}_k$

$\hat{\mathcal{V}} \leftarrow \mathcal{V}_k$

**return** $(\hat{\mathcal{U}}, \hat{\mathcal{V}})$

### 8.5.2 Partition for population-within models

When training PW models, we partitioned each subject-specific dataset following the procedure in Supplementary Discussion 8.5.1 and aggregated the individual subsets. This ensures the proportional contribution of all subjects to the population train and test set.

### 8.5.3 Partition for population-disjoint models

For PD models, we implemented another partition method with whole subjects being included either in the train or test set. To achieve the desired train:test ratio, we started with an empty test set and added a single subject, selected at random and without replacement from the remaining subject pool. We stopped adding new subjects when the actual train:set ratio exceeded the desired ratio.

## 8.6 Minibatch sampling

We trained the models by distributing the samples into minibatches, each containing $32$ samples. To avoid learning bias due to sample order, we shuffled the samples at the beginning of each training epoch. The optimal shuffling strategy depends on the type of dataset. For SS datasets, we applied a global approach where the samples were randomly shuffled before minibatch formation. For our PW and PD dataset, however, this approach proved computationally prohibitive due to its random access pattern. Specifically, with samples distributed across $91$ source files totaling approximately $400$ GB, the random access patterns would require loading and unloading different files multiple times, incurring substantial data transfer overhead that dominated training time. On the other hand, sequential sampling involves processing all samples from one subject at a time, which eliminates transfer overhead but introduces systematic bias due to minibatch homogeneity. We therefore required a sampling strategy that balances two competing objectives: source locality to minimize data transfer, and sample diversity to prevent subject-specific bias within minibatches. To meet these constraints, we implemented a cache system with a tunable capacity, and designed a stratified randomization approach that encourages cache locality.

Formally, we defined $\mathcal{S}$ as the population dataset with $|\mathcal{S}|$ total samples. The samples were aggregated from $N_D$ subject-specific sets $\mathcal{D}_j$ and organized into $N_B$ minibatches $\mathcal{B}_k$, with $j \in \{1, \ldots, N_D\}$ and $k \in \{1, \ldots, N_B\}$ denoting the subject index and minibatch index, respectively. The sampling problem reduces to finding a permutation of the samples in $\mathcal{S}$ before assigning them to minibatches. In the global randomization approach, the samples were shuffled without considering their source datasets. In contrast, our stratified approach included an additional shuffle at the subject level. The procedure is as follows:

1. We first grouped the samples based on their source datasets and maintained the natural order of the constituent datasets. We denoted this ordering as $\mathcal{S} = \bigcup_j \mathcal{D}_j$.
2. Next, we shuffled the source datasets, yielding $\mathcal{S}^{(\pi)} = \bigcup_j \mathcal{D}_{\pi(j)}$, where $\pi : \mathbb{N} \to \mathbb{N}$ is a bijective function representing the random permutation and $\pi(j)$ denotes the permuted index $j$. This step determines the accessing order of the source datasets.
3. We then combined the consecutive datasets $\mathcal{D}_{\pi(j)}$ into groups, representing the train set

as

$$\mathcal{S}^{(\pi)} = \bigcup_i \underbrace{\left(\bigcup_j \mathcal{D}_{\pi(j)}\right)}_{\mathcal{G}_i}, \text{ for } (i-1)\cdot G < j \leq i \cdot G,$$

where $\mathcal{G}_i$ denotes the group with index $i \geq 1$, and $G \in \{1, 2, \ldots, N_D\}$ is the number of datasets included in each group.

4. For each group, we concatenated the samples from all constituent datasets to yield $\mathcal{G}_i = \{s_{i,q} \mid 1 < q \leq |\mathcal{G}_i|\}$, with $s_{i,q}$ denoting the sample with relative index $q$ from group $i$, and $|\mathcal{G}_i|$ being the total number of samples. We then shuffled the samples, yielding $\mathcal{G}_i^{(\omega)} = \{s_{i,\omega(q)} \mid s_{i,q} \in \mathcal{G}_i\}$, where $\omega$ is a random permutation function, and $\omega(n)$ denotes the permuted index. This step created local randomization, ensuring that samples from different sources were interleaved within each group.
5. Finally, we concatenated the samples from all permuted groups into $\mathcal{S}^{(\pi,\omega)} = \bigcup_i \mathcal{G}_i^{(\omega)} = \{s_{\omega(\pi(n))} \mid s_n \in \mathcal{S}\}$ and assigned the samples into minibatches as

$$\mathcal{B}_k := \{s_{\omega(\pi(n))} \mid (k-1)\cdot B < n \leq k \cdot B\},$$

where $B$ is the minibatch size. Each of the resulting minibatch contained samples from approximately $G$ independent datasets (or maximum $B$ independent subjects, if $B < G$). The minibatches were then processed sequentially during training.

The hyperparameter $G$ controls the tradeoff between minibatch diversity and cache capacity, with $G = 1$ being equivalent to sequential sampling (most restrictive) and $G = N_D$ being equivalent to global randomization (most diverse). In this study, we used $G = B + 2$ and repeated steps 1–5 at every epoch, further reducing bias introduced by any specific subject combination. Our strategy, illustrated in Supplementary Fig. 30b, guaranteed that at any time there were at most $G$ independent subject-specific datasets loaded into memory. Furthermore, each source dataset was loaded exactly once and referenced in subsequent minibatches, and then unloaded from memory once all of its samples were processed.

## 8.7 Loss function

We aim to reconstruct BP waveforms with minimal point-wise error and minimal shape distortion. To achieve both objectives, we developed a loss function combining mean square error (MSE) and cosine distance (CD), i.e.

$$\mathcal{L}_n(\boldsymbol{\theta}) := \alpha \cdot \underbrace{\|\mathbf{Y}_n - f_{\boldsymbol{\theta}}(\tilde{\mathbf{X}}_n)\|^2}_{\text{MSE}} + (1-\alpha)\cdot \underbrace{\left(1 - \frac{\left\langle \mathbf{Y}_n, f_{\boldsymbol{\theta}}(\tilde{\mathbf{X}}_n)\right\rangle}{\|\mathbf{Y}_n\|\,\|f_{\boldsymbol{\theta}}\tilde{\mathbf{X}}_n)\|}\right)}_{\text{CD}}, \tag{20}$$

with $\|\cdot\|$ and $\langle\cdot,\cdot\rangle$ denoting the Euclidean norm and the scalar product along the temporal dimension, respectively, and $\alpha \in [0, 1]$ controlling the relative emphasis between both objectives. Standard MSE loss penalizes magnitude deviations but can produce jagged waveform. On the other hand, the scale-invariant CD loss, which is defined through the cosine function

$$\cos(\mathbf{Y}_n, f_{\boldsymbol{\theta}}(\tilde{\mathbf{X}}_n)) := \frac{\left\langle \mathbf{Y}_n, f_{\boldsymbol{\theta}}(\tilde{\mathbf{X}}_n)\right\rangle}{\|\mathbf{Y}_n\|\,\|f_{\boldsymbol{\theta}}(\tilde{\mathbf{X}}_n)\|}, \tag{21}$$

penalizes morphological dissimilarity between target and estimated waveforms. It is worth noting that the CD term has minimal impact on fiducial estimation models, since the two-dimensional output (DBP and SBP) lacks the sequential structure to capture. However, for waveform models, this loss term enforces the preservation of clinically relevant features like dicrotic notches and diastolic decay patterns.

We also note that the two loss terms operate at different scales: CD is bounded between 0 and 2, while MSE spans orders of magnitude larger. This scale mismatch requires careful tuning of $\alpha$ to ensure meaningful contribution from both terms. For large $\alpha$, the MSE would dominate entirely, rendering the CD term ineffective. Conversely, setting $\alpha$ too small risks overemphasizing shape alignment at the expense of magnitude accuracy, especially in later epochs when MSE has decreased substantially. In this study, we set $\alpha = 0.2$, which we found to balance both objectives across the full training trajectory. For LR models, however, we used $\alpha = 1.0$ to reflect conventional LR approach.

## 8.8 Training method

### 8.8.1 Cross-sectional study

We used batch size $B = 32$ to train all SS models. For PW and PD models, we used the mini-batch sampling strategy with batch size $B = 64$ (Supplementary Discussion 8.6). We updated the loss gradient using using the AdamW optimizer, with the initial learning rate $\gamma_0 = 5 \cdot 10^{-4}$ and PyTorch's default values for other hyperparameters.[215] To evaluate model performance during training, we defined an accuracy metric $\varrho$ as

$$\varrho(\mathbf{Y}, \hat{\mathbf{Y}}) := 0.5 \cdot r(\max_T \mathbf{Y}, \max_T \hat{\mathbf{Y}}) + 0.5 \cdot r(\min_T \mathbf{Y}, \min_T \hat{\mathbf{Y}}), \tag{22}$$

where $r(\cdot,\cdot)$ is the Pearson correlation coefficient between two quantities, and the $\min_T$ and $\max_T$ functions operate along the temporal dimensions. In other words, we extracted SBP and DBP values from the model output, determined how well they correlate with the ground truth, and used the average correlation as the model accuracy. For models that estimate the fiducial values directly, the extraction through $\min_T$ and $\max_T$ is not necessary. Instead, the SBP and DBP are extracted as the first and second position of the output vector, respectively.

We also used this accuracy metric to adapt the learning rate of the optimizer. We employed PyTorch's ReduceLROnPlateau (RLROP) scheduler, which monitors the model accuracy and reduces the learning rate when performance reaches a plateau. The RLROP scheduler operates with three key hyperparameters: a reduction factor $\kappa \in (0, 1)$ for scaling the learning rate $(\gamma \leftarrow \kappa \cdot \gamma)$, a patience window $w_r \in \mathbb{N}_{>0}$ specifying how many epochs to wait before applying the reduction, and an improvement threshold $\delta_r > 0$ defining the minimum change to consider a metric change as improvement. We set $\kappa = 0.8$ and $\delta_r = 1 \cdot 10^{-4}$ for all models, $w_r = 50$ for SS models, and $w_r = 10$ for PW and PD models.

To prevent overfitting, we implemented a convergence tracking mechanism instead of setting a fixed number of training epochs. We designed the mechanism to monitor the accuracy metric (22) and terminate the training loop when the accuracy stops improving. The tracking mechanism is controlled through four hyperparameters: an activation threshold $\varrho_{\text{min}} > 0$ specifying the minimum accuracy to start monitoring, an improvement threshold $\delta_c > 0$ defining the minimum change to be considered an improvement, a patience window $w_c \in \mathbb{N}_{>0}$ specifying how many epochs to wait before terminating the training loop and restoring the model to its previous best state, and the maximum number of epochs $\tau_{\text{max}} \in \mathbb{N}_{>0}$ if the activation threshold was never reached. We used $\varrho_{\text{min}} = 0.5$ and $\delta_c = 1 \cdot 10^{-4}$ when training all models, while setting the patience window and the maximum epochs differently for different types of models. For SS models, we set $w_c = 200$ and $\tau_{\text{max}} = 5,000$. For PW and PD models, we set $w_c = 50$ and $\tau_{\text{max}} = 500$.

### 8.8.2 Pilot longitudinal study

We selected four best-performing SS configurations (ranked by their accuracy on the cross-sectional dataset) and evaluated their generalizability on the longitudinal datasets. Each configuration corresponds to a suite of 91 SS models, from which we extracted 5 models corresponding to the 5 subjects who also participated in the longitudinal protocol. For each subject, let $\text{D}_0$ denote the cross-sectional dataset, and $\text{D}_k$ denote the longitudinal datasets from day $k$, with $k = 1, \ldots, 5$. Furthermore, let $\text{C}_0$ denote the model checkpoint trained with dataset $\text{D}_0$. We first evaluated the longitudinal stability of the model by testing the cross-sectional checkpoint $\text{C}_0$ on the datasets $\text{D}_1, \ldots, \text{D}_5$ individually. This evaluation served as the baseline values which help assess the efficacy of recalibration. We then fine-tuned the model for each day $j \geq 1$ by loading the checkpoint $\text{C}_{j-1}$ while using a portion of dataset $\text{D}_j$ from the current day (explained next). We view this step as subject-specific recalibration. After recalibration, the new model checkpoint $\text{C}_j$ is tested on the datasets $\text{D}_k$ of future days ($k > j$) to evaluate the benefit of progressive daily recalibration.

The fine-tuning step slightly differed from the approach used to train the cross-sectional SS models. The dataset for each model $\text{C}_j$ includes the previously available datasets with their partition state unchanged, while incorporating the new dataset $\text{D}_j$ having $5:5$ partition ratio without shuffling. We used AdamW optimizer with an initial learning rate of $\gamma_0 = 1 \cdot 10^{-4}$, and RLROP scheduler with the same hyperparameters as before. We also employed the convergence tracker with a patience window of $w_r = 50$ epochs and a limit of $\tau_{\text{max}} = 1,000$ training epochs.

### 8.8.3 Hardware resources

All SS models were trained on a lab workstation at the University of Utah, equipped with an NVIDIA Quadro RTX 6000 (NVIDIA Corporation, Santa Clara, CA, USA) and 256 GB of memory. PW and PD models were distributed across two computing servers: the Center for High Performance Computing (CHPC, University of Utah)[216] and the Advanced Cyberinfrastructure for Education and Research (ACER, University of Illinois Chicago).[217] We used a combined assortment of NVIDIA GPUs including H100 NVL, L40S, P40, V100, A100, RTX2080Ti, RTX4500 Ada, RTX6000 Ada, and GTX TITAN X; and the lab's dedicated GPU node hosted at ACER using four NVIDIA H100 NVL and 1 TB of DDR5 RAM memory.

## 8.9 Cross-sectional ablation study

We selected representative PW models from the CRT and CRS classes to perform an ablation study comprising three experiment types. The first two are structural ablations that isolate the contribution of specific architectural blocks; the third is a hybrid approach that targets the training strategy rather than the architecture itself. Together, the ablated configurations (A1–A4) provide a systematic decomposition of the reference architecture across depth, positional encoding design, and learning rate scheduling.

### 8.9.1 Depth reduction

The first ablation type examines architectural depth by removing layers from the reference models. Specifically, we removed one layer from the convolutional encoder, reducing its depth by one. For CRT models, we additionally collapsed the multi-stage MLP head into a single FC layer $((100T_\ell + 2m) \to T_{\text{out}})$, mapping the transformer output directly to the estimated BP vector. This MLP simplification was not applied to CRS, since its MLP head includes only a single FC layer, making it structurally equivalent to the post-ablation CRT configuration. The resulting models test whether the sequential backbone (BiLSTM + Transformer/Samba) alone can effectively learn the PVI-BP relationship without the depth contributed by the convolutional encoder or the multi-stage MLP head. Models of this type are denoted with the suffix “-A1”, and were trained using the same strategy as the reference PW models.

### 8.9.2 Positional encoder replacement

The second type examines the contribution of the BiLSTM positional encoder (PE) block by replacing it with a simpler PE. Specifically, we used the sinusoidal PE, which adds fixed position-dependent frequency patterns without any trainable parameters.[200,218] We denoted models of this type with the suffix “-A2”, and were trained using the same strategy as the reference PW models.

### 8.9.3 Training strategy adjustment

The third experiment type departs from structural ablation and instead targets the training strategy, applied exclusively to CRT models. Here, we retained the full reference architecture while replacing the RLROP scheduler with the CosineAnnealingLR during training. This directly addresses premature convergence caused by RLROP’s reactive decay, substituting it with a smooth, predictable schedule better suited to Transformer optimization. Models of this type are denoted with the suffix “-A3”. Finally, we combined the A1 depth reduction with the A3 cosine schedule, again restricted to CRT models, to test whether RLROP’s conservative decay was limiting the potential of the simplified architecture. Models of this type are denoted with the suffix “-A4”. All other training configurations remain unchanged.

# 9 Supplementary Discussion 9. Results

## 9.1 Ring sensor characterization

SEM images were taken before and after PEDOT deposition.The ENIG surface contains scratches due to manufacture imperfection. The PEDOT layer contains porous clusters that increase the effective surface area. Electrical impedance spectroscopy shows relatively consistent impedance of the PEDOT electrodes across the frequency sweep range. In contrast, ENIG electrodes have large impedance and phase distortion at low frequencies while approaching the impedance of PEDOT at higher frequencies. At the injection frequency of 50 kHz, ENIG electrodes have an impedance of $13.33 \pm 4.31$ kΩ and phase of $-13.58 \pm 9.53$ degrees, while PEDOT electrodes have an impedance of $8.75 \pm 1.33$ kΩ and phase of $-11.57 \pm 22.17$ degrees. Cyclic voltammetry (CV) characterization shows that PEDOT-coated electrodes had significantly larger charge capacitance ($331.69 \pm 576.28$ µF) than ENIG electrodes ($5.93 \pm 3.61$ µF). Both types of electrodes also underwent a 100-cycle stability test and exhibited minimal change in their charce capacitance between the first and last cycle, indicating good electrochemical stability (Fig. 1). Overall, the gold electrode provides stable, low-charge storage capacity while exhibiting conductive behavior, while the PEDOT-based electrodes exhibit stable, high CSC with capacitive behavior.

## 9.2 Computational fluid dynamics simulations

### 9.2.1 Particle transport under baseline condition

We performed the simulations using a time step size of $\delta t = 1$ ms for a total of 6,000 time steps, equivalent to 6 cardiac cycles. During simulation, the number of suspended particles within the domain $N_\text{p}$ reached a maximum of approximately $1.6 \cdot 10^6$. The results reported here were sampled over the time interval between 3 s and 6 s to ensure that transient effects from initialization have decayed and the system has fully saturated by injected particles. Fig. 2c and Fig. 2d illustrate snapshots of particle distributions within the reconstructed palmar arterial network at the peak systolic phase (5.1 s) and the corresponding diastolic phase (5.9 s), respectively.

We first noted that particle velocity spanned a range from near-zero during diastole to $\mathcal{O}(10^2)$ mm/s at peak systole (Fig. 2c ,d). During systole, the pulsatile inflow induced strong acceleration of the fluid, resulting in high particle velocity, most notably in the proximal arterial segments (Supplementary Video 1). In contrast, during diastole, particle velocity decreased substantially throughout the network as the driving pressure gradient diminished. In view of the whole system, at the first bifurcation (towards artery 5 and 6), particles experienced strong acceleration due to local geometric constriction and flow redistribution (Supplementary Video 2). We noticed that recirculation zones developed in regions of high curvature, creating partially unsaturated regions with reduced particle concentration. These recirculation zones reduced the effective flow cross-section, thereby increasing particle velocity, most notably in the downstream segments. Particularly at the early bifurcation corresponding to artery 6, we also observed incomplete particle saturation, attributable to recirculation and flow separation in-duced by the highly curved geometry. Particles at the bifurcation between arteries 3 and 4 exhibited distinct flow patterns (Fig. 2c ii and Fig. 2d ii). As fluid entered the curved bifurcation region, particle velocity decreased near the inner wall of the curvature due to viscous shear and local pressure distribution, leading to a clear separation of transport patterns between the downstream arteries.

At the two proper palmar arteries (4 and 5) along the ring finger, particles exhibited a radial velocity gradient characteristic of laminar Poiseuille flow (Fig. 2c iii and d iii, Supplementary Video 3). Overall, the particle velocity was higher near the lumen center and lower near the vessel wall due to viscous shear within the laminar boundary layer. Additionally, due to variations in geometric resistance and outlet impedance, the flow rate differed between the two arteries. During systole, the peak volumetric flow reached approximately 170 $mm^3/s$ in artery 4 and 140 $mm^3/s$ in artery 5, with up to 40% discrepancy in mean particle velocity. During diastole, the volumetric flow rate in both arteries reached 25 $mm^3/s$, accompanied by a near stagnation in particle transport throughout the entire network. The weakened flow reduced radial velocity gradients, leading to more uniform velocity distribution across the lumen cross-section. We also noted the morphological morphological discrepancy between pressure and flow rate within the same arteries due to peripheral compliance. Specifically, the pressure peak was delayed relative to the peak inflow, and did not exhibit a sharp drop immediately after systole. Instead, pressure decayed gradually during diastole, consistent with the capacitive behavior of the Windkessel model.

#### 9.2.2 Particle transport with moderate and severe stenosis

In the $75\%$ area occlusion scenario, the overall profiles for pressure and flow rate remained almost identical to the baseline scenario for both arteries 4 and 5 (Fig. 2e, f and Supplementary Video 4). In contrast, the $96\%$ occlusion scenario yielded substantial hemodynamic alterations. In artery 4, the flow waveform was significantly damped, leading to a considerable reduction in pulsatility and peak flow magnitude. Specifically, the flow amplituded decreased from $\approx$150 $mm^3/s$ in the baseline scenario to $\approx$20 $mm^3/s$ in the 96% occlusion scenario. The flow attenuation was accompanied by a pronounced phase lag due to increased impedance at the occluded region. Similarly, the mean pressure in artery 4 decreased by approximately 15 mm Hg, accompanied by reduced pulsatility and delayed systolic peak, reflecting substantial energy loss at the stenosis lesion. In contrast, artery 5 exhibited an increase in flow amplitude under severe occlusion, with amplitude reaching approximately 200 $mm^3/s$. Consistent with the elevated flow rate, the pressure profile at artery 5 was increased by 8 mm Hg throughout the entire cardiac cycle.

### 9.3 Image reconstruction algorithm

To validate our implementation, we performed forward and inverse simulations and compared our results with the open-source Electrical Impedance and Diffuse Optical Tomography Reconstruction Software (EIDORS).[219] For quantitative comparison, we used the root-mean-square

error (RMSE) defined as

$$\text{RMSE}(\mathbf{X}, \hat{\mathbf{X}}) := \frac{1}{\sqrt{N}} \|\mathbf{X} - \hat{\mathbf{X}}\|_{\text{F}}, \tag{23}$$

where $\mathbf{X}$ and $\hat{\mathbf{X}}$ are the reference and computed quantities, respectively, $N$ is the number of elements in $\mathbf{X}$, and $\|\cdot\|_{\text{F}}$ denotes the Frobenius norm. Here, $\mathbf{X}$ and $\hat{\mathbf{X}}$ can be either vectors or matrices, depending on the context.

### 9.3.1 Forward module validation results

We validated the forward solver in two scenarios: rectangular resistor model, and a circular disk model. In the first scenario, we compared the forward solution with the theoretical value. Our results show the nodal voltage $u_i$ matches the expected linear relationship $\hat{u}_i = x_i \cdot I/(A\sigma)$ with RMSE$(\mathbf{u}, \hat{\mathbf{u}}) = 3.9 \cdot 10^{-14}$ V (Supplementary Fig. 21a). The electrode voltages are $-0.1$ V and $2.1$ V for the left and right electrode, respectively, corresponding to the $0.1$ V difference accounted by equation (1d). In the second validation scenario, we compared the Jacobian from our implementation with EIDORS using the same mesh. Qualitatively, the Jacobian from our implementation exhibits a pattern with sign as expected from theory (Supplementary Fig. 21b). Specifically, the Jacobian is positive in the region between the injection and measurement pair where the lead and reciprocal field form an acute angle, and negative at regions where the fields form an obtuse angle. A comparison between our implementation with EIDORS revealed an excellent numerical agreement, with RMSE$(\mathbf{J}, \hat{\mathbf{J}}) = 7.53 \cdot 10^{-21}$ V m/S.

### 9.3.2 Inverse module validation results

We validated the inverse solver with experimental data[191] and compared the results with EIDORS. The reconstructed conductivity images from our PVI algorithm shows similar patterns to images from EIDORS (Supplementary Fig. 22a). We computed the pixel-wise difference between the two images and found a mean absolute error of $22.8$ nS/m with standard deviation of $19.3$ nS/m, and RMSE$(\Delta\boldsymbol{\sigma}, \Delta\hat{\boldsymbol{\sigma}}) = 37.2$ nS/m (Supplementary Fig. 22b). To complement the validation, we also investigated the effects of regularization on image quality. We used data from a water tank experiment and performed image reconstruction with four different regularization parameters: $\lambda = 10^{-6}$, $\lambda = 10^{-4}$, $\lambda = 10^{-3}$, and $\lambda = 10^{-2}$. We found that the images contain ringing artifacts for lower regularization values, and blurry contrast for higher regularization values (Supplementary Fig. 23).

### 9.3.3 Finger isopotential lines and sensitivity results

We simulated all injection pairs for ring with 8 and 16 electrodes and extracted the potential distribution inside the computational finger (Supplementary Fig. 4–13). While the potential distribution is a scalar field extending in all three dimensions, for clarity we simplified the visualization as follows. First, we represented the spatial distribution by extracting its isopotential surfaces, which we then reduced to a 2D cross-sectional slice, yielding the isopotential lines shown. Secondly, we only show the isopotential lines corresponding to the boundary electrode voltages, leading to an apparent symmetrical distribution; their numerical values are not necessarily symmetric. Furthermore, the electrode voltages were determined by Dirichlet boundary conditions $\pm 1$ V at the injecting electrodes. Under linear assumption, these values can be scaled and translated arbitrarily to match a specific injection current. Our analysis revealed that the isopotential lines are distorted in two ways: they clustered in the regions near the tissue interfaces (most apparent at the tendons), and refracted when transitioning across the interface (most apparent at the skin-fat interface).

The sensitivity analysis requires both the lead and the reciprocal field. We conducted the analysis for 8 electrodes and combined the injection pairs from the skip-2 and skip-1 pattern to match the EII setup used in our experimental study (Supplementary Discussion 7). Although a complete EII frame with 8 electrodes consist of 32 independent injection–measurement configurations, we selected three representative configurations. The first configuration, $(\ell_1, \ell_4)$-injection with $(\ell_8, \ell_2)$-measurement, represents the situation where all electrodes are located on the dorsal or lateral side of the finger, furthest away from the arteries. The second configuration, $(\ell_4, \ell_7)$-injection with $(\ell_3, \ell_5)$-measurement, corresponds to the instance when all electrodes are located on the palmar side of the finger and closest to the arteries. In the last configuration, $(\ell_7, \ell_2)$-injection with $(\ell_4, \ell_6)$-measurement, the injection and measurement pair are on the opposite side of the finger. Supplementary Fig. 14–16 show the results of our sensitivity analysis. We characterized the penetration depth as the spatial dimension of the $95\%$-region of the total impedance magnitude measured at the sensing electrodes, and found them to be $18.16 \times 30.95 \times 19.24$ mm$^3$, $19.24 \times 43.99 \times 17.68$ mm$^3$, and $18.86 \times 41.65 \times 16.58$ mm$^3$ for the three electrode configurations, respectively.

Additionally, we computed the contribution of each tissue to the total absolute resistance and reactance (Supplementary Fig. 17 and Supplementary Table 3). We compared the BioZ content of all tissues and found that SAT has the highest content, with absolute resistance between $33\%$–$94\%$, absolute reactance between $15\%$–$65\%$, and impedance magnitude between $32\%$–$93\%$. In contrast, muscle has the least BioZ content, with absolute resistance between $0.001\%$–$0.020\%$, absolute reactance between $0.009\%$–$0.026\%$, and impedance magnitude between $0.001\%$–$0.020\%$. We compared the BioZ content of arteries across three configurations and found that it is lowest when the injection and measurement pairs are on the dorsal side (configuration 1), with $0.290\%$, $0.511\%$ and $0.289\%$ contribution to absolute resistance, absolute reactance, and impedance magnitude, respectively. The arterial contributions are at the highest when the injection and measurement pairs are on the palmar side (configuration 2), with $5.973\%$, $2.324\%$ and $5.835\%$ contribution to absolute resistance, absolute reactance, and impedance magnitude, respectively.

### 9.3.4 Signal reconstruction of synthetic conductivity waveform

We used synthetic conductivity waveforms to validate our reconstruction algorithms with both 8 and 16 electrodes (Supplementary Fig. 18). In both configurations, the temporal evolution of blood conductivity is accurately recovered, as evidenced by the near-perfect agreement between the estimated and reference waveforms throughout the cardiac cycle. The correlation analysis and error distribution showed excellent tracking fidelity with $r^2 = 0.999$, $(p < 0.001)$ for both electrode counts. The representative PVI frames at selected time points reveal a critical difference in spatial resolution. With 8 electrodes, the reconstructed images show dif-fuse conductivity changes that lack clear anatomical correspondence to the underlying arterial structures. In contrast, 16 electrodes produce images where the spatial distribution of conductivity changes visually aligns more closely with the known positions of the ulnar and radial digital arteries.

## 9.4 Experimental data collection

### 9.4.1 Cross-sectional study

We enrolled $N = 99$ healthy subjects for the cross-sectional protocol, consisting of three phases: static, Valsalva, and cold pressor (Supplementary Discussion 7.3). We excluded $N = 3$ subjects from the study and follow up data analyses due to inability to complete any phase of the study. All remaining $N = 96$ subjects completed the static phase, of which $N = 80$ also completed the Valsalva phase, and $N = 58$ completed all three phases of the study.

Supplementary Table 4 summarizes the cohort characteristics and features from our experimental datasets. The $N = 96$ subjects included in the cross-sectional analysis had a male:female ratio of $40 : 56$, age of $28.3 \pm 8.8$ years, weight of $68.4 \pm 11.2$ kg, height of $169.8 \pm 8.6$ cm, and body mass index (BMI) of $23.7 \pm 3.6$ kg/m$^2$. Our processing pipeline detected $N_P = 315,886$ raw periods, of which $N_C = 263,929$ were classified as clean. The number of raw periods varied across subjects, from as low as $N_P = 810$ to as high as $N_P = 6,081$. The number of ML samples, i.e. sequences of $m = 5$ consecutive clean periods, was $N_S = 172,686$. The dataset has SBP of $128.6 \pm 15.2$ mm Hg, DBP of $82.4 \pm 11$ mm Hg, and heart rate of $74.1 \pm 12.8$ bpm. We also inspected the morphology of the $\mathbf{Z}^{(\text{HP})}$ and $\mathbf{S}^{(\text{HP})}$ periods and found that they both exhibit patterns similar to BP waveform: a systolic peak followed by a clear dicrotic notch and a subtle second reflection during diastole. The impedance waveforms have a peak-to-peak range of $6.2 \pm 4.8$ mΩ, and the conductivity waveform has a peak-to-peak range of $9.6 \pm 7.0$ mS/m.

### 9.4.2 Pilot longitudinal study

We recruited $N = 5$ subjects (002, 010, 070, 094, and 095) from the cross-sectional study for a follow-up pilot longitudinal study. All 5 subjects were measured 6–12 months after participating in the cross-sectional study. These participants completed the protocol involving 5 consecutive days of data collection, each consisting of 3 trials. Each trial lasted 8 minutes and included 20 seconds of Valsalva maneuver. The cohort has a male:female ratio of $3 : 2$, age of $27.2 \pm 2.6$ years, weight of $66.5 \pm 9$ kg, height of $170.7 \pm 8.3$ cm, and BMI of $23 \pm 4$ kg/m$^2$ (Supplementary Table 4). Our processing pipeline detected $N_P = 55,258$ raw periods for the longitudinal study, classified $N_C = 48,364$ as clean, and yielded $N_S = 37,948$ ML samples.

## 9.5 Machine learning

Here, we discuss the results of our ML approach. We first describe the notation used for identifying the models. In total, we trained $1,820$ subject-specific (SS) models ($20$ configurations $\times$ $91$ subjects), $20$ population-within (PW) models, and $20$ population-disjoint (PD) models. Sup-plementary Table 5 summarizes the training progress of all models. We denote the models as follows:

- All model configurations are assigned numeric identifiers. The identifiers begin with 01 and end with 20, for 20 configurations.
- Configurations are ordered at three levels. The first is based on the model class: Linear Regression (LR) models are 01–04, Multilayer Perceptron (MLP) models are 05–08, Convolutional Neural Network (CNN) models are 09–12, Convolutional Recurrent Transformer (CRT) models are 13–16, and Convolutional Recurrent Samba (CRS) models are 17–20. Secondly, the configurations are ordered within each class based on input modality: the first two using image input ($\tilde{\mathbf{X}}^{(\text{3D})}$), and the other two using BioZ input ($\tilde{\mathbf{X}}^{(\text{1D})}$). Finally, the configurations are ordered based on output modality: waveform estimation ($\hat{\mathbf{Y}}^{(\text{wave})}$), followed by fiducial estimation ($\hat{\mathbf{Y}}^{(\text{fp})}$).
- Since each model configuration was trained using different datasets and partition method, we distinguish them by adding the prefixes SS, PW and PD to the model identifier. We also note that since SS models are trained on individual subjects, the model identifier refers to the configuration as a whole, rather than for any specific subject. For example, while PW01 and PD01 each refer to a single model, SS01 refers to a collection of 91 models with the same configuration. Consequently, the number of train and test samples and the number of epochs for SS configurations reported in Supplementary Table 5 are the sum of their respective sub-models.

### 9.5.1 Dataset metrics

In what follows, we describe the performance metrics used to evaluate the models' goodness of fit. However, to determine the models' capacity to generalize, we also characterized how these metrics change with respect to dataset degradation. To enable this analysis, we constructed two metrics associated with the datasets: one to quantify their inherent quality, and the other to measure the train-test partition gap that might affect generalization.

First, to compare dataset quality across subjects, we defined the data quality index (DQI) as the proportion of raw periods passing the quality assessment criteria. To account for statistical uncertainty due to sample size, we computed the DQI as the lower bound of the Wilson score rather than the point estimate.[220] Let $\hat{p} := N_C/N_P$ denote the ratio between $N_C$ clean periods and $N_P$ total raw periods, the DQI is computed as

$$\text{DQI} := \frac{\hat{p} + \dfrac{z^2}{2N_P} - \dfrac{z^2}{N_P}\sqrt{\hat{p}(1-\hat{p}) + \dfrac{z^2}{2N_P}}}{1 + \dfrac{z^2}{N_P}}, \tag{24}$$

where $z \approx 1.96$ corresponds to the 95% confidence interval.

Second, to quantify the divergence between the train and test set, we defined the partition gap in the output, $\mathcal{W}_{\text{label}} := \mathcal{W}_1(\mathbf{Y}_{\mathcal{U}}, \mathbf{Y}_{\mathcal{V}})$. Here, $\mathcal{W}_1(\mathbf{Y}_{\mathcal{U}}, \mathbf{Y}_{\mathcal{V}})$ is the Wasserstein-1 distance between the training labels $\mathbf{Y}_{\mathcal{U}}$ and test labels $\mathbf{Y}_{\mathcal{V}}$. Formally, the Wasserstein-1 distance is defined as the solution to the linear optimization problem[221]

$$\mathcal{W}_{\text{label}} \equiv \mathcal{W}_1(\mathbf{Y}_{\mathcal{U}}, \mathbf{Y}_{\mathcal{V}}) := \underset{\boldsymbol{\Gamma}}{\text{minimize}} \sum_{i,j} (\mathbf{D} \odot \boldsymbol{\Gamma})_{i,j} \tag{25a}$$

$$\text{subject to} \sum_i \boldsymbol{\Gamma}_{i,j} = \frac{1}{|\mathcal{V}|}, \sum_j \boldsymbol{\Gamma}_{i,j} = \frac{1}{|\mathcal{U}|}, \tag{25b}$$

$$\boldsymbol{\Gamma}_{i,j} \geq 0, \ \forall\, (i,j) \in \{1, \ldots, |\mathcal{U}|\} \times \{1, \ldots, |\mathcal{V}|\}; \tag{25c}$$

where $\odot$ is the Hadamard product of two matrices with the same shape, $\boldsymbol{\Gamma} \in \mathbb{R}^{|\mathcal{V}|\times|\mathcal{U}|}$ is the joint distribution of the train and test set, $\mathbf{D} \in \mathbb{R}^{|\mathcal{V}|\times|\mathcal{U}|}$ is the pairwise distance matrix with entries defined as $\mathbf{D}_{ij} := \|\mathbf{Y}_i - \mathbf{Y}_j\|_2$, and $i \in \{1, \ldots, |\mathcal{U}|\}$ and $j \in \{1, \ldots, |\mathcal{V}|\}$ are the indices of the train and test samples, respectively.

While the DQI is inherent to the dataset, the label gap $\mathcal{W}_{\text{label}}$ is dependent on the dataset partition, which we randomized when initializing the models. Therefore, we computed $\mathcal{W}_{\text{label}}$ for all SS, PW, and PD models. We used the Python Optimal Transport package (ot, version 0.9.6)[222] to set up the linear program (25) and solved for $\mathcal{W}_{\text{label}}$. For PD and PW models, we avoided loading the entire population dataset due to memory constraint. Instead, we utilized the cache-aware sampling algorithm (Supplementary Discussion 8.6) to assemble the distance matrix $\mathbf{D}$ in blocks and stored separately before solving for $\mathcal{W}_{\text{label}}$.

### 9.5.2 Model accuracy metrics

For all models, we extracted the estimated SBP and DBP and compared with the respective true values. We employed multiple complementary metrics that capture goodness of fit: regression metrics to evaluate how well estimated trend align with true values' trend, error metrics to quantify sample-wise accuracy, and distribution metrics assess the overall statistical discrepancy. Specifically, we use the following metrics:

- Inter-subject determination coefficient and concordance coefficient: $r^2_{\text{a}}$ and $\hat{\rho}_{c,\text{a}}$, respectively. We calculated these two quantities as

$$r^2_{\text{a}} := r^2(y, \hat{y}) \qquad \text{and} \qquad \hat{\rho}_{c,\text{a}} := r(y, \hat{y}) \frac{2 s_y s_{\hat{y}}}{s_y^2 + s_{\hat{y}}^2 + (m_y - m_{\hat{y}})^2}, \tag{26}$$

  where $r(\cdot,\cdot)$ is Pearson's correlation coefficient determined from simple linear regression; $y,\ \hat{y} \in \mathbb{R}$ are the true and estimated BP values (either SBP or DBP), respectively; $m_y,\ m_{\hat{y}} \in \mathbb{R}$ are their respective mean values; and $s_y,\ s_{\hat{y}} \in \mathbb{R}$ are their respective standard deviations. While $r^2$ determines how closely the samples align with the best-fit line, $\hat{\rho}_c$ provides a complementary description and determines how closely the best-fit line aligns with the reference $45°$ line of perfect correlation.[223] Here, the true and estimated values are from the population test set, aggregated from all subject-specific test sets ($\mathcal{V} := \mathcal{V}_1 \cup \cdots \cup \mathcal{V}_N$).

- Weighted intra-subject determination coefficient and concordance coefficient: $r^2_{\text{w}}$ and $\hat{\rho}_{c,\text{w}}$, respectively, defined as the weighted average of the corresponding subject-specific regression metrics, i.e.

$$r^2_{\text{w}} := \frac{1}{|\mathcal{V}|} \sum_k^N r_k^2\, |\mathcal{V}_k| \qquad \text{and} \qquad \hat{\rho}_{c,\text{w}} := \frac{1}{|\mathcal{V}|} \sum_k^N \rho_{c,k}^2\, |\mathcal{V}_k|\,, \tag{27}$$

where $\mathcal{V}_k$ is the subject-specific test set; and $r_k^2$, $\rho_{c,k}^2$ are computed similarly to (26) with the estimated and true values from $\mathcal{V}_k$. While the aggregated regression metrics (26) quantify the overall trend of the population data, the weighted regression metrics (27) isolate within-subject performance and avoid the ecological fallacy.[224] We also note by weighting the individual metrics by the number of samples instead of simply averaging across all subjects, we acquired the least biased estimator of the true population-level metrics.[225]

- Mean error (ME) and limits of agreement (LOA) covering 95% of errors.[226] Here, we computed the lower and upper LOA as the 2.5th and 97.5th percentile of the error distribution, respectively.
- Mean and standard deviation of the absolute error (MAE±SDAE). Compared to the ME, MAE is robust against mutual cancellation between positive and negative errors.
- Cumulative probability $\mathcal{P}_\tau$ of estimations with absolute error (AE) below the threshold $\tau$ of 5, 10, and 15 mm Hg, i.e.

$$\mathcal{P}_\tau := \mathrm{P}\left(\text{AE} \leq \tau\right), \tag{28}$$

where AE $:= |y - \hat{y}|$ is the absolute error, and $\mathrm{P}$ is the probability computed from the empirical error distribution with bin size of $\mathrm{d}p = 0.1$ mm Hg.

- Distance between the estimated and true BP distributions, $\mathcal{W}_{\text{pred}}$. While the above metrics evaluate pairwise discrepancy between estimated and true BP samples, the $\mathcal{W}_{\text{pred}}$ distance quantifies the dissimilarity between the distributions, thereby assessing whether the model captures the underlying statistical properties of the data. Formally, we defined the $\mathcal{W}_{\text{pred}}$ as the Wasserstein-1 distance $\mathcal{W}_{\text{pred}} := \mathcal{W}_1(y, \hat{y})$, i.e. similar to the distribution gap $\mathcal{W}_1(\mathbf{Y}_\mathcal{U}, \mathbf{Y}_\mathcal{V})$. However, for univariate distributions, the optimization problem (25) has the following closed-form solution[227]

$$\mathcal{W}_{\text{pred}} \equiv \mathcal{W}_1(y, \hat{y}) := \int_{\mathbb{R}} \left| F_{\mathcal{Y}}(p) - F_{\hat{\mathcal{Y}}}(p) \right| \mathrm{d}p, \tag{29}$$

where $F_{\mathcal{Y}} : \mathbb{R} \to [0, 1]$ and $F_{\hat{\mathcal{Y}}} : \mathbb{R} \to [0, 1]$ are the cumulative distribution functions of the estimated and true BP values, respectively. Here, we computed the cumulative distribution function from the data histograms with discretization $\mathrm{d}p = 0.1$ mm Hg.

- In addition to the above metrics for fiducial BP values, we also report the average mean absolute error (AMAE) and average root-mean-square error (ARMSE) for models with waveform output, i.e. those with odd identifiers 01, 03,..17, 19. These metrics are defined as

$$\text{AMAE}(\mathbf{Y}, \hat{\mathbf{Y}}) := \frac{1}{|\mathcal{V}|} \sum_{n=1}^{|\mathcal{V}|} \left( \frac{1}{T} \sum_{k=1}^{T} \left| \mathbf{Y}_n[k] - \hat{\mathbf{Y}}_n[k] \right| \right); \tag{30}$$

$$\text{ARMSE}(\mathbf{Y}, \hat{\mathbf{Y}}) := \frac{1}{|\mathcal{V}|} \sum_{n=1}^{|\mathcal{V}|} \sqrt{\frac{1}{T} \sum_{k=1}^{T} \left( \mathbf{Y}_n[k] - \hat{\mathbf{Y}}_n[k] \right)^2}, \tag{31}$$

where $\mathbf{Y} \in \mathbb{R}^{|\mathcal{V}| \times T}$ and $\hat{\mathbf{Y}} \in \mathbb{R}^{|\mathcal{V}| \times T}$ are the tensors containing true and estimated BP waveforms, respectively; $|\mathcal{V}|$ is the number of test samples, $T = 50$ is the BP waveform length, $n$ is the sample index, and $k$ is the time index.

### 9.5.3 Model robustness quantification

With the dataset metrics and model accuracy metrics established, we characterized how the model performance depends on dataset quality and composition. Specifically, we examined whether SS models trained on high-quality data demonstrate better accuracy, and whether accuracy degradation correlates with train-test domain gap. For this analysis, we selected the 10 configurations with waveform outputs and, for each configuration, computed the correlation between AMAE and DQI as well as correlation between AMAE and $\mathcal{W}_{\text{label}}$. We believe this analysis provides valuable insight regarding the models' capacity to generalize on the holdout datasets.

### 9.5.4 Summary of model parameters

Supplementary Table 5 summarizes the architectural complexity and the training progress of all models. Overall, models estimating waveforms have more parameters than models estimating fiducial values. Regarding input, however, model complexity does not strictly correlate with input dimensionality. For MLP and LR models, image inputs require more neurons than BioZ inputs. On the contrary, CNN, CRT, and CRS models show the reverse trend, with more parameters for BioZ inputs than for image inputs. This is because they all have a convolutional encoder that expanded the channel dimensions while reducing the spatial resolution (Supplementary Discussion 8.4). However, since BioZ inputs do not have spatial dimensions, there was a net increase in parameter counts. The CNN model class has the most parameters, ranging between ≈100M and ≈154M, while CRS model class has the fewest number of parameters, ranging between ≈298K and ≈2.2M. The CRT model class has the most consistent number of parameters across all input and output modalities, with only 370,500 parameters difference (5%) between the smallest and largest model configuration (7,174,530 and 7,545,030 parameters, respectively). LR and MLP models, which use exclusively fully-connected layers, have the most variance in parameter counts: LR with ≈99.8% difference between the smallest and largest configuration (96K and 60M, respectively), and MLP with ≈96% difference (4.8M and 120M).

### 9.5.5 Overview of training progress

Supplementary Table 5 also shows the total number of training epochs for all model configurations and dataset composition. For PW and PD models, the training curves are shown in Supplementary Fig. 31 and 32, respectively. It is worth noting that, due to our convergence tracker's threshold-based activation, training termination does not guarantee optimal convergence (Supplementary Discussion 8.8). In contrast, we identified three termination types: non-convergence, premature, and standard. Non-convergence termination occurred when models reached the maximum epochs (5,000 for SS; and 500 for PW and PD), without achieving the threshold $\varrho_{\text{min}} = 0.5$ required to activate the convergence tracker. Examples of this termination type are most LR models across all partition strategies, e.g. PW01–PW04 (Supplementary Fig. 31a), and most PD models across all architectural classes (Supplementary Fig. 32). The second type of termination (premature) occurred when models trained for very few epochs with test accuracy barely exceeding the tracking threshold, or with learning curves lacking plateau regions. These models struggled to increase the test accuracy beyond the improvement threshold $\delta_c = 1 \cdot 10^{-4}$ and thus exhausted the tracker's patience window after activation. Examples include PW14 (Supplementary Fig. 31d), PD09 and PD11 (Supplementary Fig. 32c), and PD15 (Supplementary Fig. 32d). Finally, standard termination includes all remaining cases where models exceeded the tracking threshold and show sufficient improvement $(\delta > \delta_c)$ to repeatedly reset the patience window. The learning curves for these models typically plateau in later epochs, with some showing train-test convergence, effectively narrowing the generalization gap. Representative examples include PW11 and PW15 (Supplementary Fig. 31c and 31d, respectively).

### 9.5.6 Cross-sectional results for subject-specific models

The performance from all SS model configurations are summarized in Supplementary Table 6 and visualized in Supplementary Fig. 33–52. We noticed the following consistent patterns across model classes and configurations. First, all models estimated DBP more accurately than SBP, as evidenced by higher regression metrics ($r^2$ and $\hat{\rho}_c$) and lower error metrics (MAE and SDAE) for DBP than for SBP. This is expected since true DBP has lower physiological variance than SBP.[228] Second, models trained to estimate fiducial values outperform those trained to estimate full BP waveform, demonstrating that waveform estimation is a more challenging learning task than fiducials points alone. Lastly, models using BioZ inputs generally outperform those using image inputs, though this performance gap is narrow for transformer-based architectures (CRT and CRS), and especially negligible for the CRS class (SS17—SS20).

To compare among model classes, we first examine three representative cases: LR (establishing baseline), CNN (worst performance), and CRS (best performance). We summarize here the baseline metrics from the LR class (SS01–SS04, Supplementary Fig. 33–36): determination coefficient $0.51 \leq r^2_{\text{a}} \leq 0.74$ $(p < 0.001)$ and concordance coefficient $0.68 \leq \hat{\rho}_{c,\text{a}} \leq 0.86$ for both SBP and DBP; MAE±SDAE between $5.72 \pm 5.64$ mm Hg and $8.39 \pm 10.82$ mm Hg for SBP; and between $4.1 \pm 3.87$ mm Hg and $5.75 \pm 7.23$ mm Hg for DBP. Bland–Altman analysis for the LR class showed small bias, with $-1.41$ mm Hg $\leq$ ME $\leq -0.69$ mm Hg for SBP; and $-0.72$ mm Hg $\leq$ ME $\leq 0.07$ mm Hg for DBP. The LOA for SBP are between $[-17.93, 14.82]$ mm Hg and $[-33.18, 21.79]$ mm Hg; and the LOA for DBP are between $[-11.96, 10.57]$ mm Hg and $[-19.34, 16.17]$ mm Hg. The best LR configuration is SS04, with cumulative percentages $\mathcal{P}_5$, $\mathcal{P}_{10}$, and $\mathcal{P}_{15}$ of $57\%$, $84\%$, and $94\%$ for SBP, respectively; and $71\%$, $93\%$, and $98\%$ for DBP.

The CNN class (SS09–SS12, Supplementary Fig. 41–44) yielded the worst performance among all SS models. The best configuration of this class (SS12) achieved $r^2_{\text{a}} = 0.58$ $(p < 0.001)$, $\hat{\rho}_{c,\text{a}} = 0.34$ for SBP; and $r^2_{\text{a}} = 0.57$ $(p < 0.001)$, $\hat{\rho}_{c,\text{a}} = 0.36$ for DBP. Bland–Altman analysis revealed severe underestimation of fiducial BP values, with ME and LOA of $-24.22$, $[-43.03, -0.26]$ mm Hg and $-15.69$, $[-28.78, 1.07]$ mm Hg for SBP and DBP, respectively. The corresponding MAE±SDAE for this configuration is $24.52 \pm 9.84$ mm Hg for SBP, and $15.94 \pm 6.79$ mm Hg for DBP. Consequently, the cumulative percentages $\mathcal{P}_5$, $\mathcal{P}_{10}$, and $\mathcal{P}_{15}$ for SBP are $4\%$, $9\%$, and $17\%$, respectively; and for DBP $7\%$, $19\%$, and $42\%$, respectively. Compared to the baseline LR class, the CNN class yielded lower accuracy, higher bias and higher variance, which can be attributed to overfitting, since the per-subject sample size was insufficient to train its 100M–150M parameters.

On the opposite extreme, CRS models (SS17–SS20, Supplementary Fig. 49–52) achieved the best performance. For models with waveform outputs (SS17 and SS19), their corresponding waveform metrics are the lowest of all SS models, with AMAE = $4.09$ mm Hg and ARMSE = $4.47$ mm Hg for SS17 model, and AMAE = $4.14$ mm Hg and ARMSE = $4.53$ mm Hg for SS19 model. All models of the SS CRS class yielded SBP determination coefficient between $r^2_{\mathrm{a}} = 0.8$ $(p < 0.001)$ and $r^2_{\mathrm{a}} = 0.81$ $(p < 0.001)$, SBP concordance coefficient between $\hat{\rho}_{c,\mathrm{a}} = 0.89$ and $\hat{\rho}_{c,\mathrm{a}} = 0.9$, DBP determination coefficient between $r^2_{\mathrm{a}} = 0.81$ $(p < 0.001)$ and $r^2_{\mathrm{a}} = 0.83$ $(p < 0.001)$, and DBP concordance coefficient between $\hat{\rho}_{c,\mathrm{a}} = 0.9$ and $\hat{\rho}_{c,\mathrm{a}} = 0.91$. We found the CRS models to slightly overstimate the fiducial BP values, with ME between $0.14$ mm Hg and $0.5$ mm Hg for SBP; and between $-0.7$ mm Hg and $0.51$ mm Hg. The CRS class also achieved the tightest LOA among all SS models, with LOA between $[-12.59, 13.42]$ mm Hg and $[-13.17, 13.22]$ mm Hg for SBP; and between $[-8.53, 9.49]$ mm Hg and $[-9.72, 9.14]$ mm Hg for DBP. The corresponding MAE±SDAE metrics are also the lowest and differ by a small margin among CRS configurations. For SBP estimation, the highest and lowest AE metrics are $4.85 \pm 4.63$ mm Hg (SS19) and $4.73 \pm 4.46$ mm Hg (SS17), respectively. For DBP estimation, the highest and lowest AE metrics are $3.53 \pm 3.17$ mm Hg (SS20) and $3.4 \pm 3.01$ mm Hg (SS19), respectively. Furthermore, all SS CRS models achieve superior AE distribution compared to other models, with cumulative percentages $\mathcal{P}_5 \geq 63\%$, $\mathcal{P}_{10} \geq 89\%$, and $\mathcal{P}_{15} \geq 97\%$ for SBP; and $\mathcal{P}_5 \geq 76\%$, $\mathcal{P}_{10} \geq 96\%$, and $\mathcal{P}_{15} \geq 99\%$ for DBP.

The remaining two model classes (MLP and CRT) occupy intermediate position, with MLP models having similar performance to the baseline LR models, while CRT models achieved the second-best metrics: $0.71 \leq r^2_{\mathrm{a}} \leq 0.76$ $(p < 0.001)$, $6.09$ mm Hg $\leq$ MAE $\leq 7.41$ mm Hg for SBP; and $0.73 \leq r^2_{\mathrm{a}} \leq 0.77$ $(p < 0.001)$, $5.28$ mm Hg $\leq$ MAE $\leq 4.87$ mm Hg for DBP. Here, a notable pattern emerges when comparing their input modalities. For image-based configurations, CRT models (SS13 and SS14) substantially outperforms MLP models (SS05 and SS06); while for BioZ-based configurations, CRT models (SS15 and SS16) marginally underperformed relative to their MLP counterparts (SS07 and SS08). However, when considering intra-subject estimation through the weighted regression metrics ($r^2_{\mathrm{w}}$ and $\hat{\rho}_{c,\mathrm{w}}$), we found the CRT configurations to have the highest score, with $0.63 \leq r^2_{\mathrm{w}} \leq 0.66$, $0.55 \leq \hat{\rho}_{c,\mathrm{w}} \leq 0.62$ for SBP; and $0.59 \leq r^2_{\mathrm{w}} \leq 0.60$, $0.55 \leq \hat{\rho}_{c,\mathrm{w}} \leq 0.59$ for DBP, suggesting its capability to learn intra-subject patterns.

### 9.5.7 Cross-sectional generalizability of subject-specific models

Our SS models were trained specifically for each participant and were not meant to generalize to other participants. We therefore did not evaluate the SS models on the holdout datasets.

### 9.5.8 Dependency of subject-specific accuracy on dataset quality and partition

Supplementary Fig. 113a and 113b show the dependency of waveform accuracy on dataset quality and label gap, respectively, for SS configurations. For simple LR and MLP configurations (SS01, SS03, SS05 and SS07), we found that accuracy depends on both dataset quality and label gap, as evidenced by the negative correlation between AMAE and DQI $(-0.51 \leq r \leq -0.44,\ p < 0.05)$, and positive correlation between AMAE and $\mathcal{W}_{\mathrm{label}}$ $(0.44 \leq$ $r \leq 0.61,\ p < 0.05$). For CRS configurations (SS17 and SS19), we found weak to moderate dependency on dataset quality and label gap, with $r = -0.31$ $(p < 0.05)$ and $r = -0.07$ $(p = 0.52)$ correlation between AMAE and DQI for SS17 and SS19, respectively; and corresponding $r = 0.39$ $(p < 0.05)$ and $r = 0.5$ $(p < 0.05)$ correlation between AMAE and $\mathcal{W}_{\text{label}}$.

For CNN configurations (SS09 and SS11), we observed large variations in AMAE (between $4.77$ mm Hg and $56.8$ mm Hg) with no statistically significant correlation to dataset characteristics (AMAE vs. DQI: $r = 0.01$ $(p = 0.9)$ for SS09 and $r = 0.06$ $(p = 0.57)$ for SS11; AMAE vs. $\mathcal{W}_{\text{label}}$: $r = -0.13$ $(p = 0.22)$ for SS09 and $r = -0.09$ $(p = 0.41)$ for SS11. As discussed in Supplementary Discussion 9.5.6, this high variance in AMAE is attributable to overfitting rather than architectural robustness.

CRT configurations (SS13 and SS15) similarly showed no statistically significant correlation with DQI $(r = 0.09,\ p = 0.38;\ r = -0.02,\ p = 0.84)$ or $\mathcal{W}_{\text{label}}$ $(r = 0.09,\ p = 0.39;\ r = -0.07,\ p = 0.51)$. However, unlike CNN configurations, CRT configurations achieved lower AMAE with substantially smaller variation (between $2.46$ mm Hg and $16.64$ mm Hg) across SS sub-models. With limited sample size ($N = 91$), this non-correlation results could reflect either genuine architectural robustness or insufficient statistical power to detect underlying dependency. An alternative explanation is that CRT architecture effectively learn subject-specific patterns, which compensate for label gap and data quality variations. This interpretation is supported by the high $r^2_{\text{w}}$ achieved by the CRT class, as noted in Supplementary Discussion 9.5.6 and Supplementary Table 6.

### 9.5.9 Cross-sectional results for population-within models

Supplementary Table 7 summarizes the estimation metrics from all 20 PW configurations, and Supplementary Fig. 53–72 show their results. Based on the overall performance, we rank the model classes from worst to best as follows: LR, CNN, MLP, CRS, and CRT. Here, we discuss the performance of LR, MLP, and CRT classes as representative and compared them to their SS counterparts. We found the LR class (PW01–PW04, Supplementary Fig. 53–56) to be the worst performer of all PW models, with its best configuration (PW02) achieving a determination coefficient of $r^2_{\text{a}} = 0.05$ $(p < 0.001)$, concordance coefficient of $\hat{\rho}_{c,\text{a}} = 0.2$, ME of $-3.99$ mm Hg, LOA of $[-41.1, 27.07]$ mm Hg, and MAE±SDAE of $13.39 \pm 11.47$ mm Hg for SBP. The corresponding DBP metrics for this configuration are $r^2_{\text{a}} = 0.02$ $(p < 0.001)$, $\hat{\rho}_{c,\text{a}} = 0.14$, ME $= -2.4$ mm Hg, LOA $= [-30.23, 21.11]$ mm Hg, and MAE±SDAE $= 9.89 \pm 8.58$ mm Hg. The cumulative percentages $\mathcal{P}_5$, $\mathcal{P}_{10}$, and $\mathcal{P}_{15}$ for PW02 are $27\%$, $48\%$, and $64\%$ for SBP, respectively; and $34\%$, $61\%$, and $78\%$ for DBP, respectively. Compared to their SS counterparts (SS01–SS04), the PW LR results are substantially worse, demonstrating their inability of simple linear architecture to capture large variability in the PW dataset.

For MLP configurations (PW05–PW08, Supplementary Fig. 57–60), we found that image-based models (PW05 and PW06) outperformed BioZ-based models (PW07 and PW08). For example, PW05 configuration produces waveform metrics AMAE $= 4.62$ mm Hg and ARMSE $= 5.08$ mm Hg, while the corresponding metrics for PW07 are AMAE $= 5.07$ mm Hg and ARMSE $= 5.57$ mm Hg. When comparing PW MLP models with SS MLP models, we found the following opposing trends: image-based PW models outperformed their SS counterparts (SS05 and SS06), while Bioz-based PW models exhibited performance degradation compared to their corresponding SS models (SS07 and SS08). For example, PW05 configuration yielded $r^2_{\mathrm{a}} = 0.75$ ($p < 0.001$), $\hat{\rho}_{c,\mathrm{a}} = 0.85$, ME and LOA of $-2.08$, $[-17.94, 12.49]$ mm Hg, MAE±SDAE of $5.62 \pm 5.39$ mm Hg for SBP; and $r^2_{\mathrm{a}} = 0.77$ ($p < 0.001$), $\hat{\rho}_{c,\mathrm{a}} = 0.87$, ME and LOA of $0.17$, $[-10.21, 11.28]$ mm Hg, MAE±SDAE of $3.87 \pm 3.64$ mm Hg for DBP. For comparison, the corresponding metrics for SS05 are $r^2_{\mathrm{a}} = 0.52$ ($p < 0.001$), $\hat{\rho}_{c,\mathrm{a}} = 0.7$, ME and LOA of $-0.49$, $[-24.24, 25.08]$ mm Hg, and MAE±SDAE of $8.44 \pm 9.81$ mm Hg for SBP; and $r^2_{\mathrm{a}} = 0.53$ ($p < 0.001$), $\hat{\rho}_{c,\mathrm{a}} = 0.72$, ME and LOA of $0.35$, $[-15.82, 18.49]$ mm Hg, MAE±SDAE of $6 \pm 6.76$ mm Hg for DBP Supplementary Table 6). Here, the PW05 configuration achieved better correlation, lower bias, and also narrower LOA.

In contrast, comparing SS08 and PW08 revealed a particularly sharp degradation in performance, indicated by reduction in both correlation metrics, increased bias and increased error variance. Specifically, for SBP metrics we observed the following trends (from SS08 to PW08): $r^2_{\mathrm{a}}$ ($0.78 \rightarrow 0.7$), $\hat{\rho}_{c,\mathrm{a}}$ ($0.88 \rightarrow 0.82$), ME ($-0.47 \rightarrow -1.02$ mm Hg), LOA ($[-14.95, 13.17] \rightarrow [-17.26, 15.31]$ mm Hg), MAE ($5.19 \rightarrow 6.31$ mm Hg), and SDAE ($4.89 \rightarrow 5.37$ mm Hg). The corresponding trends in DPB metrics are: $r^2_{\mathrm{a}}$ ($0.76 \rightarrow 0.64$), $\hat{\rho}_{c,\mathrm{a}}$ ($0.87 \rightarrow 0.73$), ME ($-0.30 \rightarrow 2.93$ mm Hg), LOA ($[-11.18, 10.58] \rightarrow [-9.82, 16.38]$ mm Hg), MAE ($4.09 \rightarrow 5.61$ mm Hg), and SDAE ($3.61 \rightarrow 4.6$ mm Hg). This opposing trend along input modality and dataset composition can be attributed to the substantial difference in model size: ≈120M parameters for image-based MLP models in contrast to ≈4.8M for BioZ models (Supplementary Table 5). As a consequence, for MLP architectures with sufficient size, the PVI reconstruction pathway combined with population-scale training offers advantages over direct BioZ inputs. This relationship between model size and dataset size also extends to the CNN model class, with PW CNN configurations (PW09–PW12) outperformed the corresponding SS configurations (SS09–SS12).

We also observed an increase in performance for the PW CRT models PW13, PW15 and PW16 (Supplementary Fig. 65, 67 and 68, respectively) compared to their SS counterparts. These three PW models are also among the best performers, with $0.8 \leq r^2_{\mathrm{a}} \leq 0.85$ ($p < 0.001$), $0.88 \leq \hat{\rho}_{c,\mathrm{a}} \leq 0.91$, ME and LOA between $-1.79$, $[-14.06, 9.5]$ mm Hg and $-1.09$, $[-14.62, 12.1]$ mm Hg, and MAE±SDAE between $4.44 \pm 4.27$ mm Hg and $4.79 \pm 4.75$ mm Hg for SBP; and $0.8 \leq r^2_{\mathrm{a}} \leq 0.85$ ($p < 0.001$), $0.88 \leq \hat{\rho}_{c,\mathrm{a}} \leq 0.91$, $-0.21$ mm Hg $\leq$ ME $\leq 0.23$ mm Hg, LOA between $[-8.55, 8.45]$ mm Hg and $[-10.13, 9.76]$ mm Hg, and MAE±SDAE between $3.14 \pm 2.92$ mm Hg and $3.53 \pm 3.4$ mm Hg for DBP. The three models also achieved impressive AE distributions, with cumulative percentages $\mathcal{P}_5 \geq 64\%$, $\mathcal{P}_{10} \geq 89\%$, and $\mathcal{P}_{15} \geq 96\%$ for SBP; and $\mathcal{P}_5 \geq 77\%$, $\mathcal{P}_{10} \geq 95\%$, and $\mathcal{P}_{15} \geq 99\%$ for DBP.

We found PW14 (Supplementary Fig. 66) to be the worst performer in the CRT class, underperforming across all metrics: $r^2_{\mathrm{a}} = 0.36$ ($p < 0.001$), $\hat{\rho}_{c,\mathrm{a}} = 0.52$, ME and LOA of $-3.81$, $[-27.74, 20.08]$ mm Hg, MAE±SDAE $= 9.83 \pm 7.91$ mm Hg, $\mathcal{P}_5 = 33\%$, $\mathcal{P}_{10} = 59\%$, and $\mathcal{P}_{15} = 78\%$ for SBP; and $r^2_{\mathrm{a}} = 0.38$ ($p < 0.001$), $\hat{\rho}_{c,\mathrm{a}} = 0.59$, ME and LOA of $-2.07$, $[19.5, 15.29]$ mm Hg, and MAE±SDAE $= 6.94 \pm 6.52$ mm Hg, $\mathcal{P}_5 = 46\%$, $\mathcal{P}_{10} = 75\%$, and $\mathcal{P}_{15} = 90\%$ for DBP. The low accuracy of PW14 can be attributed to premature termination (142 epochs) due to insufficient improvement to reset the convergence tracker, whereas other PW CRT configurations (PW13, PW15, and PW16) demonstrated standard termination behavior (Supplementary Discussion 9.5.5) and converged after 218, 416, and 475 epochs, respectively.

The CRS class, established as the best performer across all SS configurations (SS17–SS20), showed modest improvement when trained on the PW datasets. The PW CRS models (PW17–PW20, Supplementary Fig. 69–72) achieved waveform metrics of $3.8$ mm Hg $\leq$ AMAE $\leq 4.17$ mm Hg and $4.11$ mm Hg $\leq$ ARMSE $\leq 4.56$ mm Hg; SBP metrics of $0.81 \leq r_{\text{a}}^2 \leq 0.84$ $(p < 0.001)$, $0.89 \leq \hat{\rho}_{c,\text{a}} \leq 0.9$, ME and LOA between $-1.2$, $[-13.28, 10.96]$ mm Hg and $-0.15$, $[-13.29, 12.48]$ mm Hg, and MAE±SDAE between $4.44 \pm 4.3$ mm Hg and $4.82 \pm 4.47$ mm Hg; and DBP metrics of $0.81 \leq r_{\text{a}}^2 \leq 0.83$ $(p < 0.001)$, $0.89 \leq \hat{\rho}_{c,\text{a}} \leq 0.9$, ME and LOA between $-0.75$, $[-10.12, 9.19]$ mm Hg and $0.40$, $[-8.54, 9.22]$ mm Hg, and MAE±SDAE between $3.28 \pm 3.04$ mm Hg and $3.57 \pm 3.24$ mm Hg. Similar to the PW CRT models, PW CRS models also achieved excellent AE distribution, with cumulative percentages $\mathcal{P}_5 \geq 63\%$, $\mathcal{P}_{10} \geq 89\%$, and $\mathcal{P}_{15} \geq 97\%$ for SBP; and $\mathcal{P}_5 \geq 75\%$, $\mathcal{P}_{10} \geq 96\%$, and $\mathcal{P}_{15} \geq 99\%$ for DBP. The performance of the PW CRS models is almost indistinguishable from PW CRT, despite the CRS class having dramatically fewer parameters.

#### 9.5.10 Cross-sectional generalizability of population-within models

To test the generalizability of our ML architectures, we used the PW model weights to perform inference on the holdout datasets. The inference results are summarized in Supplementary Table 8 and visualized in Supplementary Fig. 73–92. All models yielded similarly low accuracy, with waveform metrics of $9.85$ mm Hg $\leq$ AMAE $\leq 11.39$ mm Hg and $10.81$ mm Hg $\leq$ ARMSE $\leq 14.95$ mm Hg; SBP metrics of $r_{\text{a}}^2 \leq 0.08$ $(p < 0.001)$, $0.03 \leq \hat{\rho}_{c,\text{a}} \leq 0.19$, $-15.94$ mm Hg $\leq$ ME $\leq -7.76$ mm Hg, LOA between $[-37.19, 13.58]$ mm Hg and $[-46.79, 20.07]$ mm Hg, and MAE±SDAE between $13.36 \pm 9.53$ mm Hg and $18.75 \pm 12.48$ mm Hg; with corresponding DBP metrics of $r_{\text{a}}^2 \leq 0.12$ $(p < 0.001)$, $0.03 \leq \hat{\rho}_{c,\text{a}} \leq 0.31$, $-4.11$ mm Hg $\leq$ ME $\leq 3.08$ mm Hg, LOA between $[-19.87, 17.58]$ mm Hg and $[-27.92, 21.77]$ mm Hg, and MAE±SDAE between $8.21 \pm 5.79$ mm Hg and $11.14 \pm 8.17$ mm Hg. The cumulative percentages for all configurations are $\mathcal{P}_5 \leq 22\%$, $\mathcal{P}_{10} \leq 43\%$, and $\mathcal{P}_{15} \leq 62\%$ for SBP, and $\mathcal{P}_5 \leq 36\%$, $\mathcal{P}_{10} \leq 67\%$, and $\mathcal{P}_{15} \leq 87\%$ for DBP.

The low accuracy was uniform across all architectural choices, input modalities, and output modalities. To investigate whether this limitation is attributable to the cross-cohort BP discrepancy, we computed the $\mathcal{W}_{\text{label}}$ distance between BP waveform distributions in the PW train set and the holdout set (Supplementary Table 10). We found the distributional distance between holdout and PW train set to be between $27.31$ mm Hg and $27.63$ mm Hg, approximately $2.6$ times larger than the typical PW train-test gap (between $10.68$ and $11.06$ mm Hg). Given our SS robustness analysis (Supplementary Discussion 9.5.8) showing that several model classes (LR, MLP, CRS) exhibit statistically significant sensitivity to label gap, the observed performance degradation for these model classes is expected. For CRT models, our robustness analysis revealed no significant correlation with label gap. However, the low estimation accuracy on the holdout dataset suggests that CRT's strength in learning subject-specific patterns does not extend to cross-cohort generalization.

### 9.5.11 Cross-sectional results for population-disjoint models

We trained PD models where datasets from individual subjects were wholly assigned either to the train or test subset (Supplementary Discussion 8.5.3). The PD partition strategy is stricter than PW, as indicated by the larger train-test label gaps for PD configurations (between $16.66$ mm Hg and $25.08$ mm Hg) compared to their PW counterparts (between $10.68$ mm Hg and $11.06$ mm Hg), as shown in Supplementary Table 10. Because each PD configuration used a different random subject assignment, direct comparison between configurations conflates architectural differences with partition variability. However, comparing each PD configuration to its PW counterpart isolates the effect of partition strategy.

Supplementary Table 9 summarizes the estimation metrics from all 20 PD configurations, and Supplementary Fig. 53–112 show their results. As expected from the increased label gap, PD models showed substantial performance degradation relative to their PW counterparts (Supplementary Table 7). All PD models achieved waveform metrics AMAE between $8.67$ mm Hg and $15.90$ mm Hg, and ARMSE between $9.42$ mm Hg and $16.49$ mm Hg; SBP metrics of $r_{\text{a}}^2 \leq 0.32$ $(p < 0.05)$, $-0.32 \leq \hat{\rho}_{c,\text{a}} \leq 0.46$, ME between $-8.25$ mm Hg and $9.4$ mm Hg, LOA between $[-22.13, 22.01]$ mm Hg and $[-44.3, 38.2]$ mm Hg, MAE between $9.18$ mm Hg and $19.16$ mm Hg, and SDAE between $6.85$ mm Hg and $13.88$ mm Hg; with corresponding DBP metrics of $r_{\text{a}}^2 \leq 0.35$ $(p < 0.05)$, $-0.27 \leq \hat{\rho}_{c,\text{a}} \leq 0.44$, ME between $-5.05$ mm Hg and $6.55$ mm Hg, LOA between $[-12.11, 23.33]$ mm Hg and $[-28.98, 28.09]$ mm Hg, MAE between $7.48$ mm Hg and $13.65$ mm Hg, and SDAE between $5.85$ mm Hg and $9.02$ mm Hg. The cumulative percentages for all PD configurations are $\mathcal{P}_5 \leq 34\%$, $\mathcal{P}_{10} \leq 63\%$, and $\mathcal{P}_{15} \leq 83\%$ for SBP, and $\mathcal{P}_5 \leq 46\%$, $\mathcal{P}_{10} \leq 75\%$, and $\mathcal{P}_{15} \leq 89\%$ for DBP.

We found that most PD configurations were terminated after 500 epochs without convergence, while PD09, PD11, and PD15 were terminated prematurely at 52, 105, and 65 epochs, respectively, as shown in Supplementary Table 5 and Supplementary Fig. 32. These three models are also the best performers of all PD models. Specifically, PD09 achieved $r_{\text{a}}^2 = 0.28$ $(p < 0.05)$, $\hat{\rho}_{c,\text{a}} = 0.33$ for SBP and $r_{\text{a}}^2 = 0.35$ $(p < 0.05)$, $\hat{\rho}_{c,\text{a}} = 0.32$ for DBP; PD11 achieved $r_{\text{a}}^2 = 0.26$ $(p < 0.05)$, $\hat{\rho}_{c,\text{a}} = 0.36$ for SBP and $r_{\text{a}}^2 = 0.27$ $(p < 0.05)$, $\hat{\rho}_{c,\text{a}} = 0.41$ for DBP; and PD15 achieved $r_{\text{a}}^2 = 0.32$ $(p < 0.05)$, $\hat{\rho}_{c,\text{a}} = 0.46$ for SBP and $r_{\text{a}}^2 = 0.24$ $(p < 0.05)$, $\hat{\rho}_{c,\text{a}} = 0.44$ for DBP. However, the premature termination behavior suggests that they fortuitously encountered parameter combinations that satisfied the convergence threshold rather than reflecting genuine learning (Supplementary Discussion 9.5.5).

### 9.5.12 Cross-sectional generalizability of population-disjoint models

For PD models, the training and testing datasets are mutually exclusive. We thus view the testing results for PD models as equivalent to generalizability study and did not evaluate the models on the holdout datasets.

### 9.5.13 Longitudinal stability and frequency of recalibration of subject-specific models

Based on the performance of the SS configurations (Supplementary Discussion 9.5.6 and Supplementary Table 6), we selected the following $4$ configurations to perform a pilot assessment of frequency of recalibration: SS03, SS07, SS15, and SS17, which are the best performer from their respective model class LR, MLP, CRT, and CRS. For each configuration, we extracted $5$ subject-specific model weights corresponding to the subjects who also participated in the longitudinal study, and recalibrate them on the longitudinal datasets across $5$ days. In total, we trained $100$ longitudinal SS models ($4$ configurations $\times$ $5$ subjects $\times$ $5$ days). We evaluated their performance using the ARMSE as accuracy metric, since they are all waveform-output models.

The recalibration results are shown in Supplementary Table 11. We examine the results at progressively finer levels of granularity, beginning with overall trends averaged across all subjects and models, followed by subject-specific and model-specific patterns. Looking at the overall average performance (lower right corner of the table), we observed the following two trends. First, model performance degrades over time when evaluated on future unseen datasets without calibration. For example, at checkpoints $C_0$, ARMSE increased from $11.94$ mm Hg For $D_1$ to $22.54$ mm Hg for $D_5$. Second, progressive recalibration with data from previous days improves performance. For example, the ARMSE for dataset $D_5$ improved from $22.54$ mm Hg (achieved by $C_0$, cross-sectional checkpoint) to $20.21$ mm Hg (achieved by $C_4$ with data from previous day. Furthermore, same-day calibration significantly improved the ARMSE to $6.60$ mm Hg for $D_5$. These findings are expected, since temporal drift in physiological signals and electrode contact are expected to degrade model accuracy, while incorporating recent subject-specific data allows models to adapt to current conditions.

Examining model-averaged trends (bottom rows of the table) for individual subjects reveals substantial variability for inter-subject performance. We found that subject 095 had worse ARMSE compared to other subjects across most time points, with particularly severe degradation for $D_5$: ARMSE between $51.10$ and $59.64$ mm Hg across all calibration levels, compared to $8.34$ mm Hg and $21.56$ mm Hg for other subjects. When examining subject-averaged trends for each configuration (rightmost columns of the table), architectural differences became apparent. While all models exhibited performance degradation on future unseen data, feedforward models (SS03 and SS07) deteriorated far more severely than hybrid transformer models (SS15 and SS17). Specifically, the performance deterioration by $C_0$ from $D_1$ to $D_5$ was $12.30 \rightarrow 33.51$ mm Hg for SS03, $11.70 \rightarrow 27.84$ mm Hg for SS07, $12.21 \rightarrow 13.42$ mm Hg for SS15, and $11.56 \rightarrow 15.38$ mm Hg for SS17. Furthermore, even after recalibration, feedforward models either marginally exceeded or failed to match the baseline performance of transformer-based models. For example, the ARMSE for $D_3$ achieved by SS03 and SS07 at $C_2$ (i.e. after two days of recalibration) were $17.27$ and $13.62$ mm Hg, respectively. In contrast, the uncalibrated checkpoints $C_0$ of SS15 and SS17 achieved ARMSE of $10.60$ and $11.10$ mm Hg, respectively. Similar trends were observed across other daily datasets ($D_1, D_2, D_4, D_5$). These findings suggest that temporal modeling provides inherent robustness to longitudinal drift that cannot be easily recovered through recalibration alone by simpler architectures.

### 9.5.14 Ablation study

Based on cross-sectional performance, we selected PW15 and PW19 for the ablation study, yielding six experiments in total: PW15-A1 through PW15-A4, and PW19-A1 and PW19-A2. The training progress of our ablated models is summarized in Supplementary Table 12, and their results are shown in Supplementary Table 13 and Supplementary Fig. 114–119.

We observed that the A2 sinusoidal positional encoder (PE) replacement produced consistent deterioration or no improvement across all metrics for both PW15 and PW19. This contrasts with standard Transformer practice, where fixed sinusoidal encodings are often sufficient, and suggests that the BiLSTM PE captures temporal structure that a parameter-free encoding cannot replicate in this context. A similar pattern emerges for the A1 variant (depth ablation) on PW19. At 2.5M parameters, one of our smallest models, the depth ablation reduced parameter count to 2.3M parameters (a ≈8% compression) and produced deterioration across most metrics. This suggests the reference PW19 architecture is already near its capacity-performance optimum, leaving little room for structural compression.

In contrast, the A1 ablation had a substantially larger structural impact on PW15, reducing parameter count by ≈72% from 7.5M to 2.12M, while accelerating convergence from 416 to 324 epochs. Despite this compression, the ablated model PW15-A1 exhibited slight improvement on SBP estimation metrics. Specifically, the $\hat{\rho}_{c,\text{a}}$ and $\hat{\rho}_{c,\text{w}}$ improved from $0.91 \to 0.92$ and $0.70 \to 0.73$, respectively. Estimation bias and variance also tightened, with ME improvement from $-1.79 \to 0.26$ mm Hg and LOA improvement from $[-14.06, 9.50] \to [-11.64, 11.89]$ mm Hg. The SBP absolute error metrics followed the same trend, with MAE±SDAE improvement from $4.44 \pm 4.27 \to 4.26 \pm 4.09$ mm Hg, and corresponding gains in the cumulative distribution metrics $\mathcal{P}_5$ of $67\% \to 69\%$, $\mathcal{P}_{10}$ of $91\% \to 92\%$, and $\mathcal{P}_{15}$ of $97\% \to 98\%$. Other metrics for PW15-A1, including waveform errors and DBP metrics, either remain unchanged or showed slight deterioration compared to the reference model. The CosineAnnealingLR scheduler variant (PW15-A3) produced no meaningful improvement over the reference. However, when combined with the depth ablation, the CosineAnnealingLR scheduler variant (PW15-A4) yielded additional DBP gains: with $r^2_\text{a}$ increased from $0.85 \to 0.86$, $r^2_\text{w}$ from $0.54 \to 0.56$, $\hat{\rho}_{c,\text{a}}$ from $0.91 \to 0.92$, and $\hat{\rho}_{c,\text{w}}$ from $0.68 \to 0.69$; while maintaining comparable SBP performance.

# 10 Supplementary Discussion 10. Discussion

## 10.1 Ring sensors

The selection of electrode materials and form factor critically determines the accuracy and practicality of BioZ measurements and, in turn, cuffless BP prediction errors (see Supplementary Discussion 10.7.2 for an example of performance degradation due to electrode material). In term of signal quality, silver chloride (Ag/AgCl) electrodes with conductive wet gel are superior and therefore used as clinical gold standard. The conductive gel layer moistens the skin while providing a large effective contact area, thereby reducing the skin-electrode contact impedance. However, gel electrodes are disposable, susceptible to drying, and can cause skin irritation, making them impractical for long-term wearable use. Liquid metal electrodes (LME), such as gallium-based liquid alloys, is an emerging technology that addresses the disadvantages of wet gel electrodes. LMEs have been shown to outperform wet gel electrodes in acquiring high-fidelity ECG data.[229] Additionally, their fluidic state at room temperature enables superior conformability to irregular skin topology, making them ideal candidates for stretchable electrodes used in dynamic wearable applications.[230] However, a critical weakness of LME is the potential leakage and rapid surface oxidation of liquid metal without special protective encapsulation.[231] Furthermore, although gallium-based alloys have been shown to be biocompatible in bulk, its toxicity profile at nanoscale may be different and thus require a thorough characterization.[230] Finally, its highly sophisticated manufacturing process and availability create challenges for clinical deployment.

The limitations of wet electrode systems drive the search for dry electrode alternatives for wearable devices with practical and clinical application. Dry metal electrodes such as stainless steel offer compelling advantages including manufacturing at low cost, reusability, electronic integration, and convenient deployment without skin preparation.[125] Gold-plated and ENIG finishes provide additional corrosion resistance and biocompatibility while being cost-effective through compatibility with standard manufacturing processes for printed circuit boards.[232] However, dry metal electrodes exhibit substantially higher skin-electrode contact impedance compared to wet electrodes, especially at low measurement frequency. Therefore, post-production treatments such as deposition of PEDOT have been explored to reduce contact impedance of dry metal electrodes, especially when the electrode effective area is extremely small.[233] PEDOT is an electrochemically stable polymer that enables mixed ionic and electronic conduction at the skin-electrode contact layer, thereby reducing the contact impedance significantly compared to dry metal electrodes.[171,234] PEDOT is also biocompatible and has been implemented in wearable health monitoring applications.[235,236] However, PEDOT deposition processes, such as spin-coating and electrodeposition, are time consuming and not easily scalable at this moment in time and, importantly, the resulting layers deposited in the ring electrodes are mechanically fragile to finger insertion, thus limiting realistic long-term usage without significant degradation.

Carbon-based materials such as graphene and carbon nanotubes (CNTs) have attracted significant research interest as dry electrode materials due to their flexibility, large surface-to-volume ratios, and high electrical conductivity.[237–239] CNTs, in form of composite material with a conductive polymer, such as PEDOT-CNT, has been shown to achieve lower contact impedance than PEDOT while maintaining exceptional mechanical stability.[240–242] Graphene-based electrodes share structural similarities with CNTs but offer a unique advantage of optical transparency, thereby allowing optical and electrical monitoring.[243] Graphene electronic tattoos have been implemented in wearable BioZ devices for electrodermal activity monitoring[103] and cuffless BP monitoring.[244] However, the integration of carbon-based electrodes into wearable devices is challenging due to their complicated fabrication processes, which have not been standardized.[245] Furthermore, both materials have risk of cytotoxicity, with CNTs possessing significantly greater concern than graphene.[246] These challenges are major impediments to the use and ultimate broad adoption of carbon-based and graphene materials in commercial wearable devices despite their impressive electrical and mechanical properties.

In this study, we designed a ring platform with ENIG electrodes and benchmarked their performance against the same ENIG electrodes with PEDOT coating (Fig. 1, Supplementary Fig. 1). ENIG electrodes exhibited impedance characteristics comparable to PEDOT (low impedance magnitude with minimal phase distortion at the measurement frequency of 50 kHz), while maintaining predominantly conductive behavior suitable for bioimpedance sensing. Critically, PEDOT coatings degraded substantially after minimal use, which we noticed through visual inspection; whereas ENIG electrodes remained visually stable throughout our entire data collection phase. Unlike PEDOT, which requires specialized post-processing, ENIG finishing is integrated into well-established PCB fabrication techniques. These combined advantages position ENIG electrodes as a viable candidate for commercial wearable BioZ devices.

In addition to electrode material, the number of electrodes also requires consideration, especially for EII applications. In our design, we opted for an 8-electrode configuration over 16-electrode sensor array to balance hardware complexity while optimizing signal penetration (Supplementary Discussion 3.2). In general, the number of independent measurements is proportional to the number of electrodes, thereby increasing the information content and resolution of the reconstructed images.[77] However, the tradeoff for increased spatial resolution is lower temporal resolution, as the injection pair would have to be cycled across all electrodes in use before scanning the next frame. Using the Sciospec EIT32 system with 8-electrode configuration, we could achieve a consistent frame rate of $50 \pm 1.2$ fps. With more electrodes (e.g. at least 16), we saw a dramatic reduction of the frame rate to between $15$ and $25$ fps. Furthermore, a higher number of electrodes could lead to an overly conductive finger boundary layer that shunts the injected current along the skin-electrode interface, potentially reducing the measurement depth and therefore the ability to detect hemodynamic changes within the digital arteries.

## 10.2 Computational fluid dynamics simulations

The FEM–DEM simulation showed high-fidelity particle transport in the palmar digital arterial network in pulsatile hemodynamic condition (Fig. 2). Overall, transient particle disturbances and localized recirculation regions emerged near highly curved bifurcation segments. During systole, the rapid acceleration of the fluid phase produced strong radial velocity gradients across the lumen, enhancing particle transport and separation. In contrast, during diastole, the weakened flow field reduced particle motion and diminished velocity stratification. Our results are consistent with previous studies of particle transport in bifurcating vasculature.[247,248]

Notably, our occlusion simulation results revealed that moderate stenosis (75% area occlusion) induced only minor perturbations in pressure and flow rate due to effective redistribution within the arterial network, i.e. reduced perfusion in the occluded branch is offset by increased flow through an adjacent branch. In contrast, severe stenosis (96% area occlusion) lead to pronounced flow reduction, loss of pulsatility, and significant pressure damping across the affected branch. The resulting hemodynamic imbalance lead to compensatory increases in pressure and flow rate in adjacent vessels. While this mechanism preserved total flow, it can lead to elevated wall shear stress and increased mechanical load in the compensating vessel, potentially contributing to vascular remodeling or damage.

## 10.3 Image reconstruction algorithm

While open-source software for impedance imaging exist,[249,250] here we developed our own software specifically to provide a high-efficiency and customizable path for future implementation into resource-constrained embedded systems. Our imaging algorithm employs the one-step Gauss–Newton approach with smoothing regularization and customized finite element meshes specific to our application (Supplementary Fig. 19 and 20, Supplementary Discussion 6). A major challenge in EII is the degradation of image fidelity caused by unknown boundary geometries where the electrodes are placed.[251] However, by leveraging the rigid geometry of our ring and predefined electrode dimensions, we effectively eliminate boundary-induced artifacts, which contributes to stable and accurate forward modeling results. We validated both our forward and inverse solver against EIDORS (Supplementary Fig. 21 and 22, respectively). Under comparable initial conditions and regularization parameters, our implementation achieved negligible error, confirming numerical correctness (Supplementary Discussion 9.3).

## 10.4 Finger isopotential lines and volume impedance density

Our EQS simulation study offered a detailed look at the internal isopotential distribution lines (Supplementary Fig. 4–13), the BioZ measurement sensitivity (Supplementary Fig. 14–16), and the specific contribution of individual finger tissues (Supplementary Fig. 17 and Supplementary Table 3), which were lacking in the existing literature. The isopotential lines portray the spatial variation in potential between source and sink electrodes. Our analysis revealed strong distortion of the isopotential lines at the tissue interfaces, indicating sharp voltage gradients and electrical field discontinuity due to high conductivity contrast between adjacent tissues.[252]

Beyond characterizing internal field distributions, our simulation study also addressed the choice of electrode injection configuration. In EII, the established optimal injection pattern is the trigonometric pattern, in which currents are injected through all electrodes with appropriately scaled amplitude and sign to represent a sinusoidal pattern on the boundary.[57,60] However, the optimality of this pattern is restricted to the theoretical scenario with circular domain having concentric circular contrast regions and a continuous current distribution on the boundary. For real-world applications with non-circular domains and finite discrete electrodes, numerical simulations are required to determine the optimal bipolar injection configurations. Our investigation therefore provides a principled basis for identifying the optimal bipolar injection configuration for this realistic geometry, directly guiding the hardware design.

Furthermore, because our reconstruction methodology employs time-difference imaging, baseline impedance of time-invariant anatomical structures such as SAT or muscle reported in Supplementary Table 3 could be mathematically subtracted, leaving only the temporal variation for modeling purposes. This ensures that the conductivity signals extracted from the resulting images are purely representative of pulsatile hemodynamic variations, isolating the physiological blood conductivity and volume changes from the static finger BioZ background.

## 10.5 Signal reconstruction of synthetic conductivity waveform

We also performed numerical experiments with an anatomically accurate finger phantom and physiologically realistic conductivity waveform (Fig. 3, Supplementary Fig. 18). In an idealized simulation setting, our results revealed that both 8-electrode and 16-electrode configurations achieve sufficient morphological fidelity for conductivity waveform reconstruction, with relative errors on the order of 1% (Supplementary Discussion 9.3.4). However, with 16 electrodes, the reconstructed images showed clearer anatomical conductivity features matching the digital arteries in the finger phantom. Although these results were acquired in simulation under ideal noiseless measurement scenario, the 16-electrode reconstruction revealed the achievable upper bound when sufficient angular sampling is provided, demonstrating the feasibility of implementing EII in a wearable form factor. Furthermore, the simulation results validated our reconstruction pipeline while illustrating the tradeoff between hardware constraints and anatomical localization capability.

## 10.6 Experimental study

Our cross-sectional study collected data from $96$ subjects (Supplementary Fig. 24), representing the largest academic dataset for cuffless ring BP monitoring based on BioZ (Supplementary Table 14). This work is also the first to acquire multi-channel BioZ data at the finger using a ring, and the first to perform impedance imaging at this anatomical site for cuffless BP monitoring, establishing a new paradigm for wearable cardiovascular monitoring that balances clinical utility with practical wearability with a standard form factor. Additionally, in our study we conducted for the first time a pilot longitudinal study tracking $5$ subjects over multiple consecutive days to assess temporal stability and recalibration requirements. While limited in scale, this longitudinal component of our study can serve as a template for evaluating long-term deployment feasibility of wearable BP monitoring technology.

Beyond the primary BioZ and reference BP measurements, we also acquired ultrasound images of the finger vasculature for all subjects (Supplementary Fig. 25). While these ultrasound data were not utilized in the current EII analyses, they represent a valuable resource for future work on our image reconstruction pipeline, potentially serving as ground-truth references for ML-based reconstruction algorithms or providing subject-specific anatomical priors in the finite element meshes to improve the accuracy of our forward and inverse solvers.

## 10.7 Machine learning

We used the conductivity images as input to our ML models for cuffless BP estimation and compared their performance against direct multi-channel BioZ input. Both input modalities achieved comparable accuracy across multiple model architectures, with image-based inputs showing marginal improvements in certain configurations such as the CRS model. These results demonstrated that imaging-based approaches are competitive with direct BioZ input data for cardiovascular monitoring applications. Importantly, it is worth noting that the reconstruction algorithm employed in our implementation prioritizes computational efficiency over image quality through simplified regularization schemes. Advanced reconstruction algorithms producing high-quality images would directly translate to improved feature extraction for downstream ML models, representing a clear trajectory for future performance gains.

Our ablation study revealed that the sequential structure of the input signals demands a positional encoder with learnable weights. Furthermore, structural compression of the CRT class through depth reduction yielded marginal performance gains, indicating that the larger CRT configurations (between 7.1M and 7.5M parameters) carry architectural redundancy and these could be further optimized without sacrificing estimation fidelity. In contrast, the compact CRS models degraded under all ablations, suggesting that they already operated near their representational capacity.

### 10.7.1 Comparison to existing literature for BP estimation with ring sensors

Current finger-based cuffless BP estimation technologies predominantly focus on single-modality sensors, such as PPG or single-channel BioZ, to derive discrete SBP and DBP. A summary of these studies is given in Supplementary Table 14. Here we benchmark our imaging and modeling framework against these established methods, comparing the advantages of integrating multi-channel BioZ and imaging data to enable continuous, cuffless BP waveform prediction which had been previously unaddressed in the ring form factor.

Supplementary Table 14 also includes our results from models SS17, SS19, PW13, and PW15-A4. We selected these models for comparison as they represent the multiple dimensions of our ML approach: model architecture (CRT vs. CRS), input modality (image vs. BioZ), and dataset composition (SS vs. PW). In what follows, we compare our results with the tabulated studies on a case-by-case basis. For completeness, we first summarize the context of each study, e.g. sensing modality, model architecture, cohort size, before comparing the performance metrics. However, we note that a fair comparison is difficult to achieve due to fundamental differences between our work and the existing studies. In 6 out of 9 studies, the authors reported mean and standard deviation of errors (ME±SDE) as a measure of accuracy instead of mean and standard deviation of absolute errors (MAE±SDAE).[198,253–257] As discussed in Supplementary Discussion 9.5.2, ME can appear small due to cancellation of positive and negative error values, which could lead to overly optimistic interpretations. On the other hand, ME±SDE is most informative when reported within a Bland–Altman analysis using a reference standard, as it enables assessment of potential systematic bias in the finger-based sensor technology under evaluation.[125] Here, instead, we report a comprehensive analysis of performance metrics to quantitatively evaluate the overall performance metrics of our models.

The works from Lee et al.[253] and Kim et al.[254] are validation studies for the CART-I Plus commercial device (Sky Labs Inc., Gunpo, Korea). The CART-I Plus ring measured PPG and estimated the fiducial points through a multi-step convolutional encoder.[258] Both studies measured cohorts with mixed cardiovascular conditions: Lee et al. with $33$ subjects ($N = 15$ on medication) and Kim et al. with $N = 89$ subjects ($18$ on medication). Both studies used a estimation model that was trained on $N = 4,185$ subjects under surgery.[258] The results from Lee et al. yielded $r_\text{a}^2$, ME, and LOA of $0.77$, $1.74$ mm Hg and $[-11.38, 14.85]$ mm Hg for SBP, and $0.71$, $-3.24$ mm Hg and $[-16, 9.5]$ mm Hg for DBP, respectively. With considerably less training data, our PW13 model achieved higher regression metrics ($r_\text{a}^2$ of $0.83$ for SBP and $0.82$ for DBP), slightly smaller bias (ME of $-1.61$ mm Hg for SBP and $0.23$ mm Hg for DBP), and slightly tighter LOA ($[-14.38, 10.49]$ mm Hg for SBP and $[-9.1, 9.52]$ mm Hg for DBP). On the other hand, Kim et al. achieved slightly better regression metrics and slightly smaller bias and LOA than ours, with $r_\text{a}^2$ of $0.88$ and $0.9$, ME and LOA of $0.16$, $[-11.41, 11.72]$ and $-0.07$, $[-9.26, 9.1]$ mm Hg for SBP and DBP, respectively.

Schukraft et al.[259] evaluated the Senbiosys SBF2003 commercial device (Senbiosys SA, Neuchâtel, Switzerland). The SBF2003 estimated uncalibrated SBP and DBP using pulse wave analysis (PWA) method with 27 hand-picked features extracted from the PPG waveforms,[260] and required calibration to compute the correct offset for each subject. Because of the subject-specific calibration, we feel it is more appropriate to compare their results with our subject-specific SS17 model. They evaluated the device only on $N = 25$ subjects and achieved slightly higher regression metrics than ours: $r_\text{a}^2$ of $0.86$ and $0.88$ for SBP and DBP, respectively, compared to $0.81$ and $0.83$ achieved by our SS17 model. Their LOA is $[-11.31, 15.94]$ mm Hg for SBP, which is larger than ours ($[-12.98, 12.65]$ mm Hg); and $[-6.41, 7.39]$ mm Hg for DBP, which is tighter than ours ($[-9.09, 9.23]$ mm Hg). However, our model addresses a more complex task of full BP waveform prediction, which accounts for the performance difference while providing richer morphological interpretation.

Zhu et al.[255] developed the ringBP system for the uAita product (Aita Technology Co. Ltd., Shenzen, Guangdong, China). They used a MLP to estimate SBP and DBP from a set of 8 features, hand-picked from the PPG data and from the experimental protocol. One of the features is the height difference between two positions of the arm, which represents a complex procedure likely not being able to be performed by elderly people or at home, ultimately impact patient adherence to long term monitoring. They trained the MLP on $N = 85$ subjects with leave-one-out cross validation strategy and achieved $r_\text{a}^2$ of $0.86$ for SBP and $0.81$ for DBP, which are similar to our PW13 results ($0.83$ and $0.82$ for SBP and DBP, respectively). They achieved smaller ME but with larger LOA ($0.01, [-13.21, 13.22]$ and $0.02, [-12.27, 12.31]$ mm Hg for SBP and DBP, respectively, compared to ours ($-1.69, [-14.38, 10.49]$ and $0.23, [-9.10, 9.52]$ mm Hg for SBP and DBP, respectively).

Tang et al.[261] developed RingTool, an open-source toolkit that analyzed PPG and accelerometer data to estimate SBP and DBP among other cardiovascular parameters. For each separate estimation task, they trained and compared various neural network models, including Transformer and Mamba. Among the studies listed in Supplementary Table 14, this is the closest to our work in term of ML design and training framework. They trained the models on $N = 34$ subjects with 5-fold cross validation and reported that the Transformer architecture yielded the best performance. However, their results ($r^2_{\text{a}}$ and MAE±SDAE of $0.13$, $13.33 \pm 0.78$ mm Hg and $0.0$, $7.56 \pm 0.54$ mm Hg for SBP and DBP respectively) are worse to our PW13 model ($r^2_{\text{a}}$ listed above, with MAE±SDAE of $4.70 \pm 4.39$ mm Hg and $3.44 \pm 3.14$ mm Hg for SBP and DBP, respectively).

Fortin et al.[257] implemented the CNAP2GO ring (CNSystems Medizintechnik GmbH, Graz, Austria) that directly estimates mean arterial BP (mBP) using the volume clamp (VC) method. They validated the device on $N = 46$ subjects undergoing neurosurgery and acquired reference mBP through an arterial catheter. For comparison, here we computed mBP from our estimated SBP and DBP as mBP $= 1/3 \cdot$ SBP $+ 2/3 \cdot$ DBP. Our PW13 model achieved ME and LOA of $-0.38$, $[-10.2, 9.52]$ mm Hg for mBP estimation, which is slightly better than their results (with ME and LOA of $-1.0$, $[-14.8, 12.7]$ mm Hg). Similar to Fortin et al., Panula et al.[262] also implemented a wearable device that uses the VC method to measure mBP directly. However, their device also measured DBP and fit a subject-specific nonlinear regression model to determine the SBP from the two measured quantities. They validated their device on $N = 43$ subjects and achieved $r^2_{\text{a}}$, ME, and LOA of $0.62$, $-3.5$ mm Hg and $[-20, 13]$ mm Hg for SBP, and $0.79$, $-4$ mm Hg and $[-13, 4.6]$ mm Hg for DBP, respectively. In contrast, our SS19 model achieved better results across all relevant metrics ($r^2_{\text{a}}$, ME, and LOA of $0.8$, $0.31$ mm Hg and $[-12.59, 13.42]$ mm Hg for SBP, and $0.83$, $0.51$ mm Hg and $[-8.53, 9.49]$ mm Hg for DBP, respectively).

Ni et al.[263] proposed a wearable ring that estimated BP with eddy current. The ring used a subject-specific regression model to determine SBP and DBP with input features derived from the frequency evolution of the current. The authors validated their device with a limited cohort of $N = 10$ subjects and achieved lower regression metrics ($r^2_{\text{a}}$ of $0.77$ for SBP and $0.64$ for DBP) compared to our SS19 model ($0.8$ for SBP and $0.83$ for DBP), although with lower ME±SDAE ($3.64 \pm 2.71$ mm Hg for SBP and $3.2 \pm 2.66$ mm Hg for DBP) compared to ours ($4.85 \pm 4.63$ mm Hg for SBP and $3.4 \pm 3.01$ mm Hg). However, we do not believe their results are representative of true model performance given the limited sample size evaluated, with predominantly male participants ($8 : 2$ male:female ratio).

Finally, Sel et al.[198] implemented a ring that estimates BP from single-channel BioZ data. For each subject, they implemented two different subject-specific AdaBoost regression models to estimate SBP and DBP independently from a set of $50$ BioZ features. The authors trained and evaluated the regression models on a cohort of $N = 10$ subjects ($90\%$ male), aged between $19$ and $26$ years. They achieved ME and LOA of $0.11$, $[-10.22, 10.44]$ mm Hg for SBP and $0.11$, $[-7.48, 7.77]$ mm Hg for DBP, slightly better than our PW15-A4 model (with ME and LOA of $1.21$, $[-10.34, 12.33]$ mm Hg for SBP and $1.19$, $[-7.02, 9.81]$ mm Hg for DBP. As a metric for trend alignment, the authors reported the mean correlation coefficient across all subjects, which we denote here as $r_{\text{m}}$. To facilitate comparison, we transformed the correlation coefficient to the determination coefficient $r^2_{\text{m}}$, and compared with our weighted determination coefficient $r^2_{\text{w}}$. However, as mentioned in Supplementary Discussion 9.5.2, the averaged metric $r^2_{\text{m}}$ can be biased when not accounted for sample size and provide overly optimistic determination coefficient value. Despite this, we found that our weighted coefficient exceeded their averaged coefficient for SBP ($r^2_{\text{w}} = 0.62$ vs. $r^2_{\text{m}} = 0.58$), while it was lower than theirs for DBP ($r^2_{\text{w}} = 0.56$ vs. $r^2_{\text{m}} = 0.66$).

Beyond these performance differences, our study differed from Sel et al.[198] in several fundamental aspects. First, we evaluated on a 10-fold larger cohort with broader diversity in both age and gender ($96$ subjects, aged between $18$ and $64$ years, with $58\%$ female). Second, they trained two separate models to predict each fiducial point (one model for SBP and one model for DBP) on individual subjects. In contrast, our PW15-A4 is a single model trained to predict full BP waveforms, from which the fiducial points were extracted–a more complex prediction task. Third, Sel et al. averaged their extracted features and BP labels across $10$ consecutive periods with $50\%$ overlaps, thereby substantially reducing the variability of the target quantities. Indeed, our true SBP and DBP had standard deviations of $15.2$ mm Hg and $11.0$ mm Hg, respectively, compared to their substantially lower $9.6$ mm Hg and $8.4$ mm Hg in their study, respectively. Furthermore, the authors did not provide a detailed explanation on how they handled the overlapping windows during train:test partition, which could have inadvertently led to data leakage thus creating a scenario with overoptimistic accuracy. Finally, we also report cumulative error distributions and concordance coefficients providing clinically interpretable performance benchmarks not reported in their work; and distributional distance that helps evaluate model robustness. Considering these fundamental differences in measurement principles, modeling approach, experimental study, and model evaluation, we view our prediction task as more challenging than that of Sel et al. Despite this increased complexity of our overall approach, our technology achieves comparable, and in some cases improved, performance while predicting full BP waveforms across a broader cohort of participants.

### 10.7.2 Comparison to existing literature for cuffless BP estimation with wrist sensors

Despite the experimental differences and for completeness, we also compare our ring-based cuffless BP technology with wrist-based form factors using BioZ for cuffless BP prediction, listed in Supplementary Table 15. First, Huynh et al.[264] implemented a wrist-based sensor that determines pulse wave velocity (PWV) through multi-channel BioZ measurement at the wrist. With a cohort of $N = 15$ healthy subjects ($9 : 6$ male:female), the authors fitted a linear regression model to predict SBP and DBP from the PWV data. They achieved ME and LOA of $0.01$, $[-15.87, 15.88]$ mm Hg for SBP and $-0.06$, $[-10.76, 10.64]$ mm Hg for DBP. In contrast, our SS17 model achieved slightly larger ME than theirs, but with tighter LOA: $0.28$, $[-12.98, 12.65]$ mm Hg for SBP and $0.29$, $[-9.09, 9.23]$ mm Hg for DBP. The authors remarked that most errors “lie within the limits of agreement” and thus concluded that their estimated BP values “agree closely” with the reference device. However, this interpretation of LOA conflates its definition ($95\%$ coverage of errors) with accuracy (whether the LOA is clinically acceptable).[226] Furthermore, instead of reporting regression metrics between estimated and true BP values, the authors reported regression metrics between measured PWV and BP: $r^2_{\text{m}}$ of $0.66$ and $0.71$ for SBP and DBP, respectively. However, these values reflect sensors sensitivity while contributing little to interpreting the accuracy of BP estimation, since PWV and BP are dependent and both are manifestation of the same underlying cardiovascular dynamics.[265]

Ibrahim et al.[264] proposed a wrist-worn BioZ system with two columns of electrodes along the radial and ulnar artery and evaluated their approach on $N = 10$ healthy subjects ($7 : 3$ male:female). Similar to Sel et al.,[198] here the authors trained two separate AdaBoost re-gression models for SBP and DBP based on $50$ features extracted from the BioZ data. They reported impressive results, with $r^2_{\text{m}}$, ME, and LOA of $0.74$, $-0.02$ mm Hg, $[-5.03, 4.99]$ mm Hg for SBP and $0.59$, $-0.09$ mm Hg, $[-6.96, 6.78]$ mm Hg for DBP. These are better than our SS19 models, with $r^2_{\text{w}}$, ME and LOA of $0.5$, $0.31$ mm Hg, $[-12.59, 13.42]$ mm Hg for SBP and $0.5$, $-0.09$ mm Hg, $[-8.53, 9.49]$ mm Hg for DBP. However, their results may be overly optimistic due to several methodological factors. First, since the authors used a similar feature extraction and ML training approach to Sel et al., our earlier remarks in Supplementary Discussion 10.7.1 applies here. Second, the authors employed wet gel electrodes, which provide superior signal quality compared to our dry metal electrodes due to markedly lower skin contact impedance. We believe that these combined approaches substantially reduced the complexity of their prediction task. Subsequent works from the same group explored alternative electrode materials to address the inconvenience of wet gel electrodes and saw a sharp decline in BP prediction accuracy. Specifically, the follow-up work by Ibrahim et al.[266] used dry silver electrodes and trained a CNN autoencoder prior to BioZ feature extraction. They trained and evaluated their models on $N = 4$ subjects and yielded $r^2_{\text{m}}$, ME, and LOA of $0.62$, $0.2$ mm Hg, $[-12.54, 12.94]$ mm Hg for SBP and $0.64$, $0.5$ mm Hg, $[-9.3, 10.3]$ mm Hg for DBP. As another example, the work by Kireev et al.[244] explored graphene electronic tattoos on $N = 7$ subjects and yielded $r^2_{\text{m}}$, ME and LOA of $0.79$, $0.17$ mm Hg, $[-11.14, 11.48]$ mm Hg for SBP and $0.77$, $0.16$ mm Hg, $[-8.6, 8.9]$ mm Hg for DBP. While both studies achieved higher metrics than our SS19 model, their performance decline demonstrate the impact of transitioning from clinical-grade to wearable electrodes. Furthermore, our other methodological concerns regarding their feature extraction and ML approach, i.e. low BP variability, potential temporal leakage, and separate models for SBP and DBP predictions, still remain; leaving ambiguity in interpreting their results and making a fair comparison challenging on our end.

Finally, Crandall et al. presented a cuffless BioZ smartwatch that employs a novel signal-tagged physics-informed neural network (sPINN) for hemodynamic monitoring.[125] By embedding Navier–Stokes equations into the loss function, the authors trained the sPINN to estimate spatially- and temporally-varying BP and blood velocity fields from single-channel BioZ data. The authors performed an exhaustive evaluation of their technology in $N = 75$ healthy subjects following a protocol with static and dynamic BP measurements; as well as $N = 86$ clinically-relevant patients, including $N = 32$ patients with hypertension (HTN), $N = 22$ patients with cardiovascular diseases (CVD), and $N = 32$ patients with other conditions. The sPINN yielded $r^2_{\text{a}}$ and MAE±SDAE of $0.58$ and $7.24 \pm 6.98$ mm Hg for SBP and $0.63$ and $5.45 \pm 5.19$ mm Hg for DBP in the healthy cohort; and $0.77$ and $7.26 \pm 8.46$ mm Hg for SBP and $0.81$ and $3.89 \pm 4.63$ mm Hg for DBP in the clinic cohort. Since our dataset only comprises of healthy subjects, we compared their PW sPINN model with our PW15-A4 model, which yielded $r^2_{\text{a}}$ and MAE±SDAE of $0.86$ and $4.22 \pm 3.96$ mm Hg for SBP and $0.86$ and $3.21 \pm 2.95$ mm Hg for DBP. While the sPINN model explicitly embeds fluid dynamic principles within a ≈1M parameter architecture, our findings suggest that cuffless BP estimation with high accuracy may be achievable without imposing physics-based constraints. Specifically, a sufficiently expressive data-driven model with sequence-modeling architecture and increased complexity (e.g., ≈2.1M parameters for ablated CRT models) coupled with multi-channel sensing strategy like ours have the potential to enrich spatial information and potentially contribute to capture the underlying hemodynamic structure required for accurate inference. However, this hypothesis requires further research to investigate the prediction limits of our approach also in clinically relevant populations.

# 11 Supplementary Discussion 11. Limitations and outlook

In this study, we developed a wearable ring with peripheral vascular imaging and BP monitoring capability for managing peripheral vascular health. This development consists of multiple stages: ring sensor design and fabrication, computational fluid dynamic simulations of the palmar arteries, imaging algorithm implementation, sensitivity analysis with computational finger phantom, experimental validation in controlled laboratory conditions, and cuffless BP estimation through ML. At each stage, design decisions were made under practical constraints. In what follows, we examine the limitations specific to each research component and outline directions for future work.

Our EIS and CV characterization was limited in scope, having sampled only 10 electrodes from a production batch of 500 electrodes. Furthermore, our characterization did not include physical (such as surface deformation due to wear and tear) and chemical (such as oxidation, or corrosion due to sweat) stability of the ENIG layer over time. These characterizations would not only disentangle electrode variability from other experimental factors, but also provide uncertainty bounds for downstream stages, e.g. experiment design, data filtering and image reconstruction. Future research should include these additional studies in order to establish standardized sensor validation protocols, a prerequisite for clinical deployment.

Our study employed the FEM–DEM framework using a point-particle approximation. While this approach allowed efficient simulation of millions of particles, our model assumed one-way coupling where the particle phase did not influence the fluid phase. This assumption is appropriate for dilute suspensions but can be insufficient at higher particle concentration when accounting for higher hematocrit concentration. Future work should focus on extending the model to include two-way or four-way coupling to account for particle-fluid and particle-particle interactions. Such extensions would enable more accurate modeling of dense suspensions and collective particle dynamics. Furthermore, our model simulated red blood cells as rigid spheres and thus did not explicitly capture their deformability. Development of deformable particle models or reduced-order representations of red blood cells would improved prediction of particle transport physics necessary for further evaluation of finger flow and perfusion. Additionally, the current simulations assume rigid wall vessels. A combined numerical framework with deformability of the vascular wall would provide a more physiologically realistic representation of hand circulation and improve predictions of pulsatile hemodynamics in finger arteries. Finally, since blood conductivity contributes to BioZ signals, future studies will also incorporate conductivity models coupled with particle volume fraction fields to calculate the electrical signals measured by the ring.[125]

We identified the following four limitations with our image reconstruction algorithm. First, although our forward solver incorporates the electrode contact impedance through the CEM, we did not estimate these impedance values during the measurement. Instead, we assumed a constant contact impedance across all electrodes and at all time. Second, without prior information about the conductivity distribution, our inverse solver was initialized under homogeneous assumption, from which the Jacobian was computed and held fixed throughout the reconstruction. Third, our difference imaging approach is based on the linearized inverse problem which assumes small conductivity fluctuation localized at the arteries. However, this assumption may not hold when there is a global disturbance due to motion artifact. Finally, our reconstruction cost function, while stabilized with the smoothness regularization term, did not account for other physiological constraints such as sharp boundaries between tissues and bounded conductivity ranges. In order to improve imaging accuracy, these limitations need to be addressed holistically, which requires a thorough redesign of the reconstruction pipeline. Accuracy begins with better accounting for electrode-specific contact impedance values, either by estimating them directly during measurement, or by reformulating them as unknowns in the inverse problem. In addition, anatomical information contained in ultrasound images, such as arterial depth and radius or tissue boundaries, should be extracted and incorporated as regularization constraint or as initial assumption. By addressing these points, a more rigorous inverse solver can be implemented with spatial-temporal regularization, which will enable accurate and problem-specific reconstruction of fast hemodynamic events.

Our numerical study characterized the sensitivity and penetration depth at different EII injection patterns. However, our results are limited for a single phantom model at a single frequency with perfect conductors as electrodes. Future research should extend our analysis by including multiple phantom models with diverse anatomical composition, using a wide range of injection frequencies, and accounting for electrode impedance. In addition, the design of imaging hardware and reconstruction algorithm would benefit from in silico investigations with focus on different experimental and physiological aspects, such as the inadvertent movement of electrodes between EII frames, the time-dependent dilation of the arteries during systole, or the drift in electrode impedance due to the accumulation of sweats.

The experimental results presented here were evaluated intentionally under strictly static conditions. Since the experimental study did not include exercise or dynamic movement as part of the protocol, the robustness of the BP waveform prediction for dynamic application against motion artifacts for ambulatory application remains to be determined in future work. Also, the study was initially conceived as a cross-sectional evaluation limited to healthy volunteers and did not include patients with diagnosed diabetes or other CVD conditions. Consequently, further research is required to determine the system's accuracy and performance within the specific target populations that would benefit most from continuous BP monitoring and flow imaging. Regarding the temporal stability of the system, the longitudinal experimental study was performed on a relatively limited dataset. This small window of data constrains the interpretation of long-term sensor performance and model reliability and frequency of recalibration required, suggesting that more extensive longitudinal tracking is necessary to validate the ring's performance over weeks or months. Finally, although this study represents the largest academic BioZ investigation conducted on a multi-channel BioZ ring form factor to date, the lack of broad generalizability can be attributed to the relatively small sample size. Future work should prioritize increasing the cohort size to provide a more comprehensive representation of physiological and population values, ensuring the model remains accurate across diverse demographics.

Our ML approach was intended to evaluate five distinct model architectures trained with various input/output configurations and different dataset compositions. This breadth-first approach, while establishing a comparative landscape from which future investigation can depart, entailed significant cost in term of training time and hardware resource. Future work will focus on further cross validating these architectures, which is especially important for the PW and PD training strategies since they have larger dataset gaps. In addition, we used the same hyperparameters for all model classes to ensure methodological consistency, but they may be suboptimal for each architecture. Furthermore, our ablation study was carried out only at the structural level, i.e. by removing specific layers and/or replacing specific blocks with simpler alternatives. To complement this, future studies should also validate the design choices at a finer scale, such as the use of separate derivative channels, or stacking multiple consecutive periods, or the use of augmented features. Finally, our results revealed that CRS and CRT classes achieved the highest estimation accuracy for both SS and PW training paradigm. However, when evaluated for the PW generalization test and the PD training paradigm, all architectures exhibited deterioration in performance. These are the conditions most susceptible to the methodological limitations outlined above, i.e. insufficient cross-validation, untuned hyperparameters, and unevaluated fine-grained design choices. This ambiguity motivates a more focused investigation and present new directions for our future work.

Our full pipeline, from signal acquisition and preconditioning, to image reconstruction and ML inference, poses challenges for existing wearable ring architectures. The ARM Cortex-M7, a widely adopted microcontroller for edge ML deployment, has 384 KB of static memory and 1 MB of flash storage.[267,268] For comparison, our best-performing model class CRT with up to 8M parameters could be compressed to 8 MB or 4 MB of storage through aggressive quantization with int8 or int4 representation, respectively; which still exceeds the 1 MB budget of the Cortex-M7. Furthermore, the 50 fps throughput required for real-time continuous monitoring would not only rapidly exhaust the relatively small batteries characteristic of the ring form factor. At the current state of the art, the full pipeline can be implemented on smartphone devices with storage and memory in orders of magnitude above that of microcontrollers. Consequently, a complete co-design of custom ASICs or parallelized neural processor might be required for edge implementation, wherein the multi-channel BioZ acquisition front-end, image reconstruction, and ML inference are unified into a single architecture optimized jointly for the target ultra-low power envelope and latency constraint without compromising accuracy.[269] As an intermediate step towards this fully embedded implementation, a hybrid-approach where multi-channel BioZ data is transmitted to a smartphone via low power Bluetooth will be evaluated as part of our future research efforts.

# 12 Supplementary Discussion 12. Statistical methods

Unless specified otherwise, all post-hoc analyses of simulation and ML results were carried out in MATLAB R2024a. For linear regression analyses between true and estimated (or simulated) values, we computed the $p$-value using the studentized permutation test with 50,000 iterations.[270,271] Here, the null hypothesis is that there is no correlation between the estimated values and the true values, i.e. $r = 0$ where $r$ is the Pearson's correlation coefficient. We considered the correlation $r \neq 0$ statistically significant if $p < 0.05$. Such analyses were performed to compare synthetic vs. reconstructed conductivity waveform (Fig. 3, Supplementary Fig. 18); evaluate ML performance dependency on dataset properties (Supplementary Fig. 113); and compare estimated vs. true SBP/DBP (Supplementary Table 6–9, Supplementary Fig. 33–112, subfigures a i and b i). Additionally for ML results, we evaluated the Bland–Altman limits of agreement (LOA) of the errors between estimated and true DBP/SBP (Supplementary Table 6–9, Supplementary Fig. 33–112, subfigures a ii and b ii). Here, we computed the lower and upper LOA as the 2.5$^{\text{th}}$ and 97.5$^{\text{th}}$ percentile of the error distribution, respectively. To compute the cumulative percentage $\mathcal{P}_\tau$ of absolute errors (AE) below a threshold $\tau$ (Supplementary Table 6–9, Supplementary Fig. 33–112, subfigures a iii and b iii), we first discretized the AE distribution uniformly with $\mathrm{d}p = 0.1$ mm Hg. We also used the same procedure to determine the cumulative distribution functions of estimated and true SBP/DBP, which are required when computing the distribution gap $\mathcal{W}_{\text{pred}}$ (Supplementary Table 6–9, Supplementary Fig. 33–112, subfigures a iv and b iv).

# Supplementary Tables

## Supplementary Table 1. Geometric and Windkessel parameters for palmar arterial system boundary faces

Geometric and Windkessel parameters for palmar arterial system boundary faces.

| | | Artery identifier | | | | | |
|---|---|---|---|---|---|---|---|
| **Parameters** | **Inlet** | **1** | **2** | **3** | **4** | **5** | **6** |
| Face diameter (mm) | 3.35 | 1.45 | 1.53 | 1.36 | 1.86 | 1.77 | 2.00 |
| Face area ($mm^2$) | 8.81 | 1.64 | 1.83 | 1.46 | 2.72 | 2.46 | 3.14 |
| Proximal resistance ($\mu g/mm^4 s$) | † | 7.3 | 6.4 | 8.3 | 4.1 | 4.6 | 3.4 |
| Distal resistance ($g/mm^4 s$) | † | 280 | 250 | 320 | 170 | 190 | 150 |
| Distal capacitance ($mm^4 s^2/kg$) | † | 11 | 12 | 10 | 18 | 16 | 21 |

# Supplementary Table 2. Physical and numerical parameters for particle-laden simulation

Physical and numerical parameters for particle-laden simulation.

| Parameter | Value | Units |
|---|---|---|
| Fluid dynamic viscosity | 3.5 | mg/(mm s) |
| Fluid density | 1 | $mg/mm^3$ |
| Particle density | 1 | $mg/mm^3$ |
| Particle diameter | 50 | μm |
| Mean fluid inflow | 358 | $mm^3/s$ |
| Mean particle inflow | 22.4 | $mm^3/s$ |
| Mean hematocrit | 6.26% | Dimensionless |
| Reynolds number | $\approx$39 | Dimensionless |
| Maximum number of particles | 1,641,264 | Dimensionless |
| Time step size | 1 | ms |
| Total physical time | 6 | s |
| Sampled physical time | 3–6 | s |
| Number of processors | 180 | Dimensionless |
| Computational cost | 69 | min |

# Supplementary Table 3. Contribution of specific tissues to finger bioimpedance

Contribution of individual tissues to the total absolute resistance, absolute reactance and impedance magnitude measured at the surface of the human finger phantom, at three different ring electrode configurations. The configurations are representative of the skip-2 injection and skip-1 measurement pattern for peripheral vascular impedance imaging system with 8 electrodes. Configuration 1 uses $(\ell_1, \ell_4)$ injection pair and $(\ell_8, \ell_2)$ measurement pair. Configuration 2 uses $(\ell_4, \ell_7)$ injection pair and $(\ell_3, \ell_5)$ measurement pair. Configuration 3 uses $(\ell_7, \ell_2)$ injection pair and $(\ell_4, \ell_6)$ measurement pair. R, absolute resistance contribution; X, absolute reactance contribution; Z, impedance magnitude contribution; SAT, subcutaneous adipose tissue.

| | **Configuration 1** | | | **Configuration 2** | | | **Configuration 3** | | |
|---|---|---|---|---|---|---|---|---|---|
| Tissue | R (%) | X (%) | Z (%) | R (%) | X (%) | Z (%) | R (%) | X (%) | Z (%) |
| Arteries | 0.290 | 0.511 | 0.289 | 5.973 | 2.324 | 5.835 | 3.853 | 2.646 | 3.799 |
| Cancellous bone | 0.673 | 1.975 | 0.673 | 10.444 | 2.693 | 10.198 | 1.051 | 1.758 | 1.037 |
| Cortical bone | 4.348 | 10.896 | 4.347 | 34.407 | 17.127 | 33.639 | 3.871 | 10.527 | 3.826 |
| Muscle | 0.004 | 0.017 | 0.004 | 0.020 | 0.026 | 0.020 | 0.001 | 0.009 | 0.001 |
| SAT | 93.572 | 64.601 | 93.343 | 32.731 | 34.693 | 32.197 | 84.110 | 14.953 | 82.924 |
| Skin | 0.020 | 6.547 | 0.177 | 0.168 | 16.775 | 1.951 | 0.133 | 57.143 | 1.521 |
| Ligament | 1.093 | 15.452 | 1.167 | 16.257 | 26.362 | 16.159 | 6.983 | 12.964 | 6.893 |

# Supplementary Table 4. Summary of experimental data

Summary of demographic information and dataset features from different cohorts. The train-test and holdout (test-exclusive) datasets are from the cross-sectional study. The demographic information was taken during the data collection appointment, while the dataset features were computed from the clean periods. The augmented samples were prepared by stacking $m = 5$ consecutive clean periods. Peak-peak bioimpedance (BioZ) was extracted from the high-passed, channel-averaged BioZ signal, (i.e. $\text{BioZ}_{\text{avg}}$, Supplementary Discussion 7.6). Peak-peak conductivity was extracted from the high-passed, pixel-averaged conductivity image sequences. BioZ, bioimpedance; BP, blood pressure; bpm, beats per minute. When applicable, data are given as mean ± standard deviation.

| | | Cross-sectional | | Longitudinal |
|---|---|---|---|---|
| | | **Train-test** | **Holdout** | |
| Cohort characteristics | Subjects | 91 | 5 | 5 |
| | Male:Female | 37:54 | 3:2 | 3:2 |
| | Age (years) | 28.4±9.1 | 26.2±3 | 27.2±2.6 |
| | Weight (kg) | 68.5±11.2 | 66.5±13.2 | 66.5±9 |
| | Height (cm) | 169.8±8.5 | 169.7±13.1 | 170.7±8.3 |
| | Body mass index (kg/m$^2$) | 23.8±3.6 | 22.9±1.7 | 23.0±4.0 |
| Dataset features | Raw periods | 298,239 | 17,621 | 55,258 |
| | Clean periods | 248,774 | 15,151 | 48,364 |
| | Augmented samples | 162,367 | 10,322 | 37,948 |
| | Systolic BP (mm Hg) | 128.3±15.3 | 133±12.8 | 121.3±14.8 |
| | Diastolic BP (mm Hg) | 82.5±11.1 | 80.3±10.3 | 76.4±11 |
| | Heart rate (bpm) | 74.1±12.8 | 74.8±12.7 | 80.3±13 |
| | Peak-peak BioZ (mΩ) | 6.3±5.0 | 5.5±1.8 | 10.5±14.0 |
| | Peak-peak conductivity (mS/m) | 9.6±7.1 | 8.7±4.8 | 15.5±11.69 |

# Supplementary Table 5. Model naming conventions and training summary

Naming conventions and summary of training progress for all model configurations. The model architectures, as well as model input and output configurations, are described in Supplementary Discussion 8.4. The generic identifier (id.) goes from 1 to 20, with prefixes corresponding to a particular dataset composition: SS for subject-specific, PW population-within, and PD for population-disjoint. Here, each SS configuration corresponds to a suite of $91$ sub-models, trained on individual SS datasets. The number of train samples, test samples, and epochs for SS configurations are sums of the corresponding values from the corresponding sub-models. On the other hand, each PW configuration consists of a single model trained on the population dataset, aggregated from the $91$ SS datasets. Similarly, PD models were trained on the same population dataset, albeit with a different partition strategy than PW. id., identifier; LR, Linear Regression class; MLP, Multilayer Perceptron class; CNN, Convolutional Neural Network class; CRT, Convolutional Recurrent Transformer class; CRS, Convolutional Recurrent Samba class; SS, subject-specific; PD, population-disjoint; PW, population-within.

| Model configurations | | | | | Subject-specific models | | | | | Population-within models | | | | Population-disjoint models | | | |
|---|---|---|---|---|---|---|---|---|---|---|---|---|---|---|---|---|---|
| id. | Model class | Input mode | Output mode | Parameters | id. | Train samples | Test samples | Epochs (total) | Epochs (average) | id. | Train samples | Test samples | Epochs | id. | Train samples | Test samples | Epochs |
| 1 | LR | image | waveform | 60,000,550 | SS01 | 102,434 | 11,367 | 339,015 | 3,725 | PW01 | 102,449 | 11,377 | 500 | PD01 | 143,102 | 19,265 | 500 |
| 2 | | image | fiducial | 2,400,022 | SS02 | 102,494 | 11,382 | 342,172 | 3,760 | PW02 | 102,301 | 11,352 | 500 | PD02 | 143,585 | 18,782 | 500 |
| 3 | | bioz | waveform | 2,400,550 | SS03 | 103,848 | 11,514 | 452,267 | 4,970 | PW03 | 102,309 | 11,348 | 500 | PD03 | 144,304 | 18,063 | 500 |
| 4 | | bioz | fiducial | 96,022 | SS04 | 102,463 | 11,370 | 441,661 | 4,853 | PW04 | 102,330 | 11,357 | 500 | PD04 | 144,492 | 17,875 | 500 |
| 5 | MLP | image | waveform | 120,056,650 | SS05 | 102,205 | 11,336 | 248,417 | 2,730 | PW05 | 104,277 | 11,582 | 149 | PD05 | 144,839 | 17,528 | 500 |
| 6 | | image | fiducial | 120,051,802 | SS06 | 102,407 | 11,372 | 242,093 | 2,660 | PW06 | 101,990 | 11,327 | 229 | PD06 | 144,265 | 18,102 | 500 |
| 7 | | bioz | waveform | 4,856,650 | SS07 | 102,496 | 11,368 | 113,579 | 1,248 | PW07 | 102,398 | 11,358 | 150 | PD07 | 143,249 | 19,118 | 500 |
| 8 | | bioz | fiducial | 4,851,802 | SS08 | 102,279 | 11,350 | 113,005 | 1,242 | PW08 | 102,179 | 11,342 | 137 | PD08 | 144,368 | 17,999 | 500 |
| 9 | CNN | image | waveform | 153,754,238 | SS09 | 102,432 | 11,360 | 44,378 | 488 | PW09 | 102,611 | 11,391 | 431 | PD09 | 144,950 | 17,417 | 52 |
| 10 | | image | fiducial | 153,741,902 | SS10 | 102,437 | 11,367 | 49,492 | 544 | PW10 | 102,395 | 11,363 | 174 | PD10 | 144,028 | 18,339 | 500 |
| 11 | | bioz | waveform | 100,447,922 | SS11 | 102,336 | 11,365 | 41,640 | 458 | PW11 | 102,456 | 11,367 | 137 | PD11 | 144,863 | 17,504 | 105 |
| 12 | | bioz | fiducial | 100,435,586 | SS12 | 102,327 | 11,351 | 43,985 | 483 | PW12 | 102,023 | 11,322 | 205 | PD12 | 144,909 | 17,458 | 500 |
| 13 | CRT | image | waveform | 7,186,866 | SS13 | 102,455 | 11,372 | 25,736 | 283 | PW13 | 104,339 | 11,585 | 218 | PD13 | 144,875 | 17,492 | 500 |
| 14 | | image | fiducial | 7,174,530 | SS14 | 102,109 | 11,317 | 24,473 | 269 | PW14 | 102,189 | 11,338 | 142 | PD14 | 144,473 | 17,894 | 500 |
| 15 | | bioz | waveform | 7,545,030 | SS15 | 102,112 | 11,335 | 22,681 | 249 | PW15 | 102,315 | 11,358 | 416 | PD15 | 143,474 | 18,893 | 65 |
| 16 | | bioz | fiducial | 7,532,694 | SS16 | 102,303 | 11,349 | 17,569 | 193 | PW16 | 102,257 | 11,350 | 475 | PD16 | 142,957 | 19,410 | 500 |
| 17 | CRS | image | waveform | 2,155,122 | SS17 | 105,471 | 11,700 | 47,339 | 518 | PW17 | 102,529 | 11,388 | 173 | PD17 | 143,253 | 19,114 | 500 |
| 18 | | image | fiducial | 954,594 | SS18 | 103,126 | 11,432 | 38,102 | 419 | PW18 | 102,452 | 11,380 | 118 | PD18 | 143,176 | 19m191 | 500 |
| 19 | | bioz | waveform | 2,513,286 | SS19 | 103,909 | 11,527 | 26,928 | 296 | PW19 | 102,406 | 11,369 | 418 | PD19 | 143,850 | 18,517 | 500 |
| 20 | | bioz | fiducial | 1,312,758 | SS20 | 103,853 | 11,512 | 21,321 | 234 | PW20 | 102,371 | 11,365 | 290 | PD20 | 144,962 | 17,405 | 500 |

# Supplementary Table 6. Estimation results of subject-specific models

Training results of all subject-specific (SS) model configurations. We aggregated the results from all $91$ sub-models and computed the metrics reported here. For regression analysis, we also evaluated the intra-subject performance by computing the regression metrics (determination coefficient and concordance coefficient) on individual subjects, and reported the weighted average. id., identifier; BP, brachial blood pressure; AMAE, average mean absolute error; ARMSE, average root mean square error; $r^2_{\mathrm{a}}$ and $r^2_{\mathrm{w}}$, aggregated and weighted determination coefficient, respectively; $\hat{\rho}_{c,\mathrm{a}}$ and $\hat{\rho}_{c,\mathrm{w}}$, aggregated and weighted concordance coefficient, respectively; ME and LOA, mean and limits of agreement of errors, respectively; MAE and SDAE, mean and standard deviation of absolute errors (AE), respectively; $\mathcal{P}_5$, $\mathcal{P}_{10}$, and $\mathcal{P}_{15}$, cumulative percentage of estimations with AE within 5, 10, and 15 mm Hg, respectively; $\mathcal{W}_{\mathrm{pred}}$, Wasserstein distance between true and predicted distribution; †, waveform metrics not applicable to models estimating fiducial BP.

| | BP waveform | | Systolic BP | | | | | | | | | | | | | Diastolic BP | | | | | | | | | | | | |
|---|---|---|---|---|---|---|---|---|---|---|---|---|---|---|---|---|---|---|---|---|---|---|---|---|---|---|---|---|
| id. | AMAE (mm Hg) | ARMSE (mm Hg) | $r^2_{\mathrm{a}}$ | $r^2_{\mathrm{w}}$ | $\hat{\rho}_{c,\mathrm{a}}$ | $\hat{\rho}_{c,\mathrm{w}}$ | ME (mm Hg) | LOA (mm Hg) | MAE±SDAE (mm Hg) | $\mathcal{P}_5$ | $\mathcal{P}_{10}$ | $\mathcal{P}_{15}$ | $\mathcal{W}_{\mathrm{pred}}$ (mm Hg) | $r^2_{\mathrm{a}}$ | $r^2_{\mathrm{w}}$ | $\hat{\rho}_{c,\mathrm{a}}$ | $\hat{\rho}_{c,\mathrm{w}}$ | ME (mm Hg) | LOA (mm Hg) | MAE±SDAE (mm Hg) | $\mathcal{P}_5$ | $\mathcal{P}_{10}$ | $\mathcal{P}_{15}$ | $\mathcal{W}_{\mathrm{pred}}$ (mm Hg) |
| SS01 | 6.93 | 7.25 | 0.50 | 0.32 | 0.68 | 0.50 | -1.10 | [-27.59, 21.04] | 8.39±10.82 | 45% | 73% | 87% | 2.36 | 0.54 | 0.29 | 0.72 | 0.47 | -0.72 | [-18.27, 14.81] | 5.76±7.23 | 59% | 85% | 94% | 1.36 |
| SS02 | † | † | 0.52 | 0.33 | 0.70 | 0.49 | -1.41 | [-33.18, 21.79] | 8.60±10.40 | 45% | 73% | 86% | 2.51 | 0.54 | 0.29 | 0.72 | 0.46 | -0.61 | [-19.34, 16.17] | 5.93±6.85 | 58% | 85% | 93% | 1.55 |
| SS03 | 5.00 | 5.27 | 0.71 | 0.36 | 0.84 | 0.54 | -0.88 | [-17.92, 14.68] | 5.85±5.85 | 56% | 84% | 94% | 0.97 | 0.71 | 0.30 | 0.84 | 0.47 | 0.07 | [-11.73, 11.70] | 4.25±4.29 | 69% | 92% | 98% | 0.46 |
| SS04 | † | † | 0.74 | 0.40 | 0.86 | 0.58 | -0.69 | [-17.93, 14.82] | 5.72±5.64 | 57% | 84% | 94% | 0.70 | 0.74 | 0.33 | 0.86 | 0.50 | -0.40 | [-11.96, 10.57] | 4.10±3.87 | 71% | 93% | 98% | 0.45 |
| SS05 | 7.13 | 7.46 | 0.52 | 0.28 | 0.70 | 0.44 | -0.49 | [-24.24, 25.08] | 8.44±9.81 | 45% | 71% | 85% | 1.98 | 0.53 | 0.23 | 0.72 | 0.39 | 0.35 | [-15.82, 18.49] | 6.00±6.76 | 57% | 83% | 93% | 1.30 |
| SS06 | † | † | 0.54 | 0.32 | 0.72 | 0.48 | -0.41 | [-22.96, 24.41] | 8.12±9.92 | 46% | 73% | 86% | 2.18 | 0.54 | 0.25 | 0.72 | 0.42 | -0.27 | [-16.23, 17.03] | 5.80±6.84 | 58% | 84% | 94% | 0.54 |
| SS07 | 4.66 | 5.12 | 0.77 | 0.41 | 0.87 | 0.60 | -0.72 | [-15.39, 13.71] | 5.49±5.16 | 56% | 84% | 95% | 0.77 | 0.77 | 0.36 | 0.88 | 0.55 | 0.71 | [-9.56, 11.73] | 4.01±3.64 | 71% | 94% | 99% | 0.72 |
| SS08 | † | † | 0.78 | 0.45 | 0.88 | 0.64 | -0.47 | [-14.95, 13.17] | 5.19±4.89 | 60% | 88% | 96% | 0.55 | 0.76 | 0.33 | 0.87 | 0.53 | -0.30 | [-11.18, 10.58] | 4.09±3.61 | 70% | 93% | 99% | 0.46 |
| SS09 | 28.85 | 29.27 | 0.37 | 0.56 | 0.15 | 0.07 | -37.20 | [-64.82, -12.34] | 37.28±12.82 | 1% | 2% | 4% | 37.20 | 0.38 | 0.52 | 0.18 | 0.09 | -23.11 | [-41.78, -3.15] | 23.21±8.86 | 3% | 6% | 15% | 23.11 |
| SS10 | † | † | 0.42 | 0.54 | 0.21 | 0.12 | -31.06 | [-53.70, -3.01] | 31.23±12.41 | 3% | 6% | 11% | 31.06 | 0.40 | 0.49 | 0.24 | 0.13 | -19.52 | [-35.93, 0.73] | 19.83±8.44 | 5% | 13% | 28% | 19.52 |
| SS11 | 26.39 | 26.74 | 0.57 | 0.58 | 0.20 | 0.09 | -33.66 | [-50.82, -9.73] | 33.69±9.96 | 1% | 3% | 6% | 33.67 | 0.60 | 0.54 | 0.24 | 0.10 | -20.97 | [-33.39, -5.50] | 21.00±6.83 | 2% | 8% | 17% | 20.97 |
| SS12 | † | † | 0.58 | 0.57 | 0.34 | 0.16 | -24.22 | [-43.03, -0.26] | 24.52±9.84 | 4% | 9% | 17% | 24.23 | 0.57 | 0.52 | 0.36 | 0.16 | -15.69 | [-28.78, 1.07] | 15.95±6.79 | 7% | 19% | 42% | 15.69 |
| SS13 | 6.02 | 6.33 | 0.71 | 0.65 | 0.82 | 0.55 | -3.43 | [-18.12, 13.92] | 7.41±5.34 | 38% | 72% | 92% | 3.43 | 0.73 | 0.60 | 0.85 | 0.55 | -1.55 | [-11.78, 10.92] | 4.87±3.60 | 58% | 91% | 99% | 1.55 |
| SS14 | † | † | 0.77 | 0.65 | 0.83 | 0.57 | -4.74 | [-17.85, 10.21] | 7.04±5.02 | 40% | 75% | 94% | 4.75 | 0.78 | 0.60 | 0.84 | 0.56 | -3.12 | [-12.44, 7.36] | 4.83±3.46 | 58% | 92% | 99% | 3.15 |
| SS15 | 5.12 | 5.47 | 0.76 | 0.63 | 0.86 | 0.62 | -2.03 | [-15.99, 13.01] | 6.09±4.79 | 49% | 82% | 95% | 2.05 | 0.76 | 0.59 | 0.87 | 0.59 | -0.66 | [-11.10, 9.94] | 4.28±3.35 | 65% | 94% | 99% | 0.78 |
| SS16 | † | † | 0.75 | 0.66 | 0.84 | 0.58 | -3.85 | [-18.18, 12.14] | 6.89±5.04 | 41% | 77% | 93% | 3.87 | 0.77 | 0.60 | 0.86 | 0.56 | -2.36 | [-12.67, 8.95] | 4.76±3.51 | 59% | 92% | 99% | 2.38 |
| SS17 | 4.09 | 4.47 | 0.81 | 0.48 | 0.90 | 0.59 | 0.28 | [-12.98, 12.65] | 4.73±4.46 | 64% | 90% | 97% | 1.17 | 0.83 | 0.48 | 0.90 | 0.60 | 0.29 | [-9.09, 9.23] | 3.39±3.05 | 77% | 96% | 99% | 0.95 |
| SS18 | † | † | 0.80 | 0.48 | 0.89 | 0.61 | 0.50 | [-12.93, 13.67] | 4.85±4.60 | 63% | 89% | 97% | 1.28 | 0.82 | 0.46 | 0.90 | 0.59 | 0.39 | [-9.09, 9.62] | 3.49±3.13 | 77% | 96% | 99% | 0.82 |
| SS19 | 4.14 | 4.53 | 0.80 | 0.50 | 0.89 | 0.63 | 0.31 | [-12.59, 13.42] | 4.85±4.63 | 63% | 89% | 97% | 1.15 | 0.83 | 0.50 | 0.91 | 0.63 | 0.51 | [-8.53, 9.49] | 3.40±3.01 | 77% | 97% | 99% | 0.74 |
| SS20 | † | † | 0.80 | 0.51 | 0.89 | 0.63 | 0.14 | [-13.17, 13.22] | 4.85±4.62 | 64% | 90% | 97% | 1.01 | 0.81 | 0.48 | 0.90 | 0.61 | -0.07 | [-9.72, 9.14] | 3.53±3.17 | 76% | 97% | 99% | 0.76 |

## Supplementary Table 7. Estimation results of population-within models

Training results of all population-within (PW) model configurations. For regression analysis, we also evaluated the intra-subject performance by computing the regression metrics (determination coefficient and concordance coefficient) on individual subjects, and reported the weighted average. id., identifier; BP, brachial blood pressure; AMAE, average mean absolute error; ARMSE, average root mean square error; $r^2_{\text{a}}$ and $r^2_{\text{w}}$, aggregated and weighted determination coefficient, respectively; $\hat{\rho}_{c,\text{a}}$ and $\hat{\rho}_{c,\text{w}}$, aggregated and weighted concordance coefficient, respectively; ME and LOA, mean and limits of agreement of errors, respectively; MAE and SDAE, mean and standard deviation of absolute errors (AE), respectively; $\mathcal{P}_5$, $\mathcal{P}_{10}$, and $\mathcal{P}_{15}$, cumulative percentage of estimations with AE within 5, 10, and 15 mm Hg, respectively; $\mathcal{W}_{\text{pred}}$, Wasserstein distance between true and predicted distribution; †, waveform metrics not applicable to models estimating fiducial BP.

| | BP waveform | | Systolic BP | | | | | | | | | | | Diastolic BP | | | | | | | | | | |
|---|---|---|---|---|---|---|---|---|---|---|---|---|---|---|---|---|---|---|---|---|---|---|---|---|
| id. | AMAE (mm Hg) | ARMSE (mm Hg) | $r^2_{\text{a}}$ | $r^2_{\text{w}}$ | $\hat{\rho}_{c,\text{a}}$ | $\hat{\rho}_{c,\text{w}}$ | ME (mm Hg) | LOA (mm Hg) | MAE±SDAE (mm Hg) | $\mathcal{P}_5$ | $\mathcal{P}_{10}$ | $\mathcal{P}_{15}$ | $\mathcal{W}_{\text{pred}}$ (mm Hg) | $r^2_{\text{a}}$ | $r^2_{\text{w}}$ | $\hat{\rho}_{c,\text{a}}$ | $\hat{\rho}_{c,\text{w}}$ | ME (mm Hg) | LOA (mm Hg) | MAE±SDAE (mm Hg) | $\mathcal{P}_5$ | $\mathcal{P}_{10}$ | $\mathcal{P}_{15}$ | $\mathcal{W}_{\text{pred}}$ (mm Hg) |
| PW01 | 11.23 | 11.73 | 0.11 | 0.05 | 0.26 | 0.09 | -9.11 | [-40.49, 19.31] | 13.92±11.28 | 24% | 46% | 62% | 9.14 | 0.13 | 0.05 | 0.31 | 0.07 | -4.49 | [-26.46, 15.74] | 9.15±7.48 | 35% | 63% | 81% | 4.51 |
| PW02 | † | † | 0.05 | 0.11 | 0.20 | 0.14 | -3.99 | [-41.10, 27.07] | 13.39±11.47 | 27% | 48% | 64% | 4.39 | 0.02 | 0.08 | 0.14 | 0.06 | -2.40 | [-30.23, 21.11] | 9.89±8.58 | 34% | 61% | 78% | 2.85 |
| PW03 | 12.75 | 13.34 | 0.01 | 0.07 | 0.08 | 0.05 | -6.36 | [-46.70, 28.28] | 15.48±12.37 | 22% | 42% | 58% | 6.61 | 0.02 | 0.07 | -0.11 | -0.02 | -2.56 | [-32.27, 22.56] | 10.64±8.59 | 31% | 57% | 75% | 3.99 |
| PW04 | † | † | 0.01 | 0.07 | 0.10 | 0.07 | -3.84 | [-43.66, 30.37] | 15.00±11.81 | 22% | 42% | 59% | 4.33 | 0.02 | 0.08 | -0.11 | -0.02 | -2.04 | [-31.54, 22.81] | 10.68±8.49 | 30% | 56% | 75% | 3.78 |
| PW05 | 4.62 | 5.08 | 0.75 | 0.45 | 0.85 | 0.59 | -2.08 | [-17.94, 12.49] | 5.62±5.39 | 57% | 85% | 94% | 2.38 | 0.77 | 0.40 | 0.87 | 0.57 | 0.17 | [-10.21, 11.28] | 3.87±3.64 | 73% | 94% | 98% | 1.10 |
| PW06 | † | † | 0.72 | 0.43 | 0.83 | 0.57 | -0.73 | [-17.05, 15.05] | 5.87±5.43 | 55% | 83% | 94% | 2.11 | 0.71 | 0.35 | 0.83 | 0.50 | -0.43 | [-12.41, 11.15] | 4.42±3.93 | 66% | 92% | 98% | 1.56 |
| PW07 | 5.07 | 5.57 | 0.71 | 0.44 | 0.83 | 0.57 | 0.01 | [-16.42, 15.47] | 6.01±5.29 | 52% | 83% | 94% | 2.06 | 0.72 | 0.38 | 0.84 | 0.54 | 0.14 | [-11.10, 11.51] | 4.28±3.83 | 68% | 93% | 98% | 1.40 |
| PW08 | † | † | 0.70 | 0.44 | 0.82 | 0.56 | -1.02 | [-17.26, 15.31] | 6.31±5.37 | 49% | 80% | 93% | 2.33 | 0.64 | 0.34 | 0.73 | 0.43 | 2.93 | [-9.82, 16.38] | 5.61±4.60 | 55% | 84% | 96% | 3.50 |
| PW09 | 6.45 | 7.13 | 0.68 | 0.35 | 0.59 | 0.31 | -8.37 | [-27.83, 8.44] | 9.84±7.51 | 31% | 58% | 78% | 8.44 | 0.71 | 0.32 | 0.76 | 0.40 | -2.16 | [-15.13, 9.68] | 4.96±4.25 | 61% | 88% | 97% | 3.38 |
| PW10 | † | † | 0.28 | 0.23 | 0.41 | 0.27 | -5.89 | [-31.79, 18.58] | 10.92±8.95 | 32% | 56% | 72% | 7.00 | 0.31 | 0.18 | 0.47 | 0.24 | -2.04 | [-21.28, 15.58] | 7.27±5.90 | 45% | 73% | 88% | 3.58 |
| PW11 | 6.58 | 7.16 | 0.69 | 0.41 | 0.59 | 0.36 | -5.96 | [-27.70, 10.67] | 8.71±7.50 | 38% | 67% | 83% | 7.60 | 0.71 | 0.37 | 0.72 | 0.40 | -1.35 | [-16.33, 10.64] | 5.13±4.36 | 59% | 88% | 96% | 3.81 |
| PW12 | † | † | 0.67 | 0.38 | 0.77 | 0.48 | -2.81 | [-20.69, 14.71] | 6.96±5.99 | 46% | 76% | 90% | 3.41 | 0.67 | 0.33 | 0.78 | 0.44 | -1.01 | [-14.24, 11.45] | 4.83±4.16 | 62% | 89% | 97% | 2.07 |
| PW13 | 3.95 | 4.27 | 0.83 | 0.57 | 0.89 | 0.67 | -1.61 | [-14.38, 10.49] | 4.70±4.39 | 64% | 90% | 97% | 2.40 | 0.82 | 0.51 | 0.89 | 0.64 | 0.23 | [-9.10, 9.52] | 3.44±3.14 | 77% | 96% | 99% | 1.41 |
| PW14 | † | † | 0.36 | 0.18 | 0.52 | 0.26 | -3.81 | [-27.74, 20.08] | 9.83±7.91 | 33% | 59% | 78% | 5.10 | 0.38 | 0.17 | 0.59 | 0.24 | -2.07 | [-19.50, 15.29] | 6.94±5.62 | 46% | 75% | 90% | 2.52 |
| PW15 | 3.69 | 4.00 | 0.85 | 0.61 | 0.91 | 0.70 | -1.79 | [-14.06, 9.50] | 4.44±4.27 | 67% | 91% | 97% | 2.10 | 0.85 | 0.54 | 0.91 | 0.68 | -0.21 | [-8.55, 8.45] | 3.14±2.92 | 81% | 97% | 99% | 1.01 |
| PW16 | † | † | 0.80 | 0.51 | 0.88 | 0.66 | -1.09 | [-14.62, 12.10] | 4.79±4.75 | 65% | 89% | 96% | 1.77 | 0.80 | 0.44 | 0.89 | 0.61 | -0.04 | [-10.13, 9.76] | 3.53±3.40 | 77% | 95% | 99% | 1.01 |
| PW17 | 4.17 | 4.56 | 0.81 | 0.53 | 0.89 | 0.66 | -0.15 | [-13.29, 12.48] | 4.82±4.47 | 63% | 89% | 97% | 1.80 | 0.82 | 0.49 | 0.90 | 0.64 | 0.24 | [-8.86, 9.98] | 3.52±3.11 | 76% | 96% | 99% | 1.14 |
| PW18 | † | † | 0.82 | 0.54 | 0.90 | 0.68 | -0.51 | [-13.10, 11.91] | 4.63±4.47 | 66% | 90% | 97% | 1.60 | 0.81 | 0.47 | 0.90 | 0.62 | 0.29 | [-9.12, 9.64] | 3.48±3.14 | 77% | 96% | 99% | 1.02 |
| PW19 | 3.80 | 4.11 | 0.84 | 0.57 | 0.90 | 0.68 | -0.47 | [-13.28, 10.96] | 4.44±4.30 | 67% | 92% | 97% | 1.99 | 0.83 | 0.52 | 0.90 | 0.65 | 0.40 | [-8.54, 9.22] | 3.28±3.04 | 79% | 97% | 99% | 1.19 |
| PW20 | † | † | 0.83 | 0.55 | 0.89 | 0.67 | -1.20 | [-14.46, 10.99] | 4.65±4.43 | 65% | 90% | 97% | 2.20 | 0.81 | 0.47 | 0.89 | 0.61 | -0.75 | [-10.12, 9.19] | 3.57±3.24 | 75% | 96% | 99% | 1.47 |

# Supplementary Table 8. Generalizability of population-within models on holdout datasets

Inference results of all population-within (PW) models on the holdout datasets. For regression analysis, we also evaluated the intra-subject performance by computing the regression metrics (determination coefficient and concordance coefficient) on individual subjects, and reported the weighted average. id., identifier; BP, brachial blood pressure; AMAE, average mean absolute error; ARMSE, average root mean square error; $r^2_{\text{a}}$ and $r^2_{\text{w}}$, aggregated and weighted determination coefficient, respectively; $\hat{\rho}_{c,\text{a}}$ and $\hat{\rho}_{c,\text{w}}$, aggregated and weighted concordance coefficient, respectively; ME and LOA, mean and limits of agreement of errors, respectively; MAE and SDAE, mean and standard deviation of absolute errors (AE), respectively; $\mathcal{P}_5$, $\mathcal{P}_{10}$, and $\mathcal{P}_{15}$, cumulative percentage of estimations with AE within 5, 10, and 15 mm Hg, respectively; $\mathcal{W}_{\text{pred}}$, Wasserstein distance between true and predicted distribution; †, waveform metrics not applicable to models estimating fiducial BP.

| | BP waveform | | Systolic BP | | | | | | | | | | | Diastolic BP | | | | | | | | | | |
|---|---|---|---|---|---|---|---|---|---|---|---|---|---|---|---|---|---|---|---|---|---|---|---|---|
| id. | AMAE (mm Hg) | ARMSE (mm Hg) | $r^2_{\text{a}}$ | $r^2_{\text{w}}$ | $\hat{\rho}_{c,\text{a}}$ | $\hat{\rho}_{c,\text{w}}$ | ME (mm Hg) | LOA (mm Hg) | MAE±SDAE (mm Hg) | $\mathcal{P}_5$ | $\mathcal{P}_{10}$ | $\mathcal{P}_{15}$ | $\mathcal{W}_{\text{pred}}$ (mm Hg) | $r^2_{\text{a}}$ | $r^2_{\text{w}}$ | $\hat{\rho}_{c,\text{a}}$ | $\hat{\rho}_{c,\text{w}}$ | ME (mm Hg) | LOA (mm Hg) | MAE±SDAE (mm Hg) | $\mathcal{P}_5$ | $\mathcal{P}_{10}$ | $\mathcal{P}_{15}$ | $\mathcal{W}_{\text{pred}}$ (mm Hg) |
| PW01 | 13.94 | 14.95 | 0.02 | 0.05 | 0.09 | -0.02 | -15.94 | [-45.58, 14.96] | 18.75±12.48 | 16% | 30% | 43% | 15.95 | 0.00 | 0.04 | 0.03 | -0.07 | -4.11 | [-27.92, 21.77] | 11.14±8.17 | 28% | 52% | 70% | 4.30 |
| PW02 | † | † | 0.05 | 0.03 | 0.17 | 0.04 | -7.76 | [-35.84, 20.01] | 13.36±9.53 | 21% | 43% | 62% | 7.76 | 0.01 | 0.08 | 0.09 | -0.08 | 1.73 | [-20.22, 25.48] | 9.99±7.03 | 27% | 56% | 79% | 4.10 |
| PW03 | 11.16 | 12.14 | 0.06 | 0.01 | 0.15 | -0.03 | -12.84 | [-38.60, 12.48] | 15.33±10.74 | 20% | 38% | 54% | 12.85 | 0.03 | 0.13 | 0.16 | -0.20 | -0.96 | [-20.88, 18.57] | 9.12±5.99 | 29% | 59% | 83% | 3.70 |
| PW04 | † | † | 0.08 | 0.01 | 0.19 | -0.01 | -10.31 | [-35.89, 15.37] | 13.90±9.91 | 22% | 43% | 59% | 10.34 | 0.03 | 0.13 | 0.15 | -0.21 | -0.65 | [-20.46, 18.93] | 9.08±5.98 | 29% | 60% | 83% | 3.76 |
| PW05 | 10.61 | 11.69 | 0.02 | 0.06 | 0.10 | 0.02 | -9.60 | [-40.66, 19.41] | 14.69±10.48 | 18% | 39% | 59% | 9.60 | 0.07 | 0.05 | 0.24 | 0.12 | 0.02 | [-19.62, 21.52] | 8.48±6.32 | 35% | 65% | 85% | 3.41 |
| PW06 | † | † | 0.02 | 0.11 | 0.12 | 0.05 | -8.17 | [-36.72, 22.97] | 13.78±9.82 | 21% | 41% | 61% | 8.17 | 0.04 | 0.05 | 0.18 | 0.11 | -0.11 | [-20.56, 23.04] | 9.19±6.66 | 32% | 61% | 81% | 3.11 |
| PW07 | 11.48 | 12.38 | 0.00 | 0.14 | 0.03 | 0.11 | -11.92 | [-46.79, 20.07] | 15.58±11.92 | 20% | 39% | 56% | 11.92 | 0.05 | 0.10 | 0.22 | 0.17 | -1.90 | [-21.80, 20.55] | 9.21±6.61 | 33% | 60% | 80% | 3.36 |
| PW08 | † | † | 0.01 | 0.13 | 0.07 | 0.08 | -9.94 | [-40.84, 21.90] | 13.98±10.51 | 21% | 42% | 62% | 9.94 | 0.03 | 0.05 | 0.14 | 0.07 | 3.08 | [-17.57, 22.44] | 9.31±6.26 | 29% | 58% | 81% | 4.90 |
| PW09 | 11.39 | 12.52 | 0.02 | 0.14 | 0.06 | -0.02 | -15.89 | [-41.51, 12.05] | 17.73±10.92 | 13% | 28% | 44% | 15.90 | 0.03 | 0.06 | 0.15 | 0.05 | -1.68 | [-21.01, 18.42] | 8.91±6.08 | 31% | 61% | 83% | 4.15 |
| PW10 | † | † | 0.02 | 0.14 | 0.06 | -0.02 | -13.69 | [-37.59, 16.83] | 16.42±10.01 | 14% | 30% | 47% | 13.73 | 0.01 | 0.06 | 0.08 | 0.10 | -1.58 | [-22.38, 19.80] | 8.84±6.65 | 36% | 63% | 81% | 4.05 |
| PW11 | 10.83 | 11.93 | 0.01 | 0.15 | 0.04 | -0.02 | -14.07 | [-37.71, 13.16] | 16.28±9.79 | 13% | 29% | 48% | 14.19 | 0.00 | 0.11 | 0.04 | 0.06 | -0.51 | [-20.21, 18.96] | 8.67±6.04 | 34% | 62% | 84% | 5.43 |
| PW12 | † | † | 0.01 | 0.16 | 0.07 | -0.02 | -11.17 | [-34.95, 21.27] | 15.44±9.32 | 16% | 32% | 49% | 11.20 | 0.02 | 0.08 | 0.12 | 0.08 | -1.49 | [-21.26, 22.58] | 9.27±6.54 | 31% | 60% | 80% | 3.14 |
| PW13 | 11.04 | 11.88 | 0.04 | 0.10 | 0.14 | 0.00 | -9.44 | [-34.19, 20.42] | 14.47±9.31 | 19% | 37% | 54% | 9.45 | 0.02 | 0.08 | 0.14 | 0.07 | 0.88 | [-17.91, 26.39] | 9.50±6.57 | 28% | 59% | 83% | 3.65 |
| PW14 | † | † | 0.01 | 0.12 | 0.06 | -0.05 | -11.29 | [-38.34, 17.45] | 15.52±9.83 | 15% | 32% | 52% | 11.31 | 0.06 | 0.07 | 0.24 | 0.05 | -0.93 | [-19.00, 20.57] | 9.07±6.00 | 30% | 59% | 83% | 2.95 |
| PW15 | 9.85 | 10.81 | 0.07 | 0.18 | 0.15 | 0.17 | -11.35 | [-37.19, 13.58] | 13.87±9.83 | 21% | 41% | 60% | 11.38 | 0.05 | 0.11 | 0.19 | 0.17 | 0.71 | [-19.10, 19.18] | 8.60±5.77 | 32% | 63% | 85% | 4.42 |
| PW16 | † | † | 0.01 | 0.13 | 0.06 | 0.08 | -12.55 | [-42.83, 19.65] | 15.95±11.30 | 19% | 36% | 52% | 12.56 | 0.02 | 0.06 | 0.13 | 0.08 | -1.12 | [-22.00, 20.58] | 9.38±6.58 | 31% | 58% | 80% | 2.96 |
| PW17 | 9.95 | 10.90 | 0.03 | 0.09 | 0.11 | 0.00 | -9.23 | [-34.16, 19.15] | 13.94±9.16 | 19% | 38% | 58% | 9.36 | 0.03 | 0.08 | 0.13 | 0.16 | 0.69 | [-20.12, 20.22] | 8.73±6.15 | 33% | 63% | 82% | 4.49 |
| PW18 | † | † | 0.04 | 0.13 | 0.13 | -0.01 | -11.41 | [-40.15, 18.66] | 15.73±10.62 | 17% | 35% | 52% | 11.42 | 0.12 | 0.10 | 0.31 | 0.17 | -1.95 | [-19.87, 17.58] | 8.21±5.79 | 35% | 67% | 87% | 3.39 |
| PW19 | 10.92 | 11.88 | 0.03 | 0.18 | 0.09 | 0.07 | -14.24 | [-39.99, 12.77] | 16.48±10.72 | 17% | 33% | 47% | 14.26 | 0.02 | 0.10 | 0.12 | 0.12 | -1.38 | [-21.41, 17.71] | 9.13±6.01 | 29% | 60% | 82% | 3.84 |
| PW20 | † | † | 0.01 | 0.16 | 0.05 | 0.09 | -12.02 | [-41.29, 16.67] | 15.57±11.12 | 19% | 37% | 54% | 12.02 | 0.05 | 0.11 | 0.20 | 0.13 | -1.24 | [-19.05, 18.65] | 9.08±5.62 | 28% | 58% | 84% | 3.57 |

## Supplementary Table 9. Estimation results of population-disjoint models

Training results of all population-disjoint (PD) model configurations. For regression analysis, we also evaluated the intra-subject performance by computing the regression metrics (determination coefficient and concordance coefficient) on individual subjects, and reported the weighted average. id., identifier; BP, brachial blood pressure; AMAE, average mean absolute error; ARMSE, average root mean square error; $r^2_{\mathrm{a}}$ and $r^2_{\mathrm{w}}$, aggregated and weighted determination coefficient, respectively; $\hat{\rho}_{c,\mathrm{a}}$ and $\hat{\rho}_{c,\mathrm{w}}$, aggregated and weighted concordance coefficient, respectively; ME and LOA, mean and limits of agreement of errors, respectively; MAE and SDAE, mean and standard deviation of absolute errors (AE), respectively; $\mathcal{P}_5$, $\mathcal{P}_{10}$, and $\mathcal{P}_{15}$, cumulative percentage of estimations with AE within 5, 10, and 15 mm Hg, respectively; $\mathcal{W}_{\mathrm{pred}}$, Wasserstein distance between true and predicted distribution; †, waveform metrics not applicable to models estimating fiducial BP.

| | BP waveform | | Systolic BP | | | | | | | | | | | | | Diastolic BP | | | | | | | | | | | | |
|---|---|---|---|---|---|---|---|---|---|---|---|---|---|---|---|---|---|---|---|---|---|---|---|---|---|---|---|---|
| id. | AMAE (mm Hg) | ARMSE (mm Hg) | $r^2_\text{a}$ | $r^2_\text{w}$ | $\hat{\rho}_{c,\text{a}}$ | $\hat{\rho}_{c,\text{w}}$ | ME (mm Hg) | LOA (mm Hg) | MAE±SDAE (mm Hg) | $\mathcal{P}_5$ | $\mathcal{P}_{10}$ | $\mathcal{P}_{15}$ | $\mathcal{W}_\text{pred}$ (mm Hg) | $r^2_\text{a}$ | $r^2_\text{w}$ | $\hat{\rho}_{c,\text{a}}$ | $\hat{\rho}_{c,\text{w}}$ | ME (mm Hg) | LOA (mm Hg) | MAE±SDAE (mm Hg) | $\mathcal{P}_5$ | $\mathcal{P}_{10}$ | $\mathcal{P}_{15}$ | $\mathcal{W}_\text{pred}$ (mm Hg) |
| PD01 | 14.23 | 14.89 | 0.05 | 0.06 | 0.18 | 0.04 | 9.40 | [-25.66, 51.51] | 18.16±13.88 | 17% | 34% | 50% | 9.40 | 0.04 | 0.04 | 0.17 | 0.00 | 6.37 | [-16.79, 30.84] | 11.65±8.46 | 26% | 50% | 69% | 6.37 |
| PD02 | † | † | 0.00 | 0.02 | 0.04 | 0.04 | 4.04 | [-26.38, 44.94] | 14.64±11.74 | 22% | 42% | 61% | 4.04 | 0.03 | 0.05 | 0.15 | 0.10 | -3.00 | [-21.75, 21.14] | 9.31±6.76 | 32% | 59% | 81% | 3.19 |
| PD03 | 15.90 | 16.49 | 0.11 | 0.11 | -0.32 | 0.03 | -2.42 | [-44.30, 38.20] | 19.16±12.70 | 14% | 28% | 42% | 2.95 | 0.08 | 0.11 | -0.27 | 0.00 | 2.01 | [-28.98, 28.09] | 13.65±9.02 | 23% | 40% | 56% | 3.26 |
| PD04 | † | † | 0.01 | 0.03 | -0.07 | 0.04 | -6.56 | [-35.31, 30.94] | 15.79±10.39 | 19% | 35% | 50% | 6.89 | 0.07 | 0.04 | -0.21 | -0.04 | -2.12 | [-26.61, 23.74] | 11.00±7.91 | 29% | 51% | 70% | 4.64 |
| PD05 | 12.08 | 12.59 | 0.02 | 0.14 | 0.14 | 0.18 | -0.03 | [-29.99, 30.20] | 13.39±9.14 | 20% | 41% | 61% | 1.75 | 0.00 | 0.08 | 0.02 | 0.11 | 0.99 | [-23.35, 26.26] | 11.23±7.30 | 22% | 49% | 72% | 1.14 |
| PD06 | † | † | 0.00 | 0.07 | 0.01 | 0.12 | -6.54 | [-36.50, 31.97] | 15.79±10.35 | 17% | 34% | 51% | 8.16 | 0.02 | 0.05 | 0.11 | 0.07 | -3.39 | [-27.68, 23.58] | 11.64±7.97 | 25% | 48% | 67% | 5.37 |
| PD07 | 13.41 | 14.25 | 0.00 | 0.13 | 0.05 | 0.17 | -8.25 | [-50.06, 24.67] | 17.19±13.20 | 17% | 35% | 52% | 8.34 | 0.02 | 0.09 | 0.12 | 0.12 | -3.07 | [-28.30, 23.13] | 10.90±8.06 | 29% | 54% | 72% | 3.07 |
| PD08 | † | † | 0.08 | 0.12 | 0.25 | 0.14 | 0.96 | [-41.39, 34.64] | 15.89±11.60 | 19% | 37% | 54% | 6.70 | 0.12 | 0.08 | 0.33 | 0.10 | 1.36 | [-24.52, 27.03] | 11.10±8.00 | 28% | 51% | 70% | 3.95 |
| PD09 | 8.67 | 9.42 | 0.28 | 0.14 | 0.33 | 0.15 | -1.08 | [-26.57, 20.02] | 9.18±7.45 | 34% | 63% | 82% | 6.83 | 0.35 | 0.11 | 0.32 | 0.14 | 0.08 | [-26.36, 17.31] | 7.48±6.85 | 46% | 75% | 87% | 5.86 |
| PD10 | † | † | 0.14 | 0.13 | 0.29 | 0.14 | -3.45 | [-27.77, 24.95] | 12.56±8.04 | 21% | 41% | 63% | 7.02 | 0.04 | 0.05 | 0.14 | 0.08 | -5.05 | [-23.00, 15.40] | 9.85±6.32 | 27% | 53% | 78% | 5.07 |
| PD11 | 10.36 | 11.05 | 0.26 | 0.20 | 0.36 | 0.23 | -2.88 | [-27.57, 26.97] | 11.21±8.25 | 27% | 53% | 71% | 7.75 | 0.27 | 0.20 | 0.41 | 0.17 | 2.95 | [-16.64, 24.29] | 9.50±6.31 | 27% | 58% | 82% | 5.73 |
| PD12 | † | † | 0.09 | 0.22 | 0.28 | 0.26 | 3.12 | [-22.13, 22.01] | 9.88±6.85 | 28% | 55% | 78% | 4.55 | 0.06 | 0.16 | 0.21 | 0.21 | 1.32 | [-19.71, 20.04] | 8.11±6.28 | 40% | 66% | 83% | 3.33 |
| PD13 | 9.88 | 10.61 | 0.19 | 0.16 | 0.38 | 0.24 | 5.76 | [-17.32, 27.38] | 10.68±7.66 | 28% | 53% | 73% | 5.78 | 0.16 | 0.08 | 0.30 | 0.12 | 6.55 | [-12.11, 23.33] | 9.08±6.40 | 32% | 60% | 82% | 6.57 |
| PD14 | † | † | 0.10 | 0.04 | 0.26 | 0.09 | -0.03 | [-29.93, 30.97] | 14.19±9.01 | 17% | 37% | 56% | 7.03 | 0.12 | 0.09 | 0.29 | 0.11 | 0.25 | [-23.47, 26.07] | 10.61±7.28 | 26% | 52% | 75% | 4.36 |
| PD15 | 9.07 | 9.76 | 0.32 | 0.19 | 0.46 | 0.24 | -6.49 | [-29.87, 19.25] | 11.68±8.37 | 26% | 48% | 68% | 6.73 | 0.24 | 0.10 | 0.44 | 0.15 | -2.94 | [-21.46, 16.98] | 7.74±6.21 | 42% | 70% | 86% | 3.13 |
| PD16 | † | † | 0.14 | 0.15 | 0.32 | 0.22 | 0.05 | [-41.13, 28.72] | 12.88±10.96 | 27% | 50% | 68% | 6.51 | 0.10 | 0.12 | 0.28 | 0.18 | 3.04 | [-19.67, 22.84] | 8.29±6.43 | 38% | 68% | 84% | 3.66 |
| PD17 | 10.33 | 10.93 | 0.06 | 0.07 | 0.20 | 0.10 | -2.09 | [-31.85, 26.66] | 12.78±9.16 | 24% | 45% | 64% | 5.90 | 0.13 | 0.07 | 0.30 | 0.09 | 1.20 | [-18.32, 21.99] | 8.57±6.16 | 35% | 64% | 84% | 4.45 |
| PD18 | † | † | 0.08 | 0.05 | 0.26 | 0.11 | -4.41 | [-28.90, 25.57] | 11.68±8.44 | 25% | 49% | 70% | 4.55 | 0.21 | 0.05 | 0.42 | 0.11 | -2.38 | [-23.00, 17.87] | 8.91±6.52 | 33% | 63% | 83% | 2.64 |
| PD19 | 9.30 | 10.02 | 0.10 | 0.13 | 0.29 | 0.17 | -5.12 | [-31.89, 27.80] | 12.04±9.54 | 28% | 52% | 68% | 5.14 | 0.14 | 0.08 | 0.37 | 0.09 | -0.98 | [-18.29, 20.29] | 7.71±5.85 | 38% | 72% | 89% | 1.05 |
| PD20 | † | † | 0.03 | 0.08 | 0.14 | 0.12 | -4.70 | [-32.01, 29.32] | 13.55±9.76 | 24% | 44% | 59% | 6.40 | 0.00 | 0.09 | 0.05 | 0.09 | -0.59 | [-21.93, 22.73] | 9.56±6.86 | 32% | 58% | 77% | 1.53 |

# Supplementary Table 10. Label gap and waveform estimation accuracy for population models

Partition gap $\mathcal{W}_{\text{label}} := \mathcal{W}_1(\mathbf{Y}_\mathcal{U}, \mathbf{Y}_\mathcal{V})$ of blood pressure waveform distributions for all population model configurations and dataset compositions, and corresponding average mean absolute error (AMAE) as representative waveform estimation metric. For population-within (PW) and population-disjoint (PD) models, the sets $\mathcal{U}$ and $\mathcal{V}$ are the train and test sets used when training the models, respectively, and the AMAE was computed on the test set. For the holdout dataset, we used $\mathcal{U}$ as the corresponding PW train set and $\mathcal{V}$ as the entire holdout dataset. The computation of $\mathcal{W}_{\text{label}}$ is described Supplementary Discussion 9.5.1. The AMAE values for PW, holdout, and PD are taken from Supplementary Table 7, 8, and 9, respectively.

| | Population-within | | Holdout | | Population-disjoint | |
|---|---|---|---|---|---|---|
| id. | $\mathcal{W}_{\text{label}}$ (mm Hg) | AMAE (mm Hg) | $\mathcal{W}_{\text{label}}$ (mm Hg) | AMAE (mm Hg) | $\mathcal{W}_{\text{label}}$ (mm Hg) | AMAE (mm Hg) |
| 1 | 10.77 | 11.23 | 27.48 | 13.94 | 18.74 | 14.23 |
| 3 | 10.71 | 12.75 | 27.52 | 11.16 | 19.25 | 15.90 |
| 5 | 10.84 | 4.62 | 27.61 | 10.61 | 18.19 | 12.08 |
| 7 | 10.96 | 5.07 | 27.52 | 11.48 | 16.66 | 13.41 |
| 9 | 10.75 | 6.45 | 27.31 | 11.39 | 17.09 | 8.67 |
| 11 | 10.68 | 6.58 | 27.33 | 10.83 | 25.08 | 10.36 |
| 13 | 10.92 | 3.95 | 27.54 | 11.04 | 21.74 | 9.88 |
| 15 | 10.85 | 3.69 | 27.63 | 9.85 | 20.58 | 9.07 |
| 17 | 11.06 | 4.17 | 27.34 | 9.95 | 21.56 | 10.33 |
| 19 | 10.94 | 3.80 | 27.49 | 10.92 | 20.10 | 9.30 |

## Supplementary Table 11. Recalibration results of subject-specific models on longitudinal study

Recalibration results for subject-specific (SS) models on longitudinal datasets, evaluated using average root-mean-square error (ARMSE). Each block (separated by horizontal and vertical lines) shows the performance of an individual SS configuration for an individual subject. Each cell $(\mathrm{C}_j, \mathrm{D}_k)$ shows the ARMSE of model checkpoint $\mathrm{C}_j$ evaluated on dataset $\mathrm{D}_k$. Checkpoints $\mathrm{C}_0$ were trained with SS datasets from the cross-sectional study. Checkpoints $\mathrm{C}_j$ $(j \geq 1)$ were trained by loading $\mathrm{C}_{j-1}$ from the previous day and fine-tuning with 50% of datasets $\mathrm{D}_j$ from the current day. For a fixed row $\mathrm{C}_j$, columns from left to right shows the longitudinal stability of the model for new datasets. For a fixed column $\mathrm{D}_k$, the rows from top down shows the effectiveness of recalibration with newly available dataset. Grey cells $(k = j)$ represent same-day performance where 50% of datasets $\mathrm{D}_j$ was used to obtain checkpoints $\mathrm{C}_j$. White cells $(k > j)$ represent true longitudinal generalizability, where checkpoints $\mathrm{C}_j$ was test on future unseen datasets $\mathrm{D}_k$. Model-average and subject-average cells were determined by averaging the ARMSE from the corresponding $(\mathrm{C}_j, \mathrm{D}_k)$ cells across all $4$ models and/or all $5$ subjects. All values are given in mm Hg. Cells with dagger (†) represent invalid evaluations where $\mathrm{C}_j$ would be tested on historical data $\mathrm{D}_k$ $(k < j)$ already incorporated into the model's training. LR, Linear Regression class; MLP, Multilayer Perceptron class; CRT, Convolutional Recurrent Transformer class; CRS, Convolutional Recurrent Samba class; SS, subject-specific class.

|  | | Subject 002 | | | | | Subject 010 | | | | | Subject 070 | | | | |
|---|---|---|---|---|---|---|---|---|---|---|---|---|---|---|---|---|
| | | $D_1$ | $D_2$ | $D_3$ | $D_4$ | $D_5$ | $D_1$ | $D_2$ | $D_3$ | $D_4$ | $D_5$ | $D_1$ | $D_2$ | $D_3$ | $D_4$ | $D_5$ |
| LR (SS03) | $C_0$ | 11.15 | 13.40 | 7.88 | 13.53 | 9.52 | 7.83 | 26.24 | 14.33 | 10.07 | 14.52 | 14.59 | 11.12 | 11.49 | 10.19 | 11.05 |
| | $C_1$ | 11.47 | 13.36 | 7.92 | 13.11 | 9.62 | 5.21 | 27.41 | 17.33 | 10.43 | 15.31 | 9.17 | 9.57 | 11.93 | 11.15 | 8.36 |
| | $C_2$ | † | 9.45 | 9.21 | 13.69 | 9.38 | † | 8.89 | 8.08 | 9.14 | 11.30 | † | 7.08 | 12.37 | 11.68 | 8.13 |
| | $C_3$ | † | † | 5.49 | 13.90 | 10.37 | † | † | 5.61 | 9.68 | 10.64 | † | † | 8.82 | 11.29 | 8.39 |
| | $C_4$ | † | † | † | 6.02 | 11.77 | † | † | † | 8.60 | 11.61 | † | † | † | 10.88 | 8.37 |
| | $C_5$ | † | † | † | † | 6.58 | † | † | † | † | 9.70 | † | † | † | † | 7.98 |
| MLP (SS7) | $C_0$ | 9.15 | 12.89 | 9.22 | 11.48 | 17.82 | 6.29 | 11.62 | 8.85 | 9.24 | 14.40 | 20.70 | 22.99 | 19.53 | 10.62 | 14.87 |
| | $C_1$ | 6.11 | 11.31 | 9.39 | 19.04 | 14.53 | 4.36 | 11.82 | 8.31 | 9.25 | 17.02 | 6.95 | 14.59 | 14.77 | 13.19 | 10.10 |
| | $C_2$ | † | 8.86 | 6.87 | 19.41 | 13.72 | † | 6.53 | 10.04 | 9.34 | 17.93 | † | 5.41 | 13.22 | 13.33 | 8.92 |
| | $C_3$ | † | † | 5.57 | 20.38 | 12.37 | † | † | 3.28 | 9.88 | 14.01 | † | † | 5.70 | 12.85 | 8.29 |
| | $C_4$ | † | † | † | 3.92 | 14.00 | † | † | † | 7.11 | 12.86 | † | † | † | 8.81 | 8.34 |
| | $C_5$ | † | † | † | † | 4.59 | † | † | † | † | 7.02 | † | † | † | † | 4.79 |
| CRT (SS15) | $C_0$ | 8.22 | 11.23 | 7.58 | 9.96 | 9.83 | 16.72 | 12.14 | 12.57 | 14.66 | 16.66 | 10.14 | 13.04 | 6.25 | 12.13 | 8.82 |
| | $C_1$ | 6.37 | 10.66 | 5.28 | 13.96 | 6.61 | 7.35 | 14.84 | 11.43 | 11.35 | 9.42 | 5.97 | 8.07 | 10.02 | 14.98 | 8.74 |
| | $C_2$ | † | 9.43 | 6.40 | 16.09 | 6.35 | † | 7.95 | 7.52 | 10.88 | 7.92 | † | 5.68 | 11.57 | 16.40 | 9.50 |
| | $C_3$ | † | † | 6.43 | 15.74 | 6.61 | † | † | 6.26 | 10.97 | 7.62 | † | † | 6.28 | 14.92 | 8.27 |
| | $C_4$ | † | † | † | 7.42 | 7.58 | † | † | † | 6.80 | 7.52 | † | † | † | 12.35 | 7.11 |
| | $C_5$ | † | † | † | † | 7.90 | † | † | † | † | 7.80 | † | † | † | † | 5.85 |
| CRS (SS17) | $C_0$ | 12.85 | 12.46 | 13.99 | 10.26 | 12.61 | 11.74 | 11.46 | 8.39 | 11.20 | 11.56 | 11.79 | 16.63 | 11.39 | 12.06 | 14.31 |
| | $C_1$ | 4.58 | 11.34 | 12.18 | 8.18 | 11.32 | 3.39 | 14.69 | 8.75 | 8.64 | 8.57 | 4.10 | 13.58 | 9.77 | 12.51 | 13.31 |
| | $C_2$ | † | 4.74 | 9.53 | 8.31 | 9.94 | † | 6.19 | 9.97 | 8.59 | 8.92 | † | 4.62 | 7.42 | 11.73 | 11.75 |
| | $C_3$ | † | † | 2.78 | 8.37 | 9.54 | † | † | 2.66 | 8.35 | 7.59 | † | † | 3.13 | 11.61 | 9.85 |
| | $C_4$ | † | † | † | 3.96 | 10.21 | † | † | † | 6.33 | 8.40 | † | † | † | 7.48 | 9.52 |
| | $C_5$ | † | † | † | † | 3.17 | † | † | † | † | 6.22 | † | † | † | † | 4.10 |
| Model-average | $C_0$ | 10.34 | 12.49 | 9.67 | 11.31 | 12.45 | 10.64 | 15.37 | 11.04 | 11.29 | 14.28 | 14.30 | 15.95 | 12.17 | 11.2 5 | 12.26 |
| | $C_1$ | 7.13 | 11.67 | 8.69 | 13.57 | 10.52 | 5.08 | 17.19 | 11.46 | 9.92 | 12.58 | 6.55 | 11.45 | 11.62 | 12.96 | 10.13 |
| | $C_2$ | † | 8.12 | 8.00 | 14.38 | 9.85 | † | 7.39 | 8.90 | 9.49 | 11.52 | † | 5.70 | 11.14 | 13.29 | 9.57 |
| | $C_3$ | † | † | 5.07 | 14.60 | 9.73 | † | † | 4.45 | 9.72 | 9.96 | † | † | 5.98 | 12.67 | 8.70 |
| | $C_4$ | † | † | † | 5.33 | 10.89 | † | † | † | 7.21 | 10.10 | † | † | † | 9.88 | 8.34 |
| | $C_5$ | † | † | † | † | 5.56 | † | † | † | † | 7.69 | † | † | † | † | 5.68 |

|  | | Subject 094 | | | | | Subject 095 | | | | | Subject-average | | | | |
|---|---|---|---|---|---|---|---|---|---|---|---|---|---|---|---|---|
| | | $D_1$ | $D_2$ | $D_3$ | $D_4$ | $D_5$ | $D_1$ | $D_2$ | $D_3$ | $D_4$ | $D_5$ | $D_1$ | $D_2$ | $D_3$ | $D_4$ | $D_5$ |
| LR (SS03) | $C_0$ | 11.17 | 15.69 | 36.64 | 15.74 | 31.75 | 16.79 | 18.91 | 10.71 | 22.30 | 100.73 | 12.30 | 17.07 | 16.21 | 14.37 | 33.51 |
| | $C_1$ | 8.62 | 14.58 | 37.69 | 14.92 | 33.42 | 6.96 | 15.10 | 7.54 | 15.04 | 83.32 | 8.29 | 16.01 | 16.48 | 12.93 | 30.01 |
| | $C_2$ | † | 9.10 | 47.81 | 12.92 | 28.08 | † | 12.96 | 8.87 | 14.87 | 70.56 | † | 9.49 | 17.27 | 12.46 | 25.49 |
| | $C_3$ | † | † | 14.01 | 12.03 | 24.10 | † | † | 6.97 | 13.70 | 59.23 | † | † | 8.18 | 12.12 | 22.55 |
| | $C_4$ | † | † | † | 6.56 | 15.49 | † | † | † | 11.32 | 54.67 | † | † | † | 8.68 | 20.38 |
| | $C_5$ | † | † | † | † | 7.36 | † | † | † | † | 10.50 | † | † | † | † | 8.42 |
| MLP (SS7) | $C_0$ | 12.14 | 15.33 | 26.55 | 15.95 | 28.48 | 10.21 | 12.92 | 8.05 | 14.39 | 63.64 | 11.70 | 15.15 | 14.44 | 12.34 | 27.84 |
| | $C_1$ | 7.82 | 14.14 | 40.01 | 11.11 | 18.91 | 4.87 | 11.98 | 8.11 | 10.57 | 81.75 | 6.02 | 12.77 | 16.12 | 12.63 | 28.46 |
| | $C_2$ | † | 6.55 | 29.26 | 10.48 | 12.77 | † | 6.70 | 8.70 | 11.10 | 107.85 | † | 6.81 | 13.62 | 12.73 | 32.24 |
| | $C_3$ | † | † | 9.02 | 7.93 | 10.45 | † | † | 4.58 | 9.77 | 124.87 | † | † | 5.63 | 12.16 | 34.00 |
| | $C_4$ | † | † | † | 3.29 | 11.16 | † | † | † | 7.18 | 150.23 | † | † | † | 6.06 | 39.32 |
| | $C_5$ | † | † | † | † | 3.98 | † | † | † | † | 6.16 | † | † | † | † | 5.31 |
| CRT (SS15) | $C_0$ | 9.12 | 10.14 | 10.73 | 8.99 | 9.74 | 16.87 | 16.73 | 15.86 | 19.83 | 22.04 | 12.21 | 12.65 | 10.60 | 13.11 | 13.42 |
| | $C_1$ | 8.62 | 8.53 | 14.81 | 7.74 | 9.52 | 8.60 | 15.60 | 14.66 | 16.12 | 22.46 | 7.38 | 11.54 | 11.24 | 12.83 | 11.35 |
| | $C_2$ | † | 8.89 | 16.58 | 9.45 | 11.58 | † | 11.00 | 14.21 | 15.44 | 24.71 | † | 8.59 | 11.26 | 13.65 | 12.01 |
| | $C_3$ | † | † | 11.93 | 9.63 | 11.27 | † | † | 7.43 | 13.44 | 22.00 | † | † | 7.66 | 12.94 | 11.15 |
| | $C_4$ | † | † | † | 6.96 | 11.48 | † | † | † | 10.89 | 18.90 | † | † | † | 8.88 | 10.52 |
| | $C_5$ | † | † | † | † | 10.76 | † | † | † | † | 10.78 | † | † | † | † | 8.62 |
| CRS (SS17) | $C_0$ | 11.21 | 9.36 | 11.16 | 12.82 | 16.27 | 10.23 | 16.48 | 10.58 | 17.34 | 22.15 | 11.56 | 13.28 | 11.10 | 12.74 | 15.38 |
| | $C_1$ | 5.06 | 8.98 | 9.38 | 10.89 | 15.62 | 3.04 | 11.06 | 7.93 | 13.39 | 16.89 | 4.03 | 11.93 | 9.60 | 10.72 | 13.14 |
| | $C_2$ | † | 4.01 | 8.55 | 9.40 | 12.17 | † | 5.88 | 6.61 | 12.28 | 16.61 | † | 5.09 | 8.42 | 10.06 | 11.88 |
| | $C_3$ | † | † | 4.68 | 9.71 | 11.90 | † | † | 3.64 | 10.19 | 17.35 | † | † | 3.38 | 9.65 | 11.25 |
| | $C_4$ | † | † | † | 3.36 | 10.23 | † | † | † | 5.18 | 14.75 | † | † | † | 5.26 | 10.62 |
| | $C_5$ | † | † | † | † | 3.39 | † | † | † | † | 3.38 | † | † | † | † | 4.05 |
| Model-average | $C_0$ | 10.91 | 12.63 | 21.27 | 13.37 | 21.56 | 13.53 | 16.26 | 11.30 | 18.46 | 52.14 | 11.94 | 14.54 | 13.09 | 13.14 | 22.54 |
| | $C_1$ | 7.53 | 11.56 | 25.47 | 11.16 | 19.37 | 5.87 | 13.44 | 9.56 | 13.78 | 51.10 | 6.43 | 13.06 | 13.36 | 12.28 | 20.74 |
| | $C_2$ | † | 7.13 | 25.55 | 10.56 | 16.15 | † | 9.13 | 9.60 | 13.42 | 54.93 | † | 7.49 | 12.64 | 12.23 | 20.40 |
| | $C_3$ | † | † | 9.91 | 9.83 | 14.43 | † | † | 5.65 | 11.77 | 55.86 | † | † | 6.21 | 11.72 | 19.74 |
| | $C_4$ | † | † | † | 5.04 | 12.09 | † | † | † | 8.64 | 59.64 | † | † | † | 7.22 | 20.21 |
| | $C_5$ | † | † | † | † | 6.37 | † | † | † | † | 5.56 | † | † | † | † | 6.60 |

# Supplementary Table 12. Model training summary of ablation experiments

Model naming conventions and summary of training progress for the ablation experiments. The reference population-within models (PW15 and PW19) are highlighted in grey cells. Model PW15 (and its ablated variants) belong to the Convolutional Recurrent Transformer (CRT) class, while model PW19 (and its ablated variants) belong to the Convolutional Recurrent Samba (CRS) class (see Supplementary Discussion 8.4). The suffixes "-A1"–"-A4" denote changes to the reference models as well as adjustment to the training strategy (see Supplementary Discussion 8.9). All models were trained using the PW partition strategy, with multi-channel bioimpedance as input and full BP waveform as output. id., identifier; CRT, Convolutional Recurrent Transformer class; CRS, Convolutional Recurrent Samba class; PW, population-within.

| id. | Parameters | Train samples | Test samples | Epochs |
|---|---|---|---|---|
| PW15 | 7,545,030 | 102,315 | 11,358 | 416 |
| PW15-A1 | 2,129,226 | 102,448 | 11,372 | 324 |
| PW15-A2 | 7,301,986 | 102,445 | 11,378 | 333 |
| PW15-A3 | 7,545,030 | 102,410 | 11,372 | 403 |
| PW15-A4 | 2,129,226 | 102,265 | 11,356 | 441 |
| PW19 | 2,513,286 | 102,406 | 11,369 | 418 |
| PW19-A1 | 2,328,390 | 102,339 | 11,362 | 136 |
| PW19-A2 | 2,270,242 | 102,593 | 11,390 | 306 |

# Supplementary Table 13. Estimation results from ablated models

Results of all ablated population-within models PW15 and PW19. The reference models are highlighted in grey cells. For regression analysis, we also evaluated the intra-subject performance by computing the regression metrics (determination coefficient and concordance coefficient) on individual subjects, and reported the weighted average. id., identifier; BP, brachial blood pressure; AMAE, average mean absolute error; ARMSE, average root mean square error; $r^2_{\mathsf{a}}$ and $r^2_{\mathsf{w}}$, aggregated and weighted determination coefficient, respectively; $\hat{\rho}_{c,\mathsf{a}}$ and $\hat{\rho}_{c,\mathsf{w}}$, aggregated and weighted concordance coefficient, respectively; ME and LOA, mean and limits of agreement of errors, respectively; MAE and SDAE, mean and standard deviation of absolute errors (AE), respectively; $\mathcal{P}_5$, $\mathcal{P}_{10}$, and $\mathcal{P}_{15}$, cumulative percentage of estimations with AE within 5, 10, and 15 mm Hg, respectively; $\mathcal{W}_{\mathsf{pred}}$, Wasserstein distance between true and predicted distribution; †, waveform metrics not applicable to models estimating fiducial BP.

| id. | BP waveform | | Systolic BP | | | | | | | | | | | Diastolic BP | | | | | | | | | | |
|---|---|---|---|---|---|---|---|---|---|---|---|---|---|---|---|---|---|---|---|---|---|---|---|---|
| | AMAE (mm Hg) | ARMSE (mm Hg) | $r^2_a$ | $r^2_w$ | $\hat{\rho}_{c,a}$ | $\hat{\rho}_{c,w}$ | ME (mm Hg) | LOA (mm Hg) | MAE±SDAE (mm Hg) | $\mathcal{P}_5$ | $\mathcal{P}_{10}$ | $\mathcal{P}_{15}$ | $\mathcal{W}_{pred}$ (mm Hg) | $r^2_a$ | $r^2_w$ | $\hat{\rho}_{c,a}$ | $\hat{\rho}_{c,w}$ | ME (mm Hg) | LOA (mm Hg) | MAE±SDAE (mm Hg) | $\mathcal{P}_5$ | $\mathcal{P}_{10}$ | $\mathcal{P}_{15}$ | $\mathcal{W}_{pred}$ (mm Hg) |
| PW15 | 3.69 | 4.00 | 0.85 | 0.61 | 0.91 | 0.70 | -1.79 | [-14.06, 9.50] | 4.44±4.27 | 67% | 91% | 97% | 2.10 | 0.85 | 0.54 | 0.91 | 0.68 | -0.21 | [-8.55, 8.45] | 3.14±2.92 | 81% | 97% | 99% | 1.01 |
| PW15-A1 | 3.75 | 4.06 | 0.85 | 0.60 | 0.92 | 0.73 | 0.26 | [-11.64, 11.89] | 4.26±4.09 | 69% | 92% | 98% | 1.04 | 0.85 | 0.54 | 0.91 | 0.67 | 1.03 | [-7.62, 9.81] | 3.27±2.97 | 80% | 96% | 99% | 1.09 |
| PW15-A2 | 4.19 | 4.50 | 0.85 | 0.60 | 0.88 | 0.66 | -3.61 | [-16.29, 7.53] | 5.28±4.59 | 58% | 87% | 96% | 3.66 | 0.84 | 0.53 | 0.90 | 0.64 | -1.58 | [-10.63, 7.06] | 3.52±3.08 | 76% | 96% | 99% | 1.85 |
| PW15-A3 | 3.80 | 4.08 | 0.86 | 0.61 | 0.90 | 0.69 | -2.69 | [-14.29, 8.30] | 4.64±4.33 | 65% | 91% | 97% | 2.76 | 0.86 | 0.54 | 0.91 | 0.67 | -1.13 | [-9.39, 7.35] | 3.21±2.91 | 80% | 97% | 99% | 1.45 |
| PW15-A4 | 3.69 | 3.97 | 0.86 | 0.62 | 0.92 | 0.74 | 1.21 | [-10.34, 12.33] | 4.22±3.96 | 69% | 92% | 98% | 1.28 | 0.86 | 0.56 | 0.92 | 0.69 | 1.19 | [-7.02, 9.81] | 3.21±2.95 | 80% | 97% | 99% | 1.20 |
| PW19 | 3.80 | 4.11 | 0.84 | 0.57 | 0.90 | 0.68 | -0.47 | [-13.28, 10.96] | 4.44±4.30 | 67% | 92% | 97% | 1.99 | 0.83 | 0.52 | 0.90 | 0.65 | 0.40 | [-8.54, 9.22] | 3.28±3.04 | 79% | 97% | 99% | 1.19 |
| PW19-A1 | 4.16 | 4.52 | 0.82 | 0.53 | 0.90 | 0.67 | -0.04 | [-13.86, 12.32] | 4.69±4.48 | 65% | 90% | 97% | 0.90 | 0.81 | 0.47 | 0.89 | 0.61 | 0.93 | [-8.44, 10.70] | 3.63±3.28 | 75% | 95% | 99% | 1.36 |
| PW19-A2 | 3.81 | 4.13 | 0.84 | 0.57 | 0.91 | 0.72 | 0.15 | [-12.65, 11.76] | 4.38±4.28 | 68% | 91% | 97% | 1.00 | 0.83 | 0.51 | 0.91 | 0.66 | 0.56 | [-8.40, 9.76] | 3.31±3.10 | 79% | 96% | 99% | 0.81 |

## Supplementary Table 14. Comparison to blood pressure ring sensor studies

Comparison to related blood pressure studies using ring sensors. For our work, we selected four models (SS17, SS19, PW13, and PW15-A4) representing the combinations of model architecture (CRT vs. CRS), input modality (image vs. BioZ), and dataset composition (SS vs. PW). These models are also among the best performers within each category. The complete results are reported in Supplementary Table 6 and 7. Here, the cohort size refers to the size of the study and not necessarily the number of subjects in the training/test dataset. AdaBoost, Adaptive Boosting; BioZ, bioimpedance; BP, blood pressure; CNN, convolutional neural network; CRS, Convolutional Recurrent Samba class; CRT, Convolutional Recurrent Transformer class; DBP and SBP, diastolic and systolic blood pressure, respectively; HTN, hypertension; MAE and SDAE, mean and standard deviation of absolute errors (AE), respectively; mBP, mean arterial pressure; ME and LOA, mean and limits of agreement of errors, respectively; MLP, multilayer perceptron; MLR, multilinear regression; PPG, photoplethysmography; PVI, peripheral vascular impedance; PW, population-within; SD, standard deviation; SS, subject-specific; VC, Volume clamp; †, Information not reported; *, Coefficients of determination computed from reported Pearson's coefficient of correlation; **, LOA computed from reported ME and SDE under normality assumption.

| Reference | Cohort size (male:female) | Age | Health condition | Study protocol | Sensing modality | BP estimation model | Accuracy, $r^2$ SBP | Accuracy, $r^2$ DBP | Estimation error (mm Hg) SBP | Estimation error (mm Hg) DBP |
|---|---|---|---|---|---|---|---|---|---|---|
| Lee et al.[253] | 33 (13:20) | (Mean±SD) 52.6±10.8 | Mixed (15 medicated) | 24-hour ambulatory | PPG | CNN | 0.77* | 0.71* | (ME, LOA) 1.74, [-11.38, 14.85] | -3.24, [-16, 9.51] |
| Kim et al.[254] | 89 (42:47) | (Mean±SD) 40.1±12 | Mixed (18 medicated) | Static | PPG | CNN | 0.88* | 0.9* | (ME, LOA) 0.16 [-11.41, 11.72] | -0.07 [-9.26, 9.1] |
| Haddad et al.[272] | 6 (6:0) | (Mean±SD) 32±2.61 | Healthy | Static; dynamic | PPG | MLR | † | † | (ME, LOA) -0.28, [-15.06, 14.5] | -1.3, [-15.37, 12.77] |
| Schukraft et al.[259] | 25 (15:10) | (Mean±SD) 68.9±6.4 | Mixed (22 HTN) | Coronary angiography | PPG | MLR | 0.86* | 0.88* | (ME, LOA) 2.3, [11.31, 15.94] | 0.49, [6.41, 7.39] |
| Zhu et al.[255] | 85 (43:42) | (Range) [18–60] | Mixed | Static; ambulatory | PPG | MLP | 0.86 | 0.81 | (ME, LOA) 0.01, [-13.21, 13.22] | 0.02, [-12.27, 12.31] |
| Tang et al.[261] | 34 (14:20) | (Mean±SD) 21.24±2.62 | Healthy | Static; dynamic | PPG | Transformer | 0.13* | 0.0* | (MAE±SDAE) 13.33±0.78 | 7.56±0.54 |
| Fortin et al.[257] | 46 (19:27) | (Median; range) 36; [23–49] | Mixed | Neurosurgery; anesthesia | VC | Direct | † | † | (ME, LOA) mBP: -1.0, [-14.8, 12.7] | |
| Panula et al.[262] | 43 (30:13) | (Mean; range) 48; [24–83] | Mixed (4 medicated) | Static | VC | Direct | 0.62* | 0.79* | (ME, LOA) -3.5, [-20, 13] | -4.0, [-13, 4.6] |
| Ni et al.[263] | 10 (8:2) | (Mean±SD) 21±2 | Healthy | † | Eddy current | MLR | 0.77* | 0.62* | (MAE±SDAE) 3.64±2.71 | 3.20±2.66 |
| Sel et al.[39] | 10 (9:1) | (Median; range) 21; [19–26] | Healthy | Static; cold pressor | BioZ | AdaBoost | 0.58* | 0.66* | (ME, LOA**) 0.11, [-10.22, 10.44] | 0.11, [-7.48, 7.7] |
| This work | 96 (40:56) | (Mean±SD) 28.4±9.1 | Healthy | Static; Valsalva; cold pressor | PVI image | CRS (SS17) | 0.81 | 0.83 | (MAE±SDAE) 4.73±4.46 | 3.39±3.05 |
| | | | | | | CRT (PW13) | 0.83 | 0.82 | 4.7±4.39 | 3.44±3.14 |
| | | | | | BioZ | CRS (SS19) | 0.80 | 0.83 | 4.85±4.63 | 3.4±3.01 |
| | | | | | | CRT (PW15-A4) | 0.86 | 0.86 | 4.22±3.96 | 3.21±2.95 |

# Supplementary Table 15. Comparison to blood pressure wrist sensor studies

Comparison to related blood pressure studies using wrist bioimpedance sensors. For our work, we selected four models (SS17, SS19, PW13, and PW15-A4) representing the combinations of model architecture (CRT vs. CRS), input modality (image vs. BioZ), and dataset composition (SS vs. PW). These models are also among the best performers within each category. The complete results are reported in Supplementary Table 6 and 7. Here, the cohort size refers to the size of the study and not necessarily the number of subjects in the training/test dataset. AdaBoost, Adaptive Boosting; BioZ, bioimpedance; BP, blood pressure; CNN, convolutional neural network; CRS, Convolutional Recurrent Samba class; CRT, Convolutional Recurrent Transformer class; CVD, cardiovascular diseases; DBP and SBP, diastolic and systolic blood pressure, respectively; ENIG, electroless nickel immersion gold; HTN, hypertension; LR, simple linear regression; MAE and SDAE, mean and standard deviation of absolute errors (AE), respectively; ME and LOA, mean and limits of agreement of errors, respectively; PVI, peripheral vascular impedance; PW, population-within; PWA, pulse wave analysis; PWV, pulse wave velocity; SD, standard deviation; sPINN, signal-tagged physics-informed neural network; SS, subject-specific; †, Information not reported; *, Coefficients of determination computed from reported Pearson's coefficient of correlation; **, LOA computed from reported ME and SDE under normality assumption.

| Reference | Cohort size (male:female) | Age | Health condition | Study protocol | Electrode material | BP<br>BP estimation model | Accuracy, $r^2$ | | Estimation error (mm Hg) | |
|---|---|---|---|---|---|---|---|---|---|---|
| | | | | | | | SBP | DBP | SBP | DBP |
| Huynh et al.[273] | 15 (9:6) | (Mean±SD) 30±5 | Healthy | Static; dynamic | Copper foil | PWV; LR | † | † | (ME, LOA**) 0.01, [-15.87, 15.88] | -0.06, [-10.76, 10.64] |
| Ibrahim et al.[264] | 10 (7:3) | (Range) [18–30] | Healthy | Static; dynamic | Gel | PWA; AdaBoost | 0.74* | 0.59* | (ME, LOA**) -0.02, [-5.03, 4.99] | -0.09, [-6.96, 6.78] |
| Ibrahim et al.[266] | 4 | (Range) [20–25] | Healthy | Static; dynamic | Silver | PWA; CNN autoencoder | 0.62* | 0.64* | (ME, LOA**) 0.2, [-12.54, 12.94] | 0.5, [-9.3, 10.3] |
| Kireev et al.[244] | 7 | Mid 20s | Healthy | Static; dynamic | Graphene | PWA; AdaBoost | 0.79* | 0.77* | (ME, LOA**) 0.17, [-11.14, 11.48] | 0.16, [-8.6, 8.9] |
| Crandall et al.[125] | 75 (39:36) | (Range) [18–48] | Healthy | Static; dynamic | Stainless steel | sPINN | 0.58 | 0.63 | (MAE±SDAE) 7.24±6.98 | 5.45±5.19 |
| | 86 (41:45) | (Range) [24–93] | Patients (32 HTN, 22 CVD) | Clinic | | | 0.77 | 0.81 | 7.26±8.46 | 3.89±4.63 |
| This work | 96 (40:56) | (Mean±SD) 28.4±9.1 | Healthy | Static; Valsalva; cold pressor | ENIG | CRS (SS17) | 0.81 | 0.83 | (MAE±SDAE) 4.73±4.46 | 3.39±3.05 |
| | | | | | | CRT (PW13) | 0.83 | 0.82 | 4.7±4.39 | 3.44±3.14 |
| | | | | | | CRS (SS19) | 0.80 | 0.83 | 4.85±4.63 | 3.4±3.01 |
| | | | | | | CRT (PW15-A4) | 0.86 | 0.86 | 4.22±3.96 | 3.21±2.95 |

# Supplementary Figures

## Supplementary Fig. 1. Computer-aided design images of the electrical ring imager

Rendered model of a representative ring (size 9.5 North America) in Autodesk Fusion, showing the empty ring housing (**a**), ring with electrodes (**b**), fully assembled ring (**c**), and a sample electrode with coaxial cable (**d**). Scale bars, 5 mm.

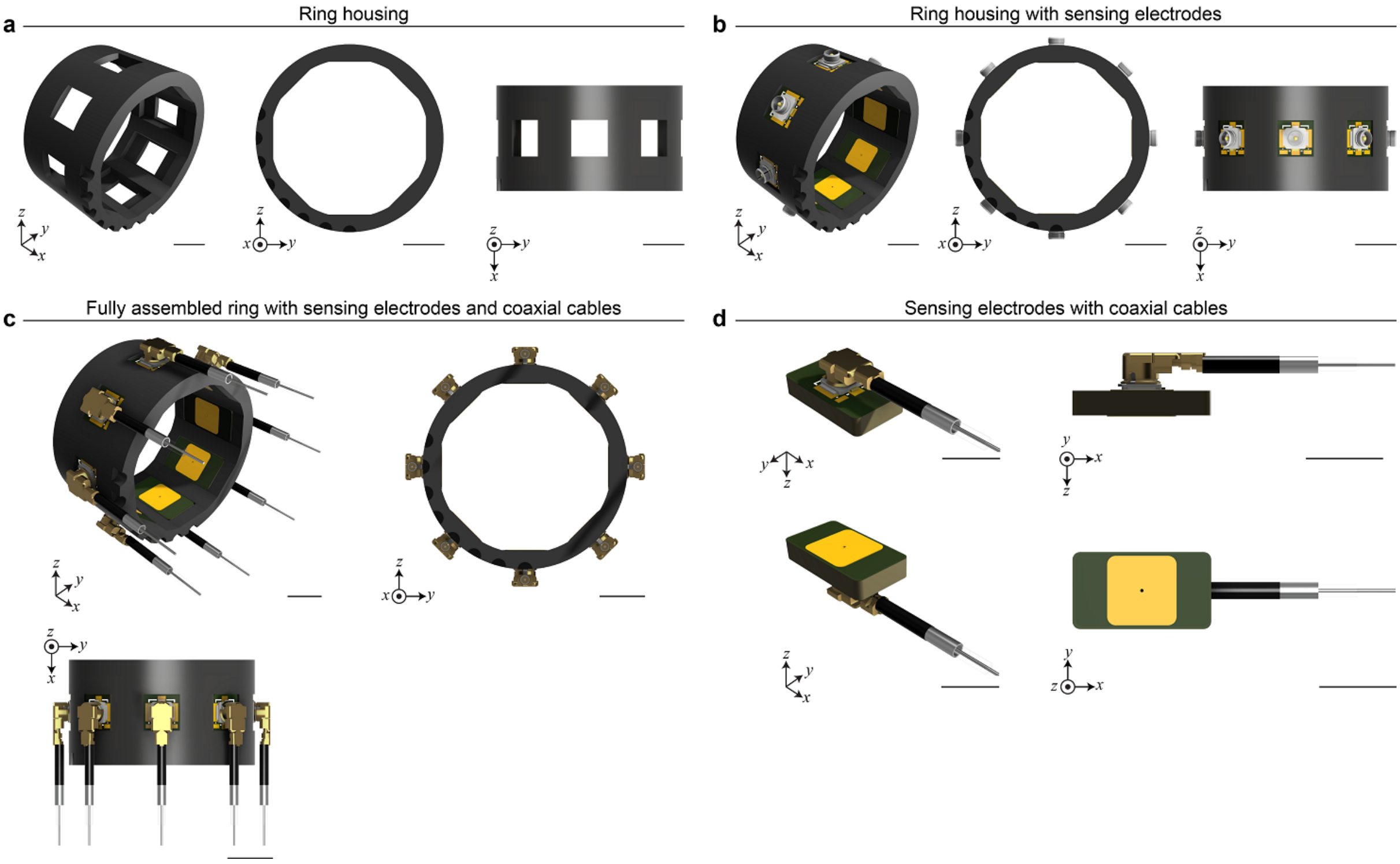

# Supplementary Fig. 2. Baseline palmar arterial model for particle-laden flow simulation

**a**, Subject-specific model of the palmar arterial network. **b**, Detail of local mesh refinement.

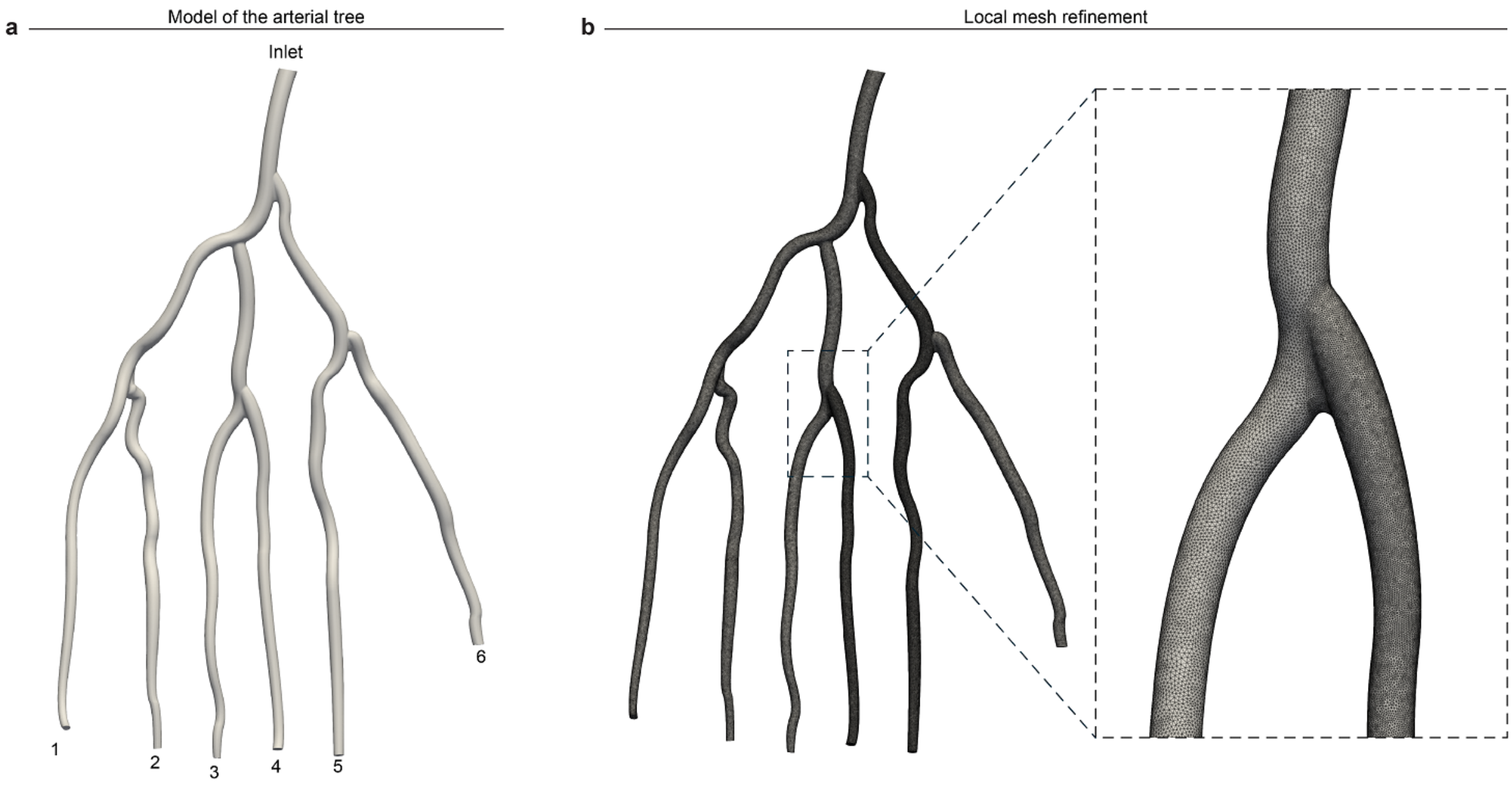

## Supplementary Fig. 3. Instantaneous hematocrit at the injection site of the particles

Scale bar, a quarter period.

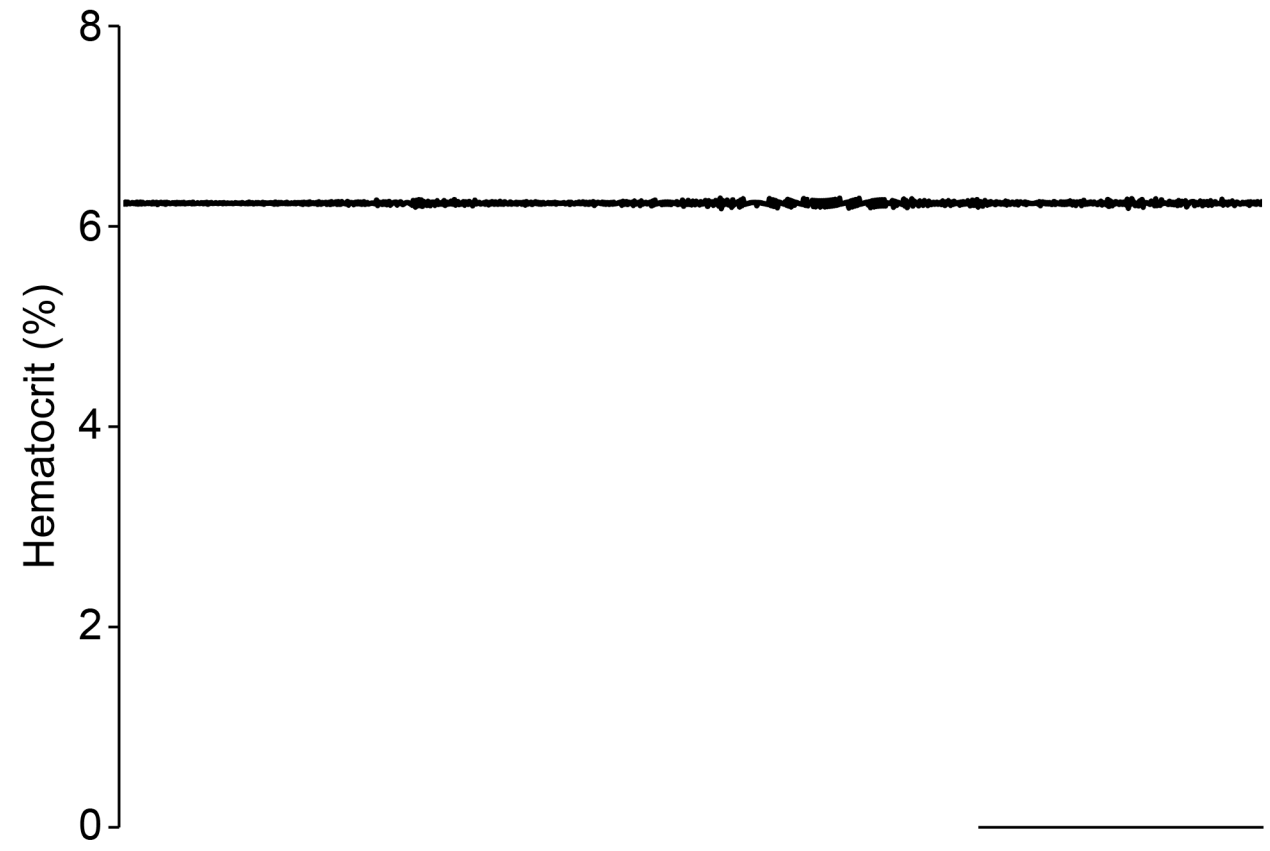

## Supplementary Fig. 4. Isopotential simulations for 8 electrodes at skip-0 and skip-1 injection pattern

Isopotential lines (black) for all injection pairs for peripheral vascular impedance (PVI) ring simulations with 8 electrodes, using the skip-0 (adjacent, **a**) and skip-1 (**b**) pattern. All numbers denote real electrical potentials in volts (V) at the corresponding electrodes. Electrodes with 1 V and -1 V are current source and sink, respectively. Scale bars, 5 mm.

**a**

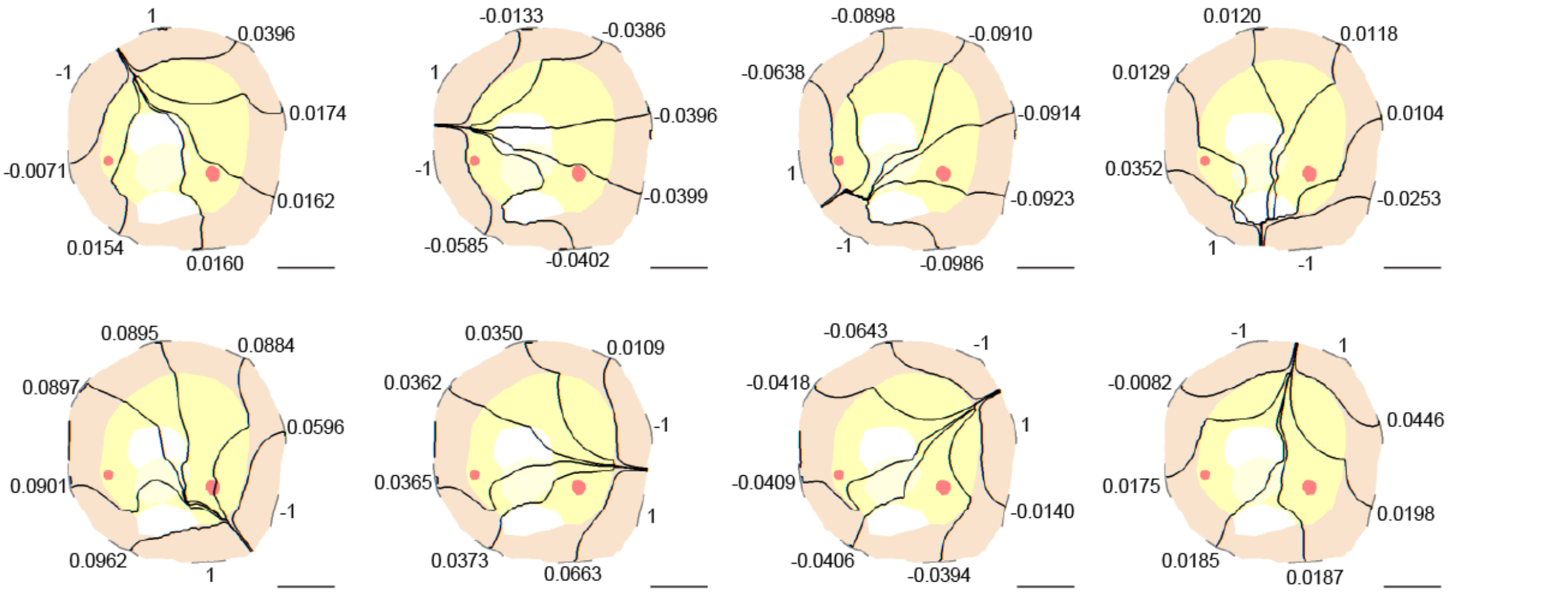

**b**

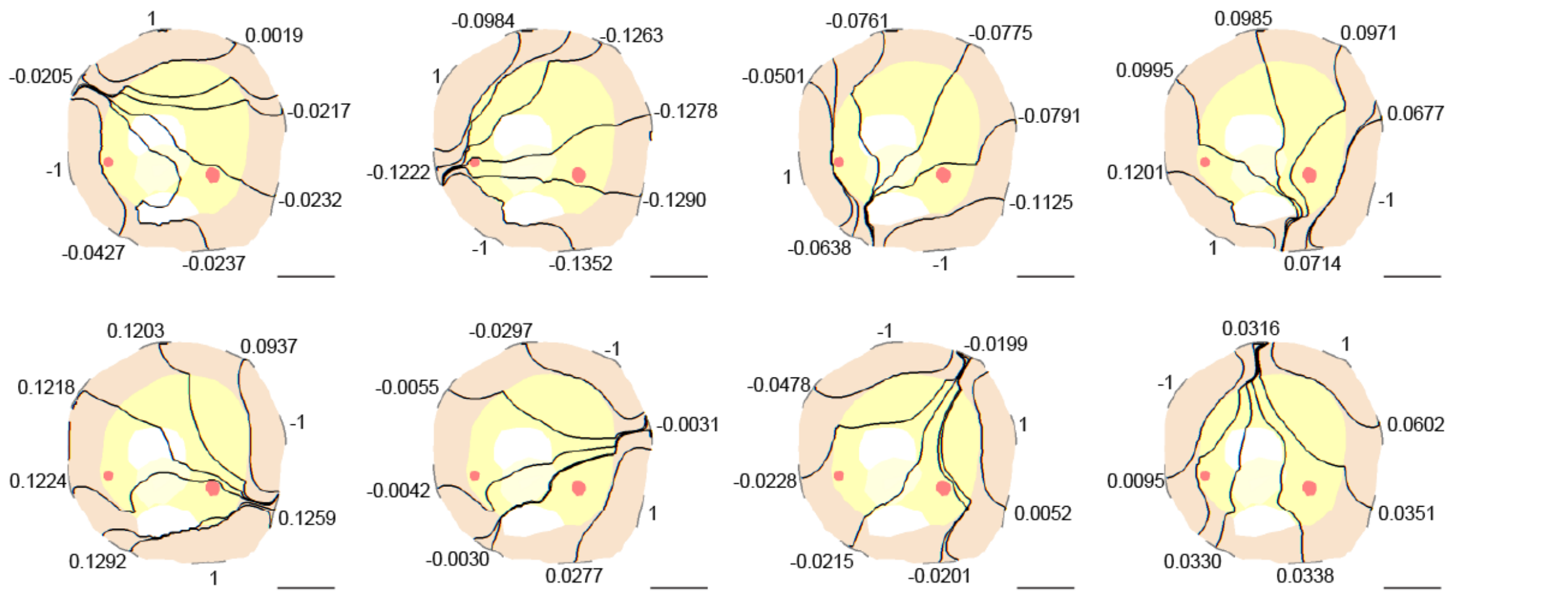

## Supplementary Fig. 5. Isopotential simulations for 8 electrodes at skip-2 and skip-3 injection pattern

Isopotential lines (black) for all injection pairs for peripheral vascular impedance (PVI) ring simulations with 8 electrodes, using the skip-2 (**a**) and skip-3 (opposite, **b**) pattern. All numbers denote real electrical potentials in volts (V) at the corresponding electrodes. Electrodes with 1 V and -1 V are current source and sink, respectively. Scale bars, 5 mm.

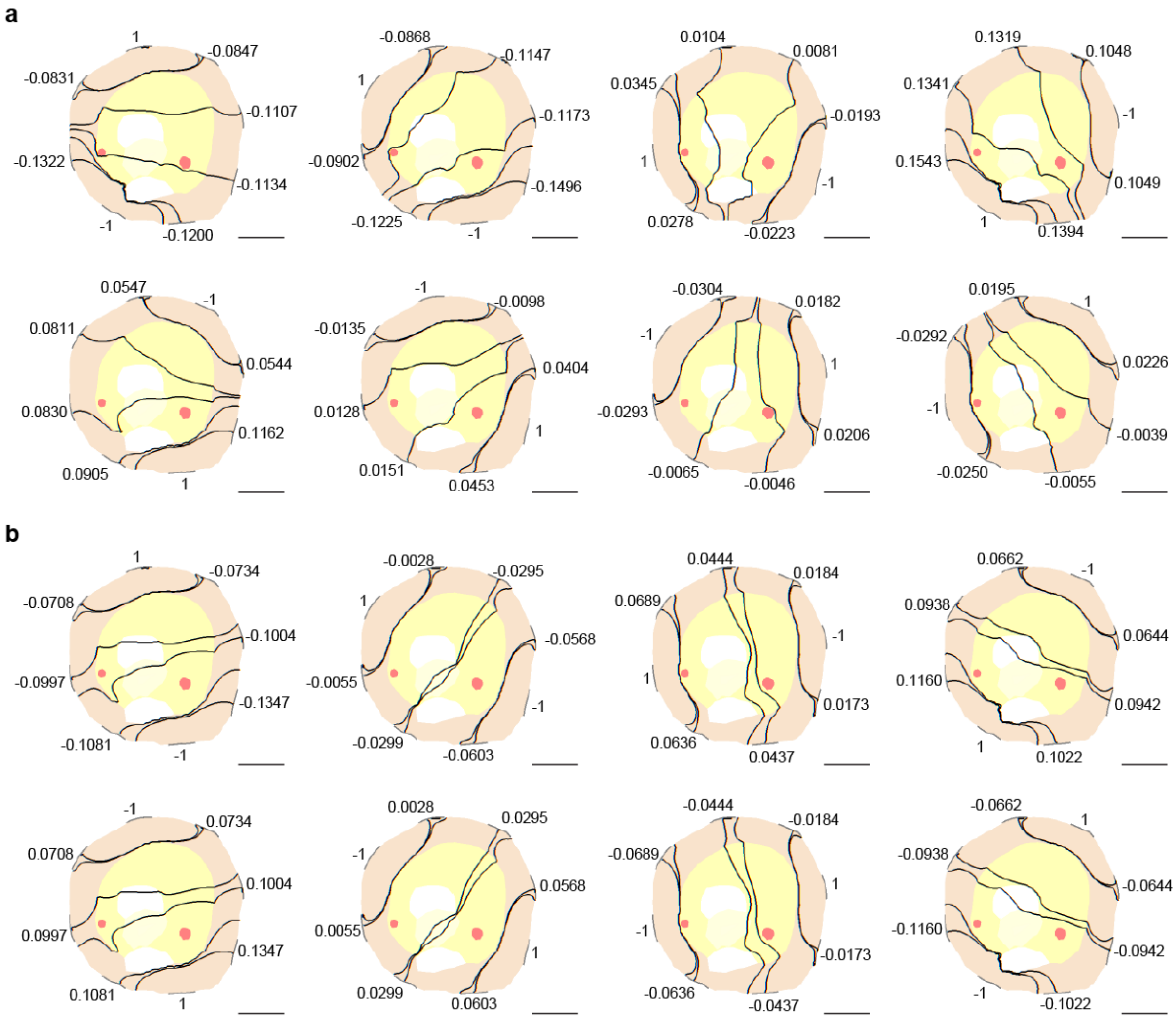

## Supplementary Fig. 6. Isopotential simulations for 16 electrodes at skip-0 injection pattern

Isopotential lines (black) for all injection pairs for peripheral vascular impedance (PVI) ring simulations with 16 electrodes, using the skip-0 (adjacent) pattern. All numbers denote real electrical potentials in volts (V) at the corresponding electrodes. Electrodes with 1 V and -1 V are current source and sink, respectively. Scale bars, 5 mm.

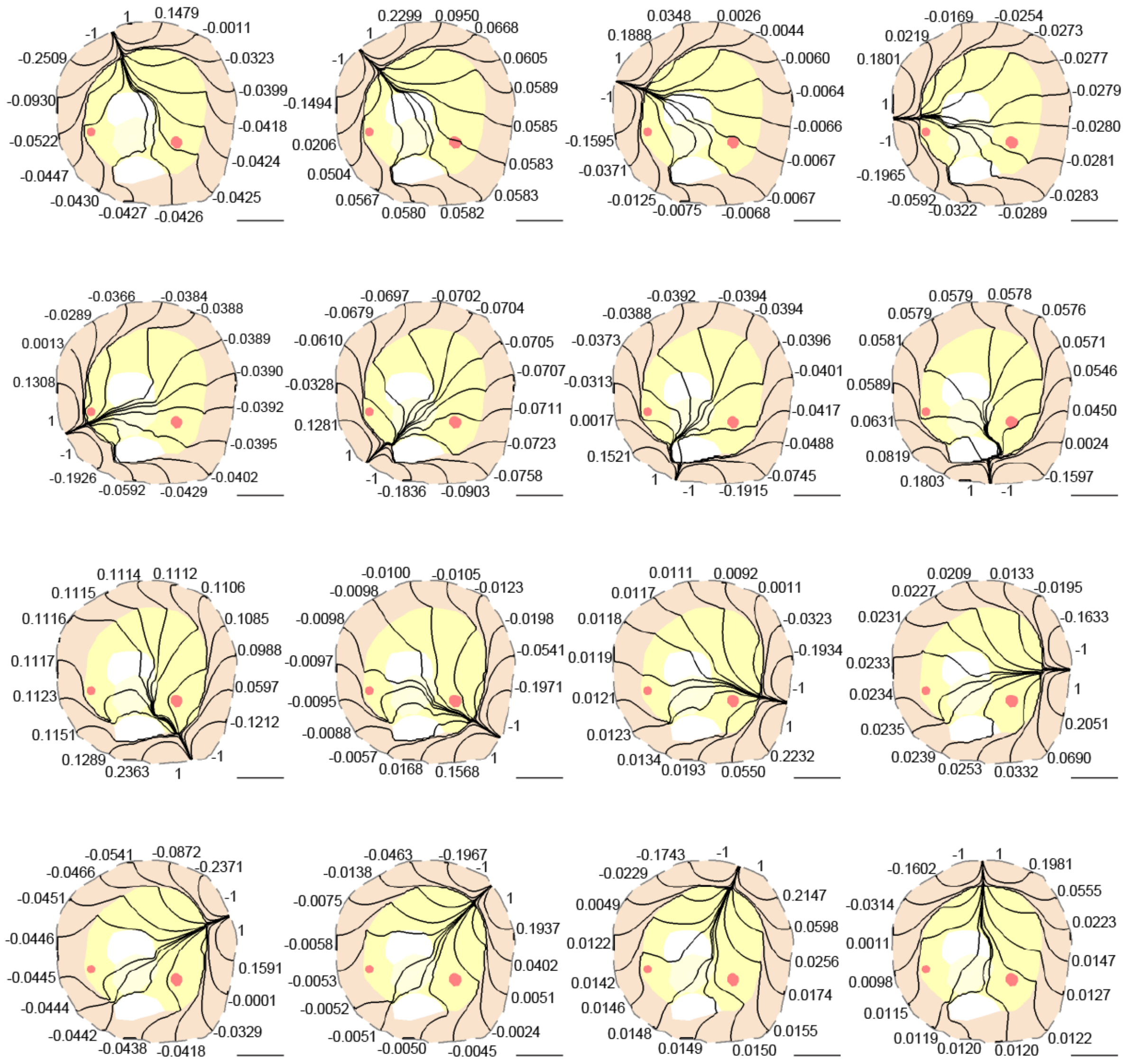

## Supplementary Fig. 7. Isopotential simulations for 16 electrodes at skip-1 injection pattern

Isopotential lines (black) for all injection pairs for peripheral vascular impedance (PVI) ring simulations with 16 electrodes, using the skip-1 pattern. All numbers denote real electrical potentials in volts (V) at the corresponding electrodes. Electrodes with 1 V and -1 V are current source and sink, respectively. Scale bars, 5 mm.

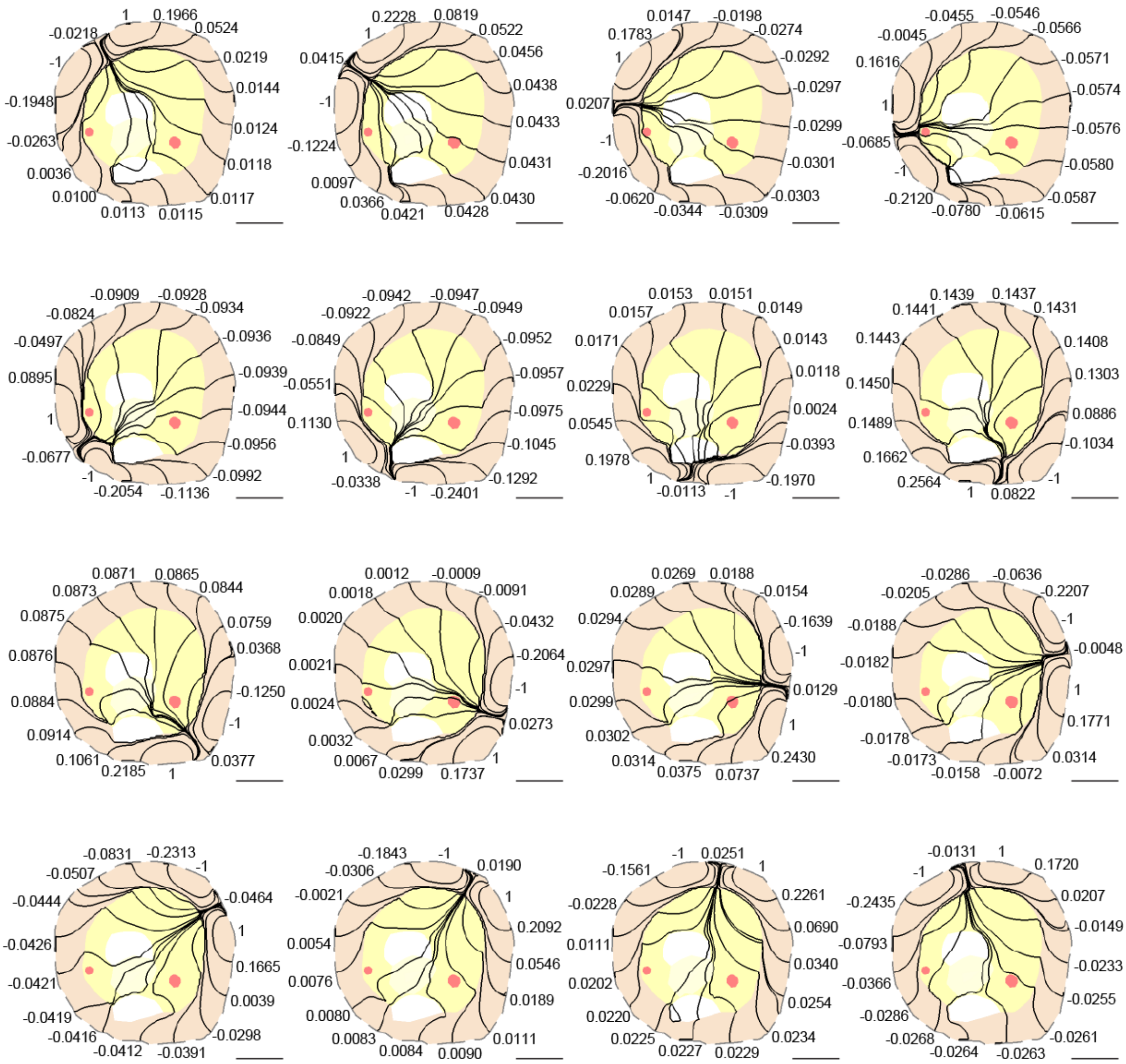

## Supplementary Fig. 8. Isopotential simulations for 16 electrodes at skip-2 injection pattern

Isopotential lines (black) for all injection pairs for peripheral vascular impedance (PVI) ring simulations with 16 electrodes, using the skip-2 pattern. All numbers denote real electrical potentials in volts (V) at the corresponding electrodes. Electrodes with 1 V and -1 V are current source and sink, respectively. Scale bars, 5 mm.

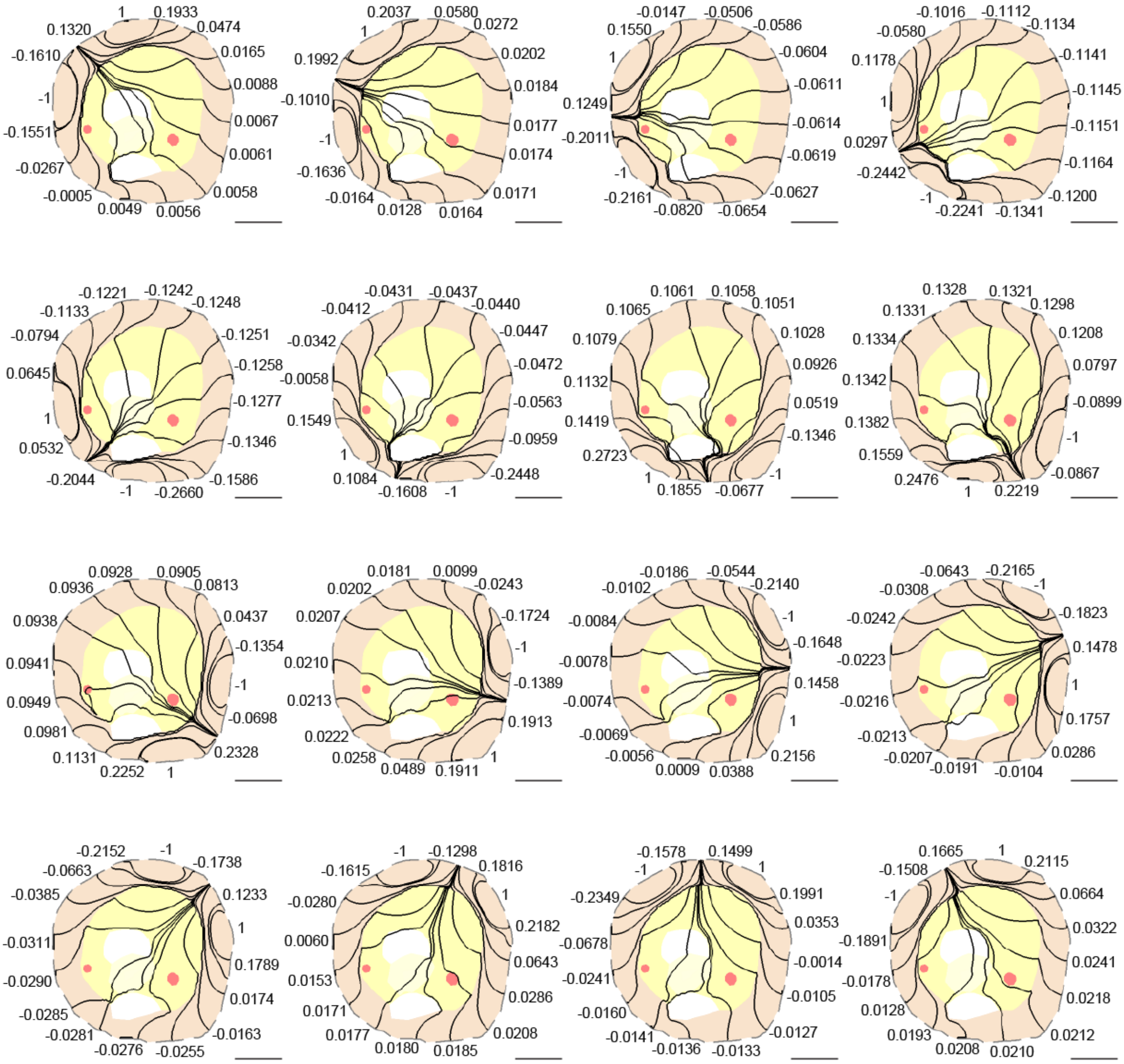

## Supplementary Fig. 9. Isopotential simulations for 16 electrodes at skip-3 injection pattern

Isopotential lines (black) for all injection pairs for peripheral vascular impedance (PVI) ring simulations with 16 electrodes, using the skip-3 pattern. All numbers denote real electrical potentials in volts (V) at the corresponding electrodes. Electrodes with 1 V and -1 V are current source and sink, respectively. Scale bars, 5 mm.

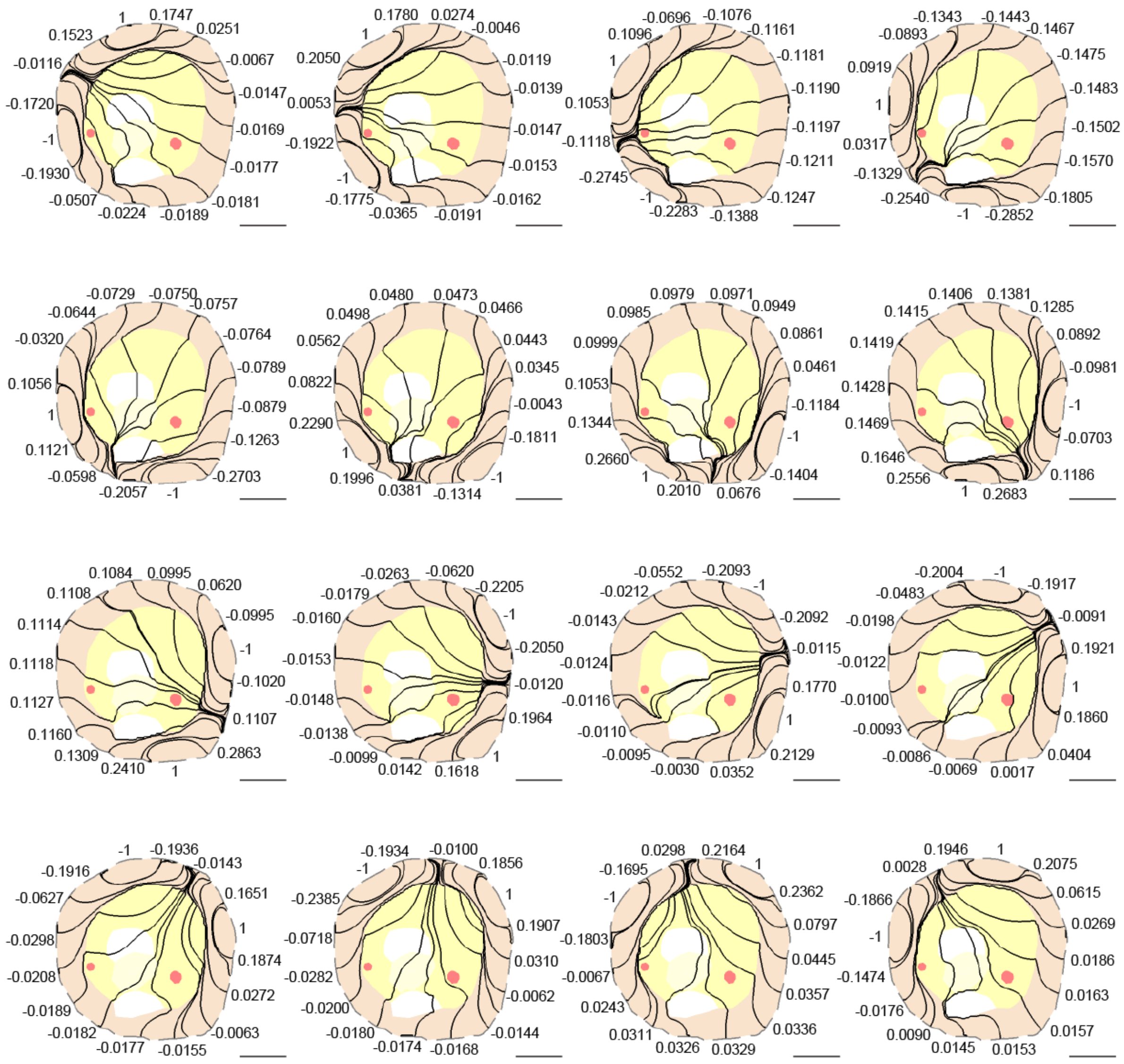

## Supplementary Fig. 10. Isopotential simulations for 16 electrodes at skip-4 injection pattern

Isopotential lines (black) for all injection pairs for peripheral vascular impedance (PVI) ring simulations with 16 electrodes, using the skip-4 pattern. All numbers denote real electrical potentials in volts (V) at the corresponding electrodes. Electrodes with 1 V and -1 V are current source and sink, respectively. Scale bars, 5 mm.

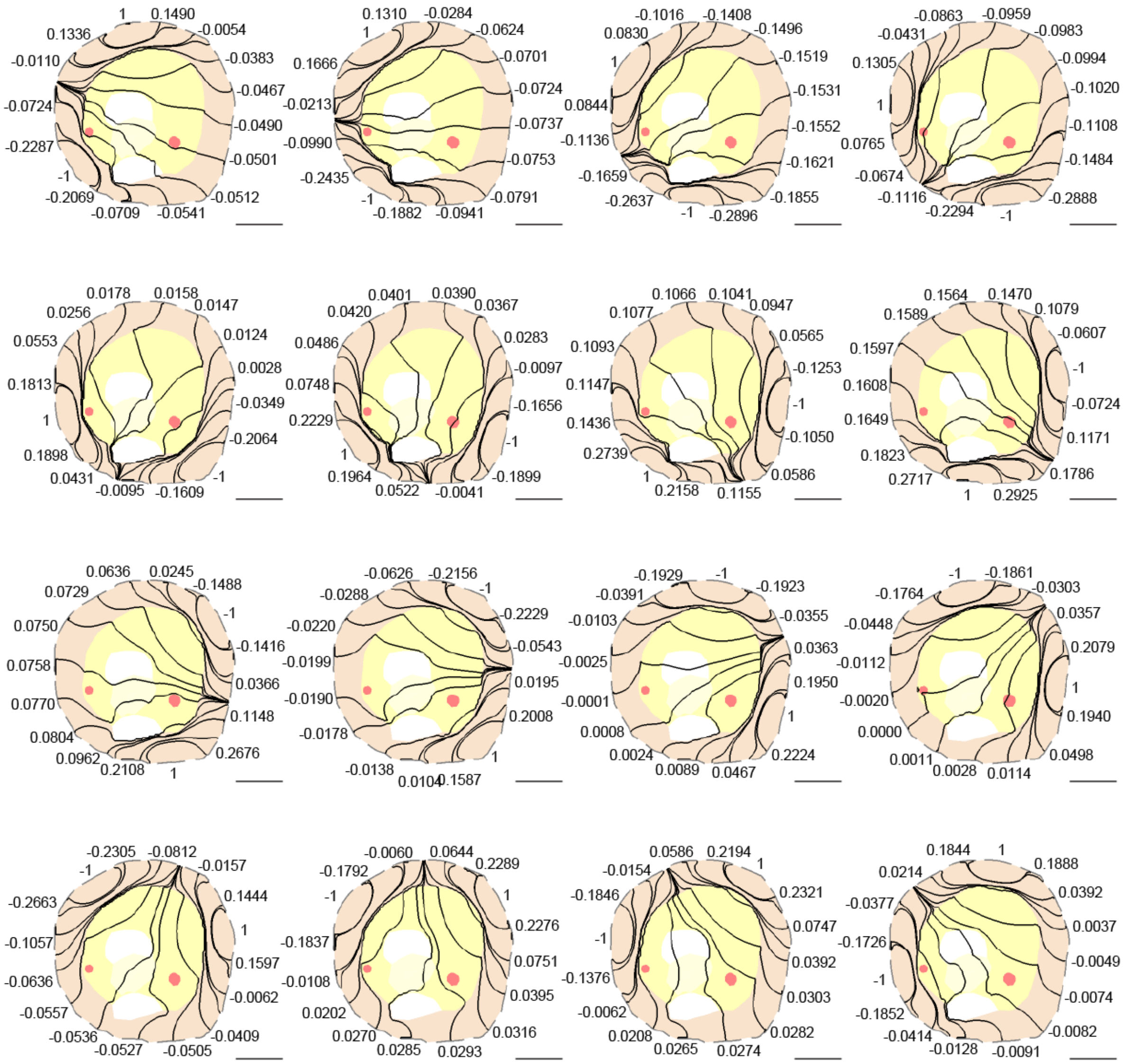

## Supplementary Fig. 11. Isopotential simulations for 16 electrodes at skip-5 injection pattern

Isopotential lines (black) for all injection pairs for peripheral vascular impedance (PVI) ring simulations with 16 electrodes, using the skip-5 pattern. All numbers denote real electrical potentials in volts (V) at the corresponding electrodes. Electrodes with 1 V and -1 V are current source and sink, respectively. Scale bars, 5 mm.

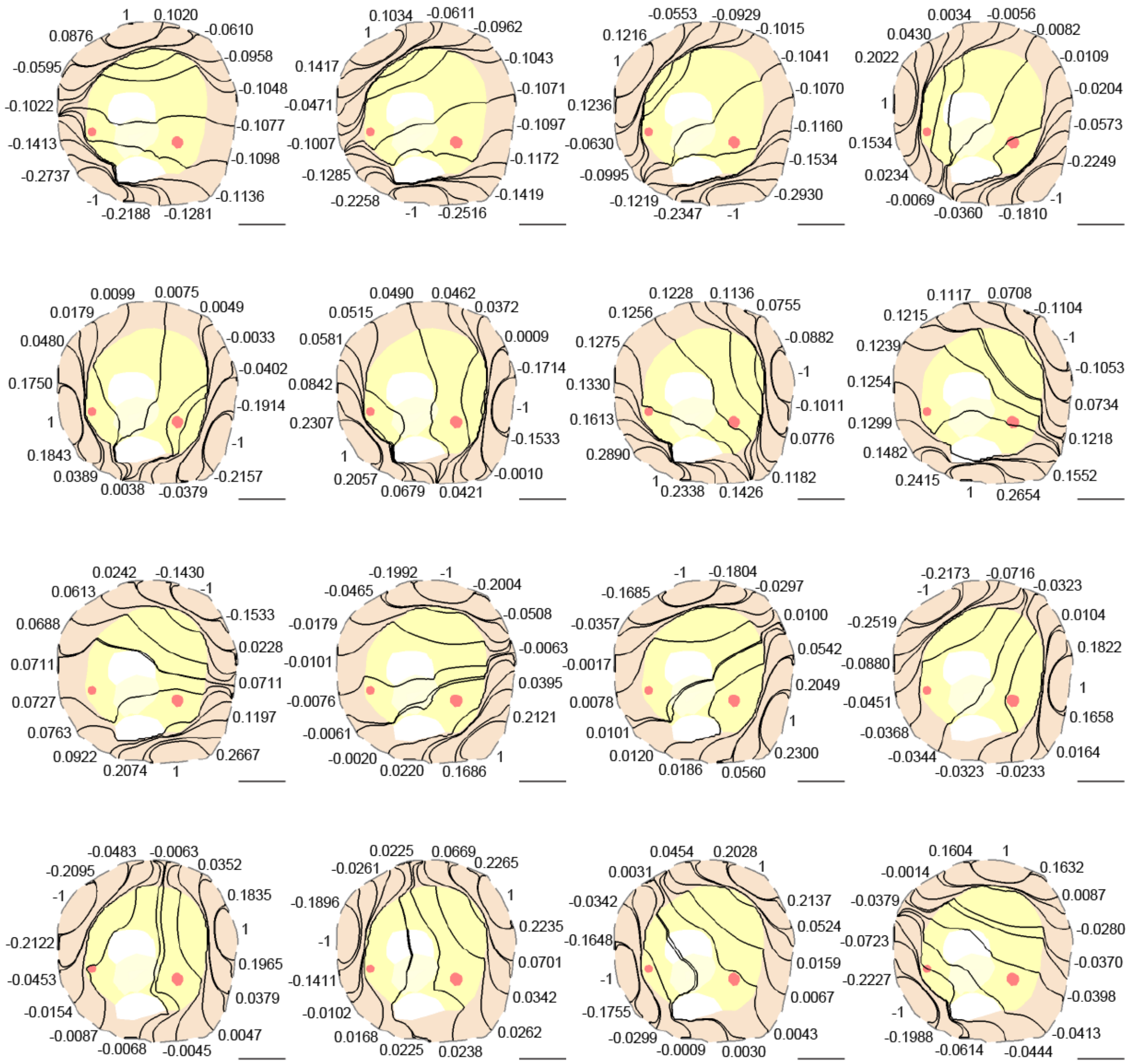

## Supplementary Fig. 12. Isopotential simulations for 16 electrodes at skip-6 injection pattern

Isopotential lines (black) for all injection pairs for peripheral vascular impedance (PVI) ring simulations with 16 electrodes, using the skip-6 pattern. All numbers denote real electrical potentials in volts (V) at the corresponding electrodes. Electrodes with 1 V and -1 V are current source and sink, respectively. Scale bars, 5 mm.

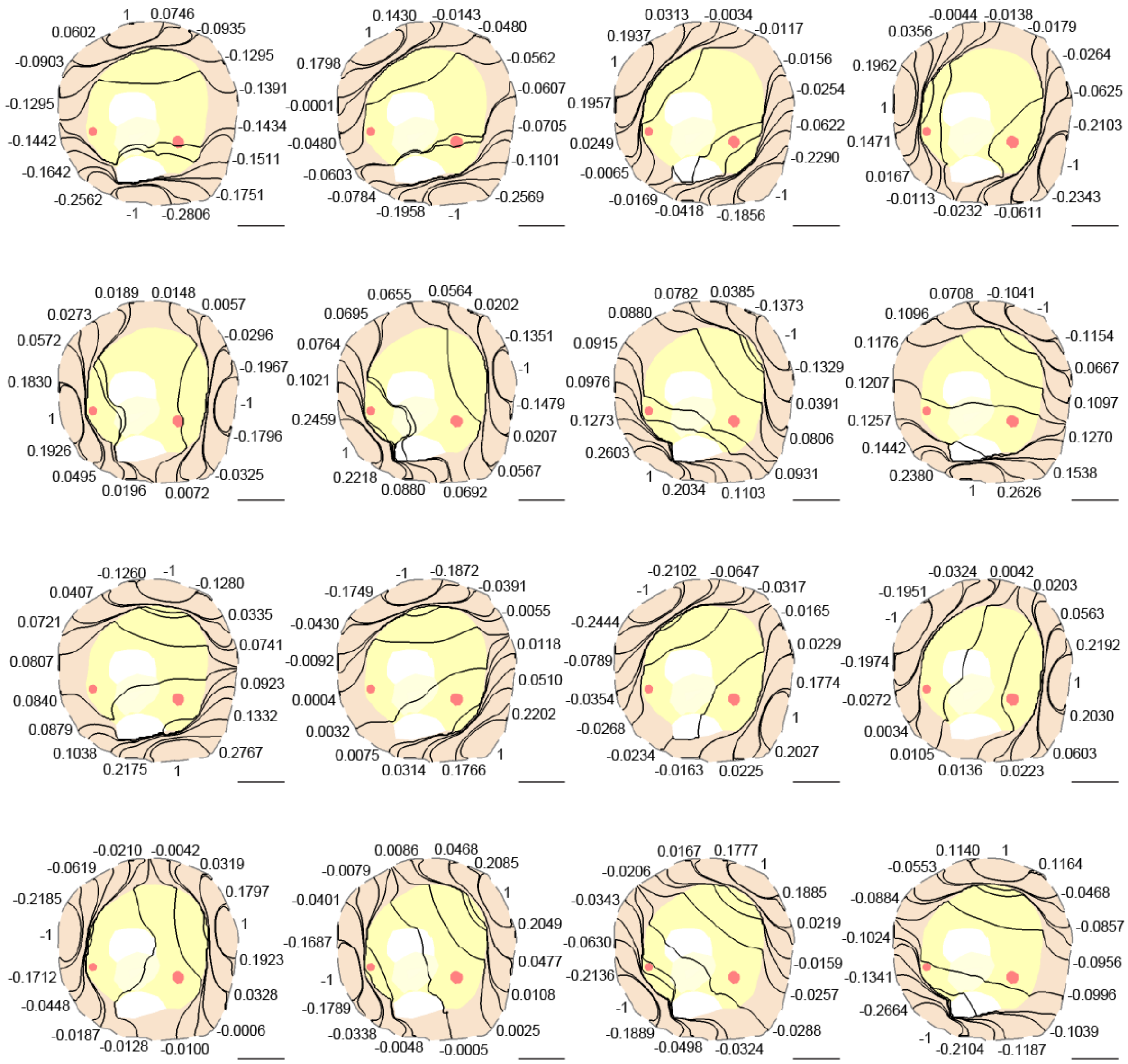

## Supplementary Fig. 13. Isopotential simulations for 16 electrodes at skip-7 injection pattern

Isopotential lines (black) for all injection pairs for peripheral vascular impedance (PVI) ring simulations with 16 electrodes, using the skip-7 (opposite) pattern. All numbers denote real electrical potentials in volts (V) at the corresponding electrodes. Electrodes with 1 V and -1 V are current source and sink, respectively. Scale bars, 5 mm.

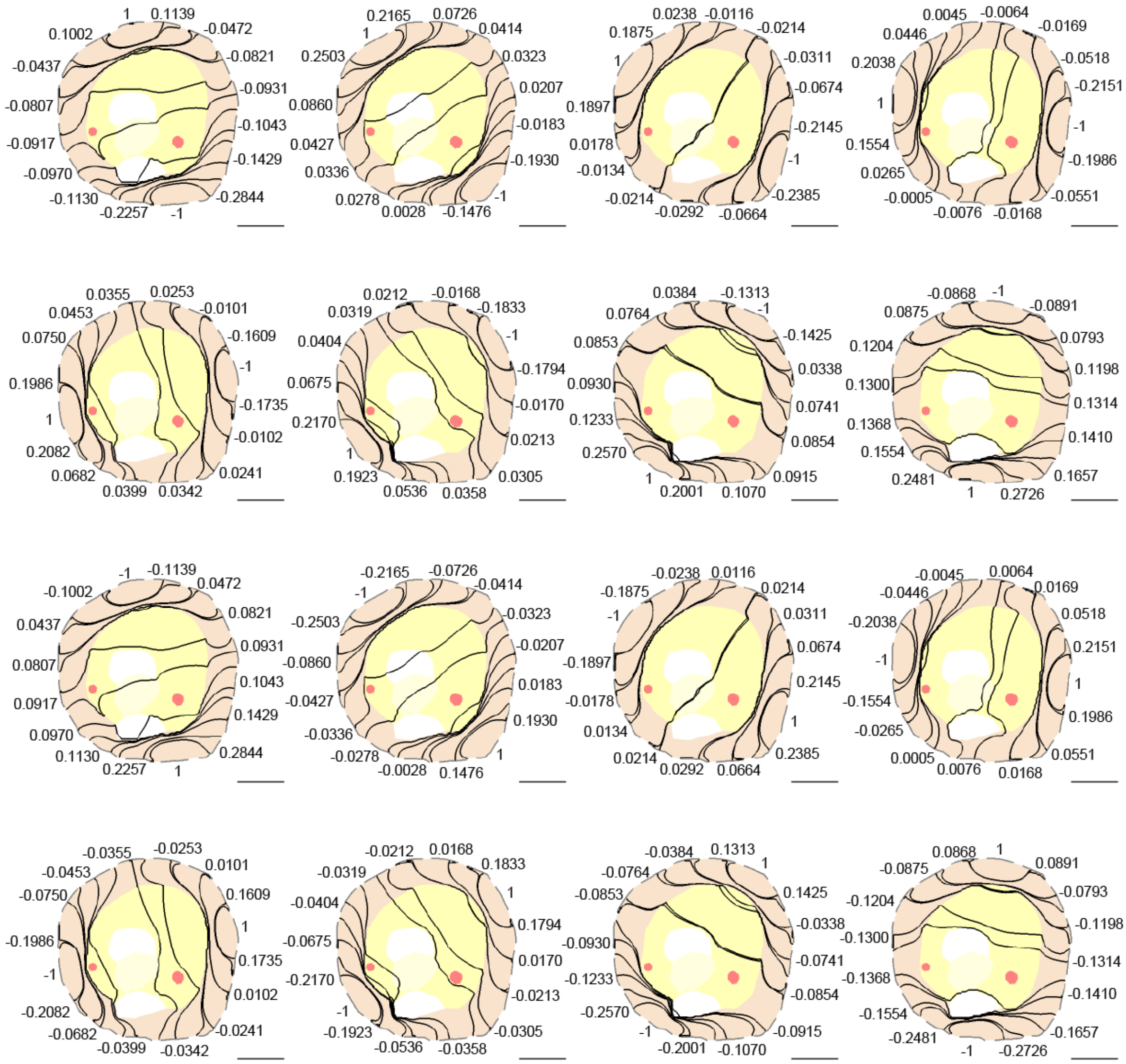

## Supplementary Fig. 14. Volume impedance analysis for 8 electrodes with $(\ell_1, \ell_4)$-injection and $(\ell_8, \ell_2)$-measurement pair

**a**, Electrode configuration. **b**, $xz$ and $yz$ slices with highest sensitivity. **c**, Impedance density distribution in the region that accounts for 95% of total impedance magnitude. VID, volume impedance density; $I^{(+,-)}$, current injection electrodes; $V^{(+,-)}$, voltage sensing electrodes; Green and blue patches, injection and measurement electrodes, respectively; Scale bars, 5 mm.

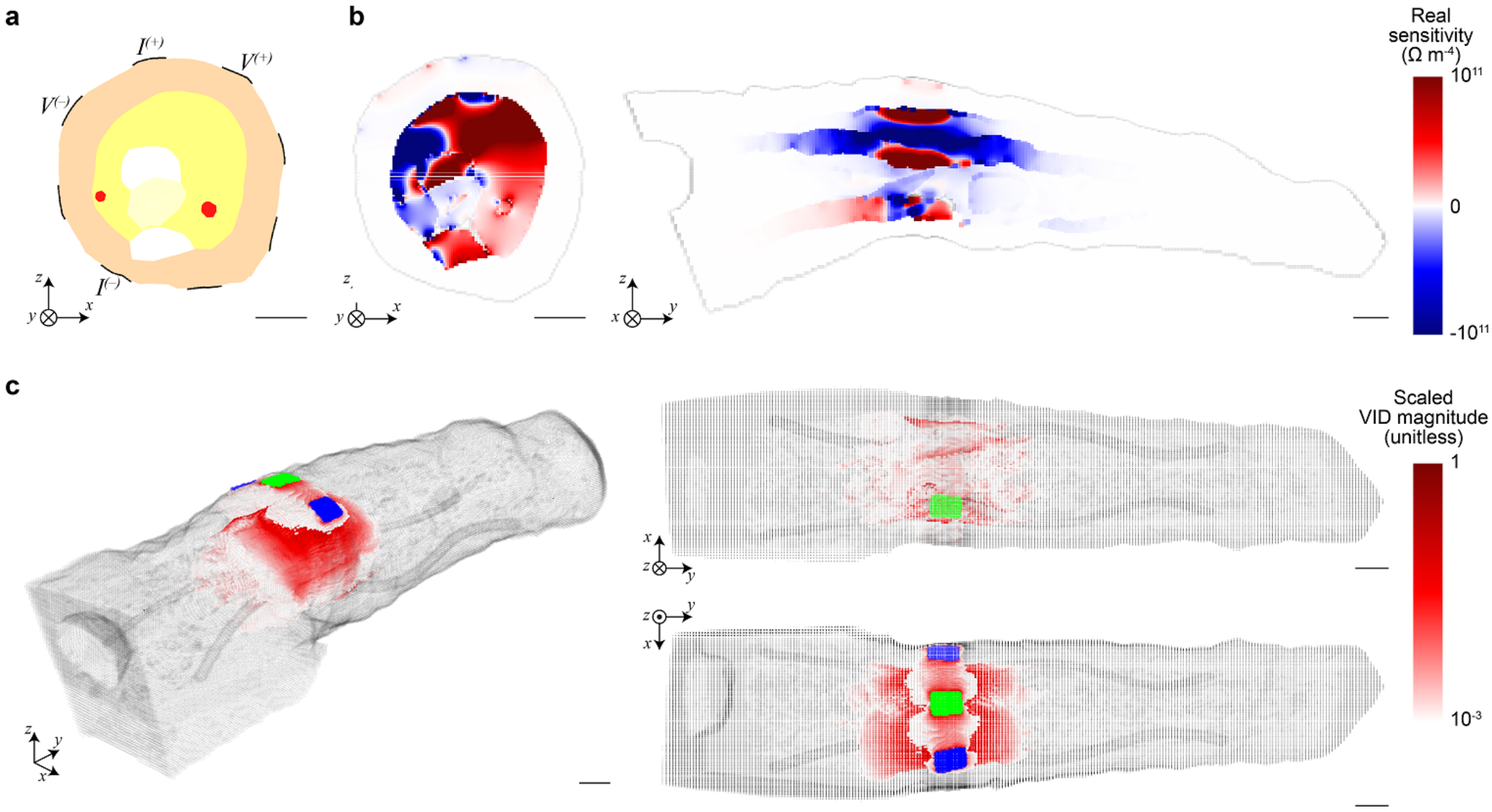

## Supplementary Fig. 15. Volume impedance analysis for 8 electrodes with $(\ell_4, \ell_7)$-injection and $(\ell_3, \ell_5)$-measurement pair

**a**, Electrode configuration. **b**, $xz$ and $yz$ slices with highest sensitivity. **c**, Impedance density distribution in the region that accounts for 95% of total impedance magnitude. VID, volume impedance density; $I^{(+,-)}$, current injection electrodes; $V^{(+,-)}$, voltage sensing electrodes; Green and blue patches, injection and measurement electrodes, respectively; Scale bars, 5 mm.

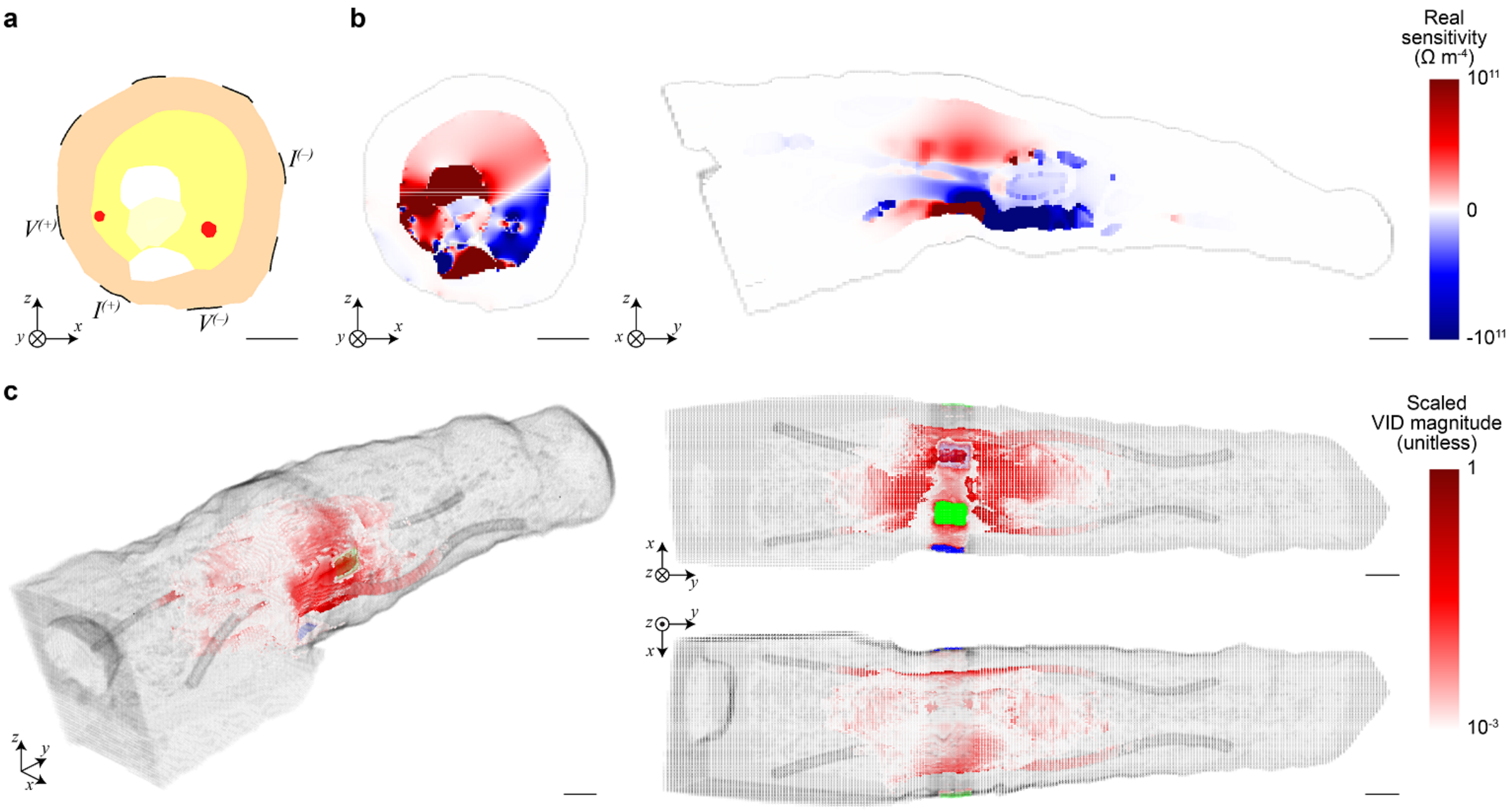

## Supplementary Fig. 16. Volume impedance analysis for 8 electrodes with $(\ell_7, \ell_2)$-injection and $(\ell_4, \ell_6)$-measurement pair

**a**, Electrode configuration. **b**, $xz$ and $yz$ slices with highest sensitivity. **c**, Impedance density distribution in the region that accounts for 95% of total impedance magnitude. VID, volume impedance density; $I^{(+,-)}$, current injection electrodes; $V^{(+,-)}$, voltage sensing electrodes; Green and blue patches, injection and measurement electrodes, respectively; Scale bars, 5 mm.

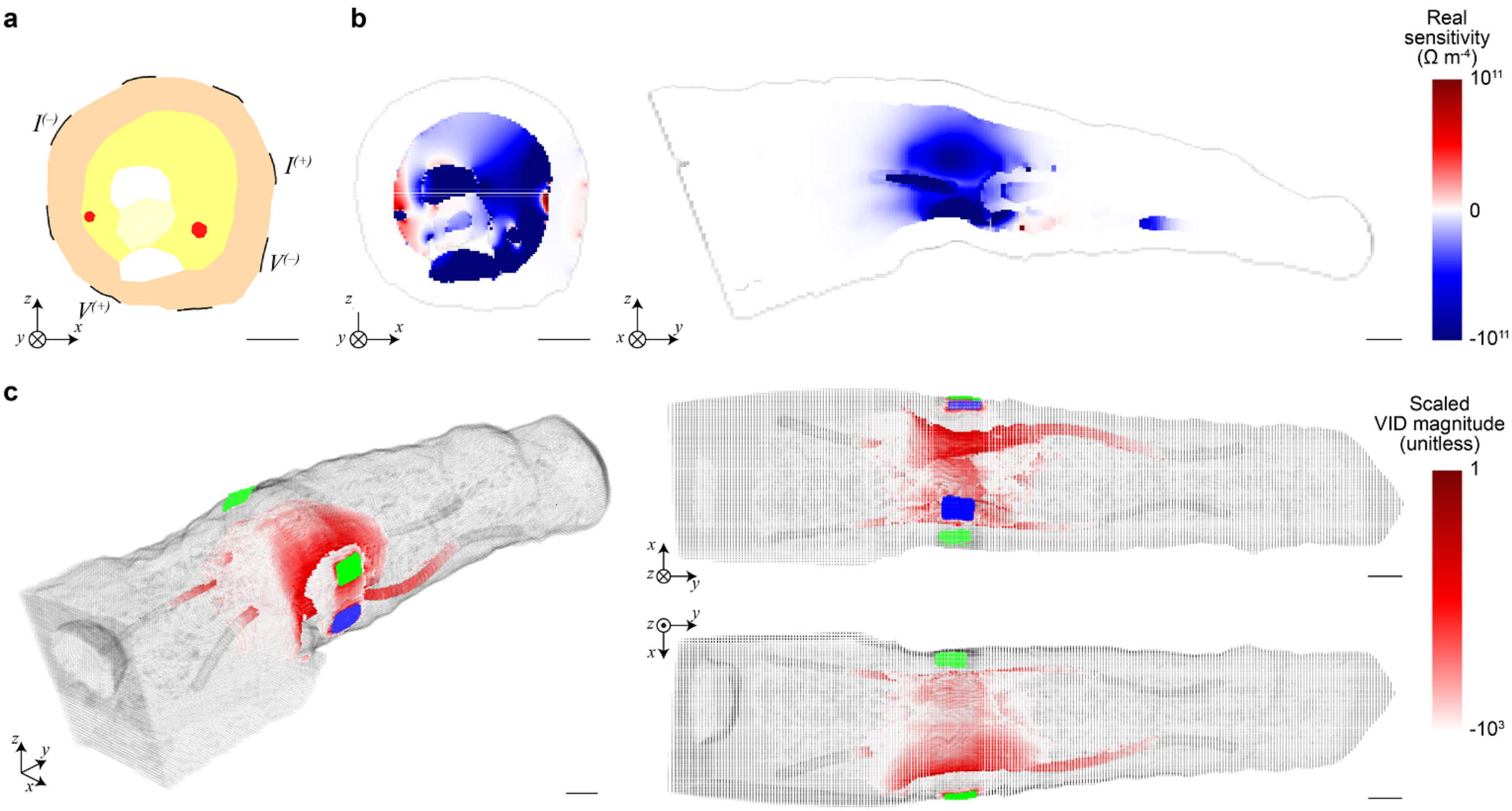

# Supplementary Fig. 17. Tissue contribution to surface resistance and reactance

Tissue contribution to surface resistance and reactance from three different configurations. Configuration 1 uses $(\ell_1, \ell_4)$ injection pair and $(\ell_8, \ell_2)$ measurement pair. Configuration 2 uses $(\ell_4, \ell_7)$ injection pair and $(\ell_3, \ell_5)$ measurement pair. Configuration 3 uses $(\ell_7, \ell_2)$ injection pair and $(\ell_4, \ell_6)$ measurement pair. The configurations are shown in Supplementary Fig. 14–16. SAT, subcutaneous adipose tissue.

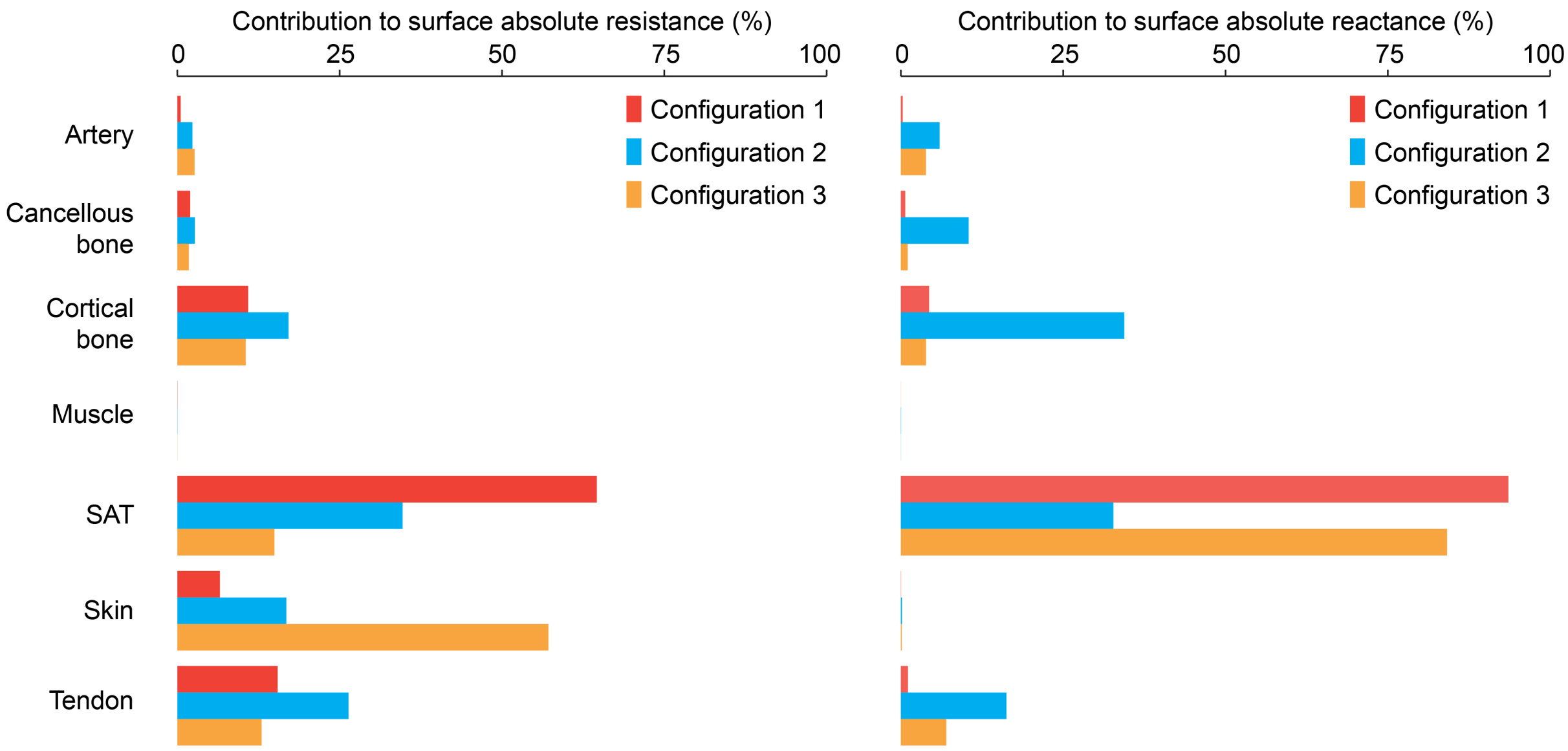

# Supplementary Fig. 18. Reconstruction of synthetic conductivity waveform with 8 and 16 electrodes

Reconstruction of synthetic conductivity waveform with 8 electrodes (**a**) and 16 electrodes (**b**). **i**. Reconstructed signal extracted at region of interest (ROI) with representative peripheral vascular impedance (PVI) images at time indices 30, 47, 70, and 100. The ROI were detected using singular value decomposition. **ii**. Correlation between reconstructed and reference conductivity. **iii**. Histogram of absolute errors between reconstructed and reference conductivity. $t$, time index; $r^2$, coefficient of determination; p, p-value; white bars in **i**, regions of interest.

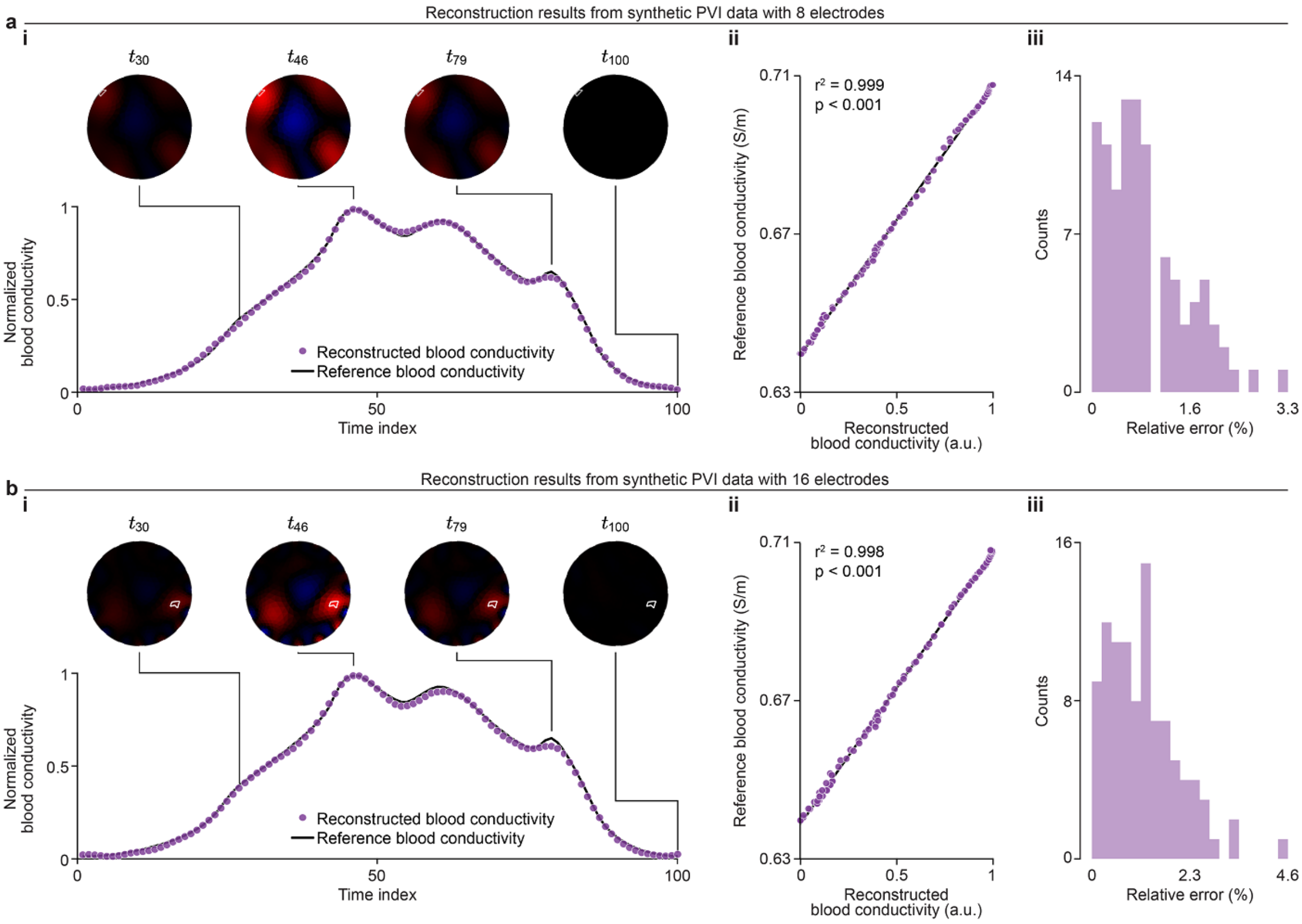

# Supplementary Fig. 19. Design of the peripheral vascular impedance imaging reconstruction software

**a**, Architecture of the peripheral vascular impedance image reconstruction software, with three main modules: meshing, forward solver, and inverse solver. The meshing module prepares a finite-element mesh from a given ring size, and computes the refined mesh as well as the Laplacian matrix for regularization. The forward solver calculates the electrode voltages for the refined mesh based on the electrode configurations and an initial value for the internal conductivity distribution. The inverse solver compares the forward solution with new measurements and updates the conductivity change accordingly. **b**, Core operations of the meshing modules. The `laplacian` operation is based on the graph degree and graph adjacency matrix. The `refine` operation adds new nodes at the midpoints of the edges, and connects the nodes to create new edges. The `mesh2img` transformation uses polygon clipping algorithms to compute intersection between all elements in the unstructured mesh and all pixels in the structured grid.

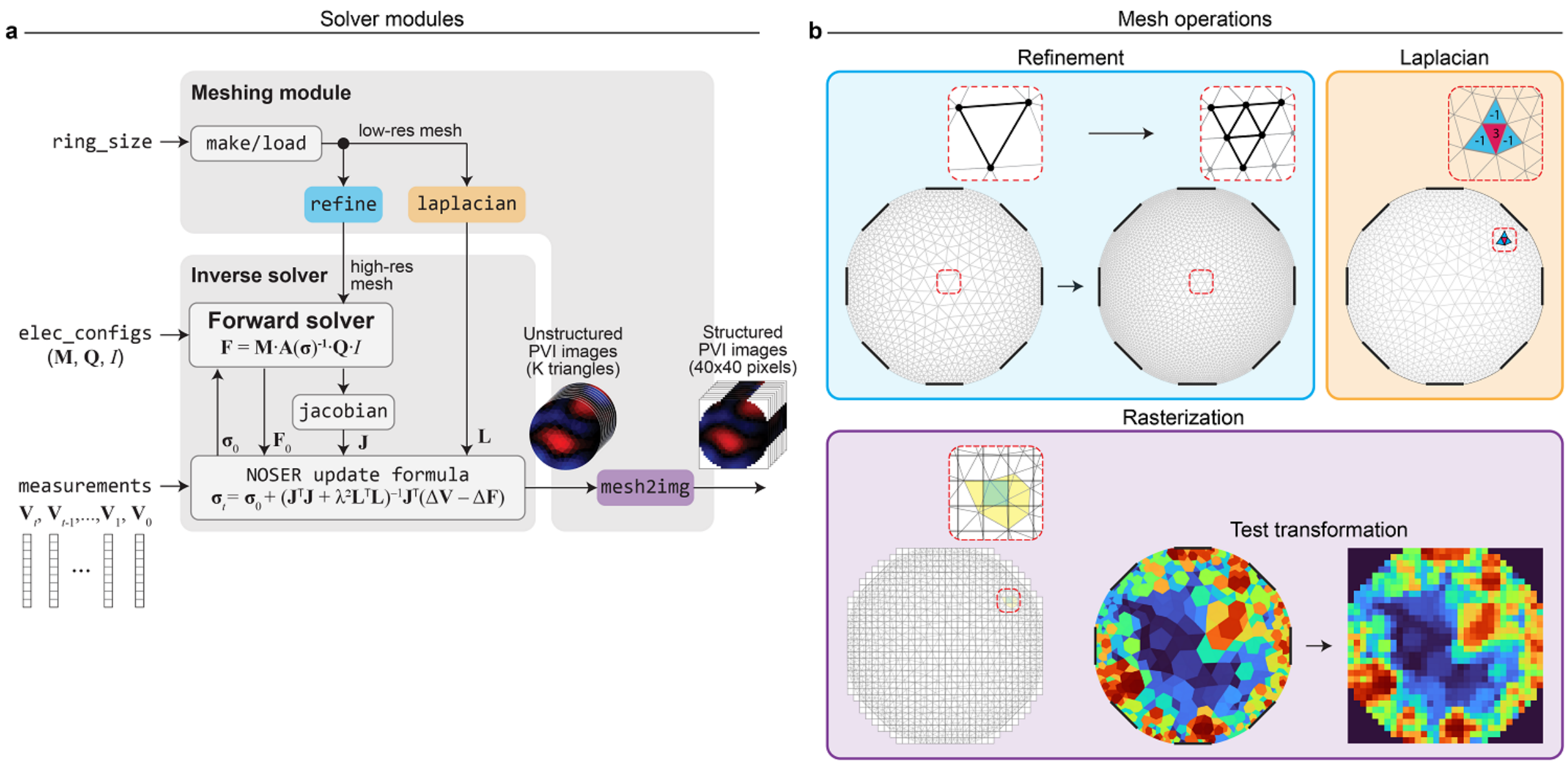

# Supplementary Fig. 20. Finite element ring meshes conformal to various ring sizes

Finite element meshes of all rings from size 6 to size 15, with overlays of the ring profiles. The thick bars on the meshes' peripheral denote electrodes. The semicircular indices on the rings' outer brim encode the ring size. N, number of nodes; K, number of elements. Scale bars, 5 mm.

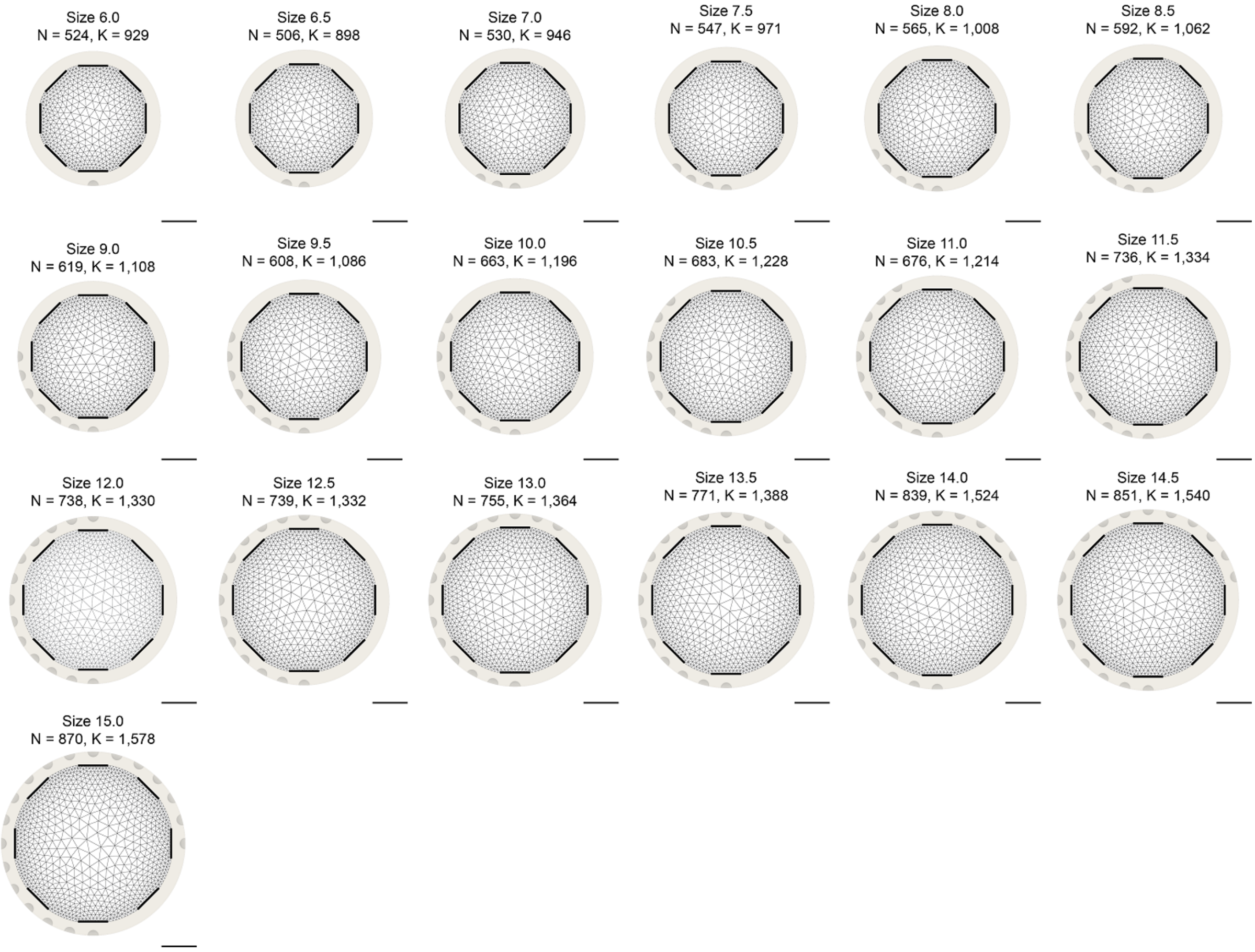

## Supplementary Fig. 21. Numerical validation of the forward solver module

**a**, Validation of the forward operator using a homogeneous rectangular resistor model. **i**, model discretization and parameters; **ii**, simulated results shows linear voltage gradient along the resistor's length. **b**, Validation of the Jacobian computation using a homogeneous disc model having 16 electrodes with $(0, 3)$-injection pair and $(0, 1)$-measurement pair. **i**, model discretization and parameters, with representative injection pair $(7, 10)$ and measurement pair $(2, 3)$; **ii**, corresponding Jacobian results; **iii**, numerical benchmark comparison with Jacobian solution from EIDORS. Scale bars, 5 cm.

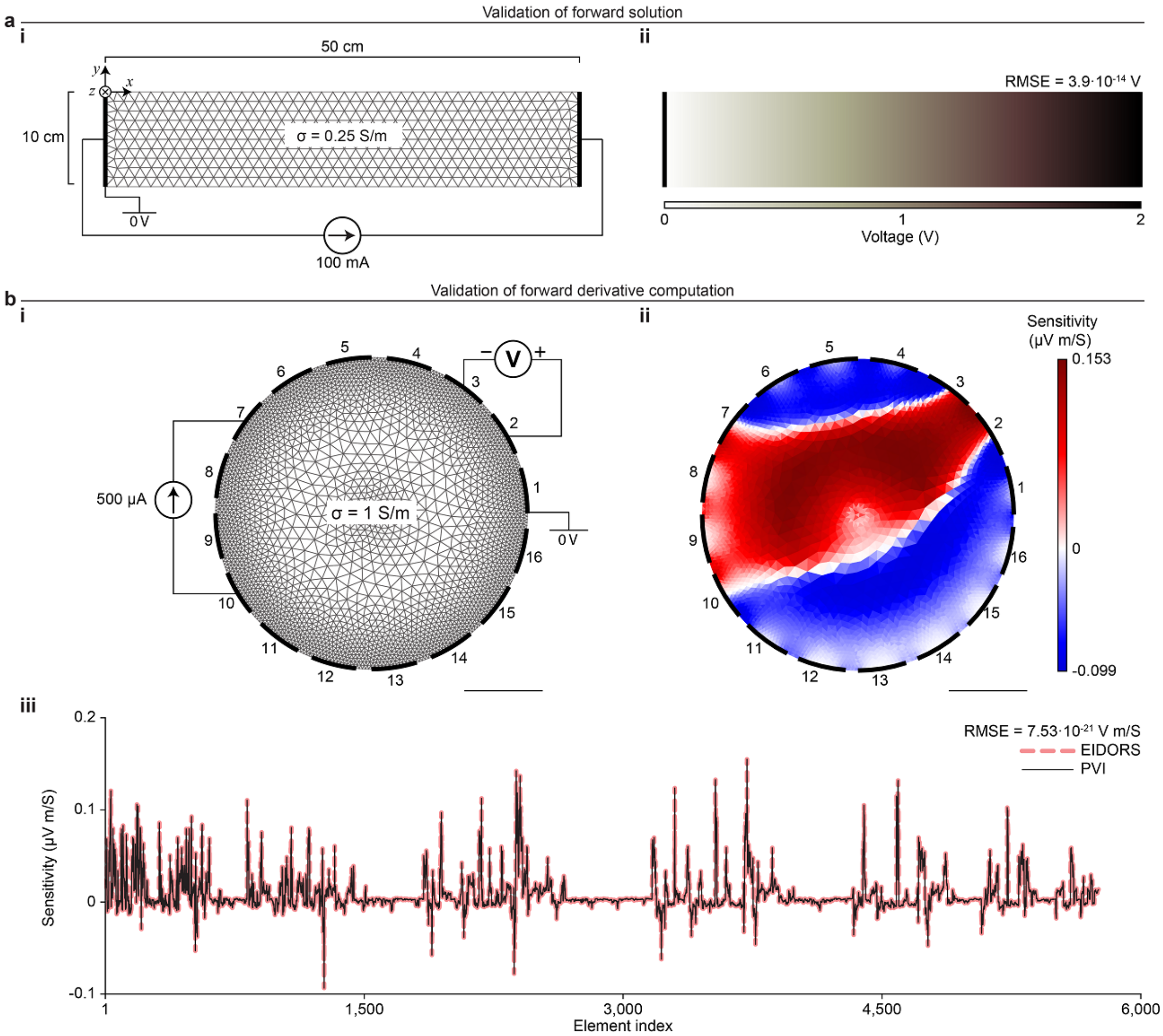

# Supplementary Fig. 22. Numerical validation of the inverse solver module

Validation of the inverse solver using open-source reference experimental data. **a**, representative conductivity images at maximum (**i**) and minimum (**ii**) conductivity change. **b**, numerical benchmark comparison of reconstructed images between our imaging algorithm and EIDORS, showing the histogram of the absolute error (**i**) and the mean conductivity signal (**ii**). Scale bar, 20 s.

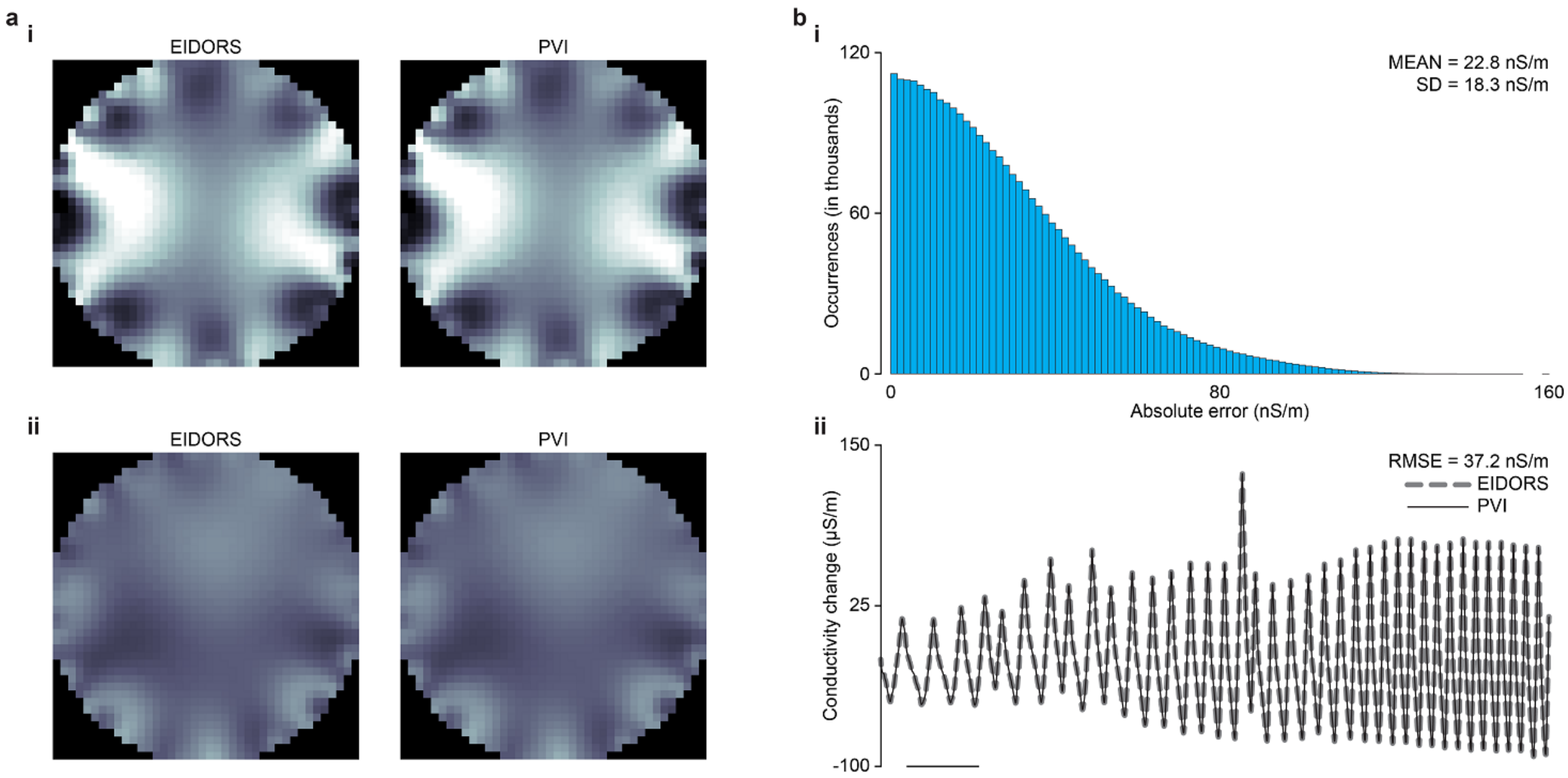

# Supplementary Fig. 23. Effects of regularization on reconstructed image quality

**a**, Experiment setup of a phantom tank containing water and a beaker at the center. **b**, Effect of regularization parameter $\lambda$ on the smoothness of the reconstructed conductivity distribution. From left to right, the regularization parameters were $\lambda = 10^{-6}$, $\lambda = 10^{-4}$, $\lambda = 10^{-3}$, and $\lambda = 10^{-2}$. With lower regularization values, the images contain artifacts due to ringing effects. For larger values, the reconstructed target becomes blurry and faint. Scale bar, 5 cm.

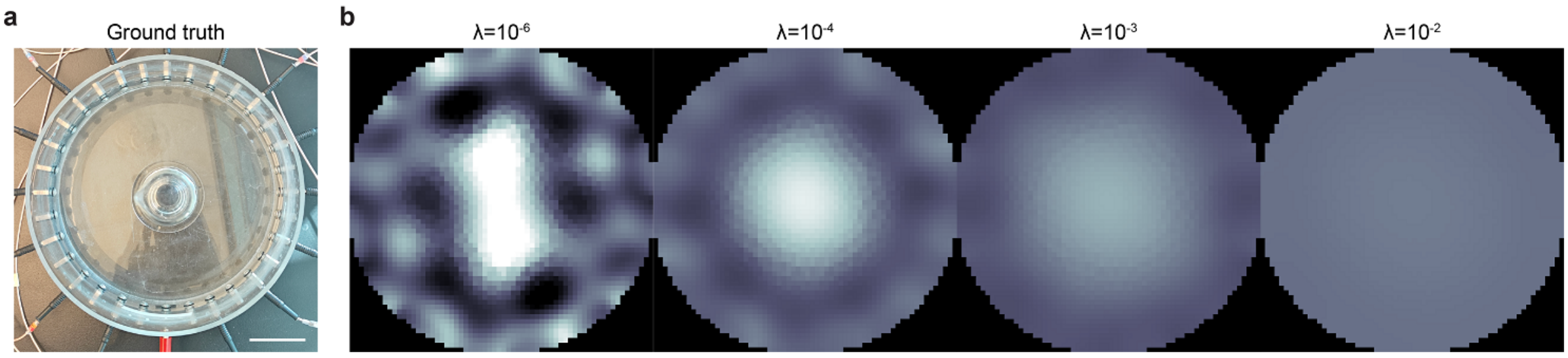

# Supplementary Fig. 24. Experimental study design

**a**, Flow diagram showing participants enrolled into the study, with a total of $N = 96$ subjects participated in the cross-sectional study, of which $N = 5$ returned for a pilot longitudinal study. **b**, Study protocol for the cross-sectional study (**i**) and longitudinal study (**ii**).

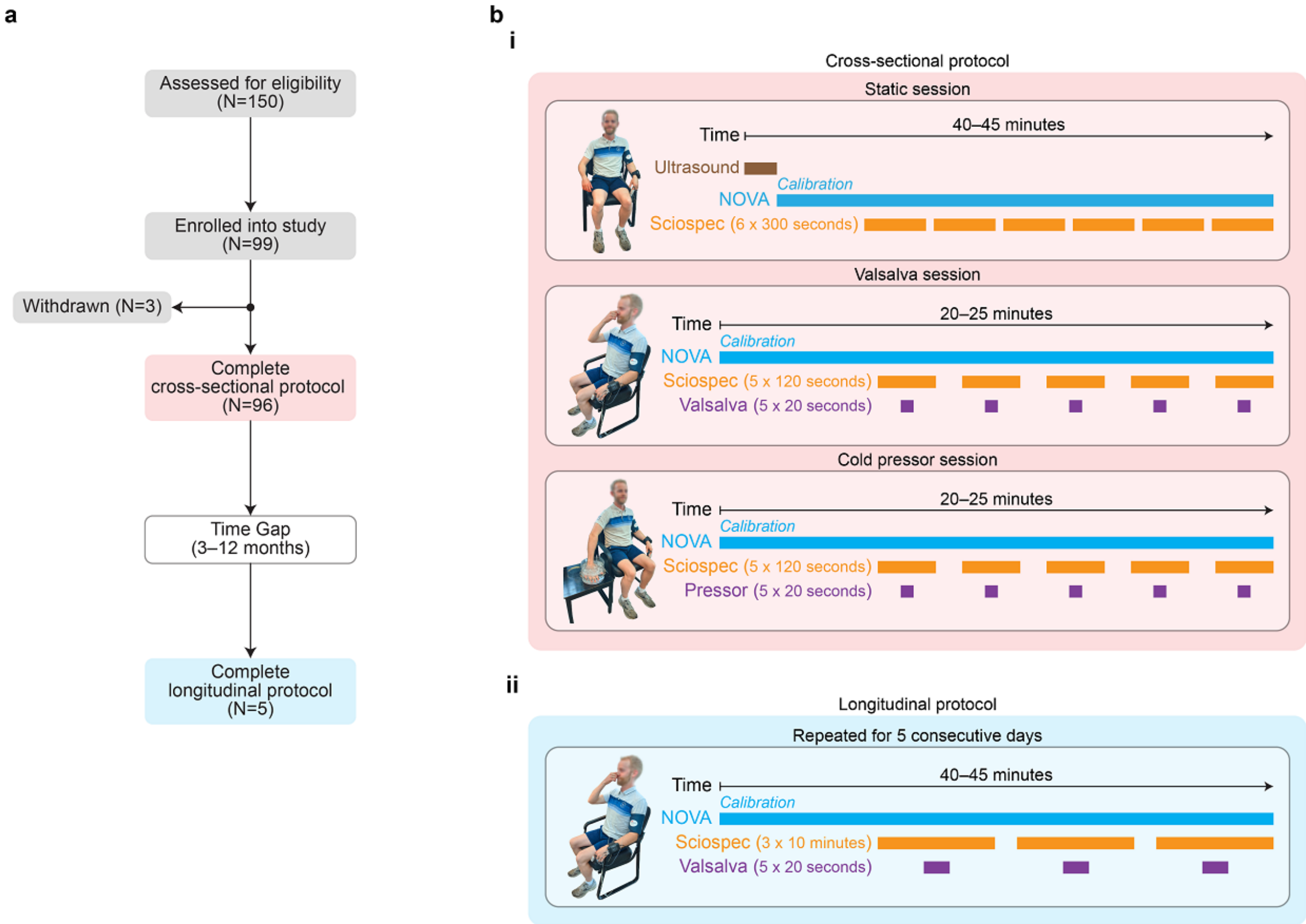

# Supplementary Fig. 25. Doppler flowmetry and B-mode imaging of the finger

Doppler flowmetry images of representative subjects from the cross-sectional cohort. The ultrasound transducer was pressed against the palmar side and on the proximal phalanx of the left index finger.

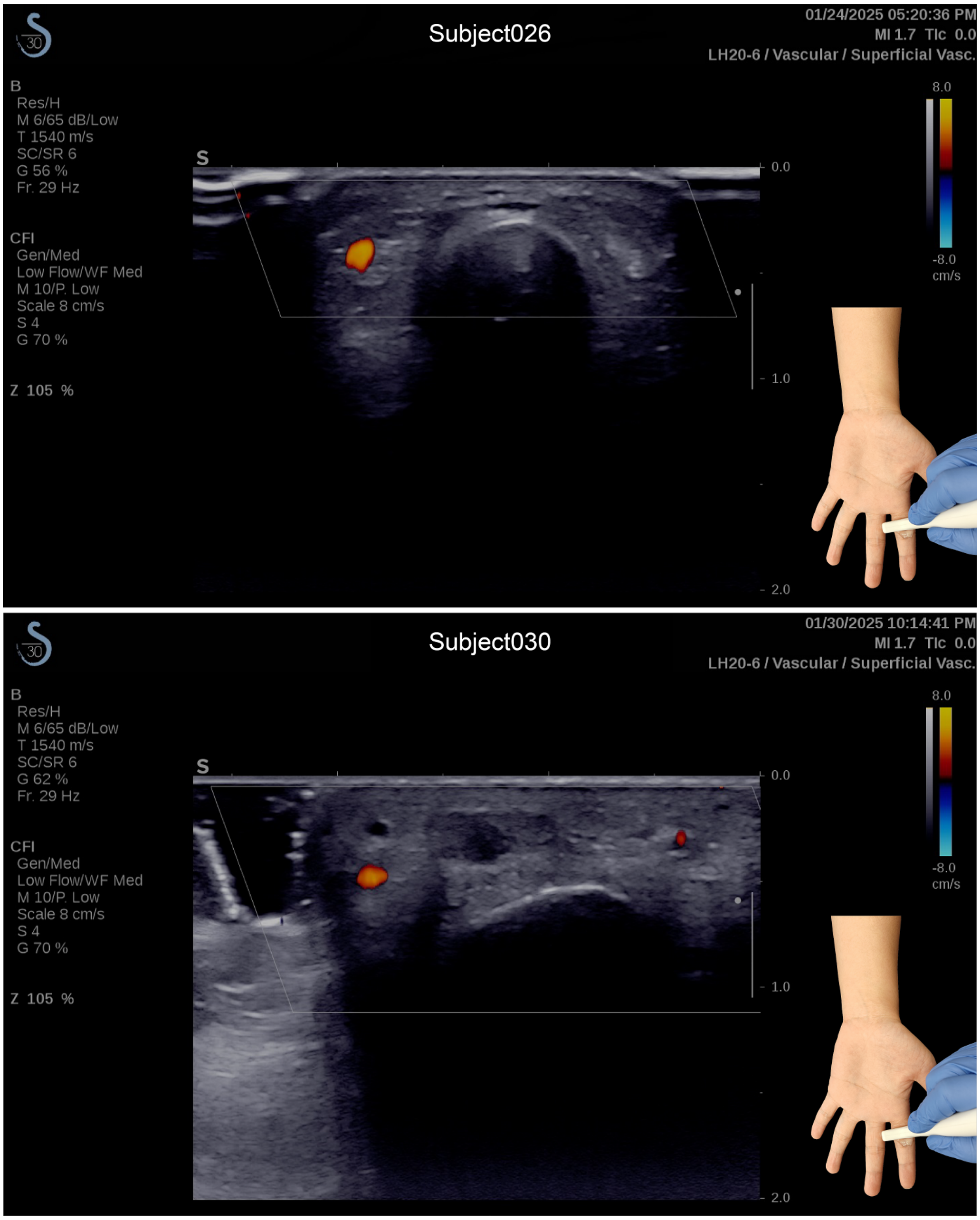

# Supplementary Fig. 26. Processing pipeline for experimental data

**a,** The first two stages include temporal filtering and alignment. Raw data was recorded with two different data acquisition systems. Brachial blood pressure (BP) and electrocardiogram (ECG) were recorded with the Nova Plus system, and multi-channel bioimpedance (BioZ) signals were recorded with the Sciospec EIT32 system connected to our ring. The data was lowpass-filtered at 5 Hz, with an additional step of detrending and channel-averaging for BioZ signals. Next, the filtered signals were passed through peak detection algorithm as preparation for alignment. The inter-beat intervals (IBI) of ECG and channel-averaged bioimpedance ($BioZ_{avg}$) signals were computed and aligned using a total least square cost. ECG and BP peak delays were used to estimate the pulse arrival time, which was then added to the IBI alignment result to yield the BP-$BioZ_{avg}$ alignment. The alignment marks were then stored in spreadsheets for later use. **b,** The later two stages in the pipeline include segmentation and signal quality assessment (SQA). BP and $BioZ_{avg}$ signals were segmented at the BP diastolic peaks and the periods where resampled to a uniform length of 50 points. In the SQA stage, $BioZ_{avg}$ periods were classified as clean based on the ensemble Mahalanobis distance, while BP periods were classified threshold applied directly on the time and signal domain. Both classification algorithms exported binary masks as output, which were then combined into a unified mask vector. Concurrently, the low-pass multi-channel BioZ signals were used for peripheral vascular impedance (PVI) image reconstruction. Both BioZ signals and PVI images were then decomposed into high-pass and low-pass components using moving average filter. The SQA mask was then applied to the BP data as well as the decomposed BioZ and PVI image data, which were then exported as tensors. $BioZ_{avg}$, detrended and channel-average bioimpedance signal; $\mathbf{P}$, blood pressure tensor; $\mathbf{Z}^{(HP)}$, impedance tensor (high-passed filtered); $\mathbf{Z}^{(LP)}$, impedance tensor (low-passed filtered); $\mathbf{S}^{(HP)}$, conductivity image tensor (high-passed filtered); $\mathbf{S}^{(LP)}$, conductivity image tensor (low-passed filtered).

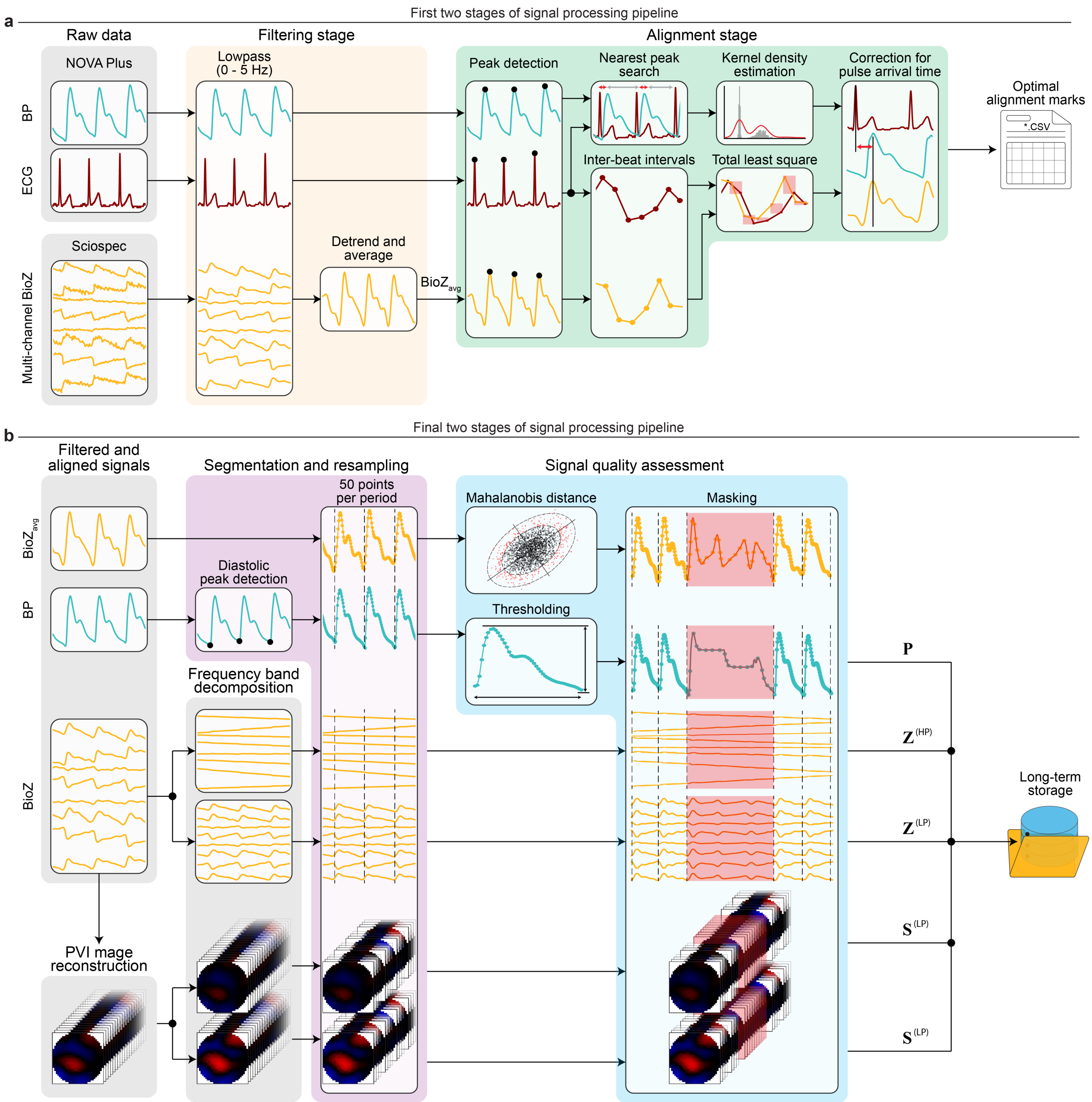
a
First two stages of signal processing pipeline
Raw data
Filtering stage
Alignment stage
NOVA Plus
BP
ECG
Sciospec
Multi-channel BioZ
Lowpass
(0 - 5 Hz)
Detrend and
average
BioZavg
Peak detection
Nearest peak
search
Kernel density
estimation
Correction for
pulse arrival time
Inter-beat intervals
Total least square
Optimal
alignment marks
*.csv
b
Final two stages of signal processing pipeline
Filtered and
aligned signals
Segmentation and resampling
Signal quality assessment
BioZavg
BP
BioZ
50 points
per period
Diastolic
peak detection
Frequency band
decomposition
PVI mage
reconstruction
Mahalanobis distance
Thresholding
Masking
P
Z (HP)
Z (LP)
S (LP)
S (LP)
Long-term
storage

# Supplementary Fig. 27. Physiological features distributions of processed datasets

Violin plots showing physiological features from all processed datasets with machine learning. BP, brachial blood pressure.

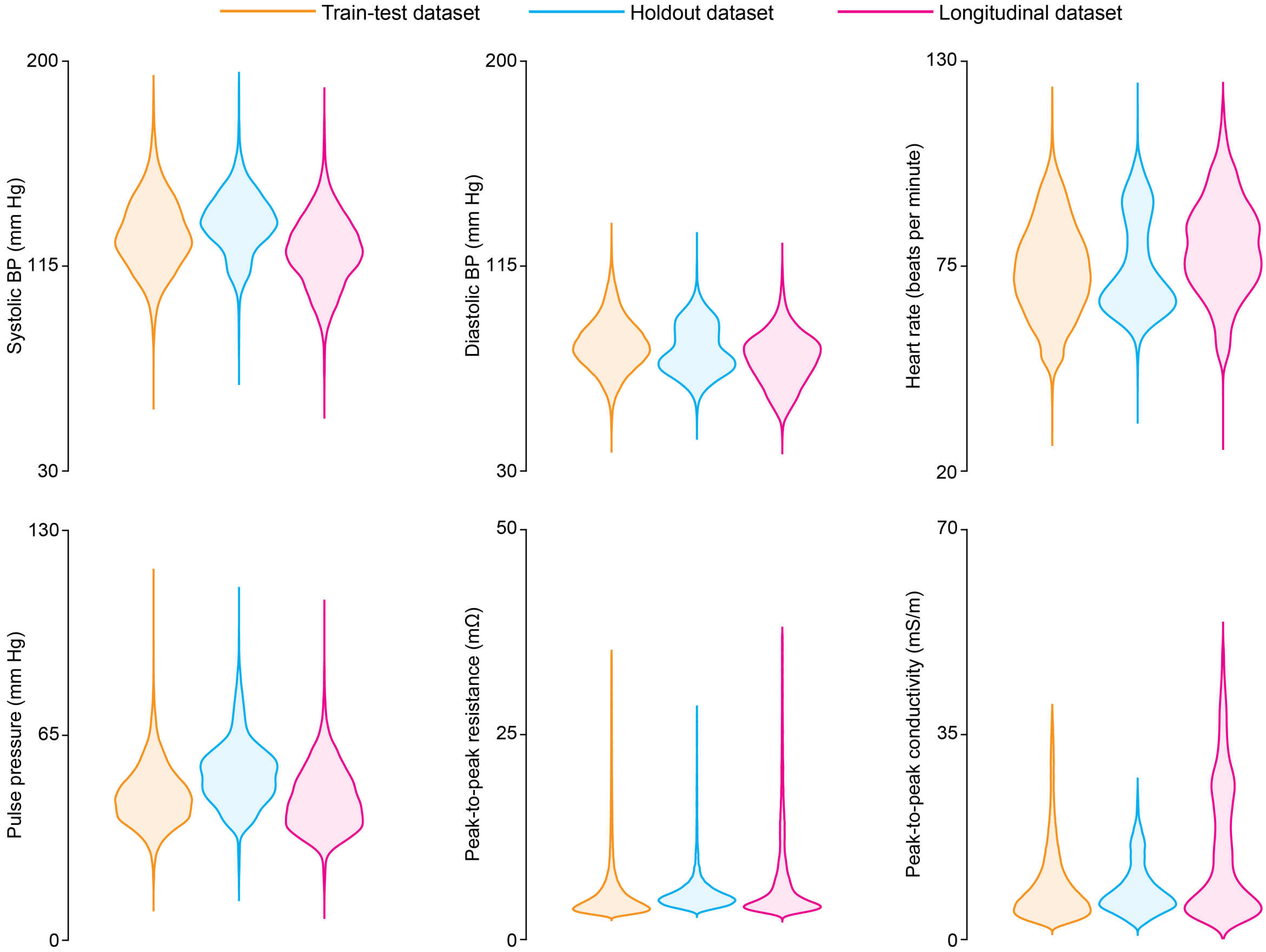

## Supplementary Fig. 28. Ensemble of reference brachial blood pressure and conductivity waveforms

Solid line, ensemble average waveform; dashed lines, standard deviation from ensemble average; scale bars, a quarter period.

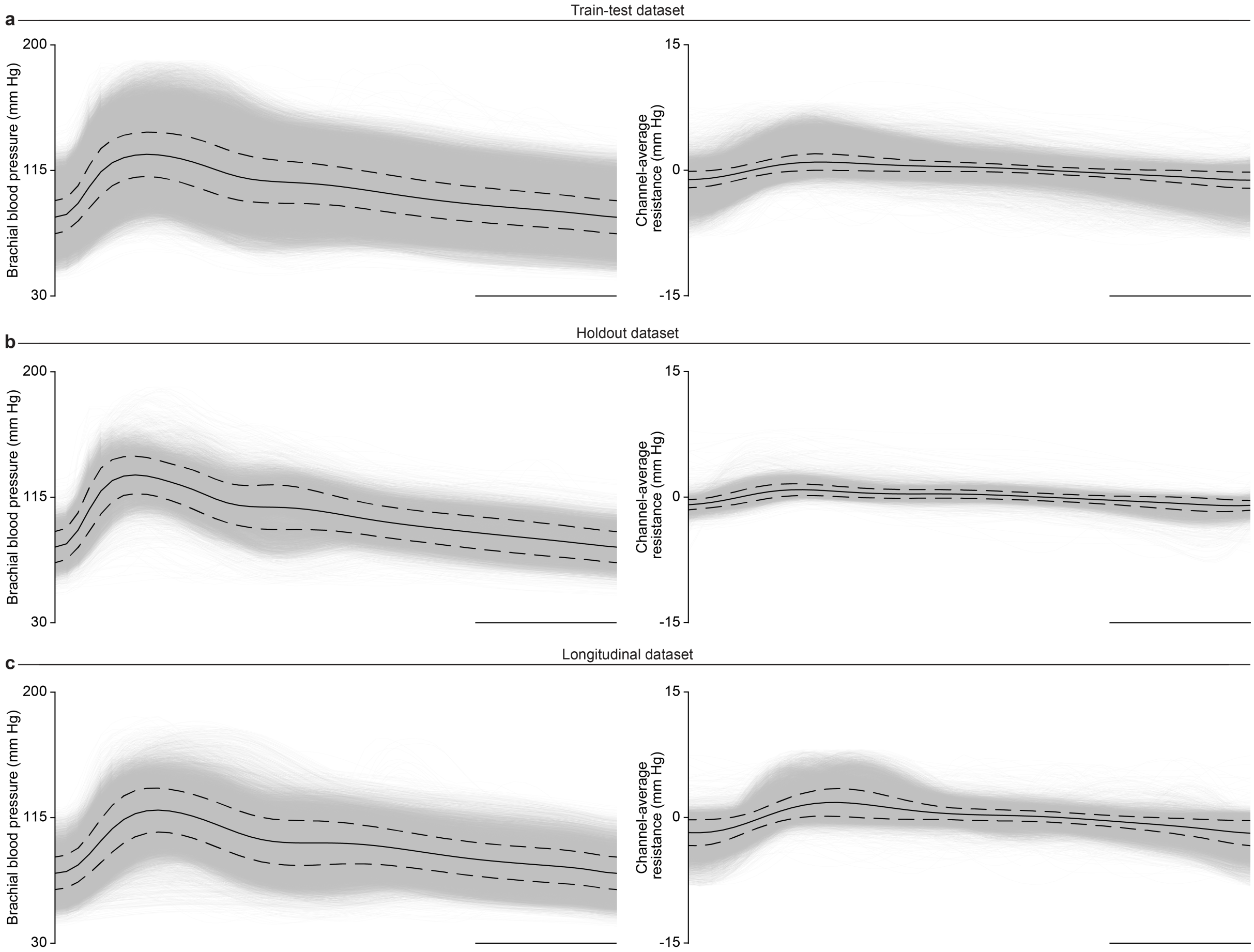

# Supplementary Fig. 29. Architecture of all machine learning model classes

**a,** Linear Regression (LR) class; **b,** Multilayer Perceptron (MLP) class; **c,** Convolutional Neural Network (CNN) class; **d,** Convolutional-Recurrent-Transformer (CRT) class; **e,** Convolutional-Recurrent-Samba (CRS) class. BiLSTM, bidirectional long short-term memory; ReLU, rectified linear unit; SiLU, sigmoid linear unit.

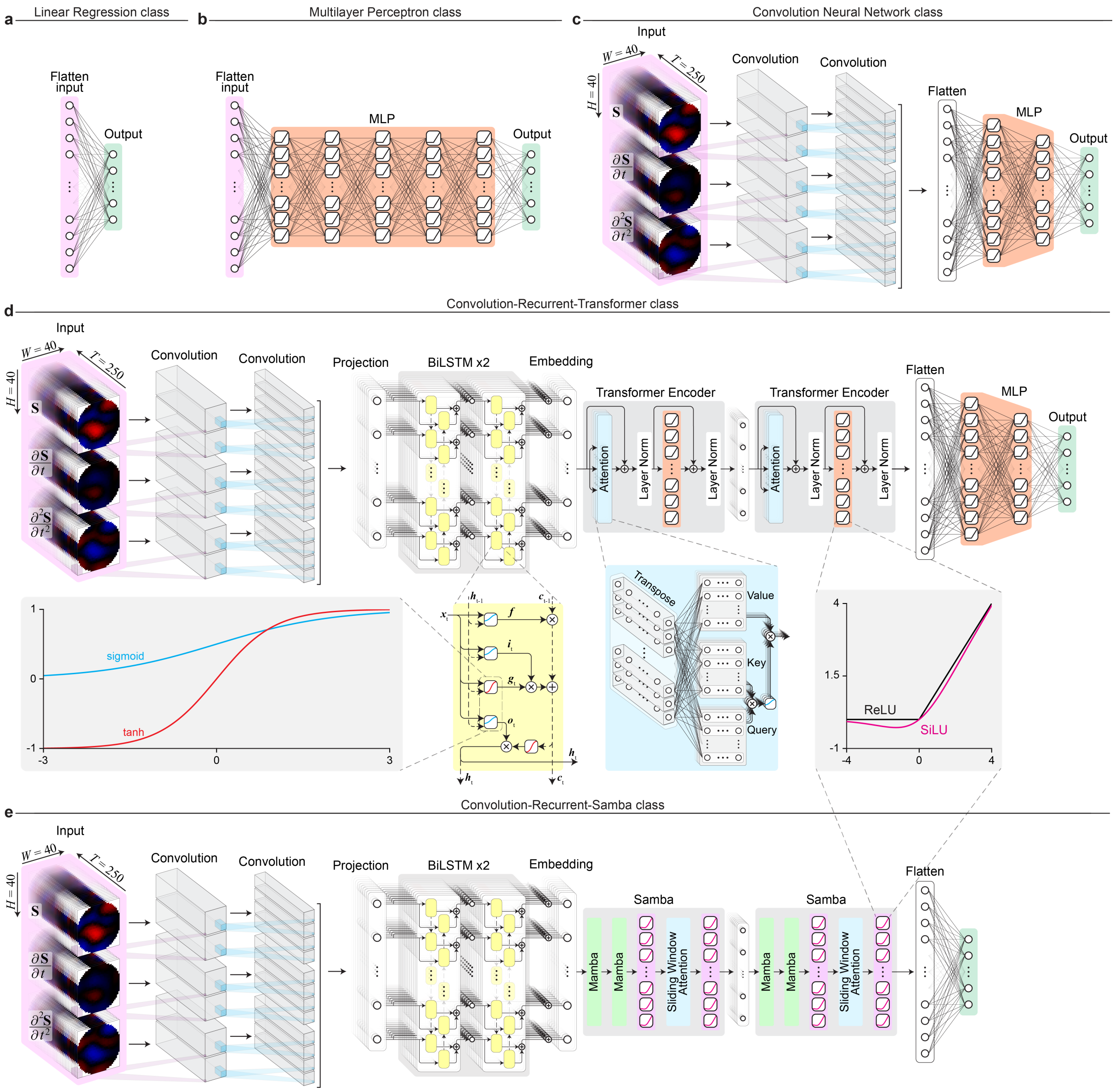

# Supplementary Fig. 30. Dataset partition and batch sampling strategy

Examples demonstrating our bipartite graph partition algorithm (**a**) and stratified batch randomization approach (**b**). In **a**, each node represents a sample built from multiple consecutive periods. Each edge indicates the overlap between the samples, with the edge colors representing the overlapping length, i.e. how many periods per overlap. At initialization, the samples were sorted based on their overlapping degree. At each iteration, a single sample is removed either from the train and test set, and the overlaps are recomputed. The process repeats until there is no overlaps in the remaining sets. In **b**, each dot represents a sample, with colors encoding the source datasets. At initialization, the samples are grouped by their source dataset. At each epoch, the source files are shuffled and grouped into clusters. The samples from each clusters are then shuffled and finally distributed into minibatches. The cluster size and minibatch size can be adjusted based on available memory. In this example, the cluster size is $3$ and minibatch size is $9$. $k$, partition iteration number; $\mathcal{U}$, train set; $\mathcal{V}$, test set; $\mathcal{S}$, aggregated (population) dataset; $\mathcal{D}$, source (subject-specific) dataset; $\mathcal{G}$, cache cluster; $\mathcal{B}$, minibatch; $\pi(\cdot)$ and $\omega(\cdot)$, permutation functions.

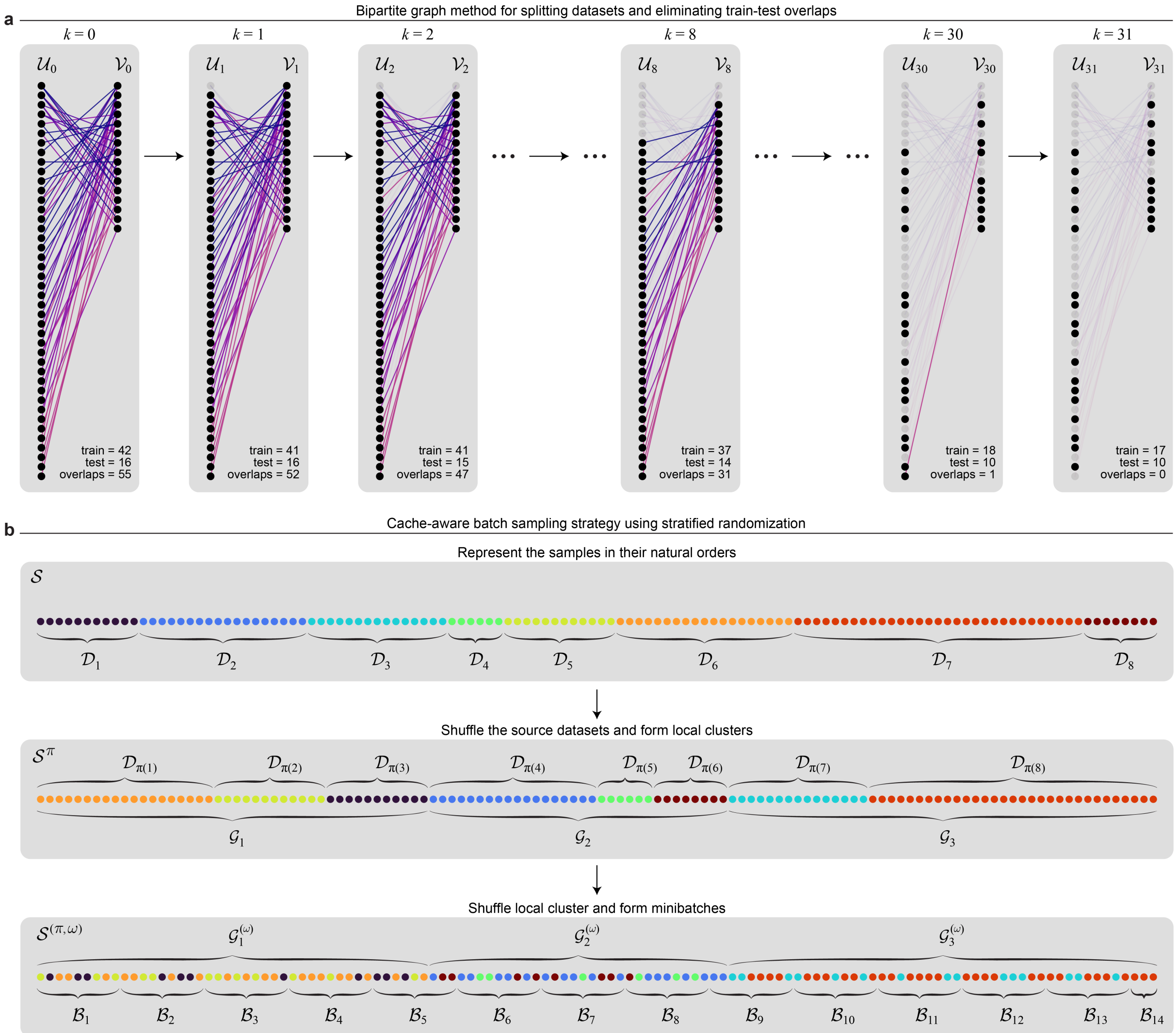

# Supplementary Fig. 31. Training curves for population-within models

Training loss and accuracy of all population-within (PW) model configurations of classes Linear Regression (**a**), Multilayer Perceptron (**b**), Convolutional Neural Network (**c**), Convolutional-Recurrent-Transformer (**d**), and Convolutional-Recurrent-Samba (**d**). The loss and accuracy are defined in Supplementary Discussion 8.7 and Supplementary Discussion 8.8, respectively. Blue solid lines, train loss and/or accuracy; orange solid lines, test loss and/or accuracy.

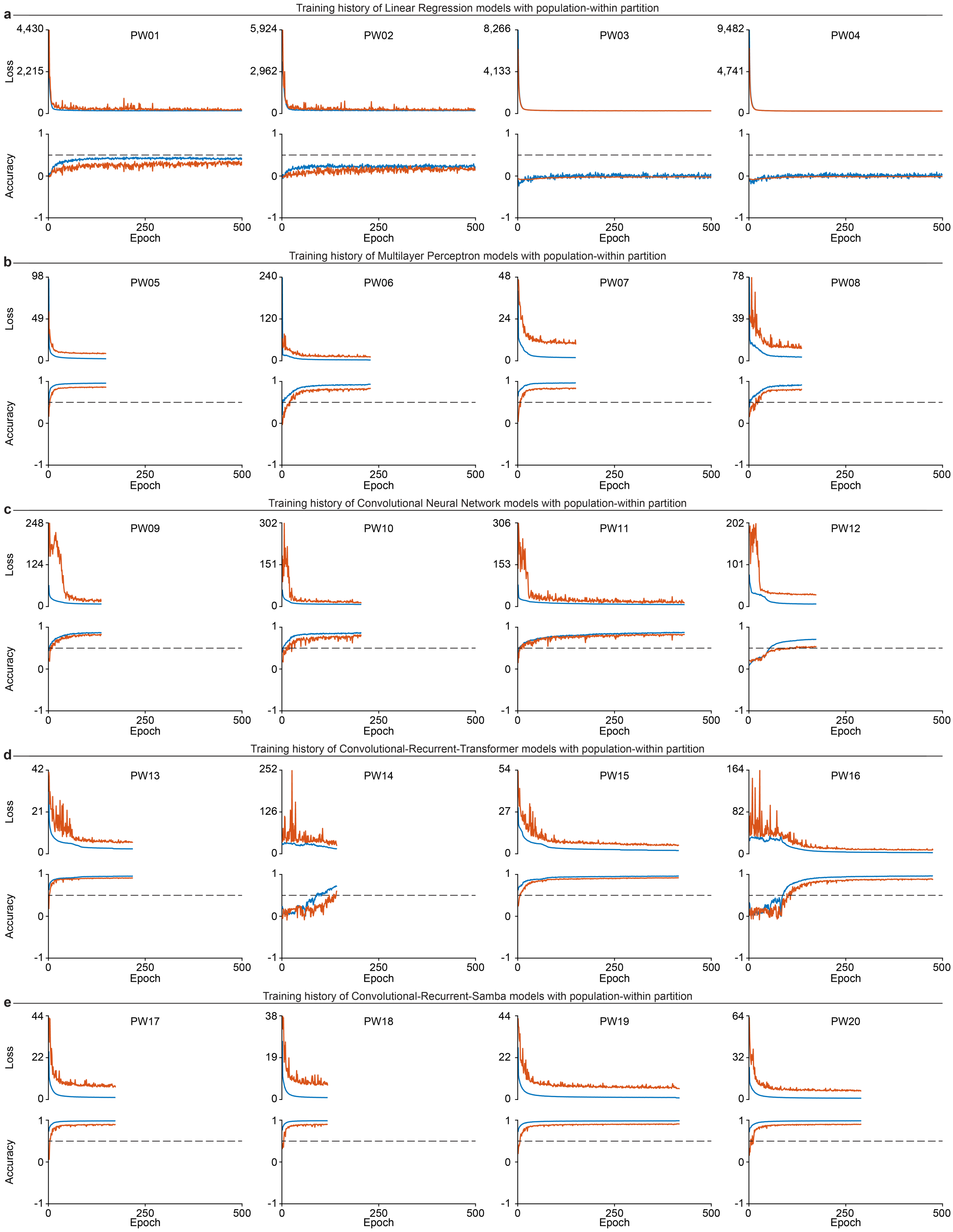
a
Training history of Linear Regression models with population-within partition
PW01
PW02
PW03
PW04
b
Training history of Multilayer Perceptron models with population-within partition
PW05
PW06
PW07
PW08
c
Training history of Convolutional Neural Network models with population-within partition
PW09
PW10
PW11
PW12
d
Training history of Convolutional-Recurrent-Transformer models with population-within partition
PW13
PW14
PW15
PW16
e
Training history of Convolutional-Recurrent-Samba models with population-within partition
PW17
PW18
PW19
PW20
Loss
Accuracy
Epoch
0
250
500

## Supplementary Fig. 32. Training curves for population-disjoint models

Training loss and accuracy of all population-disjoint (PD) model configurations of classes Linear Regression (**a**), Multilayer Perceptron (**b**), Convolutional Neural Network (**c**), Convolutional-Recurrent-Transformer (**d**), and Convolutional-Recurrent-Samba (**d**). The loss and accuracy are defined in Supplementary Discussion 8.7 and Supplementary Discussion 8.8, respectively. Blue solid lines, train loss and/or accuracy; orange solid lines, test loss and/or accuracy.

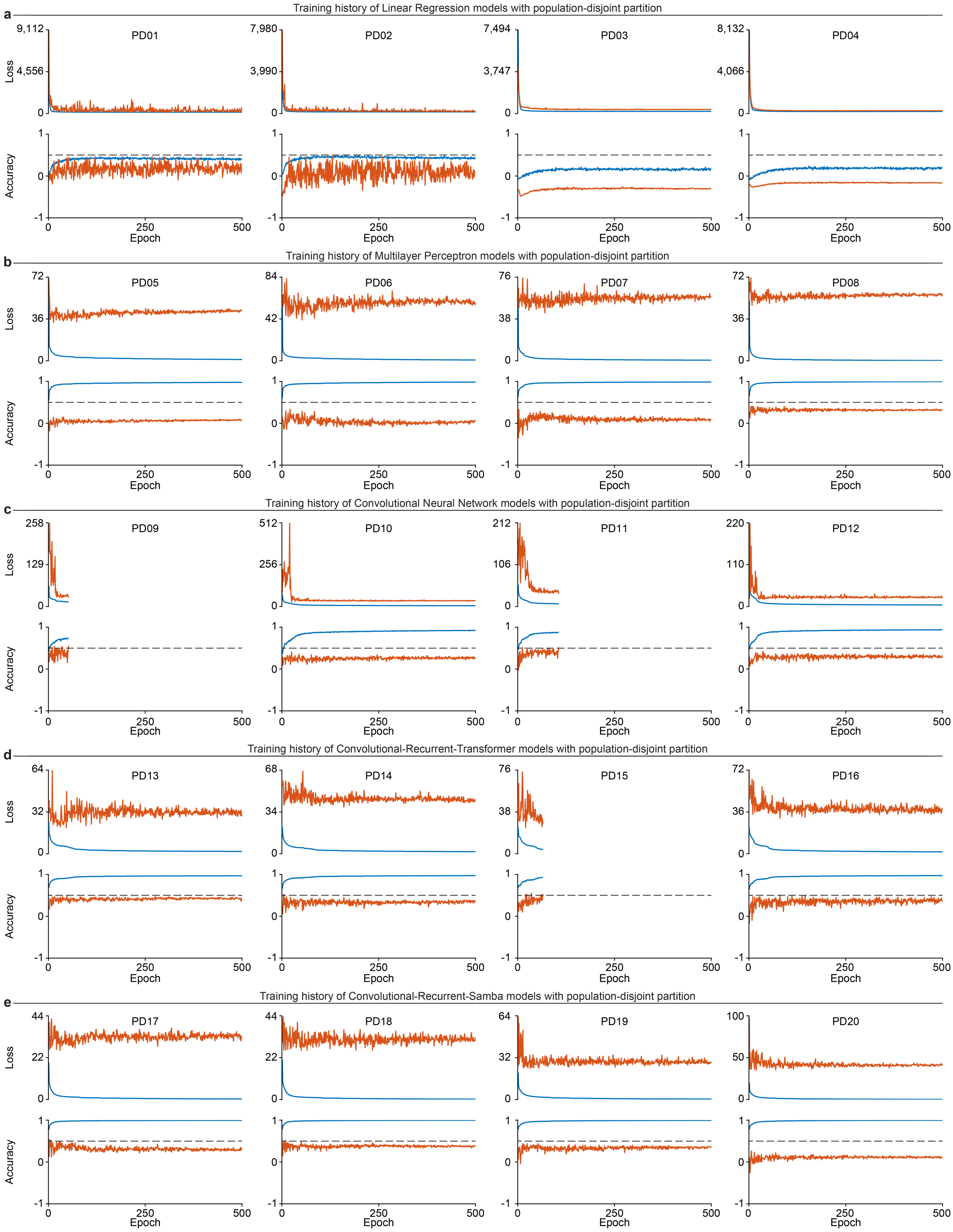
a
Training history of Linear Regression models with population-disjoint partition
PD01
PD02
PD03
PD04
b
Training history of Multilayer Perceptron models with population-disjoint partition
PD05
PD06
PD07
PD08
c
Training history of Convolutional Neural Network models with population-disjoint partition
PD09
PD10
PD11
PD12
d
Training history of Convolutional-Recurrent-Transformer models with population-disjoint partition
PD13
PD14
PD15
PD16
e
Training history of Convolutional-Recurrent-Samba models with population-disjoint partition
PD17
PD18
PD19
PD20
Loss
Accuracy
Epoch

# Supplementary Fig. 33. Estimation results of subject-specific model SS01

Aggregated results from SS01 configuration: Linear Regression class with image input and waveform output, trained with subject-specific (SS) datasets. **a,** Estimation accuracy for systolic brachial blood pressure (SBP); **b,** Estimation accuracy for diastolic brachial blood pressure (DBP); **c,** Waveform ensemble of all estimated and true brachial blood pressure (BP) periods. For **a** and **b**: **i**, correlation plots; **ii**, limits of agreement (LOA) plots; **iii**, histogram of absolute errors (AE); and **iv**, histogram of estimated and true BP distributions. For **c**: **i**, ensemble of estimated BP periods; **ii**, ensemble of true BP periods. For correlation plots: $r_{\mathrm{a}}^2$, aggregated coefficient of determination; $\hat{\rho}_{c,\mathrm{a}}$, aggregated coefficient of concordance; solid line, empirical linear regression line; dashed line, $45°$ line of perfect correlation. For LOA plots: solid line, mean of errors between estimated and true BP values; dashed lines, 2.5$^{\mathrm{th}}$ percentile (lower) and 97.5$^{\mathrm{th}}$ percentile (upper). For AE histogram plots: MAE and SDAE, mean and standard deviation of AE, respectively; $\mathcal{P}_5$, $\mathcal{P}_{10}$, and $\mathcal{P}_{15}$, cumulative percentage of estimations with AE within 5, 10, and 15 mm Hg, respectively. For fiducial histogram plots: $\mathcal{W}_{\mathrm{pred}}$, Wasserstein distance between true and estimated distribution. For ensemble plots: AMAE, average mean absolute error; ARMSE, average root mean square error; solid line, ensemble average of all periods; dashed lines, ensemble average $\pm$ standard deviation of all periods; scale bars, one-quarter of period.

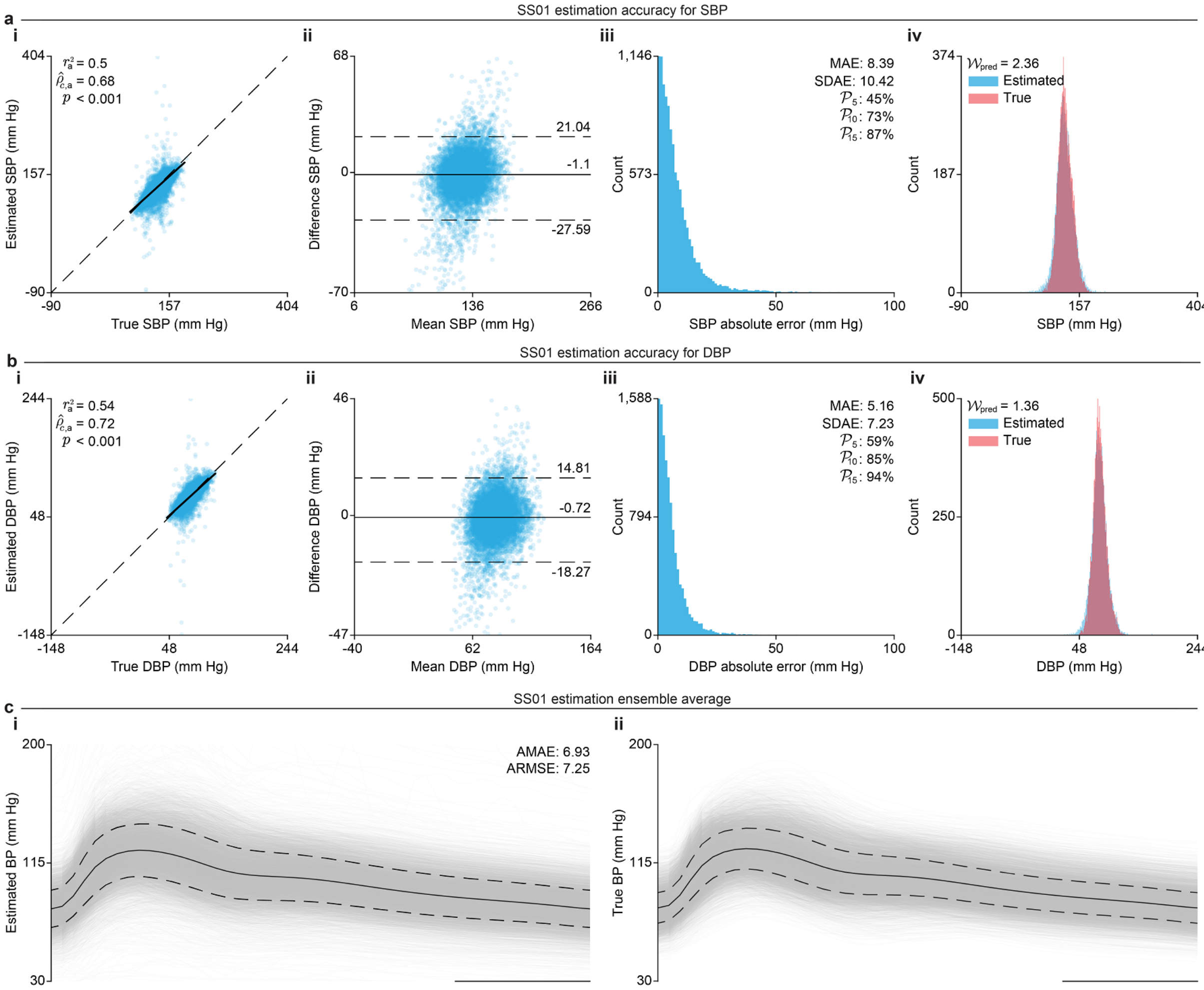

# Supplementary Fig. 34. Estimation results of subject-specific model SS02

Aggregated results from SS02 configuration: Linear Regression class with image input and fiducial output, trained with subject-specific (SS) datasets. **a,** Estimation accuracy for systolic brachial blood pressure (SBP); **b,** Estimation accuracy for diastolic brachial blood pressure (DBP); **i**, correlation plots; **ii**, limits of agreement (LOA) plots; **iii**, histogram of absolute errors (AE); and **iv**, histogram of estimated and true BP distributions. BP, blood pressure; DBP, diastolic blood pressure; SBP, systolic blood pressure. For correlation plots: $r^2_{\text{a}}$, aggregated coefficient of determination; $\hat{\rho}_{c,\text{a}}$, aggregated coefficient of concordance; solid line, empirical linear regression line; dashed line, $45°$ line of perfect correlation. For LOA plots: solid line, mean of errors between estimated and true BP values; dashed lines, 2.5$^{\text{th}}$ percentile (lower) and 97.5$^{\text{th}}$ percentile (upper). For AE histogram plots: MAE and SDAE, mean and standard deviation of AE, respectively; $\mathcal{P}_5$, $\mathcal{P}_{10}$, and $\mathcal{P}_{15}$, cumulative percentage of estimations with AE within 5, 10, and 15 mm Hg, respectively. For fiducial histogram plots: $\mathcal{W}_{\text{pred}}$, Wasserstein distance between true and estimated distribution.

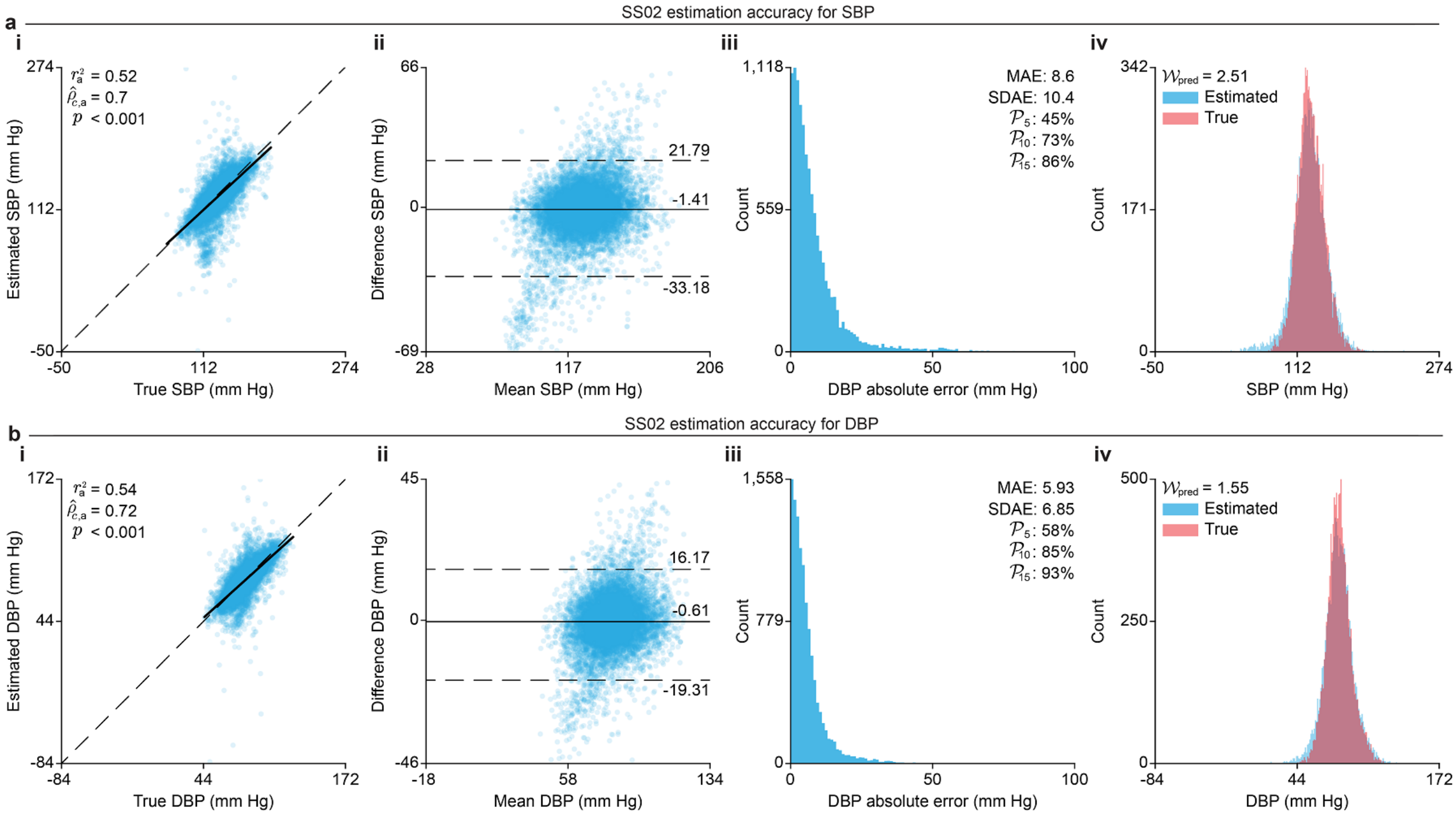

# Supplementary Fig. 35. Estimation results of subject-specific model SS03

Aggregated results from SS03 configuration: Linear Regression class with impedance input and waveform output, trained with subject-specific (SS) datasets. **a,** Estimation accuracy for systolic brachial blood pressure (SBP); **b,** Estimation accuracy for diastolic brachial blood pressure (DBP); **c,** Waveform ensemble of all estimated and true brachial blood pressure (BP) periods. For **a** and **b**: **i**, correlation plots; **ii**, limits of agreement (LOA) plots; **iii**, histogram of absolute errors (AE); and **iv**, histogram of estimated and true BP distributions. For **c**: **i**, ensemble of estimated BP periods; **ii**, ensemble of true BP periods. For correlation plots: $r_{\mathrm{a}}^2$, aggregated coefficient of determination; $\hat{\rho}_{c,\mathrm{a}}$, aggregated coefficient of concordance; solid line, empirical linear regression line; dashed line, $45°$ line of perfect correlation. For LOA plots: solid line, mean of errors between estimated and true BP values; dashed lines, 2.5$^{\mathrm{th}}$ percentile (lower) and 97.5$^{\mathrm{th}}$ percentile (upper). For AE histogram plots: MAE and SDAE, mean and standard deviation of AE, respectively; $\mathcal{P}_5$, $\mathcal{P}_{10}$, and $\mathcal{P}_{15}$, cumulative percentage of estimations with AE within 5, 10, and 15 mm Hg, respectively. For fiducial histogram plots: $\mathcal{W}_{\mathrm{pred}}$, Wasserstein distance between true and estimated distribution. For ensemble plots: AMAE, average mean absolute error; ARMSE, average root mean square error; solid line, ensemble average of all periods; dashed lines, ensemble average $\pm$ standard deviation of all periods; scale bars, one-quarter of period.

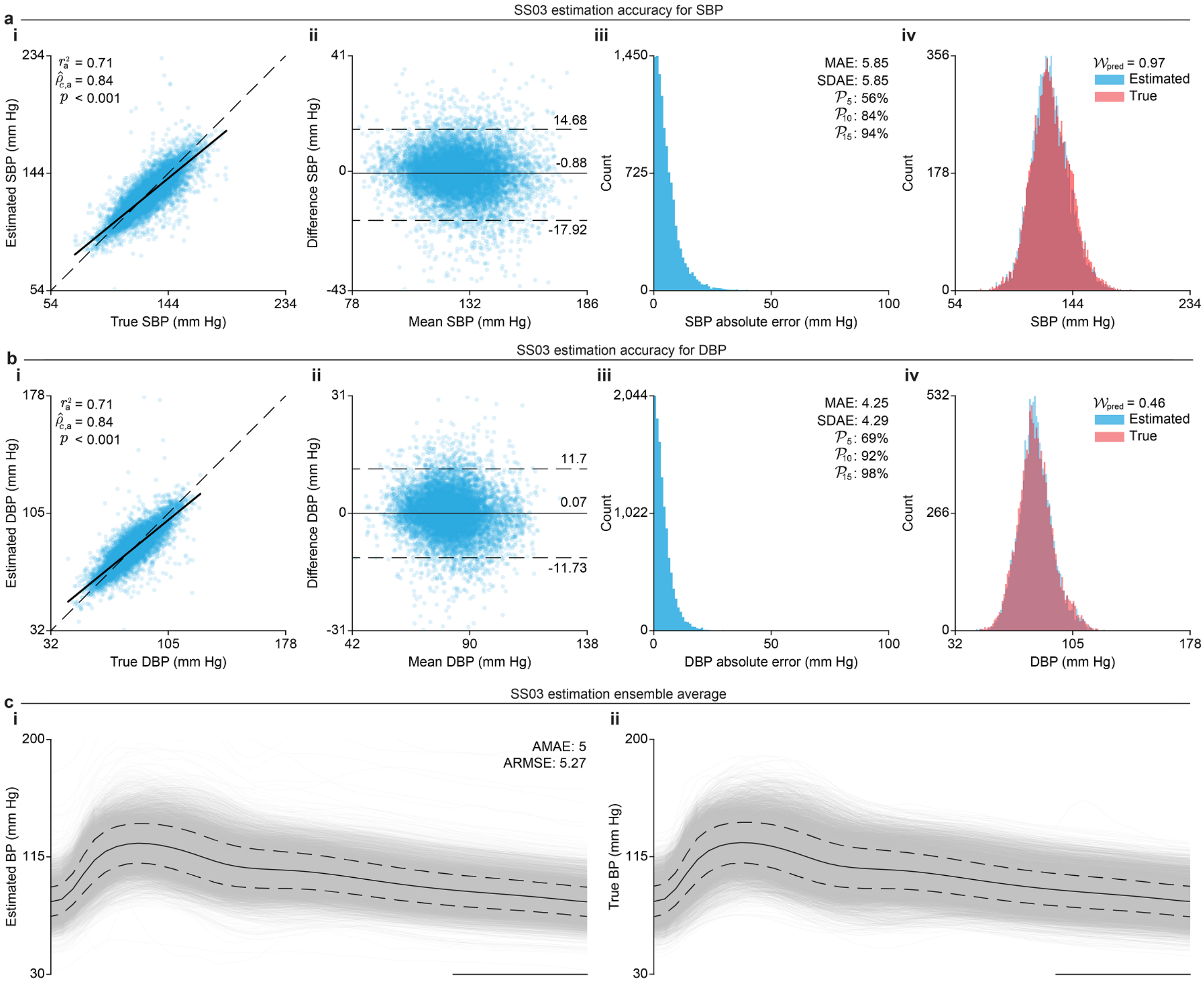

# Supplementary Fig. 36. Estimation results of subject-specific model SS04

Aggregated results from SS04 configuration: Linear Regression class with impedance input and fiducial output, trained with subject-specific (SS) datasets. **a,** Estimation accuracy for systolic brachial blood pressure (SBP); **b,** Estimation accuracy for diastolic brachial blood pressure (DBP); **i**, correlation plots; **ii**, limits of agreement (LOA) plots; **iii**, histogram of absolute errors (AE); and **iv**, histogram of estimated and true BP distributions. BP, blood pressure; DBP, diastolic blood pressure; SBP, systolic blood pressure. For correlation plots: $r_{\text{a}}^2$, aggregated coefficient of determination; $\hat{\rho}_{c,\text{a}}$, aggregated coefficient of concordance; solid line, empirical linear regression line; dashed line, $45°$ line of perfect correlation. For LOA plots: solid line, mean of errors between estimated and true BP values; dashed lines, 2.5$^{\text{th}}$ percentile (lower) and 97.5$^{\text{th}}$ percentile (upper). For AE histogram plots: MAE and SDAE, mean and standard deviation of AE, respectively; $\mathcal{P}_5$, $\mathcal{P}_{10}$, and $\mathcal{P}_{15}$, cumulative percentage of estimations with AE within 5, 10, and 15 mm Hg, respectively. For fiducial histogram plots: $\mathcal{W}_{\text{pred}}$, Wasserstein distance between true and estimated distribution.

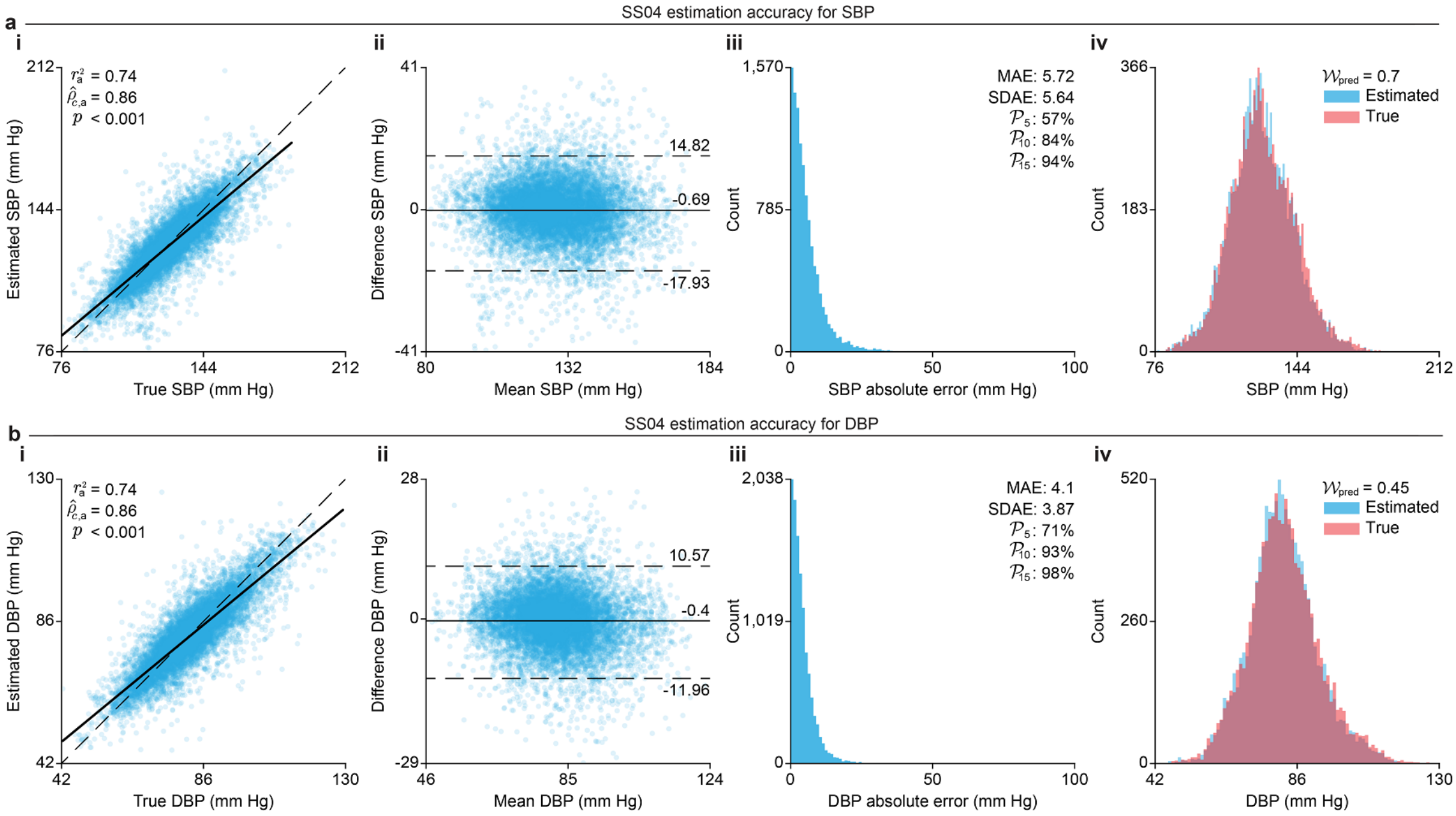

# Supplementary Fig. 37. Estimation results of subject-specific model SS05

Aggregated results from SS05 configuration: Multilayer Perceptron class with image input and waveform output, trained with subject-specific (SS) datasets. **a,** Estimation accuracy for systolic brachial blood pressure (SBP); **b,** Estimation accuracy for diastolic brachial blood pressure (DBP); **c,** Waveform ensemble of all estimated and true brachial blood pressure (BP) periods. For **a** and **b**: **i**, correlation plots; **ii**, limits of agreement (LOA) plots; **iii**, histogram of absolute errors (AE); and **iv**, histogram of estimated and true BP distributions. For **c**: **i**, ensemble of estimated BP periods; **ii**, ensemble of true BP periods. For correlation plots: $r_\text{a}^2$, aggregated coefficient of determination; $\hat{\rho}_{c,\text{a}}$, aggregated coefficient of concordance; solid line, empirical linear regression line; dashed line, $45°$ line of perfect correlation. For LOA plots: solid line, mean of errors between estimated and true BP values; dashed lines, 2.5$^{\text{th}}$ percentile (lower) and 97.5$^{\text{th}}$ percentile (upper). For AE histogram plots: MAE and SDAE, mean and standard deviation of AE, respectively; $\mathcal{P}_5$, $\mathcal{P}_{10}$, and $\mathcal{P}_{15}$, cumulative percentage of estimations with AE within 5, 10, and 15 mm Hg, respectively. For fiducial histogram plots: $\mathcal{W}_\text{pred}$, Wasserstein distance between true and estimated distribution. For ensemble plots: AMAE, average mean absolute error; ARMSE, average root mean square error; solid line, ensemble average of all periods; dashed lines, ensemble average $\pm$ standard deviation of all periods; scale bars, one-quarter of period.

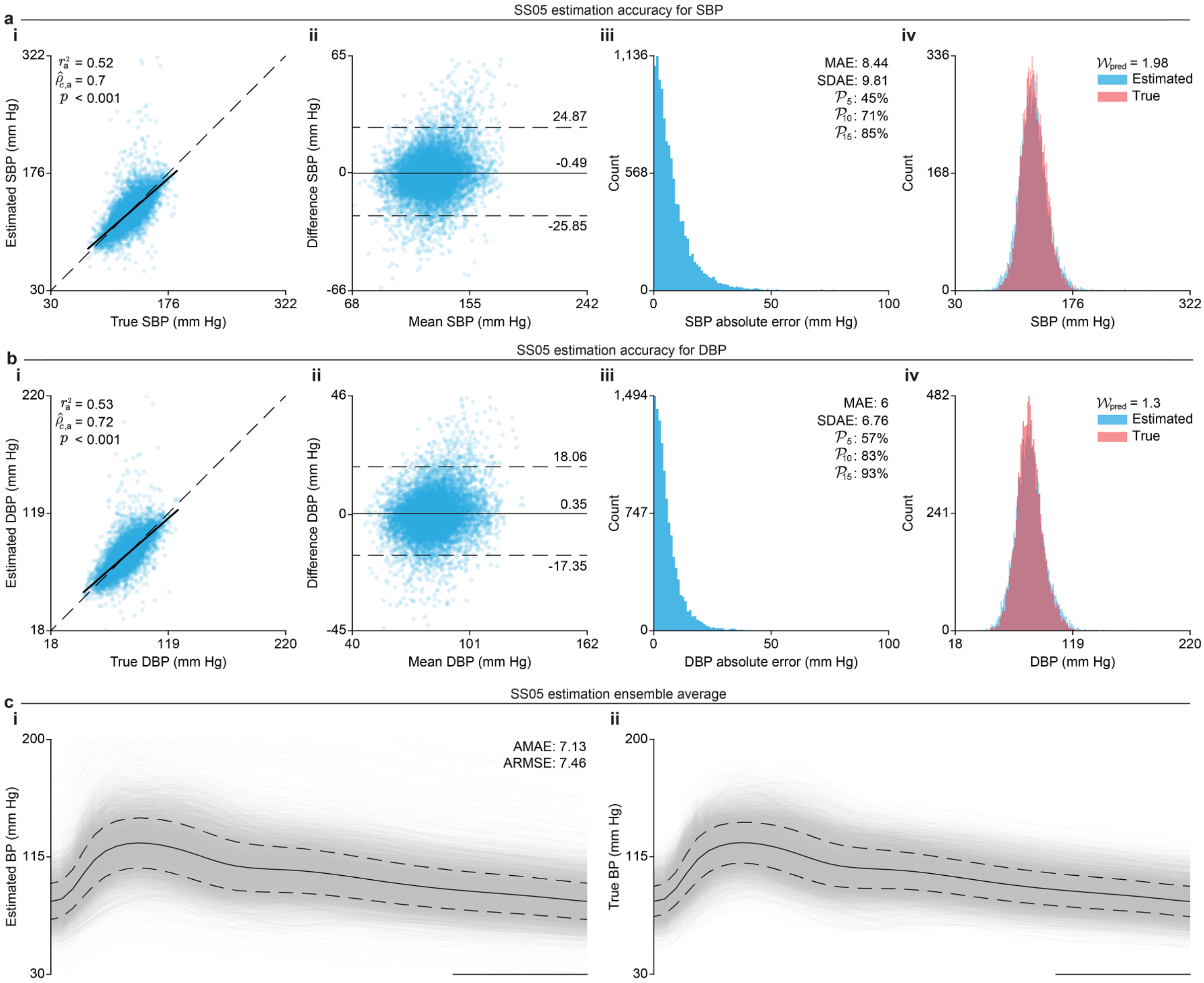

# Supplementary Fig. 38. Estimation results of subject-specific model SS06

Aggregated results from SS06 configuration: Multilayer Perceptron class with image input and fiducial output, trained with subject-specific (SS) datasets. **a,** Estimation accuracy for systolic brachial blood pressure (SBP); **b,** Estimation accuracy for diastolic brachial blood pressure (DBP); **i**, correlation plots; **ii**, limits of agreement (LOA) plots; **iii**, histogram of absolute errors (AE); and **iv**, histogram of estimated and true BP distributions. BP, blood pressure; DBP, diastolic blood pressure; SBP, systolic blood pressure. For correlation plots: $r^2_{\text{a}}$, aggregated coefficient of determination; $\hat{\rho}_{c,\text{a}}$, aggregated coefficient of concordance; solid line, empirical linear regression line; dashed line, $45°$ line of perfect correlation. For LOA plots: solid line, mean of errors between estimated and true BP values; dashed lines, 2.5$^{\text{th}}$ percentile (lower) and 97.5$^{\text{th}}$ percentile (upper). For AE histogram plots: MAE and SDAE, mean and standard deviation of AE, respectively; $\mathcal{P}_5$, $\mathcal{P}_{10}$, and $\mathcal{P}_{15}$, cumulative percentage of estimations with AE within 5, 10, and 15 mm Hg, respectively. For fiducial histogram plots: $\mathcal{W}_{\text{pred}}$, Wasserstein distance between true and estimated distribution.

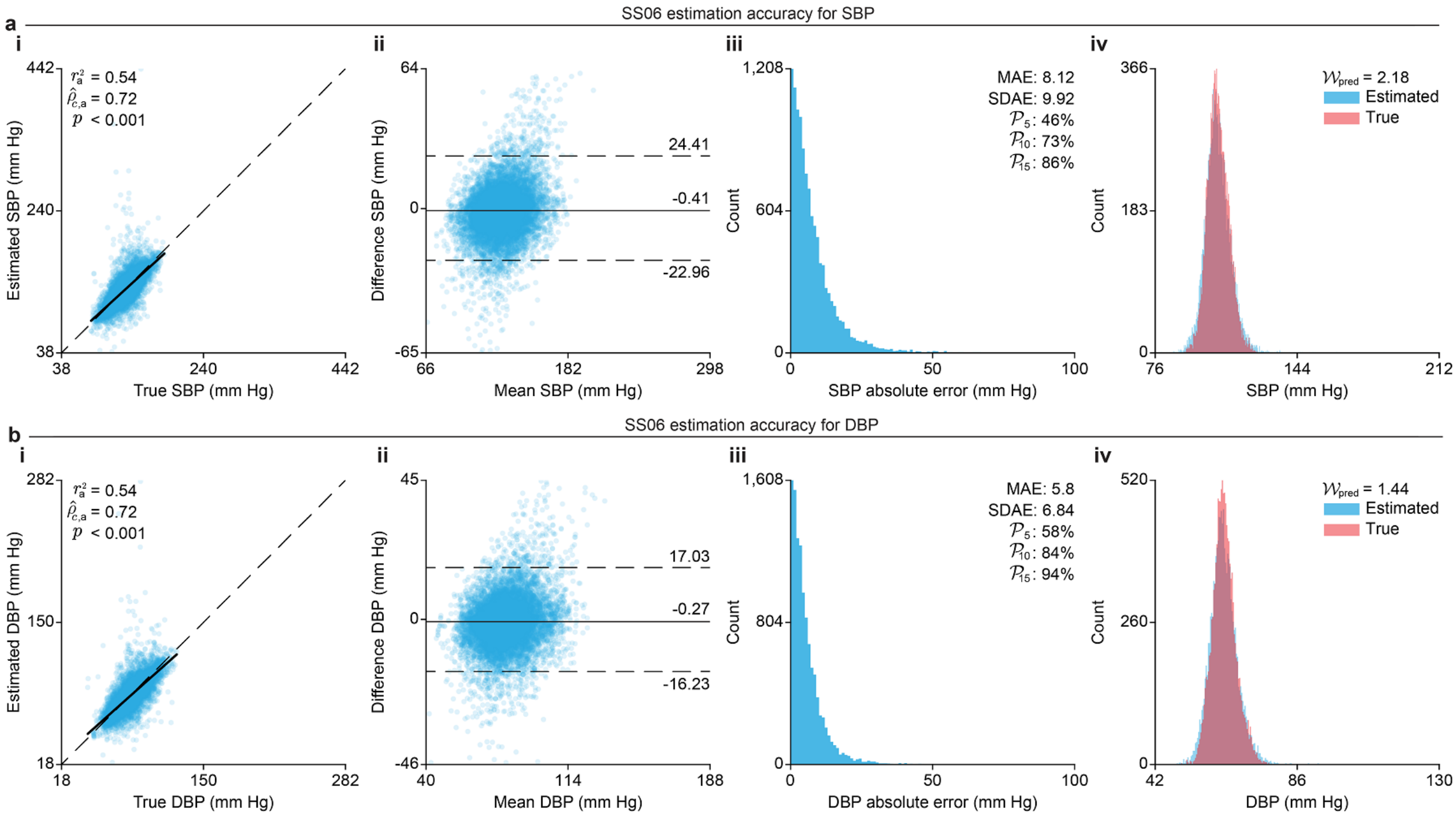

# Supplementary Fig. 39. Estimation results of subject-specific model SS07

Aggregated results from SS07 configuration: Multilayer Perceptron class with impedance input and waveform output, trained with subject-specific (SS) datasets. **a,** Estimation accuracy for systolic brachial blood pressure (SBP); **b,** Estimation accuracy for diastolic brachial blood pressure (DBP); **c,** Waveform ensemble of all estimated and true brachial blood pressure (BP) periods. For **a** and **b**: **i**, correlation plots; **ii**, limits of agreement (LOA) plots; **iii**, histogram of absolute errors (AE); and **iv**, histogram of estimated and true BP distributions. For **c**: **i**, ensemble of estimated BP periods; **ii**, ensemble of true BP periods. For correlation plots: $r_{\mathrm{a}}^2$, aggregated coefficient of determination; $\hat{\rho}_{c,\mathrm{a}}$, aggregated coefficient of concordance; solid line, empirical linear regression line; dashed line, $45°$ line of perfect correlation. For LOA plots: solid line, mean of errors between estimated and true BP values; dashed lines, 2.5$^{\mathrm{th}}$ percentile (lower) and 97.5$^{\mathrm{th}}$ percentile (upper). For AE histogram plots: MAE and SDAE, mean and standard deviation of AE, respectively; $\mathcal{P}_5$, $\mathcal{P}_{10}$, and $\mathcal{P}_{15}$, cumulative percentage of estimations with AE within 5, 10, and 15 mm Hg, respectively. For fiducial histogram plots: $\mathcal{W}_{\mathrm{pred}}$, Wasserstein distance between true and estimated distribution. For ensemble plots: AMAE, average mean absolute error; ARMSE, average root mean square error; solid line, ensemble average of all periods; dashed lines, ensemble average $\pm$ standard deviation of all periods; scale bars, one-quarter of period.

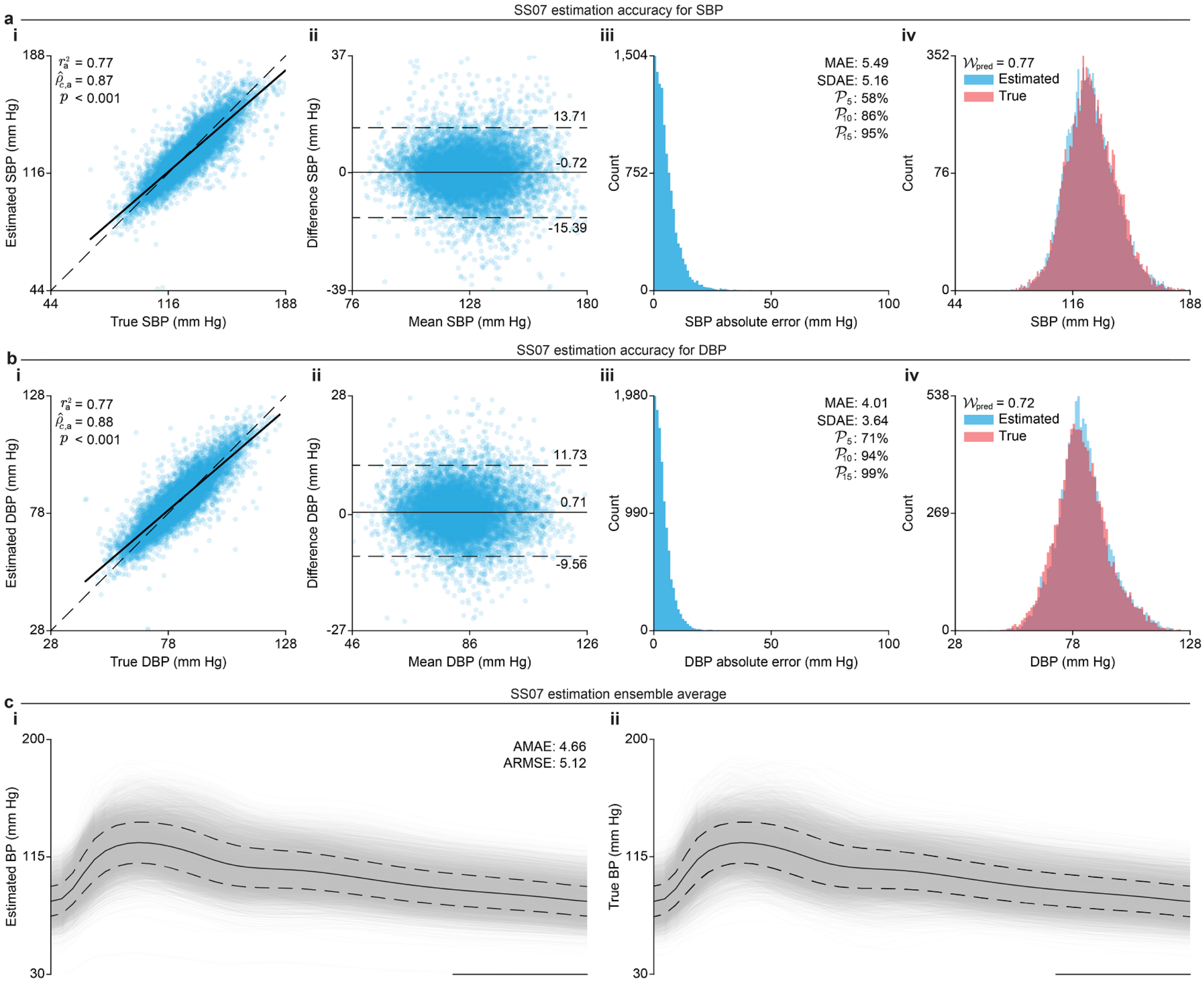

# Supplementary Fig. 40. Estimation results of subject-specific model SS08

Aggregated results from SS08 configuration: Multilayer Perceptron class with impedance input and fiducial output, trained with subject-specific (SS) datasets. **a,** Estimation accuracy for systolic brachial blood pressure (SBP); **b,** Estimation accuracy for diastolic brachial blood pressure (DBP); **i**, correlation plots; **ii**, limits of agreement (LOA) plots; **iii**, histogram of absolute errors (AE); and **iv**, histogram of estimated and true BP distributions. BP, blood pressure; DBP, diastolic blood pressure; SBP, systolic blood pressure. For correlation plots: $r_{\text{a}}^2$, aggregated coefficient of determination; $\hat{\rho}_{c,\text{a}}$, aggregated coefficient of concordance; solid line, empirical linear regression line; dashed line, $45°$ line of perfect correlation. For LOA plots: solid line, mean of errors between estimated and true BP values; dashed lines, 2.5th percentile (lower) and 97.5th percentile (upper). For AE histogram plots: MAE and SDAE, mean and standard deviation of AE, respectively; $\mathcal{P}_5$, $\mathcal{P}_{10}$, and $\mathcal{P}_{15}$, cumulative percentage of estimations with AE within 5, 10, and 15 mm Hg, respectively. For fiducial histogram plots: $\mathcal{W}_{\text{pred}}$, Wasserstein distance between true and estimated distribution.

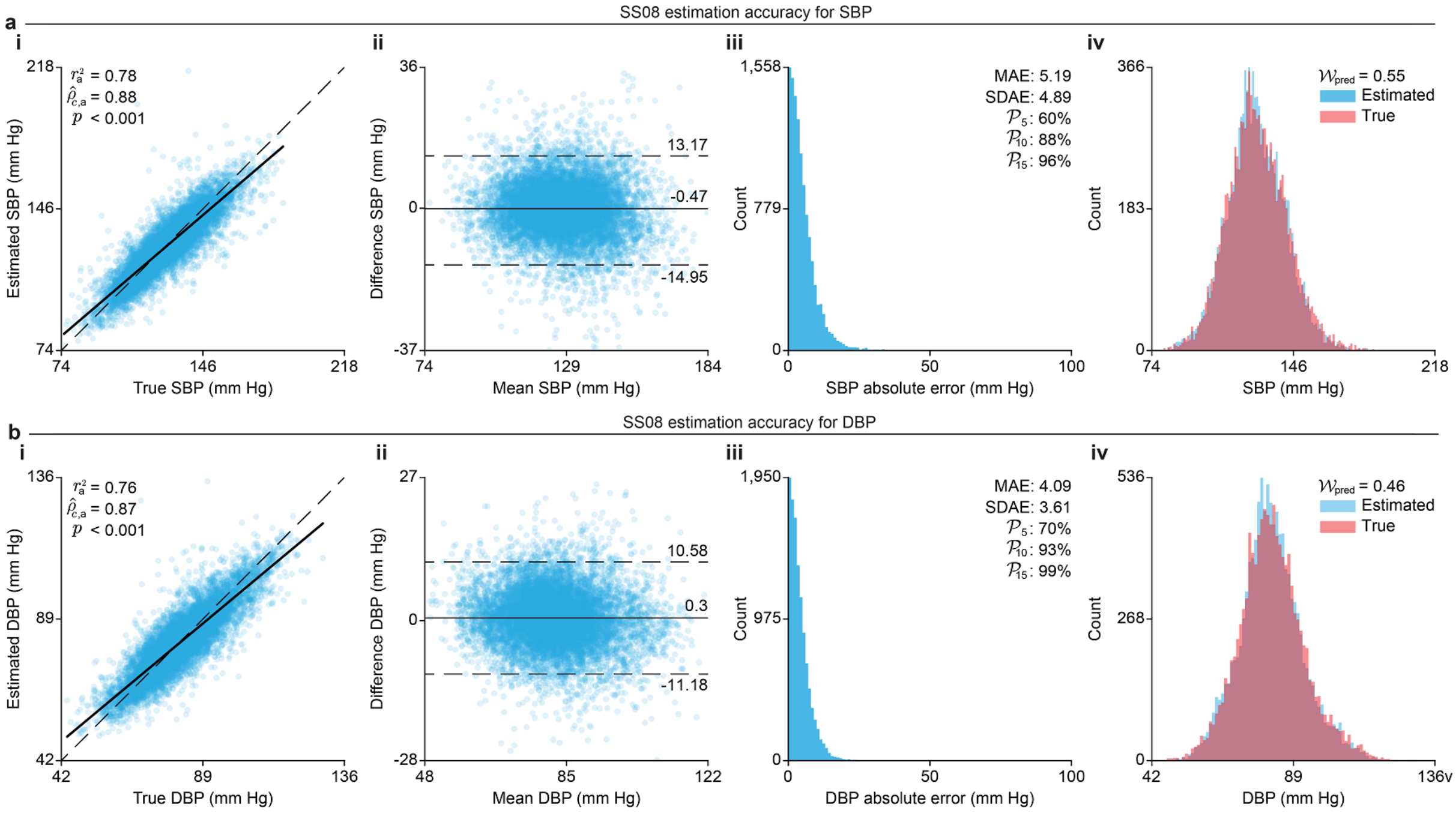

# Supplementary Fig. 41. Estimation results of subject-specific model SS09

Aggregated results from SS09 configuration: Convolutional Neural Network class with image input and waveform output, trained with subject-specific (SS) datasets. **a,** Estimation accuracy for systolic brachial blood pressure (SBP); **b,** Estimation accuracy for diastolic brachial blood pressure (DBP); **c,** Waveform ensemble of all estimated and true brachial blood pressure (BP) periods. For **a** and **b**: **i**, correlation plots; **ii**, limits of agreement (LOA) plots; **iii**, histogram of absolute errors (AE); and **iv**, histogram of estimated and true BP distributions. For **c**: **i**, ensemble of estimated BP periods; **ii**, ensemble of true BP periods. For correlation plots: $r^2_{\text{a}}$, aggregated coefficient of determination; $\hat{\rho}_{c,\text{a}}$, aggregated coefficient of concordance; solid line, empirical linear regression line; dashed line, $45°$ line of perfect correlation. For LOA plots: solid line, mean of errors between estimated and true BP values; dashed lines, 2.5$^{\text{th}}$ percentile (lower) and 97.5$^{\text{th}}$ percentile (upper). For AE histogram plots: MAE and SDAE, mean and standard deviation of AE, respectively; $\mathcal{P}_5$, $\mathcal{P}_{10}$, and $\mathcal{P}_{15}$, cumulative percentage of estimations with AE within 5, 10, and 15 mm Hg, respectively. For fiducial histogram plots: $\mathcal{W}_{\text{pred}}$, Wasserstein distance between true and estimated distribution. For ensemble plots: AMAE, average mean absolute error; ARMSE, average root mean square error; solid line, ensemble average of all periods; dashed lines, ensemble average $\pm$ standard deviation of all periods; scale bars, one-quarter of period.

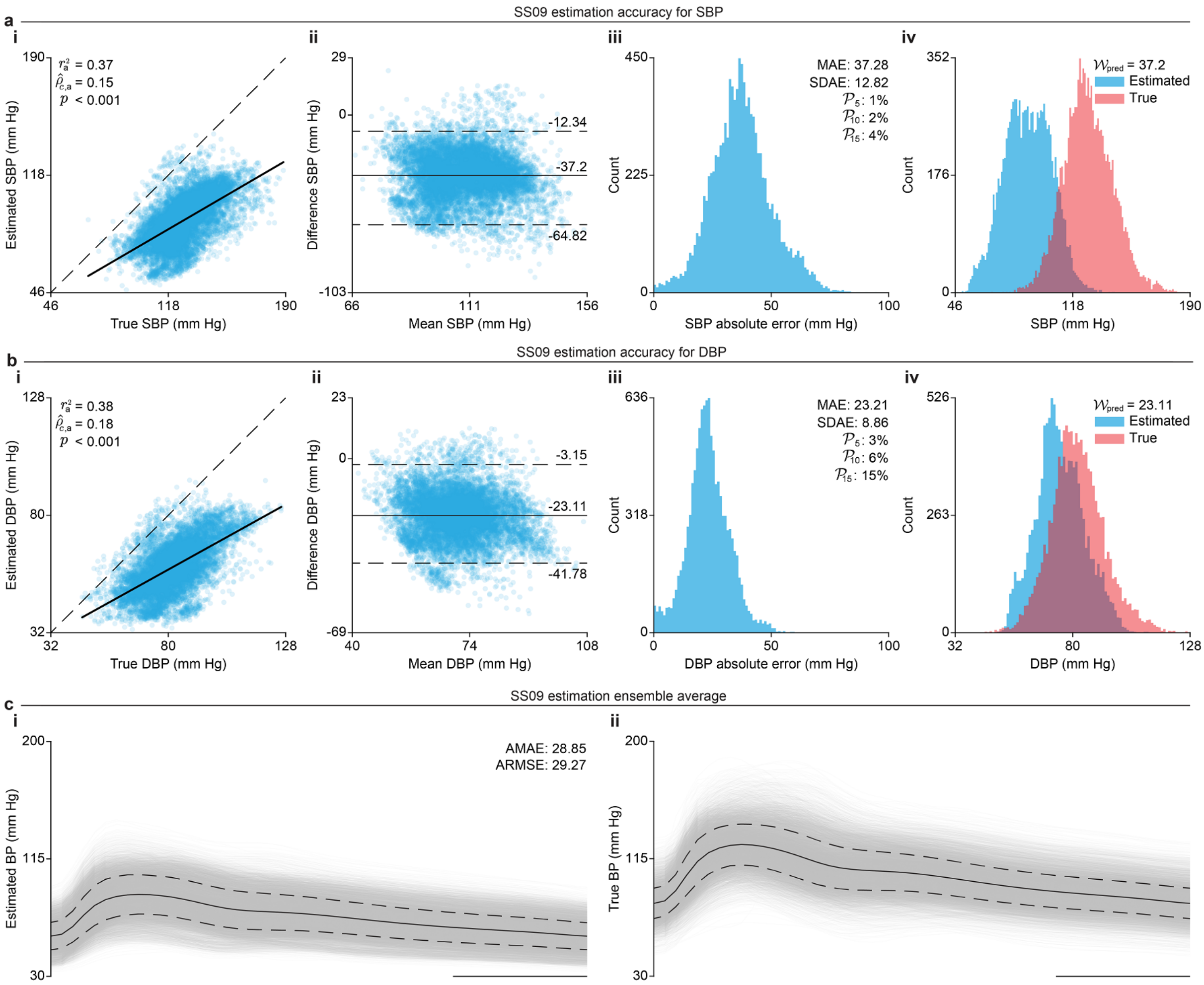

# Supplementary Fig. 42. Estimation results of subject-specific model SS10

Aggregated results from SS10 configuration: Convolutional Neural Network class with image input and fiducial output, trained with subject-specific (SS) datasets. **a,** Estimation accuracy for systolic brachial blood pressure (SBP); **b,** Estimation accuracy for diastolic brachial blood pressure (DBP); **i**, correlation plots; **ii**, limits of agreement (LOA) plots; **iii**, histogram of absolute errors (AE); and **iv**, histogram of estimated and true BP distributions. BP, blood pressure; DBP, diastolic blood pressure; SBP, systolic blood pressure. For correlation plots: $r^2_{\text{a}}$, aggregated coefficient of determination; $\hat{\rho}_{c,\text{a}}$, aggregated coefficient of concordance; solid line, empirical linear regression line; dashed line, $45°$ line of perfect correlation. For LOA plots: solid line, mean of errors between estimated and true BP values; dashed lines, 2.5$^{\text{th}}$ percentile (lower) and 97.5$^{\text{th}}$ percentile (upper). For AE histogram plots: MAE and SDAE, mean and standard deviation of AE, respectively; $\mathcal{P}_5$, $\mathcal{P}_{10}$, and $\mathcal{P}_{15}$, cumulative percentage of estimations with AE within 5, 10, and 15 mm Hg, respectively. For fiducial histogram plots: $\mathcal{W}_{\text{pred}}$, Wasserstein distance between true and estimated distribution.

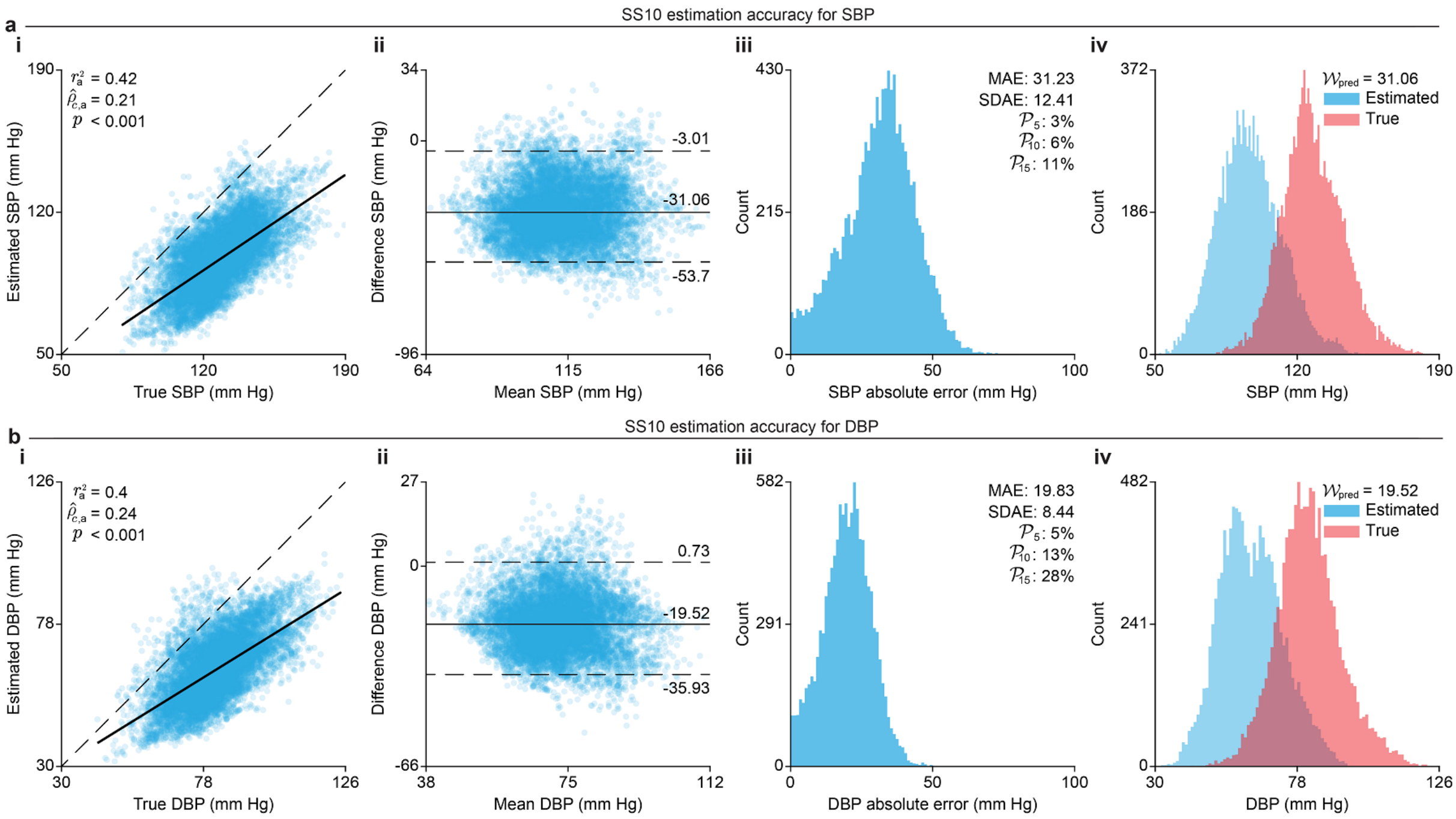

# Supplementary Fig. 43. Estimation results of subject-specific model SS11

Aggregated results from SS11 configuration: Convolutional Neural Network class with impedance input and waveform output, trained with subject-specific (SS) datasets. **a,** Estimation accuracy for systolic brachial blood pressure (SBP); **b,** Estimation accuracy for diastolic brachial blood pressure (DBP); **c,** Waveform ensemble of all estimated and true brachial blood pressure (BP) periods. For **a** and **b**: **i**, correlation plots; **ii**, limits of agreement (LOA) plots; **iii**, histogram of absolute errors (AE); and **iv**, histogram of estimated and true BP distributions. For **c**: **i**, ensemble of estimated BP periods; **ii**, ensemble of true BP periods. For correlation plots: $r^2_{\text{a}}$, aggregated coefficient of determination; $\hat{\rho}_{c,\text{a}}$, aggregated coefficient of concordance; solid line, empirical linear regression line; dashed line, $45°$ line of perfect correlation. For LOA plots: solid line, mean of errors between estimated and true BP values; dashed lines, 2.5th percentile (lower) and 97.5th percentile (upper). For AE histogram plots: MAE and SDAE, mean and standard deviation of AE, respectively; $\mathcal{P}_5$, $\mathcal{P}_{10}$, and $\mathcal{P}_{15}$, cumulative percentage of estimations with AE within 5, 10, and 15 mm Hg, respectively. For fiducial histogram plots: $\mathcal{W}_{\text{pred}}$, Wasserstein distance between true and estimated distribution. For ensemble plots: AMAE, average mean absolute error; ARMSE, average root mean square error; solid line, ensemble average of all periods; dashed lines, ensemble average $\pm$ standard deviation of all periods; scale bars, one-quarter of period.

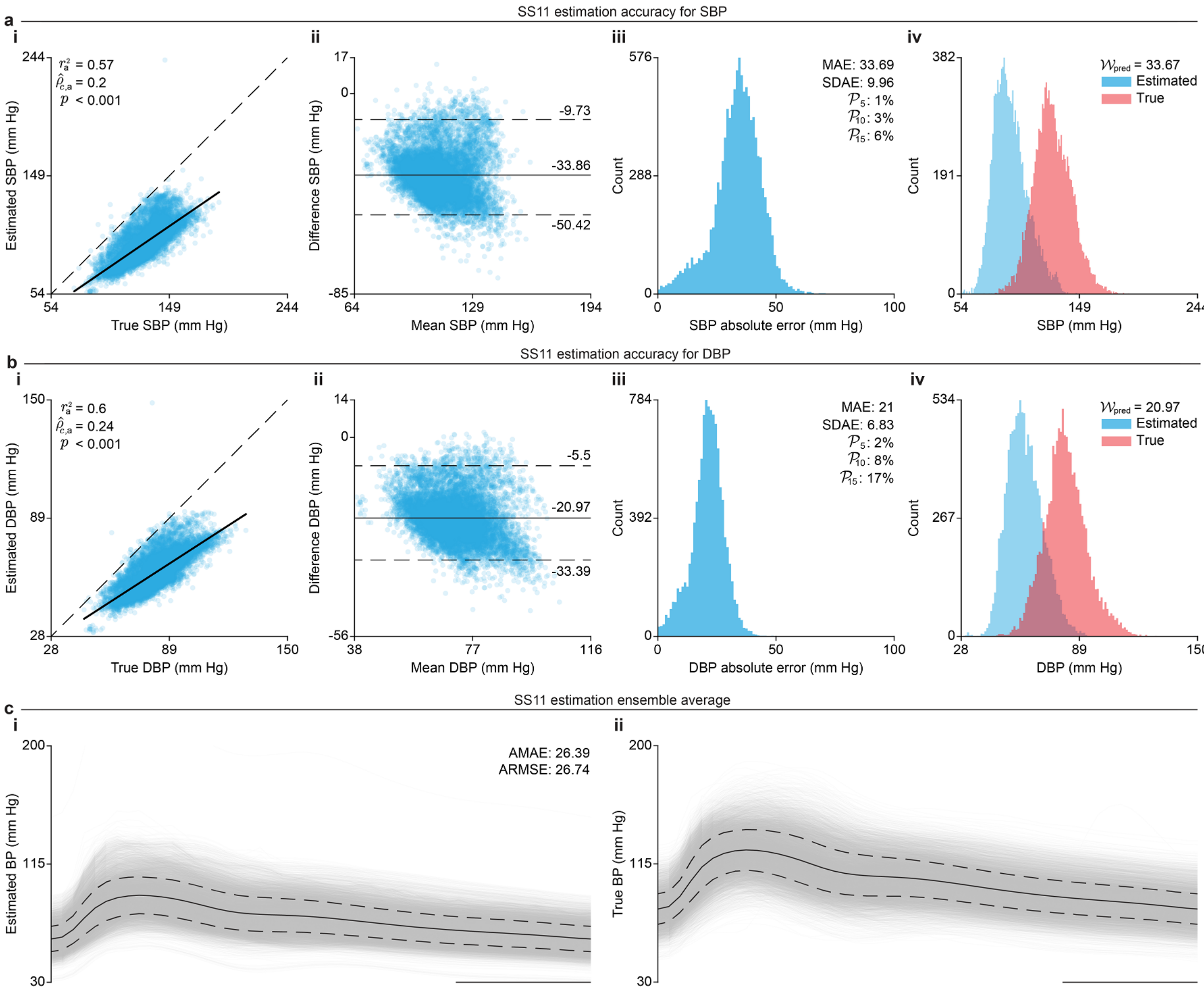

# Supplementary Fig. 44. Estimation results of subject-specific model SS12

Aggregated results from SS12 configuration: Convolutional Neural Network class with impedance input and fiducial output, trained with subject-specific (SS) datasets. **a,** Estimation accuracy for systolic brachial blood pressure (SBP); **b,** Estimation accuracy for diastolic brachial blood pressure (DBP); **i**, correlation plots; **ii**, limits of agreement (LOA) plots; **iii**, histogram of absolute errors (AE); and **iv**, histogram of estimated and true BP distributions. BP, blood pressure; DBP, diastolic blood pressure; SBP, systolic blood pressure. For correlation plots: $r^2_{\text{a}}$, aggregated coefficient of determination; $\hat{\rho}_{c,\text{a}}$, aggregated coefficient of concordance; solid line, empirical linear regression line; dashed line, $45°$ line of perfect correlation. For LOA plots: solid line, mean of errors between estimated and true BP values; dashed lines, 2.5$^{\text{th}}$ percentile (lower) and 97.5$^{\text{th}}$ percentile (upper). For AE histogram plots: MAE and SDAE, mean and standard deviation of AE, respectively; $\mathcal{P}_5$, $\mathcal{P}_{10}$, and $\mathcal{P}_{15}$, cumulative percentage of estimations with AE within 5, 10, and 15 mm Hg, respectively. For fiducial histogram plots: $\mathcal{W}_{\text{pred}}$, Wasserstein distance between true and estimated distribution.

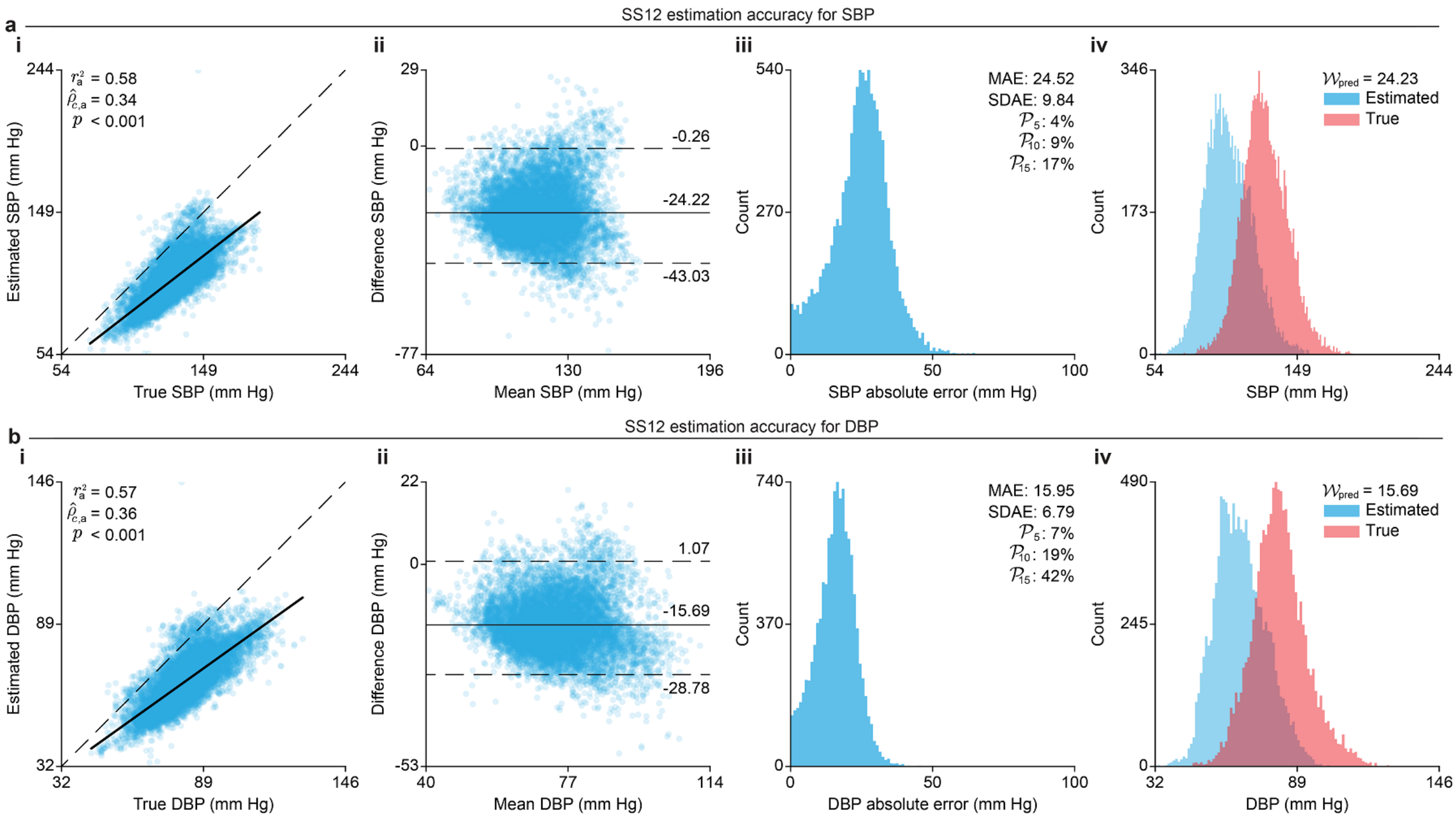

# Supplementary Fig. 45. Estimation results of subject-specific model SS13

Aggregated results from SS13 configuration: Convolutional Recurrent Transformer class with image input and waveform output, trained with subject-specific (SS) datasets. **a,** Estimation accuracy for systolic brachial blood pressure (SBP); **b,** Estimation accuracy for diastolic brachial blood pressure (DBP); **c,** Waveform ensemble of all estimated and true brachial blood pressure (BP) periods. For **a** and **b**: **i**, correlation plots; **ii**, limits of agreement (LOA) plots; **iii**, histogram of absolute errors (AE); and **iv**, histogram of estimated and true BP distributions. For **c**: **i**, ensemble of estimated BP periods; **ii**, ensemble of true BP periods. For correlation plots: $r_{\mathrm{a}}^2$, aggregated coefficient of determination; $\hat{\rho}_{c,\mathrm{a}}$, aggregated coefficient of concordance; solid line, empirical linear regression line; dashed line, $45^\circ$ line of perfect correlation. For LOA plots: solid line, mean of errors between estimated and true BP values; dashed lines, 2.5$^{\mathrm{th}}$ percentile (lower) and 97.5$^{\mathrm{th}}$ percentile (upper). For AE histogram plots: MAE and SDAE, mean and standard deviation of AE, respectively; $\mathcal{P}_5$, $\mathcal{P}_{10}$, and $\mathcal{P}_{15}$, cumulative percentage of estimations with AE within 5, 10, and 15 mm Hg, respectively. For fiducial histogram plots: $\mathcal{W}_{\mathrm{pred}}$, Wasserstein distance between true and estimated distribution. For ensemble plots: AMAE, average mean absolute error; ARMSE, average root mean square error; solid line, ensemble average of all periods; dashed lines, ensemble average $\pm$ standard deviation of all periods; scale bars, one-quarter of period.

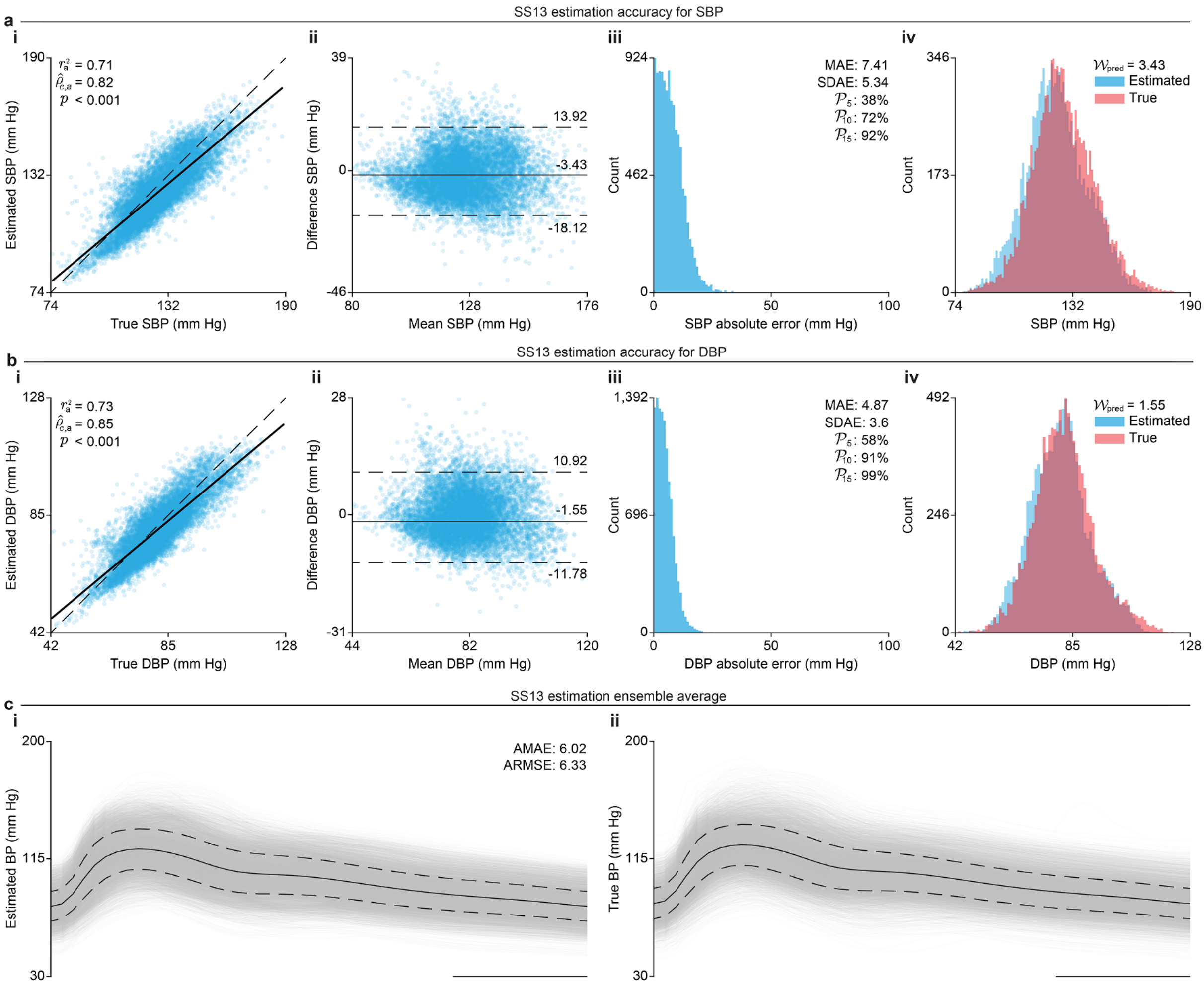

## Supplementary Fig. 46. Estimation results of subject-specific model SS14

Aggregated results from SS14 configuration: Convolutional Recurrent Transformer class with image input and fiducial output, trained with subject-specific (SS) datasets. **a,** Estimation accuracy for systolic brachial blood pressure (SBP); **b,** Estimation accuracy for diastolic brachial blood pressure (DBP); **i**, correlation plots; **ii**, limits of agreement (LOA) plots; **iii**, histogram of absolute errors (AE); and **iv**, histogram of estimated and true BP distributions. BP, blood pressure; DBP, diastolic blood pressure; SBP, systolic blood pressure. For correlation plots: $r^2_{\text{a}}$, aggregated coefficient of determination; $\hat{\rho}_{c,\text{a}}$, aggregated coefficient of concordance; solid line, empirical linear regression line; dashed line, $45°$ line of perfect correlation. For LOA plots: solid line, mean of errors between estimated and true BP values; dashed lines, 2.5$^{\text{th}}$ percentile (lower) and 97.5$^{\text{th}}$ percentile (upper). For AE histogram plots: MAE and SDAE, mean and standard deviation of AE, respectively; $\mathcal{P}_5$, $\mathcal{P}_{10}$, and $\mathcal{P}_{15}$, cumulative percentage of estimations with AE within 5, 10, and 15 mm Hg, respectively. For fiducial histogram plots: $\mathcal{W}_{\text{pred}}$, Wasserstein distance between true and estimated distribution.

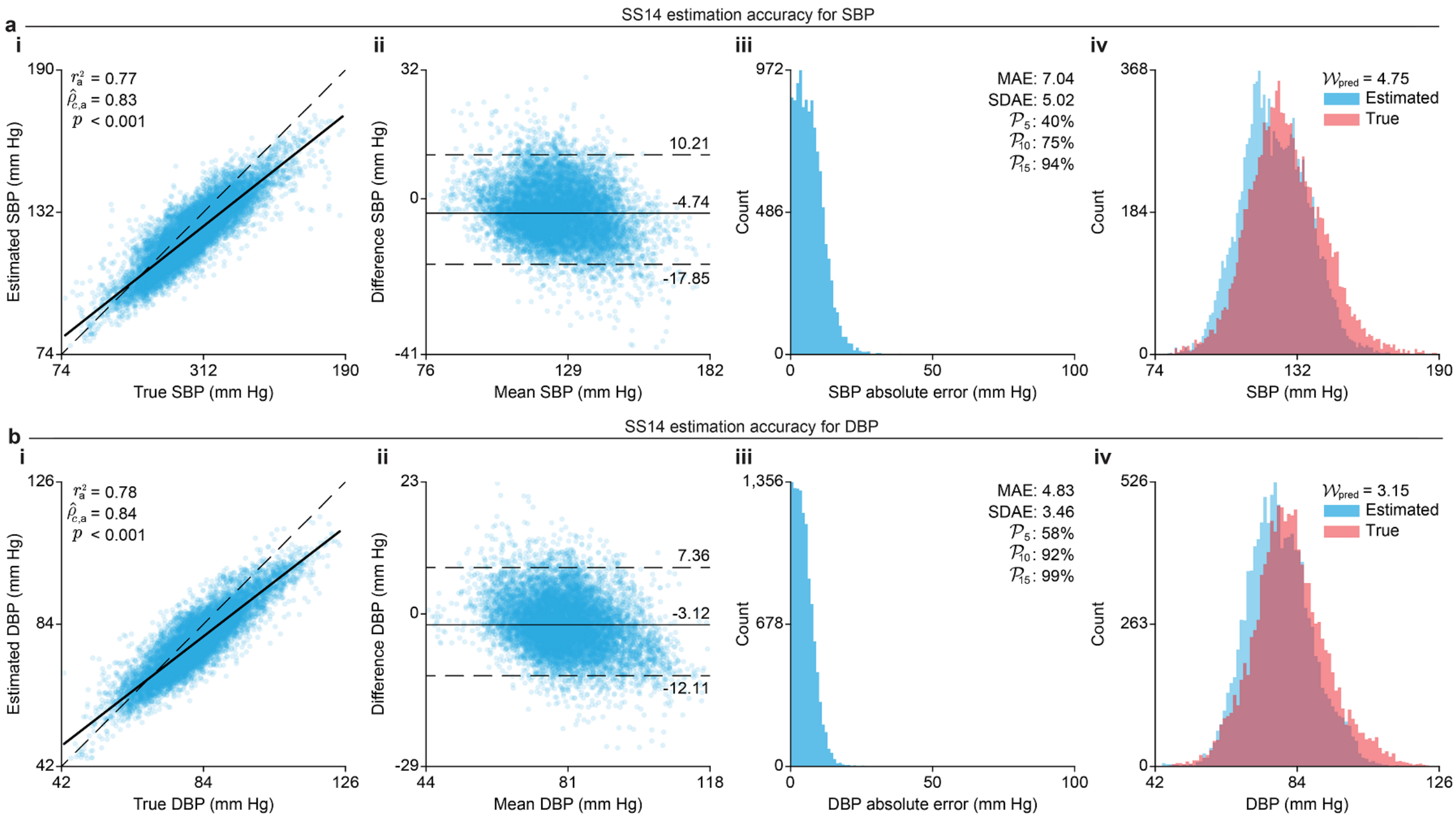

# Supplementary Fig. 47. Estimation results of subject-specific model SS15

Aggregated results from SS15 configuration: Convolutional Recurrent Transformer class with impedance input and waveform output, trained with subject-specific (SS) datasets. **a,** Estimation accuracy for systolic brachial blood pressure (SBP); **b,** Estimation accuracy for diastolic brachial blood pressure (DBP); **c,** Waveform ensemble of all estimated and true brachial blood pressure (BP) periods. For **a** and **b**: **i**, correlation plots; **ii**, limits of agreement (LOA) plots; **iii**, histogram of absolute errors (AE); and **iv**, histogram of estimated and true BP distributions. For **c**: **i**, ensemble of estimated BP periods; **ii**, ensemble of true BP periods. For correlation plots: $r_{\text{a}}^2$, aggregated coefficient of determination; $\hat{\rho}_{c,\text{a}}$, aggregated coefficient of concordance; solid line, empirical linear regression line; dashed line, $45°$ line of perfect correlation. For LOA plots: solid line, mean of errors between estimated and true BP values; dashed lines, 2.5$^{\text{th}}$ percentile (lower) and 97.5$^{\text{th}}$ percentile (upper). For AE histogram plots: MAE and SDAE, mean and standard deviation of AE, respectively; $\mathcal{P}_5$, $\mathcal{P}_{10}$, and $\mathcal{P}_{15}$, cumulative percentage of estimations with AE within 5, 10, and 15 mm Hg, respectively. For fiducial histogram plots: $\mathcal{W}_{\text{pred}}$, Wasserstein distance between true and estimated distribution. For ensemble plots: AMAE, average mean absolute error; ARMSE, average root mean square error; solid line, ensemble average of all periods; dashed lines, ensemble average $\pm$ standard deviation of all periods; scale bars, one-quarter of period.

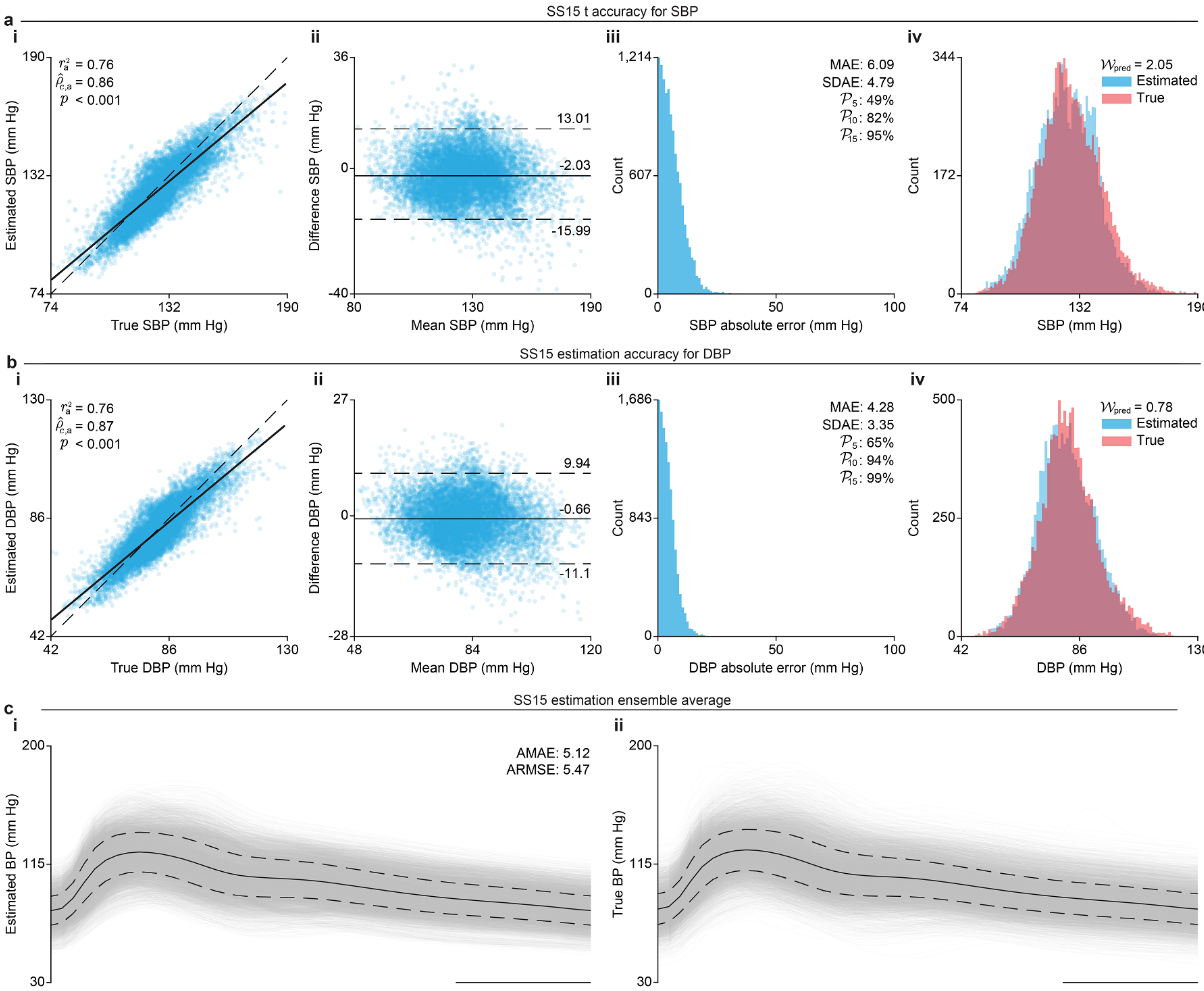

# Supplementary Fig. 48. Estimation results of subject-specific model SS16

Aggregated results from SS16 configuration: Convolutional Recurrent Transformer class with impedance input and fiducial output, trained with subject-specific (SS) datasets. **a,** Estimation accuracy for systolic brachial blood pressure (SBP); **b,** Estimation accuracy for diastolic brachial blood pressure (DBP); **i**, correlation plots; **ii**, limits of agreement (LOA) plots; **iii**, histogram of absolute errors (AE); and **iv**, histogram of estimated and true BP distributions. BP, blood pressure; DBP, diastolic blood pressure; SBP, systolic blood pressure. For correlation plots: $r_{\text{a}}^2$, aggregated coefficient of determination; $\hat{\rho}_{c,\text{a}}$, aggregated coefficient of concordance; solid line, empirical linear regression line; dashed line, $45°$ line of perfect correlation. For LOA plots: solid line, mean of errors between estimated and true BP values; dashed lines, 2.5$^{\text{th}}$ percentile (lower) and 97.5$^{\text{th}}$ percentile (upper). For AE histogram plots: MAE and SDAE, mean and standard deviation of AE, respectively; $\mathcal{P}_5$, $\mathcal{P}_{10}$, and $\mathcal{P}_{15}$, cumulative percentage of estimations with AE within 5, 10, and 15 mm Hg, respectively. For fiducial histogram plots: $\mathcal{W}_{\text{pred}}$, Wasserstein distance between true and estimated distribution.

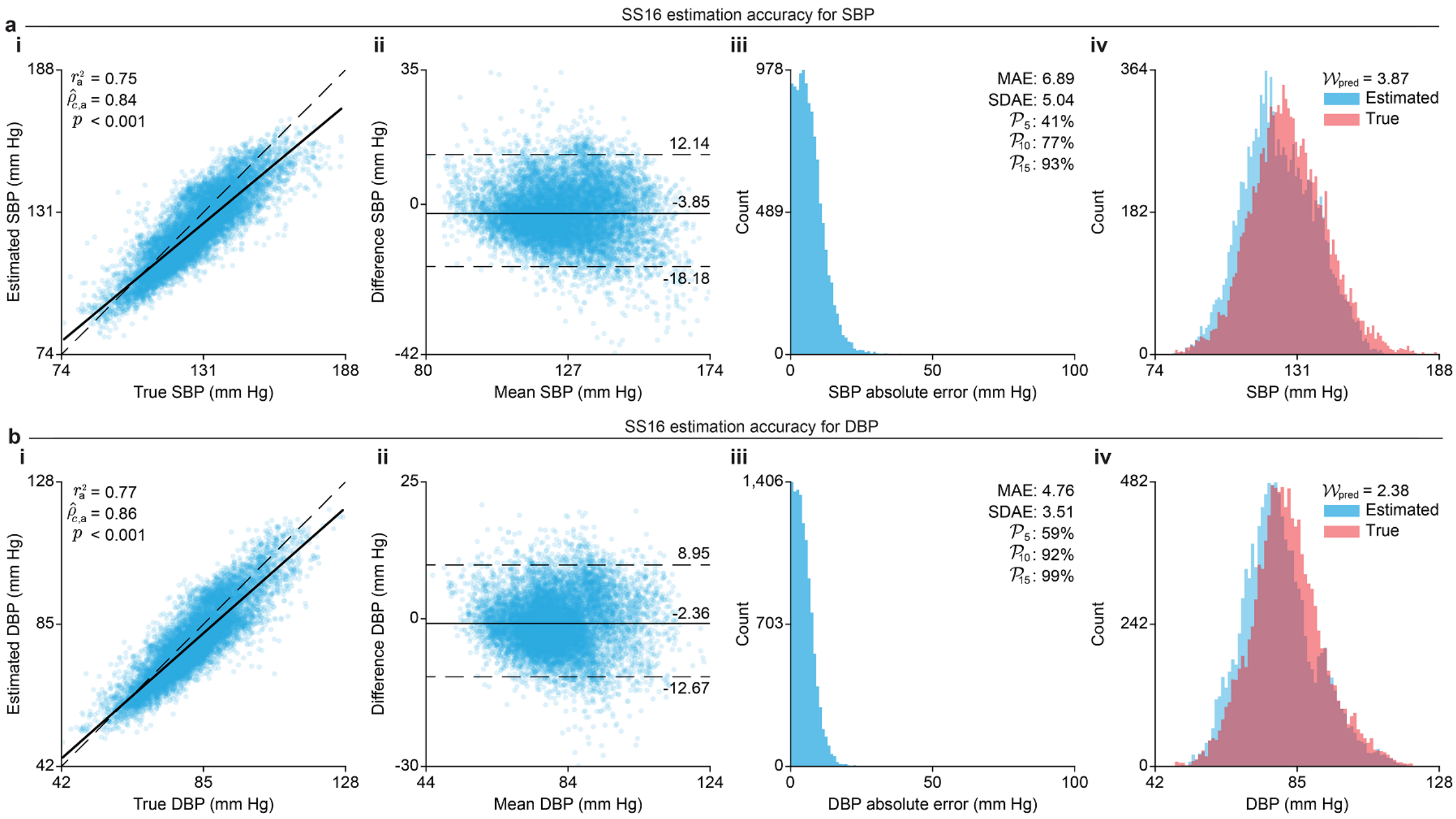

# Supplementary Fig. 49. Estimation results of subject-specific model SS17

Aggregated results from SS17 configuration: Convolutional Recurrent Samba class with image input and waveform output, trained with subject-specific (SS) datasets. **a,** Estimation accuracy for systolic brachial blood pressure (SBP); **b,** Estimation accuracy for diastolic brachial blood pressure (DBP); **c,** Waveform ensemble of all estimated and true brachial blood pressure (BP) periods. For **a** and **b**: **i**, correlation plots; **ii**, limits of agreement (LOA) plots; **iii**, histogram of absolute errors (AE); and **iv**, histogram of estimated and true BP distributions. For **c**: **i**, ensemble of estimated BP periods; **ii**, ensemble of true BP periods. For correlation plots: $r^2_{\mathrm{a}}$, aggregated coefficient of determination; $\hat{\rho}_{c,\mathrm{a}}$, aggregated coefficient of concordance; solid line, empirical linear regression line; dashed line, $45°$ line of perfect correlation. For LOA plots: solid line, mean of errors between estimated and true BP values; dashed lines, 2.5$^{\mathrm{th}}$ percentile (lower) and 97.5$^{\mathrm{th}}$ percentile (upper). For AE histogram plots: MAE and SDAE, mean and standard deviation of AE, respectively; $\mathcal{P}_5$, $\mathcal{P}_{10}$, and $\mathcal{P}_{15}$, cumulative percentage of estimations with AE within 5, 10, and 15 mm Hg, respectively. For fiducial histogram plots: $\mathcal{W}_{\mathrm{pred}}$, Wasserstein distance between true and estimated distribution. For ensemble plots: AMAE, average mean absolute error; ARMSE, average root mean square error; solid line, ensemble average of all periods; dashed lines, ensemble average $\pm$ standard deviation of all periods; scale bars, one-quarter of period.

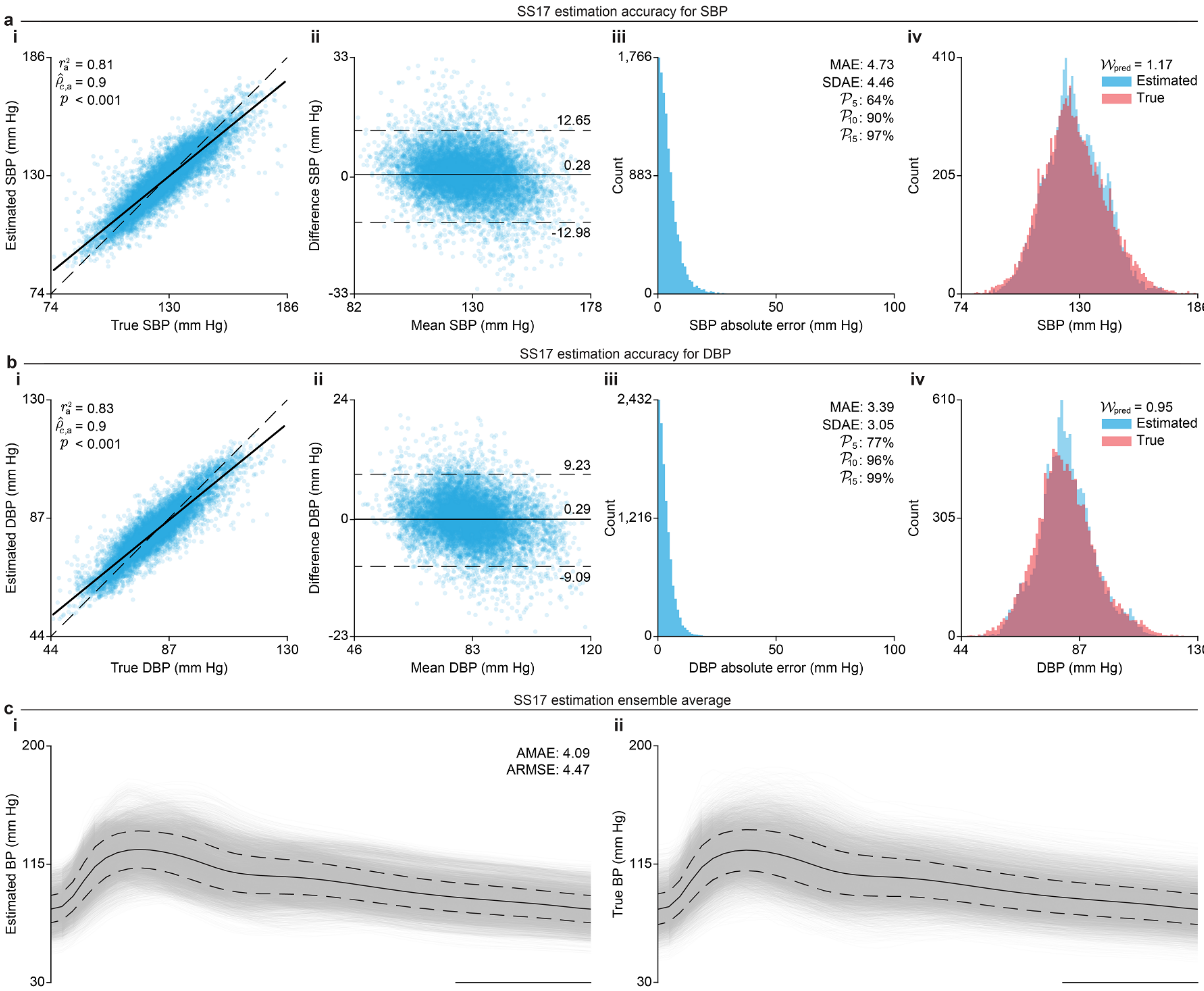

# Supplementary Fig. 50. Estimation results of subject-specific model SS18

Aggregated results from SS18 configuration: Convolutional Recurrent Samba class with image input and fiducial output, trained with subject-specific (SS) datasets. **a,** Estimation accuracy for systolic brachial blood pressure (SBP); **b,** Estimation accuracy for diastolic brachial blood pressure (DBP); **i**, correlation plots; **ii**, limits of agreement (LOA) plots; **iii**, histogram of absolute errors (AE); and **iv**, histogram of estimated and true BP distributions. BP, blood pressure; DBP, diastolic blood pressure; SBP, systolic blood pressure. For correlation plots: $r_{\text{a}}^2$, aggregated coefficient of determination; $\hat{\rho}_{c,\text{a}}$, aggregated coefficient of concordance; solid line, empirical linear regression line; dashed line, $45°$ line of perfect correlation. For LOA plots: solid line, mean of errors between estimated and true BP values; dashed lines, 2.5th percentile (lower) and 97.5th percentile (upper). For AE histogram plots: MAE and SDAE, mean and standard deviation of AE, respectively; $\mathcal{P}_5$, $\mathcal{P}_{10}$, and $\mathcal{P}_{15}$, cumulative percentage of estimations with AE within 5, 10, and 15 mm Hg, respectively. For fiducial histogram plots: $\mathcal{W}_{\text{pred}}$, Wasserstein distance between true and estimated distribution.

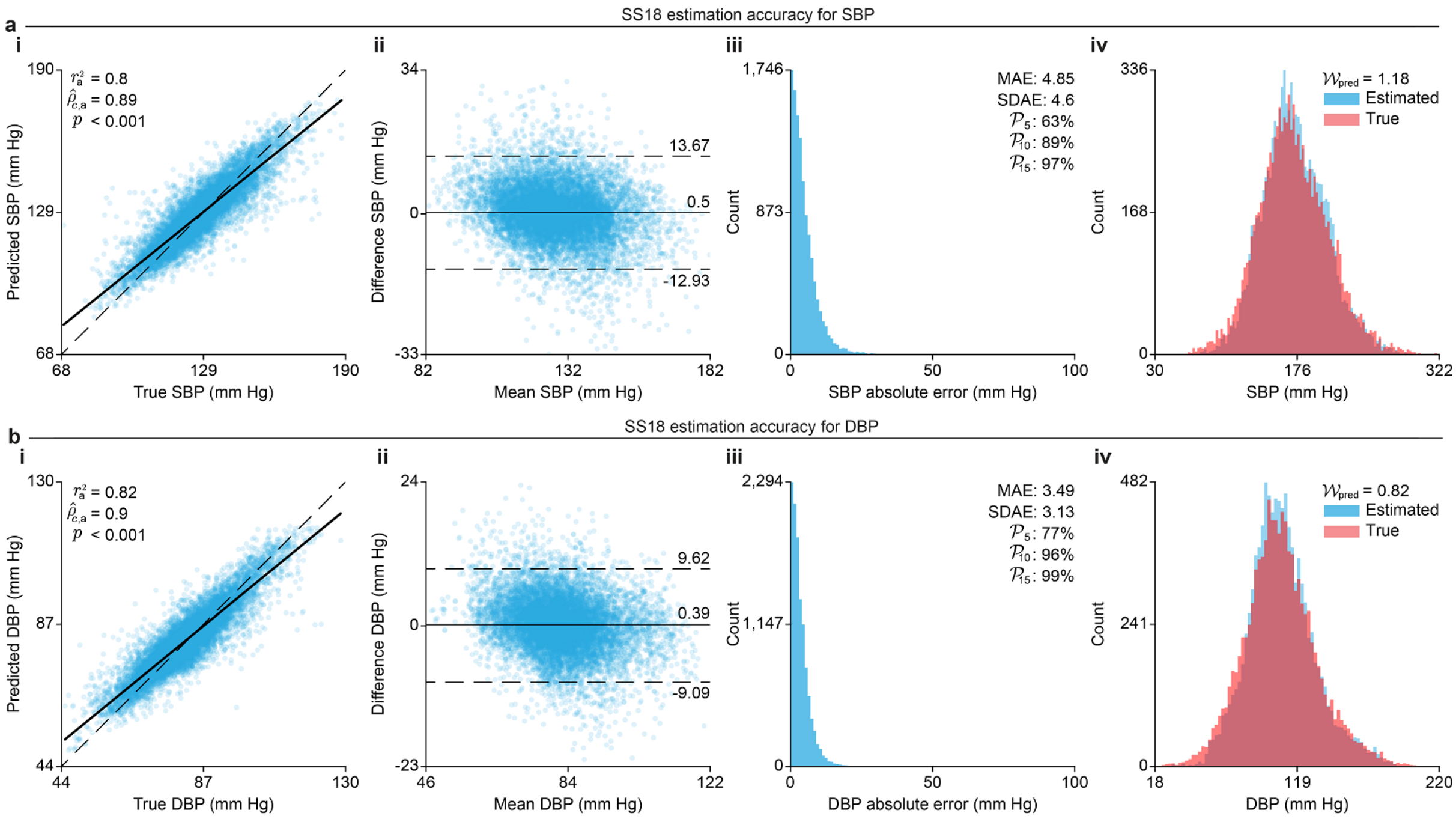

# Supplementary Fig. 51. Estimation results of subject-specific model SS19

Aggregated results from SS19 configuration: Convolutional Recurrent Samba class with impedance input and waveform output, trained with subject-specific (SS) datasets. **a,** Estimation accuracy for systolic brachial blood pressure (SBP); **b,** Estimation accuracy for diastolic brachial blood pressure (DBP); **c,** Waveform ensemble of all estimated and true brachial blood pressure (BP) periods. For **a** and **b**: **i**, correlation plots; **ii**, limits of agreement (LOA) plots; **iii**, histogram of absolute errors (AE); and **iv**, histogram of estimated and true BP distributions. For **c**: **i**, ensemble of estimated BP periods; **ii**, ensemble of true BP periods. For correlation plots: $r_{\text{a}}^2$, aggregated coefficient of determination; $\hat{\rho}_{c,\text{a}}$, aggregated coefficient of concordance; solid line, empirical linear regression line; dashed line, $45°$ line of perfect correlation. For LOA plots: solid line, mean of errors between estimated and true BP values; dashed lines, 2.5$^{\text{th}}$ percentile (lower) and 97.5$^{\text{th}}$ percentile (upper). For AE histogram plots: MAE and SDAE, mean and standard deviation of AE, respectively; $\mathcal{P}_5$, $\mathcal{P}_{10}$, and $\mathcal{P}_{15}$, cumulative percentage of estimations with AE within 5, 10, and 15 mm Hg, respectively. For fiducial histogram plots: $\mathcal{W}_{\text{pred}}$, Wasserstein distance between true and estimated distribution. For ensemble plots: AMAE, average mean absolute error; ARMSE, average root mean square error; solid line, ensemble average of all periods; dashed lines, ensemble average $\pm$ standard deviation of all periods; scale bars, one-quarter of period.

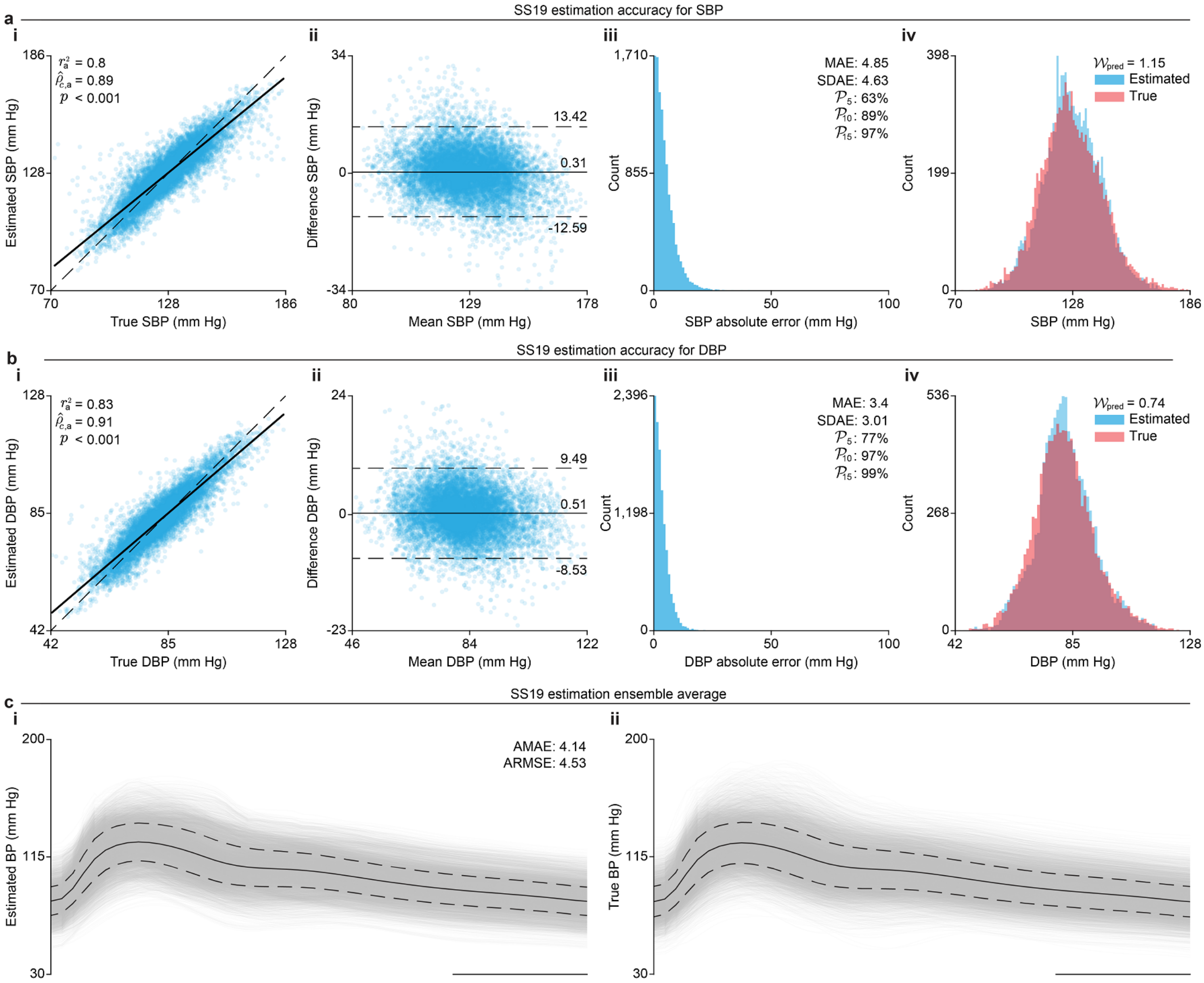

# Supplementary Fig. 52. Estimation results of subject-specific model SS20

Aggregated results from SS20 configuration: Convolutional Recurrent Samba class with impedance input and fiducial output, trained with subject-specific (SS) datasets. **a,** Estimation accuracy for systolic brachial blood pressure (SBP); **b,** Estimation accuracy for diastolic brachial blood pressure (DBP); **i**, correlation plots; **ii**, limits of agreement (LOA) plots; **iii**, histogram of absolute errors (AE); and **iv**, histogram of estimated and true BP distributions. BP, blood pressure; DBP, diastolic blood pressure; SBP, systolic blood pressure. For correlation plots: $r_{\text{a}}^2$, aggregated coefficient of determination; $\hat{\rho}_{c,\text{a}}$, aggregated coefficient of concordance; solid line, empirical linear regression line; dashed line, $45°$ line of perfect correlation. For LOA plots: solid line, mean of errors between estimated and true BP values; dashed lines, 2.5$^{\text{th}}$ percentile (lower) and 97.5$^{\text{th}}$ percentile (upper). For AE histogram plots: MAE and SDAE, mean and standard deviation of AE, respectively; $\mathcal{P}_5$, $\mathcal{P}_{10}$, and $\mathcal{P}_{15}$, cumulative percentage of estimations with AE within 5, 10, and 15 mm Hg, respectively. For fiducial histogram plots: $\mathcal{W}_{\text{pred}}$, Wasserstein distance between true and estimated distribution.

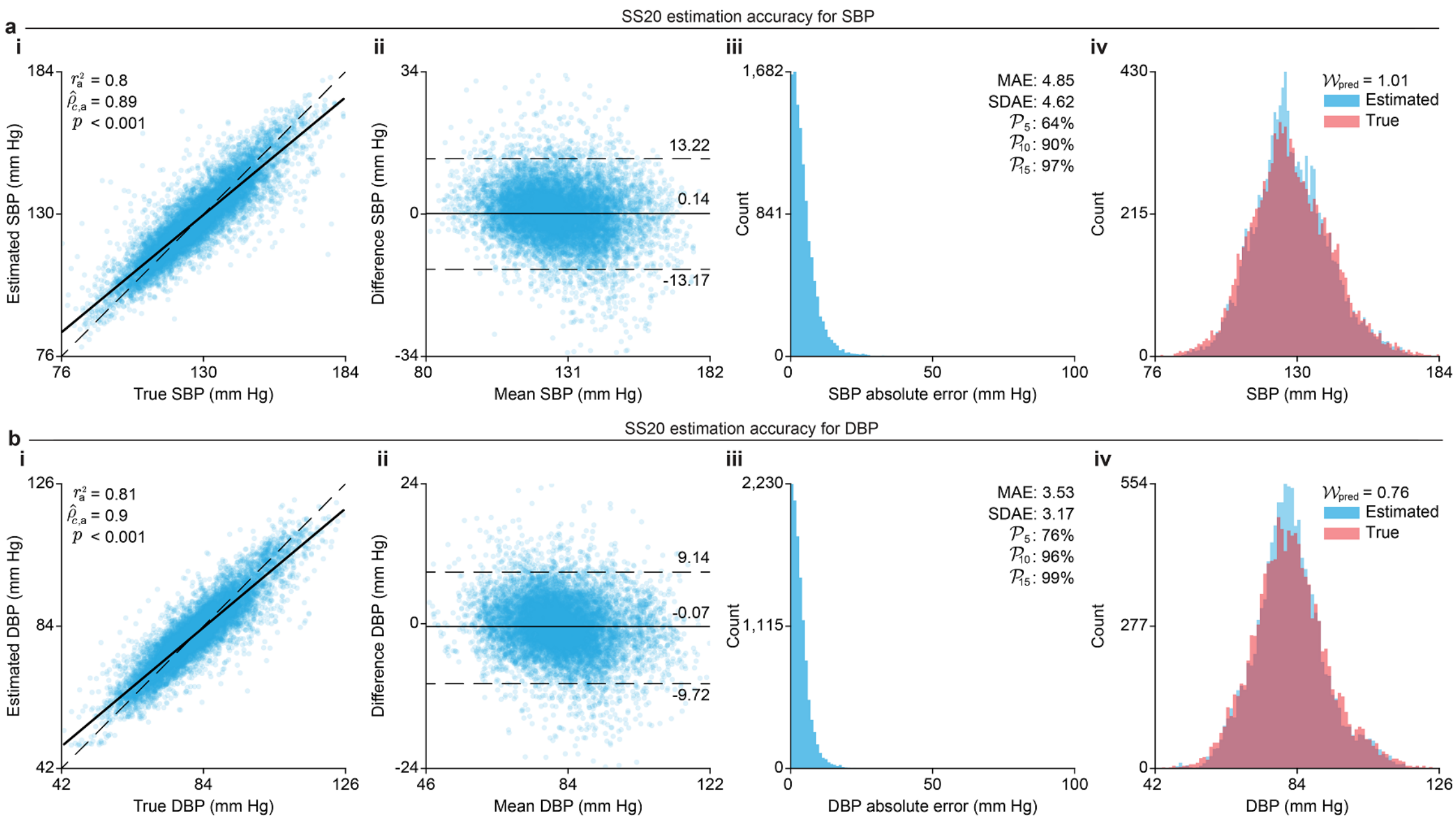

# Supplementary Fig. 53. Estimation results of population-within model PW01

Aggregated results from PW01 configuration: Linear Regression class with image input and waveform output, trained with the population-within (PW) partition. **a,** Estimation accuracy for systolic brachial blood pressure (SBP); **b,** Estimation accuracy for diastolic brachial blood pressure (DBP); **c,** Waveform ensemble of all estimated and true brachial blood pressure (BP) periods. For **a** and **b**: **i**, correlation plots; **ii**, limits of agreement (LOA) plots; **iii**, histogram of absolute errors (AE); and **iv**, histogram of estimated and true BP distributions. For **c**: **i**, ensemble of estimated BP periods; **ii**, ensemble of true BP periods. For correlation plots: $r_{\mathrm{a}}^2$, aggregated coefficient of determination; $\hat{\rho}_{c,\mathrm{a}}$, aggregated coefficient of concordance; solid line, empirical linear regression line; dashed line, $45°$ line of perfect correlation. For LOA plots: solid line, mean of errors between estimated and true BP values; dashed lines, 2.5$^{\text{th}}$ percentile (lower) and 97.5$^{\text{th}}$ percentile (upper). For AE histogram plots: MAE and SDAE, mean and standard deviation of AE, respectively; $\mathcal{P}_5$, $\mathcal{P}_{10}$, and $\mathcal{P}_{15}$, cumulative percentage of estimations with AE within 5, 10, and 15 mm Hg, respectively. For fiducial histogram plots: $\mathcal{W}_{\text{pred}}$, Wasserstein distance between true and estimated distribution. For ensemble plots: AMAE, average mean absolute error; ARMSE, average root mean square error; solid line, ensemble average of all periods; dashed lines, ensemble average $\pm$ standard deviation of all periods; scale bars, one-quarter of period.

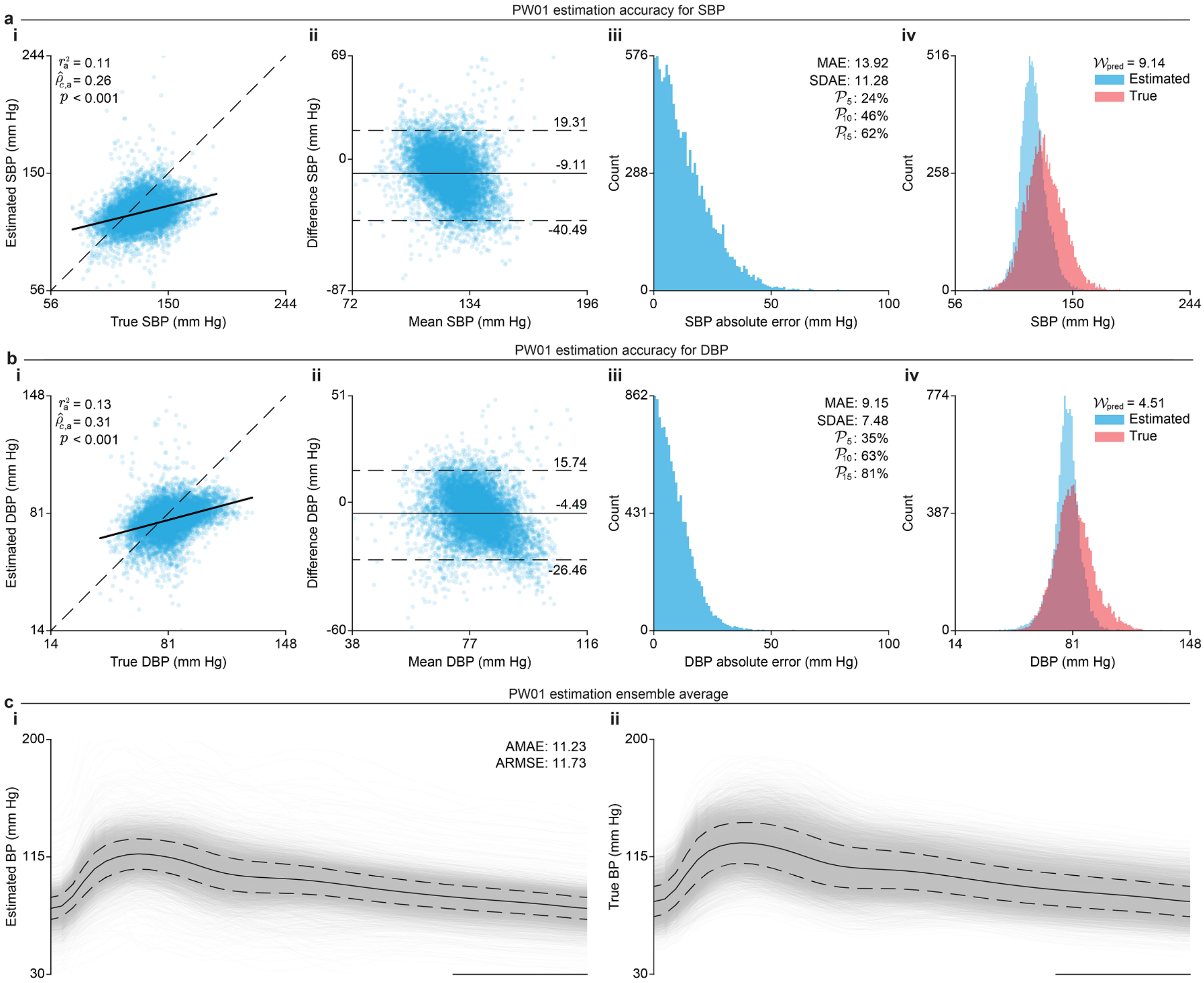

# Supplementary Fig. 54. Estimation results of population-within model PW02

Aggregated results from PW02 configuration: Linear Regression class with image input and fiducial output, trained with the population-within (PW) partition. **a,** Estimation accuracy for systolic brachial blood pressure (SBP); **b,** Estimation accuracy for diastolic brachial blood pressure (DBP); **i**, correlation plots; **ii**, limits of agreement (LOA) plots; **iii**, histogram of absolute errors (AE); and **iv**, histogram of estimated and true BP distributions. BP, blood pressure; DBP, diastolic blood pressure; SBP, systolic blood pressure. For correlation plots: $r_{\text{a}}^2$, aggregated coefficient of determination; $\hat{\rho}_{c,\text{a}}$, aggregated coefficient of concordance; solid line, empirical linear regression line; dashed line, $45°$ line of perfect correlation. For LOA plots: solid line, mean of errors between estimated and true BP values; dashed lines, 2.5$^{\text{th}}$ percentile (lower) and 97.5$^{\text{th}}$ percentile (upper). For AE histogram plots: MAE and SDAE, mean and standard deviation of AE, respectively; $\mathcal{P}_5$, $\mathcal{P}_{10}$, and $\mathcal{P}_{15}$, cumulative percentage of estimations with AE within 5, 10, and 15 mm Hg, respectively. For fiducial histogram plots: $\mathcal{W}_{\text{pred}}$, Wasserstein distance between true and estimated distribution.

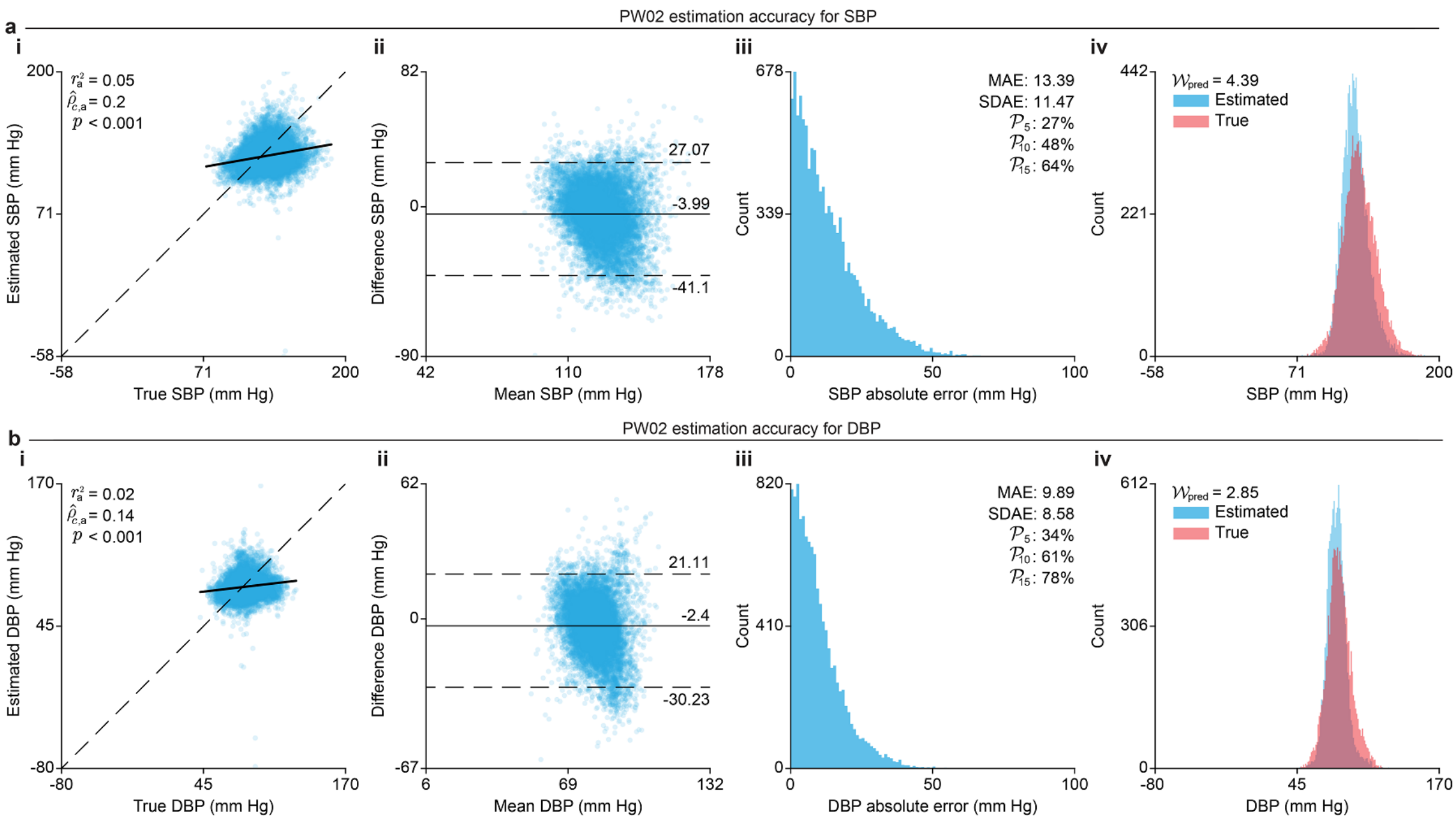

# Supplementary Fig. 55. Estimation results of population-within model PW03

Aggregated results from PW03 configuration: Linear Regression class with impedance input and waveform output, trained with the population-within (PW) partition. **a,** Estimation accuracy for systolic brachial blood pressure (SBP); **b,** Estimation accuracy for diastolic brachial blood pressure (DBP); **c,** Waveform ensemble of all estimated and true brachial blood pressure (BP) periods. For **a** and **b**: **i**, correlation plots; **ii**, limits of agreement (LOA) plots; **iii**, histogram of absolute errors (AE); and **iv**, histogram of estimated and true BP distributions. For **c**: **i**, ensemble of estimated BP periods; **ii**, ensemble of true BP periods. For correlation plots: $r_{\mathrm{a}}^2$, aggregated coefficient of determination; $\hat{\rho}_{c,\mathrm{a}}$, aggregated coefficient of concordance; solid line, empirical linear regression line; dashed line, $45°$ line of perfect correlation. For LOA plots: solid line, mean of errors between estimated and true BP values; dashed lines, 2.5$^{\text{th}}$ percentile (lower) and 97.5$^{\text{th}}$ percentile (upper). For AE histogram plots: MAE and SDAE, mean and standard deviation of AE, respectively; $\mathcal{P}_5$, $\mathcal{P}_{10}$, and $\mathcal{P}_{15}$, cumulative percentage of estimations with AE within 5, 10, and 15 mm Hg, respectively. For fiducial histogram plots: $\mathcal{W}_{\text{pred}}$, Wasserstein distance between true and estimated distribution. For ensemble plots: AMAE, average mean absolute error; ARMSE, average root mean square error; solid line, ensemble average of all periods; dashed lines, ensemble average $\pm$ standard deviation of all periods; scale bars, one-quarter of period.

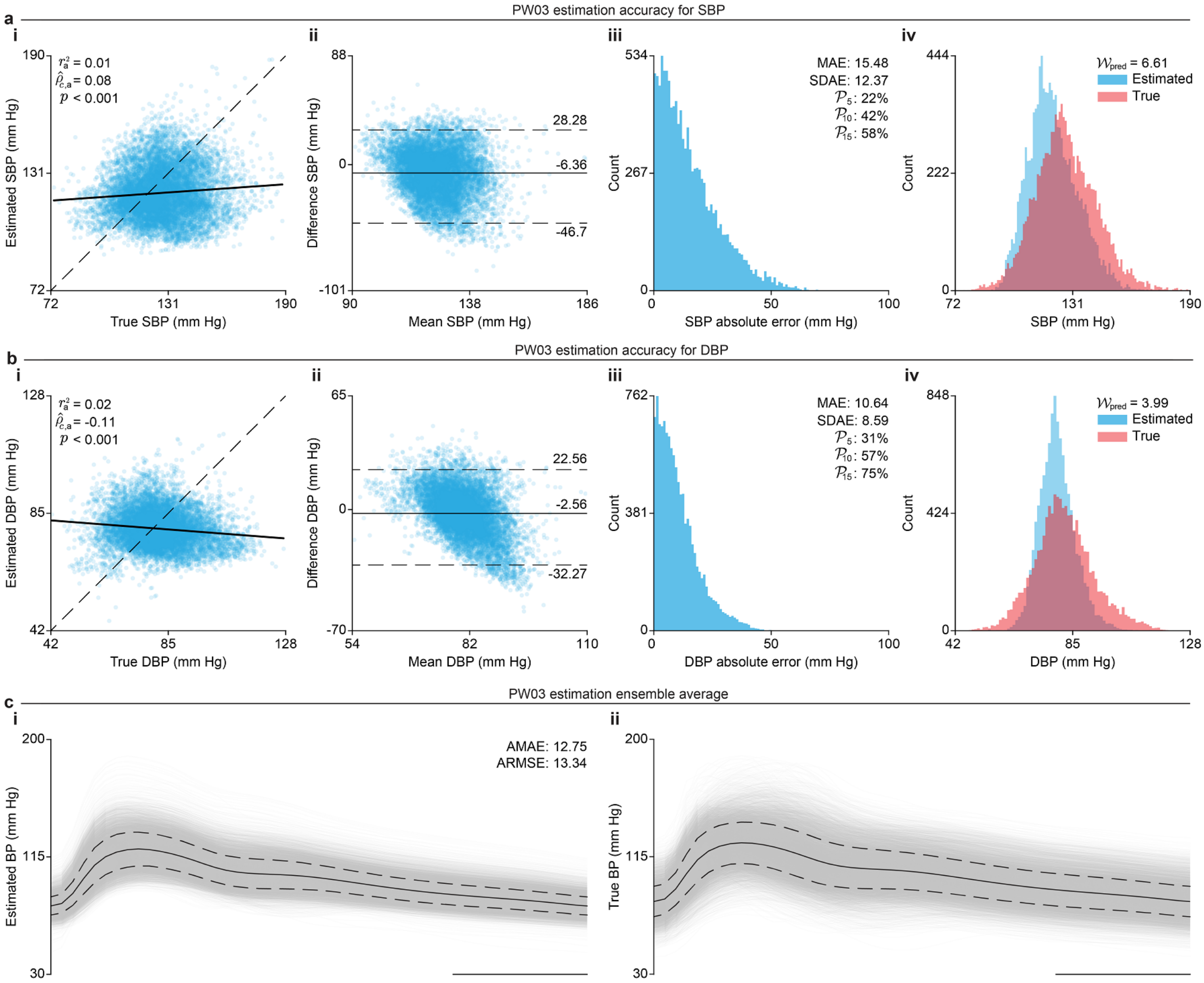

# Supplementary Fig. 56. Estimation results of population-within model PW04

Aggregated results from PW04 configuration: Linear Regression class with impedance input and fiducial output, trained with the population-within (PW) partition. **a,** Estimation accuracy for systolic brachial blood pressure (SBP); **b,** Estimation accuracy for diastolic brachial blood pressure (DBP); **i**, correlation plots; **ii**, limits of agreement (LOA) plots; **iii**, histogram of absolute errors (AE); and **iv**, histogram of estimated and true BP distributions. BP, blood pressure; DBP, diastolic blood pressure; SBP, systolic blood pressure. For correlation plots: $r_{\mathrm{a}}^2$, aggregated coefficient of determination; $\hat{\rho}_{c,\mathrm{a}}$, aggregated coefficient of concordance; solid line, empirical linear regression line; dashed line, $45°$ line of perfect correlation. For LOA plots: solid line, mean of errors between estimated and true BP values; dashed lines, 2.5$^{\text{th}}$ percentile (lower) and 97.5$^{\text{th}}$ percentile (upper). For AE histogram plots: MAE and SDAE, mean and standard deviation of AE, respectively; $\mathcal{P}_5$, $\mathcal{P}_{10}$, and $\mathcal{P}_{15}$, cumulative percentage of estimations with AE within 5, 10, and 15 mm Hg, respectively. For fiducial histogram plots: $\mathcal{W}_{\text{pred}}$, Wasserstein distance between true and estimated distribution.

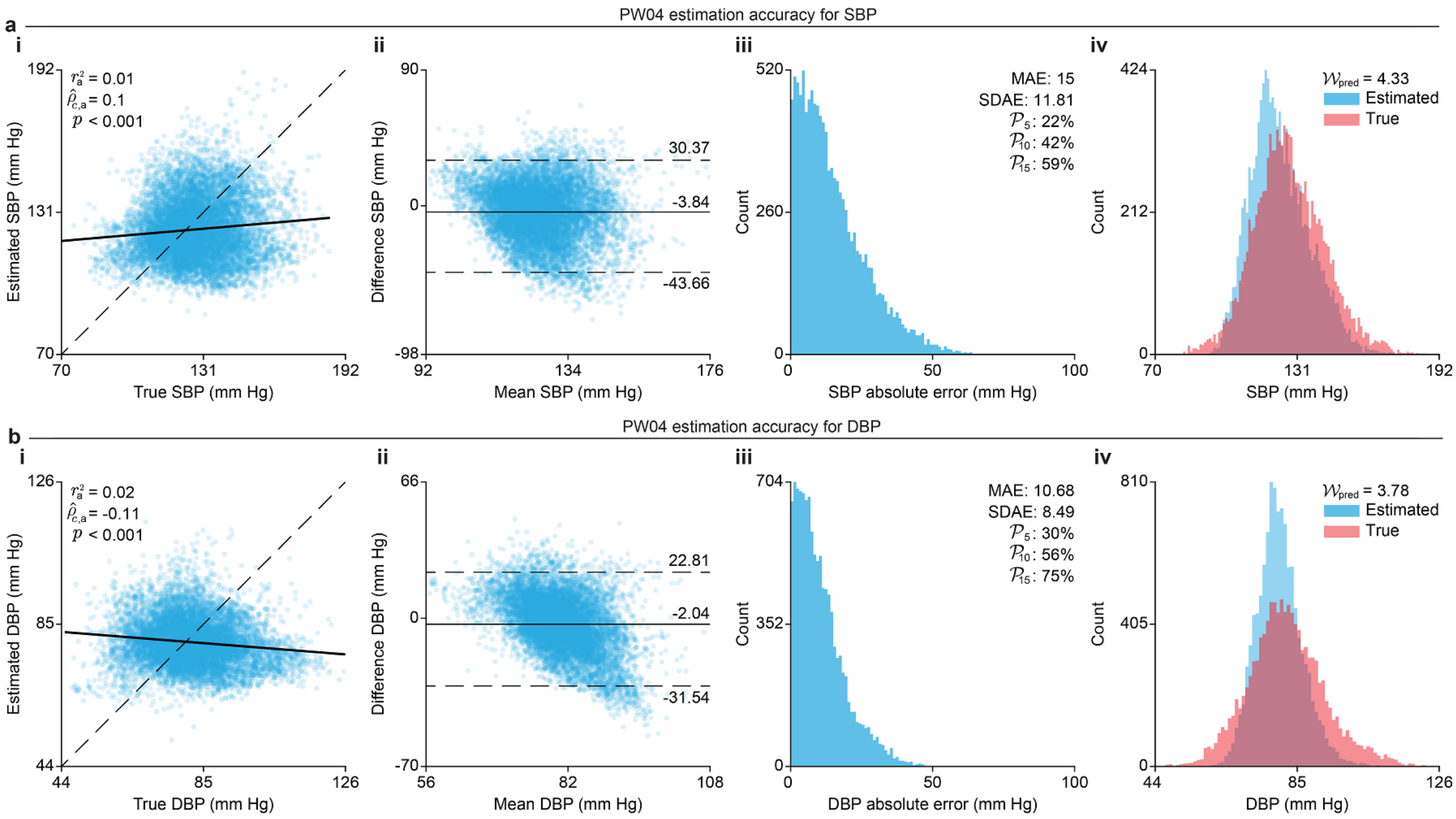

# Supplementary Fig. 57. Estimation results of population-within model PW05

Aggregated results from PW05 configuration: Multilayer Perceptron class with image input and waveform output, trained with the population-within (PW) partition. **a,** Estimation accuracy for systolic brachial blood pressure (SBP); **b,** Estimation accuracy for diastolic brachial blood pressure (DBP); **c,** Waveform ensemble of all estimated and true brachial blood pressure (BP) periods. For **a** and **b**: **i**, correlation plots; **ii**, limits of agreement (LOA) plots; **iii**, histogram of absolute errors (AE); and **iv**, histogram of estimated and true BP distributions. For **c**: **i**, ensemble of estimated BP periods; **ii**, ensemble of true BP periods. For correlation plots: $r^2_{\text{a}}$, aggregated coefficient of determination; $\hat{\rho}_{c,\text{a}}$, aggregated coefficient of concordance; solid line, empirical linear regression line; dashed line, $45°$ line of perfect correlation. For LOA plots: solid line, mean of errors between estimated and true BP values; dashed lines, 2.5$^{\text{th}}$ percentile (lower) and 97.5$^{\text{th}}$ percentile (upper). For AE histogram plots: MAE and SDAE, mean and standard deviation of AE, respectively; $\mathcal{P}_5$, $\mathcal{P}_{10}$, and $\mathcal{P}_{15}$, cumulative percentage of estimations with AE within 5, 10, and 15 mm Hg, respectively. For fiducial histogram plots: $\mathcal{W}_{\text{pred}}$, Wasserstein distance between true and estimated distribution. For ensemble plots: AMAE, average mean absolute error; ARMSE, average root mean square error; solid line, ensemble average of all periods; dashed lines, ensemble average $\pm$ standard deviation of all periods; scale bars, one-quarter of period.

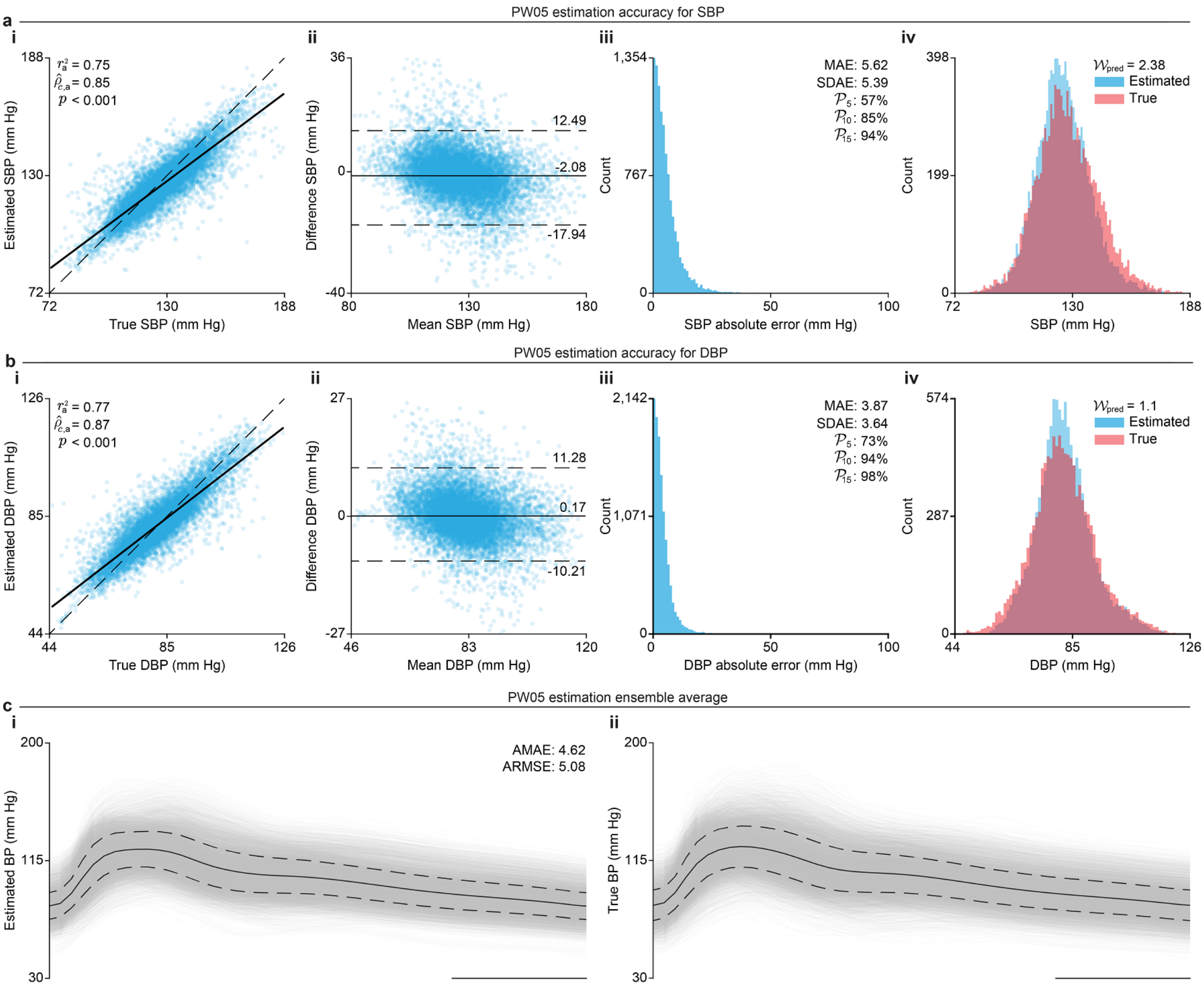

# Supplementary Fig. 58. Estimation results of population-within model PW06

Aggregated results from PW06 configuration: Multilayer Perceptron class with image input and fiducial output, trained with the population-within (PW) partition. **a,** Estimation accuracy for systolic brachial blood pressure (SBP); **b,** Estimation accuracy for diastolic brachial blood pressure (DBP); **i**, correlation plots; **ii**, limits of agreement (LOA) plots; **iii**, histogram of absolute errors (AE); and **iv**, histogram of estimated and true BP distributions. BP, blood pressure; DBP, diastolic blood pressure; SBP, systolic blood pressure. For correlation plots: $r_{\mathrm{a}}^2$, aggregated coefficient of determination; $\hat{\rho}_{c,\mathrm{a}}$, aggregated coefficient of concordance; solid line, empirical linear regression line; dashed line, $45°$ line of perfect correlation. For LOA plots: solid line, mean of errors between estimated and true BP values; dashed lines, 2.5$^{\mathrm{th}}$ percentile (lower) and 97.5$^{\mathrm{th}}$ percentile (upper). For AE histogram plots: MAE and SDAE, mean and standard deviation of AE, respectively; $\mathcal{P}_5$, $\mathcal{P}_{10}$, and $\mathcal{P}_{15}$, cumulative percentage of estimations with AE within 5, 10, and 15 mm Hg, respectively. For fiducial histogram plots: $\mathcal{W}_{\mathrm{pred}}$, Wasserstein distance between true and estimated distribution.

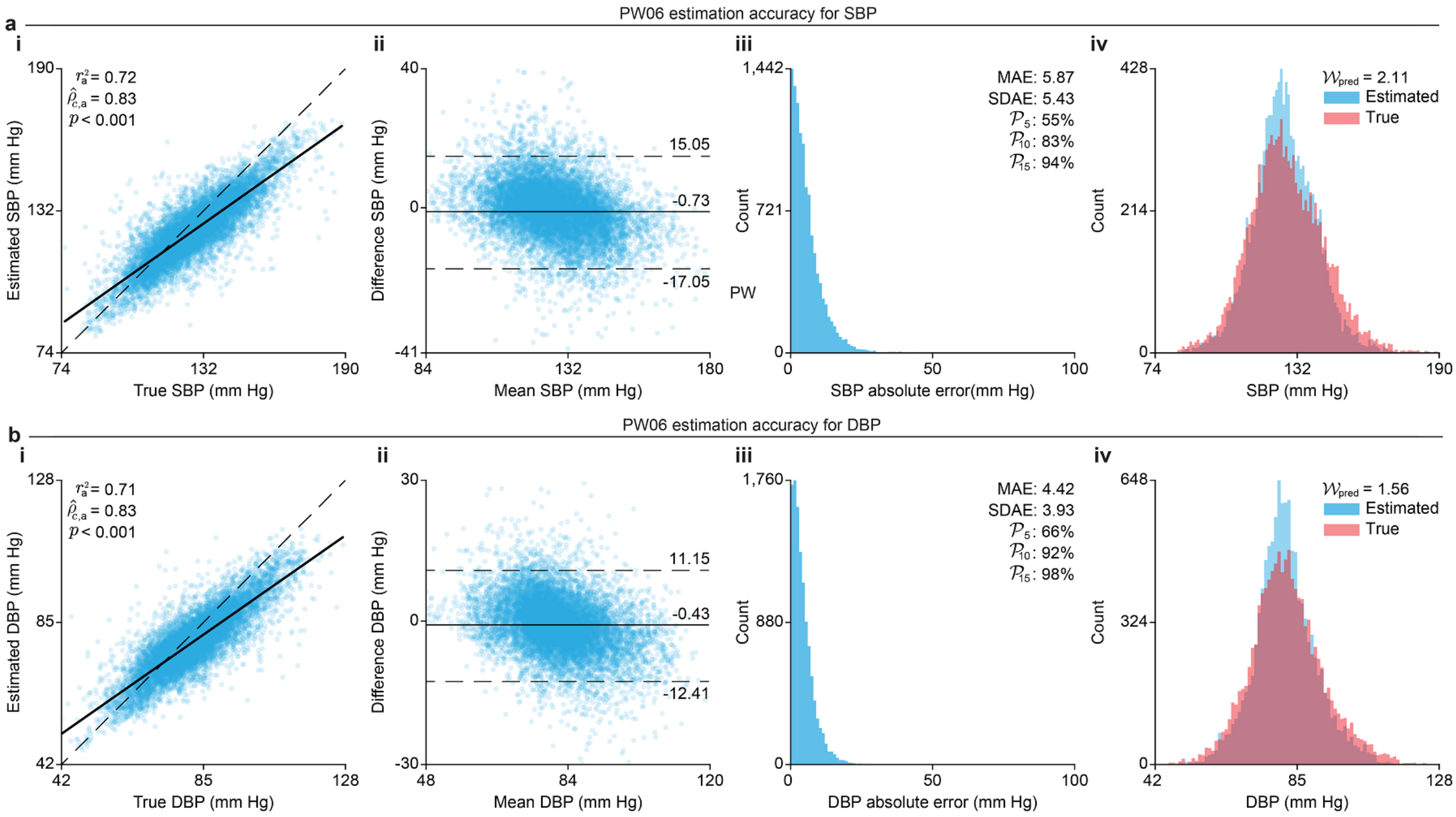

# Supplementary Fig. 59. Estimation results of population-within model PW07

Aggregated results from PW07 configuration: Multilayer Perceptron class with impedance input and waveform output, trained with the population-within (PW) partition. **a,** Estimation accuracy for systolic brachial blood pressure (SBP); **b,** Estimation accuracy for diastolic brachial blood pressure (DBP); **c,** Waveform ensemble of all estimated and true brachial blood pressure (BP) periods. For **a** and **b**: **i**, correlation plots; **ii**, limits of agreement (LOA) plots; **iii**, histogram of absolute errors (AE); and **iv**, histogram of estimated and true BP distributions. For **c**: **i**, ensemble of estimated BP periods; **ii**, ensemble of true BP periods. For correlation plots: $r^2_{\text{a}}$, aggregated coefficient of determination; $\hat{\rho}_{c,\text{a}}$, aggregated coefficient of concordance; solid line, empirical linear regression line; dashed line, $45°$ line of perfect correlation. For LOA plots: solid line, mean of errors between estimated and true BP values; dashed lines, 2.5$^{\text{th}}$ percentile (lower) and 97.5$^{\text{th}}$ percentile (upper). For AE histogram plots: MAE and SDAE, mean and standard deviation of AE, respectively; $\mathcal{P}_5$, $\mathcal{P}_{10}$, and $\mathcal{P}_{15}$, cumulative percentage of estimations with AE within 5, 10, and 15 mm Hg, respectively. For fiducial histogram plots: $\mathcal{W}_{\text{pred}}$, Wasserstein distance between true and estimated distribution. For ensemble plots: AMAE, average mean absolute error; ARMSE, average root mean square error; solid line, ensemble average of all periods; dashed lines, ensemble average $\pm$ standard deviation of all periods; scale bars, one-quarter of period.

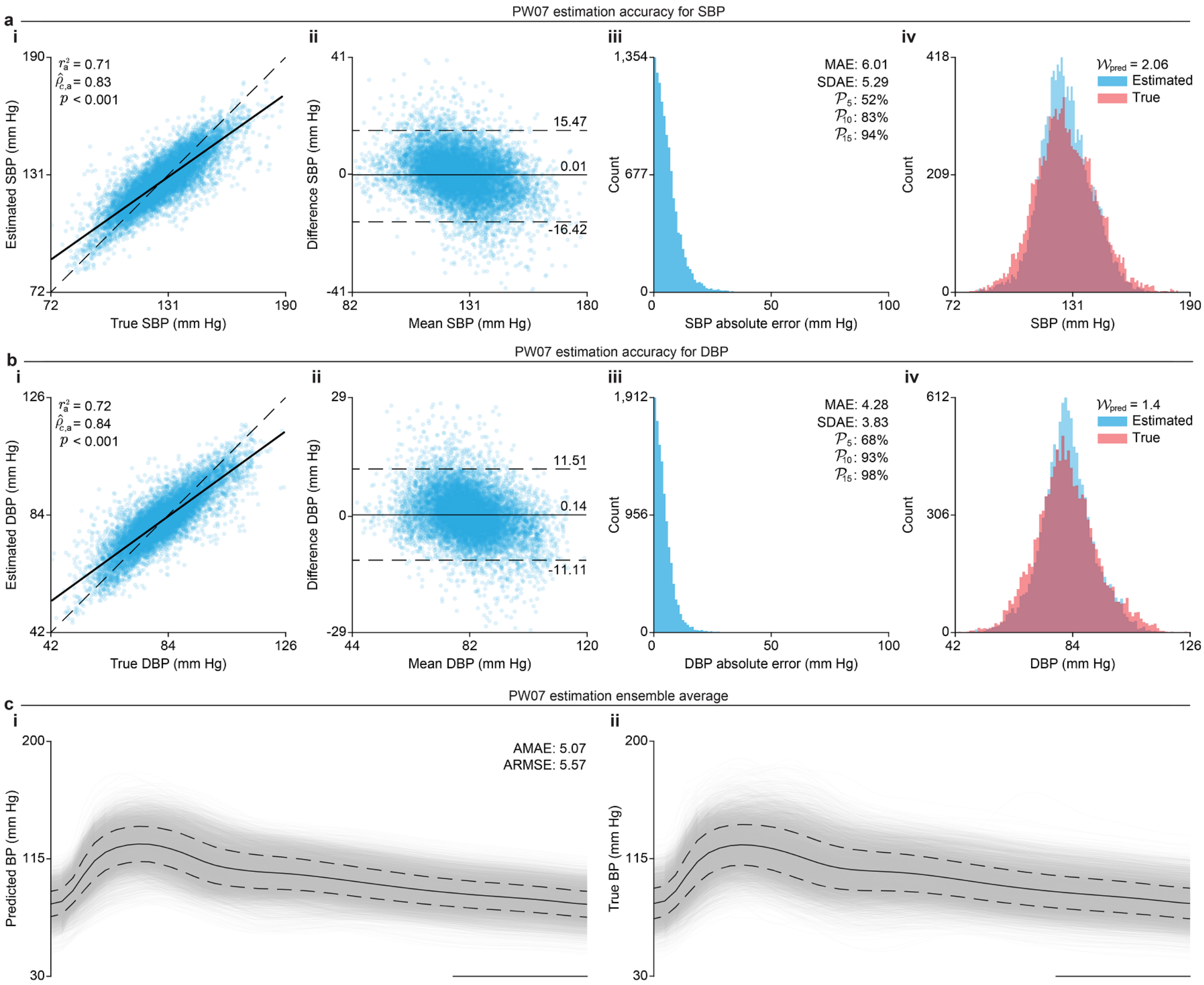

# Supplementary Fig. 60. Estimation results of population-within model PW08

Aggregated results from PW08 configuration: Multilayer Perceptron class with impedance input and fiducial output, trained with the population-within (PW) partition. **a,** Estimation accuracy for systolic brachial blood pressure (SBP); **b,** Estimation accuracy for diastolic brachial blood pressure (DBP); **i**, correlation plots; **ii**, limits of agreement (LOA) plots; **iii**, histogram of absolute errors (AE); and **iv**, histogram of estimated and true BP distributions. BP, blood pressure; DBP, diastolic blood pressure; SBP, systolic blood pressure. For correlation plots: $r^2_\mathrm{a}$, aggregated coefficient of determination; $\hat{\rho}_{c,\mathrm{a}}$, aggregated coefficient of concordance; solid line, empirical linear regression line; dashed line, $45°$ line of perfect correlation. For LOA plots: solid line, mean of errors between estimated and true BP values; dashed lines, 2.5$^{\mathrm{th}}$ percentile (lower) and 97.5$^{\mathrm{th}}$ percentile (upper). For AE histogram plots: MAE and SDAE, mean and standard deviation of AE, respectively; $\mathcal{P}_5$, $\mathcal{P}_{10}$, and $\mathcal{P}_{15}$, cumulative percentage of estimations with AE within 5, 10, and 15 mm Hg, respectively. For fiducial histogram plots: $\mathcal{W}_{\mathrm{pred}}$, Wasserstein distance between true and estimated distribution.

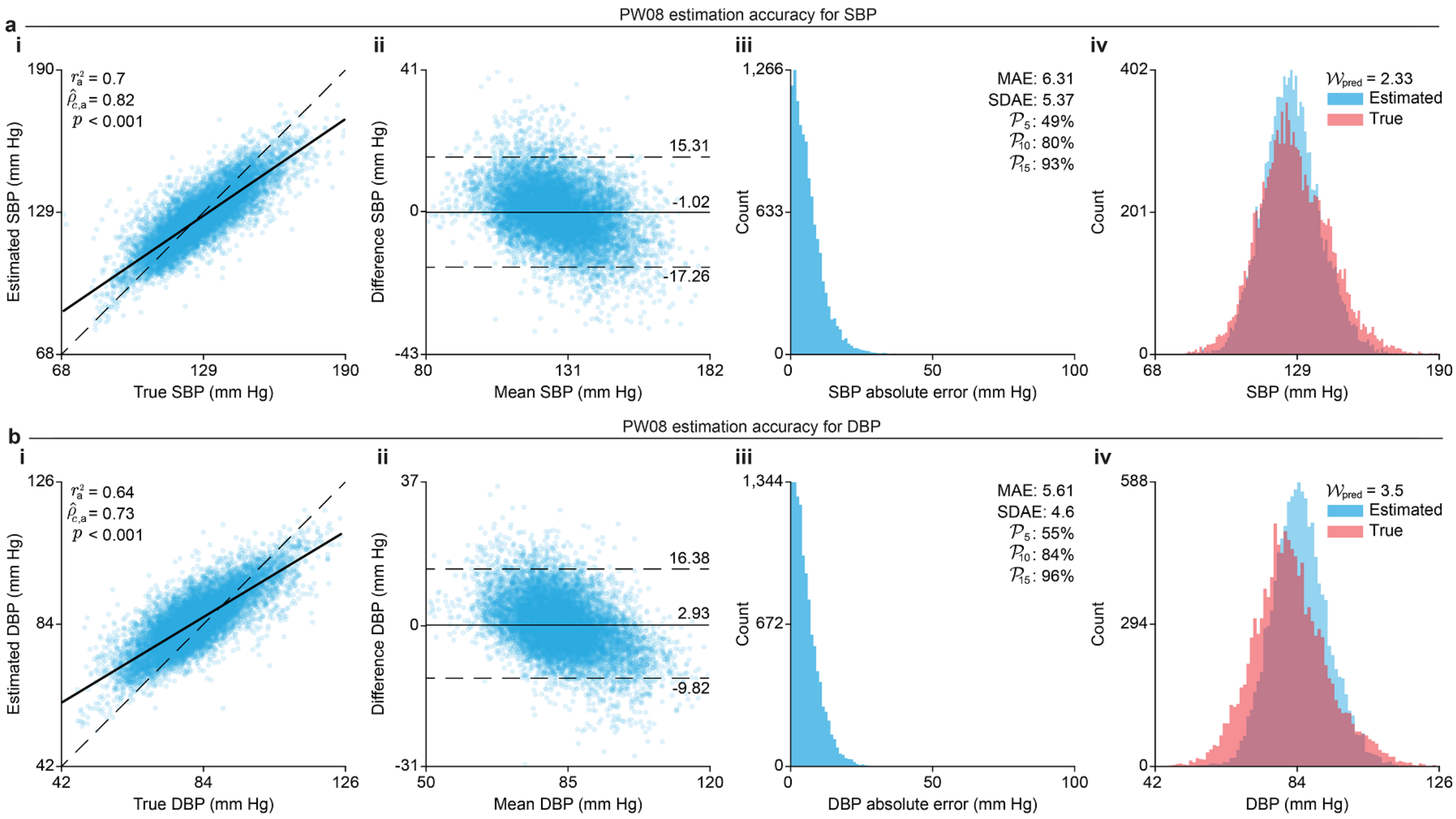

# Supplementary Fig. 61. Estimation results of population-within model PW09

Aggregated results from PW09 configuration: Convolutional Neural Network class with image input and waveform output, trained with the population-within (PW) partition. **a,** Estimation accuracy for systolic brachial blood pressure (SBP); **b,** Estimation accuracy for diastolic brachial blood pressure (DBP); **c,** Waveform ensemble of all estimated and true brachial blood pressure (BP) periods. For **a** and **b**: **i**, correlation plots; **ii**, limits of agreement (LOA) plots; **iii**, histogram of absolute errors (AE); and **iv**, histogram of estimated and true BP distributions. For **c**: **i**, ensemble of estimated BP periods; **ii**, ensemble of true BP periods. For correlation plots: $r_{\mathrm{a}}^2$, aggregated coefficient of determination; $\hat{\rho}_{c,\mathrm{a}}$, aggregated coefficient of concordance; solid line, empirical linear regression line; dashed line, $45°$ line of perfect correlation. For LOA plots: solid line, mean of errors between estimated and true BP values; dashed lines, 2.5th percentile (lower) and 97.5th percentile (upper). For AE histogram plots: MAE and SDAE, mean and standard deviation of AE, respectively; $\mathcal{P}_5$, $\mathcal{P}_{10}$, and $\mathcal{P}_{15}$, cumulative percentage of estimations with AE within 5, 10, and 15 mm Hg, respectively. For fiducial histogram plots: $\mathcal{W}_{\mathrm{pred}}$, Wasserstein distance between true and estimated distribution. For ensemble plots: AMAE, average mean absolute error; ARMSE, average root mean square error; solid line, ensemble average of all periods; dashed lines, ensemble average $\pm$ standard deviation of all periods; scale bars, one-quarter of period.

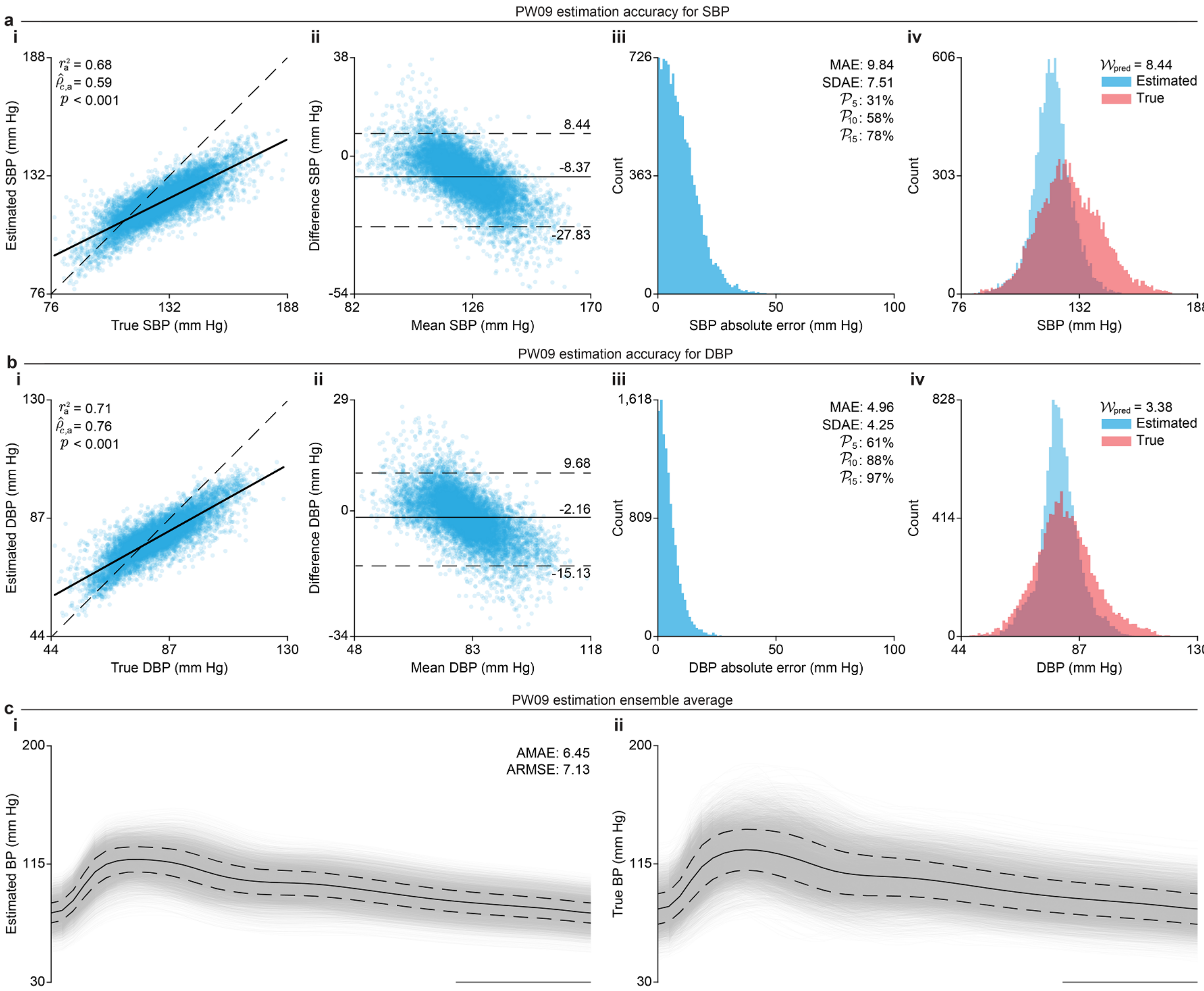

# Supplementary Fig. 62. Estimation results of population-within model PW10

Aggregated results from PW10 configuration: Convolutional Neural Network class with image input and fiducial output, trained with the population-within (PW) partition. **a,** Estimation accuracy for systolic brachial blood pressure (SBP); **b,** Estimation accuracy for diastolic brachial blood pressure (DBP); **i**, correlation plots; **ii**, limits of agreement (LOA) plots; **iii**, histogram of absolute errors (AE); and **iv**, histogram of estimated and true BP distributions. BP, blood pressure; DBP, diastolic blood pressure; SBP, systolic blood pressure. For correlation plots: $r^2_{\text{a}}$, aggregated coefficient of determination; $\hat{\rho}_{c,\text{a}}$, aggregated coefficient of concordance; solid line, empirical linear regression line; dashed line, $45°$ line of perfect correlation. For LOA plots: solid line, mean of errors between estimated and true BP values; dashed lines, 2.5$^{\text{th}}$ percentile (lower) and 97.5$^{\text{th}}$ percentile (upper). For AE histogram plots: MAE and SDAE, mean and standard deviation of AE, respectively; $\mathcal{P}_5$, $\mathcal{P}_{10}$, and $\mathcal{P}_{15}$, cumulative percentage of estimations with AE within 5, 10, and 15 mm Hg, respectively. For fiducial histogram plots: $\mathcal{W}_{\text{pred}}$, Wasserstein distance between true and estimated distribution.

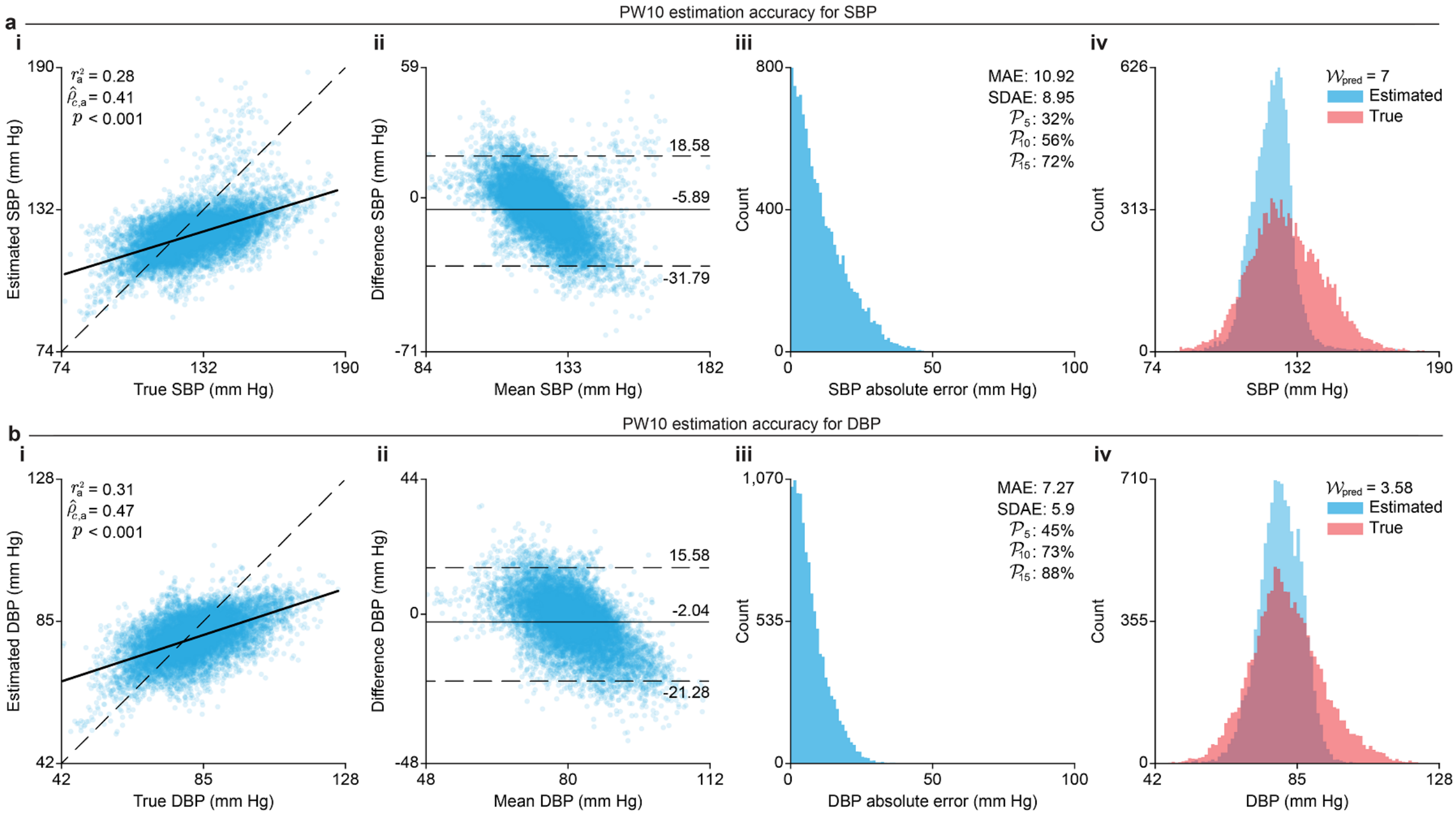

# Supplementary Fig. 63. Estimation results of population-within model PW11

Aggregated results from PW11 configuration: Convolutional Neural Network class with impedance input and waveform output, trained with the population-within (PW) partition. **a,** Estimation accuracy for systolic brachial blood pressure (SBP); **b,** Estimation accuracy for diastolic brachial blood pressure (DBP); **c,** Waveform ensemble of all estimated and true brachial blood pressure (BP) periods. For **a** and **b**: **i**, correlation plots; **ii**, limits of agreement (LOA) plots; **iii**, histogram of absolute errors (AE); and **iv**, histogram of estimated and true BP distributions. For **c**: **i**, ensemble of estimated BP periods; **ii**, ensemble of true BP periods. For correlation plots: $r_{\text{a}}^2$, aggregated coefficient of determination; $\hat{\rho}_{c,\text{a}}$, aggregated coefficient of concordance; solid line, empirical linear regression line; dashed line, $45°$ line of perfect correlation. For LOA plots: solid line, mean of errors between estimated and true BP values; dashed lines, 2.5$^{\text{th}}$ percentile (lower) and 97.5$^{\text{th}}$ percentile (upper). For AE histogram plots: MAE and SDAE, mean and standard deviation of AE, respectively; $\mathcal{P}_5$, $\mathcal{P}_{10}$, and $\mathcal{P}_{15}$, cumulative percentage of estimations with AE within 5, 10, and 15 mm Hg, respectively. For fiducial histogram plots: $\mathcal{W}_{\text{pred}}$, Wasserstein distance between true and estimated distribution. For ensemble plots: AMAE, average mean absolute error; ARMSE, average root mean square error; solid line, ensemble average of all periods; dashed lines, ensemble average $\pm$ standard deviation of all periods; scale bars, one-quarter of period.

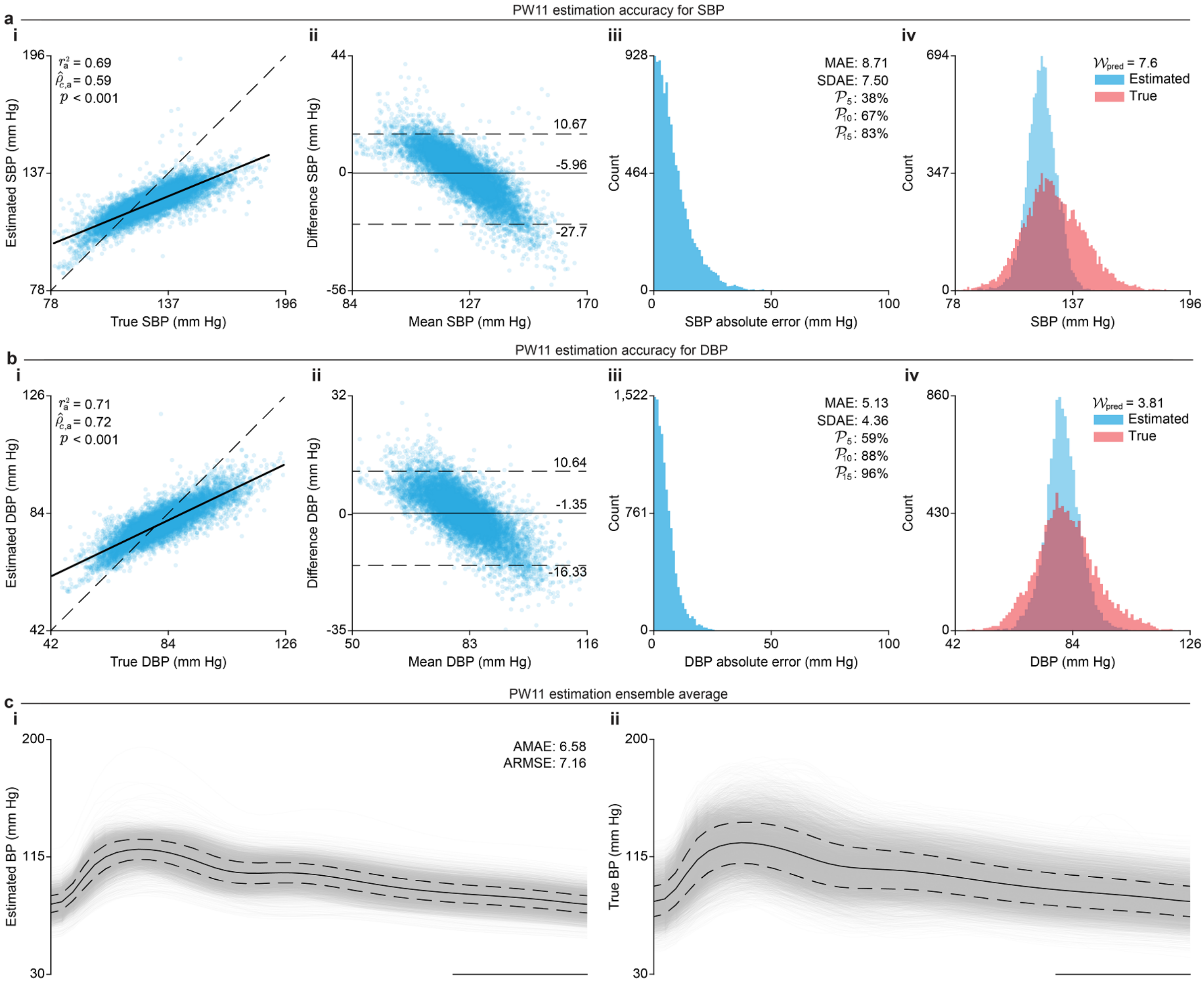

# Supplementary Fig. 64. Estimation results of population-within model PW12

Aggregated results from PW12 configuration: Convolutional Neural Network class with impedance input and fiducial output, trained with the population-within (PW) partition. **a,** Estimation accuracy for systolic brachial blood pressure (SBP); **b,** Estimation accuracy for diastolic brachial blood pressure (DBP); **i**, correlation plots; **ii**, limits of agreement (LOA) plots; **iii**, histogram of absolute errors (AE); and **iv**, histogram of estimated and true BP distributions. BP, blood pressure; DBP, diastolic blood pressure; SBP, systolic blood pressure. For correlation plots: $r_{\text{a}}^2$, aggregated coefficient of determination; $\hat{\rho}_{c,\text{a}}$, aggregated coefficient of concordance; solid line, empirical linear regression line; dashed line, $45°$ line of perfect correlation. For LOA plots: solid line, mean of errors between estimated and true BP values; dashed lines, 2.5$^{\text{th}}$ percentile (lower) and 97.5$^{\text{th}}$ percentile (upper). For AE histogram plots: MAE and SDAE, mean and standard deviation of AE, respectively; $\mathcal{P}_5$, $\mathcal{P}_{10}$, and $\mathcal{P}_{15}$, cumulative percentage of estimations with AE within 5, 10, and 15 mm Hg, respectively. For fiducial histogram plots: $\mathcal{W}_{\text{pred}}$, Wasserstein distance between true and estimated distribution.

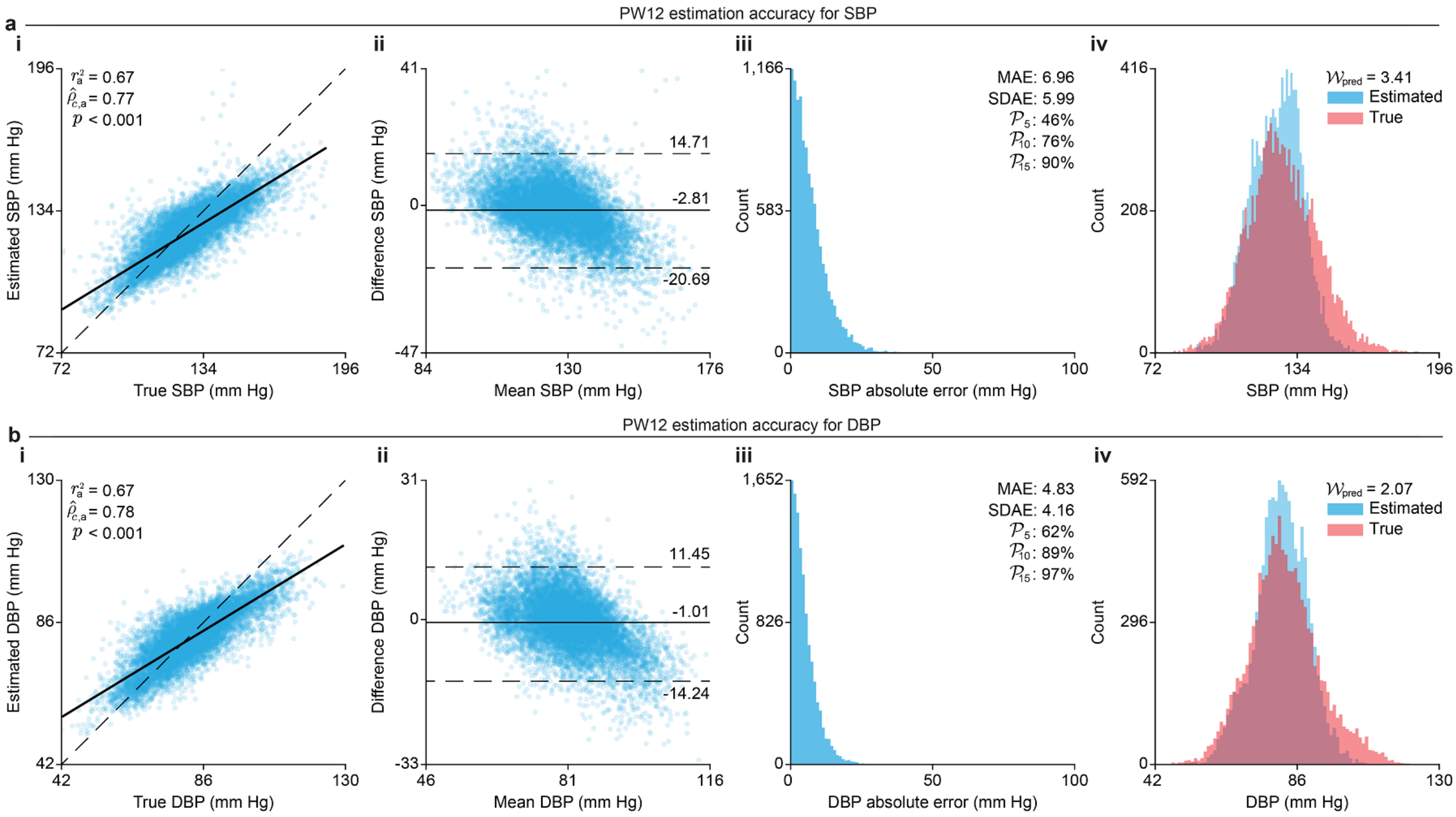

# Supplementary Fig. 65. Estimation results of population-within model PW13

Aggregated results from PW13 configuration: Convolutional Recurrent Transformer class with image input and waveform output, trained with the population-within (PW) partition. **a,** Estimation accuracy for systolic brachial blood pressure (SBP); **b,** Estimation accuracy for diastolic brachial blood pressure (DBP); **c,** Waveform ensemble of all estimated and true brachial blood pressure (BP) periods. For **a** and **b**: **i**, correlation plots; **ii**, limits of agreement (LOA) plots; **iii**, histogram of absolute errors (AE); and **iv**, histogram of estimated and true BP distributions. For **c**: **i**, ensemble of estimated BP periods; **ii**, ensemble of true BP periods. For correlation plots: $r^2_{\text{a}}$, aggregated coefficient of determination; $\hat{\rho}_{c,\text{a}}$, aggregated coefficient of concordance; solid line, empirical linear regression line; dashed line, $45°$ line of perfect correlation. For LOA plots: solid line, mean of errors between estimated and true BP values; dashed lines, 2.5th percentile (lower) and 97.5th percentile (upper). For AE histogram plots: MAE and SDAE, mean and standard deviation of AE, respectively; $\mathcal{P}_5$, $\mathcal{P}_{10}$, and $\mathcal{P}_{15}$, cumulative percentage of estimations with AE within 5, 10, and 15 mm Hg, respectively. For fiducial histogram plots: $\mathcal{W}_{\text{pred}}$, Wasserstein distance between true and estimated distribution. For ensemble plots: AMAE, average mean absolute error; ARMSE, average root mean square error; solid line, ensemble average of all periods; dashed lines, ensemble average $\pm$ standard deviation of all periods; scale bars, one-quarter of period.

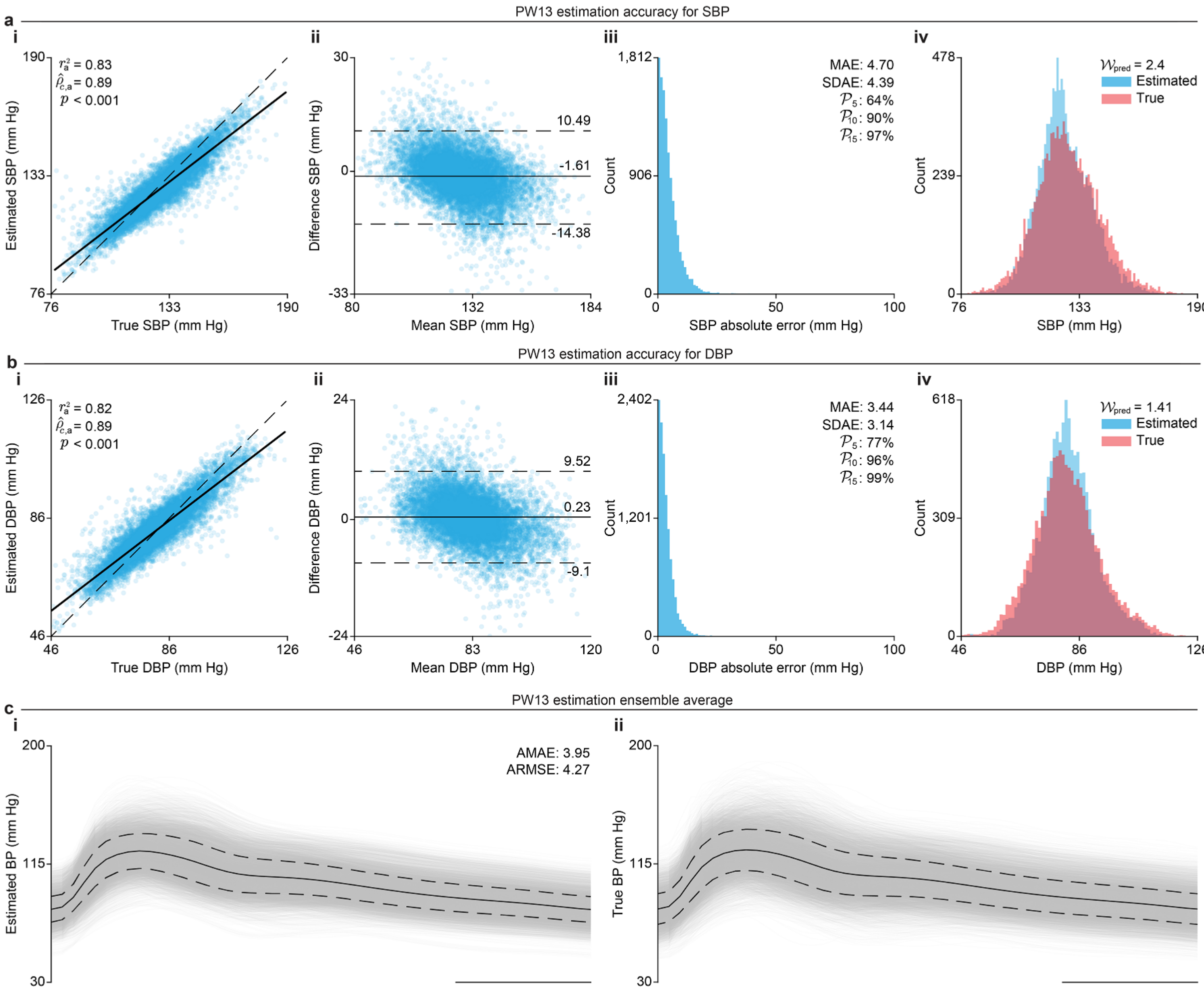

# Supplementary Fig. 66. Estimation results of population-within model PW14

Aggregated results from PW14 configuration: Convolutional Recurrent Transformer class with image input and fiducial output, trained with the population-within (PW) partition. **a,** Estimation accuracy for systolic brachial blood pressure (SBP); **b,** Estimation accuracy for diastolic brachial blood pressure (DBP); **i**, correlation plots; **ii**, limits of agreement (LOA) plots; **iii**, histogram of absolute errors (AE); and **iv**, histogram of estimated and true BP distributions. BP, blood pressure; DBP, diastolic blood pressure; SBP, systolic blood pressure. For correlation plots: $r_{\text{a}}^2$, aggregated coefficient of determination; $\hat{\rho}_{c,\text{a}}$, aggregated coefficient of concordance; solid line, empirical linear regression line; dashed line, $45°$ line of perfect correlation. For LOA plots: solid line, mean of errors between estimated and true BP values; dashed lines, 2.5th percentile (lower) and 97.5th percentile (upper). For AE histogram plots: MAE and SDAE, mean and standard deviation of AE, respectively; $\mathcal{P}_5$, $\mathcal{P}_{10}$, and $\mathcal{P}_{15}$, cumulative percentage of estimations with AE within 5, 10, and 15 mm Hg, respectively. For fiducial histogram plots: $\mathcal{W}_{\text{pred}}$, Wasserstein distance between true and estimated distribution.

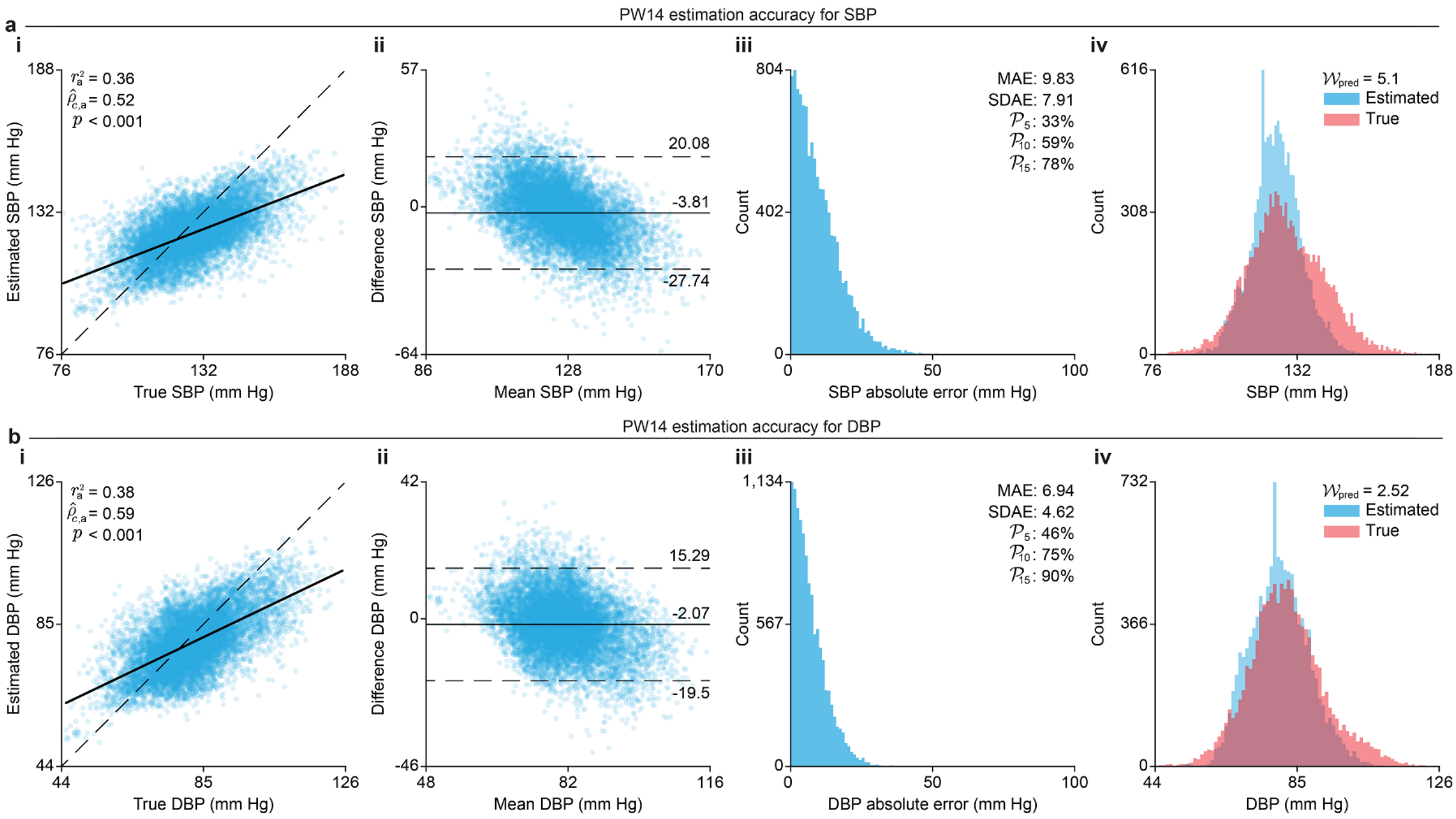

# Supplementary Fig. 67. Estimation results of population-within model PW15

Aggregated results from PW15 configuration: Convolutional Recurrent Transformer class with impedance input and waveform output, trained with the population-within (PW) partition. **a,** Estimation accuracy for systolic brachial blood pressure (SBP); **b,** Estimation accuracy for diastolic brachial blood pressure (DBP); **c,** Waveform ensemble of all estimated and true brachial blood pressure (BP) periods. For **a** and **b**: **i**, correlation plots; **ii**, limits of agreement (LOA) plots; **iii**, histogram of absolute errors (AE); and **iv**, histogram of estimated and true BP distributions. For **c**: **i**, ensemble of estimated BP periods; **ii**, ensemble of true BP periods. For correlation plots: $r^2_{\text{a}}$, aggregated coefficient of determination; $\hat{\rho}_{c,\text{a}}$, aggregated coefficient of concordance; solid line, empirical linear regression line; dashed line, $45°$ line of perfect correlation. For LOA plots: solid line, mean of errors between estimated and true BP values; dashed lines, 2.5$^{\text{th}}$ percentile (lower) and 97.5$^{\text{th}}$ percentile (upper). For AE histogram plots: MAE and SDAE, mean and standard deviation of AE, respectively; $\mathcal{P}_5$, $\mathcal{P}_{10}$, and $\mathcal{P}_{15}$, cumulative percentage of estimations with AE within 5, 10, and 15 mm Hg, respectively. For fiducial histogram plots: $\mathcal{W}_{\text{pred}}$, Wasserstein distance between true and estimated distribution. For ensemble plots: AMAE, average mean absolute error; ARMSE, average root mean square error; solid line, ensemble average of all periods; dashed lines, ensemble average $\pm$ standard deviation of all periods; scale bars, one-quarter of period.

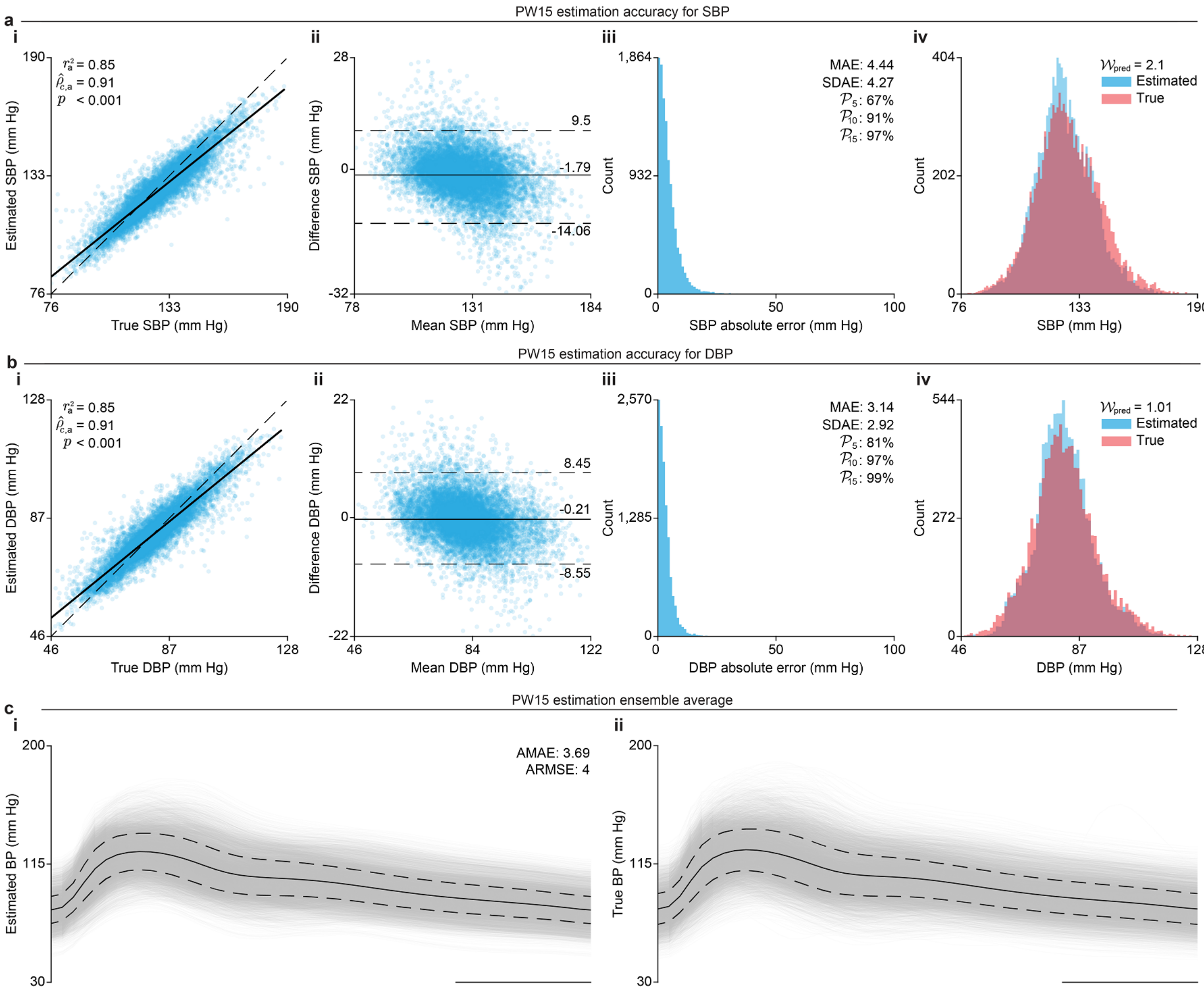

## Supplementary Fig. 68. Estimation results of population-within model PW16

Aggregated results from PW16 configuration: Convolutional Recurrent Transformer class with impedance input and fiducial output, trained with the population-within (PW) partition. **a,** Estimation accuracy for systolic brachial blood pressure (SBP); **b,** Estimation accuracy for diastolic brachial blood pressure (DBP); **i**, correlation plots; **ii**, limits of agreement (LOA) plots; **iii**, histogram of absolute errors (AE); and **iv**, histogram of estimated and true BP distributions. BP, blood pressure; DBP, diastolic blood pressure; SBP, systolic blood pressure. For correlation plots: $r^2_{\text{a}}$, aggregated coefficient of determination; $\hat{\rho}_{c,\text{a}}$, aggregated coefficient of concordance; solid line, empirical linear regression line; dashed line, $45°$ line of perfect correlation. For LOA plots: solid line, mean of errors between estimated and true BP values; dashed lines, 2.5th percentile (lower) and 97.5th percentile (upper). For AE histogram plots: MAE and SDAE, mean and standard deviation of AE, respectively; $\mathcal{P}_5$, $\mathcal{P}_{10}$, and $\mathcal{P}_{15}$, cumulative percentage of estimations with AE within 5, 10, and 15 mm Hg, respectively. For fiducial histogram plots: $\mathcal{W}_{\text{pred}}$, Wasserstein distance between true and estimated distribution.

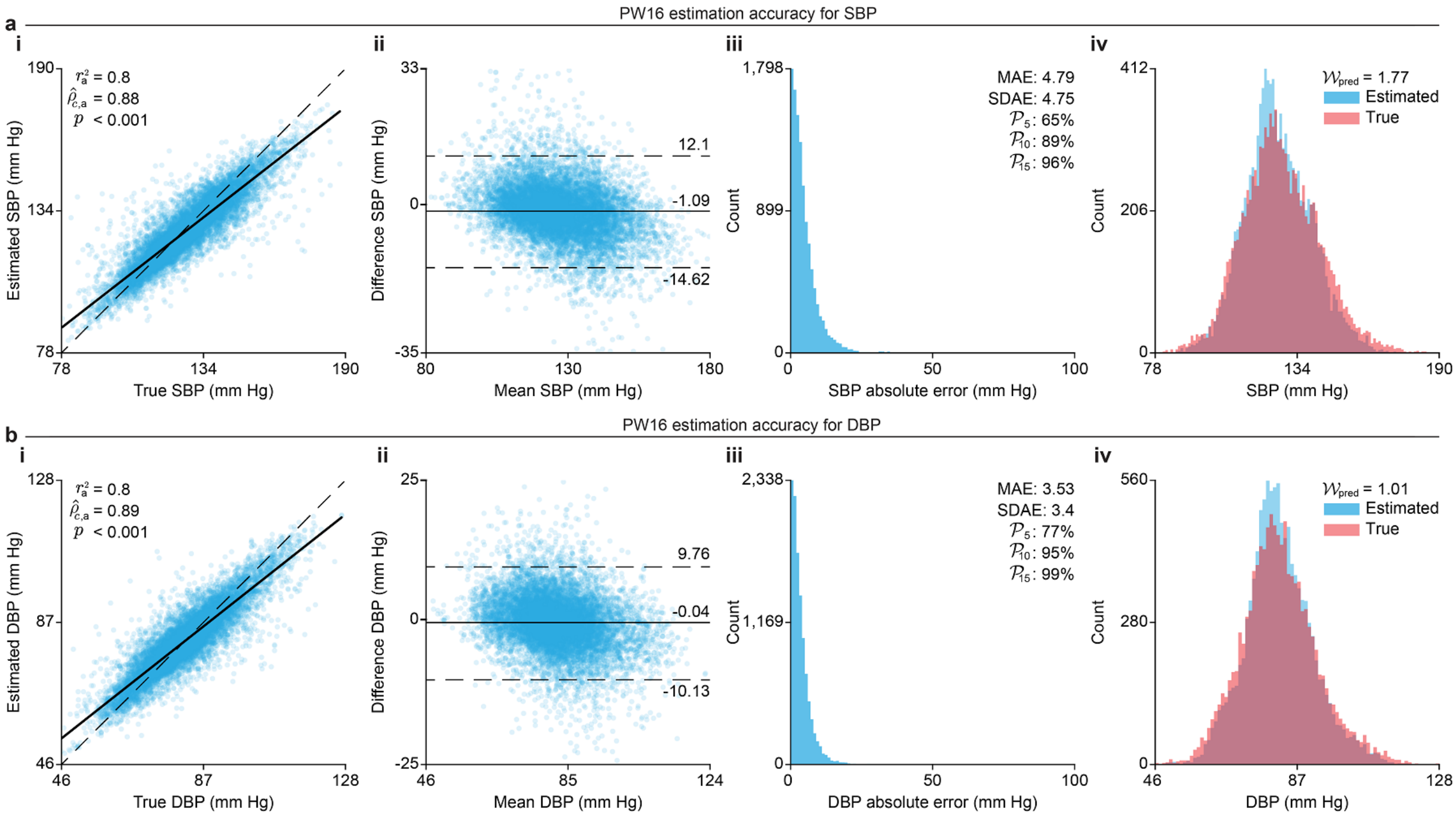

# Supplementary Fig. 69. Estimation results of population-within model PW17

Aggregated results from PW17 configuration: Convolutional Recurrent Samba class with image input and waveform output, trained with the population-within (PW) partition. **a,** Estimation accuracy for systolic brachial blood pressure (SBP); **b,** Estimation accuracy for diastolic brachial blood pressure (DBP); **c,** Waveform ensemble of all estimated and true brachial blood pressure (BP) periods. For **a** and **b**: **i**, correlation plots; **ii**, limits of agreement (LOA) plots; **iii**, histogram of absolute errors (AE); and **iv**, histogram of estimated and true BP distributions. For **c**: **i**, ensemble of estimated BP periods; **ii**, ensemble of true BP periods. For correlation plots: $r_{\mathrm{a}}^2$, aggregated coefficient of determination; $\hat{\rho}_{c,\mathrm{a}}$, aggregated coefficient of concordance; solid line, empirical linear regression line; dashed line, $45°$ line of perfect correlation. For LOA plots: solid line, mean of errors between estimated and true BP values; dashed lines, 2.5$^{\text{th}}$ percentile (lower) and 97.5$^{\text{th}}$ percentile (upper). For AE histogram plots: MAE and SDAE, mean and standard deviation of AE, respectively; $\mathcal{P}_5$, $\mathcal{P}_{10}$, and $\mathcal{P}_{15}$, cumulative percentage of estimations with AE within 5, 10, and 15 mm Hg, respectively. For fiducial histogram plots: $\mathcal{W}_{\text{pred}}$, Wasserstein distance between true and estimated distribution. For ensemble plots: AMAE, average mean absolute error; ARMSE, average root mean square error; solid line, ensemble average of all periods; dashed lines, ensemble average $\pm$ standard deviation of all periods; scale bars, one-quarter of period.

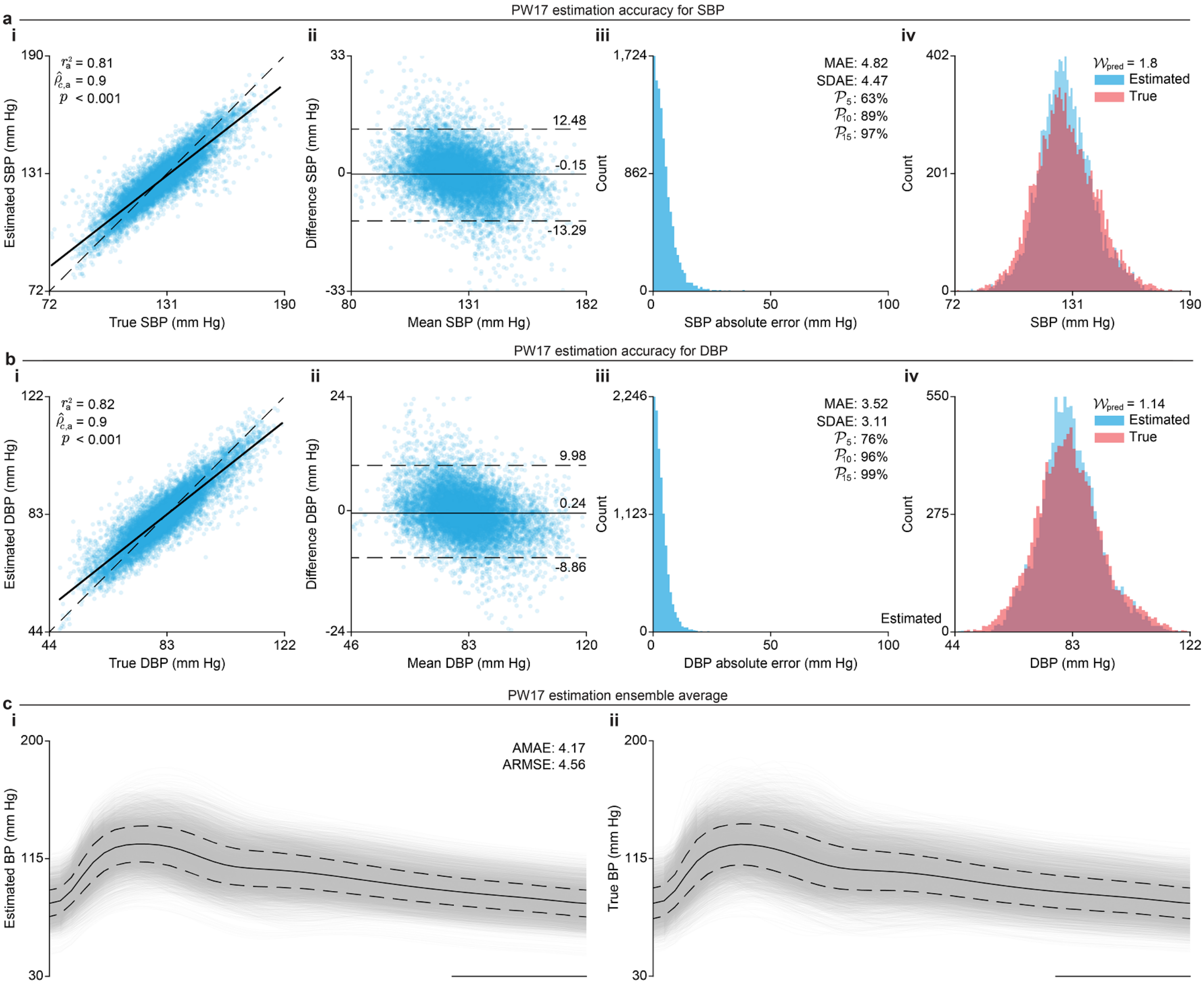

# Supplementary Fig. 70. Estimation results of population-within model PW18

Aggregated results from PW18 configuration: Convolutional Recurrent Samba class with image input and fiducial output, trained with the population-within (PW) partition. **a,** Estimation accuracy for systolic brachial blood pressure (SBP); **b,** Estimation accuracy for diastolic brachial blood pressure (DBP); **i**, correlation plots; **ii**, limits of agreement (LOA) plots; **iii**, histogram of absolute errors (AE); and **iv**, histogram of estimated and true BP distributions. BP, blood pressure; DBP, diastolic blood pressure; SBP, systolic blood pressure. For correlation plots: $r_{\text{a}}^2$, aggregated coefficient of determination; $\hat{\rho}_{c,\text{a}}$, aggregated coefficient of concordance; solid line, empirical linear regression line; dashed line, $45°$ line of perfect correlation. For LOA plots: solid line, mean of errors between estimated and true BP values; dashed lines, 2.5th percentile (lower) and 97.5th percentile (upper). For AE histogram plots: MAE and SDAE, mean and standard deviation of AE, respectively; $\mathcal{P}_5$, $\mathcal{P}_{10}$, and $\mathcal{P}_{15}$, cumulative percentage of estimations with AE within 5, 10, and 15 mm Hg, respectively. For fiducial histogram plots: $\mathcal{W}_{\text{pred}}$, Wasserstein distance between true and estimated distribution.

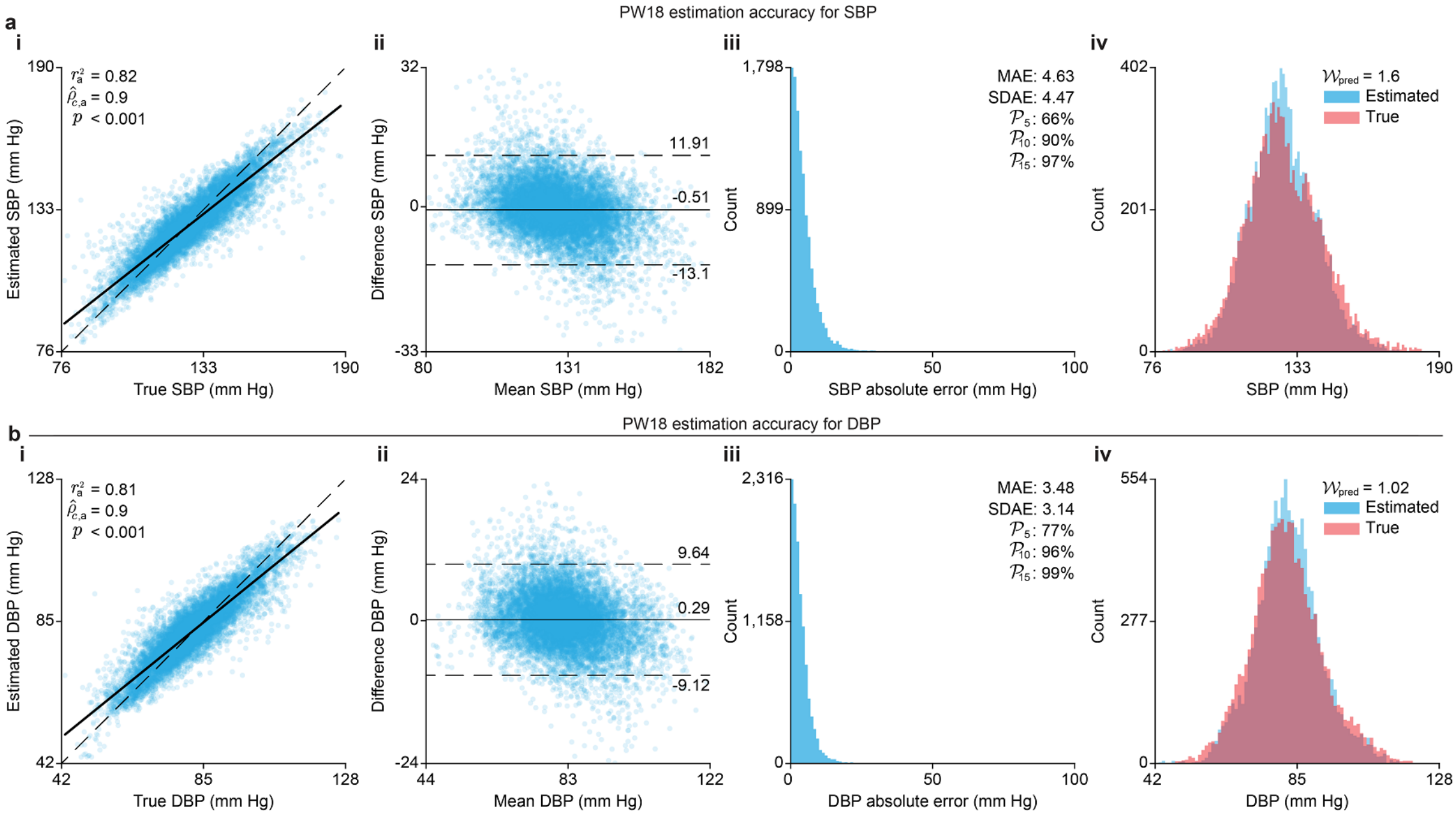

# Supplementary Fig. 71. Estimation results of population-within model PW19

Aggregated results from PW19 configuration: Convolutional Recurrent Samba class with impedance input and waveform output, trained with the population-within (PW) partition. **a,** Estimation accuracy for systolic brachial blood pressure (SBP); **b,** Estimation accuracy for diastolic brachial blood pressure (DBP); **c,** Waveform ensemble of all estimated and true brachial blood pressure (BP) periods. For **a** and **b**: **i**, correlation plots; **ii**, limits of agreement (LOA) plots; **iii**, histogram of absolute errors (AE); and **iv**, histogram of estimated and true BP distributions. For **c**: **i**, ensemble of estimated BP periods; **ii**, ensemble of true BP periods. For correlation plots: $r^2_{\mathrm{a}}$, aggregated coefficient of determination; $\hat{\rho}_{c,\mathrm{a}}$, aggregated coefficient of concordance; solid line, empirical linear regression line; dashed line, $45°$ line of perfect correlation. For LOA plots: solid line, mean of errors between estimated and true BP values; dashed lines, 2.5th percentile (lower) and 97.5th percentile (upper). For AE histogram plots: MAE and SDAE, mean and standard deviation of AE, respectively; $\mathcal{P}_5$, $\mathcal{P}_{10}$, and $\mathcal{P}_{15}$, cumulative percentage of estimations with AE within 5, 10, and 15 mm Hg, respectively. For fiducial histogram plots: $\mathcal{W}_{\mathrm{pred}}$, Wasserstein distance between true and estimated distribution. For ensemble plots: AMAE, average mean absolute error; ARMSE, average root mean square error; solid line, ensemble average of all periods; dashed lines, ensemble average $\pm$ standard deviation of all periods; scale bars, one-quarter of period.

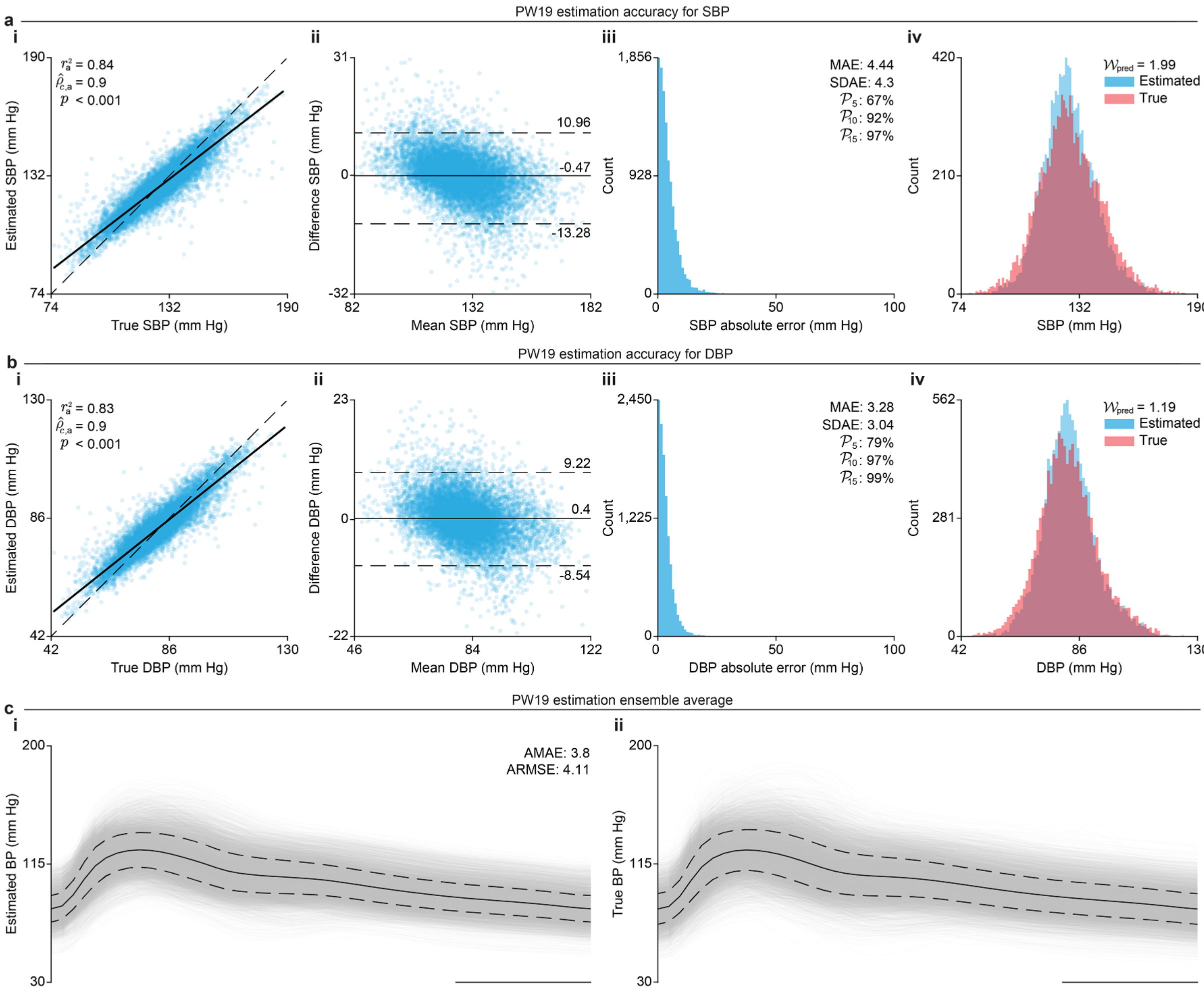

# Supplementary Fig. 72. Estimation results of population-within model PW20

Aggregated results from PW20 configuration: Convolutional Recurrent Samba class with impedance input and fiducial output, trained with the population-within (PW) partition. **a,** Estimation accuracy for systolic brachial blood pressure (SBP); **b,** Estimation accuracy for diastolic brachial blood pressure (DBP); **i**, correlation plots; **ii**, limits of agreement (LOA) plots; **iii**, histogram of absolute errors (AE); and **iv**, histogram of estimated and true BP distributions. BP, blood pressure; DBP, diastolic blood pressure; SBP, systolic blood pressure. For correlation plots: $r^2_{\text{a}}$, aggregated coefficient of determination; $\hat{\rho}_{c,\text{a}}$, aggregated coefficient of concordance; solid line, empirical linear regression line; dashed line, $45°$ line of perfect correlation. For LOA plots: solid line, mean of errors between estimated and true BP values; dashed lines, 2.5$^{\text{th}}$ percentile (lower) and 97.5$^{\text{th}}$ percentile (upper). For AE histogram plots: MAE and SDAE, mean and standard deviation of AE, respectively; $\mathcal{P}_5$, $\mathcal{P}_{10}$, and $\mathcal{P}_{15}$, cumulative percentage of estimations with AE within 5, 10, and 15 mm Hg, respectively. For fiducial histogram plots: $\mathcal{W}_{\text{pred}}$, Wasserstein distance between true and estimated distribution.

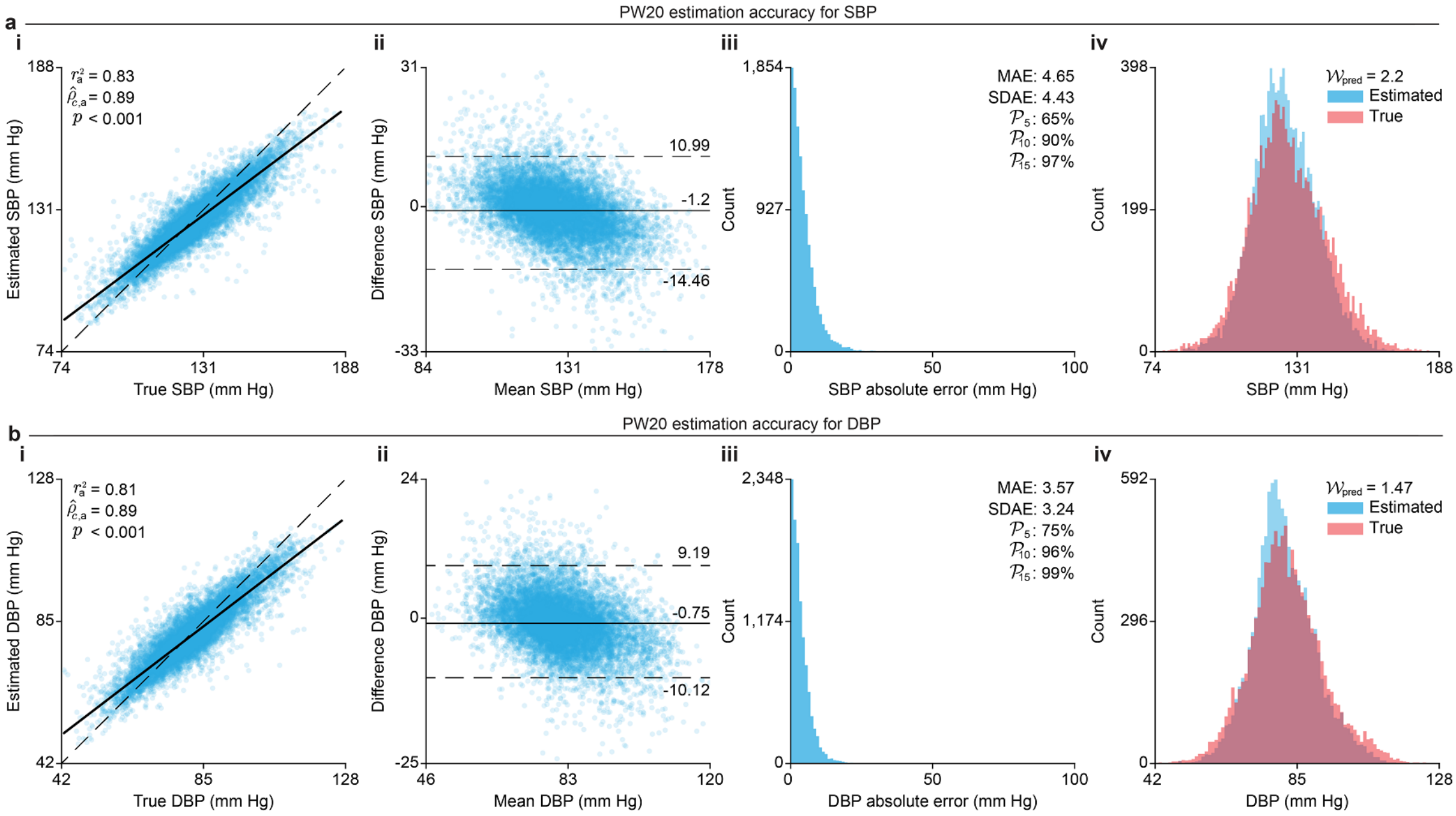

# Supplementary Fig. 73. Generalizability of population-within model PW01

Aggregated results from PW01 configuration: Linear Regression class with image input and waveform output, trained with the population-within (PW) partition, and inferred on the holdout datasets. **a,** Estimation accuracy for systolic brachial blood pressure (SBP); **b,** Estimation accuracy for diastolic brachial blood pressure (DBP); **c,** Waveform ensemble of all estimated and true brachial blood pressure (BP) periods. For **a** and **b**: **i**, correlation plots; **ii**, limits of agreement (LOA) plots; **iii**, histogram of absolute errors (AE); and **iv**, histogram of estimated and true BP distributions. For **c**: **i**, ensemble of estimated BP periods; **ii**, ensemble of true BP periods. For correlation plots: $r_\mathrm{a}^2$, aggregated coefficient of determination; $\hat{\rho}_{c,\mathrm{a}}$, aggregated coefficient of concordance; solid line, empirical linear regression line; dashed line, $45°$ line of perfect correlation. For LOA plots: solid line, mean of errors between estimated and true BP values; dashed lines, 2.5$^{\mathrm{th}}$ percentile (lower) and 97.5$^{\mathrm{th}}$ percentile (upper). For AE histogram plots: MAE and SDAE, mean and standard deviation of AE, respectively; $\mathcal{P}_5$, $\mathcal{P}_{10}$, and $\mathcal{P}_{15}$, cumulative percentage of estimations with AE within 5, 10, and 15 mm Hg, respectively. For fiducial histogram plots: $\mathcal{W}_\mathrm{pred}$, Wasserstein distance between true and estimated distribution. For ensemble plots: AMAE, average mean absolute error; ARMSE, average root mean square error; solid line, ensemble average of all periods; dashed lines, ensemble average $\pm$ standard deviation of all periods; scale bars, one-quarter of period.

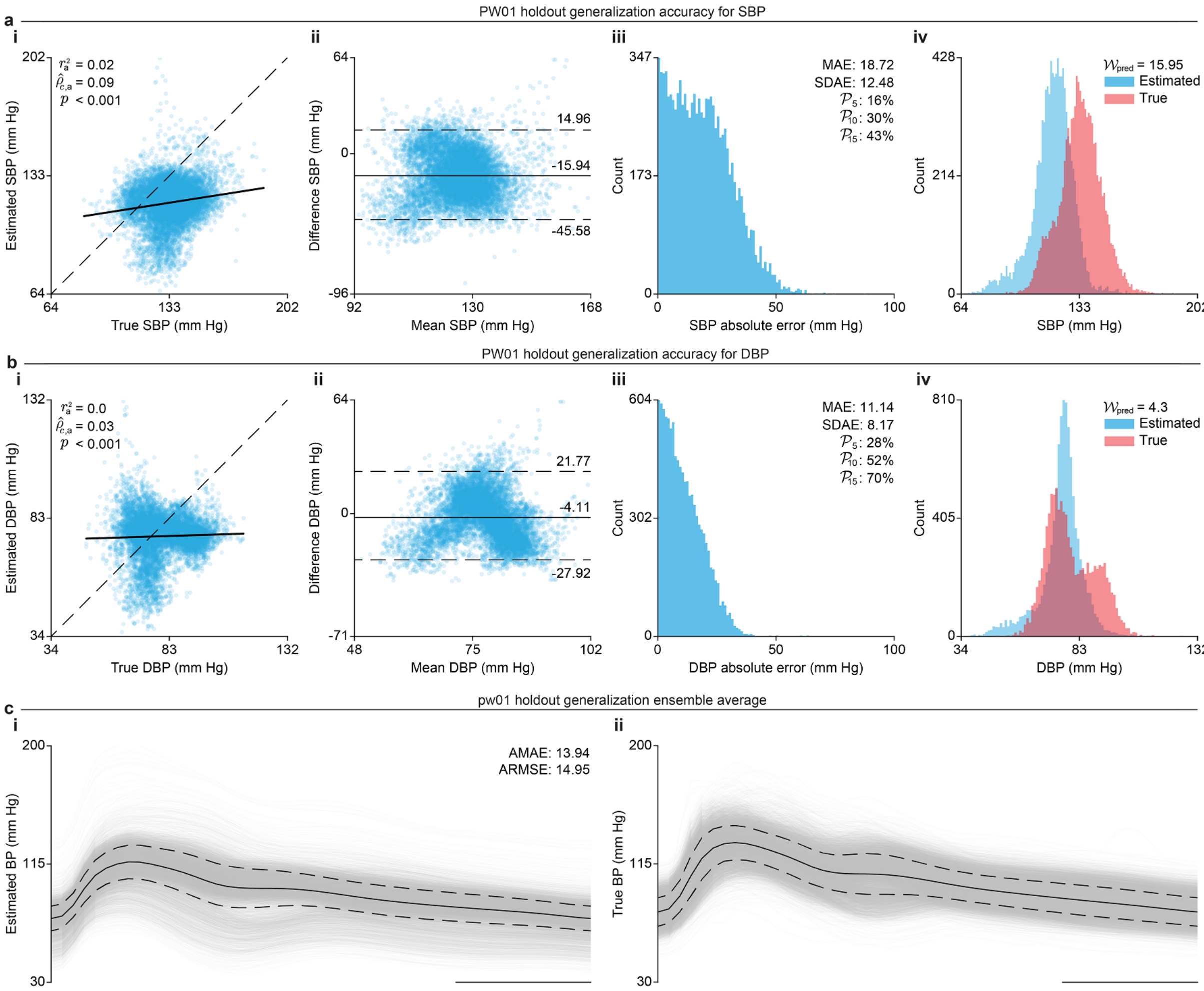

# Supplementary Fig. 74. Generalizability of population-within model PW02

Aggregated results from PW02 configuration: Linear Regression class with image input and fiducial output, trained with the population-within (PW) partition, and inferred on the holdout datasets. **a,** Estimation accuracy for systolic brachial blood pressure (SBP); **b,** Estimation accuracy for diastolic brachial blood pressure (DBP); **i**, correlation plots; **ii**, limits of agreement (LOA) plots; **iii**, histogram of absolute errors (AE); and **iv**, histogram of estimated and true BP distributions. BP, blood pressure; DBP, diastolic blood pressure; SBP, systolic blood pressure. For correlation plots: $r_{\text{a}}^2$, aggregated coefficient of determination; $\hat{\rho}_{c,\text{a}}$, aggregated coefficient of concordance; solid line, empirical linear regression line; dashed line, $45°$ line of perfect correlation. For LOA plots: solid line, mean of errors between estimated and true BP values; dashed lines, 2.5$^{\text{th}}$ percentile (lower) and 97.5$^{\text{th}}$ percentile (upper). For AE histogram plots: MAE and SDAE, mean and standard deviation of AE, respectively; $\mathcal{P}_5$, $\mathcal{P}_{10}$, and $\mathcal{P}_{15}$, cumulative percentage of estimations with AE within 5, 10, and 15 mm Hg, respectively. For fiducial histogram plots: $\mathcal{W}_{\text{pred}}$, Wasserstein distance between true and estimated distribution.

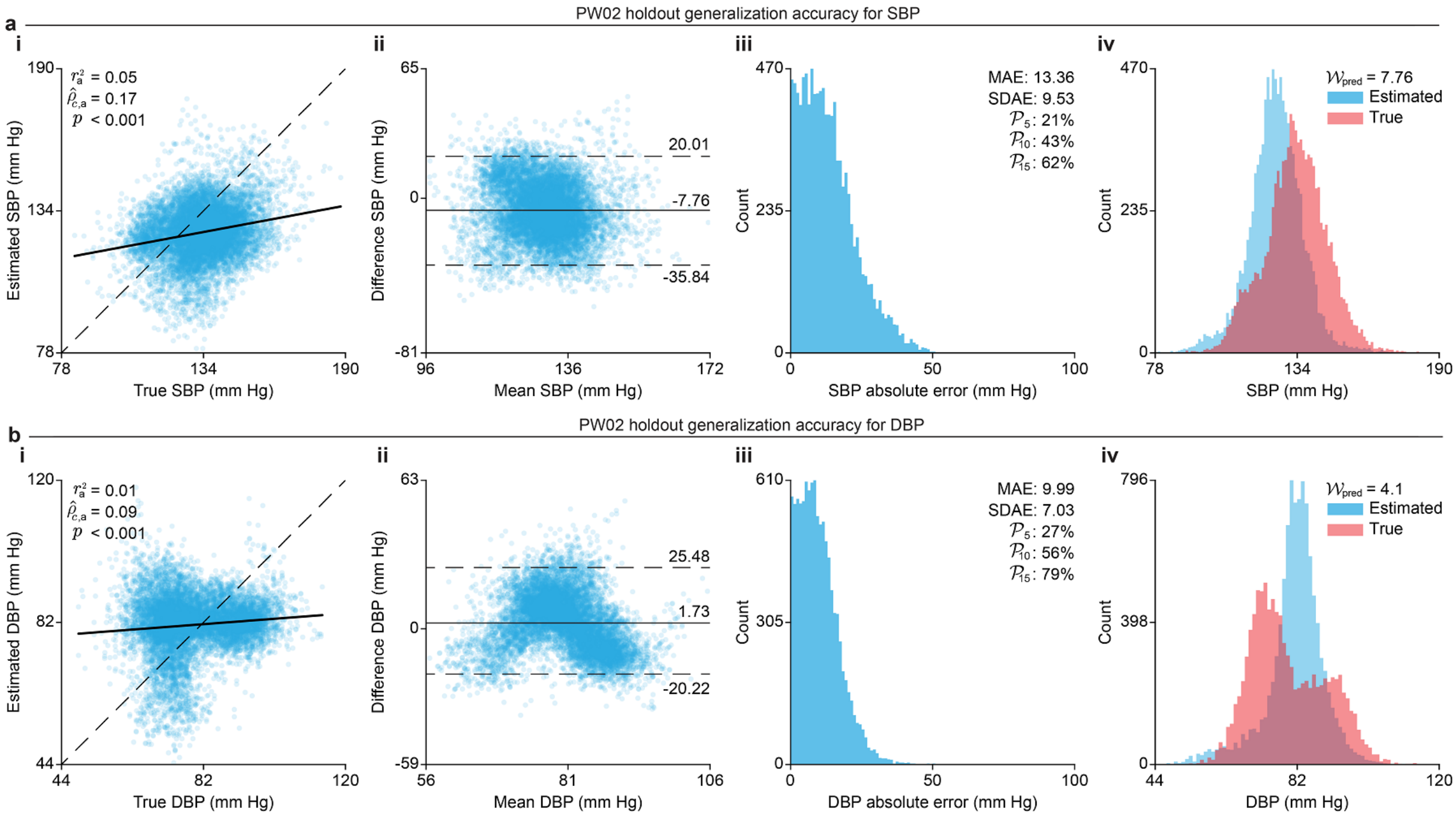

# Supplementary Fig. 75. Generalizability of population-within model PW03

Aggregated results from PW03 configuration: Linear Regression class with impedance input and waveform output, trained with the population-within (PW) partition, and inferred on the holdout datasets. **a,** Estimation accuracy for systolic brachial blood pressure (SBP); **b,** Estimation accuracy for diastolic brachial blood pressure (DBP); **c,** Waveform ensemble of all estimated and true brachial blood pressure (BP) periods. For **a** and **b**: **i**, correlation plots; **ii**, limits of agreement (LOA) plots; **iii**, histogram of absolute errors (AE); and **iv**, histogram of estimated and true BP distributions. For **c**: **i**, ensemble of estimated BP periods; **ii**, ensemble of true BP periods. For correlation plots: $r_{\text{a}}^2$, aggregated coefficient of determination; $\hat{\rho}_{c,\text{a}}$, aggregated coefficient of concordance; solid line, empirical linear regression line; dashed line, $45°$ line of perfect correlation. For LOA plots: solid line, mean of errors between estimated and true BP values; dashed lines, 2.5$^{\text{th}}$ percentile (lower) and 97.5$^{\text{th}}$ percentile (upper). For AE histogram plots: MAE and SDAE, mean and standard deviation of AE, respectively; $\mathcal{P}_5$, $\mathcal{P}_{10}$, and $\mathcal{P}_{15}$, cumulative percentage of estimations with AE within 5, 10, and 15 mm Hg, respectively. For fiducial histogram plots: $\mathcal{W}_{\text{pred}}$, Wasserstein distance between true and estimated distribution. For ensemble plots: AMAE, average mean absolute error; ARMSE, average root mean square error; solid line, ensemble average of all periods; dashed lines, ensemble average $\pm$ standard deviation of all periods; scale bars, one-quarter of period.

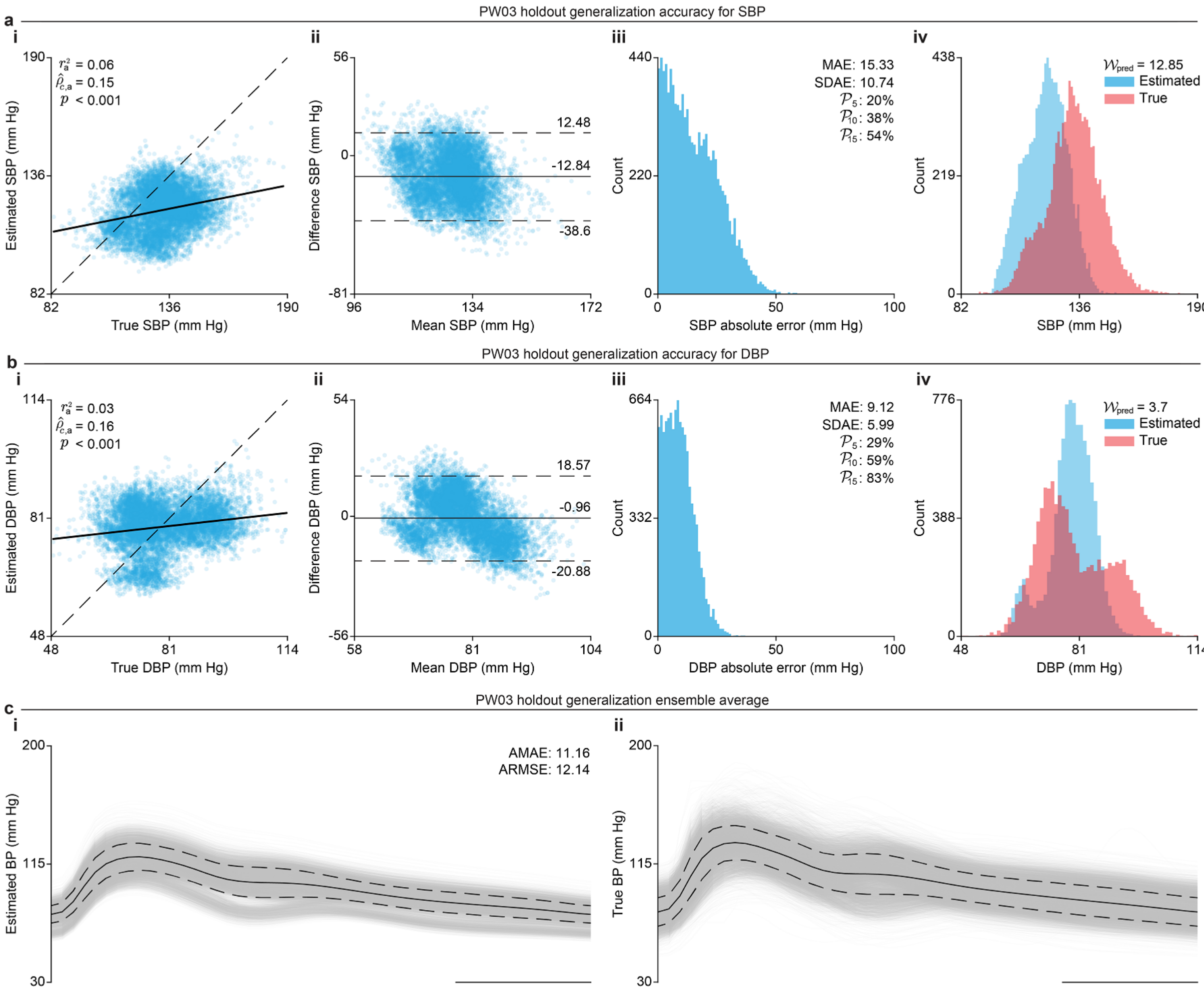

# Supplementary Fig. 76. Generalizability of population-within model PW04

Aggregated results from PW04 configuration: Linear Regression class with impedance input and fiducial output, trained with the population-within (PW) partition, and inferred on the holdout datasets. **a,** Estimation accuracy for systolic brachial blood pressure (SBP); **b,** Estimation accuracy for diastolic brachial blood pressure (DBP); **i**, correlation plots; **ii**, limits of agreement (LOA) plots; **iii**, histogram of absolute errors (AE); and **iv**, histogram of estimated and true BP distributions. BP, blood pressure; DBP, diastolic blood pressure; SBP, systolic blood pressure. For correlation plots: $r_{\mathrm{a}}^2$, aggregated coefficient of determination; $\hat{\rho}_{c,\mathrm{a}}$, aggregated coefficient of concordance; solid line, empirical linear regression line; dashed line, $45°$ line of perfect correlation. For LOA plots: solid line, mean of errors between estimated and true BP values; dashed lines, 2.5$^{\text{th}}$ percentile (lower) and 97.5$^{\text{th}}$ percentile (upper). For AE histogram plots: MAE and SDAE, mean and standard deviation of AE, respectively; $\mathcal{P}_5$, $\mathcal{P}_{10}$, and $\mathcal{P}_{15}$, cumulative percentage of estimations with AE within 5, 10, and 15 mm Hg, respectively. For fiducial histogram plots: $\mathcal{W}_{\text{pred}}$, Wasserstein distance between true and estimated distribution.

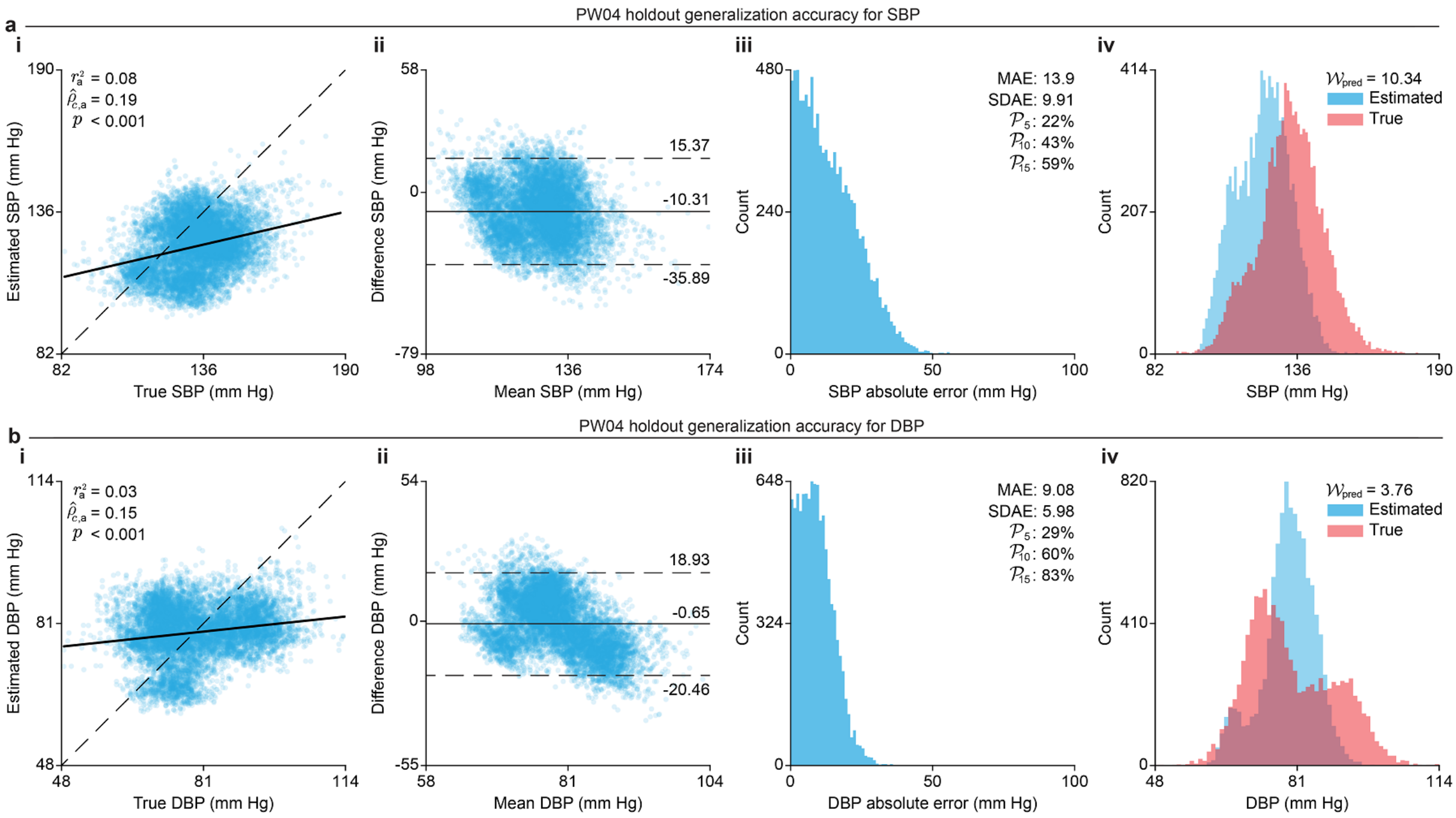

# Supplementary Fig. 77. Generalizability of population-within model PW05

Aggregated results from PW05 configuration: Multilayer Perceptron class with image input and waveform output, trained with the population-within (PW) partition, and inferred on the holdout datasets. **a,** Estimation accuracy for systolic brachial blood pressure (SBP); **b,** Estimation accuracy for diastolic brachial blood pressure (DBP); **c,** Waveform ensemble of all estimated and true brachial blood pressure (BP) periods. For **a** and **b**: **i**, correlation plots; **ii**, limits of agreement (LOA) plots; **iii**, histogram of absolute errors (AE); and **iv**, histogram of estimated and true BP distributions. For **c**: **i**, ensemble of estimated BP periods; **ii**, ensemble of true BP periods. For correlation plots: $r^2_{\mathrm{a}}$, aggregated coefficient of determination; $\hat{\rho}_{c,\mathrm{a}}$, aggregated coefficient of concordance; solid line, empirical linear regression line; dashed line, $45°$ line of perfect correlation. For LOA plots: solid line, mean of errors between estimated and true BP values; dashed lines, 2.5$^{\mathrm{th}}$ percentile (lower) and 97.5$^{\mathrm{th}}$ percentile (upper). For AE histogram plots: MAE and SDAE, mean and standard deviation of AE, respectively; $\mathcal{P}_5$, $\mathcal{P}_{10}$, and $\mathcal{P}_{15}$, cumulative percentage of estimations with AE within 5, 10, and 15 mm Hg, respectively. For fiducial histogram plots: $\mathcal{W}_{\mathrm{pred}}$, Wasserstein distance between true and estimated distribution. For ensemble plots: AMAE, average mean absolute error; ARMSE, average root mean square error; solid line, ensemble average of all periods; dashed lines, ensemble average $\pm$ standard deviation of all periods; scale bars, one-quarter of period.

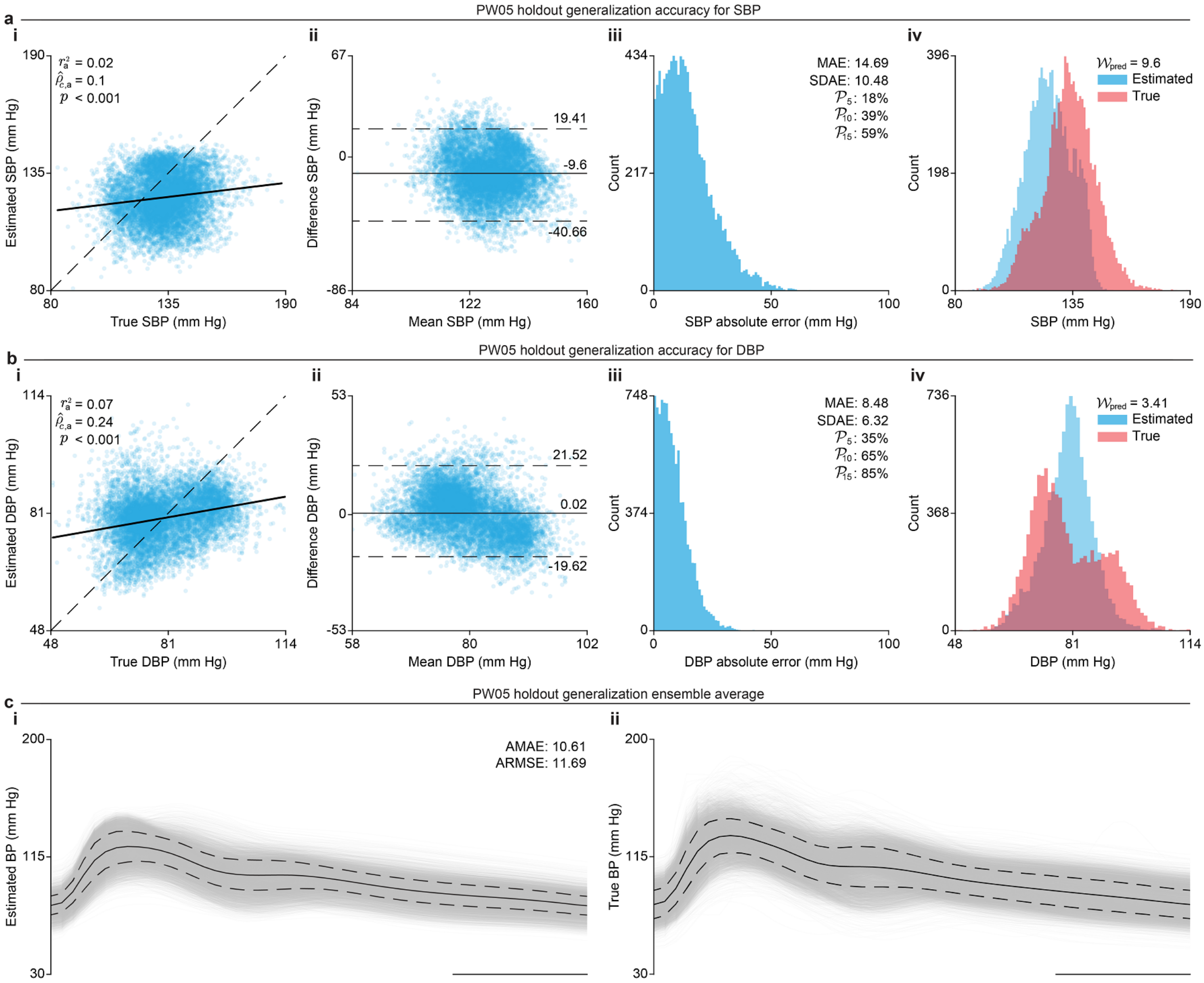

# Supplementary Fig. 78. Generalizability of population-within model PW06

Aggregated results from PW06 configuration: Multilayer Perceptron class with image input and fiducial output, trained with the population-within (PW) partition, and inferred on the holdout datasets. **a,** Estimation accuracy for systolic brachial blood pressure (SBP); **b,** Estimation accuracy for diastolic brachial blood pressure (DBP); **i**, correlation plots; **ii**, limits of agreement (LOA) plots; **iii**, histogram of absolute errors (AE); and **iv**, histogram of estimated and true BP distributions. BP, blood pressure; DBP, diastolic blood pressure; SBP, systolic blood pressure. For correlation plots: $r_{\mathrm{a}}^2$, aggregated coefficient of determination; $\hat{\rho}_{c,\mathrm{a}}$, aggregated coefficient of concordance; solid line, empirical linear regression line; dashed line, $45°$ line of perfect correlation. For LOA plots: solid line, mean of errors between estimated and true BP values; dashed lines, 2.5$^{\mathrm{th}}$ percentile (lower) and 97.5$^{\mathrm{th}}$ percentile (upper). For AE histogram plots: MAE and SDAE, mean and standard deviation of AE, respectively; $\mathcal{P}_5$, $\mathcal{P}_{10}$, and $\mathcal{P}_{15}$, cumulative percentage of estimations with AE within 5, 10, and 15 mm Hg, respectively. For fiducial histogram plots: $\mathcal{W}_{\mathrm{pred}}$, Wasserstein distance between true and estimated distribution.

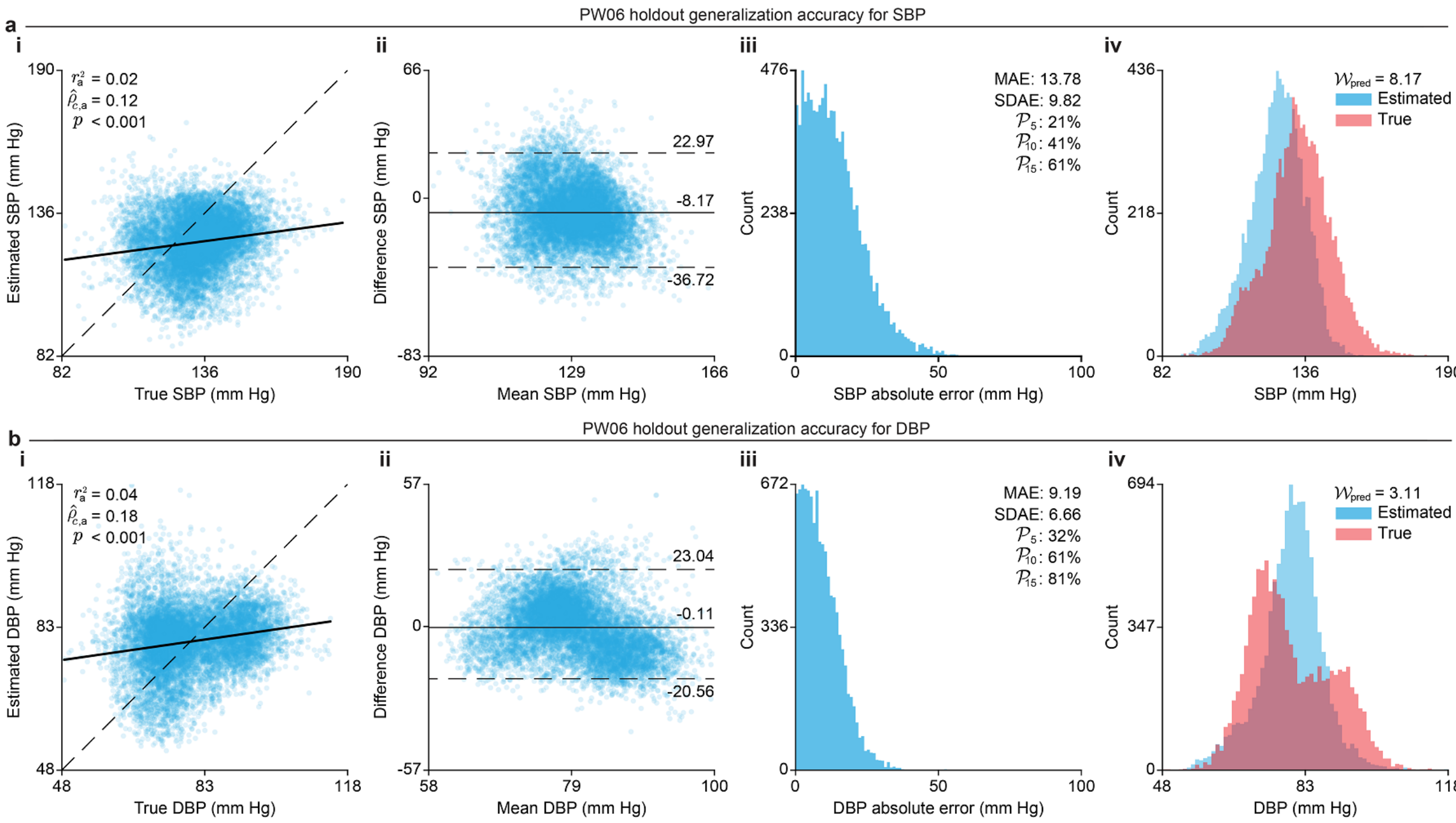

# Supplementary Fig. 79. Generalizability of population-within model PW07

Aggregated results from PW07 configuration: Multilayer Perceptron class with impedance input and waveform output, trained with the population-within (PW) partition, and inferred on the holdout datasets. **a,** Estimation accuracy for systolic brachial blood pressure (SBP); **b,** Estimation accuracy for diastolic brachial blood pressure (DBP); **c,** Waveform ensemble of all estimated and true brachial blood pressure (BP) periods. For **a** and **b**: **i**, correlation plots; **ii**, limits of agreement (LOA) plots; **iii**, histogram of absolute errors (AE); and **iv**, histogram of estimated and true BP distributions. For **c**: **i**, ensemble of estimated BP periods; **ii**, ensemble of true BP periods. For correlation plots: $r^2_{\mathrm{a}}$, aggregated coefficient of determination; $\hat{\rho}_{c,\mathrm{a}}$, aggregated coefficient of concordance; solid line, empirical linear regression line; dashed line, $45°$ line of perfect correlation. For LOA plots: solid line, mean of errors between estimated and true BP values; dashed lines, 2.5$^{\text{th}}$ percentile (lower) and 97.5$^{\text{th}}$ percentile (upper). For AE histogram plots: MAE and SDAE, mean and standard deviation of AE, respectively; $\mathcal{P}_5$, $\mathcal{P}_{10}$, and $\mathcal{P}_{15}$, cumulative percentage of estimations with AE within 5, 10, and 15 mm Hg, respectively. For fiducial histogram plots: $\mathcal{W}_{\text{pred}}$, Wasserstein distance between true and estimated distribution. For ensemble plots: AMAE, average mean absolute error; ARMSE, average root mean square error; solid line, ensemble average of all periods; dashed lines, ensemble average $\pm$ standard deviation of all periods; scale bars, one-quarter of period.

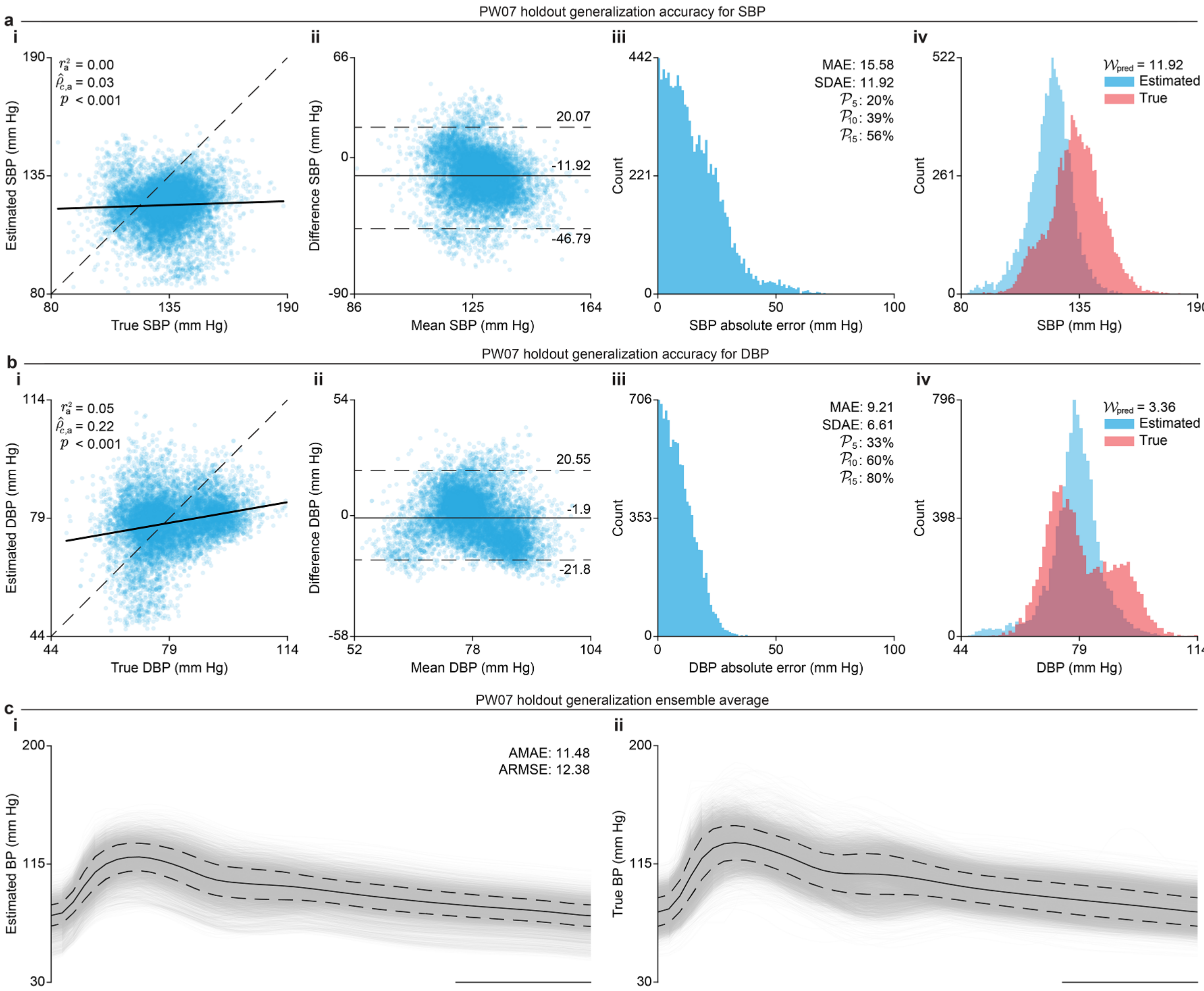

# Supplementary Fig. 80. Generalizability of population-within model PW08

Aggregated results from PW08 configuration: Multilayer Perceptron class with impedance input and fiducial output, trained with the population-within (PW) partition, and inferred on the holdout datasets. **a,** Estimation accuracy for systolic brachial blood pressure (SBP); **b,** Estimation accuracy for diastolic brachial blood pressure (DBP); **i**, correlation plots; **ii**, limits of agreement (LOA) plots; **iii**, histogram of absolute errors (AE); and **iv**, histogram of estimated and true BP distributions. BP, blood pressure; DBP, diastolic blood pressure; SBP, systolic blood pressure. For correlation plots: $r_{\mathrm{a}}^2$, aggregated coefficient of determination; $\hat{\rho}_{c,\mathrm{a}}$, aggregated coefficient of concordance; solid line, empirical linear regression line; dashed line, $45°$ line of perfect correlation. For LOA plots: solid line, mean of errors between estimated and true BP values; dashed lines, 2.5$^{\mathrm{th}}$ percentile (lower) and 97.5$^{\mathrm{th}}$ percentile (upper). For AE histogram plots: MAE and SDAE, mean and standard deviation of AE, respectively; $\mathcal{P}_5$, $\mathcal{P}_{10}$, and $\mathcal{P}_{15}$, cumulative percentage of estimations with AE within 5, 10, and 15 mm Hg, respectively. For fiducial histogram plots: $\mathcal{W}_{\mathrm{pred}}$, Wasserstein distance between true and estimated distribution.

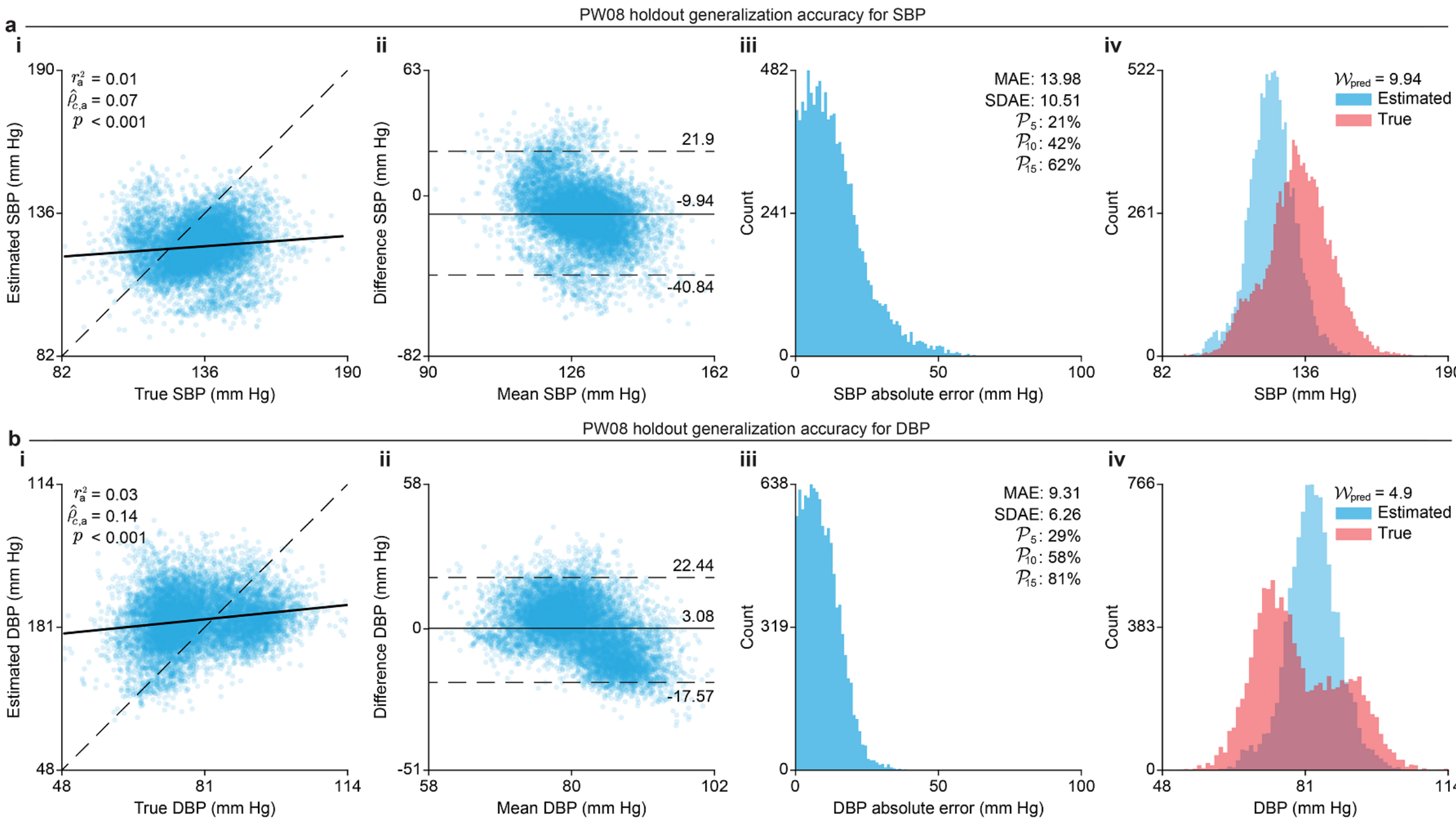

# Supplementary Fig. 81. Generalizability of population-within model PW09

Aggregated results from PW09 configuration: Convolutional Neural Network class with image input and waveform output, trained with the population-within (PW) partition, and inferred on the holdout datasets. **a,** Estimation accuracy for systolic brachial blood pressure (SBP); **b,** Estimation accuracy for diastolic brachial blood pressure (DBP); **c,** Waveform ensemble of all estimated and true brachial blood pressure (BP) periods. For **a** and **b**: **i**, correlation plots; **ii**, limits of agreement (LOA) plots; **iii**, histogram of absolute errors (AE); and **iv**, histogram of estimated and true BP distributions. For **c**: **i**, ensemble of estimated BP periods; **ii**, ensemble of true BP periods. For correlation plots: $r_{\mathrm{a}}^2$, aggregated coefficient of determination; $\hat{\rho}_{c,\mathrm{a}}$, aggregated coefficient of concordance; solid line, empirical linear regression line; dashed line, $45°$ line of perfect correlation. For LOA plots: solid line, mean of errors between estimated and true BP values; dashed lines, 2.5$^{\text{th}}$ percentile (lower) and 97.5$^{\text{th}}$ percentile (upper). For AE histogram plots: MAE and SDAE, mean and standard deviation of AE, respectively; $\mathcal{P}_5$, $\mathcal{P}_{10}$, and $\mathcal{P}_{15}$, cumulative percentage of estimations with AE within 5, 10, and 15 mm Hg, respectively. For fiducial histogram plots: $\mathcal{W}_{\mathrm{pred}}$, Wasserstein distance between true and estimated distribution. For ensemble plots: AMAE, average mean absolute error; ARMSE, average root mean square error; solid line, ensemble average of all periods; dashed lines, ensemble average $\pm$ standard deviation of all periods; scale bars, one-quarter of period.

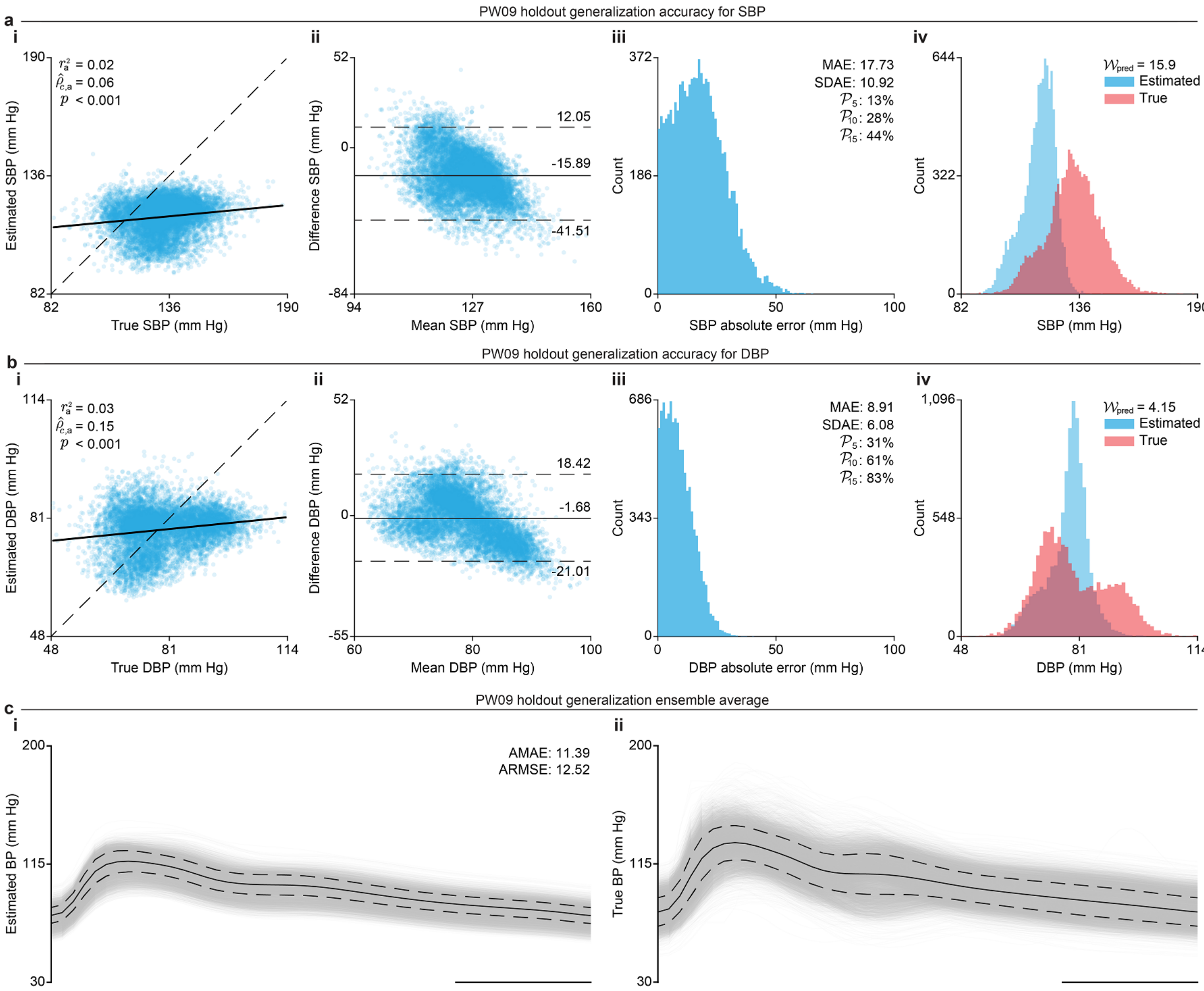

# Supplementary Fig. 82. Generalizability of population-within model PW10

Aggregated results from PW10 configuration: Convolutional Neural Network class with image input and fiducial output, trained with the population-within (PW) partition, and inferred on the holdout datasets. **a,** Estimation accuracy for systolic brachial blood pressure (SBP); **b,** Estimation accuracy for diastolic brachial blood pressure (DBP); **i**, correlation plots; **ii**, limits of agreement (LOA) plots; **iii**, histogram of absolute errors (AE); and **iv**, histogram of estimated and true BP distributions. BP, blood pressure; DBP, diastolic blood pressure; SBP, systolic blood pressure. For correlation plots: $r_{\mathrm{a}}^2$, aggregated coefficient of determination; $\hat{\rho}_{c,\mathrm{a}}$, aggregated coefficient of concordance; solid line, empirical linear regression line; dashed line, $45°$ line of perfect correlation. For LOA plots: solid line, mean of errors between estimated and true BP values; dashed lines, 2.5$^{\mathrm{th}}$ percentile (lower) and 97.5$^{\mathrm{th}}$ percentile (upper). For AE histogram plots: MAE and SDAE, mean and standard deviation of AE, respectively; $\mathcal{P}_5$, $\mathcal{P}_{10}$, and $\mathcal{P}_{15}$, cumulative percentage of estimations with AE within 5, 10, and 15 mm Hg, respectively. For fiducial histogram plots: $\mathcal{W}_{\mathrm{pred}}$, Wasserstein distance between true and estimated distribution.

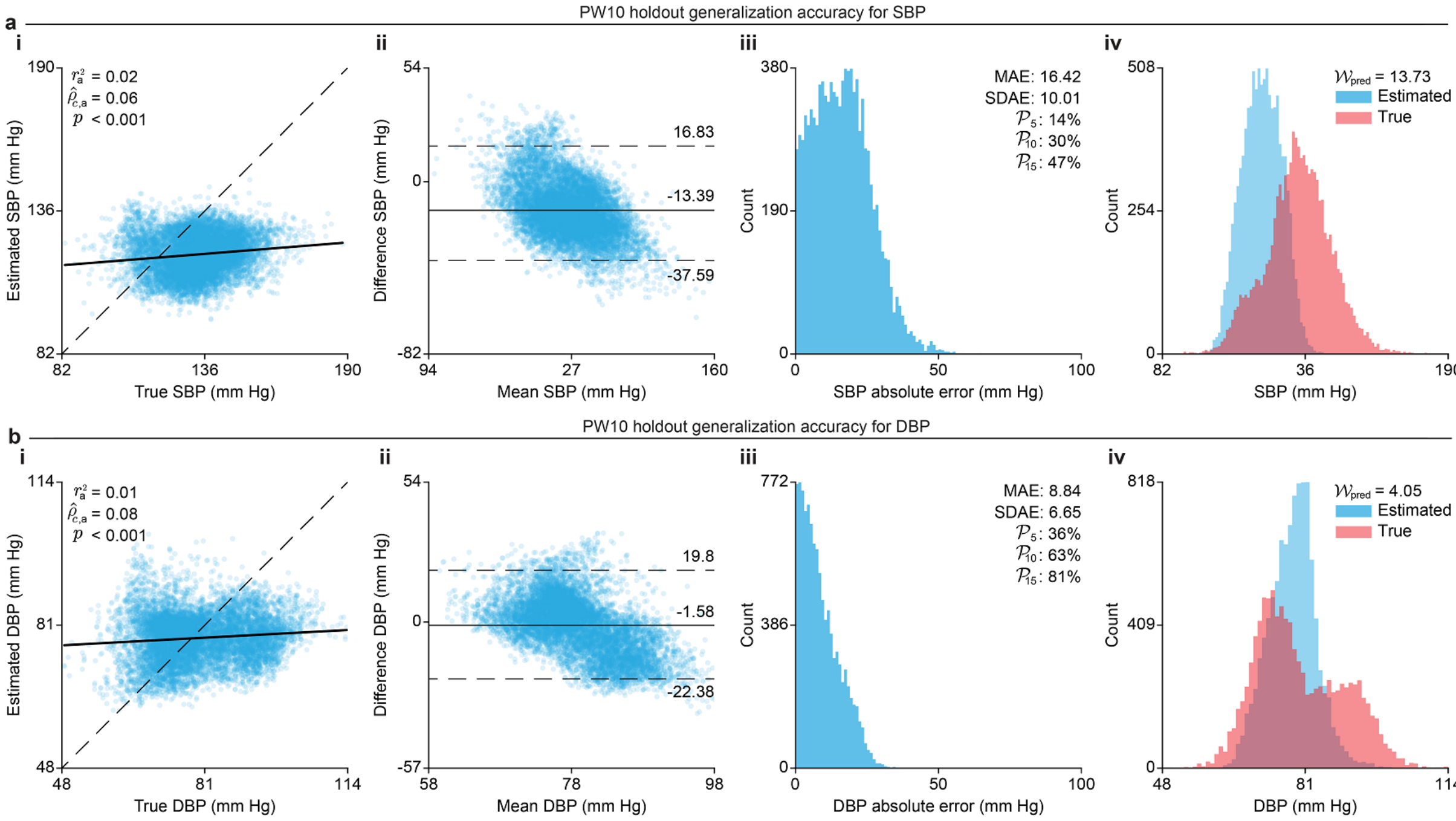

# Supplementary Fig. 83. Generalizability of population-within model PW11

Aggregated results from PW11 configuration: Convolutional Neural Network class with impedance input and waveform output, trained with the population-within (PW) partition, and inferred on the holdout datasets. **a,** Estimation accuracy for systolic brachial blood pressure (SBP); **b,** Estimation accuracy for diastolic brachial blood pressure (DBP); **c,** Waveform ensemble of all estimated and true brachial blood pressure (BP) periods. For **a** and **b**: **i**, correlation plots; **ii**, limits of agreement (LOA) plots; **iii**, histogram of absolute errors (AE); and **iv**, histogram of estimated and true BP distributions. For **c**: **i**, ensemble of estimated BP periods; **ii**, ensemble of true BP periods. For correlation plots: $r_\mathrm{a}^2$, aggregated coefficient of determination; $\hat{\rho}_{c,\mathrm{a}}$, aggregated coefficient of concordance; solid line, empirical linear regression line; dashed line, $45°$ line of perfect correlation. For LOA plots: solid line, mean of errors between estimated and true BP values; dashed lines, 2.5$^{\mathrm{th}}$ percentile (lower) and 97.5$^{\mathrm{th}}$ percentile (upper). For AE histogram plots: MAE and SDAE, mean and standard deviation of AE, respectively; $\mathcal{P}_5$, $\mathcal{P}_{10}$, and $\mathcal{P}_{15}$, cumulative percentage of estimations with AE within 5, 10, and 15 mm Hg, respectively. For fiducial histogram plots: $\mathcal{W}_{\mathrm{pred}}$, Wasserstein distance between true and estimated distribution. For ensemble plots: AMAE, average mean absolute error; ARMSE, average root mean square error; solid line, ensemble average of all periods; dashed lines, ensemble average $\pm$ standard deviation of all periods; scale bars, one-quarter of period.

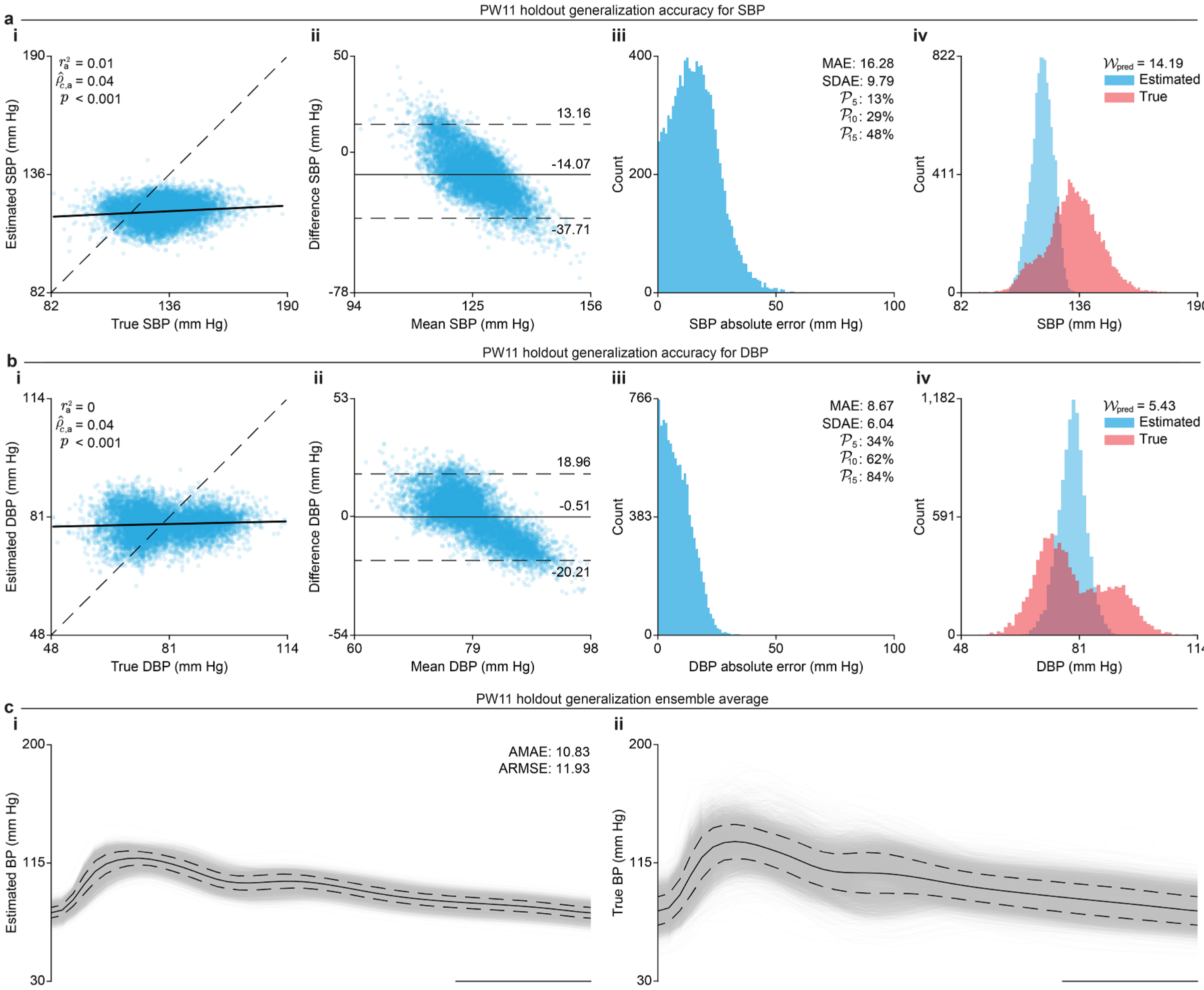

# Supplementary Fig. 84. Generalizability of population-within model PW12

Aggregated results from PW12 configuration: Convolutional Neural Network class with impedance input and fiducial output, trained with the population-within (PW) partition, and inferred on the holdout datasets. **a,** Estimation accuracy for systolic brachial blood pressure (SBP); **b,** Estimation accuracy for diastolic brachial blood pressure (DBP); **i**, correlation plots; **ii**, limits of agreement (LOA) plots; **iii**, histogram of absolute errors (AE); and **iv**, histogram of estimated and true BP distributions. BP, blood pressure; DBP, diastolic blood pressure; SBP, systolic blood pressure. For correlation plots: $r_{\mathrm{a}}^2$, aggregated coefficient of determination; $\hat{\rho}_{c,\mathrm{a}}$, aggregated coefficient of concordance; solid line, empirical linear regression line; dashed line, $45°$ line of perfect correlation. For LOA plots: solid line, mean of errors between estimated and true BP values; dashed lines, 2.5$^{\mathrm{th}}$ percentile (lower) and 97.5$^{\mathrm{th}}$ percentile (upper). For AE histogram plots: MAE and SDAE, mean and standard deviation of AE, respectively; $\mathcal{P}_5$, $\mathcal{P}_{10}$, and $\mathcal{P}_{15}$, cumulative percentage of estimations with AE within 5, 10, and 15 mm Hg, respectively. For fiducial histogram plots: $\mathcal{W}_{\mathrm{pred}}$, Wasserstein distance between true and estimated distribution.

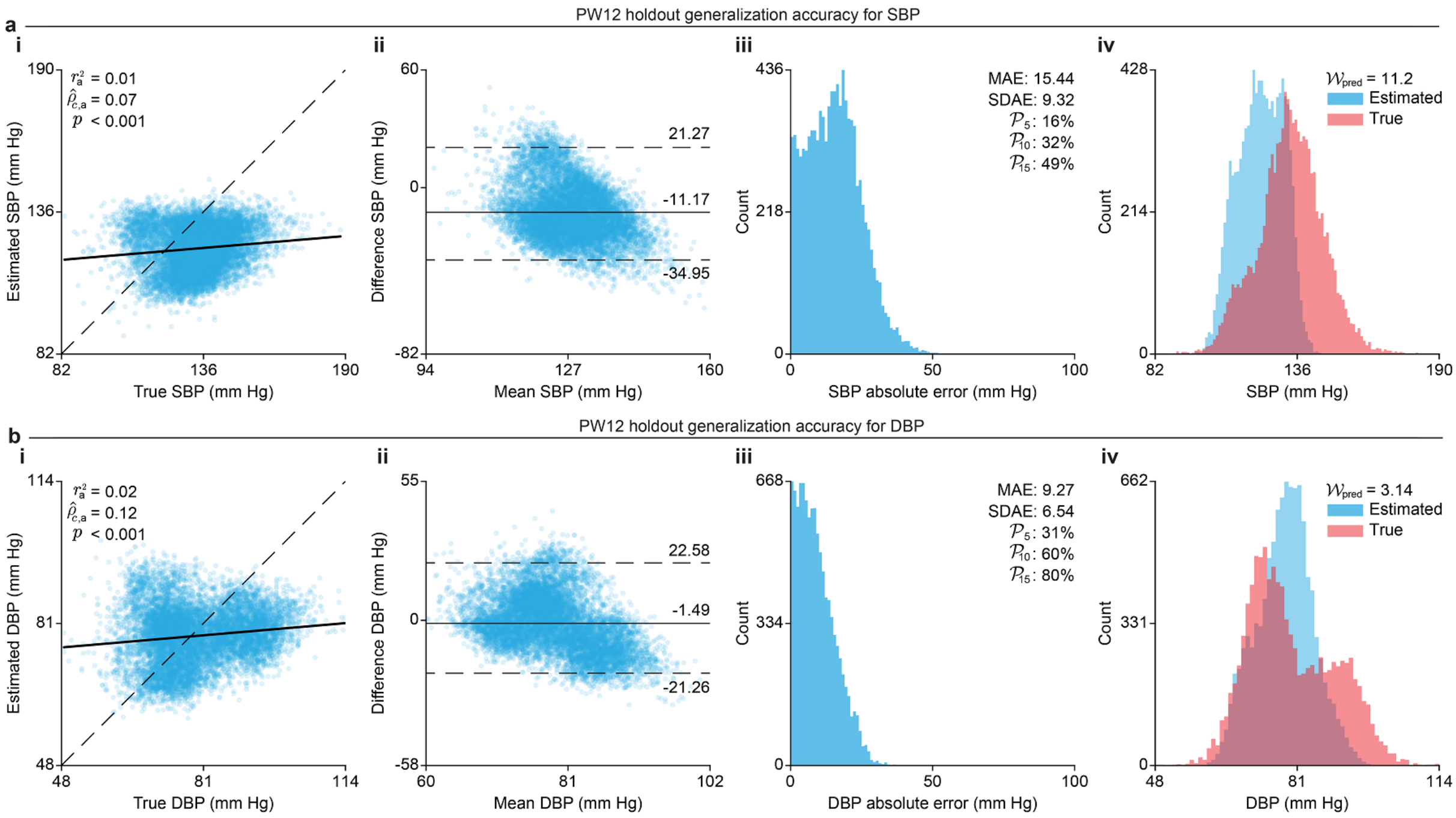

# Supplementary Fig. 85. Generalizability of population-within model PW13

Aggregated results from PW13 configuration: Convolutional Recurrent Transformer class with image input and waveform output, trained with the population-within (PW) partition, and inferred on the holdout datasets. **a,** Estimation accuracy for systolic brachial blood pressure (SBP); **b,** Estimation accuracy for diastolic brachial blood pressure (DBP); **c,** Waveform ensemble of all estimated and true brachial blood pressure (BP) periods. For **a** and **b**: **i**, correlation plots; **ii**, limits of agreement (LOA) plots; **iii**, histogram of absolute errors (AE); and **iv**, histogram of estimated and true BP distributions. For **c**: **i**, ensemble of estimated BP periods; **ii**, ensemble of true BP periods. For correlation plots: $r_{\text{a}}^2$, aggregated coefficient of determination; $\hat{\rho}_{c,\text{a}}$, aggregated coefficient of concordance; solid line, empirical linear regression line; dashed line, $45°$ line of perfect correlation. For LOA plots: solid line, mean of errors between estimated and true BP values; dashed lines, 2.5$^{\text{th}}$ percentile (lower) and 97.5$^{\text{th}}$ percentile (upper). For AE histogram plots: MAE and SDAE, mean and standard deviation of AE, respectively; $\mathcal{P}_5$, $\mathcal{P}_{10}$, and $\mathcal{P}_{15}$, cumulative percentage of estimations with AE within 5, 10, and 15 mm Hg, respectively. For fiducial histogram plots: $\mathcal{W}_{\text{pred}}$, Wasserstein distance between true and estimated distribution. For ensemble plots: AMAE, average mean absolute error; ARMSE, average root mean square error; solid line, ensemble average of all periods; dashed lines, ensemble average $\pm$ standard deviation of all periods; scale bars, one-quarter of period.

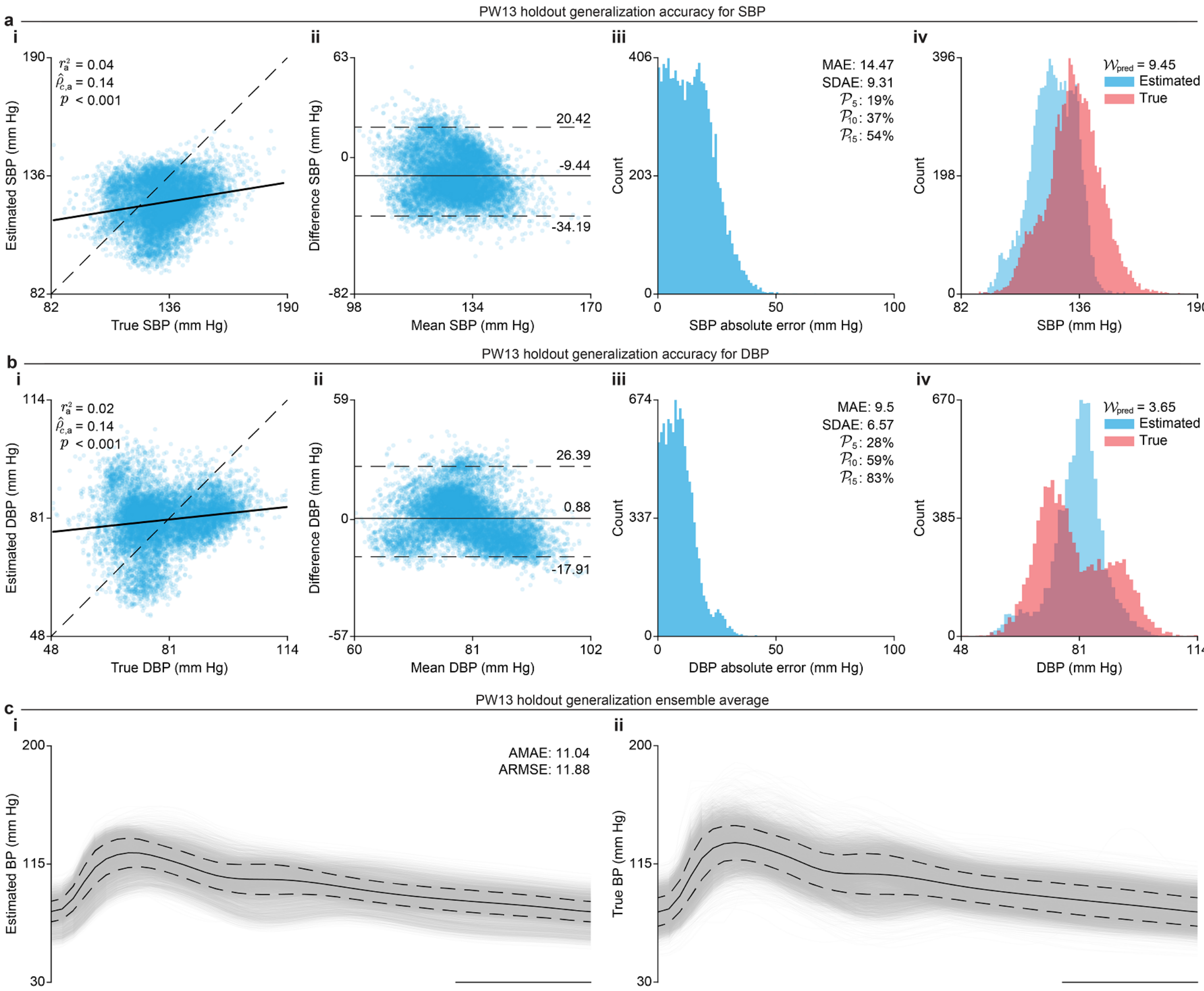

# Supplementary Fig. 86. Generalizability of population-within model PW14

Aggregated results from PW14 configuration: Convolutional Recurrent Transformer class with image input and fiducial output, trained with the population-within (PW) partition, and inferred on the holdout datasets. **a,** Estimation accuracy for systolic brachial blood pressure (SBP); **b,** Estimation accuracy for diastolic brachial blood pressure (DBP); **i**, correlation plots; **ii**, limits of agreement (LOA) plots; **iii**, histogram of absolute errors (AE); and **iv**, histogram of estimated and true BP distributions. BP, blood pressure; DBP, diastolic blood pressure; SBP, systolic blood pressure. For correlation plots: $r_{\mathrm{a}}^2$, aggregated coefficient of determination; $\hat{\rho}_{c,\mathrm{a}}$, aggregated coefficient of concordance; solid line, empirical linear regression line; dashed line, $45°$ line of perfect correlation. For LOA plots: solid line, mean of errors between estimated and true BP values; dashed lines, 2.5$^{\mathrm{th}}$ percentile (lower) and 97.5$^{\mathrm{th}}$ percentile (upper). For AE histogram plots: MAE and SDAE, mean and standard deviation of AE, respectively; $\mathcal{P}_5$, $\mathcal{P}_{10}$, and $\mathcal{P}_{15}$, cumulative percentage of estimations with AE within 5, 10, and 15 mm Hg, respectively. For fiducial histogram plots: $\mathcal{W}_{\mathrm{pred}}$, Wasserstein distance between true and estimated distribution.

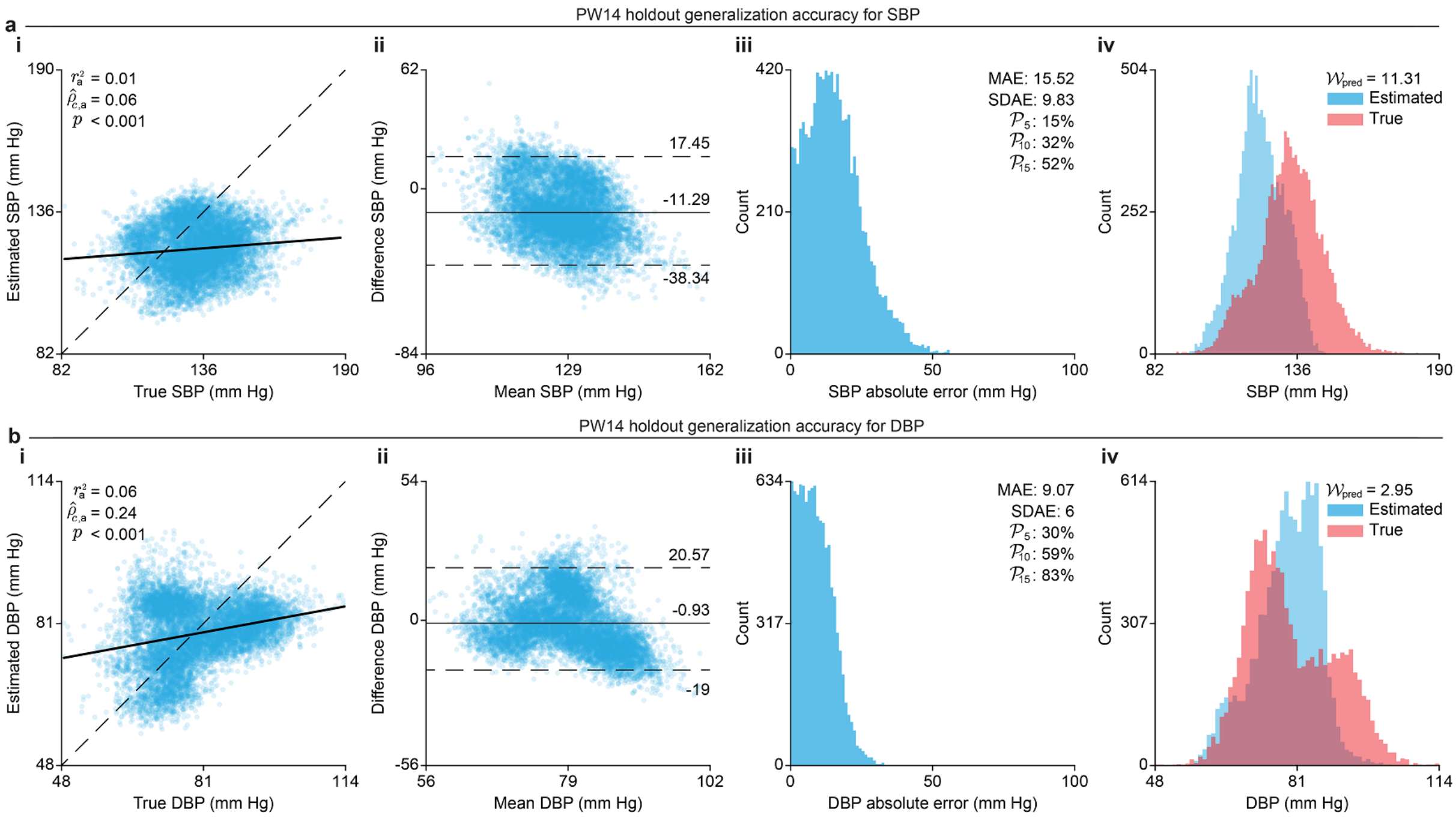

# Supplementary Fig. 87. Generalizability of population-within model PW15

Aggregated results from PW15 configuration: Convolutional Recurrent Transformer class with impedance input and waveform output, trained with the population-within (PW) partition, and inferred on the holdout datasets. **a,** Estimation accuracy for systolic brachial blood pressure (SBP); **b,** Estimation accuracy for diastolic brachial blood pressure (DBP); **c,** Waveform ensemble of all estimated and true brachial blood pressure (BP) periods. For **a** and **b**: **i**, correlation plots; **ii**, limits of agreement (LOA) plots; **iii**, histogram of absolute errors (AE); and **iv**, histogram of estimated and true BP distributions. For **c**: **i**, ensemble of estimated BP periods; **ii**, ensemble of true BP periods. For correlation plots: $r_{\mathrm{a}}^2$, aggregated coefficient of determination; $\hat{\rho}_{c,\mathrm{a}}$, aggregated coefficient of concordance; solid line, empirical linear regression line; dashed line, $45°$ line of perfect correlation. For LOA plots: solid line, mean of errors between estimated and true BP values; dashed lines, 2.5$^{\mathrm{th}}$ percentile (lower) and 97.5$^{\mathrm{th}}$ percentile (upper). For AE histogram plots: MAE and SDAE, mean and standard deviation of AE, respectively; $\mathcal{P}_5$, $\mathcal{P}_{10}$, and $\mathcal{P}_{15}$, cumulative percentage of estimations with AE within 5, 10, and 15 mm Hg, respectively. For fiducial histogram plots: $\mathcal{W}_{\mathrm{pred}}$, Wasserstein distance between true and estimated distribution. For ensemble plots: AMAE, average mean absolute error; ARMSE, average root mean square error; solid line, ensemble average of all periods; dashed lines, ensemble average $\pm$ standard deviation of all periods; scale bars, one-quarter of period.

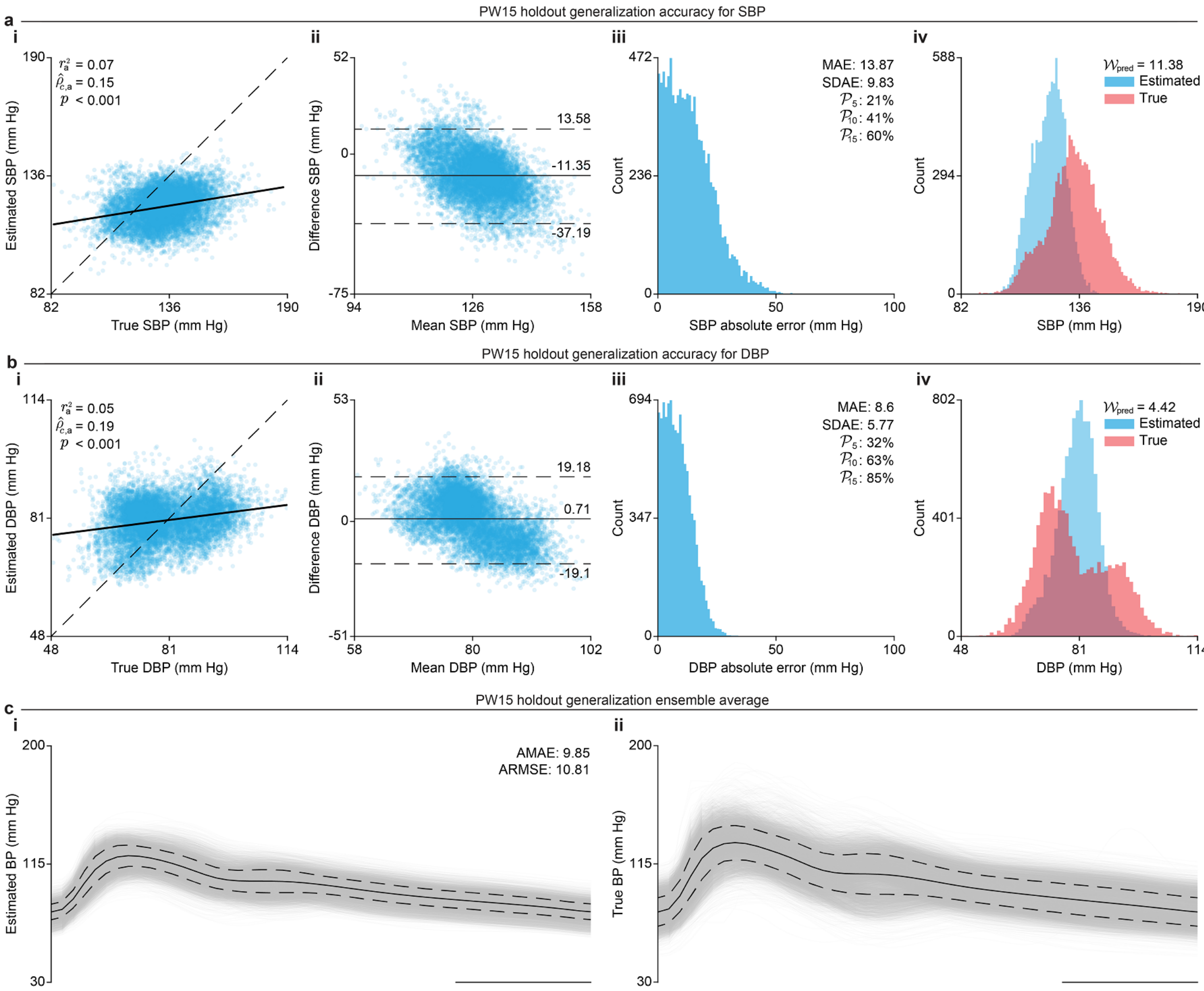

# Supplementary Fig. 88. Generalizability of population-within model PW16

Aggregated results from PW16 configuration: Convolutional Recurrent Transformer class with impedance input and fiducial output, trained with the population-within (PW) partition, and inferred on the holdout datasets. **a,** Estimation accuracy for systolic brachial blood pressure (SBP); **b,** Estimation accuracy for diastolic brachial blood pressure (DBP); **i**, correlation plots; **ii**, limits of agreement (LOA) plots; **iii**, histogram of absolute errors (AE); and **iv**, histogram of estimated and true BP distributions. BP, blood pressure; DBP, diastolic blood pressure; SBP, systolic blood pressure. For correlation plots: $r_{\text{a}}^2$, aggregated coefficient of determination; $\hat{\rho}_{c,\text{a}}$, aggregated coefficient of concordance; solid line, empirical linear regression line; dashed line, $45°$ line of perfect correlation. For LOA plots: solid line, mean of errors between estimated and true BP values; dashed lines, 2.5$^{\text{th}}$ percentile (lower) and 97.5$^{\text{th}}$ percentile (upper). For AE histogram plots: MAE and SDAE, mean and standard deviation of AE, respectively; $\mathcal{P}_5$, $\mathcal{P}_{10}$, and $\mathcal{P}_{15}$, cumulative percentage of estimations with AE within 5, 10, and 15 mm Hg, respectively. For fiducial histogram plots: $\mathcal{W}_{\text{pred}}$, Wasserstein distance between true and estimated distribution.

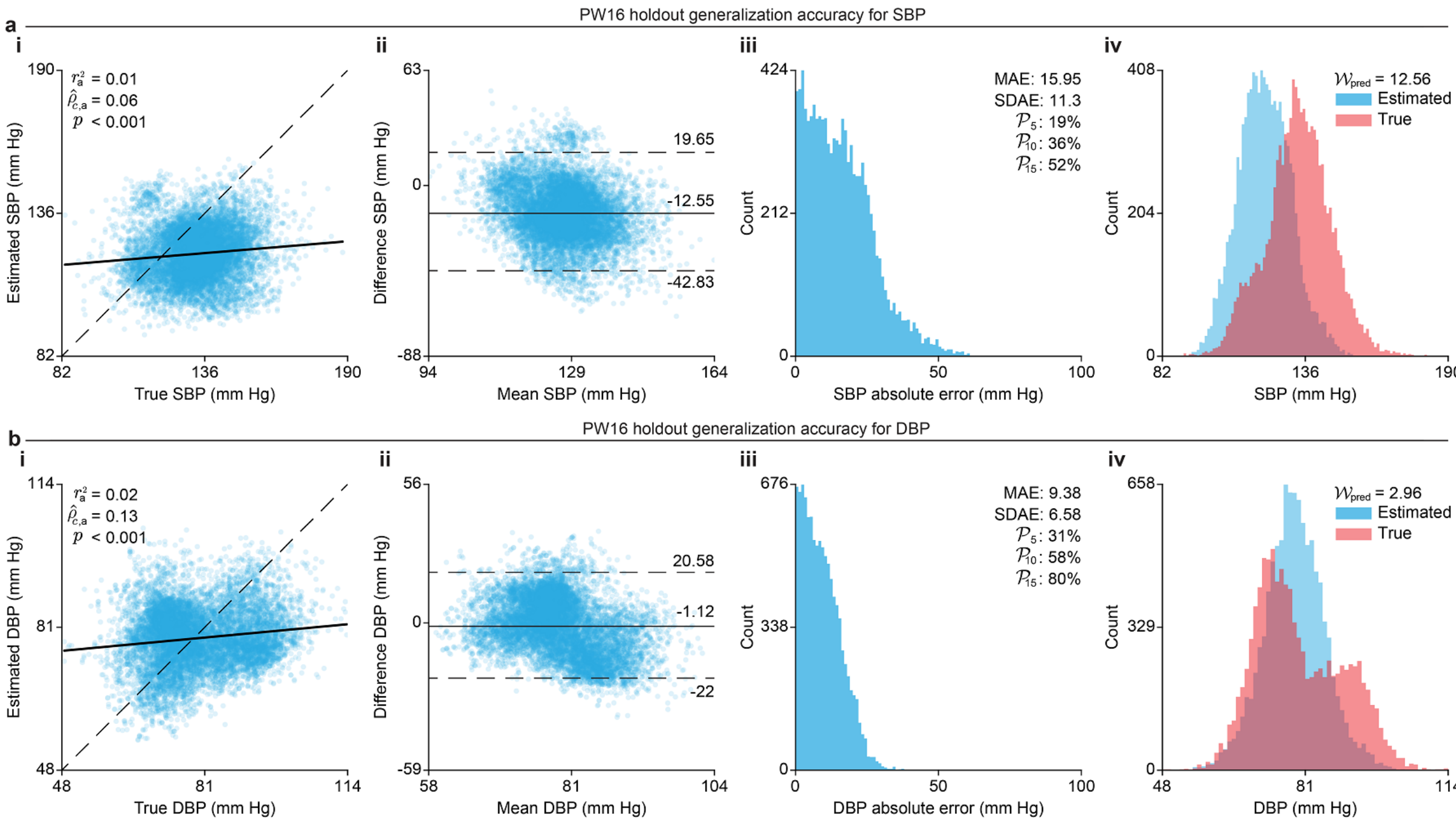

# Supplementary Fig. 89. Generalizability of population-within model PW17

Aggregated results from PW17 configuration: Convolutional Recurrent Samba class with image input and waveform output, trained with the population-within (PW) partition, and inferred on the holdout datasets. **a,** Estimation accuracy for systolic brachial blood pressure (SBP); **b,** Estimation accuracy for diastolic brachial blood pressure (DBP); **c,** Waveform ensemble of all estimated and true brachial blood pressure (BP) periods. For **a** and **b**: **i**, correlation plots; **ii**, limits of agreement (LOA) plots; **iii**, histogram of absolute errors (AE); and **iv**, histogram of estimated and true BP distributions. For **c**: **i**, ensemble of estimated BP periods; **ii**, ensemble of true BP periods. For correlation plots: $r_{\text{a}}^2$, aggregated coefficient of determination; $\hat{\rho}_{c,\text{a}}$, aggregated coefficient of concordance; solid line, empirical linear regression line; dashed line, $45°$ line of perfect correlation. For LOA plots: solid line, mean of errors between estimated and true BP values; dashed lines, 2.5$^{\text{th}}$ percentile (lower) and 97.5$^{\text{th}}$ percentile (upper). For AE histogram plots: MAE and SDAE, mean and standard deviation of AE, respectively; $\mathcal{P}_5$, $\mathcal{P}_{10}$, and $\mathcal{P}_{15}$, cumulative percentage of estimations with AE within 5, 10, and 15 mm Hg, respectively. For fiducial histogram plots: $\mathcal{W}_{\text{pred}}$, Wasserstein distance between true and estimated distribution. For ensemble plots: AMAE, average mean absolute error; ARMSE, average root mean square error; solid line, ensemble average of all periods; dashed lines, ensemble average $\pm$ standard deviation of all periods; scale bars, one-quarter of period.

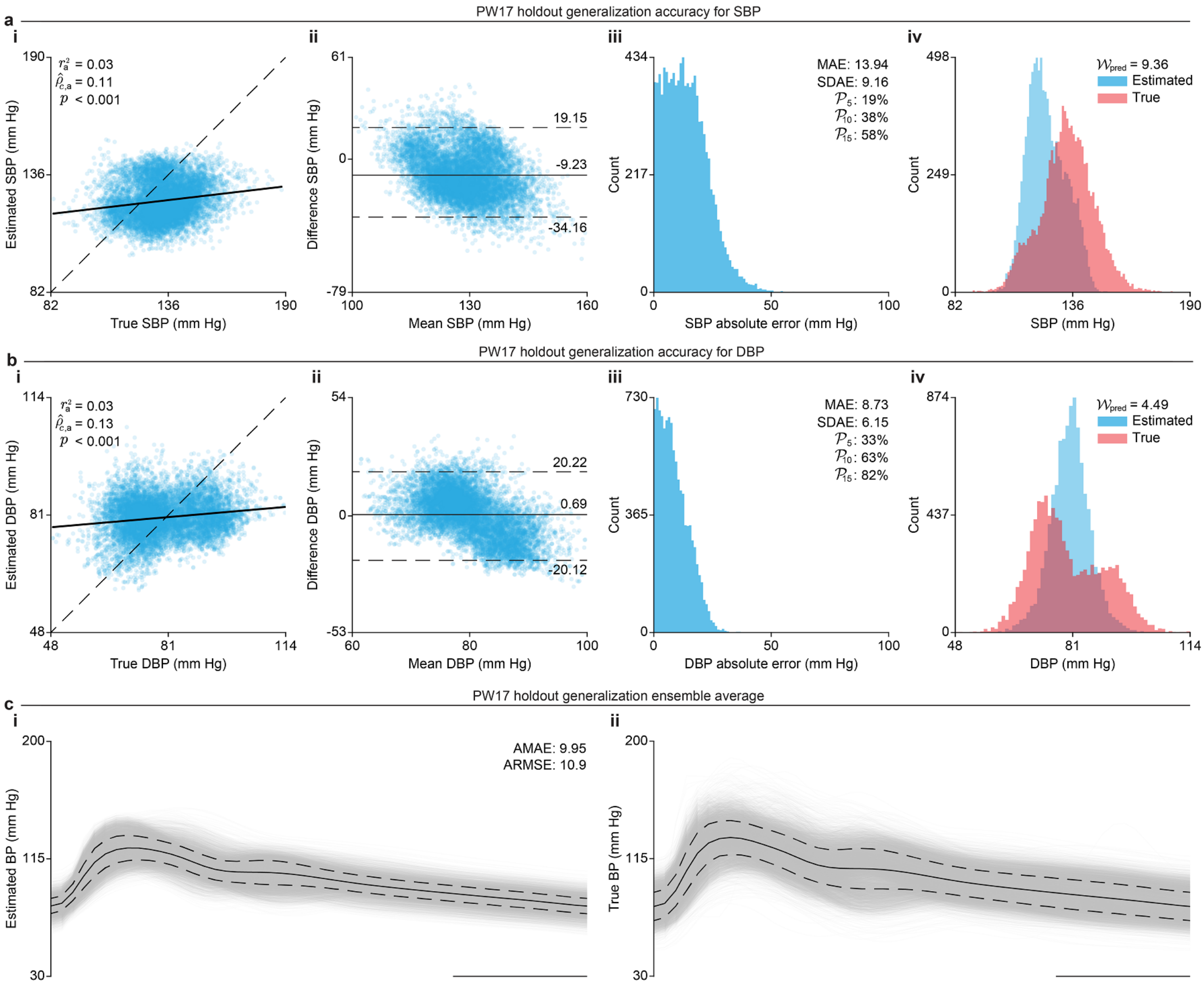

# Supplementary Fig. 90. Generalizability of population-within model PW18

Aggregated results from PW18 configuration: Convolutional Recurrent Samba class with image input and fiducial output, trained with the population-within (PW) partition, and inferred on the holdout datasets. **a,** Estimation accuracy for systolic brachial blood pressure (SBP); **b,** Estimation accuracy for diastolic brachial blood pressure (DBP); **i**, correlation plots; **ii**, limits of agreement (LOA) plots; **iii**, histogram of absolute errors (AE); and **iv**, histogram of estimated and true BP distributions. BP, blood pressure; DBP, diastolic blood pressure; SBP, systolic blood pressure. For correlation plots: $r_{\text{a}}^2$, aggregated coefficient of determination; $\hat{\rho}_{c,\text{a}}$, aggregated coefficient of concordance; solid line, empirical linear regression line; dashed line, $45°$ line of perfect correlation. For LOA plots: solid line, mean of errors between estimated and true BP values; dashed lines, 2.5$^{\text{th}}$ percentile (lower) and 97.5$^{\text{th}}$ percentile (upper). For AE histogram plots: MAE and SDAE, mean and standard deviation of AE, respectively; $\mathcal{P}_5$, $\mathcal{P}_{10}$, and $\mathcal{P}_{15}$, cumulative percentage of estimations with AE within 5, 10, and 15 mm Hg, respectively. For fiducial histogram plots: $\mathcal{W}_{\text{pred}}$, Wasserstein distance between true and estimated distribution.

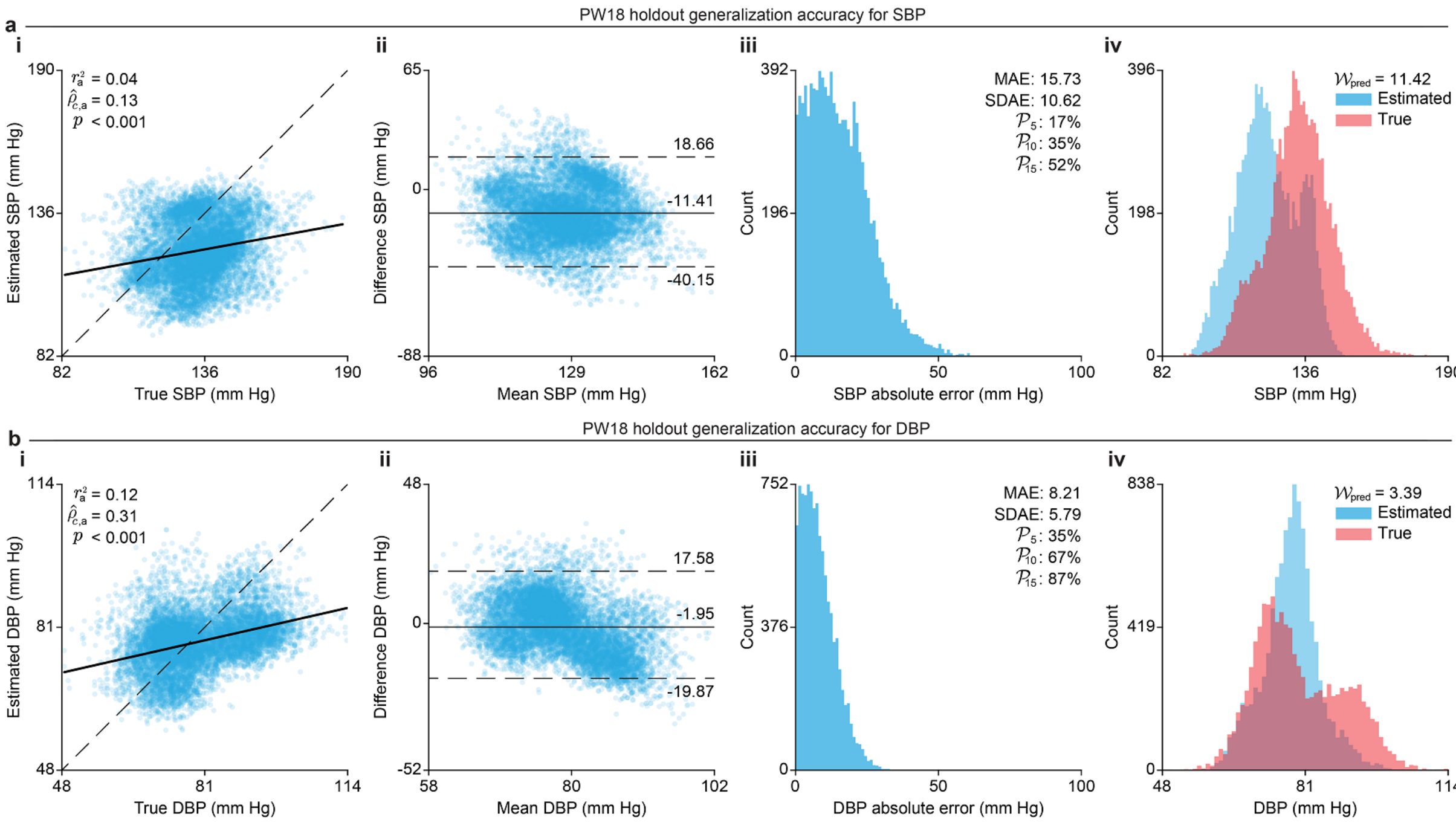

# Supplementary Fig. 91. Generalizability of population-within model PW19

Aggregated results from PW19 configuration: Convolutional Recurrent Samba class with impedance input and waveform output, trained with the population-within (PW) partition, and inferred on the holdout datasets. **a,** Estimation accuracy for systolic brachial blood pressure (SBP); **b,** Estimation accuracy for diastolic brachial blood pressure (DBP); **c,** Waveform ensemble of all estimated and true brachial blood pressure (BP) periods. For **a** and **b**: **i**, correlation plots; **ii**, limits of agreement (LOA) plots; **iii**, histogram of absolute errors (AE); and **iv**, histogram of estimated and true BP distributions. For **c**: **i**, ensemble of estimated BP periods; **ii**, ensemble of true BP periods. For correlation plots: $r_a^2$, aggregated coefficient of determination; $\hat{\rho}_{c,a}$, aggregated coefficient of concordance; solid line, empirical linear regression line; dashed line, $45°$ line of perfect correlation. For LOA plots: solid line, mean of errors between estimated and true BP values; dashed lines, 2.5$^{th}$ percentile (lower) and 97.5$^{th}$ percentile (upper). For AE histogram plots: MAE and SDAE, mean and standard deviation of AE, respectively; $\mathcal{P}_5$, $\mathcal{P}_{10}$, and $\mathcal{P}_{15}$, cumulative percentage of estimations with AE within 5, 10, and 15 mm Hg, respectively. For fiducial histogram plots: $\mathcal{W}_{pred}$, Wasserstein distance between true and estimated distribution. For ensemble plots: AMAE, average mean absolute error; ARMSE, average root mean square error; solid line, ensemble average of all periods; dashed lines, ensemble average $\pm$ standard deviation of all periods; scale bars, one-quarter of period.

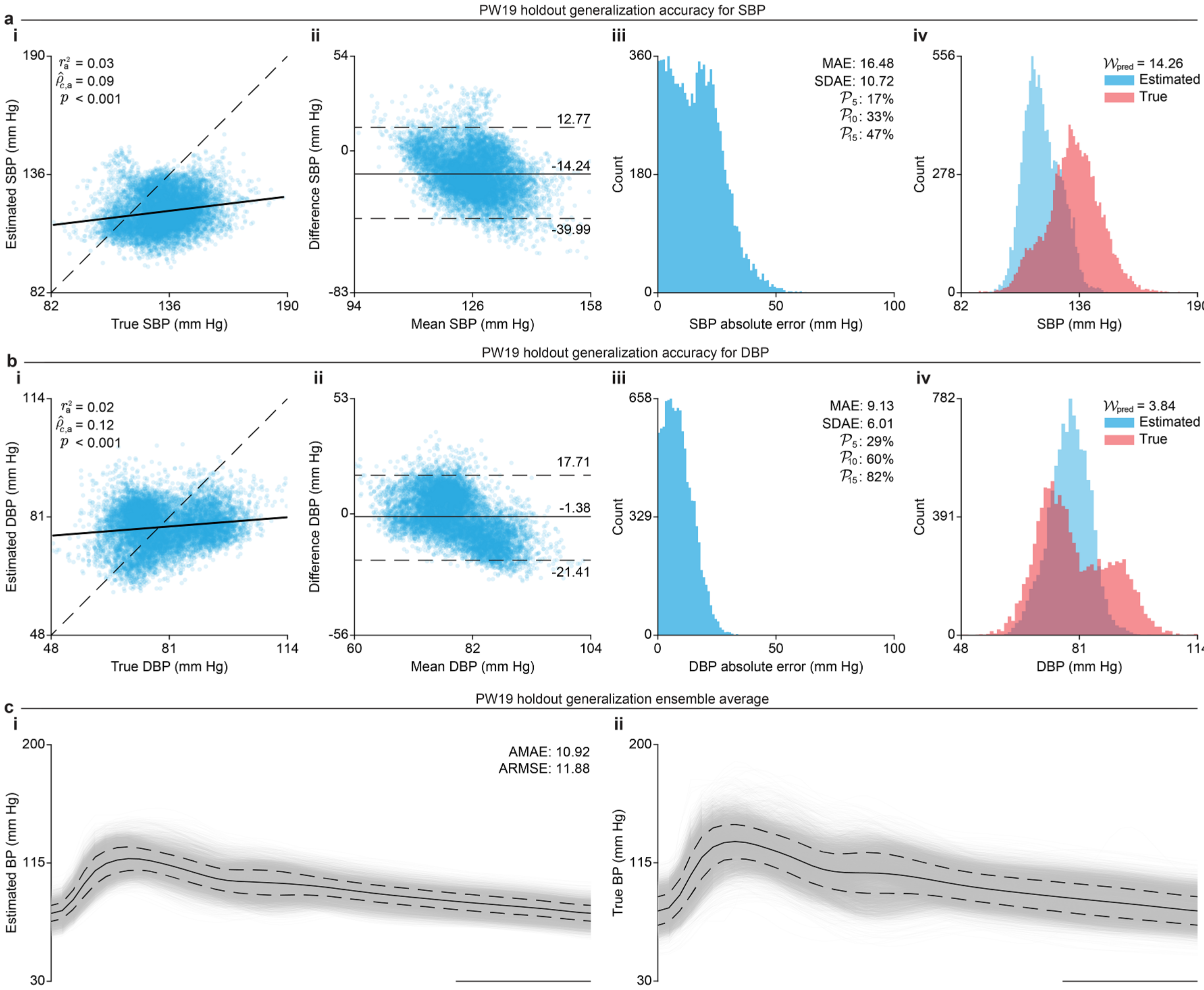

# Supplementary Fig. 92. Generalizability of population-within model PW20

Aggregated results from PW20 configuration: Convolutional Recurrent Samba class with impedance input and fiducial output, trained with the population-within (PW) partition, and inferred on the holdout datasets. **a,** Estimation accuracy for systolic brachial blood pressure (SBP); **b,** Estimation accuracy for diastolic brachial blood pressure (DBP); **i**, correlation plots; **ii**, limits of agreement (LOA) plots; **iii**, histogram of absolute errors (AE); and **iv**, histogram of estimated and true BP distributions. BP, blood pressure; DBP, diastolic blood pressure; SBP, systolic blood pressure. For correlation plots: $r_{\text{a}}^2$, aggregated coefficient of determination; $\hat{\rho}_{c,\text{a}}$, aggregated coefficient of concordance; solid line, empirical linear regression line; dashed line, $45°$ line of perfect correlation. For LOA plots: solid line, mean of errors between estimated and true BP values; dashed lines, 2.5$^{\text{th}}$ percentile (lower) and 97.5$^{\text{th}}$ percentile (upper). For AE histogram plots: MAE and SDAE, mean and standard deviation of AE, respectively; $\mathcal{P}_5$, $\mathcal{P}_{10}$, and $\mathcal{P}_{15}$, cumulative percentage of estimations with AE within 5, 10, and 15 mm Hg, respectively. For fiducial histogram plots: $\mathcal{W}_{\text{pred}}$, Wasserstein distance between true and estimated distribution.

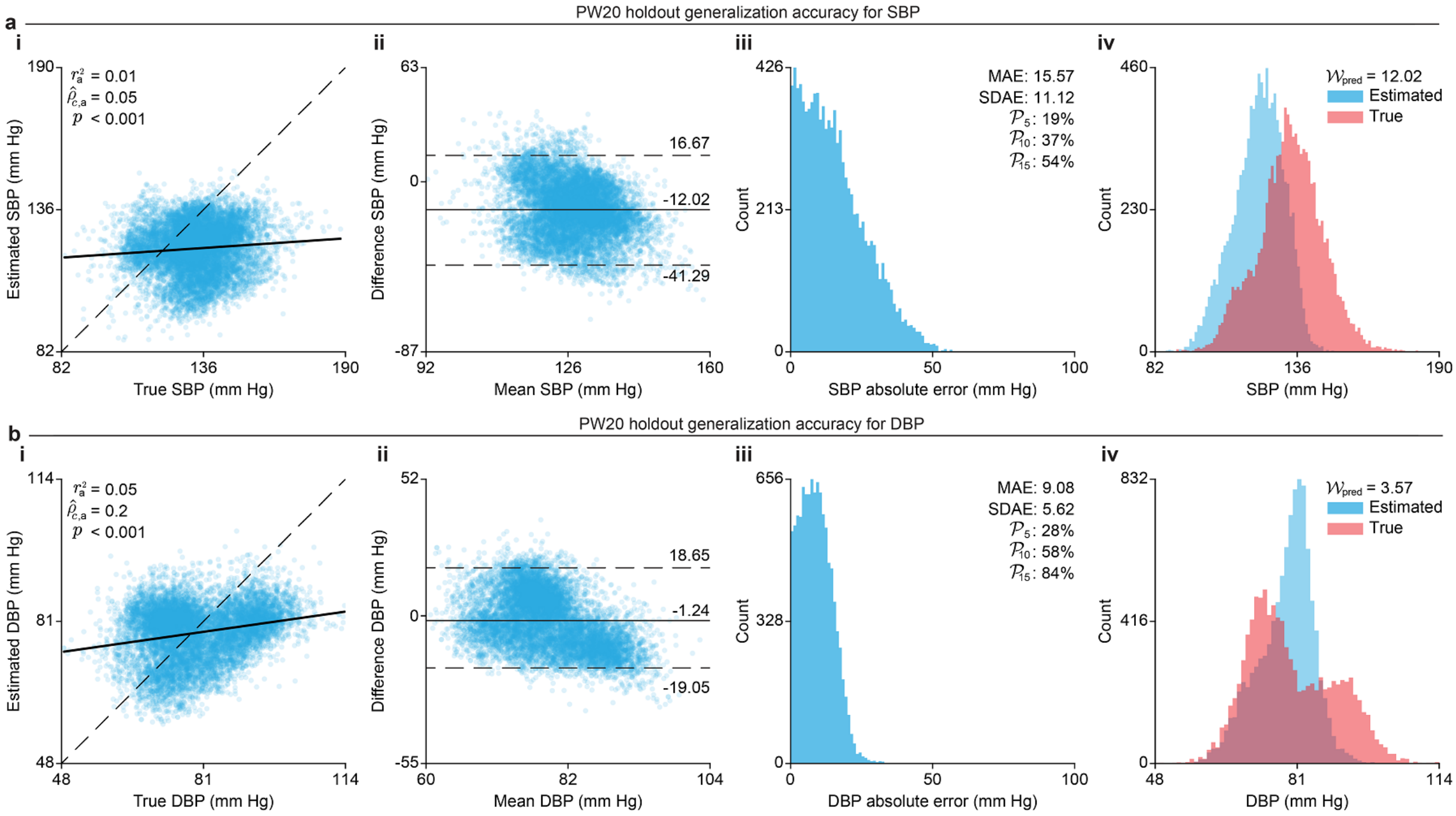

# Supplementary Fig. 93. Estimation results of population-disjoint model PD01

Aggregated results from PD01 configuration: Linear Regression class with image input and waveform output, trained with the population-disjoint (PD) partition. **a,** Estimation accuracy for systolic brachial blood pressure (SBP); **b,** Estimation accuracy for diastolic brachial blood pressure (DBP); **c,** Waveform ensemble of all estimated and true brachial blood pressure (BP) periods. For **a** and **b**: **i**, correlation plots; **ii**, limits of agreement (LOA) plots; **iii**, histogram of absolute errors (AE); and **iv**, histogram of estimated and true BP distributions. For **c**: **i**, ensemble of estimated BP periods; **ii**, ensemble of true BP periods. For correlation plots: $r^2_{\mathrm{a}}$, aggregated coefficient of determination; $\hat{\rho}_{c,\mathrm{a}}$, aggregated coefficient of concordance; solid line, empirical linear regression line; dashed line, $45°$ line of perfect correlation. For LOA plots: solid line, mean of errors between estimated and true BP values; dashed lines, 2.5$^{\text{th}}$ percentile (lower) and 97.5$^{\text{th}}$ percentile (upper). For AE histogram plots: MAE and SDAE, mean and standard deviation of AE, respectively; $\mathcal{P}_5$, $\mathcal{P}_{10}$, and $\mathcal{P}_{15}$, cumulative percentage of estimations with AE within 5, 10, and 15 mm Hg, respectively. For fiducial histogram plots: $\mathcal{W}_{\text{pred}}$, Wasserstein distance between true and estimated distribution. For ensemble plots: AMAE, average mean absolute error; ARMSE, average root mean square error; solid line, ensemble average of all periods; dashed lines, ensemble average $\pm$ standard deviation of all periods; scale bars, one-quarter of period.

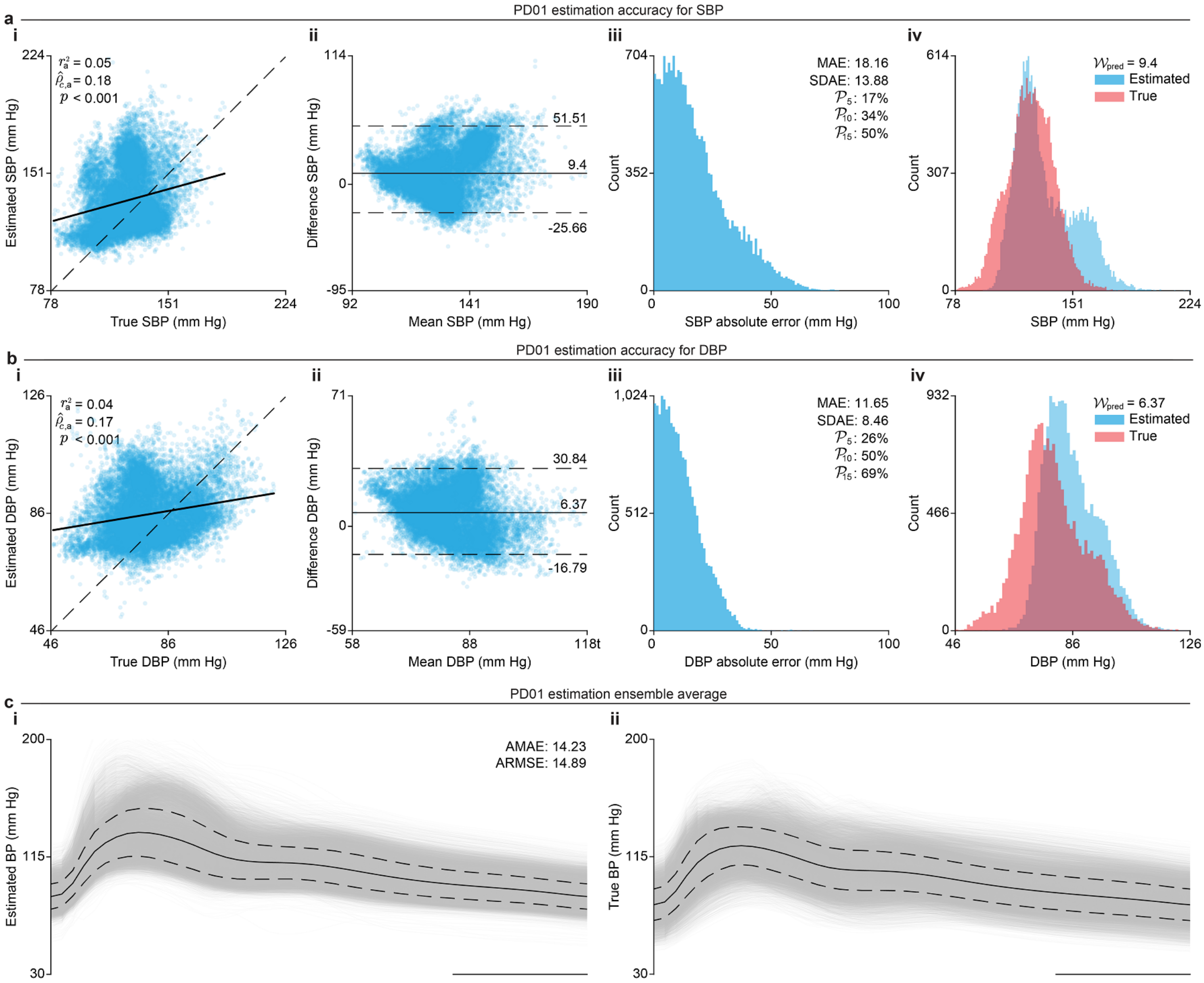

# Supplementary Fig. 94. Estimation results of population-disjoint model PD02

Aggregated results from PD02 configuration: Linear Regression class with image input and fiducial output, trained with the population-disjoint (PD) partition. **a,** Estimation accuracy for systolic brachial blood pressure (SBP); **b,** Estimation accuracy for diastolic brachial blood pressure (DBP); **i**, correlation plots; **ii**, limits of agreement (LOA) plots; **iii**, histogram of absolute errors (AE); and **iv**, histogram of estimated and true BP distributions. BP, blood pressure; DBP, diastolic blood pressure; SBP, systolic blood pressure. For correlation plots: $r_{\text{a}}^2$, aggregated coefficient of determination; $\hat{\rho}_{c,\text{a}}$, aggregated coefficient of concordance; solid line, empirical linear regression line; dashed line, $45°$ line of perfect correlation. For LOA plots: solid line, mean of errors between estimated and true BP values; dashed lines, 2.5th percentile (lower) and 97.5th percentile (upper). For AE histogram plots: MAE and SDAE, mean and standard deviation of AE, respectively; $\mathcal{P}_5$, $\mathcal{P}_{10}$, and $\mathcal{P}_{15}$, cumulative percentage of estimations with AE within 5, 10, and 15 mm Hg, respectively. For fiducial histogram plots: $\mathcal{W}_{\text{pred}}$, Wasserstein distance between true and estimated distribution.

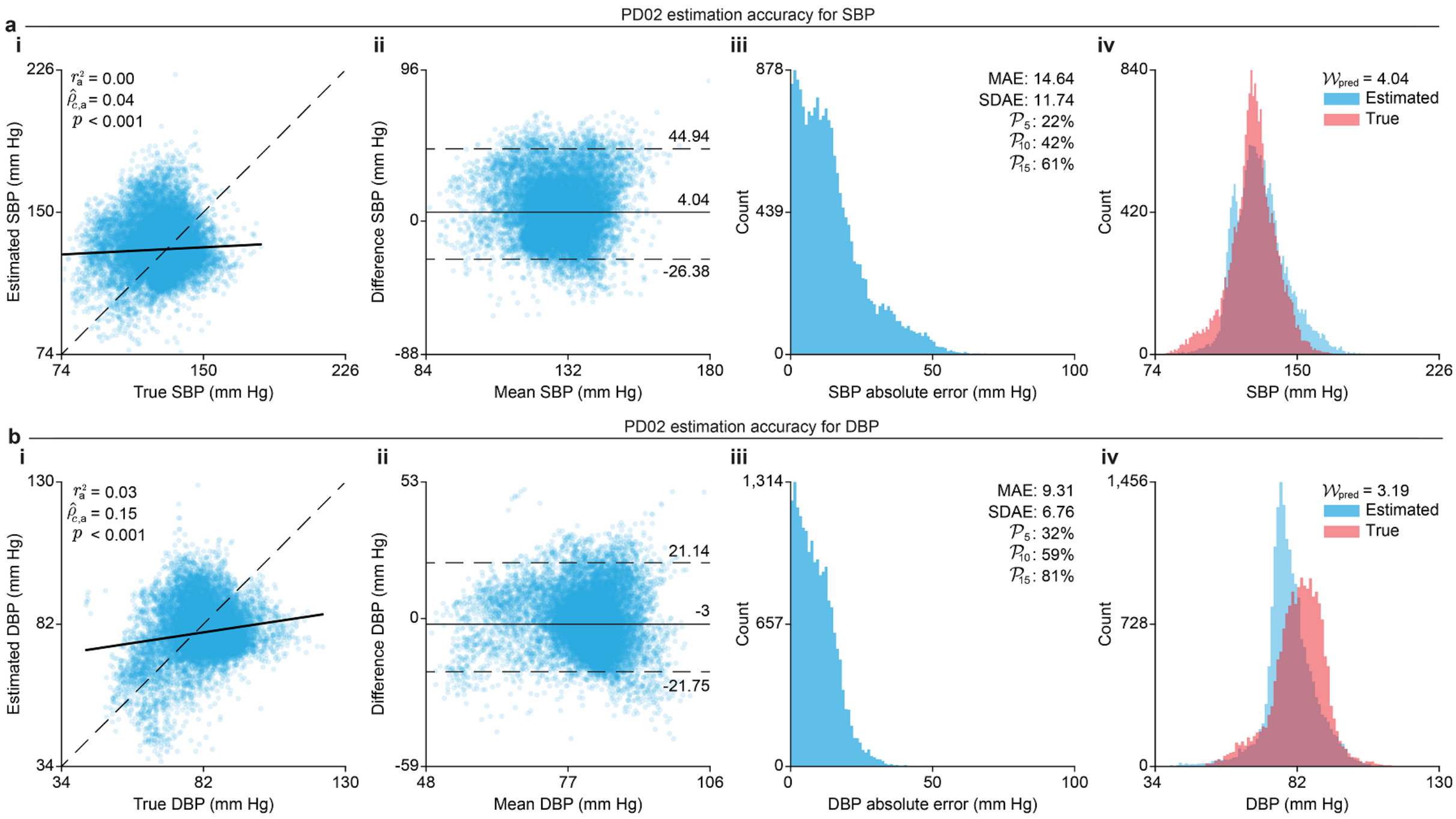

# Supplementary Fig. 95. Estimation results of population-disjoint model PD03

Aggregated results from PD03 configuration: Linear Regression class with impedance input and waveform output, trained with the population-disjoint (PD) partition. **a,** Estimation accuracy for systolic brachial blood pressure (SBP); **b,** Estimation accuracy for diastolic brachial blood pressure (DBP); **c,** Waveform ensemble of all estimated and true brachial blood pressure (BP) periods. For **a** and **b**: **i**, correlation plots; **ii**, limits of agreement (LOA) plots; **iii**, histogram of absolute errors (AE); and **iv**, histogram of estimated and true BP distributions. For **c**: **i**, ensemble of estimated BP periods; **ii**, ensemble of true BP periods. For correlation plots: $r^2_{\text{a}}$, aggregated coefficient of determination; $\hat{\rho}_{c,\text{a}}$, aggregated coefficient of concordance; solid line, empirical linear regression line; dashed line, $45°$ line of perfect correlation. For LOA plots: solid line, mean of errors between estimated and true BP values; dashed lines, 2.5$^{\text{th}}$ percentile (lower) and 97.5$^{\text{th}}$ percentile (upper). For AE histogram plots: MAE and SDAE, mean and standard deviation of AE, respectively; $\mathcal{P}_5$, $\mathcal{P}_{10}$, and $\mathcal{P}_{15}$, cumulative percentage of estimations with AE within 5, 10, and 15 mm Hg, respectively. For fiducial histogram plots: $\mathcal{W}_{\text{pred}}$, Wasserstein distance between true and estimated distribution. For ensemble plots: AMAE, average mean absolute error; ARMSE, average root mean square error; solid line, ensemble average of all periods; dashed lines, ensemble average $\pm$ standard deviation of all periods; scale bars, one-quarter of period.

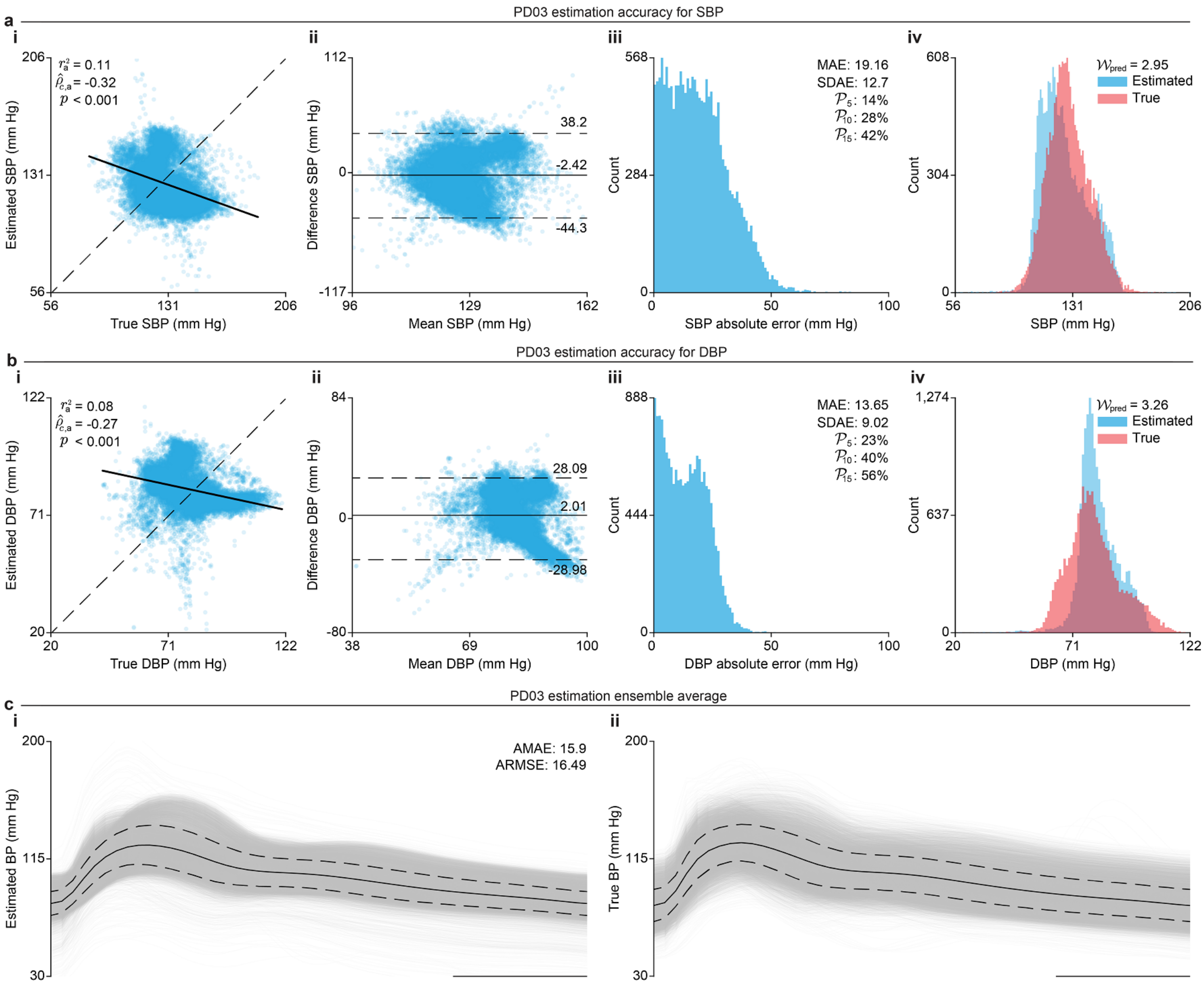

# Supplementary Fig. 96. Estimation results of population-disjoint model PD04

Aggregated results from PD04 configuration: Linear Regression class with impedance input and fiducial output, trained with the population-disjoint (PD) partition. **a,** Estimation accuracy for systolic brachial blood pressure (SBP); **b,** Estimation accuracy for diastolic brachial blood pressure (DBP); **i**, correlation plots; **ii**, limits of agreement (LOA) plots; **iii**, histogram of absolute errors (AE); and **iv**, histogram of estimated and true BP distributions. BP, blood pressure; DBP, diastolic blood pressure; SBP, systolic blood pressure. For correlation plots: $r_{\text{a}}^2$, aggregated coefficient of determination; $\hat{\rho}_{c,\text{a}}$, aggregated coefficient of concordance; solid line, empirical linear regression line; dashed line, $45°$ line of perfect correlation. For LOA plots: solid line, mean of errors between estimated and true BP values; dashed lines, 2.5th percentile (lower) and 97.5th percentile (upper). For AE histogram plots: MAE and SDAE, mean and standard deviation of AE, respectively; $\mathcal{P}_5$, $\mathcal{P}_{10}$, and $\mathcal{P}_{15}$, cumulative percentage of estimations with AE within 5, 10, and 15 mm Hg, respectively. For fiducial histogram plots: $\mathcal{W}_{\text{pred}}$, Wasserstein distance between true and estimated distribution.

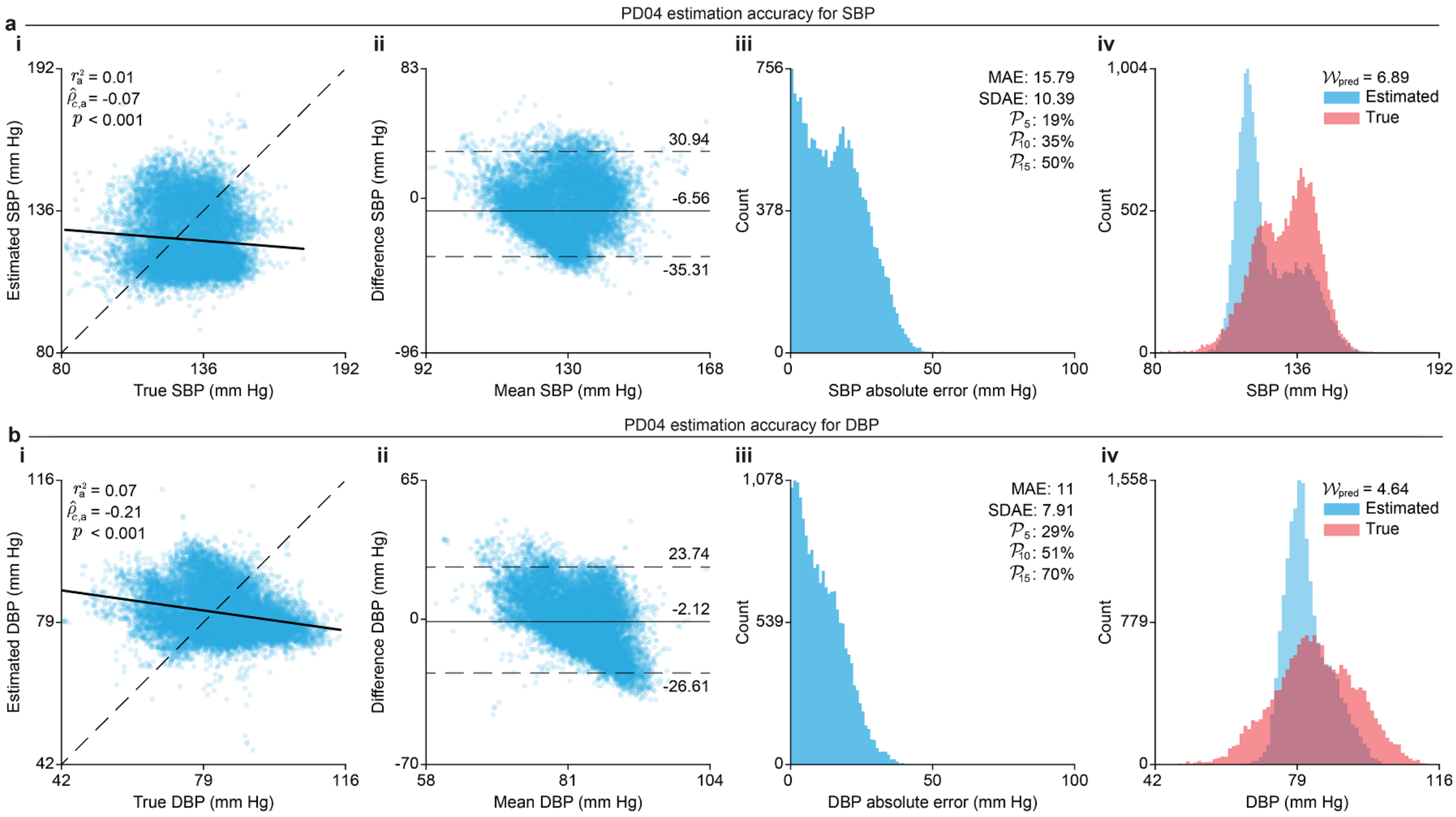

# Supplementary Fig. 97. Estimation results of population-disjoint model PD05

Aggregated results from PD05 configuration: Multilayer Perceptron class with image input and waveform output, trained with the population-disjoint (PD) partition. **a,** Estimation accuracy for systolic brachial blood pressure (SBP); **b,** Estimation accuracy for diastolic brachial blood pressure (DBP); **c,** Waveform ensemble of all estimated and true brachial blood pressure (BP) periods. For **a** and **b**: **i**, correlation plots; **ii**, limits of agreement (LOA) plots; **iii**, histogram of absolute errors (AE); and **iv**, histogram of estimated and true BP distributions. For **c**: **i**, ensemble of estimated BP periods; **ii**, ensemble of true BP periods. For correlation plots: $r_{\mathrm{a}}^2$, aggregated coefficient of determination; $\hat{\rho}_{c,\mathrm{a}}$, aggregated coefficient of concordance; solid line, empirical linear regression line; dashed line, $45°$ line of perfect correlation. For LOA plots: solid line, mean of errors between estimated and true BP values; dashed lines, 2.5$^{\text{th}}$ percentile (lower) and 97.5$^{\text{th}}$ percentile (upper). For AE histogram plots: MAE and SDAE, mean and standard deviation of AE, respectively; $\mathcal{P}_5$, $\mathcal{P}_{10}$, and $\mathcal{P}_{15}$, cumulative percentage of estimations with AE within 5, 10, and 15 mm Hg, respectively. For fiducial histogram plots: $\mathcal{W}_{\text{pred}}$, Wasserstein distance between true and estimated distribution. For ensemble plots: AMAE, average mean absolute error; ARMSE, average root mean square error; solid line, ensemble average of all periods; dashed lines, ensemble average $\pm$ standard deviation of all periods; scale bars, one-quarter of period.

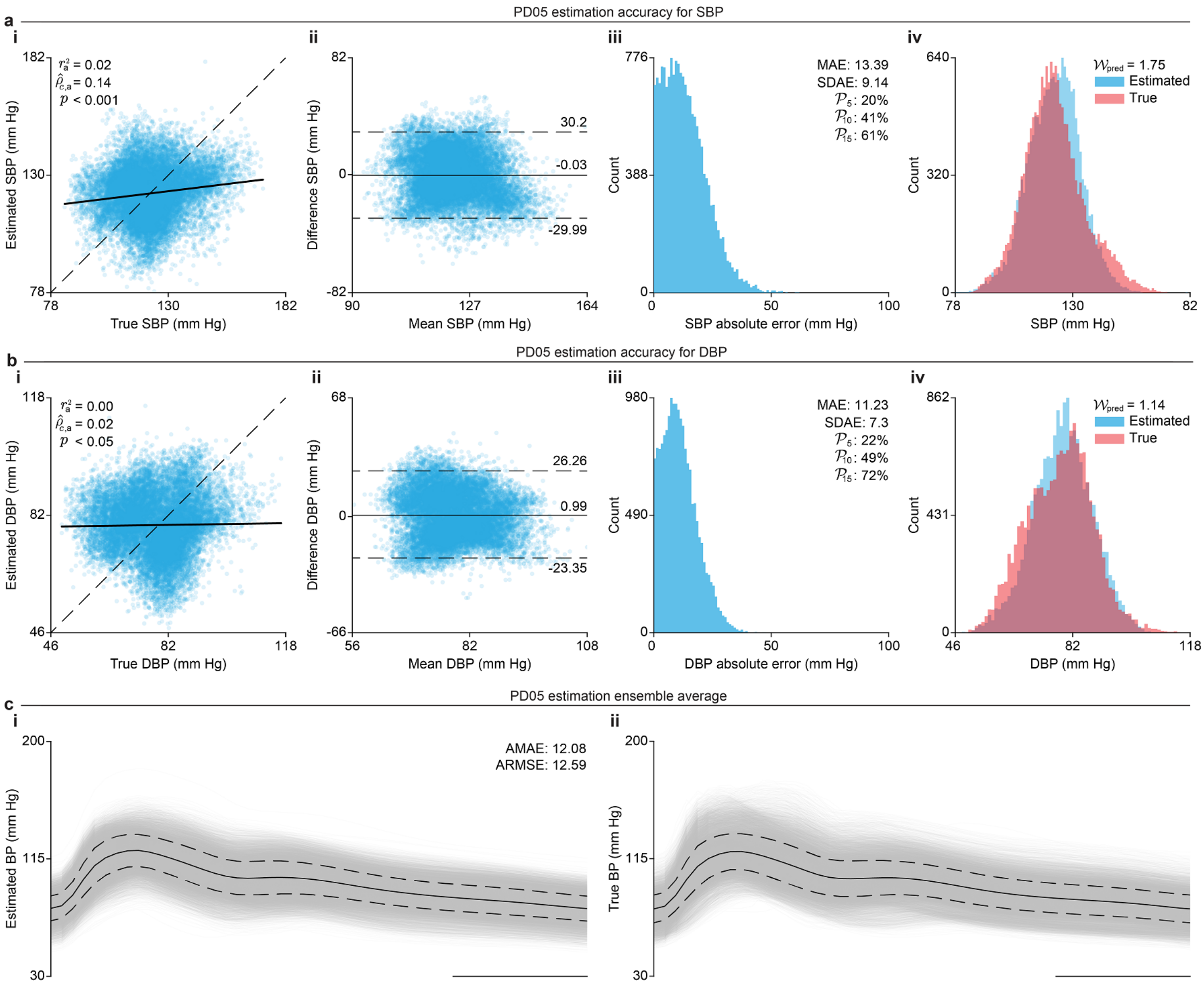

# Supplementary Fig. 98. Estimation results of population-disjoint model PD06

Aggregated results from PD06 configuration: Multilayer Perceptron class with image input and fiducial output, trained with the population-disjoint (PD) partition. **a,** Estimation accuracy for systolic brachial blood pressure (SBP); **b,** Estimation accuracy for diastolic brachial blood pressure (DBP); **i**, correlation plots; **ii**, limits of agreement (LOA) plots; **iii**, histogram of absolute errors (AE); and **iv**, histogram of estimated and true BP distributions. BP, blood pressure; DBP, diastolic blood pressure; SBP, systolic blood pressure. For correlation plots: $r_{\text{a}}^2$, aggregated coefficient of determination; $\hat{\rho}_{c,\text{a}}$, aggregated coefficient of concordance; solid line, empirical linear regression line; dashed line, $45°$ line of perfect correlation. For LOA plots: solid line, mean of errors between estimated and true BP values; dashed lines, 2.5$^{\text{th}}$ percentile (lower) and 97.5$^{\text{th}}$ percentile (upper). For AE histogram plots: MAE and SDAE, mean and standard deviation of AE, respectively; $\mathcal{P}_5$, $\mathcal{P}_{10}$, and $\mathcal{P}_{15}$, cumulative percentage of estimations with AE within 5, 10, and 15 mm Hg, respectively. For fiducial histogram plots: $\mathcal{W}_{\text{pred}}$, Wasserstein distance between true and estimated distribution.

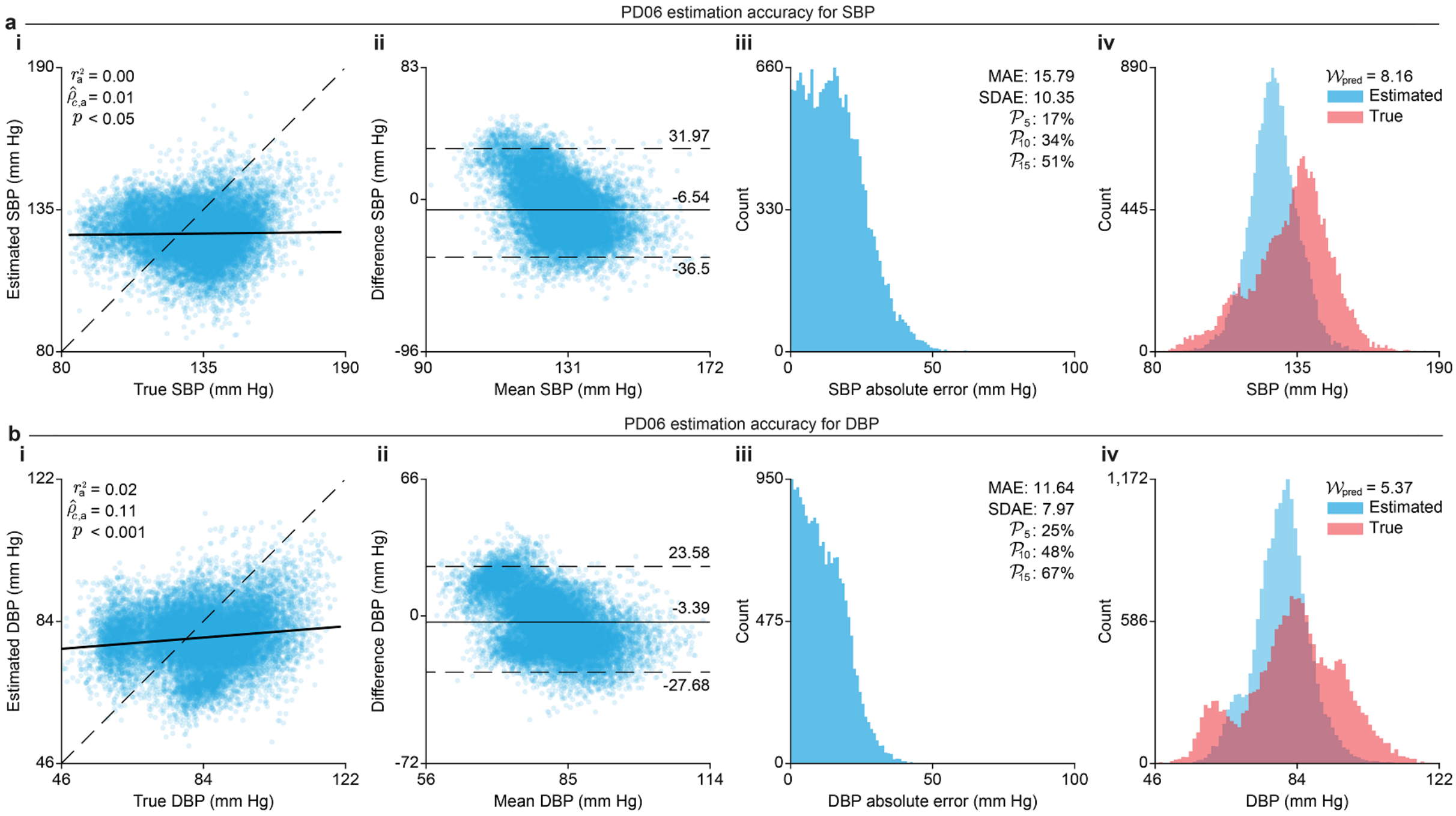

# Supplementary Fig. 99. Estimation results of population-disjoint model PD07

Aggregated results from PD07 configuration: Multilayer Perceptron class with impedance input and waveform output, trained with the population-disjoint (PD) partition. **a,** Estimation accuracy for systolic brachial blood pressure (SBP); **b,** Estimation accuracy for diastolic brachial blood pressure (DBP); **c,** Waveform ensemble of all estimated and true brachial blood pressure (BP) periods. For **a** and **b**: **i**, correlation plots; **ii**, limits of agreement (LOA) plots; **iii**, histogram of absolute errors (AE); and **iv**, histogram of estimated and true BP distributions. For **c**: **i**, ensemble of estimated BP periods; **ii**, ensemble of true BP periods. For correlation plots: $r^2_{\mathrm{a}}$, aggregated coefficient of determination; $\hat{\rho}_{c,\mathrm{a}}$, aggregated coefficient of concordance; solid line, empirical linear regression line; dashed line, $45°$ line of perfect correlation. For LOA plots: solid line, mean of errors between estimated and true BP values; dashed lines, 2.5[th] percentile (lower) and 97.5[th] percentile (upper). For AE histogram plots: MAE and SDAE, mean and standard deviation of AE, respectively; $\mathcal{P}_5$, $\mathcal{P}_{10}$, and $\mathcal{P}_{15}$, cumulative percentage of estimations with AE within 5, 10, and 15 mm Hg, respectively. For fiducial histogram plots: $\mathcal{W}_{\mathrm{pred}}$, Wasserstein distance between true and estimated distribution. For ensemble plots: AMAE, average mean absolute error; ARMSE, average root mean square error; solid line, ensemble average of all periods; dashed lines, ensemble average $\pm$ standard deviation of all periods; scale bars, one-quarter of period.

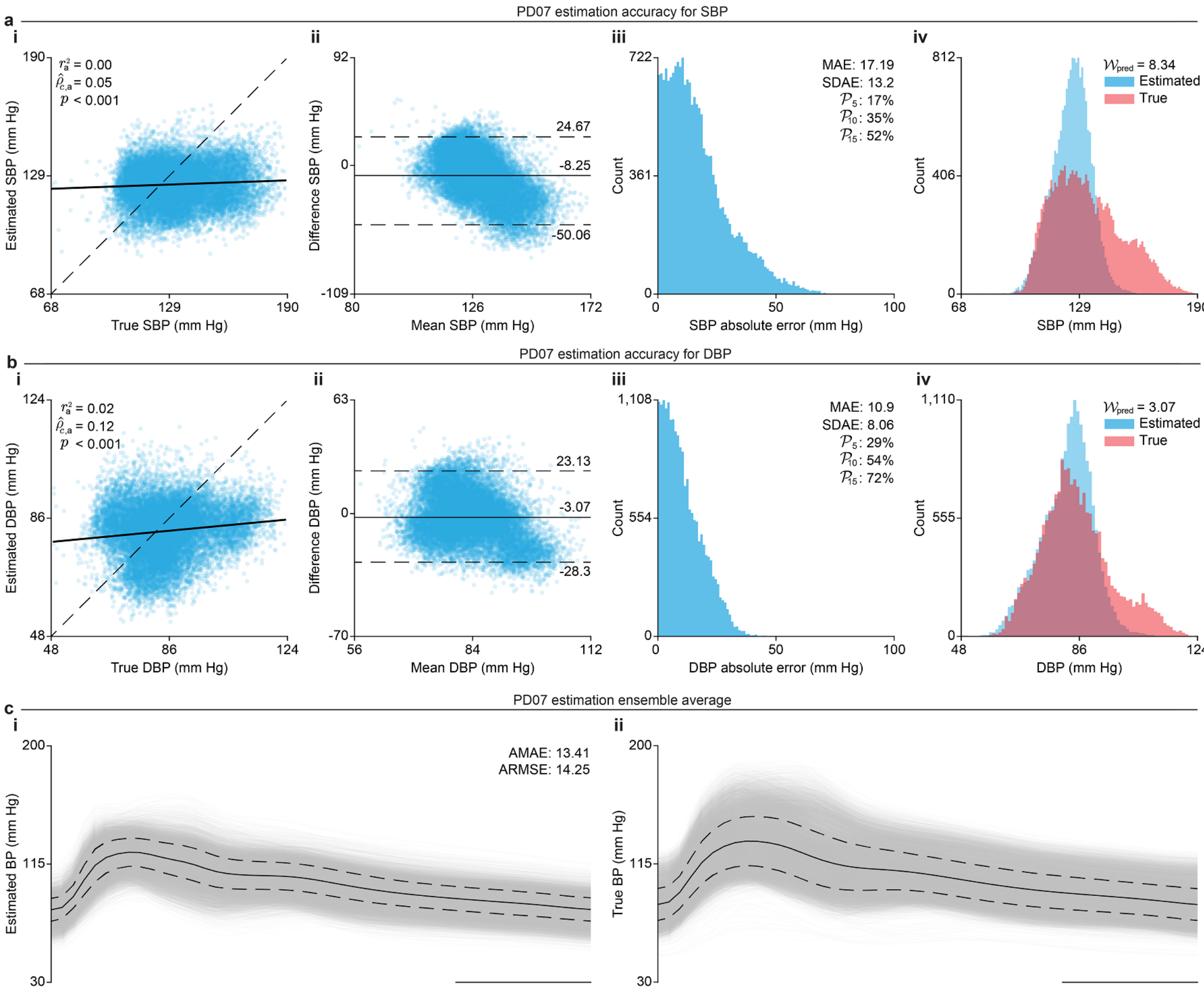

# Supplementary Fig. 100. Estimation results of population-disjoint model PD08

Aggregated results from PD08 configuration: Multilayer Perceptron class with impedance input and fiducial output, trained with the population-disjoint (PD) partition. **a,** Estimation accuracy for systolic brachial blood pressure (SBP); **b,** Estimation accuracy for diastolic brachial blood pressure (DBP); **i**, correlation plots; **ii**, limits of agreement (LOA) plots; **iii**, histogram of absolute errors (AE); and **iv**, histogram of estimated and true BP distributions. BP, blood pressure; DBP, diastolic blood pressure; SBP, systolic blood pressure. For correlation plots: $r_{\text{a}}^2$, aggregated coefficient of determination; $\hat{\rho}_{c,\text{a}}$, aggregated coefficient of concordance; solid line, empirical linear regression line; dashed line, $45°$ line of perfect correlation. For LOA plots: solid line, mean of errors between estimated and true BP values; dashed lines, 2.5$^{\text{th}}$ percentile (lower) and 97.5$^{\text{th}}$ percentile (upper). For AE histogram plots: MAE and SDAE, mean and standard deviation of AE, respectively; $\mathcal{P}_5$, $\mathcal{P}_{10}$, and $\mathcal{P}_{15}$, cumulative percentage of estimations with AE within 5, 10, and 15 mm Hg, respectively. For fiducial histogram plots: $\mathcal{W}_{\text{pred}}$, Wasserstein distance between true and estimated distribution.

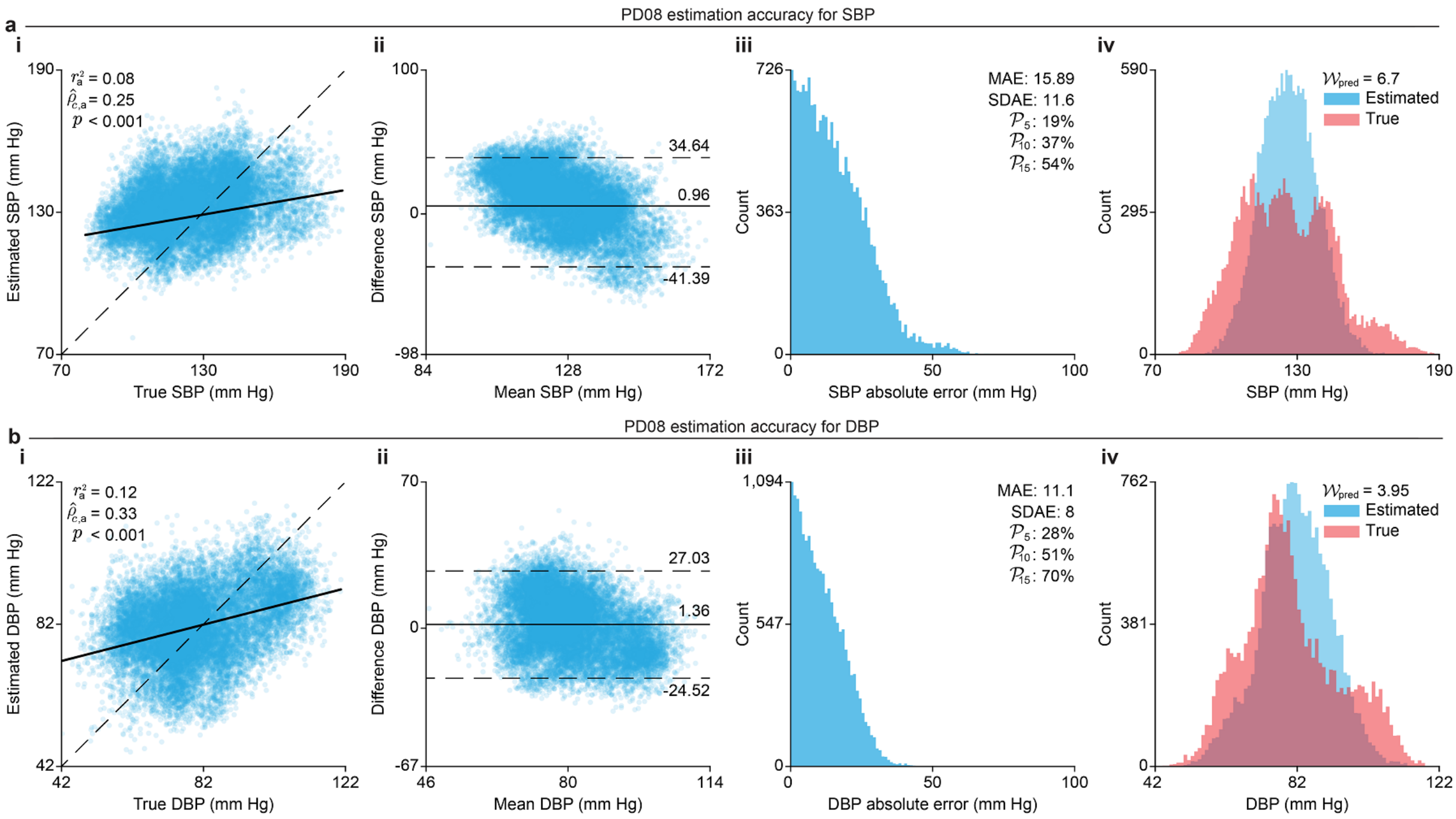

# Supplementary Fig. 101. Estimation results of population-disjoint model PD09

Aggregated results from PD09 configuration: Convolutional Neural Network class with image input and waveform output, trained with the population-disjoint (PD) partition. **a,** Estimation accuracy for systolic brachial blood pressure (SBP); **b,** Estimation accuracy for diastolic brachial blood pressure (DBP); **c,** Waveform ensemble of all estimated and true brachial blood pressure (BP) periods. For **a** and **b**: **i**, correlation plots; **ii**, limits of agreement (LOA) plots; **iii**, histogram of absolute errors (AE); and **iv**, histogram of estimated and true BP distributions. For **c**: **i**, ensemble of estimated BP periods; **ii**, ensemble of true BP periods. For correlation plots: $r^2_{\mathrm{a}}$, aggregated coefficient of determination; $\hat{\rho}_{c,\mathrm{a}}$, aggregated coefficient of concordance; solid line, empirical linear regression line; dashed line, $45°$ line of perfect correlation. For LOA plots: solid line, mean of errors between estimated and true BP values; dashed lines, 2.5$^{\text{th}}$ percentile (lower) and 97.5$^{\text{th}}$ percentile (upper). For AE histogram plots: MAE and SDAE, mean and standard deviation of AE, respectively; $\mathcal{P}_5$, $\mathcal{P}_{10}$, and $\mathcal{P}_{15}$, cumulative percentage of estimations with AE within 5, 10, and 15 mm Hg, respectively. For fiducial histogram plots: $\mathcal{W}_{\text{pred}}$, Wasserstein distance between true and estimated distribution. For ensemble plots: AMAE, average mean absolute error; ARMSE, average root mean square error; solid line, ensemble average of all periods; dashed lines, ensemble average $\pm$ standard deviation of all periods; scale bars, one-quarter of period.

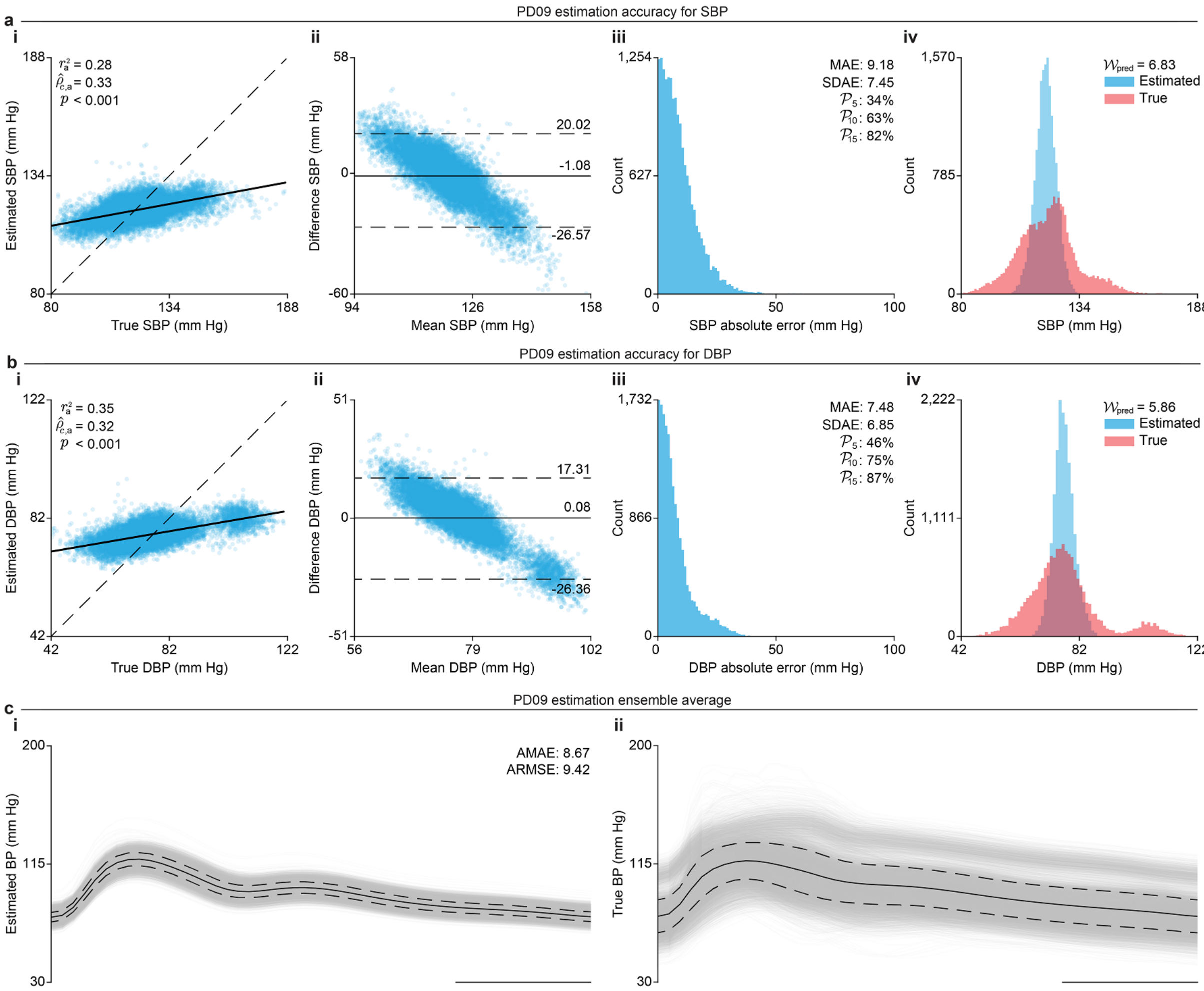

# Supplementary Fig. 102. Estimation results of population-disjoint model PD10

Aggregated results from PD10 configuration: Convolutional Neural Network class with image input and fiducial output, trained with the population-disjoint (PD) partition. **a,** Estimation accuracy for systolic brachial blood pressure (SBP); **b,** Estimation accuracy for diastolic brachial blood pressure (DBP); **i**, correlation plots; **ii**, limits of agreement (LOA) plots; **iii**, histogram of absolute errors (AE); and **iv**, histogram of estimated and true BP distributions. BP, blood pressure; DBP, diastolic blood pressure; SBP, systolic blood pressure. For correlation plots: $r_{\text{a}}^2$, aggregated coefficient of determination; $\hat{\rho}_{c,\text{a}}$, aggregated coefficient of concordance; solid line, empirical linear regression line; dashed line, $45°$ line of perfect correlation. For LOA plots: solid line, mean of errors between estimated and true BP values; dashed lines, 2.5th percentile (lower) and 97.5th percentile (upper). For AE histogram plots: MAE and SDAE, mean and standard deviation of AE, respectively; $\mathcal{P}_5$, $\mathcal{P}_{10}$, and $\mathcal{P}_{15}$, cumulative percentage of estimations with AE within 5, 10, and 15 mm Hg, respectively. For fiducial histogram plots: $\mathcal{W}_{\text{pred}}$, Wasserstein distance between true and estimated distribution.

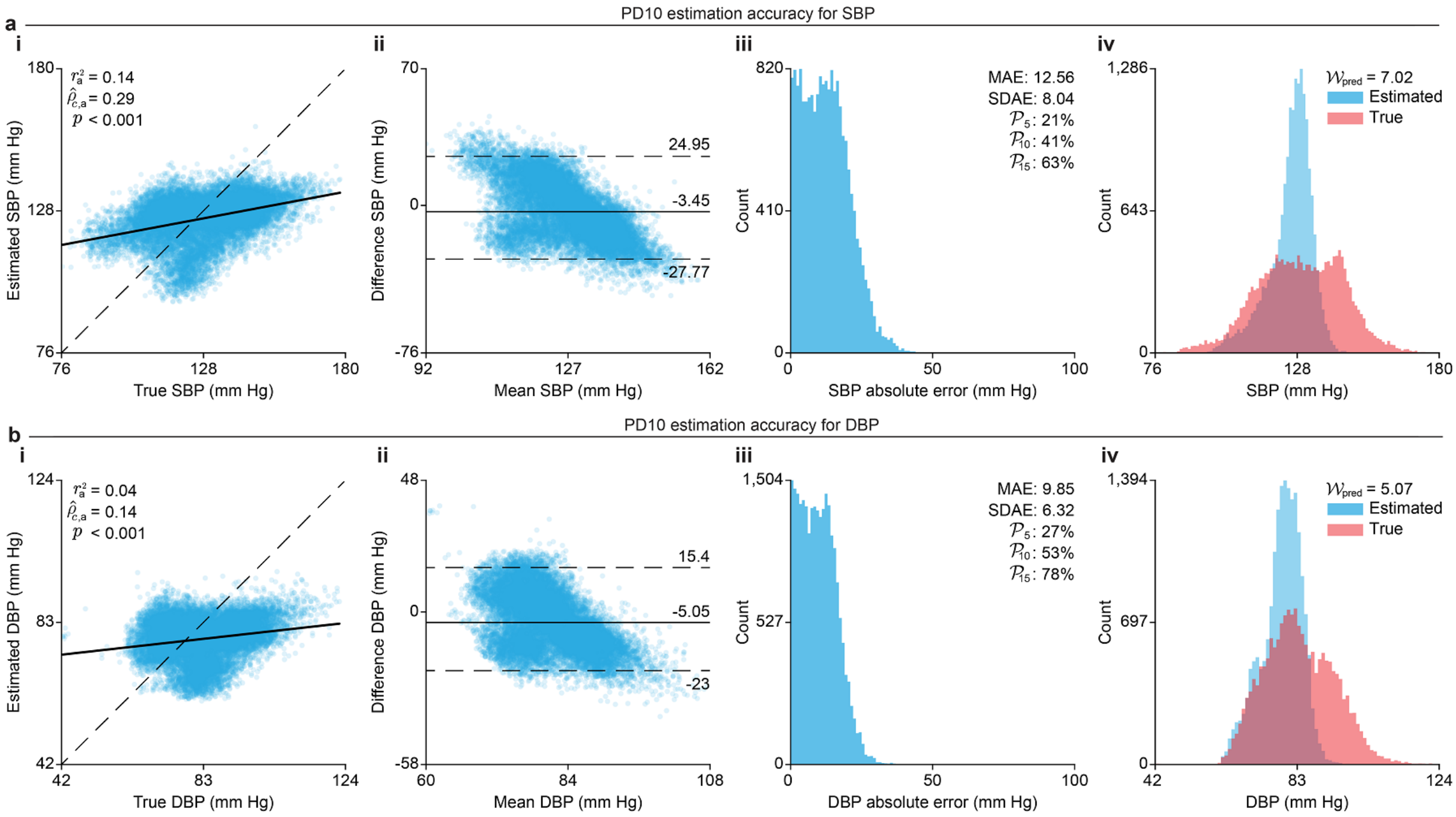

# Supplementary Fig. 103. Estimation results of population-disjoint model PD11

Aggregated results from PD11 configuration: Convolutional Neural Network class with impedance input and waveform output, trained with the population-disjoint (PD) partition. **a,** Estimation accuracy for systolic brachial blood pressure (SBP); **b,** Estimation accuracy for diastolic brachial blood pressure (DBP); **c,** Waveform ensemble of all estimated and true brachial blood pressure (BP) periods. For **a** and **b**: **i**, correlation plots; **ii**, limits of agreement (LOA) plots; **iii**, histogram of absolute errors (AE); and **iv**, histogram of estimated and true BP distributions. For **c**: **i**, ensemble of estimated BP periods; **ii**, ensemble of true BP periods. For correlation plots: $r^2_{\mathrm{a}}$, aggregated coefficient of determination; $\hat{\rho}_{c,\mathrm{a}}$, aggregated coefficient of concordance; solid line, empirical linear regression line; dashed line, $45°$ line of perfect correlation. For LOA plots: solid line, mean of errors between estimated and true BP values; dashed lines, 2.5$^{\mathrm{th}}$ percentile (lower) and 97.5$^{\mathrm{th}}$ percentile (upper). For AE histogram plots: MAE and SDAE, mean and standard deviation of AE, respectively; $\mathcal{P}_5$, $\mathcal{P}_{10}$, and $\mathcal{P}_{15}$, cumulative percentage of estimations with AE within 5, 10, and 15 mm Hg, respectively. For fiducial histogram plots: $\mathcal{W}_{\mathrm{pred}}$, Wasserstein distance between true and estimated distribution. For ensemble plots: AMAE, average mean absolute error; ARMSE, average root mean square error; solid line, ensemble average of all periods; dashed lines, ensemble average $\pm$ standard deviation of all periods; scale bars, one-quarter of period.

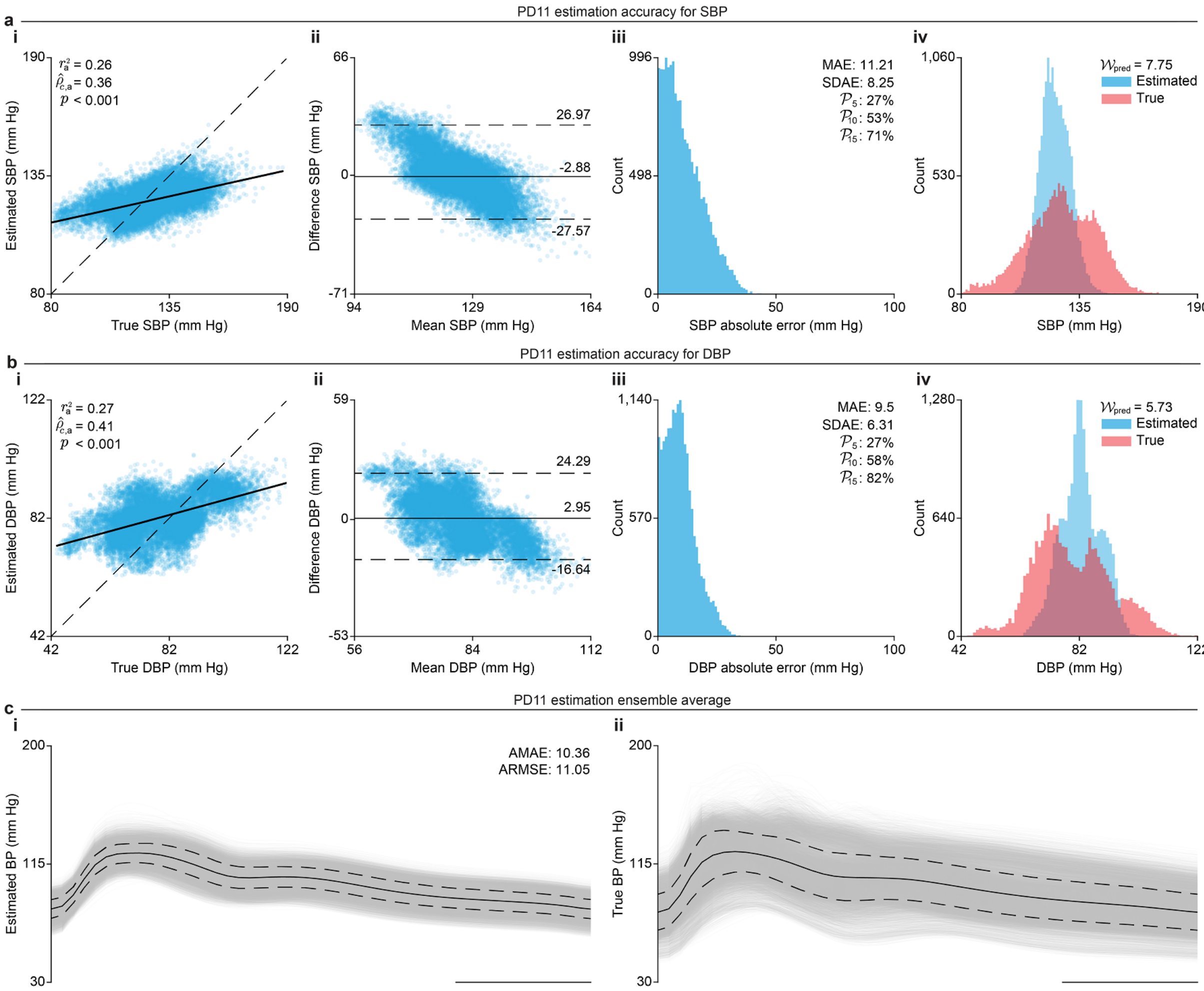

# Supplementary Fig. 104. Estimation results of population-disjoint model PD12

Aggregated results from PD12 configuration: Convolutional Neural Network class with impedance input and fiducial output, trained with the population-disjoint (PD) partition. **a,** Estimation accuracy for systolic brachial blood pressure (SBP); **b,** Estimation accuracy for diastolic brachial blood pressure (DBP); **i**, correlation plots; **ii**, limits of agreement (LOA) plots; **iii**, histogram of absolute errors (AE); and **iv**, histogram of estimated and true BP distributions. BP, blood pressure; DBP, diastolic blood pressure; SBP, systolic blood pressure. For correlation plots: $r^2_{\text{a}}$, aggregated coefficient of determination; $\hat{\rho}_{c,\text{a}}$, aggregated coefficient of concordance; solid line, empirical linear regression line; dashed line, $45°$ line of perfect correlation. For LOA plots: solid line, mean of errors between estimated and true BP values; dashed lines, 2.5$^{\text{th}}$ percentile (lower) and 97.5$^{\text{th}}$ percentile (upper). For AE histogram plots: MAE and SDAE, mean and standard deviation of AE, respectively; $\mathcal{P}_5$, $\mathcal{P}_{10}$, and $\mathcal{P}_{15}$, cumulative percentage of estimations with AE within 5, 10, and 15 mm Hg, respectively. For fiducial histogram plots: $\mathcal{W}_{\text{pred}}$, Wasserstein distance between true and estimated distribution.

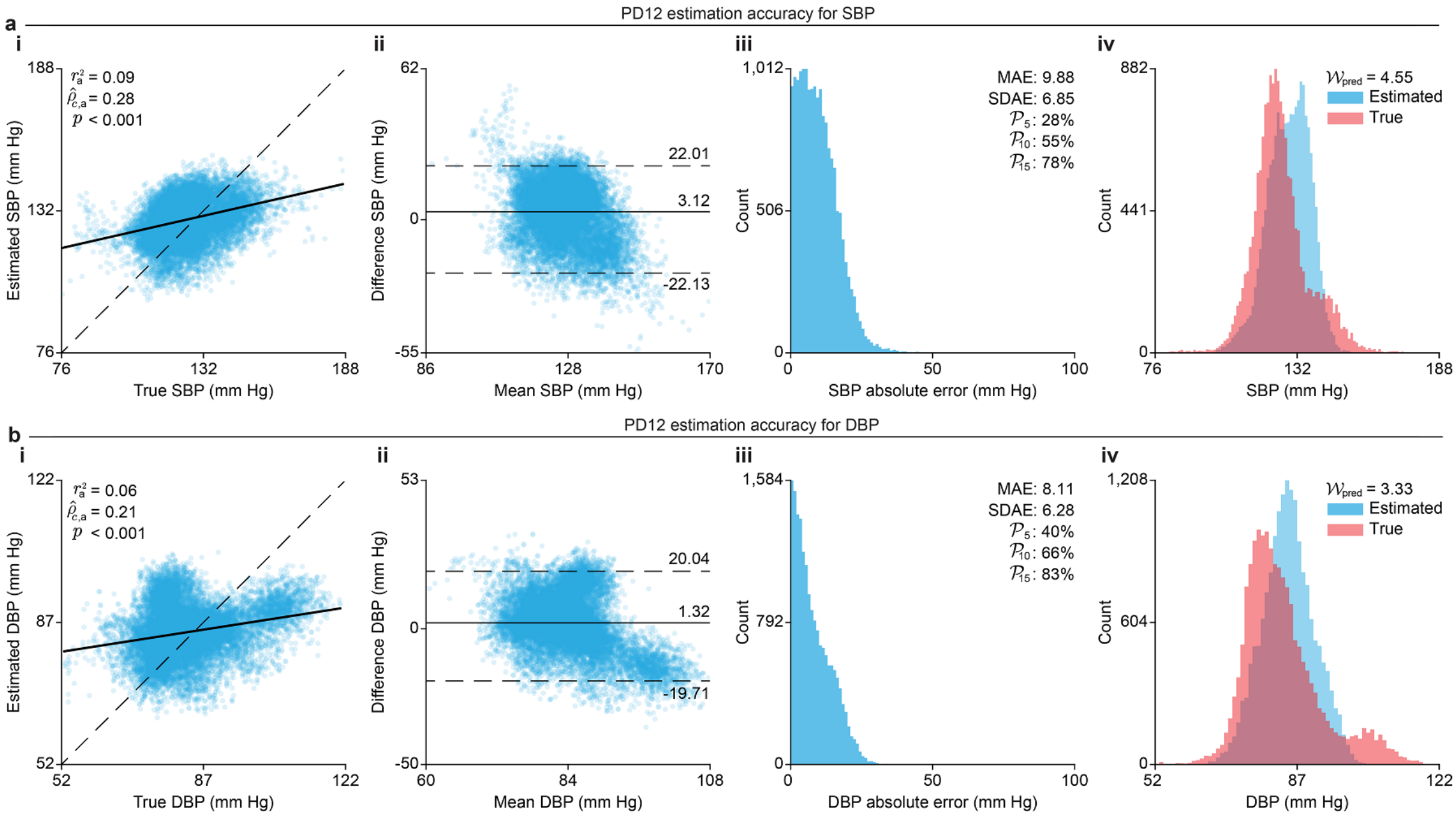

# Supplementary Fig. 105. Estimation results of population-disjoint model PD13

Aggregated results from PD13 configuration: Convolutional Recurrent Transformer class with image input and waveform output, trained with the population-disjoint (PD) partition. **a,** Estimation accuracy for systolic brachial blood pressure (SBP); **b,** Estimation accuracy for diastolic brachial blood pressure (DBP); **c,** Waveform ensemble of all estimated and true brachial blood pressure (BP) periods. For **a** and **b**: **i**, correlation plots; **ii**, limits of agreement (LOA) plots; **iii**, histogram of absolute errors (AE); and **iv**, histogram of estimated and true BP distributions. For **c**: **i**, ensemble of estimated BP periods; **ii**, ensemble of true BP periods. For correlation plots: $r_{\mathrm{a}}^2$, aggregated coefficient of determination; $\hat{\rho}_{c,\mathrm{a}}$, aggregated coefficient of concordance; solid line, empirical linear regression line; dashed line, $45°$ line of perfect correlation. For LOA plots: solid line, mean of errors between estimated and true BP values; dashed lines, 2.5[th] percentile (lower) and 97.5[th] percentile (upper). For AE histogram plots: MAE and SDAE, mean and standard deviation of AE, respectively; $\mathcal{P}_5$, $\mathcal{P}_{10}$, and $\mathcal{P}_{15}$, cumulative percentage of estimations with AE within 5, 10, and 15 mm Hg, respectively. For fiducial histogram plots: $\mathcal{W}_{\mathrm{pred}}$, Wasserstein distance between true and estimated distribution. For ensemble plots: AMAE, average mean absolute error; ARMSE, average root mean square error; solid line, ensemble average of all periods; dashed lines, ensemble average $\pm$ standard deviation of all periods; scale bars, one-quarter of period.

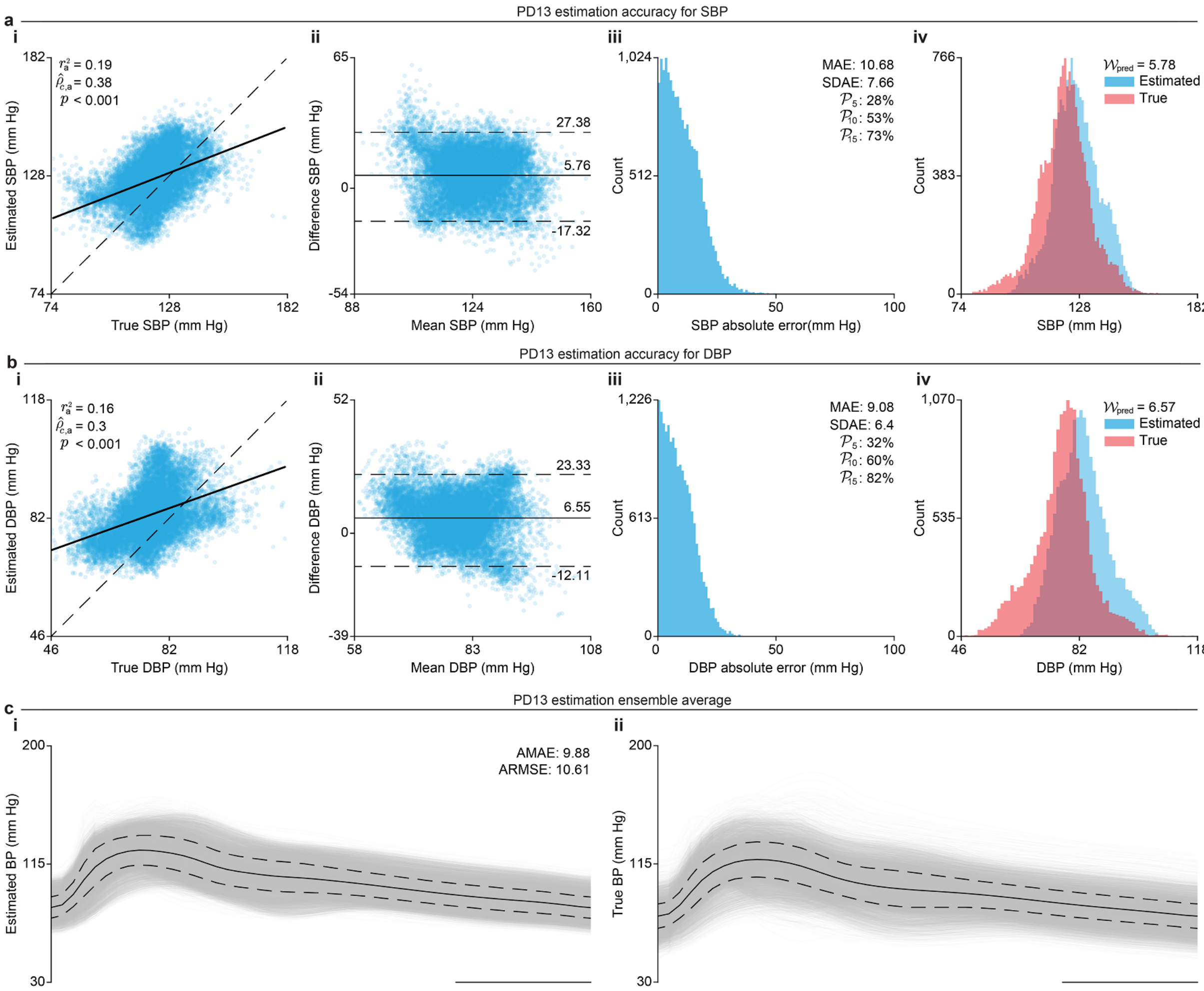

# Supplementary Fig. 106. Estimation results of population-disjoint model PD14

Aggregated results from PD14 configuration: Convolutional Recurrent Transformer class with image input and fiducial output, trained with the population-disjoint (PD) partition. **a,** Estimation accuracy for systolic brachial blood pressure (SBP); **b,** Estimation accuracy for diastolic brachial blood pressure (DBP); **i**, correlation plots; **ii**, limits of agreement (LOA) plots; **iii**, histogram of absolute errors (AE); and **iv**, histogram of estimated and true BP distributions. BP, blood pressure; DBP, diastolic blood pressure; SBP, systolic blood pressure. For correlation plots: $r_{\text{a}}^2$, aggregated coefficient of determination; $\hat{\rho}_{c,\text{a}}$, aggregated coefficient of concordance; solid line, empirical linear regression line; dashed line, $45°$ line of perfect correlation. For LOA plots: solid line, mean of errors between estimated and true BP values; dashed lines, 2.5$^{\text{th}}$ percentile (lower) and 97.5$^{\text{th}}$ percentile (upper). For AE histogram plots: MAE and SDAE, mean and standard deviation of AE, respectively; $\mathcal{P}_5$, $\mathcal{P}_{10}$, and $\mathcal{P}_{15}$, cumulative percentage of estimations with AE within 5, 10, and 15 mm Hg, respectively. For fiducial histogram plots: $\mathcal{W}_{\text{pred}}$, Wasserstein distance between true and estimated distribution.

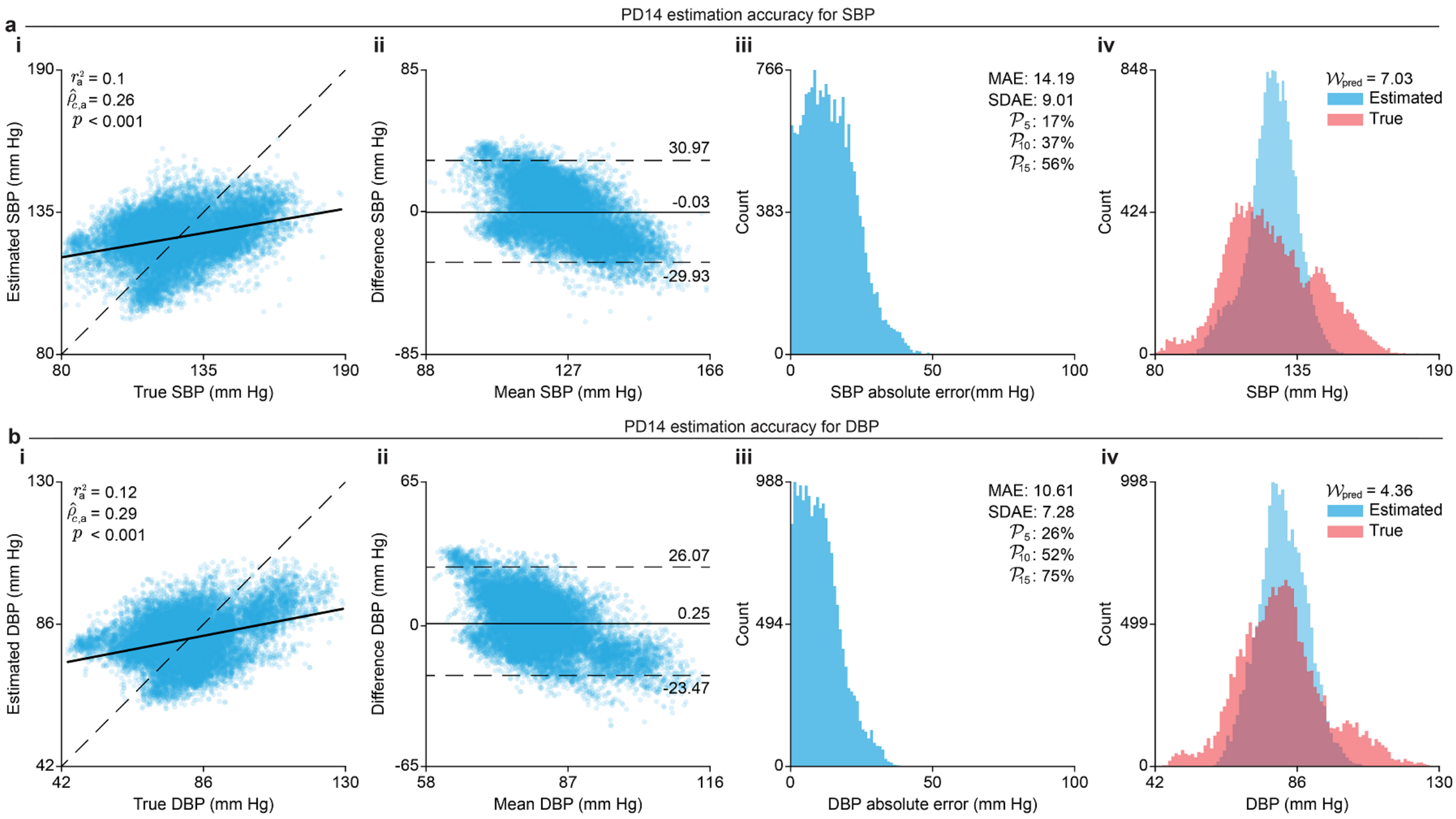

# Supplementary Fig. 107. Estimation results of population-disjoint model PD15

Aggregated results from PD15 configuration: Convolutional Recurrent Transformer class with impedance input and waveform output, trained with the population-disjoint (PD) partition. **a,** Estimation accuracy for systolic brachial blood pressure (SBP); **b,** Estimation accuracy for diastolic brachial blood pressure (DBP); **c,** Waveform ensemble of all estimated and true brachial blood pressure (BP) periods. For **a** and **b**: **i**, correlation plots; **ii**, limits of agreement (LOA) plots; **iii**, histogram of absolute errors (AE); and **iv**, histogram of estimated and true BP distributions. For **c**: **i**, ensemble of estimated BP periods; **ii**, ensemble of true BP periods. For correlation plots: $r^2_{\mathrm{a}}$, aggregated coefficient of determination; $\hat{\rho}_{c,\mathrm{a}}$, aggregated coefficient of concordance; solid line, empirical linear regression line; dashed line, $45°$ line of perfect correlation. For LOA plots: solid line, mean of errors between estimated and true BP values; dashed lines, 2.5$^{\mathrm{th}}$ percentile (lower) and 97.5$^{\mathrm{th}}$ percentile (upper). For AE histogram plots: MAE and SDAE, mean and standard deviation of AE, respectively; $\mathcal{P}_5$, $\mathcal{P}_{10}$, and $\mathcal{P}_{15}$, cumulative percentage of estimations with AE within 5, 10, and 15 mm Hg, respectively. For fiducial histogram plots: $\mathcal{W}_{\mathrm{pred}}$, Wasserstein distance between true and estimated distribution. For ensemble plots: AMAE, average mean absolute error; ARMSE, average root mean square error; solid line, ensemble average of all periods; dashed lines, ensemble average $\pm$ standard deviation of all periods; scale bars, one-quarter of period.

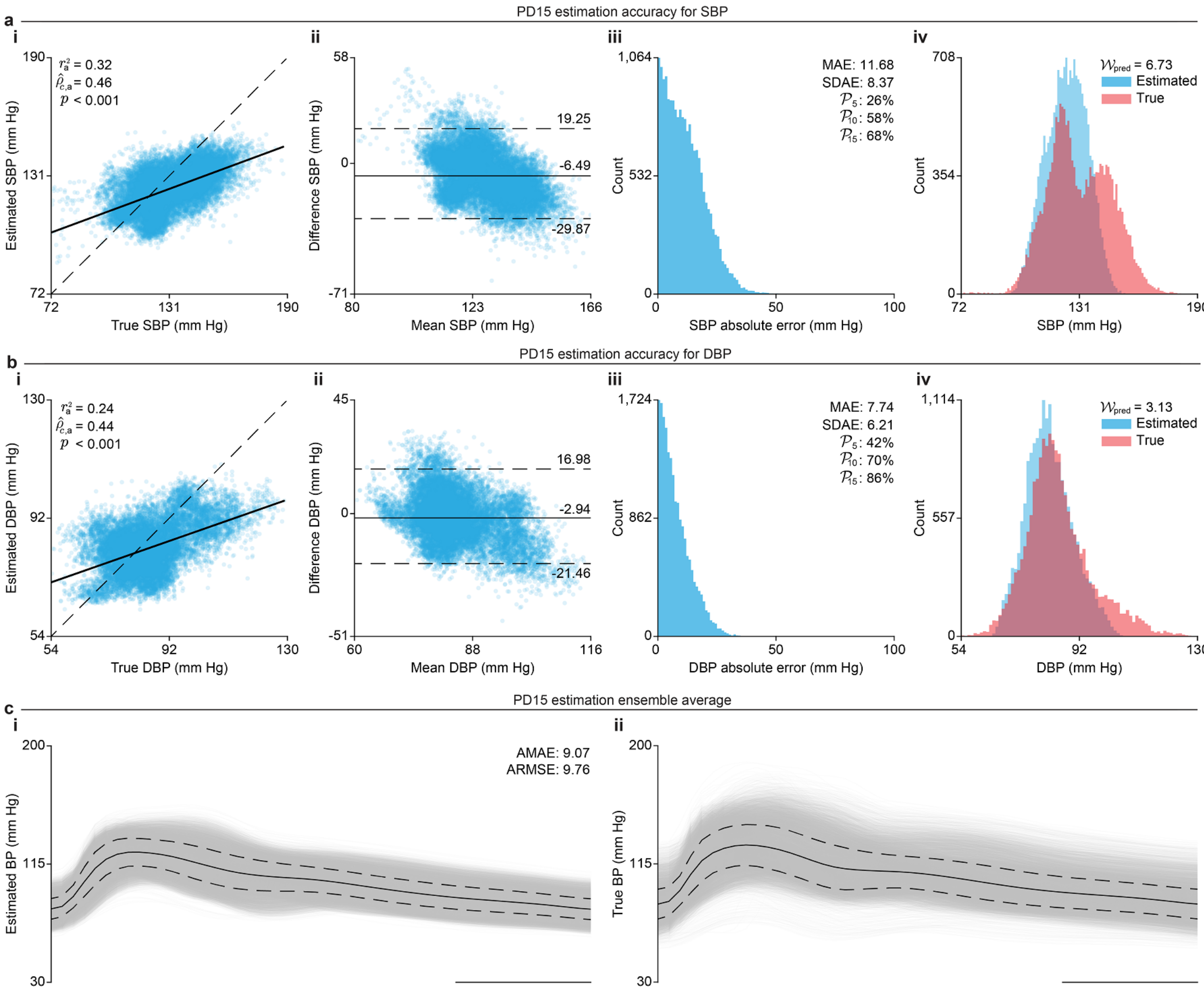

# Supplementary Fig. 108. Estimation results of population-disjoint model PD16

Aggregated results from PD16 configuration: Convolutional Recurrent Transformer class with impedance input and fiducial output, trained with the population-disjoint (PD) partition. **a,** Estimation accuracy for systolic brachial blood pressure (SBP); **b,** Estimation accuracy for diastolic brachial blood pressure (DBP); **i**, correlation plots; **ii**, limits of agreement (LOA) plots; **iii**, histogram of absolute errors (AE); and **iv**, histogram of estimated and true BP distributions. BP, blood pressure; DBP, diastolic blood pressure; SBP, systolic blood pressure. For correlation plots: $r_{\text{a}}^2$, aggregated coefficient of determination; $\hat{\rho}_{c,\text{a}}$, aggregated coefficient of concordance; solid line, empirical linear regression line; dashed line, $45°$ line of perfect correlation. For LOA plots: solid line, mean of errors between estimated and true BP values; dashed lines, 2.5th percentile (lower) and 97.5th percentile (upper). For AE histogram plots: MAE and SDAE, mean and standard deviation of AE, respectively; $\mathcal{P}_5$, $\mathcal{P}_{10}$, and $\mathcal{P}_{15}$, cumulative percentage of estimations with AE within 5, 10, and 15 mm Hg, respectively. For fiducial histogram plots: $\mathcal{W}_{\text{pred}}$, Wasserstein distance between true and estimated distribution.

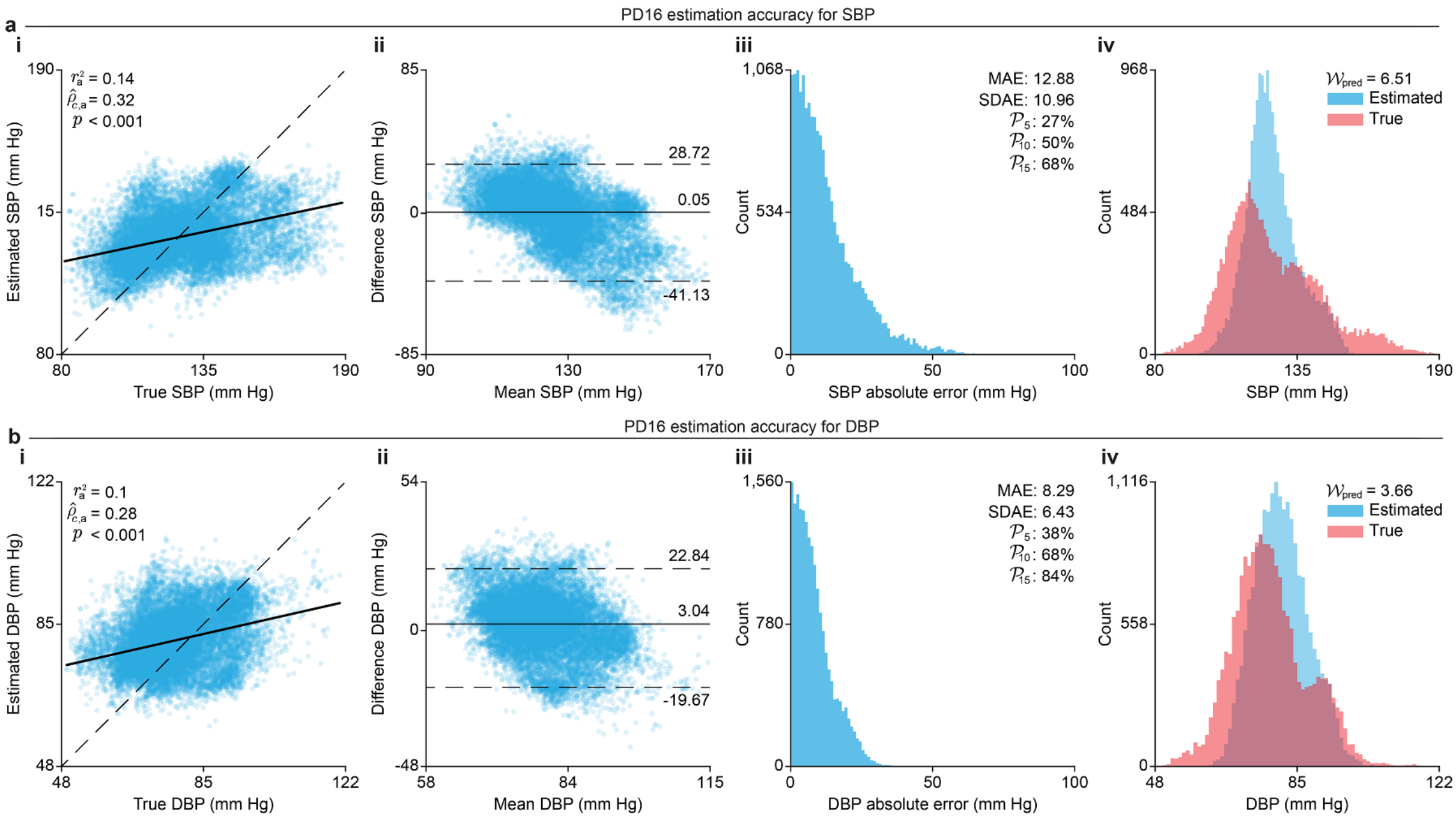

# Supplementary Fig. 109. Estimation results of population-disjoint model PD17

Aggregated results from PD17 configuration: Convolutional Recurrent Samba class with image input and waveform output, trained with the population-disjoint (PD) partition. **a,** Estimation accuracy for systolic brachial blood pressure (SBP); **b,** Estimation accuracy for diastolic brachial blood pressure (DBP); **c,** Waveform ensemble of all estimated and true brachial blood pressure (BP) periods. For **a** and **b**: **i**, correlation plots; **ii**, limits of agreement (LOA) plots; **iii**, histogram of absolute errors (AE); and **iv**, histogram of estimated and true BP distributions. For **c**: **i**, ensemble of estimated BP periods; **ii**, ensemble of true BP periods. For correlation plots: $r^2_{\text{a}}$, aggregated coefficient of determination; $\hat{\rho}_{c,\text{a}}$, aggregated coefficient of concordance; solid line, empirical linear regression line; dashed line, $45°$ line of perfect correlation. For LOA plots: solid line, mean of errors between estimated and true BP values; dashed lines, 2.5$^{\text{th}}$ percentile (lower) and 97.5$^{\text{th}}$ percentile (upper). For AE histogram plots: MAE and SDAE, mean and standard deviation of AE, respectively; $\mathcal{P}_5$, $\mathcal{P}_{10}$, and $\mathcal{P}_{15}$, cumulative percentage of estimations with AE within 5, 10, and 15 mm Hg, respectively. For fiducial histogram plots: $\mathcal{W}_{\text{pred}}$, Wasserstein distance between true and estimated distribution. For ensemble plots: AMAE, average mean absolute error; ARMSE, average root mean square error; solid line, ensemble average of all periods; dashed lines, ensemble average $\pm$ standard deviation of all periods; scale bars, one-quarter of period.

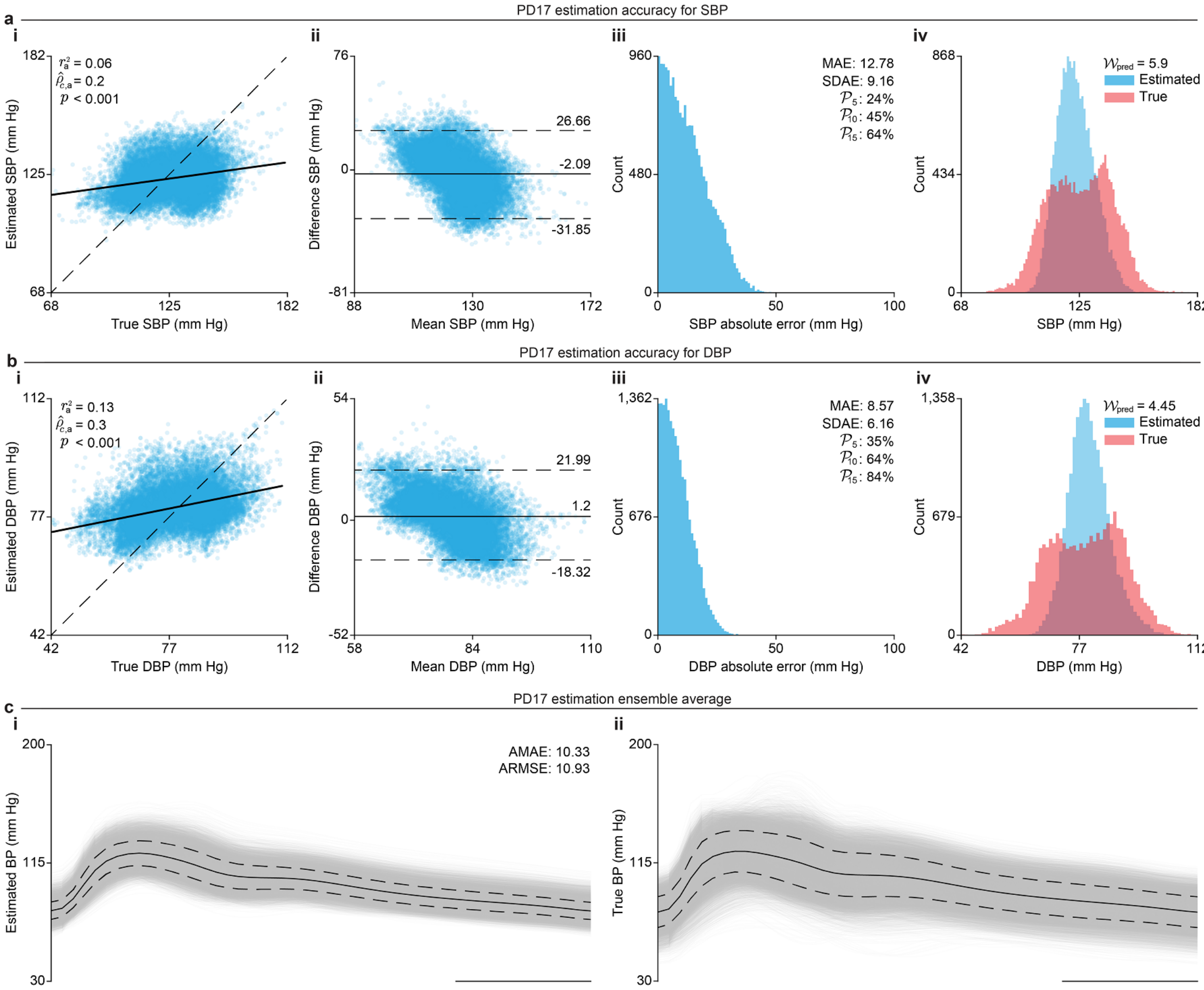

# Supplementary Fig. 110. Estimation results of population-disjoint model PD18

Aggregated results from PD18 configuration: Convolutional Recurrent Samba class with image input and fiducial output, trained with the population-disjoint (PD) partition. **a,** Estimation accuracy for systolic brachial blood pressure (SBP); **b,** Estimation accuracy for diastolic brachial blood pressure (DBP); **i**, correlation plots; **ii**, limits of agreement (LOA) plots; **iii**, histogram of absolute errors (AE); and **iv**, histogram of estimated and true BP distributions. BP, blood pressure; DBP, diastolic blood pressure; SBP, systolic blood pressure. For correlation plots: $r_a^2$, aggregated coefficient of determination; $\hat{\rho}_{c,a}$, aggregated coefficient of concordance; solid line, empirical linear regression line; dashed line, $45°$ line of perfect correlation. For LOA plots: solid line, mean of errors between estimated and true BP values; dashed lines, 2.5$^{th}$ percentile (lower) and 97.5$^{th}$ percentile (upper). For AE histogram plots: MAE and SDAE, mean and standard deviation of AE, respectively; $\mathcal{P}_5$, $\mathcal{P}_{10}$, and $\mathcal{P}_{15}$, cumulative percentage of estimations with AE within 5, 10, and 15 mm Hg, respectively. For fiducial histogram plots: $\mathcal{W}_{pred}$, Wasserstein distance between true and estimated distribution.

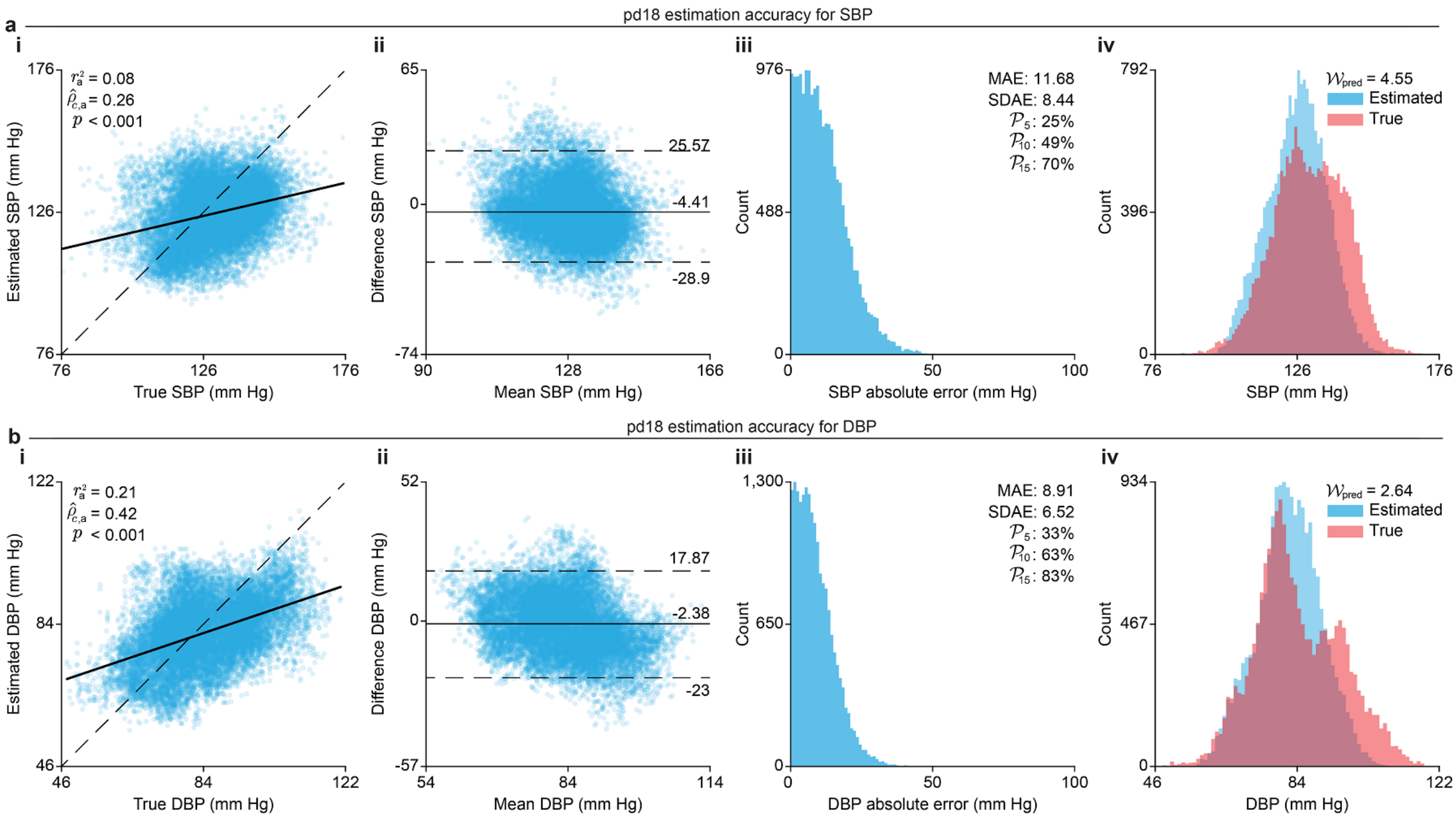

# Supplementary Fig. 111. Estimation results of population-disjoint model PD19

Aggregated results from PD19 configuration: Convolutional Recurrent Samba class with impedance input and waveform output, trained with the population-disjoint (PD) partition. **a,** Estimation accuracy for systolic brachial blood pressure (SBP); **b,** Estimation accuracy for diastolic brachial blood pressure (DBP); **c,** Waveform ensemble of all estimated and true brachial blood pressure (BP) periods. For **a** and **b**: **i**, correlation plots; **ii**, limits of agreement (LOA) plots; **iii**, histogram of absolute errors (AE); and **iv**, histogram of estimated and true BP distributions. For **c**: **i**, ensemble of estimated BP periods; **ii**, ensemble of true BP periods. For correlation plots: $r_{\text{a}}^2$, aggregated coefficient of determination; $\hat{\rho}_{c,\text{a}}$, aggregated coefficient of concordance; solid line, empirical linear regression line; dashed line, $45°$ line of perfect correlation. For LOA plots: solid line, mean of errors between estimated and true BP values; dashed lines, 2.5$^{\text{th}}$ percentile (lower) and 97.5$^{\text{th}}$ percentile (upper). For AE histogram plots: MAE and SDAE, mean and standard deviation of AE, respectively; $\mathcal{P}_5$, $\mathcal{P}_{10}$, and $\mathcal{P}_{15}$, cumulative percentage of estimations with AE within 5, 10, and 15 mm Hg, respectively. For fiducial histogram plots: $\mathcal{W}_{\text{pred}}$, Wasserstein distance between true and estimated distribution. For ensemble plots: AMAE, average mean absolute error; ARMSE, average root mean square error; solid line, ensemble average of all periods; dashed lines, ensemble average $\pm$ standard deviation of all periods; scale bars, one-quarter of period.

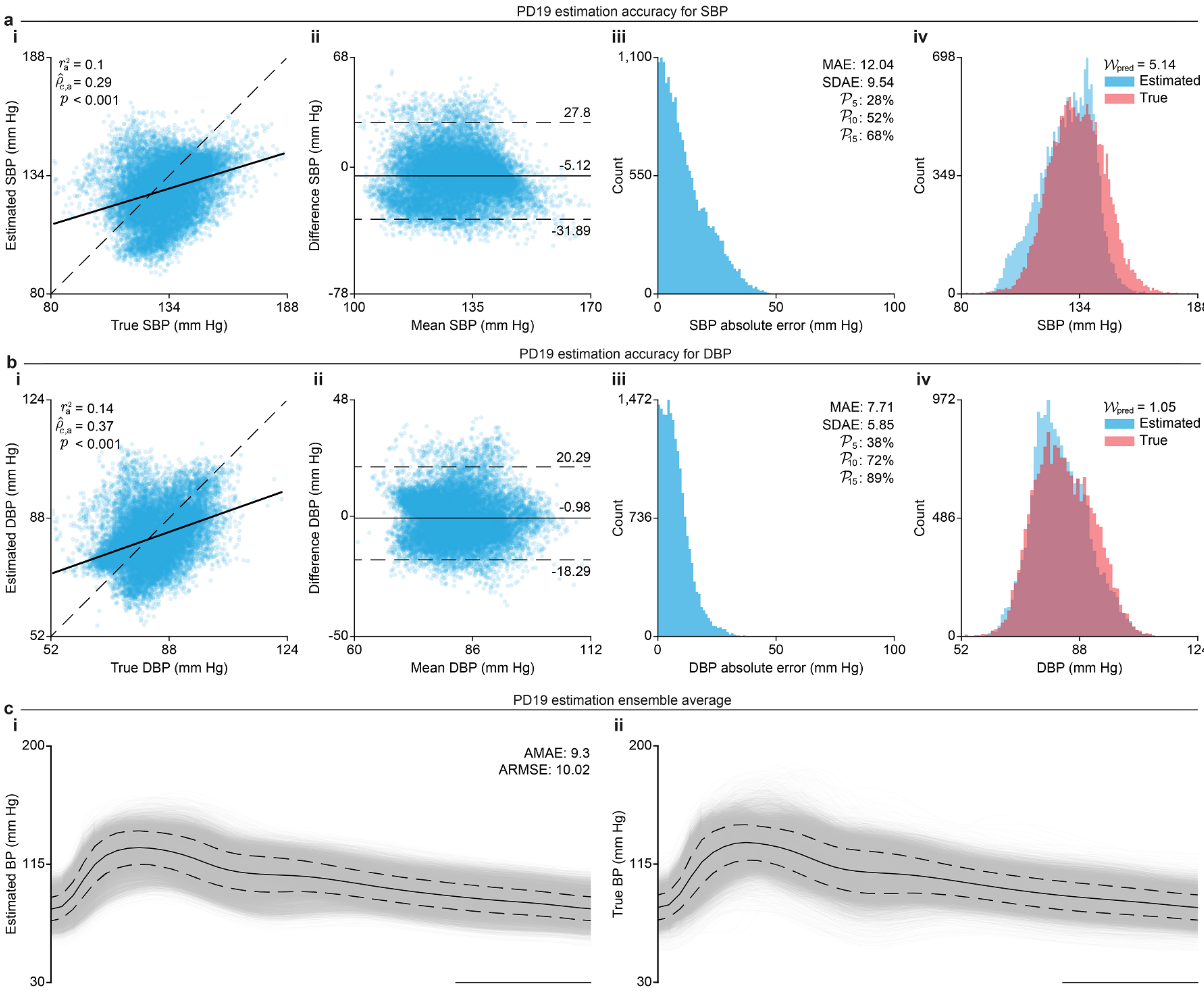

## Supplementary Fig. 112. Estimation results of population-disjoint model PD20

Aggregated results from PD20 configuration: Convolutional Recurrent Samba class with impedance input and fiducial output, trained with the population-disjoint (PD) partition. **a,** Estimation accuracy for systolic brachial blood pressure (SBP); **b,** Estimation accuracy for diastolic brachial blood pressure (DBP); **i**, correlation plots; **ii**, limits of agreement (LOA) plots; **iii**, histogram of absolute errors (AE); and **iv**, histogram of estimated and true BP distributions. BP, blood pressure; DBP, diastolic blood pressure; SBP, systolic blood pressure. For correlation plots: $r^2_{\text{a}}$, aggregated coefficient of determination; $\hat{\rho}_{c,\text{a}}$, aggregated coefficient of concordance; solid line, empirical linear regression line; dashed line, $45°$ line of perfect correlation. For LOA plots: solid line, mean of errors between estimated and true BP values; dashed lines, 2.5$^{\text{th}}$ percentile (lower) and 97.5$^{\text{th}}$ percentile (upper). For AE histogram plots: MAE and SDAE, mean and standard deviation of AE, respectively; $\mathcal{P}_5$, $\mathcal{P}_{10}$, and $\mathcal{P}_{15}$, cumulative percentage of estimations with AE within 5, 10, and 15 mm Hg, respectively. For fiducial histogram plots: $\mathcal{W}_{\text{pred}}$, Wasserstein distance between true and estimated distribution.

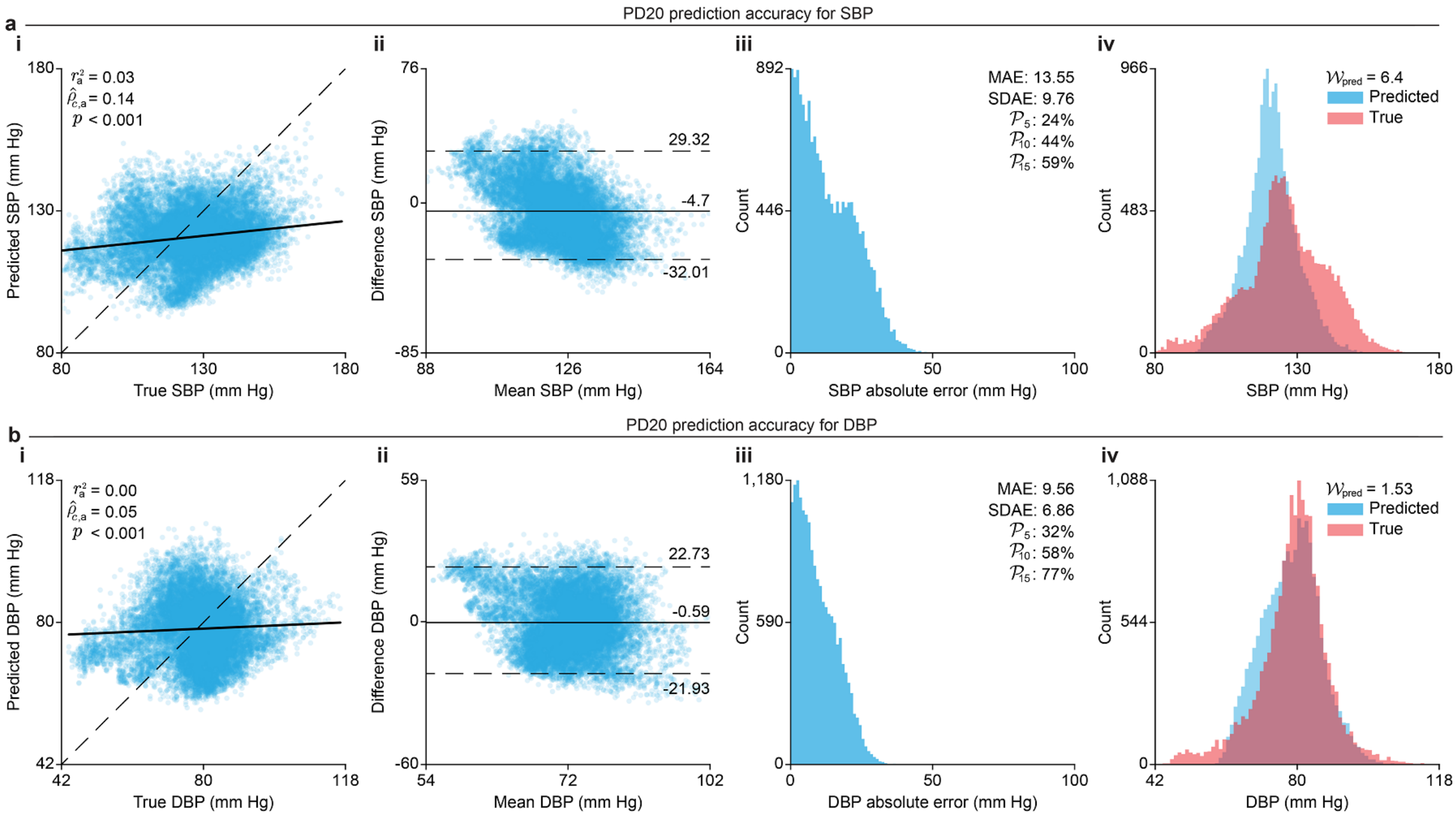

# Supplementary Fig. 113. Robustness of waveform models

**a,** Dependency of average mean absolute error (AMAE) on data quality index (DQI). **b,** Dependency of AMAE on label partition gap ($\mathcal{W}_{\text{label}}$). The quantities DQI and $\mathcal{W}_{\text{label}}$ are described in Supplementary Discussion 9.5.1. Here, each plot contains $91$ points for the $91$ subject-specific (SS) sub-models. Linear Regression models (SS01 and SS03) and Multilayer Perceptron models (SS05 and SS07) exhibit negative correlation between AMAE and DQI, and positive correlation between AMAE and $\mathcal{W}_{\text{label}}$. Convolutional-Recurrent-Samba models (SS17 and SS19) exhibited weak to moderate dependency with similar trends. Convolutional Neural Network models (SS09 and SS11) show large variation in AMAE across all SS sub-models, with no evidence of AMAE dependency on either DQI or $\mathcal{W}_{\text{label}}$. Similarly, Convolutional-Recurrent-Transformer models (SS13 and SS15) showed no statistically significant dependency to either dataset properties, though with smaller variance in AMAE across SS sub-models. r, Pearson's correlation coefficient; p, p-value; solid line, linear regression line; shade area, 95% confidence interval.

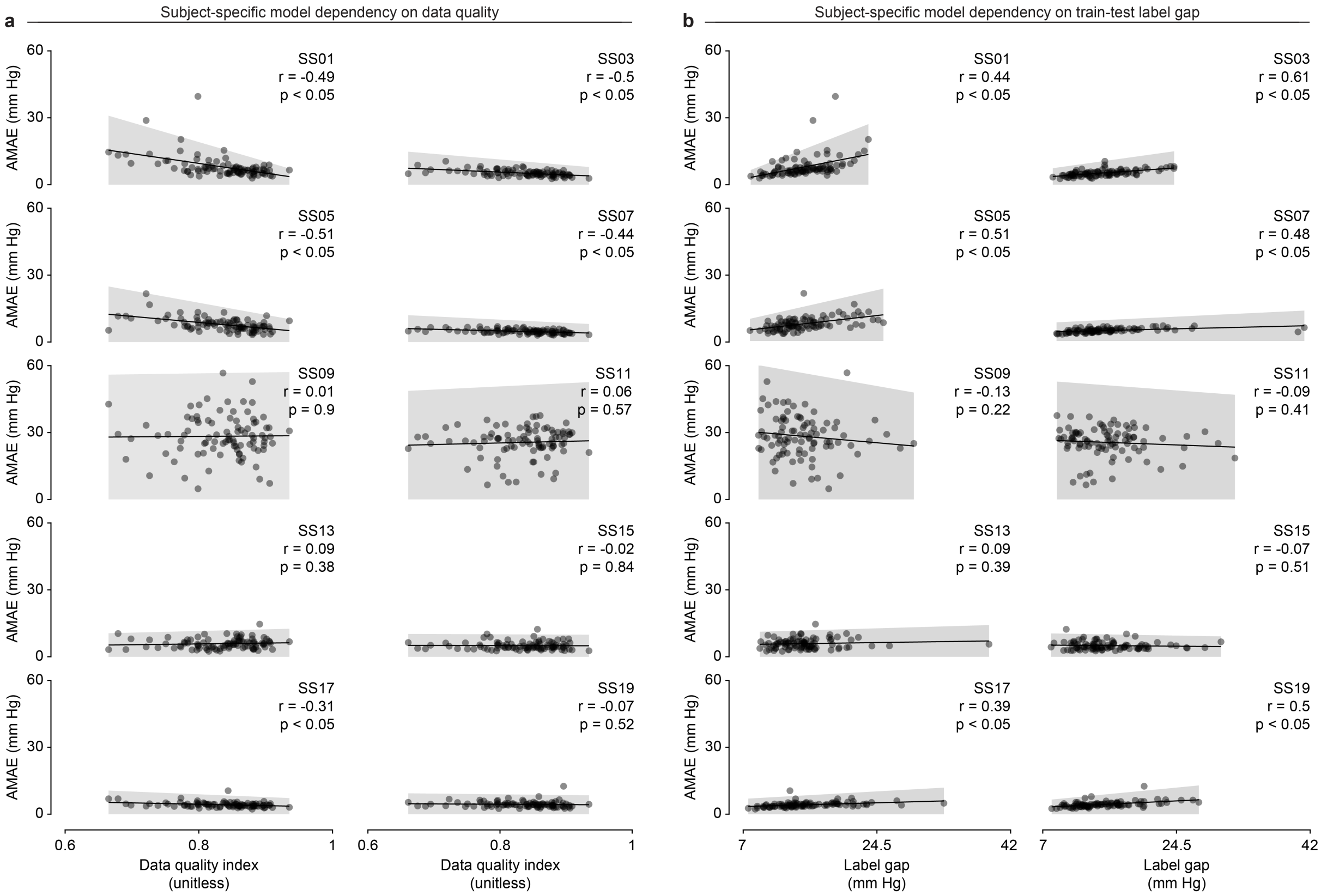

# Supplementary Fig. 114. Estimation results of ablated population-within model PW15-A1

Aggregated results from PW15-A1 configuration: Convolutional Recurrent Transformer class with impedance input and waveform output, trained with the population-within (PW) partition. **a,** Estimation accuracy for systolic brachial blood pressure (SBP); **b,** Estimation accuracy for diastolic brachial blood pressure (DBP); **c,** Waveform ensemble of all estimated and true brachial blood pressure (BP) periods. For **a** and **b**: **i**, correlation plots; **ii**, limits of agreement (LOA) plots; **iii**, histogram of absolute errors (AE); and **iv**, histogram of estimated and true BP distributions. For **c**: **i**, ensemble of estimated BP periods; **ii**, ensemble of true BP periods. For correlation plots: $r_{\mathrm{a}}^2$, aggregated coefficient of determination; $\hat{\rho}_{c,\mathrm{a}}$, aggregated coefficient of concordance; solid line, empirical linear regression line; dashed line, $45°$ line of perfect correlation. For LOA plots: solid line, mean of errors between estimated and true BP values; dashed lines, 2.5$^{\text{th}}$ percentile (lower) and 97.5$^{\text{th}}$ percentile (upper). For AE histogram plots: MAE and SDAE, mean and standard deviation of AE, respectively; $\mathcal{P}_5$, $\mathcal{P}_{10}$, and $\mathcal{P}_{15}$, cumulative percentage of estimations with AE within 5, 10, and 15 mm Hg, respectively. For fiducial histogram plots: $\mathcal{W}_{\text{pred}}$, Wasserstein distance between true and estimated distribution. For ensemble plots: AMAE, average mean absolute error; ARMSE, average root mean square error; solid line, ensemble average of all periods; dashed lines, ensemble average $\pm$ standard deviation of all periods; scale bars, one-quarter of period.

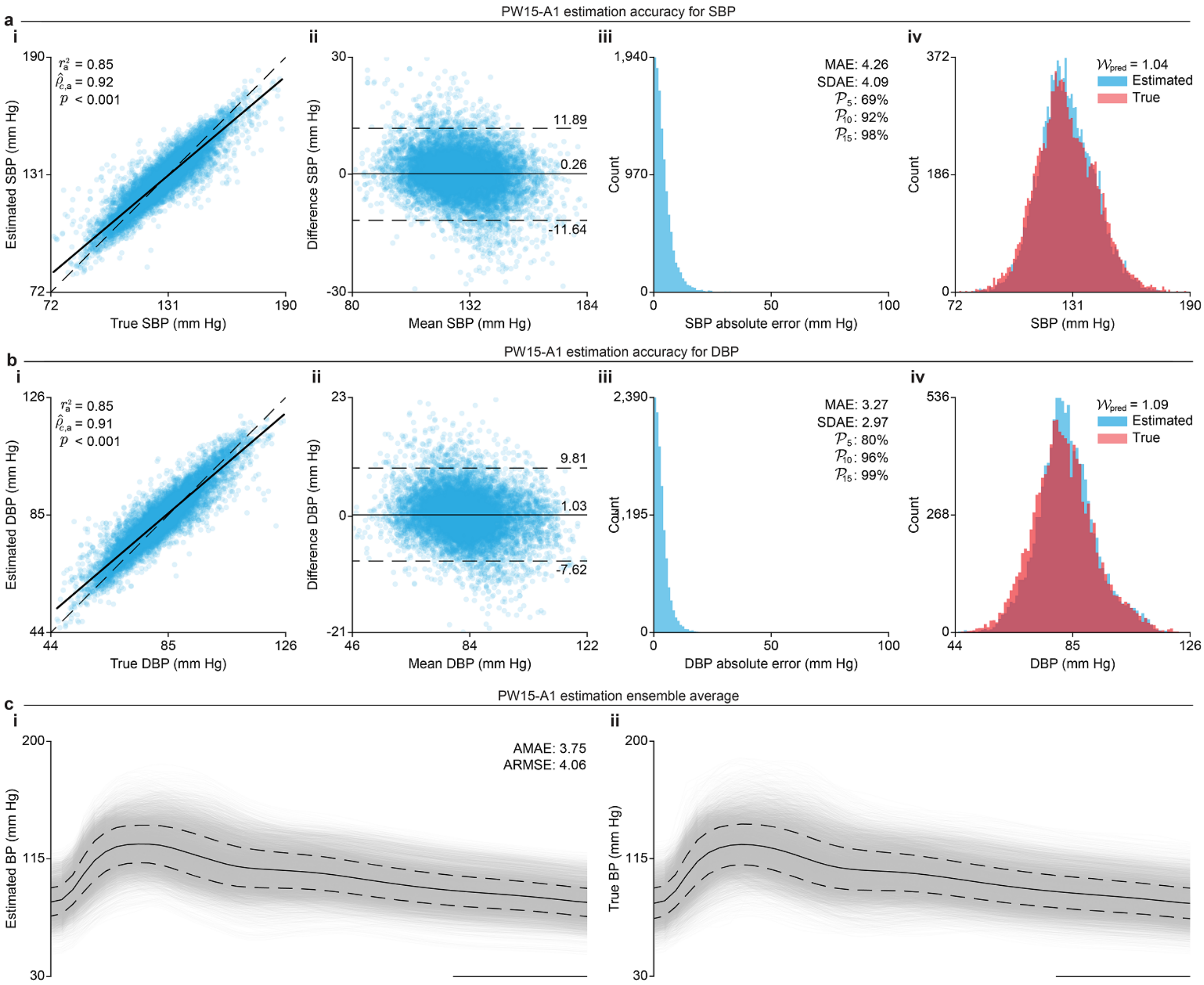

## Supplementary Fig. 115. Estimation results of ablated population-within model PW15-A2

Aggregated results from PW15-A2 configuration: Convolutional Recurrent Transformer class with impedance input and waveform output, trained with the population-within (PW) partition. **a,** Estimation accuracy for systolic brachial blood pressure (SBP); **b,** Estimation accuracy for diastolic brachial blood pressure (DBP); **c,** Waveform ensemble of all estimated and true brachial blood pressure (BP) periods. For **a** and **b**: **i**, correlation plots; **ii**, limits of agreement (LOA) plots; **iii**, histogram of absolute errors (AE); and **iv**, histogram of estimated and true BP distributions. For **c**: **i**, ensemble of estimated BP periods; **ii**, ensemble of true BP periods. For correlation plots: $r_{\mathrm{a}}^2$, aggregated coefficient of determination; $\hat{\rho}_{c,\mathrm{a}}$, aggregated coefficient of concordance; solid line, empirical linear regression line; dashed line, $45°$ line of perfect correlation. For LOA plots: solid line, mean of errors between estimated and true BP values; dashed lines, 2.5$^{\mathrm{th}}$ percentile (lower) and 97.5$^{\mathrm{th}}$ percentile (upper). For AE histogram plots: MAE and SDAE, mean and standard deviation of AE, respectively; $\mathcal{P}_5$, $\mathcal{P}_{10}$, and $\mathcal{P}_{15}$, cumulative percentage of estimations with AE within 5, 10, and 15 mm Hg, respectively. For fiducial histogram plots: $\mathcal{W}_{\mathrm{pred}}$, Wasserstein distance between true and estimated distribution. For ensemble plots: AMAE, average mean absolute error; ARMSE, average root mean square error; solid line, ensemble average of all periods; dashed lines, ensemble average $\pm$ standard deviation of all periods; scale bars, one-quarter of period.

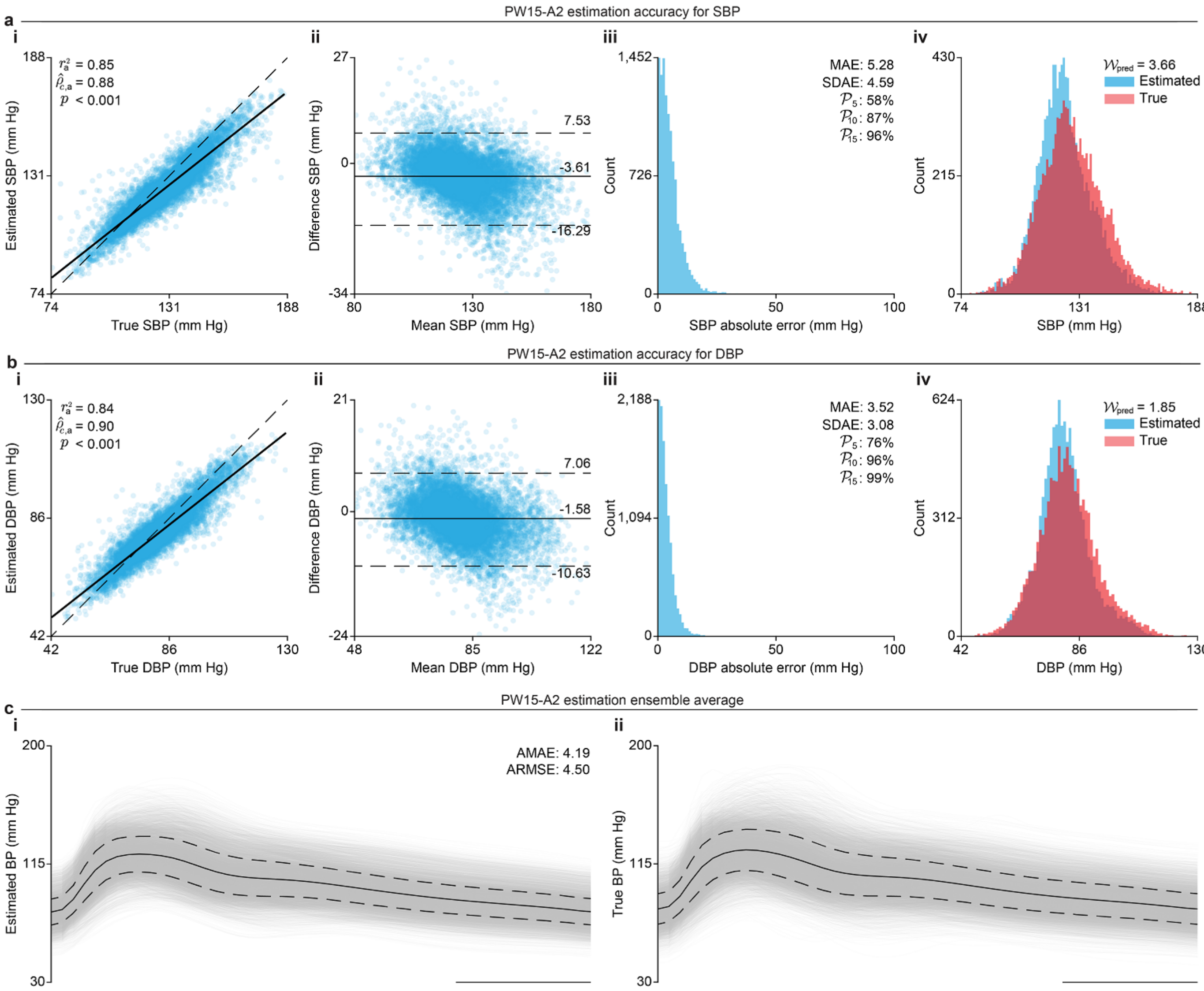

## Supplementary Fig. 116. Estimation results of ablated population-within model PW15-A3

Aggregated results from PW15-A3 configuration: Convolutional Recurrent Transformer class with impedance input and waveform output, trained with the population-within (PW) partition. **a,** Estimation accuracy for systolic brachial blood pressure (SBP); **b,** Estimation accuracy for diastolic brachial blood pressure (DBP); **c,** Waveform ensemble of all estimated and true brachial blood pressure (BP) periods. For **a** and **b**: **i**, correlation plots; **ii**, limits of agreement (LOA) plots; **iii**, histogram of absolute errors (AE); and **iv**, histogram of estimated and true BP distributions. For **c**: **i**, ensemble of estimated BP periods; **ii**, ensemble of true BP periods. For correlation plots: $r_{\mathrm{a}}^2$, aggregated coefficient of determination; $\hat{\rho}_{c,\mathrm{a}}$, aggregated coefficient of concordance; solid line, empirical linear regression line; dashed line, $45°$ line of perfect correlation. For LOA plots: solid line, mean of errors between estimated and true BP values; dashed lines, 2.5$^{\mathrm{th}}$ percentile (lower) and 97.5$^{\mathrm{th}}$ percentile (upper). For AE histogram plots: MAE and SDAE, mean and standard deviation of AE, respectively; $\mathcal{P}_5$, $\mathcal{P}_{10}$, and $\mathcal{P}_{15}$, cumulative percentage of estimations with AE within 5, 10, and 15 mm Hg, respectively. For fiducial histogram plots: $\mathcal{W}_{\mathrm{pred}}$, Wasserstein distance between true and estimated distribution. For ensemble plots: AMAE, average mean absolute error; ARMSE, average root mean square error; solid line, ensemble average of all periods; dashed lines, ensemble average ± standard deviation of all periods; scale bars, one-quarter of period.

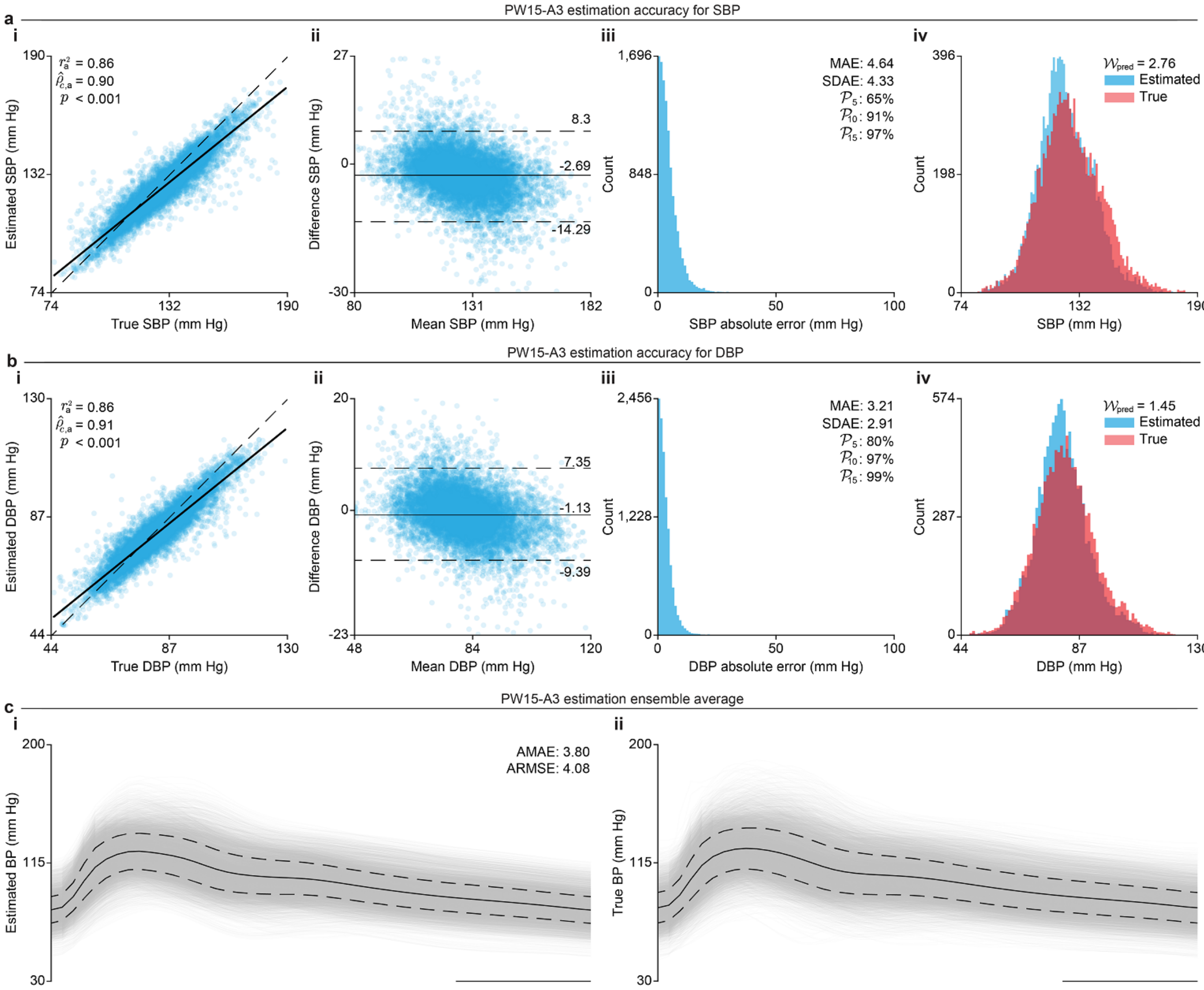

# Supplementary Fig. 117. Estimation results of ablated population-within model PW15-A4

Aggregated results from PW15-A4 configuration: Convolutional Recurrent Transformer class with impedance input and waveform output, trained with the population-within (PW) partition. **a,** Estimation accuracy for systolic brachial blood pressure (SBP); **b,** Estimation accuracy for diastolic brachial blood pressure (DBP); **c,** Waveform ensemble of all estimated and true brachial blood pressure (BP) periods. For **a** and **b**: **i**, correlation plots; **ii**, limits of agreement (LOA) plots; **iii**, histogram of absolute errors (AE); and **iv**, histogram of estimated and true BP distributions. For **c**: **i**, ensemble of estimated BP periods; **ii**, ensemble of true BP periods. For correlation plots: $r_{\mathrm{a}}^2$, aggregated coefficient of determination; $\hat{\rho}_{c,\mathrm{a}}$, aggregated coefficient of concordance; solid line, empirical linear regression line; dashed line, $45°$ line of perfect correlation. For LOA plots: solid line, mean of errors between estimated and true BP values; dashed lines, 2.5$^{\mathrm{th}}$ percentile (lower) and 97.5$^{\mathrm{th}}$ percentile (upper). For AE histogram plots: MAE and SDAE, mean and standard deviation of AE, respectively; $\mathcal{P}_5$, $\mathcal{P}_{10}$, and $\mathcal{P}_{15}$, cumulative percentage of estimations with AE within 5, 10, and 15 mm Hg, respectively. For fiducial histogram plots: $\mathcal{W}_{\mathrm{pred}}$, Wasserstein distance between true and estimated distribution. For ensemble plots: AMAE, average mean absolute error; ARMSE, average root mean square error; solid line, ensemble average of all periods; dashed lines, ensemble average $\pm$ standard deviation of all periods; scale bars, one-quarter of period.

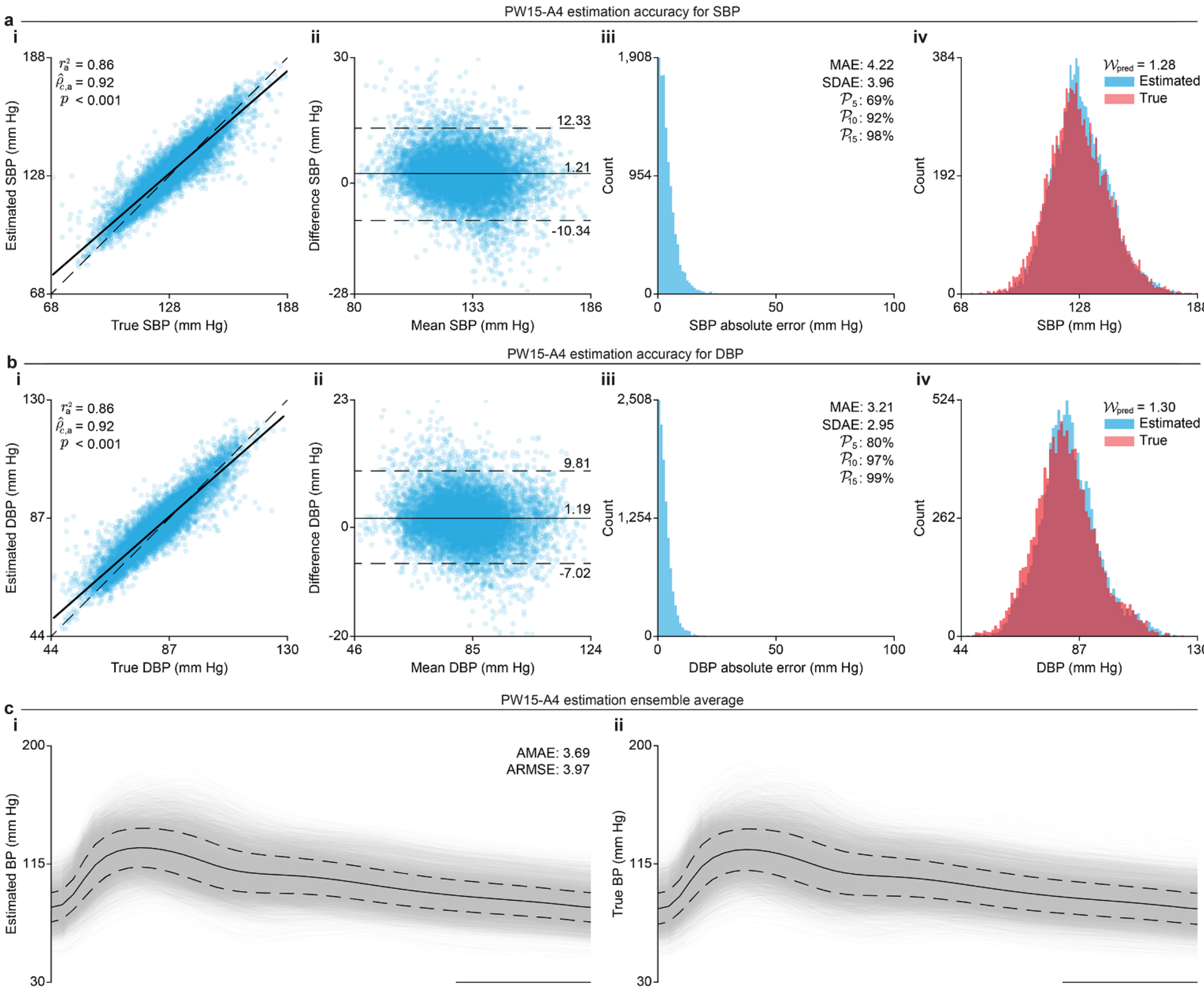

## Supplementary Fig. 118. Estimation results of ablated population-within model PW19-A1

Aggregated results from PW19-A1 configuration: Convolutional Recurrent Samba class with impedance input and waveform output, trained with the population-within (PW) partition. **a,** Estimation accuracy for systolic brachial blood pressure (SBP); **b,** Estimation accuracy for diastolic brachial blood pressure (DBP); **c,** Waveform ensemble of all estimated and true brachial blood pressure (BP) periods. For **a** and **b**: **i**, correlation plots; **ii**, limits of agreement (LOA) plots; **iii**, histogram of absolute errors (AE); and **iv**, histogram of estimated and true BP distributions. For **c**: **i**, ensemble of estimated BP periods; **ii**, ensemble of true BP periods. For correlation plots: $r^2_{\text{a}}$, aggregated coefficient of determination; $\hat{\rho}_{c,\text{a}}$, aggregated coefficient of concordance; solid line, empirical linear regression line; dashed line, $45°$ line of perfect correlation. For LOA plots: solid line, mean of errors between estimated and true BP values; dashed lines, 2.5$^{\text{th}}$ percentile (lower) and 97.5$^{\text{th}}$ percentile (upper). For AE histogram plots: MAE and SDAE, mean and standard deviation of AE, respectively; $\mathcal{P}_5$, $\mathcal{P}_{10}$, and $\mathcal{P}_{15}$, cumulative percentage of estimations with AE within 5, 10, and 15 mm Hg, respectively. For fiducial histogram plots: $\mathcal{W}_{\text{pred}}$, Wasserstein distance between true and estimated distribution. For ensemble plots: AMAE, average mean absolute error; ARMSE, average root mean square error; solid line, ensemble average of all periods; dashed lines, ensemble average $\pm$ standard deviation of all periods; scale bars, one-quarter of period.

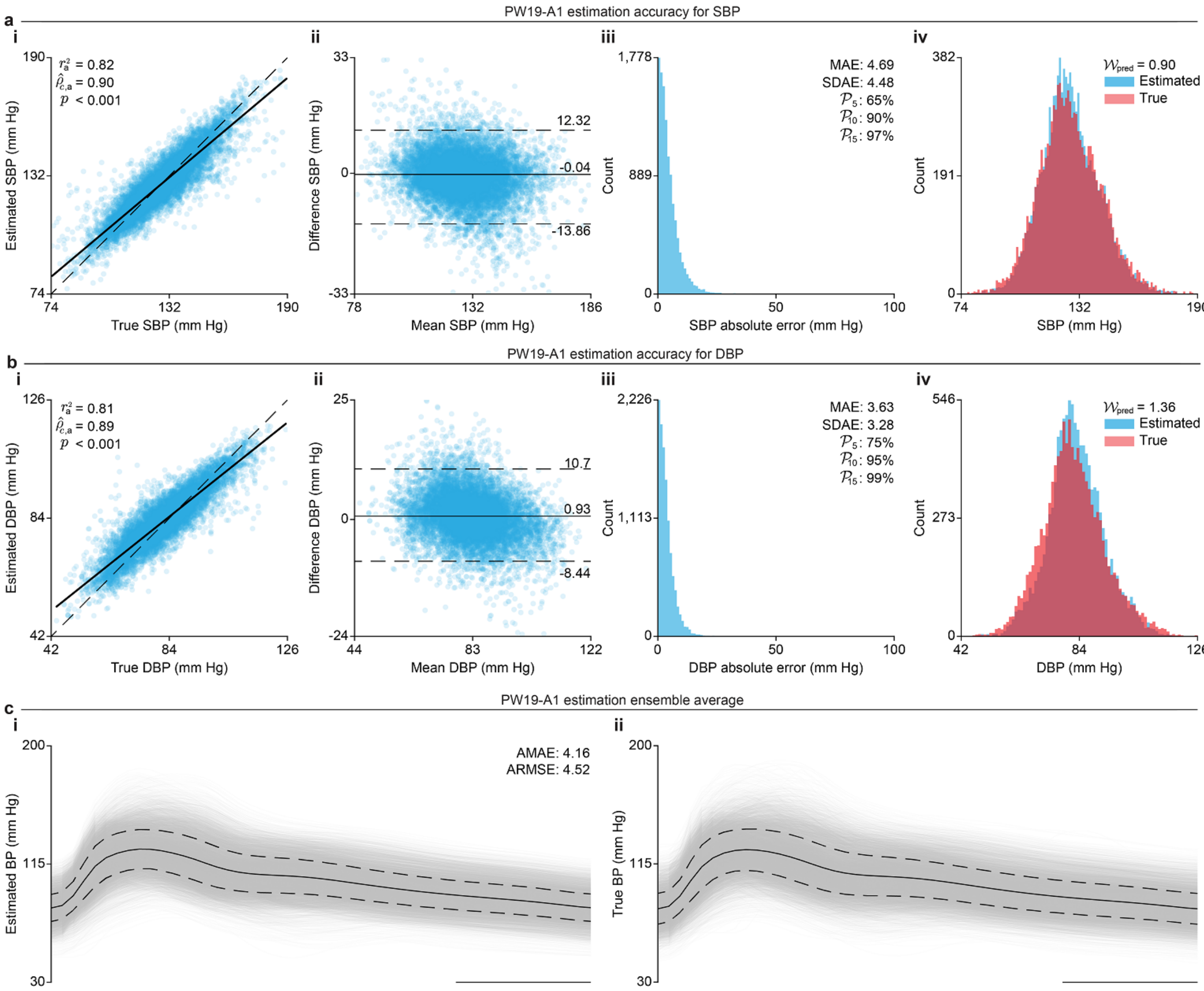

# Supplementary Fig. 119. Estimation results of ablated population-within model PW19-A1

Aggregated results from PW19-A2 configuration: Convolutional Recurrent Samba class with impedance input and waveform output, trained with the population-within (PW) partition. **a,** Estimation accuracy for systolic brachial blood pressure (SBP); **b,** Estimation accuracy for diastolic brachial blood pressure (DBP); **c,** Waveform ensemble of all estimated and true brachial blood pressure (BP) periods. For **a** and **b**: **i**, correlation plots; **ii**, limits of agreement (LOA) plots; **iii**, histogram of absolute errors (AE); and **iv**, histogram of estimated and true BP distributions. For **c**: **i**, ensemble of estimated BP periods; **ii**, ensemble of true BP periods. For correlation plots: $r_{\mathrm{a}}^2$, aggregated coefficient of determination; $\hat{\rho}_{c,\mathrm{a}}$, aggregated coefficient of concordance; solid line, empirical linear regression line; dashed line, $45°$ line of perfect correlation. For LOA plots: solid line, mean of errors between estimated and true BP values; dashed lines, 2.5$^{\text{th}}$ percentile (lower) and 97.5$^{\text{th}}$ percentile (upper). For AE histogram plots: MAE and SDAE, mean and standard deviation of AE, respectively; $\mathcal{P}_5$, $\mathcal{P}_{10}$, and $\mathcal{P}_{15}$, cumulative percentage of estimations with AE within 5, 10, and 15 mm Hg, respectively. For fiducial histogram plots: $\mathcal{W}_{\text{pred}}$, Wasserstein distance between true and estimated distribution. For ensemble plots: AMAE, average mean absolute error; ARMSE, average root mean square error; solid line, ensemble average of all periods; dashed lines, ensemble average $\pm$ standard deviation of all periods; scale bars, one-quarter of period.

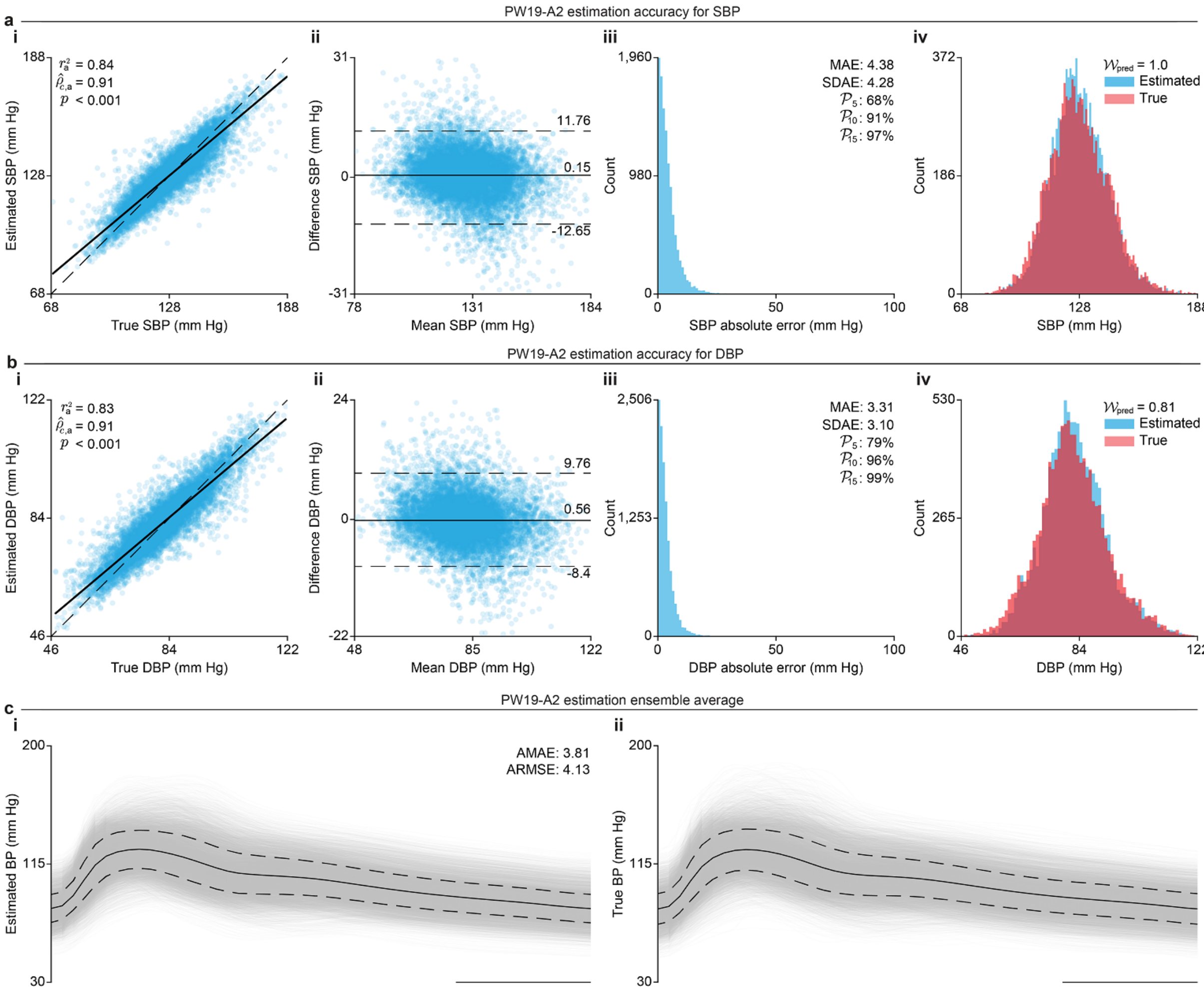

# Supplementary Videos

### Instructions to view Supplementary Videos in PDF

To view the embedded Supplementary Videos, Adobe Acrobat Reader is required. By default, multimedia content is disabled in Adobe Reader. To enable video playback:

1. Open the Supplementary Information PDF in Adobe Reader.
2. When prompted with a security warning, select "*Trust this document always*" under the Options button.

## Supplementary Video 1. Particle transport in the palmar arterial network

Particle motion over three cardiac cycles in the palmar arterial network, showing flow patterns (**a**) and particle velocity magnitude (**b**). Scale bars, 10 mm.

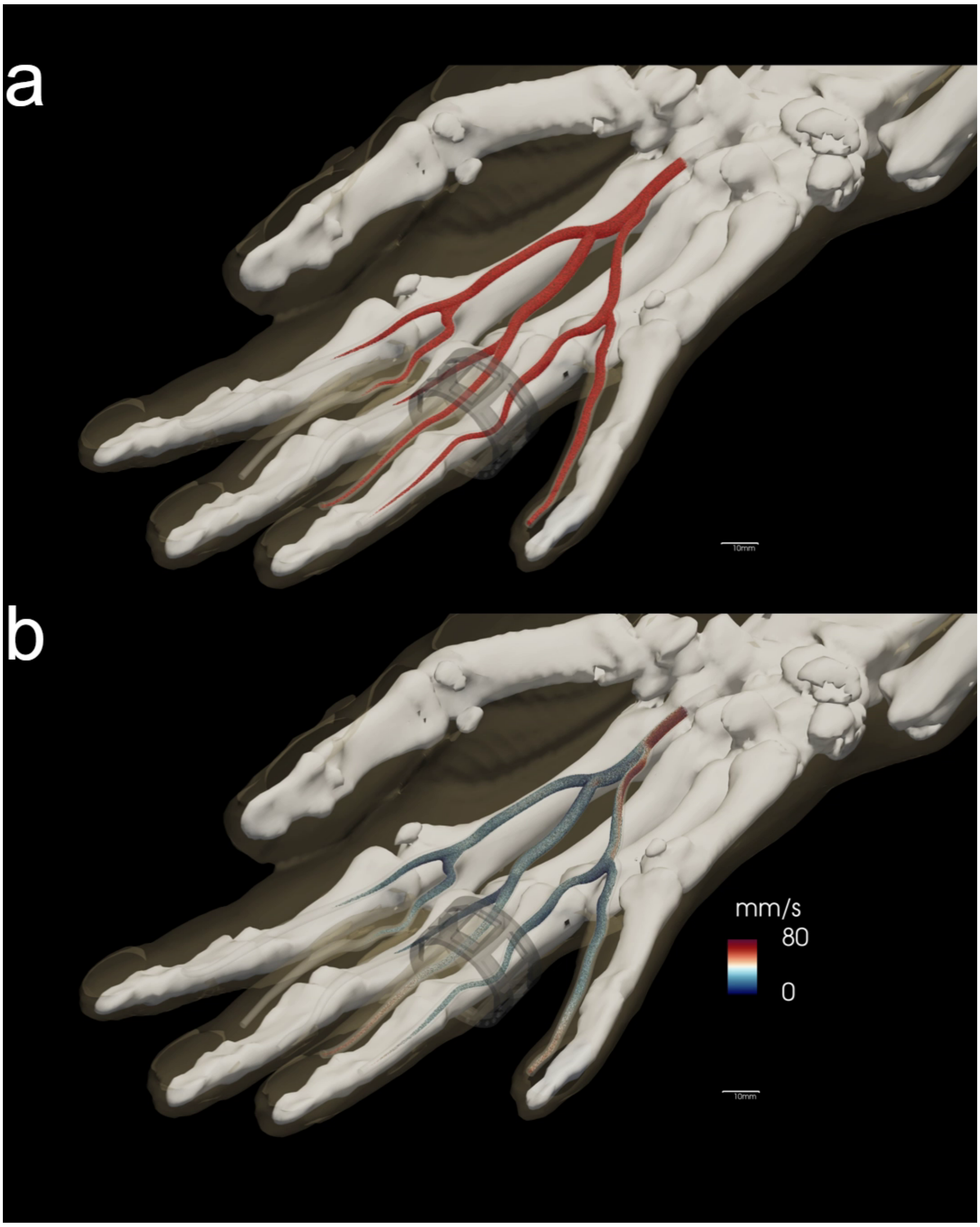

## Supplementary Video 2. Velocity field of particle transport within the isolated palmar arterial network

Particle motion over three cardiac cycles across the isolated arterial tree, showing flow patterns (**a**) and particle velocity magnitude (**b**). Scale bars, 10 mm.

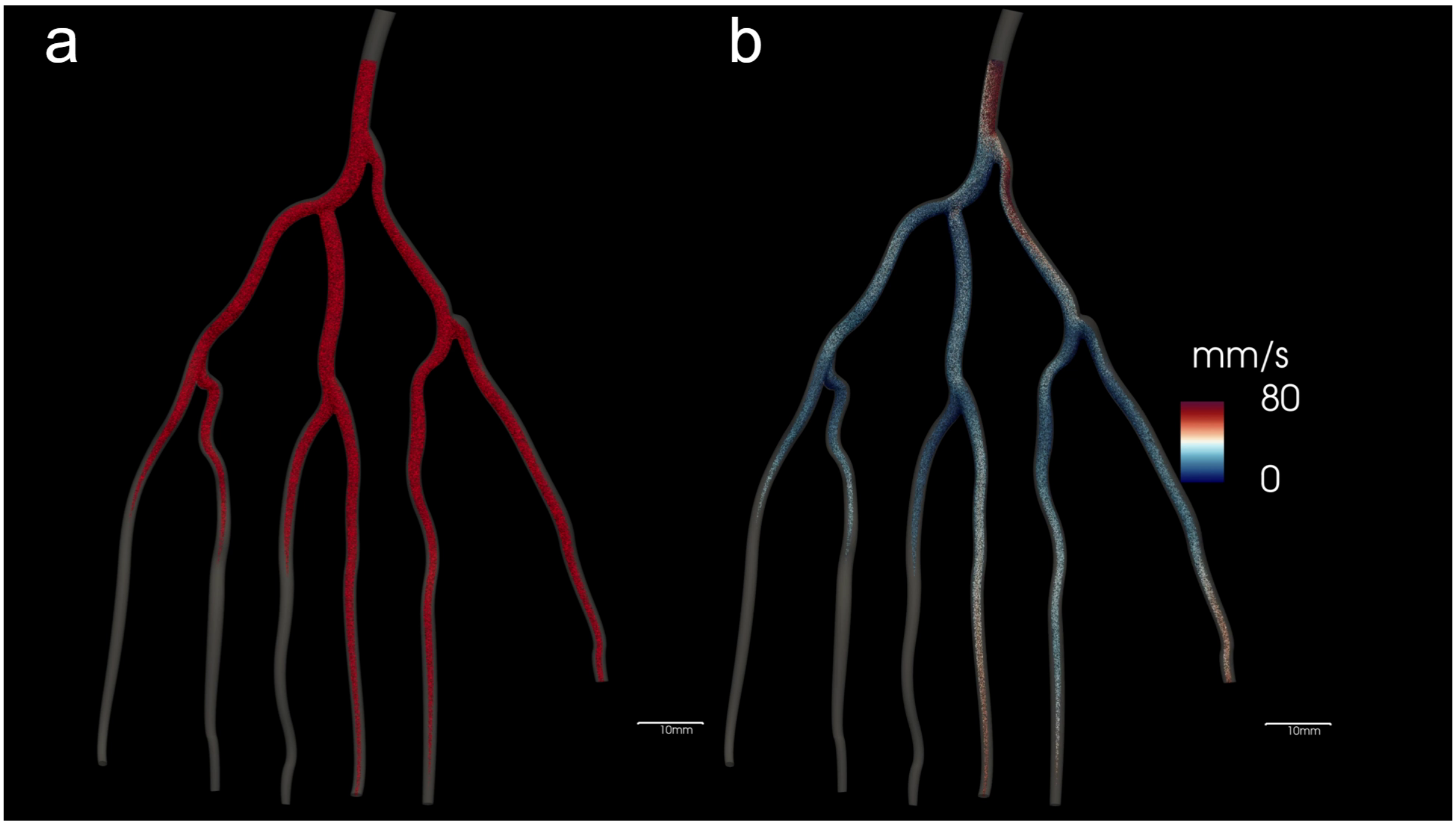

## Supplementary Video 3. Particle dynamics in the ring finger digital arteries

Particle motion over three cardiac cycles at the ring finger, showing flow patterns (**a**) and particle velocity magnitude (**b**). The particle speed exhibits laminar structure with clear radial velocity gradients influenced by local confinement. Scale bars, 1 mm.

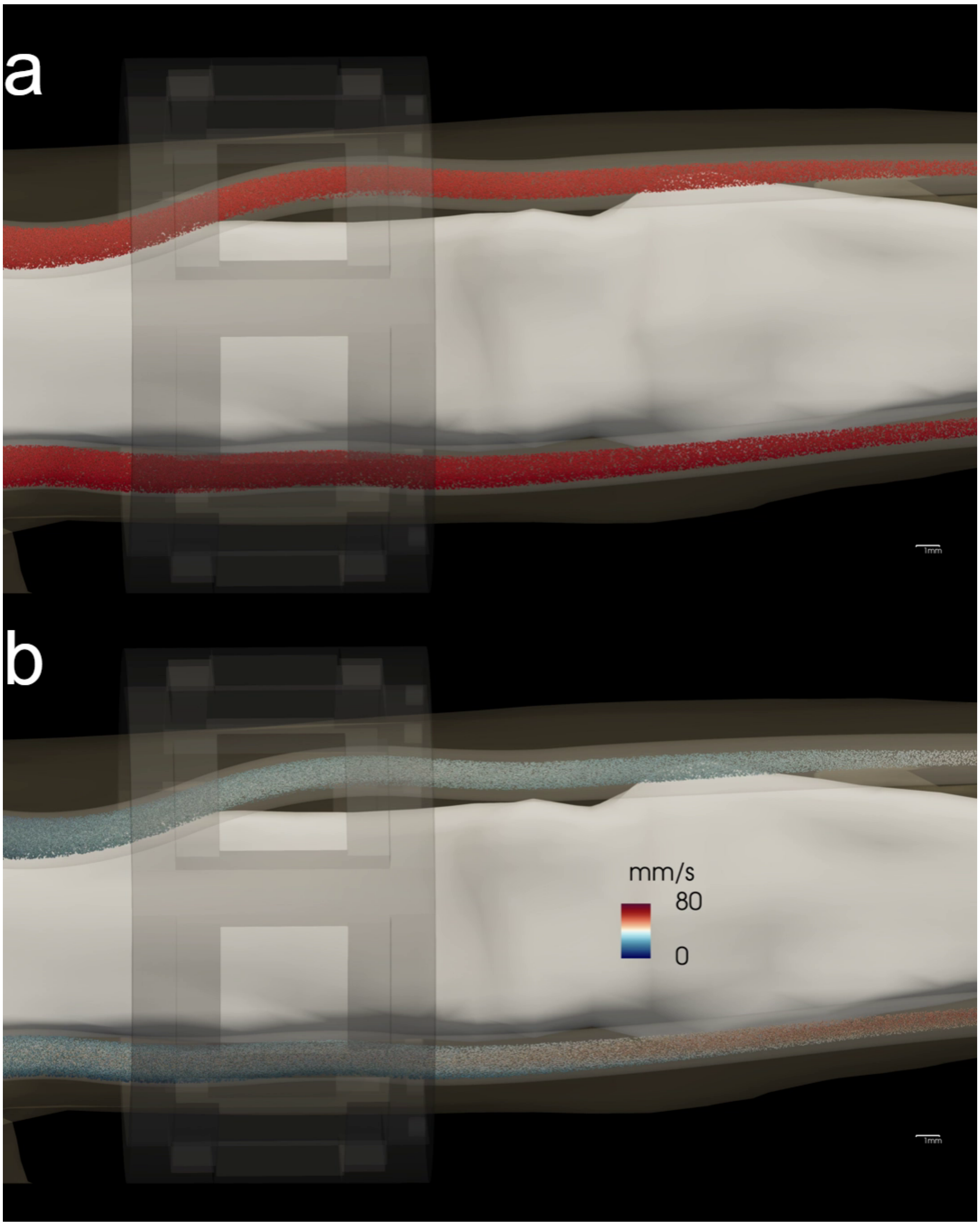

## Supplementary Video 4. Effect of stenosis severity on simulated hemodynamics in the diabetic hand vasculature

Representative hemodynamic simulations of the diabetic hand arterial tree with a stenotic lesion positioned before the digital arterial bifurcation. Simulations are shown for baseline (**a**), 75% occlusion (**b**), and 96% occlusion (**d**). Color mapping indicates blood-flow velocity in mm/s. Increasing stenosis severity is associated with a more pronounced local velocity change at the lesion and altered downstream flow distribution in the distal branches. Scale bar, 10 mm.

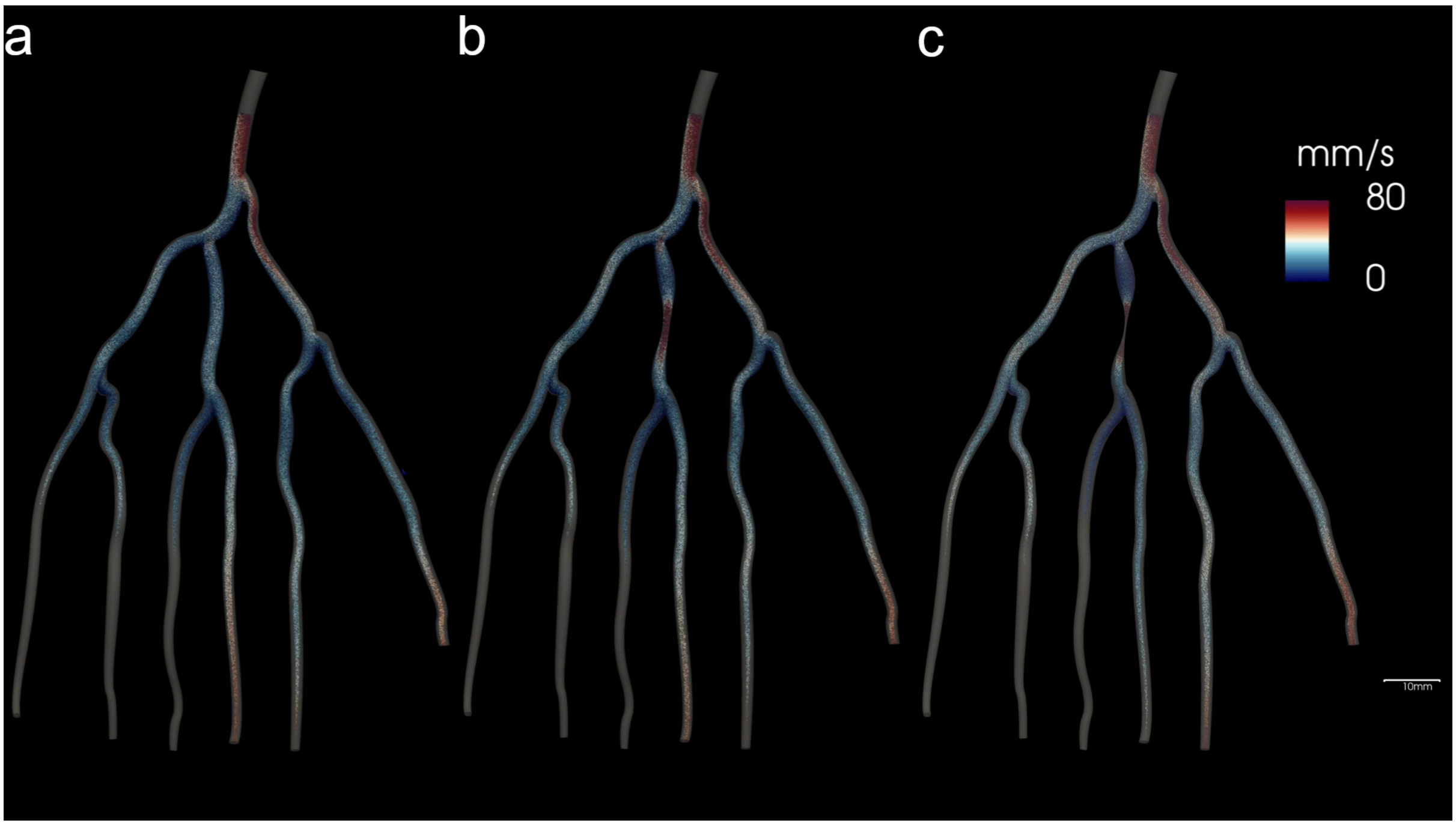

# Supplementary Video 5. Hyperparameter tuning for phantom tank experiment

Phantom tank experiment, showing the experiment setup (left) and the reconstruction quality at different regularization hyperparameter values (right). The experiment was performed with 16 electrodes, using an opposite injection pattern and adjacent measurement pattern. For small hyperparameter values ($\lambda = 0.001$ and $\lambda = 0.01$, the reconstruction algorithm was sensitive to small fluctuation of the water surface and produced ringing patterns around the targeted object. For $\lambda = 0.1$ and $\lambda = 1$, the reconstructed objects became blurry, as expected from the smoothing constraint. However, at $\lambda = 1$, the algorithm could only detect the object when it was close to the boundary. When the object was positioned at the center, its reconstruction was indistinguishable from the fluctuation of the water surface. At $\lambda = 100$, the reconstruction are dominated by smoothness constraint and produced images with uniform color. At the extreme value of $\lambda = 100$, the pattern became erratic, as the algorithm did not find a globally optimal solution.

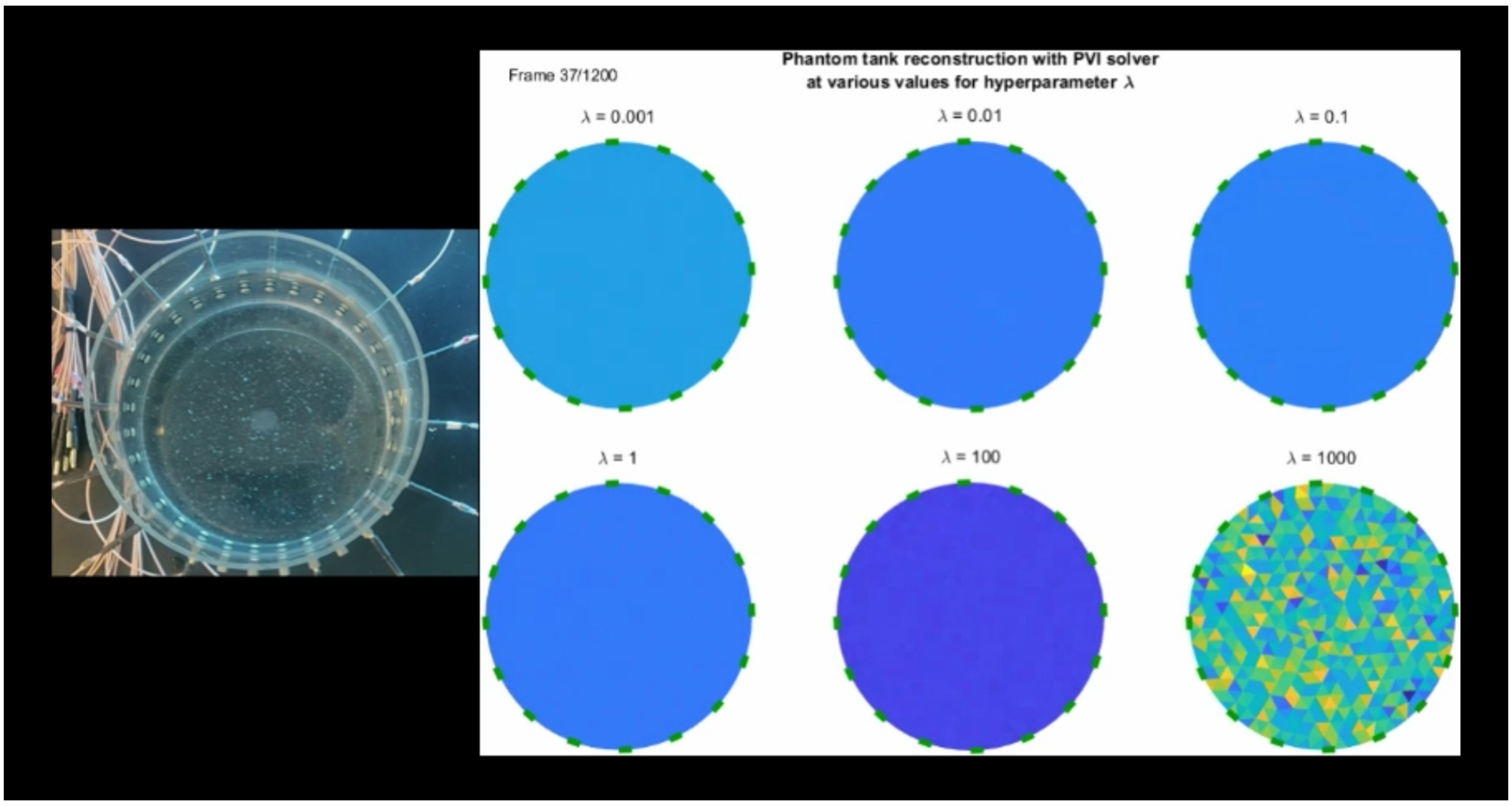

# Supplementary Video 6. Doppler flowmetry of the finger radial artery

Doppler flowmetry of the finger radial artery showing peak systolic velocity of 20 cm/s.

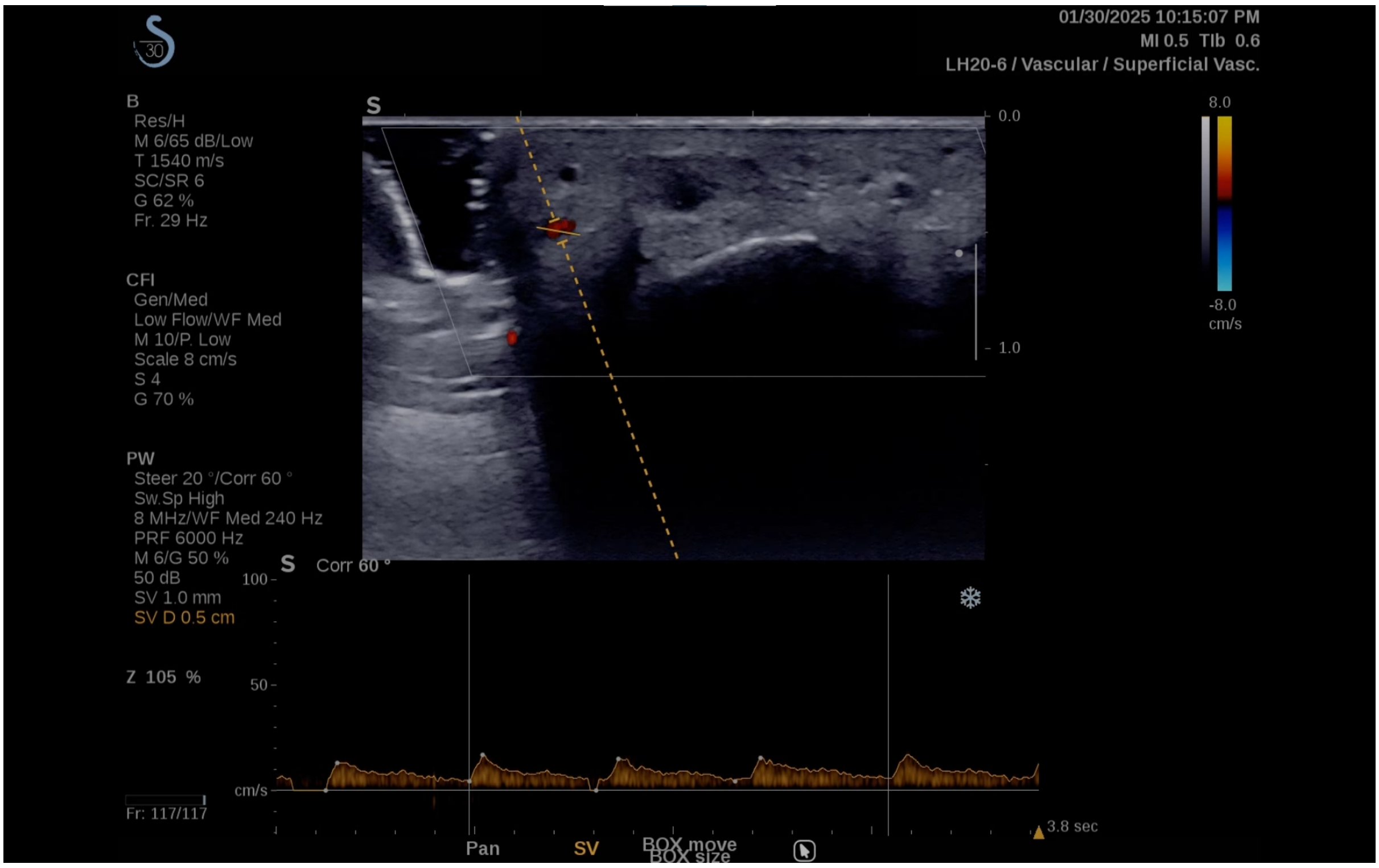

## Supplementary Video 7. Reconstruction of finger conductivity images and signal

Reconstructed conductivity data collected from a representative participant, showing the finite element mesh (top left), the $40 \times 40$ image sequence (top right), and the extracted conductivity signal (bottom). The signal was extracted and averaged from the region of interested (ROI) marked as green triangles in the mesh.

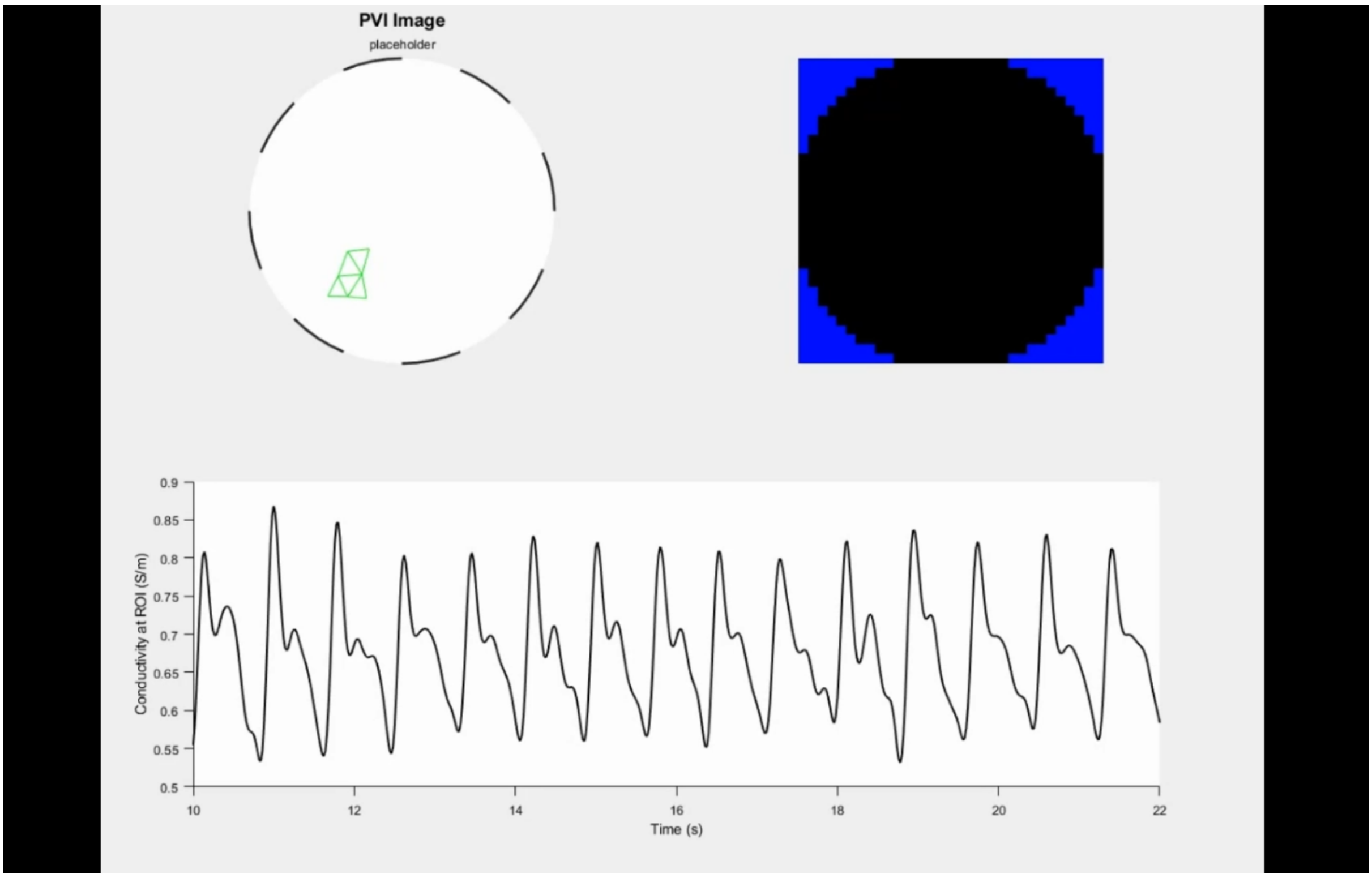

# Supplementary bibliography